\documentclass[aps,pra,onecolumn,superscriptaddress,nofootinbib,a4paper]{revtex4}

\usepackage[left=0.76in, right=0.76in, bottom=1in, top=1in]{geometry}

\usepackage{silence}
\usepackage{fancyhdr}
\fancypagestyle{header}{
  \fancyhf{}%
  \fancyhead[R]{\hyperref[toc]{\thepage}}%
}
\usepackage{appendix}

\usepackage{color}
\usepackage{xcolor}
\usepackage{graphicx}
\usepackage{amsfonts}
\usepackage{amsthm}
\usepackage{amssymb}
\usepackage{amsmath}
\usepackage{physics}
\usepackage{complexity}
\definecolor{blueviolet}{rgb}{0.2, 0.2, 0.6}
\definecolor{webgreen}{rgb}{0,.5,0}
\definecolor{webbrown}{rgb}{.6,0,0}
\usepackage[pdftex,
  bookmarks=true,
  colorlinks=true, 
  urlcolor=webbrown,
  linkcolor=blueviolet, 
  citecolor=webgreen,
  pdfstartpage=1,
  pdfstartview={XYZ},  
  bookmarksopen=true,
  bookmarksopenlevel=1,
  pdfpagemode=UseOutlines,
  pdfview=XYZ,
  ]{hyperref}
\usepackage{cleveref}
\usepackage{algorithm}      
\usepackage{algpseudocode}  
\algrenewcommand\algorithmicrequire{\textbf{Input:}}   
\algrenewcommand\algorithmicensure{\textbf{Output:}}   
\usepackage{tcolorbox}

\usepackage{aliascnt}

\newcommand{\revise}[1]{{
#1}}

\newtheorem{theorem}{Theorem}
\crefname{theorem}{Theorem}{Theorems}
\Crefname{theorem}{Theorem}{Theorems}

\newaliascnt{lemma}{theorem}
\newtheorem{lemma}[lemma]{Lemma}
\aliascntresetthe{lemma}
\crefname{lemma}{Lemma}{Lemmas}
\Crefname{lemma}{Lemma}{Lemmas}

\newaliascnt{corollary}{theorem}
\newtheorem{corollary}[corollary]{Corollary}
\aliascntresetthe{corollary}
\crefname{corollary}{Corollary}{Corollaries}
\Crefname{corollary}{Corollary}{Corollaries}

\newaliascnt{definition}{theorem}
\newtheorem{definition}[definition]{Definition}
\aliascntresetthe{definition}
\crefname{definition}{Definition}{Definitions}
\Crefname{definition}{Definition}{Definitions}

\newaliascnt{remark}{theorem}

\aliascntresetthe{remark}
\crefname{remark}{Remark}{Remarks}
\Crefname{remark}{Remark}{Remarks}

\newaliascnt{task}{theorem}
\newtheorem{task}[task]{Task}
\aliascntresetthe{task}
\crefname{task}{Task}{Tasks}
\Crefname{task}{Task}{Tasks}

\renewcommand{\poly}{\mathrm{poly}}
\renewcommand{\polylog}{\mathrm{polylog}}
\renewcommand{\E}{\mathbb{E}}
\newcommand{\bit}{\{0, 1\}}
\newcommand{\disc}{\mathrm{disc}}
\newcommand{\unif}{\mathrm{Uniform}}
\newcommand{\floor}[1]{\left\lfloor#1\right\rfloor}
\newcommand{\ceil}[1]{\left\lceil#1\right\rceil}

\newcommand{\argmax}{\mathrm{argmax}}
\newcommand{\diag}{\mathrm{diag}}
\newcommand{\Var}{\mathrm{Var}}
\newcommand{\Cov}{\mathrm{Cov}}
\newcommand{\sgn}{\mathrm{sgn}}
\newcommand{\sth}{\,:\,}
\newcommand{\lr}[1]{\left(#1\right)}

\newcommand{\cO}{\mathcal{O}}
\newcommand{\cU}{\mathcal{U}}
\newcommand{\cV}{\mathcal{V}}

\NewDocumentEnvironment{eqsplit}{b}{%
    \begin{equation}%
    \begin{split}%
        #1
    \end{split}%
    \end{equation}%
}{}

\makeatletter

\newcommand{\apptocfile}{atoc}

\let\apptoc@orig@appendix\appendix
\renewcommand{\appendix}{%
  \apptoc@orig@appendix
  \let\apptoc@orig@addtocontents\addtocontents
  \long\def\addtocontents##1##2{%
    \def\apptoc@ext{##1}%
    \def\apptoc@toc{toc}%
    \ifx\apptoc@ext\apptoc@toc
      \apptoc@orig@addtocontents{\apptocfile}{##2}%
    \else
      \apptoc@orig@addtocontents{##1}{##2}%
    \fi
  }%
}

\newcommand{\appendixtableofcontents}{%
  \begingroup
    \setcounter{tocdepth}{3}%
    \phantomsection
    \let\addcontentsline\@gobblethree
    \section*{Contents \& Roadmap}%
    \pdfbookmark[1]{Appendices}{apxcontents}%
    \@starttoc{\apptocfile}%
  \endgroup
}

\makeatother

\begin{document}
\pagestyle{header}

\title{Exponential quantum advantage in processing massive classical data}

\author{Haimeng Zhao}
\email[Corresponding author: ]{haimeng@caltech.edu}
\affiliation{California Institute of Technology, Pasadena, California 91125, USA}
\affiliation{Google Quantum AI, Venice, California 90291, USA}

\author{Alexander Zlokapa}
\affiliation{Massachusetts Institute of Technology, Cambridge, Massachusetts 02139, USA}

\author{Hartmut Neven}
\affiliation{Google Quantum AI, Venice, California 90291, USA}

\author{Ryan Babbush}
\affiliation{Google Quantum AI, Venice, California 90291, USA}

\author{John~Preskill}
\affiliation{California Institute of Technology, Pasadena, California 91125, USA}
\affiliation{Oratomic, Pasadena, California 91125, USA}

\author{Jarrod R. McClean}
\affiliation{Google Quantum AI, Venice, California 90291, USA}

\author{Hsin-Yuan Huang}
\email[Corresponding author: ]{hhuang@oratomic.com}
\email[]{hsinyuan@caltech.edu}
\affiliation{Oratomic, Pasadena, California 91125, USA}
\affiliation{California Institute of Technology, Pasadena, California 91125, USA}

\begin{abstract}
Broadly applicable quantum advantage, particularly in classical data processing and machine learning, has been a fundamental open problem. In this work, we prove that a small quantum computer of polylogarithmic size can perform large-scale classification and dimension reduction on massive classical data by processing samples on the fly, whereas any classical machine achieving the same prediction performance requires exponentially larger size. Furthermore, classical machines that are exponentially larger yet below the required size need superpolynomially more samples and time. We \revise{provide evidence for} these quantum advantages in real-world applications, including single-cell RNA sequencing and movie review sentiment analysis, demonstrating four to six orders of magnitude reduction in size with fewer than $60$ logical qubits. These quantum advantages are enabled by quantum oracle sketching, an algorithm for accessing the classical world in quantum superposition using only random classical data samples. Combined with classical shadows, our algorithm circumvents the data loading and readout bottleneck to construct succinct classical models from massive classical data, a task provably impossible for any classical machine that is not exponentially larger than the quantum machine. These quantum advantages persist even when classical machines are granted unlimited time or if $\mathsf{BPP} = \mathsf{BQP}$, and rely only on the correctness of quantum mechanics. Together, our results establish machine learning on classical data as a broad and natural domain of quantum advantage and a fundamental test of quantum mechanics at the complexity frontier.
\end{abstract}

\maketitle

\section{Introduction}

\begin{figure}[t]
    \centering
    \includegraphics[width=1\linewidth]{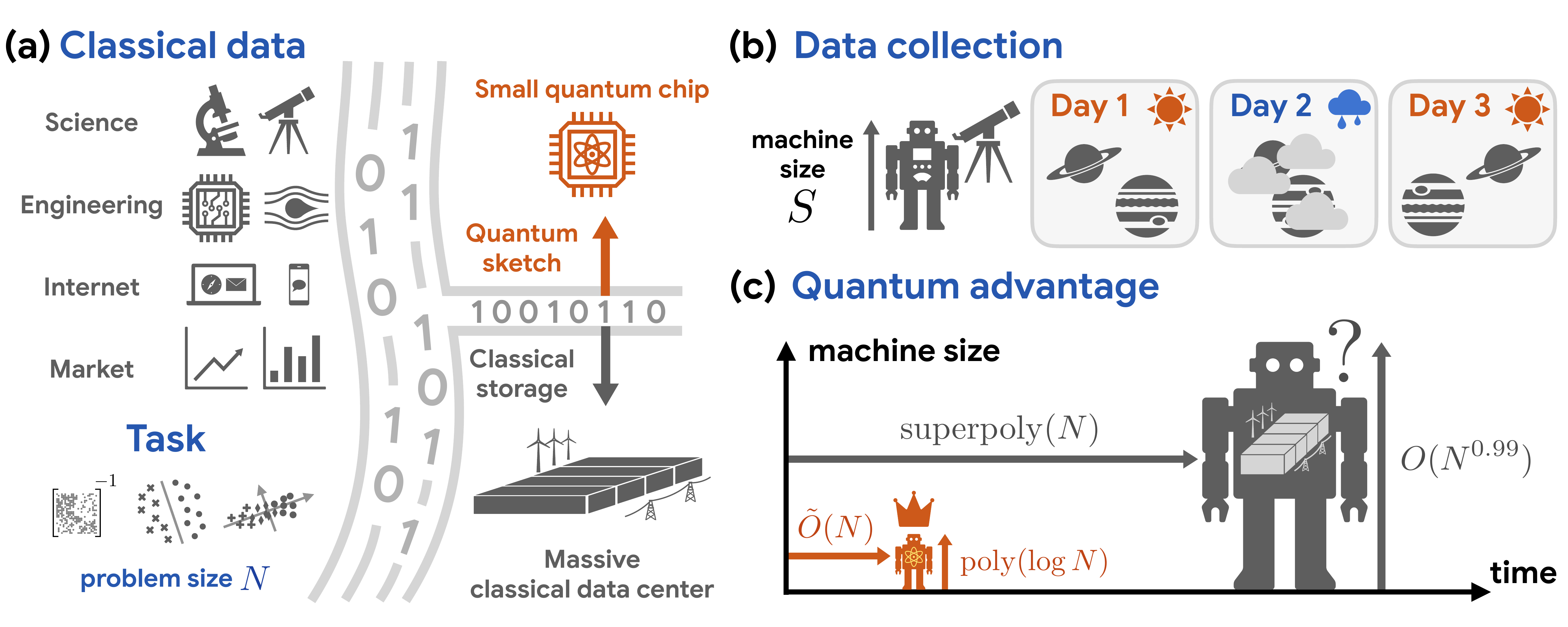}
    \caption{\textbf{Overview of quantum advantage in processing massive classical data.}
    \textbf{(a)} We prove that a quantum computer can outperform exponentially larger classical machines in a wide range of classical data processing tasks, including solving linear systems, classification, and dimension reduction.
    \textbf{(b)}
    Our quantum algorithm enables coherent quantum queries to the noisy and evolving classical world.
    \textbf{(c)}
    For various classical data processing tasks with problem size $N$, a $\poly(\log N)$-size quantum machine can succeed in $\tilde{O}(N)$ time using quantum oracle sketching.
    In contrast, we prove that any classical machine, even with exponentially larger size $O(N^{0.99})$, cannot solve the same task unless given time superpolynomial in $N$.
    This exponential quantum advantage relies only on the principle of quantum superposition, independent of any computational complexity conjectures.
    }
    \label{fig:overview}
\end{figure}

The search for useful quantum advantage has been a formidable challenge. Despite decades of effort since early foundational advances~\cite{feynman1986quantum,shor1994algorithms}, compelling end-to-end advantages with real-world impact have been established only for a few specialized tasks, most notably cryptanalysis and quantum simulation~\cite{babbush2025grand,preskill2025beyond}. These advantages rely on particular structures \cite{kitaev1995quantum,lloyd1996universal,aaronson2009need} exploitable by quantum machines which rarely arise in broader applications. Consequently, quantum computers are often seen as powerful but specialized devices, and whether their advantages extend beyond these narrow domains remains a central open question.

Classical data processing and machine learning represent perhaps the most compelling test of this question, for we are macroscopic creatures in an effectively classical world. The scale of data generated across science, industry, and everyday computation has grown at an extraordinary pace, and the need to distill useful information from it has driven the overwhelming success of modern machine learning. Numerous quantum algorithms have been developed under the banner of quantum machine learning and quantum linear algebra~\cite{dalzell2025quantum,gilyen2019quantum,martyn2021grand,biamonte2017quantum} with the hope of bringing significant speedup to this domain, but their end-to-end advantage remains largely unclear~\cite{aaronson2015read}. 

The central challenge is that these algorithms require access to classical data in superposition, modeled as \emph{quantum oracle queries}, which is fundamentally in tension with the classical access the world provides. Existing attempts to resolve this tension proceed by storing the entire dataset in a quantum random access memory (QRAM)~\cite{giovannetti2008quantum,hann2021resilience,dalzell2025distillation}, but maintaining such coherent access incurs significant overhead in fault-tolerance and control~\cite{jaques2025qram}, to the point where the classical co-processors required to sustain a QRAM could often be repurposed to solve the tasks directly. 
Skepticism about the practical impact of quantum processing on machine learning~\cite{schuld2022quantum,cerezo2023does,gil2024relation} has been reinforced by several factors: proposed quantum speedups often rely on highly contrived problems~\cite{zhao2025entanglement,gao2018quantum,gao2022enhancing,anschuetz2023interpretable,anschuetz2026arbitrary,zhang2024quantum,liu2021rigorous,gyurik2023exponential,huang2025generative}, numerous proposed algorithms have been dequantized~\cite{tang2019quantum,tang2021quantum,tang2023quantum}, and training of quantum variational models can be difficult in practice~\cite{peruzzo2014variational,mcclean2016theory,cerezo2021variational,cerezo2022challenges,du2025quantum,mcclean2018barren,cerezo2021cost,wang2021noise,larocca2025barren,anschuetz2022quantum}.
We provide a review of related works in \Cref{sec:rel-work}.

This challenge points to a more fundamental tension between machine size and data scale. Even for classical machines, processing massive data has become a critical bottleneck across science and technology, from machine learning~\cite{kaplan2020scaling,fedus2022switch,gholami2024ai,strubell2019energy} and single-cell RNA sequencing~\cite{svensson2018exponential,lahnemann2020eleven,tsuyuzaki2020benchmarking} to particle colliders~\cite{hep2019roadmap,bejaralonso2020hllhc,khachatryan2016search} and astronomical sky surveys~\cite{dewdney2009square}. Classical streaming, sketching, and online learning algorithms~\cite{muthukrishnan2005data,alon1996space,flajolet1985probabilistic,misra1982finding,cormode2005improved,woodruff2014sketching,clarkson2009numerical,andoni2020streaming,mitliagkas2013memory,shalev2025online} offer partial relief by processing data samples on the fly, using incrementally updated models without storing the entire dataset. However, in reducing machine size these techniques also sacrifice prediction accuracy. This raises a foundational question: 
\begin{quote}\emph{Can small quantum machines leverage the exponential dimensionality of quantum Hilbert spaces\\ to learn and predict better than exponentially larger classical machines?} 
\end{quote}
At first glance, such a space advantage may appear to be unlikely. Holevo's bound shows that only $n$ classical bits can be stored in an $n$-qubit state~\cite{holevo1973bounds}, and prior results imply that a large space advantage is not possible when the entire dataset is stored~\cite{watrous1999space,watrous2003complexity}. 
\revise{
Existing demonstrations of quantum space advantage are limited in application scope~\cite{le2006exponential,gavinsky2007exponential,jain2014space,kallaugher2022quantum,kallaugher2024exponential,kallaugher2025design,gilboa2024exponential,niroula2025realization,kretschmer2025demonstrating}; some rely on adversarially chosen data ordering and become classically easy under random sampling \cite{saxena2023streaming}, and none addresses natural machine learning tasks.
}

In this work, we resolve this question by proving exponential space advantages in a variety of classical data processing tasks.
We prove that a small quantum computer of $\poly(\log N)$ size can perform large-scale classification, dimension reduction, and linear system solving on massive classical data by processing classical data samples on the fly, whereas any classical machine achieving the same performance requires exponentially larger size or superpolynomially more samples and time, as summarized in \Cref{fig:overview}. 
\revise{
We illustrate the empirical scale of the space advantage through numerical simulations on real-world datasets; for movie review sentiment analysis and single-cell RNA sequencing, we provide evidence of reductions in size by six orders of magnitude compared to classical sparse-matrix and QRAM-based algorithms, and by four orders of magnitude compared to the tested classical streaming algorithms at matched terminal performance, all while using fewer than $60$ logical qubits.
}

Our results are enabled by a new framework, \emph{quantum oracle sketching}, which resolves the tension between quantum query access and classical data. Rather than storing the full dataset, this algorithm constructs coherent quantum queries from streaming classical data samples. Each sample is processed once and then immediately discarded, as a sequence of carefully designed quantum operations incrementally builds an approximate oracle that can be used within quantum algorithms. The required number of classical data samples scales quadratically with the number of quantum queries used. We prove the optimality of this quadratic scaling, which arises from the quadratic relationship between quantum amplitudes and probabilities governed by the Born rule.

Quantum oracle sketching natively handles noisy and correlated data with generic distributions and data structures. We use it to construct state preparation unitaries of vectors and block encodings of matrices. Combined with classical shadow tomography~\cite{huang2020predicting} for efficient readout, our quantum algorithms enable the construction of compact and accurate classical models from massive classical data, a task provably impossible for any classical machine that is not exponentially larger. We provide a \href{https://github.com/haimengzhao/quantum-oracle-sketching}{code} implementation in JAX and benchmark its performance via numerical simulation.

We prove classical hardness by establishing a fundamental relation between machine size and query complexity. For any oracle problem with exponential classical query complexity, if it can be solved in polynomial quantum space with slightly better than quadratic quantum query advantage, then there is a corresponding learning task using random classical data samples for which any classical machine requires exponentially larger size than the quantum machine to perform the task successfully. This quantum advantage is information-theoretic and unconditional, relying only on the validity of quantum mechanics; it persists even if classical machines are granted unlimited time or if $\mathsf{BPP} = \mathsf{BQP}$. Consequently, experimental confirmation or disproof of our results would provide a fundamental test of quantum mechanics at the complexity frontier~\cite{preskill2012quantum,kretschmer2025demonstrating}, analogous to how Bell inequalities~\cite{bell1966problem} test quantum nonlocality~\cite{clauser1969proposed,clauser1978bell}.

\section{Main results}

\begin{figure}
    \centering
    \includegraphics[width=1\linewidth]{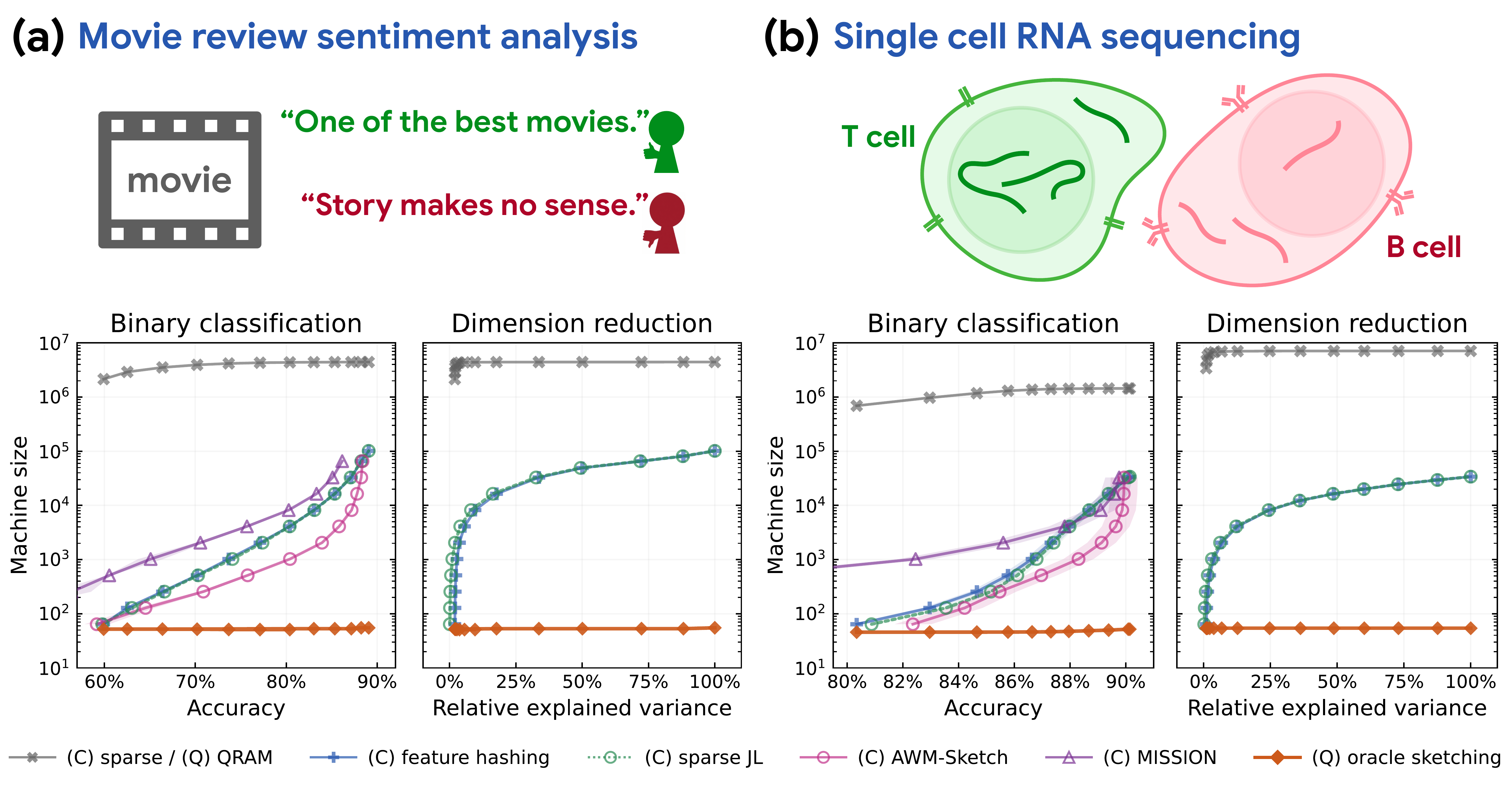}
    \caption{\revise{
    \textbf{Evidence for quantum advantages in real-world datasets from numerical simulation.}
    We perform binary classification and dimension reduction for
    \textbf{(a)}
    sentiment analysis of movie reviews from the Internet Movie Database (IMDb) \cite{maas2011learning}
    and 
    \textbf{(b)}
    single-cell RNA sequencing analysis of peripheral blood mononuclear cells (PBMC) \cite{zheng2017massively}.
    We compare the machine size required to reach a certain prediction performance using different families of classical (C) and quantum (Q) algorithms, including quantum oracle sketching (orange), quantum algorithms using QRAM (gray), classical sparse-matrix algorithms (gray), and classical streaming algorithms (blue \cite{weinberger2009feature}, green \cite{kane2014sparser}, magenta \cite{tai2018sketching}, and purple \cite{aghazadeh2018mission}).
    For each algorithm, we truncate the retained information to plot the trade-off between machine size and performance, with standard error indicated by the shaded region.
    Machine size is defined as the maximal number of fundamental memory units consumed throughout the entire computation: logical qubits for quantum and floating-point numbers for classical.
    Performance is quantified by the $5$-fold cross validation accuracy averaged over random category pairs for classification, and by the explained variance relative to the untruncated baseline for dimension reduction.
    More details are given in \Cref{sec:additional-numerics}.
    }
    }
    \label{fig:numerics}
\end{figure}

We begin by modeling how a machine processes massive amounts of classical data generated in the real world, which are typically random and noisy. The machine observes one classical data sample~$z$ at a time, processes the sample, updates its memory, and moves on to the next. Each data sample may represent, for example, a point of a function $z=(x, f(x))$ obtained from experimental observations, an entry $z=(i, j, A_{ij})$ of a matrix describing a dynamical system, or a feature vector and label $z=(i, \vec{x}_i, y_i)$ from user activity on the internet. After $M$ time steps, the machine has processed samples $z_1, \ldots, z_M$. The goal is to use these samples to learn properties of the underlying \emph{data generation process} $\mathcal{D}$, ranging from basic statistics to complex models such as classifying cells based on RNA sequences or predicting the next word in a sentence or the next frame of a video.

In realistic settings, the data generation process may evolve over time, introducing time-varying features and correlations across multiple time scales, such as the ubiquitous $1/f$ noise in electronic devices or weather-dependent fluctuations in astronomical observations (\Cref{fig:overview}(b)). In such a dynamic process, each data point $z$ is sampled from a distribution depending on the current \emph{situation}, which evolves as time progresses. We formalize this setting as a hierarchical model in \Cref{sec:data-access}. 

To quantify this correlation structure, we introduce two parameters. The \emph{refreshing time} $\tau$ is the time scale beyond which samples become effectively uncorrelated, and the \emph{repetition number} $R$ is the maximum expected number of times a sample is repeated within a window of $\tau$ time steps. A larger $R$ implies greater redundancy in that the same information is repeated multiple times, requiring proportionally more samples to gather sufficient information to solve a task. For simplicity, we assume $R = O(1)$ in the main text. All of our results extend to general $R$, where sample complexity scales linearly with $R$.

One of the most fundamental tasks in science and engineering is solving linear systems, a key subroutine in regression, optimization, and differential equations. As dataset sizes, sampling rates, or precision requirements increase, the dimension $N$ of these systems can become extremely large. Consider, for example, a power grid in an integrated circuit or civil infrastructure network, modeled as a graph with a massive number of nodes and edges. According to Ohm's law and Kirchhoff's laws, measurements of resistances and voltages yield a sparse and well-conditioned $N$-dimensional linear system $A\vec{x}=\vec{b}$, where $A\in \mathbb{R}^{N\times N}$ encodes the resistances, $\vec{b}\in \mathbb{R}^N$ the voltage values, and $\vec{x}$ the unknown currents. The accessible data are of the form $z=(i, j, A_{ij}, k, b_k)$, where $A_{ij}$ is a resistance measurement and $b_k$ a voltage measurement. To avoid overheating in critical components, we want to estimate the heat dissipation in the network. By Joule's law, this is given by a quadratic form $\vec{x}^T \mathcal{M}\vec{x}$, where $\mathcal{M}\in \mathbb{R}^{N\times N}$ specifies the components of interest.

In \Cref{sec:linear-system}, we formally define this task as estimating the normalized value of an efficiently measurable quadratic form to some error, given access to sampled data $z=(i, j, A_{ij}, k, b_k)$ from a sparse and well-conditioned linear system $A\vec{x}=\vec{b}$. Here $i, j$ specifies a uniformly random non-zero entry of $A$ and $k$ specifies a uniformly random component of $\vec{b}$. For this linear system task, we prove the following, where the exponent $0.99$ for the classical machine size can be replaced by any constant less than one.

\begin{theorem}[Solving linear systems; formalized in \Cref{thm:q-adv-linear-sys}]
    Using $\tilde{O}(N)$ samples, a quantum computer with $\poly(\log N)$ size can solve the linear system task with dimension $N$, whereas any classical machine with $O(N^{0.99})$ size cannot.
\end{theorem}

\revise{
In realistic scenarios, fluctuating noise or changing environmental conditions may cause a linear system to evolve over time while the property we wish to estimate remains approximately fixed~\cite{monticelli2000electric,abur2004power}. 
We formalize this setting as the \emph{dynamic} linear system task in \Cref{sec:linear-system}, in which the linear system changes every $\tau$ samples while approximately preserving the target property. 
For this task, we further establish the following superpolynomial quantum advantage in sample efficiency.
}

\begin{theorem}[Solving dynamic linear systems; formalized in \Cref{thm:q-adv-linear-sys-dynamic}]
    A quantum computer of $\poly(\log N)$ size can use $\tilde O(N)$ samples to solve the dynamic linear system task with dimension $N$ and sufficient $\tau=\tilde O(N)$, whereas any classical machine with $O(N^{0.99})$ size requires $\mathrm{superpoly}(N)$ samples.
\end{theorem}

The applications of computation extend far beyond traditional science and engineering, encompassing a vast array of problems in pattern recognition and artificial intelligence, where machines autonomously learn from data to perform desired tasks. A canonical example is \emph{classification}.
Consider a dataset of $N$ items, each represented by a sparse $D$-dimensional feature vector $\vec{x}_i \in \mathbb{R}^D$ and a binary label $y_i \in \{+1, -1\}$. Together, these form a training dataset with feature matrix $X \in \mathbb{R}^{N \times D}$ and label vector $\vec{y} \in \mathbb{R}^N$. The accessible data consist of samples $z = (i,\, \vec{x}_i,\, y_i)$, where $\vec{x}_i$ is a uniformly random row of $X$ and $y_i$ is the corresponding label.

This framework captures many real-world tasks. In \emph{sentiment analysis}, for instance, millions of users on e-commerce or video streaming platforms write reviews of products or movies; each vector $\vec{x}_i$ encodes a review, and each label $y_i \in \{+1,-1\}$ indicates whether the review is positive or negative. Accurately classifying such reviews helps businesses make informed commercial decisions. A standard approach to classification is the \emph{least-squares support vector machine} (LS-SVM), also known as the ridge classifier. It identifies a hyperplane decision boundary with normal vector $\vec{w} \in \mathbb{R}^D$ by minimizing the regularized loss $\mathcal{L}(\vec{w}) \;=\; \|X\vec{w} - \vec{y}\|_2^2 + \lambda\|\vec{w}\|_2^2,$ where $\lambda > 0$ is the $\ell_2$ regularization parameter. The predicted label of a test sample $\vec{x}'$ is then given by $\operatorname{sgn}(\vec{x}' \cdot \vec{w})$. A test sample is said to be \emph{classifiable} if its margin from the decision boundary, $\frac{|\vec{x}' \cdot \vec{w}|}{\|\vec{w}\|_2},$ is bounded away from zero.

In \Cref{sec:binary-classification}, we formalize this as the following learning task: given sample access to data $z = (i,\, \vec{x}_i,\, y_i)$ drawn from a sparse training set $(X, \vec{y})$ that is well-conditioned after regularization and admits a non-trivial regularized fit to the labels, predict the label of any sparse and classifiable test vector $\vec{x}'$ according to the LS-SVM decision rule. For a binary classification task of dimension $N \times D$, we establish the following quantum advantage.

\begin{theorem}[Classification; formalized in \Cref{thm:q-adv-bin-classify}]
    Using $\tilde{O}(N)$ samples, a quantum computer with $\poly(\log D)$ size can solve the binary classification task with $N$ items and $D$ features, whereas any classical machine with $O(D^{0.99})$ size cannot.
\end{theorem}

\revise{
In realistic scenarios, changing user behaviors or language habits may cause the accessible training data to evolve over time, while the classification rule we wish to model remains approximately fixed~\cite{widmer1996learning,gama2014survey}. 
We formalize this more challenging setting as the \emph{dynamic} binary classification task in \Cref{sec:binary-classification}, in which the training dataset is refreshed every $\tau$ samples while approximately preserving the underlying classification rule. For this task, we prove the following superpolynomial quantum advantage in sample efficiency.
}

\begin{theorem}[Dynamic classification; formalized in \Cref{thm:q-adv-bin-classify-dynamic}]
    A quantum computer with $\poly(\log D)$ size can use $\tilde O(N)$ samples to solve the dynamic binary classification task with $N$ items, $D$ features, and sufficient $\tau=\tilde O(N)$, whereas any classical machine with $O(D^{0.99})$ size requires $\mathrm{superpoly}(N)$ samples.
\end{theorem}

Supervised tasks such as classification require labels to guide the learning process, which may be inaccessible or prohibitively expensive to obtain, particularly when the goal is to discover unknown patterns latent in the data. Such scenarios call for \emph{unsupervised} learning approaches. A canonical example is \emph{dimension reduction}, which aims to distill high-dimensional data into a small number of explanatory variables so that hidden structure may be revealed. 

Consider $N$ items, each associated with a sparse $D$-dimensional feature vector $\vec{x}_i \in \mathbb{R}^D$. A prominent use case arises in modern biology, where experimental advances enable the generation of large volumes of unlabeled data. In \emph{single-cell RNA sequencing} (scRNA-seq), for instance, each item is a cell encoded as a sparse vector $\vec{x}_i \in \mathbb{R}^D$ representing its gene expression profile. The accessible data are of the form $z = (i, \vec{x}_i)$, forming a dataset $X \in \mathbb{R}^{N \times D}$. To discover unknown cell types or developmental trajectories hidden within this high-dimensional gene expression space, one applies \emph{principal component analysis} (PCA), which identifies the direction $\vec{w} \in \mathbb{R}^D$, $\|\vec{w}\| = 1$, of maximum variance $\vec{w}^{\top} X^{\top} X \vec{w}$, yielding low-dimensional representations $\xi(\vec{x}_i) = \vec{x}_i \cdot \vec{w}$ of the cells that distinguish, for example, T~cells from B~cells while filtering out noise. In many cases, a good initial estimate of the principal component is available by choosing an important feature with non-vanishing overlap with the true principal component; we refer to this initial estimate as a \emph{guiding vector} $\vec{g}$. Furthermore, the principal component is typically prominent enough that the spectral gap $\Delta$ between the largest and second-largest eigenvalues of $X^{\top}X$ does not vanish~\cite{newman2005power}.

In \Cref{sec:dimension-reduction}, we formalize dimension reduction as the task of estimating the low-dimensional representation $\xi(\vec{x}')$ of any sparse test vector $\vec{x}'$, given a guiding vector $\vec{g}$ and sample access to data $z = (i, \vec{x}_i)$ drawn from a sparse, spectrally gapped dataset $X$, where $\vec{x}_i$ is a uniformly random row of $X$. For a dimension reduction task of dimension $N \times D$, we prove the following quantum advantage.

\begin{theorem}[Dimension reduction; formalized in \Cref{thm:q-adv-dim-reduc}]
    Using $\tilde{O}(N)$ samples, a quantum computer with $\poly(\log D)$ size can solve the dimension reduction task with $N$ items and $D$ features, whereas any classical machine with $O(D^{0.99})$ size cannot.
\end{theorem}

\revise{
Analogously to the previous applications, changing experimental conditions, donors, or cell populations may cause the accessible data to evolve over time while the principal component of interest remains approximately fixed~\cite{haghverdi2018batch,johnson2007adjusting}. 
We formalize this setting as the \emph{dynamic} dimension reduction task. For this task, we prove the following superpolynomial advantage in sample efficiency.
}

\begin{theorem}[Dynamic dimension reduction; formalized in \Cref{thm:q-adv-dim-reduc-dynamic}]
    A quantum computer with $\poly(\log D)$ size can use $\tilde O(N)$ samples to solve the dynamic dimension reduction task with $N$ items, $D$ features, and sufficient $\tau=\tilde O(N)$, whereas any classical machine with $O(D^{0.99})$ size requires $\mathrm{superpoly}(N)$ samples.
\end{theorem}

\revise{
We conduct classical numerical simulations to provide evidence for these quantum advantages in real-world applications, including sentiment analysis of movie reviews from the Internet Movie Database (IMDb)~\cite{maas2011learning} and single-cell RNA sequencing (scRNA-seq) analysis of peripheral blood mononuclear cells (PBMCs)~\cite{zheng2017massively}. 
To handle limited memory in realistic scenarios, we truncate the feature dimension of each sample using feature hashing, randomly merging distinct words in text or genes in RNA sequences into fewer buckets \cite{weinberger2009feature}. 
Varying the number of retained features traces the trade-off between machine size and task performance.
In addition to this feature hashing method, we also consider several alternative streaming and randomized sketching techniques~\cite{kane2014sparser,tai2018sketching,aghazadeh2018mission}, described in \Cref{sec:additional-numerics}.

In \Cref{fig:numerics}, we compare the machine size required to achieve a given performance in binary classification and dimension reduction using quantum oracle sketching (orange), classical streaming algorithms (blue, green, magenta, and purple), classical sparse-matrix algorithms (gray), and QRAM-based quantum algorithms (gray). All algorithms perform the same task: predicting the label or one-dimensional representation of a single arbitrary test point. Performance is measured by $5$-fold cross-validation accuracy, averaged over random category pairs, for classification, and by the explained variance of the first principal component relative to the untruncated baseline for dimension reduction. Machine size is defined as the maximum memory used throughout the computation, measured in logical qubits for quantum processors and stored floating-point numbers for classical machines.


To isolate the dependence on machine size, we allow all algorithms to have sufficient samples and computation time to reach their converged performance under each machine size budget.
As detailed in \Cref{sec:add-num-datasets}, the convergence guarantees of the quantum algorithms allow us to predict their performance without simulating deep quantum circuits on around $60$ logical qubits.
For simplicity, we conservatively underestimate the memory consumption for the classical algorithms.
We note that these classical baselines are general-purpose algorithms for performing LS-SVM and PCA with provable guarantees; comparisons with dataset-specific heuristics or other methods are left to future work, as their performance requires extensive empirical study.

The results demonstrate that, to achieve high performance, the memory required by quantum oracle sketching remains small and nearly constant, whereas that of the tested classical and QRAM-based approaches grows exponentially. This provides evidence for quantum advantages of four to six orders of magnitude using fewer than $60$ logical qubits. In \Cref{sec:additional-numerics}, we provide further details and additional experiments demonstrating the broad applicability of this advantage on datasets spanning social media topic analysis and pharmaceutical drug discovery.
}

\section{Origin of Advantage}

\begin{figure}
    \centering
    \includegraphics[width=1\linewidth]{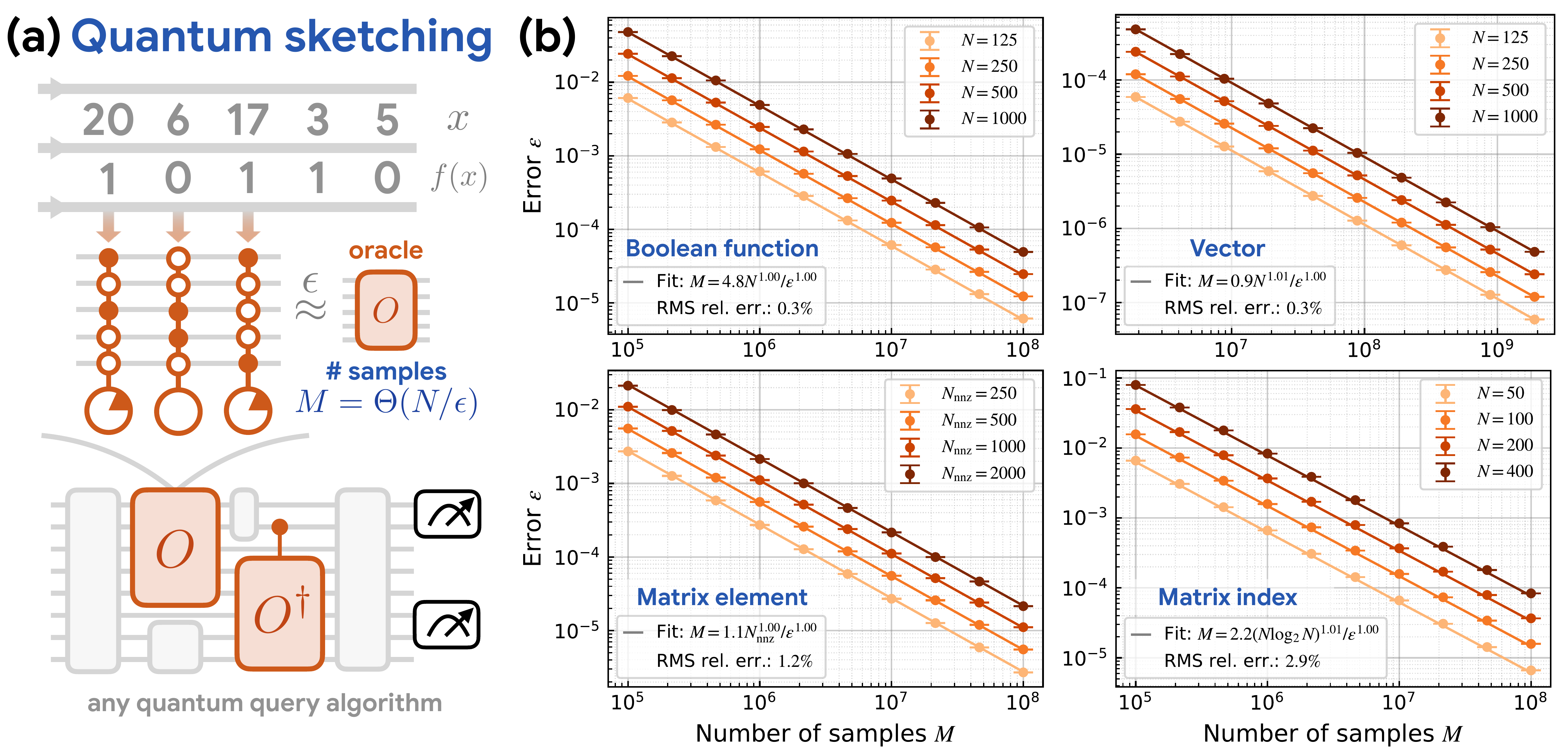}
    \caption{
    \textbf{Access the classical world in superposition with quantum oracle sketching.}
    \textbf{(a)}
    An example of making quantum coherent query to a Boolean function using its classical data $(x, f(x))$ with quantum oracle sketching.
    Upon receiving each classical sample $(x, f(x))$, we apply a multi-controlled phase gate $\exp(i\theta \ketbra{x}), \theta \propto f(x)$.
    With $M=\Theta(N/\epsilon)$ samples, the resulting random unitary channel approximates the phase oracle $O: \ket{x}\to (-1)^{f(x)}\ket{x}$ of $f$ to $\epsilon$ error in diamond distance.
    This allows us to instantiate oracle queries in any quantum query algorithm that extracts the desired property of $f$.
    \textbf{(b)}
    Numerical experiments benchmarking the number of samples $M$ needed to approximate various oracle queries to $\epsilon$ operator norm error of the expected unitary, which upper bounds the diamond distance error up to constants, as a proxy.
    We consider oracles of Boolean functions, state preparation unitaries of any vectors, and the sparse matrix element and index oracles of any sparse matrices.
    We use $N$ to denote the domain size of Boolean functions, the dimension of vectors, and the dimension of square matrices.
    $N_{\mathrm{nnz}}$ represents the number of non-zero elements in a sparse matrix.
    The solid lines represent the fitted sample complexity scaling, with fitted parameters and root-mean-squared relative errors (RMS rel. err.) listed.
    }
    \label{fig:benchmark}
\end{figure}

\subsection{Quantum Oracle Sketching}

We prove the main results by introducing \emph{quantum oracle sketching}, a quantum data-loading algorithm that resolves the fundamental tension between the quantum coherent queries we need and the classical data access the world provides. It allows us to access the classical world in quantum superposition using only classical data samples, without the overhead of storing the entire dataset.
We achieve this by applying a sequence of incremental quantum rotations using fresh data samples on the fly.
Each data sample is processed once and then immediately discarded without being stored.
With the oracles instantiated by quantum oracle sketching, we can execute any quantum query algorithm, including various quantum linear algebra algorithms~\cite{costa2022optimal,lin2020near,gilyen2019quantum}, to prepare small quantum states that encode the solutions to the application tasks.
Finally, we apply a variant of classical shadow tomography~\cite{huang2020predicting} to efficiently extract classical outputs, thereby solving the application tasks in an end-to-end fashion.

The idea of applying incremental quantum rotations based on data samples is reminiscent of classical streaming algorithms, which also update a model on the fly without storing the full dataset. 
However, the coherence requirement of quantum computation may seem at first to rule out this approach; won't the randomness and entropy in the data, continuously pumped into the quantum machine, cause it to decohere quickly?

To be more precise, suppose we have $M$ data samples $z_1, \ldots, z_M$ uniformly drawn from $N$ possibilities. 
For each sample $z_t$ we apply a small rotation $\exp(ih_{z_t}/M)$ driven by some Hamiltonian $h_{z_t}$ with constant operator norm.
One might hope that the resulting evolution approximates evolution governed by the expected Hamiltonian $\exp(i\E[h_z])\approx \mathrm{exp}(i\sum_{t} h_{z_t}/M)$. Unfortunately, results from studies of randomized Hamiltonian simulation show that the error compounds to $\epsilon \sim N^2/M$ in general~\cite{campbell2019random,chen2021concentration,kimmel2017hamiltonian}.
Hence the randomness in the data destroys the coherence of the quantum machine, incurring a large error, unless the number of samples is at least $M\sim N^2$.
However, as we discuss in \Cref{sec:q-alg-iid}, consuming $N^2$ samples to generate just a single quantum query eliminates any potential quantum advantage.

To overcome decoherence while retaining advantage, we design the incremental rotations so that their contributions accumulate coherently. The general mechanism can be illustrated by sketching the phase oracle of a Boolean function $f: [N]\to \bit$, summarized in \Cref{fig:benchmark}(a).
Consider a sequence of $M$ independent samples $z_t=(x_t, f(x_t)), t=1, \ldots, M$, where $x_t$ is sampled uniformly from $[N]=\{1, \ldots, N\}$ with distribution $p(x)=1/N$.
Our goal is to implement the phase oracle
\begin{equation}
    O: \ket{x}\to (-1)^{f(x)}\ket{x}.
\end{equation}
To achieve this goal, upon receiving each sample $(x_t, f(x_t))$, we apply a multi-controlled phase gate
\begin{equation}
	V_t = \exp(i\tau f(x_t)\ketbra{x_t}/M)
\end{equation}
for some $\tau$ to be chosen later, which applies a small phase rotation to the basis state $\ket{x_t}$ and leaves all other basis states unchanged.
After processing all $M$ samples, the resulting unitary is
\begin{equation}
	V = \prod_{t=1}^M V_t =  \exp(i\tau \sum_{t=1}^M f(x_t)\ketbra{x_t}/M) = \sum_{x=1}^N \exp(i\tau m_x f(x)) \ketbra{x},
\end{equation}
where  $m_x=\sum_{t} 1[x_t=x]/M$ is the empirical frequency of $x$.

As the sample size $M$ grows, the frequency $m_x$ concentrates around the probability $p(x)=1/N$.
Therefore, choosing the evolution time to be $\tau=\pi N$ so that $\tau m_x\approx \tau p(x)=\tau/N=\pi$, the random gate sequence $V$ approaches the phase oracle as desired:
\begin{equation}
	\sum_{x=1}^N\exp(i\tau p(x) f(x))\ketbra{x} = \sum_{x=1}^N\exp(i\pi f(x))\ketbra{x}=\sum_{x=1}^N (-1)^{f(x)}\ketbra{x}=O.
\end{equation}
In \Cref{sec:q-alg}, we rigorously prove that the error of this procedure scales as $\epsilon\sim N/M$, substantially better than the $\epsilon\sim N^2/M$ scaling for generic Hamiltonians.
The key point is that Hamiltonians associated with distinct values of $x$ act on mutually orthogonal subspaces, which prevents errors from accumulating across different basis states. 
Consequently, $M=\Theta(N/\epsilon)$ samples in this procedure suffice to construct an $\epsilon$-error approximation in diamond distance (\Cref{thm:q-oracle-sketch-iid}).
The inverse and controlled oracles can be constructed similarly by negating $\tau$ or adding control to each gate.

The above procedure allows us to execute any quantum query algorithm by sketching each oracle query with classical data samples.
To run an algorithm making $Q$ queries with total error $\epsilon$, we set the error of each oracle sketch to be $\epsilon/Q$.
Since each query consumes $\Theta(N/(\epsilon/Q))$ samples, we need 
\begin{equation}
M=Q\cdot\Theta(N/(\epsilon/Q))=\Theta(NQ^2/\epsilon)
\end{equation}
samples in total.
In \Cref{sec:q-alg-opt}, we prove that this sample complexity is optimal. The quadratic dependence on $Q$ is the necessary price to pay for converting classical samples into coherent quantum queries, mirroring the relationship between quantum amplitudes and probabilities in the Born rule.

\revise{
This pedagogical Boolean phase-oracle example illustrates the core mechanism behind quantum oracle sketching, but does not directly capture the noisy, correlated data and linear algebra data structures encountered in machine learning applications.
To address these applications, we generalize quantum oracle sketching beyond simple data distributions and orthogonal Hamiltonians, by developing a suite of algorithmic and theoretical tools in \Cref{sec:q-alg-ext}.

To accommodate more realistic data distributions, we prove that quantum oracle sketching naturally handles noisy and correlated data with time-varying features, only at the cost of increasing its sample complexity by the factor $R$, the repetition number of the data generation process.
This factor is intuitive: if each sample is repeated $R$ times, then we need $R$ times more samples to obtain the same amount of new information. We then extend the method to unknown, non-uniform data distributions using the quantum singular value transformation (QSVT).

For linear algebra data structures, we extend quantum oracle sketching in \Cref{sec:q-alg-linear-algebra} to construct the state preparation unitaries of arbitrary $N$-dimensional vectors, which we call \emph{quantum state sketching}, as well as sparse oracles and block encodings of $N$-dimensional sparse matrices.
These extensions are enabled by combining a suite of algorithmic techniques including QSVT, in-place binary search, oblivious amplitude amplification, and a signed randomized Hadamard transform with pseudorandom functions.
Because they involve non-orthogonal quantum rotations, establishing their performance requires a significantly more sophisticated variance analysis. These results yield the following guarantee.

\begin{theorem}[Quantum oracle sketching; formalized in \Cref{thm:q-oracle-sketch-corr,lem:block-encoding,thm:q-state-sketch}]
    A quantum machine can emulate $Q$ queries to the above oracles of size $N$ to $\epsilon$ total error in the diamond distance with $\poly(\log N)$ space overhead using $\tilde O(NQ^2/\epsilon)$ classical samples.
\end{theorem}
}

In \Cref{fig:benchmark}(b), we benchmark the empirical sample complexity of quantum oracle sketching on various data structures.
We generate random Boolean functions, unit vectors, and sparse matrices, use quantum oracle sketching to approximate their oracle queries, and calculate the average error, for a wide range of dimensions and sample sizes.
We report the operator norm error of the expected unitary, which upper bounds the diamond distance up to constants, as a proxy.
To isolate the performance of quantum oracle sketching without the overhead of QSVT, we report results for vectors without amplitude amplification and assume random row access to construct matrix row index oracles as in the binary classification and dimension reduction tasks.
The results show accurate agreement with theoretical predictions, highlighted by the favorable constants and exponents in the sample complexity extracted by least-squares fit, with root-mean-squared relative errors all below $3\%$.
Further details are provided in \Cref{sec:additional-numerics}.

\subsection{Interferometric Classical Shadows}

With the data successfully loaded into the quantum computer, the remaining challenge is to efficiently read out classical results.
In particular, we need to retain the sign structures that are necessary in applications such as SVM, where the sign of the inner product between the weight vector and the data vector determines the predicted label.
To this end, we develop the \emph{interferometric classical shadow} algorithm in \Cref{lem:interf-classical-shadow}, which combines the idea of the Hadamard test with the efficient offline prediction capability of classical shadow tomography.
This method allows us to construct a completely classical model that can compute predictions for an arbitrary number of sparse test inputs.

\revise{
\begin{theorem}[Efficient readout; formalized in \Cref{lem:interf-classical-shadow,thm:bin-classify-upper,thm:dim-reduc-upper}]
    A classical model of $\poly(\log D)$ size, constructed from classical shadows of the solution state, suffices to predict the label of any sparse, classifiable test vector for classification, or the one-dimensional representation of any sparse test vector for dimension reduction.
\end{theorem}
}

Remarkably, this shows that quantum technology can compress relevant information in classical data into a very compact classical representation without sacrificing accuracy, which would be impossible with classical machines alone unless provided with exponentially larger memory. This achievement does not violate the Holevo bound; rather it exploits the structure of a task to enable highly efficient extraction of useful outputs.

\subsection{Classical Hardness}

Having shown that quantum machines excel at these tasks, it remains to prove classical hardness to rigorously establish quantum advantage.
We do so by connecting advantage in machine size to separation in query complexity.
In particular, we consider the task of estimating a property of some Boolean function $f: [N]\to \bit$ using noisy query data, which we formalize as Noisy Oracle Property Estimation (NOPE) in \Cref{sec:cl-hard-noisy-oracle-property-est}.
Its complexity is characterized by the classical query complexity $Q_C$ of the target property, which captures the amount of information about $f$ that a classical machine has to extract to estimate the property.

To obtain a lower bound on the space required by a classical machine, we consider a NOPE task such that each piece of useful information is scattered across blocks of $N$ noisy samples. To extract it, the machine must store the noisy information from one block in its memory of size $S$ and combine it with the next. Consequently, after $M$ samples, the machine can carry at most $S$ bits across the $O(M/N)$ block boundaries, retrieving at most $S\cdot O(M/N)$ bits of information. To solve the task, this must be at least $Q_C$, yielding a tradeoff between space and sample size for any classical machine:
\begin{equation}
MS\geq \Omega(NQ_C) .
\end{equation}
We formally prove this lower bound in \Cref{sec:cl-hard-sample-space-lb} using communication complexity tools \cite{goos2015deterministic,goos2016rectangles,goos2020query,anshu2020query,chattopadhyay2021query,yang2024communication,fei2025multi}.
Combining this bound with the sample complexity $M=\Theta(NQ^2)$ of quantum oracle sketching yields the following fundamental relation between space advantage and oracle query separation.
\begin{theorem}[Space advantage from oracle separation; formalized in \Cref{thm:classical-lower-bound}]
\label{thm:memory-adv-from-oracle-separation}
    To solve a NOPE task with quantum query complexity $Q$ and classical query complexity $Q_C$ using the same number of samples needed by quantum machines, any classical machine must be of size at least $\Omega(Q_C/Q^2)$.
\end{theorem}

When the NOPE task has exponentially large classical query complexity and can be solved in $\poly(\log N)$ quantum space with a super-quadratic quantum query advantage, \Cref{thm:memory-adv-from-oracle-separation} unconditionally implies a corresponding exponential quantum space advantage. For example, if $Q_C=\Theta(N)$ and $Q=O(N^{0.49})$, then any classical algorithm using the same number of samples must have memory size at least $S\geq \Omega(N^{0.02})$. 
As a concrete instance, consider the NOPE task of estimating the Forrelation property \cite{aaronson2015forrelation,bansal2021kforrelation}, which exhibits the maximal average-case query separation: $Q=O(1)$ and $Q_C=\Omega(N^{1-\zeta})$ for any constant $\zeta>0$.
This immediately yields a large space separation between quantum and classical algorithms. 

From this space advantage result, we can derive a superpolynomial advantage in sample and time complexity. By developing a learning version of the XOR lemma \cite{yao1982theory,unger2009probabilistic,assadi2021graph} in \Cref{sec:cl-hard-bootstrap}, we prove that classical machines without sufficient space are fundamentally unable to track dynamically evolving data distributions and therefore require superpolynomially more samples. 
Via reductions, we also prove that the classical hardness of NOPE implies classical hardness for our target applications. Specifically, we prove in \Cref{sec:cl-hard-app} that tasks such as predicting labels or obtaining low-dimensional representations of test data are BQP-hard. Thus any algorithm that solves these tasks can be used to simulate the quantum circuit that solves NOPE. Because we have already established that NOPE is classically hard, the classical hardness of these other applications follows.

\revise{
Beyond the superpolynomial sample and time advantages we rigorously prove for the dynamic tasks, in \Cref{app:best-known-classical}, we provide computational complexity evidence for analogous advantages in static tasks as well.
In particular, when limited to only $\poly(\log N)$ space, the best known classical algorithm for solving the static linear system task has runtime $\mathrm{superpoly}(N)$ \cite{borodin1977relating,aaronson2016complexity,csanky1976fast,berkowitz1984computing}.
Improving this runtime to $\poly(N)$ time, even with $O(N^{1-\zeta})$ space for any constant $\zeta>0$, likely requires overcoming barriers in long-standing open problems in complexity theory \cite{liu2026numerical,ta2013inverting,papakonstantinou2014depth,savitch1970relationships,barnes1998sublinear}.
In contrast, quantum oracle sketching solves the task in $\poly(\log N)$ space and $\tilde O(N)$ time, which is superpolynomially faster.
Note that we exclude the comparison against this $\poly(\log N)$ space method in our numerical experiments because it requires $\mathrm{superpoly}(N)$ samples.
}

\section{Discussion}

In this work, we establish classical data processing and machine learning as a broad domain of exponential quantum advantage, extending the reach of quantum computation beyond specialized tasks.
We demonstrate this advantage across fundamental applications including solving linear systems, binary classification, and dimension reduction.
Notably, our exponential separation reaches the ultimate limit allowed by quantum mechanics, since a classical machine with $\tilde{O}(N)$ size can always store the full dataset and simulate a $\poly(\log N)$-qubit quantum computation with only $\poly(\log N)$ space overhead \cite{watrous1999space,watrous2003complexity}.

Our results are enabled by introducing quantum oracle sketching and interferometric classical shadows, which together circumvent the data loading and readout bottlenecks. These techniques allow the construction of exponentially compact classical models of massive classical data which could not be achieved efficiently without the use of quantum technology. 
The existence of such compact classical models demonstrates that, in useful tasks such as classification and dimension reduction, the relevant information has sufficient structure to permit its efficient extraction without violating the generic Holevo bound.

\revise{
While our unconditional classical lower bounds are proved using carefully constructed hard instances, our quantum algorithms provably solve the tasks efficiently for any dataset satisfying the promises. Any classical algorithm that works generally must in particular solve these hard instances and is therefore subject to our classical lower bounds.
 This distinguishes our results from previous quantum advantages based on specially constructed problems, where the quantum algorithms exploit the particular structure of the hard instances and do not extend to general inputs.
}

Our numerical experiments support the practical relevance of the quantum advantages, showing orders-of-magnitude memory savings with fewer than $60$ logical qubits.
Extrapolating this scaling, while ignoring the same $\tilde O(N)$ runtime overhead, suggests that relatively small quantum devices with hundreds of logical qubits could outperform even extremely large classical systems, barring unforeseen limitations to quantum mechanics.

\revise{
Beyond space advantages, we also establish superpolynomial advantage in sample and time complexity for dynamic tasks, showing that classical algorithms with insufficient memory are fundamentally unable to track evolving data distributions.
In the context of large-scale scientific experiments, where a gigantic volume of data is generated in each experimental run~\cite{bejaralonso2020hllhc,dewdney2009square,lahnemann2020eleven}, our results show that a small quantum machine can solve the tasks within a few runs, whereas any classical machine must either have exponentially more memory to hold the entire data volume, or repeat the experiment superpolynomially many times to achieve comparable performance.
}

\revise{The runtime of quantum oracle sketching is dominated by the $\tilde{O}(N)$ data loading time, which is unavoidable since each gate has only a constant number of degrees of freedom, while processing each sample requires only $\poly(\log N)$ time.
Characterizing its end-to-end sample and time complexity via circuit-level simulation is an important future direction.
When compared to the best known classical algorithm with memory size at most $O(N^{0.99})$, quantum oracle sketching is superpolynomially faster, even for the static tasks.
If allowed to use $\Omega(N)$ memory, however, classical algorithms have asymptotically comparable runtime, and may be faster in practice because classical gates operate at higher speeds.
The wall-clock runtime also depends on the data generation rate; if the quantum processor can keep pace with the incoming data stream, then the overall runtime will be limited by the data generation rate rather than the quantum processing speed. 
}

Notably, quantum oracle sketching is largely composed of commuting operations, suggesting significant opportunities for parallelization which might dramatically reduce the wall-clock runtime. 
Exploring such parallel implementations, potentially
through hardware-software co-design and optimized quantum error correction architectures, is an important direction for future work~\cite{zhou2025opportunities}.

We have presented quantum oracle sketching as a general classical-to-quantum converter that enables quantum queries to the classical world with provable guarantees even in the worst-case.
There are substantial opportunities to extend and optimize this framework for practical tasks, both rigorously and heuristically.
For instance, one may enhance its empirical performance by introducing trainable and variational components, or by integrating it into hybrid quantum-classical data processing pipelines.

While we have focused here on linear systems, classification, and dimension reduction, we expect similar quantum advantages to arise from quantum oracle sketching applied to a broader spectrum of real-world tasks, including solving ordinary and partial differential equations, large-scale optimization, signal processing, and communication.
Furthermore, since exponential space advantage follows generically from super-quadratic query separation, it may extend to problems previously thought to be dequantized, such as recommendation systems, where polynomial quantum speedups persist but memory bottlenecks can be critical.

From a fundamental physics perspective, our results are information-theoretic and unconditional, relying solely on the principle of quantum superposition, independent of any computational conjectures. This quantum space advantage persists even if classical machines are granted unbounded computation time, or even in the unlikely scenario where polynomial-time quantum computation is no more powerful than classical (i.e., $\mathrm{BPP} = \mathrm{BQP}$), in which case there would be no superpolynomial time advantage in quantum simulation or cryptanalysis.

Consequently, experimental confirmation or disproof of our results would serve as a fundamental test of quantum mechanics at the complexity frontier~\cite{preskill2012quantum}, probing the physical reality of exponentially large Hilbert spaces~\cite{poulin2011quantum}. 
This is analogous to how Bell inequalities test quantum nonlocality, or how particle colliders and cosmological observations probe the Standard Model at the energy frontier.
We believe that the prospect of enabling new applications of quantum devices to massive classical data, while also probing the physical limits of quantum mechanics, marks the beginning of an exciting new frontier in science and technology.

\section*{Acknowledgments}
We thank Dolev Bluvstein, Isaac Chuang, Ronald de Wolf, Dar Gilboa, Siddhartha Jain, Stephen Jordan, Robbie King, Ruohan Shen, Umesh Vazirani, and Jason Wang for insightful discussions. We are grateful to Richard Allen, Soonwon Choi, and Angus Lowe for bringing to our attention a mathematical error in a prior work on randomized Hamiltonian simulation. H.Z. was a Student Researcher at Google Quantum AI when part of this work was done. A.Z. is supported by a Hertz Fellowship. J.P. acknowledges support from the U.S. Department of Energy, Office of Science, Accelerated Research in Quantum Computing, Fundamental Algorithmic Research toward Quantum Utility (FAR-Qu), and the National Science Foundation (PHY-2317110). H.H. acknowledges support from the Broadcom Innovation Fund. H.Z., J.P., and H.H. acknowledge support from the U.S. Department of Energy, Office of Science, National Quantum Information Science Research Centers, Quantum Systems Accelerator. The Institute for Quantum Information and Matter is an NSF Physics Frontiers Center (PHY-2317110).

\clearpage
\bibliography{References}

@book{vershynin2018high,
  title={High-dimensional probability: An introduction with applications in data science},
  author={Vershynin, Roman},
  volume={47},
  year={2018},
  publisher={Cambridge University Press}
}

@article{costa2022optimal,
  title={Optimal scaling quantum linear-systems solver via discrete adiabatic theorem},
  author={Costa, Pedro CS and An, Dong and Sanders, Yuval R and Su, Yuan and Babbush, Ryan and Berry, Dominic W},
  journal={PRX Quantum},
  volume={3},
  number={4},
  pages={040303},
  year={2022},
  publisher={APS}
}

@article{morales2024quantum,
  title={Quantum linear system solvers: A survey of algorithms and applications},
  author={Morales, Mauro ES and Pira, Lirand{\"e} and Schleich, Philipp and Koor, Kelvin and Costa, Pedro and An, Dong and Aspuru-Guzik, Al{\'a}n and Lin, Lin and Rebentrost, Patrick and Berry, Dominic W},
  journal={arXiv preprint arXiv:2411.02522},
  year={2024}
}

@article{harrow2009quantum,
  title={Quantum algorithm for linear systems of equations},
  author={Harrow, Aram W and Hassidim, Avinatan and Lloyd, Seth},
  journal={Physical Review Letters},
  volume={103},
  number={15},
  pages={150502},
  year={2009},
  publisher={APS}
}

@inproceedings{bansal2021kforrelation,
  title={k-Forrelation optimally separates quantum and classical query complexity},
  author={Bansal, Nikhil and Sinha, Makrand},
  booktitle={Proceedings of the 53rd Annual ACM SIGACT Symposium on Theory of Computing},
  pages={1303--1316},
  year={2021}
}

@inproceedings{aaronson2015forrelation,
  title={Forrelation: A problem that optimally separates quantum from classical computing},
  author={Aaronson, Scott and Ambainis, Andris},
  booktitle={Proceedings of the forty-seventh annual ACM symposium on theory of computing},
  pages={307--316},
  year={2015}
}

@article{nakaji2024high,
  title={High-order randomized compiler for hamiltonian simulation},
  author={Nakaji, Kouhei and Bagherimehrab, Mohsen and Aspuru-Guzik, Al{\'a}n},
  journal={PRX Quantum},
  volume={5},
  number={2},
  pages={020330},
  year={2024},
  publisher={APS}
}

@article{martyn2021grand,
  title={Grand unification of quantum algorithms},
  author={Martyn, John M and Rossi, Zane M and Tan, Andrew K and Chuang, Isaac L},
  journal={PRX Quantum},
  volume={2},
  number={4},
  pages={040203},
  year={2021},
  publisher={APS}
}

@inproceedings{gilyen2019quantum,
  title={Quantum singular value transformation and beyond: exponential improvements for quantum matrix arithmetics},
  author={Gily{\'e}n, Andr{\'a}s and Su, Yuan and Low, Guang Hao and Wiebe, Nathan},
  booktitle={Proceedings of the 51st annual ACM SIGACT symposium on theory of computing},
  pages={193--204},
  year={2019}
}

@article{kimmel2017hamiltonian,
  title={Hamiltonian simulation with optimal sample complexity},
  author={Kimmel, Shelby and Lin, Cedric Yen-Yu and Low, Guang Hao and Ozols, Maris and Yoder, Theodore J},
  journal={npj Quantum Information},
  volume={3},
  number={1},
  pages={13},
  year={2017},
  publisher={Nature Publishing Group UK London}
}

@article{canonne2022short,
  title={A short note on an inequality between KL and TV},
  author={Canonne, Cl{\'e}ment L},
  journal={arXiv preprint arXiv:2202.07198},
  year={2022}
}

@article{angrisani2023unifying,
  title={A unifying framework for differentially private quantum algorithms},
  author={Angrisani, Armando and Doosti, Mina and Kashefi, Elham},
  journal={arXiv preprint arXiv:2307.04733},
  year={2023}
}

@book{watrous2018theory,
  title={The theory of quantum information},
  author={Watrous, John},
  year={2018},
  publisher={Cambridge University Press}
}

@inproceedings{lee2008direct,
  title={A direct product theorem for discrepancy},
  author={Lee, Troy and Shraibman, Adi and {\v{S}}palek, Robert},
  booktitle={2008 23rd Annual IEEE Conference on Computational Complexity},
  pages={71--80},
  year={2008},
  organization={IEEE}
}

@article{chen2021concentration,
  title={Concentration for random product formulas},
  author={Chen, Chi-Fang and Huang, Hsin-Yuan and Kueng, Richard and Tropp, Joel A},
  journal={PRX Quantum},
  volume={2},
  number={4},
  pages={040305},
  year={2021},
  publisher={APS}
}

@article{campbell2019random,
  title={Random compiler for fast Hamiltonian simulation},
  author={Campbell, Earl},
  journal={Physical Review Letters},
  volume={123},
  number={7},
  pages={070503},
  year={2019},
  publisher={APS}
}

@article{chattopadhyay2021query,
  title={Query-to-communication lifting using low-discrepancy gadgets},
  author={Chattopadhyay, Arkadev and Filmus, Yuval and Koroth, Sajin and Meir, Or and Pitassi, Toniann},
  journal={SIAM Journal on Computing},
  volume={50},
  number={1},
  pages={171--210},
  year={2021},
  publisher={SIAM}
}

@article{rebentrost2014quantum,
  title={Quantum support vector machine for big data classification},
  author={Rebentrost, Patrick and Mohseni, Masoud and Lloyd, Seth},
  journal={Physical Review Letters},
  volume={113},
  number={13},
  pages={130503},
  year={2014},
  publisher={APS}
}

@article{lloyd2014quantum,
  title={Quantum principal component analysis},
  author={Lloyd, Seth and Mohseni, Masoud and Rebentrost, Patrick},
  journal={Nature Physics},
  volume={10},
  number={9},
  pages={631--633},
  year={2014},
  publisher={Nature Publishing Group UK London}
}

@article{aaronson2015read,
  title={Read the fine print},
  author={Aaronson, Scott},
  journal={Nature Physics},
  volume={11},
  number={4},
  pages={291--293},
  year={2015},
  publisher={Nature Publishing Group UK London}
}

@article{somma2025quantum,
  title={Quantum algorithm for linear matrix equations},
  author={Somma, Rolando D and Low, Guang Hao and Berry, Dominic W and Babbush, Ryan},
  journal={arXiv preprint arXiv:2508.02822},
  year={2025}
}

@article{diakonikolas2009bounded,
  title={Bounded independence fools degree-2 threshold functions},
  author={Diakonikolas, Ilias and Kane, Daniel M and Nelson, Jelani},
  journal={arXiv preprint arXiv:0911.3389},
  year={2009}
}

@book{kitaev2002classical,
  title={Classical and quantum computation},
  author={Kitaev, Alexei Yu and Shen, Alexander and Vyalyi, Mikhail N},
  volume={47},
  year={2002},
  publisher={American Mathematical Soc.}
}

@article{hastings2017turning,
  title={Turning gate synthesis errors into incoherent errors},
  author={Hastings, Matthew B},
  journal={Quantum Information \& Computation},
  volume={17},
  number={5-6},
  pages={488--494},
  year={2017},
  publisher={Rinton Press, Incorporated Paramus, NJ}
}

@inproceedings{bernstein1993quantum,
  title={Quantum complexity theory},
  author={Bernstein, Ethan and Vazirani, Umesh},
  booktitle={Proceedings of the twenty-fifth annual ACM symposium on theory of computing},
  pages={11--20},
  year={1993}
}

@article{aharonov2003simple,
  title={{A simple proof that Toffoli and Hadamard are quantum universal}},
  author={Aharonov, Dorit},
  journal={arXiv preprint quant-ph/0301040},
  year={2003}
}

@article{montanaro2015quantum,
  title={Quantum speedup of Monte Carlo methods},
  author={Montanaro, Ashley},
  journal={Proceedings of the Royal Society A: Mathematical, Physical and Engineering Sciences},
  volume={471},
  number={2181},
  pages={20150301},
  year={2015},
  publisher={The Royal Society}
}

@article{brassard2000quantum,
  title={Quantum amplitude amplification and estimation},
  author={Brassard, Gilles and Hoyer, Peter and Mosca, Michele and Tapp, Alain},
  journal={arXiv preprint quant-ph/0005055},
  year={2000}
}

@article{huang2020predicting,
  title={Predicting many properties of a quantum system from very few measurements},
  author={Huang, Hsin-Yuan and Kueng, Richard and Preskill, John},
  journal={Nature Physics},
  volume={16},
  number={10},
  pages={1050--1057},
  year={2020},
  publisher={Nature Publishing Group UK London}
}

@article{johnstone2001distribution,
  title={On the distribution of the largest eigenvalue in principal components analysis},
  author={Johnstone, Iain M},
  journal={The Annals of Statistics},
  volume={29},
  number={2},
  pages={295--327},
  year={2001},
  publisher={Institute of Mathematical Statistics}
}

@article{newman2005power,
  title={Power laws, Pareto distributions and Zipf's law},
  author={Newman, Mark EJ},
  journal={Contemporary physics},
  volume={46},
  number={5},
  pages={323--351},
  year={2005},
  publisher={Taylor \& Francis}
}

@article{lin2020near,
  title={Near-optimal ground state preparation},
  author={Lin, Lin and Tong, Yu},
  journal={Quantum},
  volume={4},
  pages={372},
  year={2020},
  publisher={Verein zur F{\"o}rderung des Open Access Publizierens in den Quantenwissenschaften}
}

@article{huang2022quantum,
  title={Quantum advantage in learning from experiments},
  author={Huang, Hsin-Yuan and Broughton, Michael and Cotler, Jordan and Chen, Sitan and Li, Jerry and Mohseni, Masoud and Neven, Hartmut and Babbush, Ryan and Kueng, Richard and Preskill, John and others},
  journal={Science},
  volume={376},
  number={6598},
  pages={1182--1186},
  year={2022},
  publisher={American Association for the Advancement of Science}
}

@inproceedings{chen2022exponential,
  title={Exponential separations between learning with and without quantum memory},
  author={Chen, Sitan and Cotler, Jordan and Huang, Hsin-Yuan and Li, Jerry},
  booktitle={2021 IEEE 62nd Annual Symposium on Foundations of Computer Science (FOCS)},
  pages={574--585},
  year={2022},
  organization={IEEE}
}

@article{aharonov2022quantum,
  title={Quantum algorithmic measurement},
  author={Aharonov, Dorit and Cotler, Jordan and Qi, Xiao-Liang},
  journal={Nature Communications},
  volume={13},
  number={1},
  pages={887},
  year={2022},
  publisher={Nature Publishing Group UK London}
}

@article{oh2024entanglement,
  title={Entanglement-enabled advantage for learning a bosonic random displacement channel},
  author={Oh, Changhun and Chen, Senrui and Wong, Yat and Zhou, Sisi and Huang, Hsin-Yuan and Nielsen, Jens AH and Liu, Zheng-Hao and Neergaard-Nielsen, Jonas S and Andersen, Ulrik L and Jiang, Liang and others},
  journal={Physical Review Letters},
  volume={133},
  number={23},
  pages={230604},
  year={2024},
  publisher={APS}
}

@article{liu2025quantum,
  title={Quantum learning advantage on a scalable photonic platform},
  author={Liu, Zheng-Hao and Brunel, Romain and {\O}stergaard, Emil EB and Cordero, Oscar and Chen, Senrui and Wong, Yat and Nielsen, Jens AH and Bregnsbo, Axel B and Zhou, Sisi and Huang, Hsin-Yuan and others},
  journal={Science},
  volume={389},
  number={6767},
  pages={1332--1335},
  year={2025},
  publisher={American Association for the Advancement of Science}
}

@article{allen2025quantum,
  title={Quantum Computing Enhanced Sensing},
  author={Allen, Richard R and Machado, Francisco and Chuang, Isaac L and Huang, Hsin-Yuan and Choi, Soonwon},
  journal={arXiv preprint arXiv:2501.07625},
  year={2025}
}

@article{chen2024tight,
  title={Tight bounds on Pauli channel learning without entanglement},
  author={Chen, Senrui and Oh, Changhun and Zhou, Sisi and Huang, Hsin-Yuan and Jiang, Liang},
  journal={Physical Review Letters},
  volume={132},
  number={18},
  pages={180805},
  year={2024},
  publisher={APS}
}

@article{gao2022enhancing,
  title={Enhancing generative models via quantum correlations},
  author={Gao, Xun and Anschuetz, Eric R and Wang, Sheng-Tao and Cirac, J Ignacio and Lukin, Mikhail D},
  journal={Physical Review X},
  volume={12},
  number={2},
  pages={021037},
  year={2022},
  publisher={APS}
}

@article{anschuetz2026arbitrary,
  title={Arbitrary polynomial separations in trainable quantum machine learning},
  author={Anschuetz, Eric R and Gao, Xun},
  journal={Quantum},
  volume={10},
  pages={1976},
  year={2026},
  publisher={Verein zur F{\"o}rderung des Open Access Publizierens in den Quantenwissenschaften}
}

@article{anschuetz2023interpretable,
  title={Interpretable quantum advantage in neural sequence learning},
  author={Anschuetz, Eric R and Hu, Hong-Ye and Huang, Jin-Long and Gao, Xun},
  journal={PRX Quantum},
  volume={4},
  number={2},
  pages={020338},
  year={2023},
  publisher={APS}
}

@article{zhang2024quantum,
  title={Quantum-classical separations in shallow-circuit-based learning with and without noises},
  author={Zhang, Zhihan and Gong, Weiyuan and Li, Weikang and Deng, Dong-Ling},
  journal={Communications Physics},
  volume={7},
  number={1},
  pages={290},
  year={2024},
  publisher={Nature Publishing Group UK London}
}

@article{zhao2025entanglement,
  title={Entanglement-induced provable and robust quantum learning advantages},
  author={Zhao, Haimeng and Deng, Dong-Ling},
  journal={npj Quantum Information},
  volume={11},
  number={1},
  pages={127},
  year={2025},
  publisher={Nature Publishing Group UK London}
}

@article{gao2018quantum,
  title={A quantum machine learning algorithm based on generative models},
  author={Gao, Xun and Zhang, Z-Y and Duan, L-M},
  journal={Science advances},
  volume={4},
  number={12},
  pages={eaat9004},
  year={2018},
  publisher={American Association for the Advancement of Science}
}

@article{liu2021rigorous,
  title={A rigorous and robust quantum speed-up in supervised machine learning},
  author={Liu, Yunchao and Arunachalam, Srinivasan and Temme, Kristan},
  journal={Nature Physics},
  volume={17},
  number={9},
  pages={1013--1017},
  year={2021},
  publisher={Nature Publishing Group UK London}
}

@article{gyurik2023exponential,
  title={Exponential separations between classical and quantum learners},
  author={Gyurik, Casper and Dunjko, Vedran},
  journal={arXiv preprint arXiv:2306.16028},
  year={2023}
}

@article{huang2025generative,
  title={Generative quantum advantage for classical and quantum problems},
  author={Huang, Hsin-Yuan and Broughton, Michael and Eassa, Norhan and Neven, Hartmut and Babbush, Ryan and McClean, Jarrod R},
  journal={arXiv preprint arXiv:2509.09033},
  year={2025}
}

@article{cerezo2022challenges,
  title={Challenges and opportunities in quantum machine learning},
  author={Cerezo, Marco and Verdon, Guillaume and Huang, Hsin-Yuan and Cincio, Lukasz and Coles, Patrick J},
  journal={Nature Computational science},
  volume={2},
  number={9},
  pages={567--576},
  year={2022},
  publisher={Nature Publishing Group US New York}
}

@article{cerezo2021variational,
  title={Variational quantum algorithms},
  author={Cerezo, Marco and Arrasmith, Andrew and Babbush, Ryan and Benjamin, Simon C and Endo, Suguru and Fujii, Keisuke and McClean, Jarrod R and Mitarai, Kosuke and Yuan, Xiao and Cincio, Lukasz and others},
  journal={Nature Reviews Physics},
  volume={3},
  number={9},
  pages={625--644},
  year={2021},
  publisher={Nature Publishing Group UK London}
}

@article{du2025quantum,
  title={Quantum machine learning: A hands-on tutorial for machine learning practitioners and researchers},
  author={Du, Yuxuan and Wang, Xinbiao and Guo, Naixu and Yu, Zhan and Qian, Yang and Zhang, Kaining and Hsieh, Min-Hsiu and Rebentrost, Patrick and Tao, Dacheng},
  journal={arXiv preprint arXiv:2502.01146},
  year={2025}
}

@article{mcclean2018barren,
  title={Barren plateaus in quantum neural network training landscapes},
  author={McClean, Jarrod R and Boixo, Sergio and Smelyanskiy, Vadim N and Babbush, Ryan and Neven, Hartmut},
  journal={Nature Communications},
  volume={9},
  number={1},
  pages={4812},
  year={2018},
  publisher={Nature Publishing Group UK London}
}

@article{cerezo2021cost,
  title={Cost function dependent barren plateaus in shallow parametrized quantum circuits},
  author={Cerezo, Marco and Sone, Akira and Volkoff, Tyler and Cincio, Lukasz and Coles, Patrick J},
  journal={Nature Communications},
  volume={12},
  number={1},
  pages={1791},
  year={2021},
  publisher={Nature Publishing Group UK London}
}

@article{wang2021noise,
  title={Noise-induced barren plateaus in variational quantum algorithms},
  author={Wang, Samson and Fontana, Enrico and Cerezo, Marco and Sharma, Kunal and Sone, Akira and Cincio, Lukasz and Coles, Patrick J},
  journal={Nature Communications},
  volume={12},
  number={1},
  pages={6961},
  year={2021},
  publisher={Nature Publishing Group UK London}
}

@article{larocca2025barren,
  title={Barren plateaus in variational quantum computing},
  author={Larocca, Martin and Thanasilp, Supanut and Wang, Samson and Sharma, Kunal and Biamonte, Jacob and Coles, Patrick J and Cincio, Lukasz and McClean, Jarrod R and Holmes, Zo{\"e} and Cerezo, Marco},
  journal={Nature Reviews Physics},
  pages={1--16},
  year={2025},
  publisher={Nature Publishing Group UK London}
}

@article{anschuetz2022quantum,
  title={Quantum variational algorithms are swamped with traps},
  author={Anschuetz, Eric R and Kiani, Bobak T},
  journal={Nature Communications},
  volume={13},
  number={1},
  pages={7760},
  year={2022},
  publisher={Nature Publishing Group UK London}
}

@book{deeplearningbook,
    title={Deep Learning},
    author={Ian Goodfellow and Yoshua Bengio and Aaron Courville},
    publisher={MIT Press},
    note={\url{http://www.deeplearningbook.org}},
    year={2016}
}

@article{zhao2024learning,
  title={Learning quantum states and unitaries of bounded gate complexity},
  author={Zhao, Haimeng and Lewis, Laura and Kannan, Ishaan and Quek, Yihui and Huang, Hsin-Yuan and Caro, Matthias C},
  journal={PRX Quantum},
  volume={5},
  number={4},
  pages={040306},
  year={2024},
  publisher={APS}
}

@article{schuld2022quantum,
  title={Is quantum advantage the right goal for quantum machine learning?},
  author={Schuld, Maria and Killoran, Nathan},
  journal={PRX Quantum},
  volume={3},
  number={3},
  pages={030101},
  year={2022},
  publisher={APS}
}

@article{cerezo2023does,
  title={Does provable absence of barren plateaus imply classical simulability? Or, why we need to rethink variational quantum computing},
  author={Cerezo, Marco and Larocca, Martin and Garc{\'\i}a-Mart{\'\i}n, Diego and Diaz, Nelson L and Braccia, Paolo and Fontana, Enrico and Rudolph, Manuel S and Bermejo, Pablo and Ijaz, Aroosa and Thanasilp, Supanut and others},
  journal={arXiv preprint arXiv:2312.09121},
  year={2023}
}

@article{gil2024relation,
  title={On the relation between trainability and dequantization of variational quantum learning models},
  author={Gil-Fuster, Elies and Gyurik, Casper and P{\'e}rez-Salinas, Adri{\'a}n and Dunjko, Vedran},
  journal={arXiv preprint arXiv:2406.07072},
  year={2024}
}

@book{dalzell2025quantum,
    place={Cambridge}, 
    title={Quantum Algorithms: A Survey of Applications and End-to-end Complexities}, 
    publisher={Cambridge University Press}, 
    author={Dalzell, Alexander M. and McArdle, Sam and Berta, Mario and Bienias, Przemyslaw and Chen, Chi-Fang and Gilyén, András and Hann, Connor T. and Kastoryano, Michael J. and Khabiboulline, Emil T. and Kubica, Aleksander and et al.}, 
    year={2025}
}

@article{lloyd2018quantum,
  title={Quantum generative adversarial learning},
  author={Lloyd, Seth and Weedbrook, Christian},
  journal={Physical Review Letters},
  volume={121},
  number={4},
  pages={040502},
  year={2018},
  publisher={APS}
}

@article{leyton2008quantum,
  title={A quantum algorithm to solve nonlinear differential equations},
  author={Leyton, Sarah K and Osborne, Tobias J},
  journal={arXiv preprint arXiv:0812.4423},
  year={2008}
}

@article{montanaro2016quantum,
  title={Quantum algorithms and the finite element method},
  author={Montanaro, Ashley and Pallister, Sam},
  journal={Physical Review A},
  volume={93},
  number={3},
  pages={032324},
  year={2016},
  publisher={APS}
}

@article{clader2013preconditioned,
  title={Preconditioned quantum linear system algorithm},
  author={Clader, B David and Jacobs, Bryan C and Sprouse, Chad R},
  journal={arXiv preprint arXiv:1301.2340},
  year={2013}
}

@article{chakrabarti2020quantum,
  title={Quantum algorithms and lower bounds for convex optimization},
  author={Chakrabarti, Shouvanik and Childs, Andrew M and Li, Tongyang and Wu, Xiaodi},
  journal={Quantum},
  volume={4},
  pages={221},
  year={2020},
  publisher={Verein zur F{\"o}rderung des Open Access Publizierens in den Quantenwissenschaften}
}

@article{van2020convex,
  title={Convex optimization using quantum oracles},
  author={van Apeldoorn, Joran and Gily{\'e}n, Andr{\'a}s and Gribling, Sander and de Wolf, Ronald},
  journal={Quantum},
  volume={4},
  pages={220},
  year={2020},
  publisher={Verein zur F{\"o}rderung des Open Access Publizierens in den Quantenwissenschaften}
}

@article{jaques2025qram,
  title={{QRAM: A survey and critique}},
  author={Jaques, Samuel and Rattew, Arthur G},
  journal={Quantum},
  volume={9},
  pages={1922},
  year={2025},
  publisher={Verein zur F{\"o}rderung des Open Access Publizierens in den Quantenwissenschaften}
}

@article{giovannetti2008quantum,
  title={Quantum random access memory},
  author={Giovannetti, Vittorio and Lloyd, Seth and Maccone, Lorenzo},
  journal={Physical Review Letters},
  volume={100},
  number={16},
  pages={160501},
  year={2008},
  publisher={APS}
}

@article{dalzell2025distillation,
  title={{A distillation-teleportation protocol for fault-tolerant QRAM}},
  author={Dalzell, Alexander M and Gily{\'e}n, Andr{\'a}s and Hann, Connor T and McArdle, Sam and Salton, Grant and Nguyen, Quynh T and Kubica, Aleksander and Brand{\~a}o, Fernando GSL},
  journal={arXiv preprint arXiv:2505.20265},
  year={2025}
}

@article{babbush2018encoding,
  title={{Encoding electronic spectra in quantum circuits with linear T complexity}},
  author={Babbush, Ryan and Gidney, Craig and Berry, Dominic W and Wiebe, Nathan and McClean, Jarrod and Paler, Alexandru and Fowler, Austin and Neven, Hartmut},
  journal={Physical Review X},
  volume={8},
  number={4},
  pages={041015},
  year={2018},
  publisher={APS}
}

@book{tang2023quantum,
  title={Quantum machine learning without any quantum},
  author={Tang, Ewin},
  year={2023},
  publisher={University of Washington}
}

@inproceedings{tang2019quantum,
  title={A quantum-inspired classical algorithm for recommendation systems},
  author={Tang, Ewin},
  booktitle={Proceedings of the 51st annual ACM SIGACT symposium on theory of computing},
  pages={217--228},
  year={2019}
}

@article{tang2021quantum,
  title={Quantum principal component analysis only achieves an exponential speedup because of its state preparation assumptions},
  author={Tang, Ewin},
  journal={Physical Review Letters},
  volume={127},
  number={6},
  pages={060503},
  year={2021},
  publisher={APS}
}

@article{zhou2018achieving,
  title={Achieving the Heisenberg limit in quantum metrology using quantum error correction},
  author={Zhou, Sisi and Zhang, Mengzhen and Preskill, John and Jiang, Liang},
  journal={Nature Communications},
  volume={9},
  number={1},
  pages={78},
  year={2018},
  publisher={Nature Publishing Group UK London}
}

@article{huang2025vast,
  title={The vast world of quantum advantage},
  author={Huang, Hsin-Yuan and Choi, Soonwon and McClean, Jarrod R and Preskill, John},
  journal={arXiv preprint arXiv:2508.05720},
  year={2025}
}

@inproceedings{regev2008impossibility,
  title={Impossibility of a quantum speed-up with a faulty oracle},
  author={Regev, Oded and Schiff, Liron},
  booktitle={International Colloquium on Automata, Languages, and Programming},
  pages={773--781},
  year={2008},
  organization={Springer}
}

@article{chen2023complexity,
  title={The complexity of {NISQ}},
  author={Chen, Sitan and Cotler, Jordan and Huang, Hsin-Yuan and Li, Jerry},
  journal={Nature Communications},
  volume={14},
  number={1},
  pages={6001},
  year={2023},
  publisher={Nature Publishing Group UK London}
}

@inproceedings{grover1996fast,
  title={A fast quantum mechanical algorithm for database search},
  author={Grover, Lov K},
  booktitle={Proceedings of the twenty-eighth annual ACM symposium on theory of computing},
  pages={212--219},
  year={1996}
}

@article{holevo1973bounds,
  title={Bounds for the quantity of information transmitted by a quantum communication channel},
  author={Holevo, Alexander Semenovich},
  journal={Problemy Peredachi Informatsii},
  volume={9},
  number={3},
  pages={3--11},
  year={1973},
  publisher={Russian Academy of Sciences, Branch of Informatics, Computer Equipment and~…}
}

@article{babbush2025grand,
  title={The grand challenge of quantum applications},
  author={Babbush, Ryan and King, Robbie and Boixo, Sergio and Huggins, William and Khattar, Tanuj and Low, Guang Hao and McClean, Jarrod R and O'Brien, Thomas and Rubin, Nicholas C},
  journal={arXiv preprint arXiv:2511.09124},
  year={2025}
}

@inproceedings{shor1994algorithms,
  title={Algorithms for quantum computation: discrete logarithms and factoring},
  author={Shor, Peter W},
  booktitle={Proceedings 35th annual symposium on foundations of computer science},
  pages={124--134},
  year={1994},
  organization={Ieee}
}

@article{kitaev1995quantum,
  title={Quantum measurements and the Abelian stabilizer problem},
  author={Kitaev, Alexei},
  journal={arXiv preprint quant-ph/9511026},
  year={1995}
}

@article{feynman1986quantum,
  title={Quantum mechanical computers},
  author={Feynman, Richard P},
  journal={Foundations of physics},
  volume={16},
  number={6},
  pages={507--531},
  year={1986},
  publisher={Kluwer Academic Publishers-Plenum Publishers New York}
}

@article{bravyi2018quantum,
  title={Quantum advantage with shallow circuits},
  author={Bravyi, Sergey and Gosset, David and K{\"o}nig, Robert},
  journal={Science},
  volume={362},
  number={6412},
  pages={308--311},
  year={2018},
  publisher={American Association for the Advancement of Science}
}

@article{bravyi2020quantum,
  title={Quantum advantage with noisy shallow circuits},
  author={Bravyi, Sergey and Gosset, David and K{\"o}nig, Robert and Tomamichel, Marco},
  journal={Nature Physics},
  volume={16},
  number={10},
  pages={1040--1045},
  year={2020},
  publisher={Nature Publishing Group UK London}
}

@article{bell1966problem,
  title={On the problem of hidden variables in quantum mechanics},
  author={Bell, John S},
  journal={Reviews of Modern physics},
  volume={38},
  number={3},
  pages={447},
  year={1966},
  publisher={APS}
}

@inproceedings{alon1996space,
  title={The space complexity of approximating the frequency moments},
  author={Alon, Noga and Matias, Yossi and Szegedy, Mario},
  booktitle={Proceedings of the twenty-eighth annual ACM symposium on theory of computing},
  pages={20--29},
  year={1996}
}

@article{flajolet1985probabilistic,
  title={Probabilistic counting algorithms for data base applications},
  author={Flajolet, Philippe and Martin, G Nigel},
  journal={Journal of computer and system sciences},
  volume={31},
  number={2},
  pages={182--209},
  year={1985},
  publisher={Elsevier}
}

@article{woodruff2014sketching,
  title={Sketching as a tool for numerical linear algebra},
  author={Woodruff, David P and others},
  journal={Foundations and Trends{\textregistered} in Theoretical Computer Science},
  volume={10},
  number={1--2},
  pages={1--157},
  year={2014},
  publisher={Now Publishers, Inc.}
}

@article{muthukrishnan2005data,
  title={Data streams: Algorithms and applications},
  author={Muthukrishnan, Shanmugavelayutham and others},
  journal={Foundations and Trends{\textregistered} in Theoretical Computer Science},
  volume={1},
  number={2},
  pages={117--236},
  year={2005},
  publisher={Now Publishers, Inc.}
}

@article{misra1982finding,
  title={Finding repeated elements},
  author={Misra, Jayadev and Gries, David},
  journal={Science of computer programming},
  volume={2},
  number={2},
  pages={143--152},
  year={1982},
  publisher={Elsevier}
}

@article{cormode2005improved,
  title={An improved data stream summary: the count-min sketch and its applications},
  author={Cormode, Graham and Muthukrishnan, Shan},
  journal={Journal of Algorithms},
  volume={55},
  number={1},
  pages={58--75},
  year={2005},
  publisher={Elsevier}
}

@inproceedings{clarkson2009numerical,
  title={Numerical linear algebra in the streaming model},
  author={Clarkson, Kenneth L and Woodruff, David P},
  booktitle={Proceedings of the forty-first annual ACM symposium on theory of computing},
  pages={205--214},
  year={2009}
}

@article{andoni2020streaming,
  title={Streaming complexity of svms},
  author={Andoni, Alexandr and Burns, Collin and Li, Yi and Mahabadi, Sepideh and Woodruff, David P},
  journal={arXiv preprint arXiv:2007.03633},
  year={2020}
}

@article{mitliagkas2013memory,
  title={Memory limited, streaming PCA},
  author={Mitliagkas, Ioannis and Caramanis, Constantine and Jain, Prateek},
  journal={Advances in neural information processing systems},
  volume={26},
  year={2013}
}

@inproceedings{braverman2024new,
  title={A New Information Complexity Measure for Multi-pass Streaming with Applications},
  author={Braverman, Mark and Garg, Sumegha and Li, Qian and Wang, Shuo and Woodruff, David P and Zhang, Jiapeng},
  booktitle={Proceedings of the 56th Annual ACM Symposium on Theory of Computing},
  pages={1781--1792},
  year={2024}
}

@inproceedings{lovett2023streaming,
  title={Streaming Lower Bounds and Asymmetric Set-Disjointness},
  author={Lovett, Shachar and Zhang, Jiapeng},
  booktitle={2023 IEEE 64th Annual Symposium on Foundations of Computer Science (FOCS)},
  pages={871--882},
  year={2023},
  organization={IEEE}
}

@inproceedings{dershowitz2021communication,
  title={The communication complexity of multiparty set disjointness under product distributions},
  author={Dershowitz, Nachum and Oshman, Rotem and Roth, Tal},
  booktitle={Proceedings of the 53rd Annual ACM SIGACT Symposium on Theory of Computing},
  pages={1194--1207},
  year={2021}
}

@article{bar2004information,
  title={An information statistics approach to data stream and communication complexity},
  author={Bar-Yossef, Ziv and Jayram, Thathachar S and Kumar, Ravi and Sivakumar, D},
  journal={Journal of Computer and System Sciences},
  volume={68},
  number={4},
  pages={702--732},
  year={2004},
  publisher={Elsevier}
}

@article{anshu2020query,
  title={On query-to-communication lifting for adversary bounds},
  author={Anshu, Anurag and Ben-David, Shalev and Kundu, Srijita},
  journal={arXiv preprint arXiv:2012.03415},
  year={2020}
}

@inproceedings{goos2015deterministic,
  title={Deterministic communication vs. partition number},
  author={G{\"o}{\"o}s, Mika and Pitassi, Toniann and Watson, Thomas},
  booktitle={2015 IEEE 56th Annual Symposium on Foundations of Computer Science},
  pages={1077--1088},
  year={2015},
  organization={IEEE}
}

@article{goos2016rectangles,
  title={Rectangles are nonnegative juntas},
  author={Goos, Mika and Lovett, Shachar and Meka, Raghu and Watson, Thomas and Zuckerman, David},
  journal={SIAM Journal on Computing},
  volume={45},
  number={5},
  pages={1835--1869},
  year={2016},
  publisher={SIAM}
}

@article{goos2020query,
  title={Query-to-communication lifting for BPP},
  author={Goos, Mika and Pitassi, Toniann and Watson, Thomas},
  journal={SIAM Journal on Computing},
  volume={49},
  number={4},
  pages={FOCS17--441},
  year={2020},
  publisher={SIAM}
}

@inproceedings{yang2024communication,
  title={Communication lower bounds for collision problems via density increment arguments},
  author={Yang, Guangxu and Zhang, Jiapeng},
  booktitle={Proceedings of the 56th Annual ACM Symposium on Theory of Computing},
  pages={630--639},
  year={2024}
}

@article{fei2025multi,
  title={Multi-Pass Streaming Lower Bounds for Approximating Max-Cut},
  author={Fei, Yumou and Minzer, Dor and Wang, Shuo},
  journal={arXiv preprint arXiv:2503.23404},
  year={2025}
}

@article{raz2018fast,
  title={Fast learning requires good memory: A time-space lower bound for parity learning},
  author={Raz, Ran},
  journal={Journal of the ACM (JACM)},
  volume={66},
  number={1},
  pages={1--18},
  year={2018},
  publisher={ACM New York, NY, USA}
}

@inproceedings{raz2017time,
  title={A time-space lower bound for a large class of learning problems},
  author={Raz, Ran},
  booktitle={2017 IEEE 58th Annual Symposium on Foundations of Computer Science (FOCS)},
  pages={732--742},
  year={2017},
  organization={IEEE}
}

@inproceedings{dinur2024time,
  title={Time-space lower bounds for bounded-error computation in the random-query model},
  author={Dinur, Itai},
  booktitle={Proceedings of the 2024 Annual ACM-SIAM Symposium on Discrete Algorithms (SODA)},
  pages={2900--2915},
  year={2024},
  organization={SIAM}
}

@article{garg2021memory,
  title={Memory-sample lower bounds for learning parity with noise},
  author={Garg, Sumegha and Kothari, Pravesh K and Liu, Pengda and Raz, Ran},
  journal={arXiv preprint arXiv:2107.02320},
  year={2021}
}

@inproceedings{liu2023memory,
  title={Memory-sample lower bounds for learning with classical-quantum hybrid memory},
  author={Liu, Qipeng and Raz, Ran and Zhan, Wei},
  booktitle={Proceedings of the 55th Annual ACM Symposium on Theory of Computing},
  pages={1097--1110},
  year={2023}
}

@phdthesis{zhan2023randomness,
  title={Randomness and quantumness in space-bounded computation},
  author={Zhan, Wei},
  year={2023},
  school={Princeton University}
}

@inproceedings{assadi2021graph,
  title={Graph streaming lower bounds for parameter estimation and property testing via a streaming XOR lemma},
  author={Assadi, Sepehr and N, Vishvajeet},
  booktitle={Proceedings of the 53rd Annual ACM SIGACT Symposium on Theory of Computing},
  pages={612--625},
  year={2021}
}

@inproceedings{yao1982theory,
  title={Theory and application of trapdoor functions},
  author={Yao, Andrew C},
  booktitle={Proceedings of the 23rd Annual Symposium on Foundations of Computer Science},
  pages={80--91},
  year={1982}
}

@article{bera2023generalized,
  title={A Generalized Quantum Branching Program},
  author={Bera, Debajyoti and Sapv, Tharrmashastha},
  journal={arXiv preprint arXiv:2307.11395},
  year={2023}
}

@inproceedings{ablayev2001computational,
  title={On computational power of quantum branching programs},
  author={Ablayev, Farid and Gainutdinova, Aida and Karpinski, Marek},
  booktitle={International Symposium on Fundamentals of Computation Theory},
  pages={59--70},
  year={2001},
  organization={Springer}
}

@inproceedings{nakanishi2000ordered,
  title={Ordered quantum branching programs are more powerful than ordered probabilistic branching programs under a bounded-width restriction},
  author={Nakanishi, Masaki and Hamaguchi, Kiyoharu and Kashiwabara, Toshinobu},
  booktitle={International Computing and Combinatorics Conference},
  pages={467--476},
  year={2000},
  organization={Springer}
}

@inproceedings{kallaugher2024exponential,
  title={Exponential quantum space advantage for approximating maximum directed cut in the streaming model},
  author={Kallaugher, John and Parekh, Ojas and Voronova, Nadezhda},
  booktitle={Proceedings of the 56th Annual ACM Symposium on Theory of Computing},
  pages={1805--1815},
  year={2024}
}

@inproceedings{kallaugher2025design,
  title={How to Design a Quantum Streaming Algorithm Without Knowing Anything About Quantum Computing},
  author={Kallaugher, John and Parekh, Ojas and Voronova, Nadezhda},
  booktitle={2025 Symposium on Simplicity in Algorithms (SOSA)},
  pages={9--45},
  year={2025},
  organization={SIAM}
}

@inproceedings{gavinsky2007exponential,
  title={Exponential separations for one-way quantum communication complexity, with applications to cryptography},
  author={Gavinsky, Dmitry and Kempe, Julia and Kerenidis, Iordanis and Raz, Ran and de Wolf, Ronald},
  booktitle={Proceedings of the thirty-ninth annual ACM symposium on theory of computing},
  pages={516--525},
  year={2007}
}

@inproceedings{kallaugher2022quantum,
  title={A quantum advantage for a natural streaming problem},
  author={Kallaugher, John},
  booktitle={2021 IEEE 62nd Annual Symposium on Foundations of Computer Science (FOCS)},
  pages={897--908},
  year={2022},
  organization={IEEE}
}

@inproceedings{le2006exponential,
  title={Exponential separation of quantum and classical online space complexity},
  author={Le Gall, Fran{\c{c}}ois},
  booktitle={Proceedings of the eighteenth annual ACM symposium on Parallelism in algorithms and architectures},
  pages={67--73},
  year={2006}
}

@article{jain2014space,
  title={The space complexity of recognizing well-parenthesized expressions in the streaming model: the index function revisited},
  author={Jain, Rahul and Nayak, Ashwin},
  journal={IEEE Transactions on Information Theory},
  volume={60},
  number={10},
  pages={6646--6668},
  year={2014},
  publisher={IEEE}
}

@inproceedings{raz1999exponential,
  title={Exponential separation of quantum and classical communication complexity},
  author={Raz, Ran},
  booktitle={Proceedings of the thirty-first annual ACM symposium on theory of computing},
  pages={358--367},
  year={1999}
}

@article{gilboa2024exponential,
  title={Exponential quantum communication advantage in distributed inference and learning},
  author={Gilboa, Dar and Michaeli, Hagay and Soudry, Daniel and McClean, Jarrod},
  journal={Advances in Neural Information Processing Systems},
  volume={37},
  pages={30425--30473},
  year={2024}
}

@article{kretschmer2025demonstrating,
  title={Demonstrating an unconditional separation between quantum and classical information resources},
  author={Kretschmer, William and Grewal, Sabee and DeCross, Matthew and Gerber, Justin A and Gilmore, Kevin and Gresh, Dan and Hunter-Jones, Nicholas and Mayer, Karl and Neyenhuis, Brian and Hayes, David and others},
  journal={arXiv preprint arXiv:2509.07255},
  year={2025}
}

@article{niroula2025realization,
  title={Realization of a Quantum Streaming Algorithm on Long-lived Trapped-ion Qubits},
  author={Niroula, Pradeep and Chakrabarti, Shouvanik and Kordonowy, Steven and Kumar, Niraj and Omanakuttan, Sivaprasad and Perlin, Michael A and Allman, MS and Campora III, JP and Chernoguzov, Alex and Cooper, Samuel F and others},
  journal={arXiv preprint arXiv:2511.03689},
  year={2025}
}

@inproceedings{kerenidis2017quantum,
  title={Quantum Recommendation Systems},
  author={Kerenidis, Iordanis and Prakash, Anupam},
  booktitle={8th Innovations in Theoretical Computer Science Conference (ITCS 2017)},
  pages={49--1},
  year={2017},
  organization={Schloss Dagstuhl--Leibniz-Zentrum f{\"u}r Informatik}
}

@misc{jax2018github,
  author = {James Bradbury and Roy Frostig and Peter Hawkins and Matthew James Johnson and Chris Leary and Dougal Maclaurin and George Necula and Adam Paszke and Jake Vander{P}las and Skye Wanderman-{M}ilne and Qiao Zhang},
  title = {{JAX}: composable transformations of {P}ython+{N}um{P}y programs},
  url = {http://github.com/jax-ml/jax},
  version = {0.3.13},
  year = {2018},
}

@article{biamonte2017quantum,
  title={Quantum machine learning},
  author={Biamonte, Jacob and Wittek, Peter and Pancotti, Nicola and Rebentrost, Patrick and Wiebe, Nathan and Lloyd, Seth},
  journal={Nature},
  volume={549},
  number={7671},
  pages={195--202},
  year={2017},
  publisher={Nature Publishing Group UK London}
}

@inproceedings{aaronson2024qubit,
  title={A Qubit, a Coin, and an Advice String Walk into a Relational Problem},
  author={Aaronson, Scott and Buhrman, Harry and Kretschmer, William},
  booktitle={15th Innovations in Theoretical Computer Science Conference (ITCS 2024)},
  pages={1--1},
  year={2024},
  organization={Schloss Dagstuhl--Leibniz-Zentrum f{\"u}r Informatik}
}

@article{shalev2025online,
  title={Online learning and online convex optimization},
  author={Shalev-Shwartz, Shai},
  journal={Foundations and Trends{\textregistered} in Machine Learning},
  volume={4},
  number={2},
  pages={107--194},
  year={2025},
  publisher={Emerald Publishing Limited Boston—Delft}
}

@article{zheng2017massively,
  title={Massively parallel digital transcriptional profiling of single cells},
  author={Zheng, Grace XY and Terry, Jessica M and Belgrader, Phillip and Ryvkin, Paul and Bent, Zachary W and Wilson, Ryan and Ziraldo, Solongo B and Wheeler, Tobias D and McDermott, Geoff P and Zhu, Junjie and others},
  journal={Nature Communications},
  volume={8},
  number={1},
  pages={14049},
  year={2017},
  publisher={Nature Publishing Group UK London}
}

@inproceedings{maas2011learning,
  title={Learning word vectors for sentiment analysis},
  author={Maas, Andrew and Daly, Raymond E and Pham, Peter T and Huang, Dan and Ng, Andrew Y and Potts, Christopher},
  booktitle={Proceedings of the 49th annual meeting of the association for computational linguistics: Human language technologies},
  pages={142--150},
  year={2011}
}

@article{lloyd1996universal,
  title={Universal quantum simulators},
  author={Lloyd, Seth},
  journal={Science},
  volume={273},
  number={5278},
  pages={1073--1078},
  year={1996},
  publisher={American Association for the Advancement of Science}
}

@article{preskill2025beyond,
  title={Beyond {NISQ}: The megaquop machine},
  author={Preskill, John},
  journal={ACM Transactions on Quantum Computing},
  volume={6},
  number={3},
  pages={1--7},
  year={2025},
  publisher={ACM New York, NY}
}

@article{arunachalam2015robustness,
  title={On the robustness of bucket brigade quantum RAM},
  author={Arunachalam, Srinivasan and Gheorghiu, Vlad and Jochym-O’Connor, Tomas and Mosca, Michele and Srinivasan, Priyaa Varshinee},
  journal={New Journal of Physics},
  volume={17},
  number={12},
  pages={123010},
  year={2015},
  publisher={IOP Publishing}
}

@article{wulf1995hitting,
  title={Hitting the memory wall: Implications of the obvious},
  author={Wulf, Wm A and McKee, Sally A},
  journal={ACM SIGARCH computer architecture news},
  volume={23},
  number={1},
  pages={20--24},
  year={1995},
  publisher={ACM New York, NY, USA}
}

@article{gholami2024ai,
  title={Ai and memory wall},
  author={Gholami, Amir and Yao, Zhewei and Kim, Sehoon and Hooper, Coleman and Mahoney, Michael W and Keutzer, Kurt},
  journal={IEEE Micro},
  volume={44},
  number={3},
  pages={33--39},
  year={2024},
  publisher={IEEE}
}

@article{fedus2022switch,
  title={Switch transformers: Scaling to trillion parameter models with simple and efficient sparsity},
  author={Fedus, William and Zoph, Barret and Shazeer, Noam},
  journal={Journal of Machine Learning Research},
  volume={23},
  number={120},
  pages={1--39},
  year={2022}
}

@article{kaplan2020scaling,
  title={Scaling laws for neural language models},
  author={Kaplan, Jared and McCandlish, Sam and Henighan, Tom and Brown, Tom B and Chess, Benjamin and Child, Rewon and Gray, Scott and Radford, Alec and Wu, Jeffrey and Amodei, Dario},
  journal={arXiv preprint arXiv:2001.08361},
  year={2020}
}

@article{dao2022flashattention,
  title={{FlashAttention: Fast and memory-efficient exact attention with IO-awareness}},
  author={Dao, Tri and Fu, Dan and Ermon, Stefano and Rudra, Atri and R{\'e}, Christopher},
  journal={Advances in neural information processing systems},
  volume={35},
  pages={16344--16359},
  year={2022}
}

@article{dettmers2023qlora,
  title={{QLoRA: Efficient finetuning of quantized LLMs}},
  author={Dettmers, Tim and Pagnoni, Artidoro and Holtzman, Ari and Zettlemoyer, Luke},
  journal={Advances in neural information processing systems},
  volume={36},
  pages={10088--10115},
  year={2023}
}

@article{dettmers2022gpt3,
  title={{GPT3.int8(): 8-bit matrix multiplication for transformers at scale}},
  author={Dettmers, Tim and Lewis, Mike and Belkada, Younes and Zettlemoyer, Luke},
  journal={Advances in neural information processing systems},
  volume={35},
  pages={30318--30332},
  year={2022}
}

@inproceedings{rajbhandari2020zero,
  title={Zero: Memory optimizations toward training trillion parameter models},
  author={Rajbhandari, Samyam and Rasley, Jeff and Ruwase, Olatunji and He, Yuxiong},
  booktitle={SC20: International Conference for High Performance Computing, Networking, Storage and Analysis},
  pages={1--16},
  year={2020},
  organization={IEEE}
}

@inproceedings{kwon2023efficient,
  title={Efficient memory management for large language model serving with pagedattention},
  author={Kwon, Woosuk and Li, Zhuohan and Zhuang, Siyuan and Sheng, Ying and Zheng, Lianmin and Yu, Cody Hao and Gonzalez, Joseph and Zhang, Hao and Stoica, Ion},
  booktitle={Proceedings of the 29th symposium on operating systems principles},
  pages={611--626},
  year={2023}
}

@inproceedings{ainslie2023gqa,
  title={{GQA: Training Generalized Multi-Query Transformer Models from Multi-Head Checkpoints}},
  author={Ainslie, Joshua and Lee-Thorp, James and de Jong, Michiel and Zemlyanskiy, Yury and Lebron, Federico and Sanghai, Sumit},
  booktitle={Proceedings of the 2023 Conference on Empirical Methods in Natural Language Processing},
  pages={4895--4901},
  year={2023}
}

@article{jain2020checkmate,
  title={Checkmate: Breaking the memory wall with optimal tensor rematerialization},
  author={Jain, Paras and Jain, Ajay and Nrusimha, Aniruddha and Gholami, Amir and Abbeel, Pieter and Gonzalez, Joseph and Keutzer, Kurt and Stoica, Ion},
  journal={Proceedings of Machine Learning and Systems},
  volume={2},
  pages={497--511},
  year={2020}
}

@article{hep2019roadmap,
  title={A Roadmap for HEP Software and Computing R\&D for the 2020s},
  author={HEP Software Foundation and Albrecht, Johannes and Alves Jr, Antonio Augusto and Amadio, Guilherme and Andronico, Giuseppe and Anh-Ky, Nguyen and Aphecetche, Laurent and Apostolakis, John and Asai, Makoto and Atzori, Luca and others},
  journal={Computing and software for big science},
  volume={3},
  number={1},
  pages={7},
  year={2019},
  publisher={Springer}
}

@article{dewdney2009square,
  title={The square kilometre array},
  author={Dewdney, Peter E and Hall, Peter J and Schilizzi, Richard T and Lazio, T Joseph LW},
  journal={Proceedings of the IEEE},
  volume={97},
  number={8},
  pages={1482--1496},
  year={2009},
  publisher={IEEE}
}

@article{svensson2018exponential,
  title={Exponential scaling of single-cell RNA-seq in the past decade},
  author={Svensson, Valentine and Vento-Tormo, Roser and Teichmann, Sarah A},
  journal={Nature Protocols},
  volume={13},
  number={4},
  pages={599--604},
  year={2018},
  publisher={Nature Publishing Group UK London}
}

@article{lahnemann2020eleven,
  title={Eleven grand challenges in single-cell data science},
  author={L{\"a}hnemann, David and K{\"o}ster, Johannes and Szczurek, Ewa and McCarthy, Davis J and Hicks, Stephanie C and Robinson, Mark D and Vallejos, Catalina A and Campbell, Kieran R and Beerenwinkel, Niko and Mahfouz, Ahmed and others},
  journal={Genome biology},
  volume={21},
  number={1},
  pages={31},
  year={2020},
  publisher={Springer}
}

@article{tsuyuzaki2020benchmarking,
  title={Benchmarking principal component analysis for large-scale single-cell RNA-sequencing},
  author={Tsuyuzaki, Koki and Sato, Hiroyuki and Sato, Kenta and Nikaido, Itoshi},
  journal={Genome biology},
  volume={21},
  number={1},
  pages={9},
  year={2020},
  publisher={Springer}
}

@inproceedings{perlman2007data,
  title={Data exploration of turbulence simulations using a database cluster},
  author={Perlman, Eric and Burns, Randal and Li, Yi and Meneveau, Charles},
  booktitle={Proceedings of the 2007 ACM/IEEE Conference on Supercomputing},
  pages={1--11},
  year={2007}
}

@inproceedings{feng2008multigrid,
  title={{Multigrid on GPU: Tackling power grid analysis on parallel SIMT platforms}},
  author={Feng, Zhuo and Li, Peng},
  booktitle={2008 IEEE/ACM International Conference on Computer-Aided Design},
  pages={647--654},
  year={2008},
  organization={IEEE}
}

@article{kozhaya2002multigrid,
  title={A multigrid-like technique for power grid analysis},
  author={Kozhaya, Joseph N and Nassif, Sani R and Najm, Farid N},
  journal={IEEE Transactions on Computer-Aided Design of Integrated Circuits and Systems},
  volume={21},
  number={10},
  pages={1148--1160},
  year={2002},
  publisher={IEEE}
}

@techreport{bejaralonso2020hllhc,
  title       = {{High-Luminosity Large Hadron Collider (HL-LHC): Technical design report}},
  author      = {B\'{e}jar Alonso, I. and others (Eds.)},
  institution = {CERN},
  year        = {2020},
  number      = {CERN-2020-010},
  address     = {Geneva},
  series      = {CERN Yellow Reports: Monographs},
  doi         = {10.23731/CYRM-2020-0010},
  isbn        = {978-92-9083-556-5}
}

@article{kirkpatrick2017overcoming,
  title={Overcoming catastrophic forgetting in neural networks},
  author={Kirkpatrick, James and Pascanu, Razvan and Rabinowitz, Neil and Veness, Joel and Desjardins, Guillaume and Rusu, Andrei A and Milan, Kieran and Quan, John and Ramalho, Tiago and Grabska-Barwinska, Agnieszka and others},
  journal={Proceedings of the national academy of sciences},
  volume={114},
  number={13},
  pages={3521--3526},
  year={2017},
  publisher={National Academy of Sciences}
}

@article{hemati2014dynamic,
  title={Dynamic mode decomposition for large and streaming datasets},
  author={Hemati, Maziar S and Williams, Matthew O and Rowley, Clarence W},
  journal={Physics of Fluids},
  volume={26},
  number={11},
  year={2014},
  publisher={AIP Publishing}
}

@article{khachatryan2016search,
  title={Search for narrow resonances in dijet final states at s= 8 TeV with the novel CMS technique of data scouting},
  author={Khachatryan, Vardan and Sirunyan, Albert M and Tumasyan, Armen and Adam, Wolfgang and Asilar, E and Bergauer, Thomas and Brandstetter, Johannes and Brondolin, Erica and Dragicevic, Marko and Er{\"o}, Janos and others},
  journal={Physical Review Letters},
  volume={117},
  number={3},
  pages={031802},
  year={2016},
  publisher={APS}
}

@article{clauser1969proposed,
  title={Proposed experiment to test local hidden-variable theories},
  author={Clauser, John F and Horne, Michael A and Shimony, Abner and Holt, Richard A},
  journal={Physical Review Letters},
  volume={23},
  number={15},
  pages={880},
  year={1969},
  publisher={APS}
}

@article{clauser1978bell,
  title={Bell's theorem. Experimental tests and implications},
  author={Clauser, John F and Shimony, Abner},
  journal={Reports on Progress in Physics},
  volume={41},
  number={12},
  pages={1881},
  year={1978},
  publisher={IOP Publishing}
}

@article{watrous1999space,
  title={Space-bounded quantum complexity},
  author={Watrous, John},
  journal={Journal of Computer and System Sciences},
  volume={59},
  number={2},
  pages={281--326},
  year={1999},
  publisher={Elsevier}
}

@article{watrous2003complexity,
  title={On the complexity of simulating space-bounded quantum computations},
  author={Watrous, John},
  journal={computational complexity},
  volume={12},
  number={1},
  pages={48--84},
  year={2003},
  publisher={Springer}
}

@article{luccioni2023estimating,
  title={Estimating the carbon footprint of bloom, a 176b parameter language model},
  author={Luccioni, Alexandra Sasha and Viguier, Sylvain and Ligozat, Anne-Laure},
  journal={Journal of machine learning research},
  volume={24},
  number={253},
  pages={1--15},
  year={2023}
}

@inproceedings{strubell2019energy,
  title={Energy and policy considerations for deep learning in NLP},
  author={Strubell, Emma and Ganesh, Ananya and McCallum, Andrew},
  booktitle={Proceedings of the 57th annual meeting of the association for computational linguistics},
  pages={3645--3650},
  year={2019}
}

@article{zhou2025opportunities,
  title={Opportunities in full-stack design of low-overhead fault-tolerant quantum computation},
  author={Zhou, Hengyun and Cain, Madelyn and Lukin, Mikhail D},
  journal={Nature Computational Science},
  volume={5},
  number={12},
  pages={1110--1119},
  year={2025},
  publisher={Nature Publishing Group}
}

@article{preskill2012quantum,
  title={Quantum computing and the entanglement frontier},
  author={Preskill, John},
  journal={arXiv preprint arXiv:1203.5813},
  year={2012}
}

@techreport{joachims1996probabilistic,
  title={A Probabilistic Analysis of the Rocchio Algorithm with TFIDF for Text Categorization.},
  author={Joachims, Thorsten},
  institution={Carnegie Mellon University},
  year={1996}
}

@article{weston2003feature,
  title={Feature selection and transduction for prediction of molecular bioactivity for drug design},
  author={Weston, Jason and Perez-Cruz, Fernando and Bousquet, Olivier and Chapelle, Olivier and Elisseeff, Andre and Sch{\"o}lkopf, Bernhard},
  journal={Bioinformatics},
  volume={19},
  number={6},
  pages={764--771},
  year={2003},
  publisher={Oxford University Press}
}

@article{poulin2011quantum,
  title={Quantum simulation of time-dependent Hamiltonians and the convenient illusion of Hilbert space},
  author={Poulin, David and Qarry, Angie and Somma, Rolando and Verstraete, Frank},
  journal={Physical Review Letters},
  volume={106},
  number={17},
  pages={170501},
  year={2011},
  publisher={APS}
}

@article{Bergen2020,
    title = {Generalizing RNA velocity to transient cell states through dynamical modeling},
    volume = {38},
    ISSN = {1546-1696},
    url = {http://dx.doi.org/10.1038/s41587-020-0591-3},
    DOI = {10.1038/s41587-020-0591-3},
    number = {12},
    journal = {Nature Biotechnology},
    publisher = {Springer Science and Business Media LLC},
    author = {Bergen, Volker and Lange, Marius and Peidli, Stefan and Wolf, F. Alexander and Theis, Fabian J.},
    year = {2020},
    month = aug,
    pages = {1408-1414}
}

@article{scikit-learn,
  title={Scikit-learn: Machine Learning in {P}ython},
  author={Pedregosa, F. and Varoquaux, G. and Gramfort, A. and Michel, V.
          and Thirion, B. and Grisel, O. and Blondel, M. and Prettenhofer, P.
          and Weiss, R. and Dubourg, V. and Vanderplas, J. and Passos, A. and
          Cournapeau, D. and Brucher, M. and Perrot, M. and Duchesnay, E.},
  journal={Journal of Machine Learning Research},
  volume={12},
  pages={2825--2830},
  year={2011}
}

@misc{dorothea_169,
  author       = {Guyon, Isabelle and Gunn, Steve and Ben-Hur, Asa and Dror, Gideon},
  title        = {{Dorothea}},
  year         = {2004},
  howpublished = {UCI Machine Learning Repository},
  note         = {{DOI}: https://doi.org/10.24432/C5NK6X}
}

@article{dong2021efficient,
  title={Efficient phase-factor evaluation in quantum signal processing},
  author={Dong, Yulong and Meng, Xiang and Whaley, K Birgitta and Lin, Lin},
  journal={Physical Review A},
  volume={103},
  number={4},
  pages={042419},
  year={2021},
  publisher={APS}
}

@article{peruzzo2014variational,
  title={A variational eigenvalue solver on a photonic quantum processor},
  author={Peruzzo, Alberto and McClean, Jarrod and Shadbolt, Peter and Yung, Man-Hong and Zhou, Xiao-Qi and Love, Peter J and Aspuru-Guzik, Al{\'a}n and O’brien, Jeremy L},
  journal={Nature Communications},
  volume={5},
  number={1},
  pages={4213},
  year={2014},
  publisher={Nature Publishing Group UK London}
}

@article{mcclean2016theory,
  title={The theory of variational hybrid quantum-classical algorithms},
  author={McClean, Jarrod R and Romero, Jonathan and Babbush, Ryan and Aspuru-Guzik, Al{\'a}n},
  journal={New Journal of Physics},
  volume={18},
  number={2},
  pages={023023},
  year={2016},
  publisher={IOP Publishing}
}

@inproceedings{unger2009probabilistic,
  title={A probabilistic inequality with applications to threshold direct-product theorems},
  author={Unger, Falk},
  booktitle={2009 50th Annual IEEE Symposium on Foundations of Computer Science},
  pages={221--229},
  year={2009},
  organization={IEEE}
}

@article{hann2021resilience,
  title={Resilience of quantum random access memory to generic noise},
  author={Hann, Connor T and Lee, Gideon and Girvin, SM and Jiang, Liang},
  journal={PRX Quantum},
  volume={2},
  number={2},
  pages={020311},
  year={2021},
  publisher={APS}
}

@article{harrow2020small,
  title={Small quantum computers and large classical data sets},
  author={Harrow, Aram W},
  journal={arXiv preprint arXiv:2004.00026},
  year={2020}
}

@article{aaronson2009need,
  title={The need for structure in quantum speedups},
  author={Aaronson, Scott and Ambainis, Andris},
  journal={arXiv preprint arXiv:0911.0996},
  year={2009}
}

@article{ashhab2022quantum,
  title={Quantum state preparation protocol for encoding classical data into the amplitudes of a quantum information processing register's wave function},
  author={Ashhab, Sahel},
  journal={Physical Review Research},
  volume={4},
  number={1},
  pages={013091},
  year={2022},
  publisher={APS}
}

@article{wang2023variational,
  title={Variational quantum regression algorithm with encoded data structure},
  author={Wang, C-C Joseph and Bennink, Ryan S},
  journal={arXiv preprint arXiv:2307.03334},
  year={2023}
}

@article{jerbi2024shadows,
  title={Shadows of quantum machine learning},
  author={Jerbi, Sofiene and Gyurik, Casper and Marshall, Simon C and Molteni, Riccardo and Dunjko, Vedran},
  journal={Nature Communications},
  volume={15},
  number={1},
  pages={5676},
  year={2024},
  publisher={Nature Publishing Group UK London}
}

@article{schreiber2023classical,
  title={Classical surrogates for quantum learning models},
  author={Schreiber, Franz J and Eisert, Jens and Meyer, Johannes Jakob},
  journal={Physical Review Letters},
  volume={131},
  number={10},
  pages={100803},
  year={2023},
  publisher={APS}
}

@article{kane2014sparser,
  title={Sparser johnson-lindenstrauss transforms},
  author={Kane, Daniel M and Nelson, Jelani},
  journal={Journal of the ACM (JACM)},
  volume={61},
  number={1},
  pages={1--23},
  year={2014},
  publisher={ACM New York, NY, USA}
}

@article{guo2024quantum,
  title={Quantum Transformer: Accelerating model inference via quantum linear algebra},
  author={Guo, Naixu and Yu, Zhan and Choi, Matthew and Han, Yizhan and Agrawal, Aman and Nakaji, Kouhei and Aspuru-Guzik, Al{\'a}n and Rebentrost, Patrick},
  journal={arXiv preprint arXiv:2402.16714},
  year={2024}
}

@article{smaldone2025hybrid,
  title={A hybrid Transformer architecture with a quantized self-attention mechanism applied to molecular generation},
  author={Smaldone, Anthony M and Shee, Yu and Kyro, Gregory W and Farag, Marwa H and Chandani, Zohim and Kyoseva, Elica and Batista, Victor S},
  journal={Journal of chemical theory and computation},
  volume={21},
  number={10},
  pages={5143--5154},
  year={2025},
  publisher={ACS Publications}
}

@article{zlokapa2021quantum,
  title={A quantum algorithm for training wide and deep classical neural networks},
  author={Zlokapa, Alexander and Neven, Hartmut and Lloyd, Seth},
  journal={arXiv preprint arXiv:2107.09200},
  year={2021}
}

@inproceedings{yao1977probabilistic,
  title={Probabilistic computations: Toward a unified measure of complexity},
  author={Yao, Andrew Chi-Chin},
  booktitle={18th Annual Symposium on Foundations of Computer Science (sfcs 1977)},
  pages={222--227},
  year={1977},
  organization={IEEE}
}

@inproceedings{saxena2023streaming,
  title={Streaming complexity of CSPs with randomly ordered constraints},
  author={Saxena, Raghuvansh R and Singer, Noah and Sudan, Madhu and Velusamy, Santhoshini},
  booktitle={Proceedings of the 2023 Annual ACM-SIAM Symposium on Discrete Algorithms (SODA)},
  pages={4083--4103},
  year={2023},
  organization={SIAM}
}

@article{monticelli2000electric,
  title={Electric power system state estimation},
  author={Monticelli, Alcir},
  journal={Proceedings of the IEEE},
  volume={88},
  number={2},
  pages={262--282},
  year={2000},
  publisher={IEEE}
}

@book{abur2004power,
  title={Power system state estimation: theory and implementation},
  author={Abur, Ali and Exposito, Antonio Gomez},
  year={2004},
  publisher={CRC press}
}

@article{widmer1996learning,
  title={Learning in the presence of concept drift and hidden contexts},
  author={Widmer, Gerhard and Kubat, Miroslav},
  journal={Machine learning},
  volume={23},
  number={1},
  pages={69--101},
  year={1996},
  publisher={Springer}
}

@article{gama2014survey,
  title={A survey on concept drift adaptation},
  author={Gama, Jo{\~a}o and {\v{Z}}liobait{\.e}, Indr{\.e} and Bifet, Albert and Pechenizkiy, Mykola and Bouchachia, Abdelhamid},
  journal={ACM computing surveys (CSUR)},
  volume={46},
  number={4},
  pages={1--37},
  year={2014},
  publisher={ACM New York, NY, USA}
}

@article{johnson2007adjusting,
  title={Adjusting batch effects in microarray expression data using empirical Bayes methods},
  author={Johnson, W Evan and Li, Cheng and Rabinovic, Ariel},
  journal={Biostatistics},
  volume={8},
  number={1},
  pages={118--127},
  year={2007},
  publisher={Oxford University Press}
}

@article{haghverdi2018batch,
  title={Batch effects in single-cell RNA-sequencing data are corrected by matching mutual nearest neighbors},
  author={Haghverdi, Laleh and Lun, Aaron TL and Morgan, Michael D and Marioni, John C},
  journal={Nature biotechnology},
  volume={36},
  number={5},
  pages={421--427},
  year={2018},
  publisher={Nature Publishing Group}
}

@article{borodin1977relating,
  title={On relating time and space to size and depth},
  author={Borodin, Allan},
  journal={SIAM journal on computing},
  volume={6},
  number={4},
  pages={733--744},
  year={1977},
  publisher={SIAM}
}

@article{csanky1976fast,
  title={Fast parallel matrix inversion algorithms},
  author={Csanky, Laszlo},
  journal={SIAM Journal on Computing},
  volume={5},
  number={4},
  pages={618--623},
  year={1976},
  publisher={SIAM}
}

@article{berkowitz1984computing,
  title={On computing the determinant in small parallel time using a small number of processors},
  author={Berkowitz, Stuart J},
  journal={Information processing letters},
  volume={18},
  number={3},
  pages={147--150},
  year={1984},
  publisher={Elsevier}
}

@article{aaronson2016complexity,
  title={Complexity-theoretic foundations of quantum supremacy experiments},
  author={Aaronson, Scott and Chen, Lijie},
  journal={arXiv preprint arXiv:1612.05903},
  year={2016}
}

@inproceedings{papakonstantinou2014depth,
  title={The depth irreducibility hypothesis},
  author={Papakonstantinou, Periklis},
  booktitle={Electronic Colloquium on Computational Complexity},
  year={2014}
}

@inproceedings{ta2013inverting,
  title={Inverting well conditioned matrices in quantum logspace},
  author={Ta-Shma, Amnon},
  booktitle={Proceedings of the forty-fifth annual ACM symposium on Theory of computing},
  pages={881--890},
  year={2013}
}

@inproceedings{liu2026numerical,
  title={Numerical Linear Algebra in Linear Space},
  author={Liu, Yiping and Nguyen, Hoai-An and Yang, Junzhao},
  booktitle={Proceedings of the 2026 Annual ACM-SIAM Symposium on Discrete Algorithms (SODA)},
  pages={6392--6435},
  year={2026},
  organization={SIAM}
}

@article{savitch1970relationships,
  title={Relationships between nondeterministic and deterministic tape complexities},
  author={Savitch, Walter J},
  journal={Journal of computer and system sciences},
  volume={4},
  number={2},
  pages={177--192},
  year={1970},
  publisher={Elsevier}
}

@article{barnes1998sublinear,
  title={A sublinear space, polynomial time algorithm for directed st connectivity},
  author={Barnes, Greg and Buss, Jonathan F and Ruzzo, Walter L and Schieber, Baruch},
  journal={SIAM Journal on Computing},
  volume={27},
  number={5},
  pages={1273--1282},
  year={1998},
  publisher={SIAM}
}

@inproceedings{tai2018sketching,
  title={Sketching linear classifiers over data streams},
  author={Tai, Kai Sheng and Sharan, Vatsal and Bailis, Peter and Valiant, Gregory},
  booktitle={Proceedings of the 2018 international conference on management of data},
  pages={757--772},
  year={2018}
}

@inproceedings{aghazadeh2018mission,
  title={Mission: Ultra large-scale feature selection using count-sketches},
  author={Aghazadeh, Amirali and Spring, Ryan and LeJeune, Daniel and Dasarathy, Gautam and Shrivastava, Anshumali and others},
  booktitle={International conference on machine learning},
  pages={80--88},
  year={2018},
  organization={PMLR}
}

@inproceedings{weinberger2009feature,
  title={Feature hashing for large scale multitask learning},
  author={Weinberger, Kilian and Dasgupta, Anirban and Langford, John and Smola, Alex and Attenberg, Josh},
  booktitle={Proceedings of the 26th annual international conference on machine learning},
  pages={1113--1120},
  year={2009}
}
\bibliographystyle{unsrt}

\clearpage
\renewcommand*\appendixpagename{Appendices}
\appendix
\crefalias{section}{appendix}
\crefalias{subsection}{subappendix}
\crefalias{subsubsection}{subsubappendix}
\appendixpage
\appendixtableofcontents
\label{toc}

\counterwithin{theorem}{section}

\aliascntresetthe{lemma}
\aliascntresetthe{corollary}
\aliascntresetthe{definition}
\aliascntresetthe{remark}
\aliascntresetthe{task}

\renewcommand{\theHtheorem}{\theHsection.\arabic{theorem}}
\renewcommand{\theHlemma}{\theHtheorem}
\renewcommand{\theHcorollary}{\theHtheorem}
\renewcommand{\theHdefinition}{\theHtheorem}
\renewcommand{\theHremark}{\theHtheorem}
\renewcommand{\theHtask}{\theHtheorem}

\begin{center}
    \textbf{Roadmap}
\end{center}

We begin with a roadmap that helps the readers navigate through the appendices.
For convenience, clicking the page number in the header of any page will return the reader to this roadmap.
In \Cref{sec:additional-numerics}, we provide additional numerical experiments and the details.
The remainder of the appendices are organized into the following three themes.

\vspace{0.5em}

\textbf{Background and setup.}
We start with an overview and discussion of related works and our contributions in \Cref{sec:rel-work}.
Then we set the stage by introducing our models of data access and computation in \Cref{sec:model}.

\vspace{0.5em}

\textbf{Theoretical foundations.}
\Cref{sec:q-alg,sec:cl-hard} are devoted to establishing the rigorous theoretical foundations of processing massive classical data on small quantum computers.
This includes two parts.
\begin{enumerate}
    \item
    \emph{Quantum algorithms (\Cref{sec:q-alg}).}
    In \Cref{sec:q-alg-iid}, we begin by introducing the most intuitive and simplest version of quantum oracle sketching for IID data of Boolean functions.
    In particular, we address the apparent trap of decoherence and how we circumvent it.
    Then we establish its optimality, and extend it to handle more general data distributions and data structures such as matrices and vectors.

    \item 
    \emph{Classical hardness (\Cref{sec:cl-hard}).}
    We develop the machinery for proving classical hardness by introducing the Noisy Oracle Property Estimation (NOPE) task and its dynamic variant.
    Then we prove their classical hardness and connect them to the various applications.
\end{enumerate}

\textbf{Applications.}
In \Cref{sec:app}, we apply our theoretical tools to the real-world applications including linear systems (\Cref{fig:linear-sys}), binary classification (\Cref{fig:binary-classification}), and dimension reduction (\Cref{fig:dim-reduct}).
Each application starts with a self-contained introduction of the problem and the statements of our main quantum advantage results.
They are written without any quantum jargon and are intended for a general audience.
Detailed proofs follow, utilizing the tools we develop in the theoretical foundations.

\textbf{Suggested reading routes.} 
Our figures are designed to provide intuitive understanding without going into the details.
We suggest starting with \Cref{fig:linear-sys,fig:binary-classification,fig:dim-reduct} for the applications and quantum advantage results, then \Cref{fig:access-model} for the formal setup.
\Cref{fig:qos} provides an overview of quantum oracle sketching and \Cref{fig:hardness} illustrates the intuition behind the classical hardness proofs.

\begin{figure}
    \centering
    \includegraphics[width=1\linewidth]{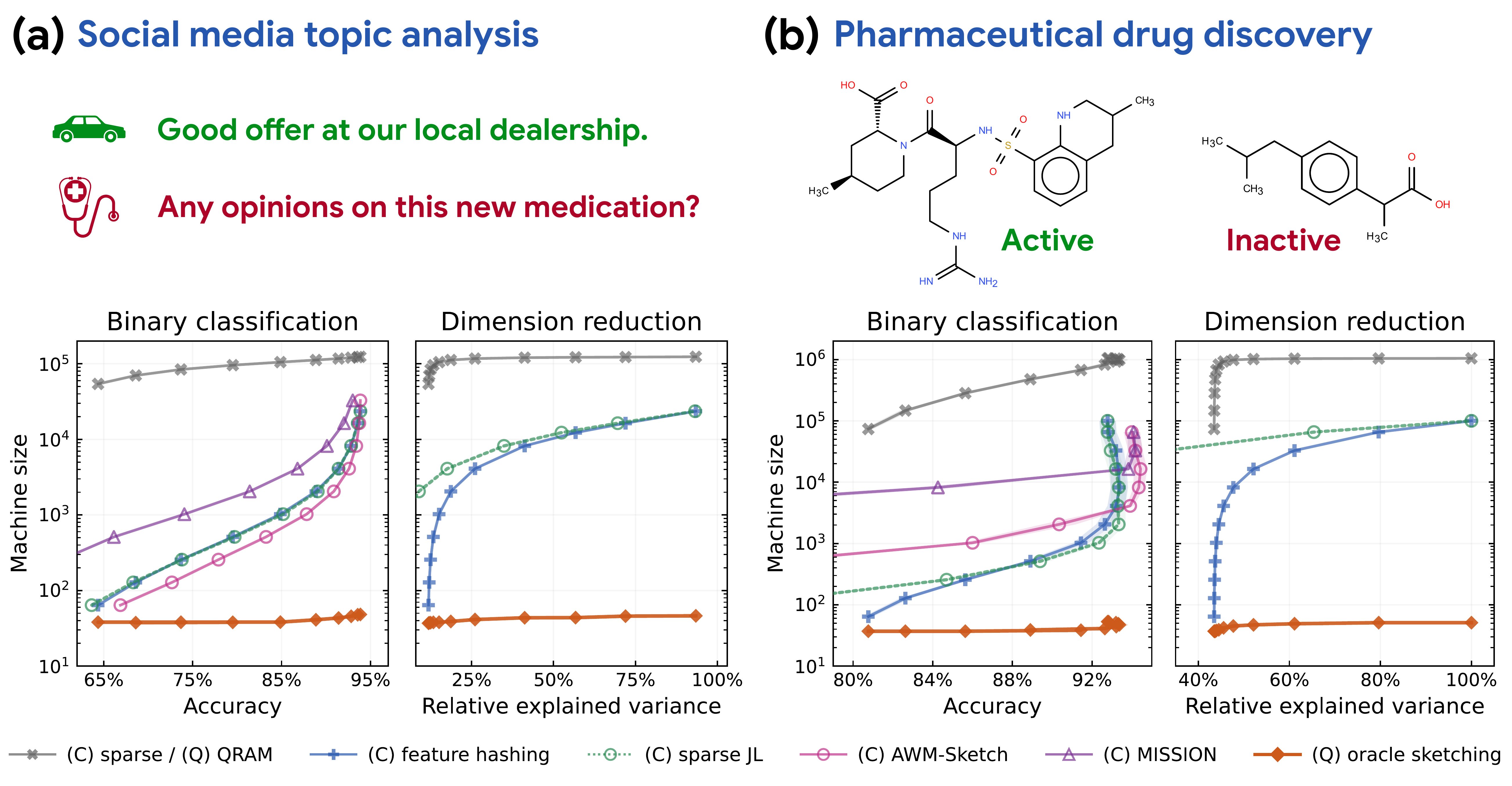}
    \caption{
    \revise{
    \textbf{Evidence for quantum advantages in additional real-world datasets from numerical simulation.}
    We perform binary classification and dimension reduction for
    \textbf{(a)}
    topic analysis of posts from 20 newsgroups \cite{joachims1996probabilistic}
    and 
    \textbf{(b)}
    chemical compound data for Thrombin binding \cite{weston2003feature}.
    We compare the machine size required to reach a certain prediction performance using different families of classical (C) and quantum (Q) algorithms, including quantum oracle sketching (orange), quantum algorithms using QRAM (gray), classical sparse-matrix algorithms (gray), and classical streaming algorithms (blue \cite{weinberger2009feature}, green \cite{kane2014sparser}, magenta \cite{tai2018sketching}, and purple \cite{aghazadeh2018mission}).
    For each algorithm, we truncate the retained information to plot the trade-off between machine size and performance, with standard error indicated by the shaded region.
    Machine size is defined as the maximal number of fundamental memory units consumed throughout the entire computation: logical qubits for quantum and floating-point numbers for classical.
    Performance is quantified by the $5$-fold cross validation accuracy averaged over random category pairs for classification, and by the explained variance relative to the untruncated baseline for dimension reduction.
    More details are given in \Cref{sec:additional-numerics}.
    }
    }
    \label{fig:more-numerics}
\end{figure}

For more details, we suggest the following routes depending on the reader's interest:

\begin{enumerate}
    \item 
    \emph{For general readers:} 
    Start with the applications in \Cref{sec:app}, which are further supported by numerical experiments in \Cref{fig:numerics,fig:more-numerics}.
    \Cref{sec:rel-work} provides more context on the various challenges in reaching quantum advantages in useful classical tasks and a semi-technical overview of how we tackle them.

    \item 
    \emph{For algorithm designers:}
    Proceed to the quantum algorithms in \Cref{sec:q-alg} and in particular \Cref{sec:q-alg-linear-algebra} for applications related to linear algebra.
    The quantum algorithm parts of \Cref{sec:app} are also good templates for applying the algorithms to potential applications.
    On \href{https://github.com/haimengzhao/quantum-oracle-sketching}{GitHub}, we provide JAX \cite{jax2018github} implementation of quantum oracle sketching and QSVT that supports GPU/TPU and automatic differentiation for variational training, with details explained in \Cref{sec:additional-numerics}.

    \item 
    \emph{For readers interested in the rigorous proofs:}
    The formal setup is detailed in \Cref{sec:model} and the sample complexity and variance analysis techniques for noisy and correlated linear algebra data are developed in \Cref{sec:q-alg-ext,sec:q-alg-linear-algebra}. 
    Theoretical computer scientists may be further interested in our technical contributions in the classical lower bound proofs (\Cref{sec:cl-hard}), including the sample-space lower bounds via simulation, learning XOR lemma with derandomization, and the BQP-hardness of useful applications.
\end{enumerate}

\textbf{Notations.}
Throughout this work, we use the standard notations of asymptotics. 
For two positive functions $f(N)$ and $g(N)$, $f(N) = O(g(N))$ if there exist $N_0, C>0$ such that $\forall N>N_0, f(N)\leq Cg(N)$. 
$f(N)=\Omega(g(N))$ if $g(N)=O(f(N))$. 
$f(N)=\Theta(g(N))$ if $f(N)=O(g(N))$ and $f(N)=\Omega(g(N))$. 
$f(N)=o(g(N))$ if $\lim_{N\to\infty} f(N)/g(N)=0$.
$f(N)=\omega(g(N))$ if $g(N)=o(f(N))$.
We add a tilde (e.g., $\tilde O(f(N))$) to omit factors that scale polynomially with $\log f(N)$.
We also use $\poly(N)=N^{O(1)}$ to denote functions of $N$ that scale polynomially with $N$ and use $\polylog(N)$ to denote $\poly(\log(N))$.
Similarly, $\mathrm{superpoly(N)}$ means $N^{\omega(1)}$.
All logarithms are base $2$ and all error parameters $\epsilon\in (0, 1)$ unless stated otherwise.
We say a function of $N$ is non-vanishing if it is at least $\Omega(1)$.

\newpage

\section{Additional numerical experiments}
\label{sec:additional-numerics}

In this section, we provide more details of the numerical experiments presented in the main text, along with additional numerical experiments.
We begin with additional evidence of our quantum space advantages in real-world datasets.
Then we move on to the benchmarking of quantum oracle sketching.
Note that in all numerical experiments, we exclude the low-space classical algorithm described in \Cref{app:best-known-classical} because it has runtime $\mathrm{superpoly}(N)$.

\subsection{Quantum advantages in realistic datasets}
\label{sec:add-num-datasets}

\revise{To provide empirical evidence for the exponential quantum space advantages, we conduct numerical simulations on four real-world datasets across diverse application domains: sentiment analysis of movie reviews (\Cref{fig:numerics}(a)), single-cell RNA sequences analysis (\Cref{fig:numerics}(b)), social media topic analysis (\Cref{fig:more-numerics}(a)), and pharmaceutical drug discovery (\Cref{fig:more-numerics}(b)).}
The specific datasets we consider are the standard benchmark datasets in the respective domains.
For movie review sentiment analysis, we use the movie review dataset from the Internet Movie Database (IMDb) \cite{maas2011learning} available at \href{https://ai.stanford.edu/~amaas/data/sentiment/}{this website}.
For single-cell RNA sequencing, we use the single-cell RNA sequences of 68k peripheral blood mononuclear cells (PBMC) \cite{zheng2017massively} commonly known as the PBMC68k or Zheng68k dataset available through the scVelo package \cite{Bergen2020}.
For social media topic analysis, we use the posts from 20 newsgroups known as the 20Newsgroup dataset \cite{joachims1996probabilistic} available through the scikit-learn package \cite{scikit-learn}.
For pharmaceutical drug discovery, we use the chemical compound dataset Dorothea \cite{dorothea_169} with labels indicating binding ability to Thrombin available at \href{https://archive.ics.uci.edu/dataset/169/dorothea}{the UCI Machine Learning Repository}.
For raw text data in IMDb and 20Newsgroup, we construct the features by the standard TF-IDF method with English stop words as default in scikit-learn.
The other two datasets are used as provided.

\revise{
To mimic the realistic scenarios where we do not have enough memory to store the whole dataset, we truncate the feature dimension by randomized feature hashing, which randomly merges distinct features to reduce the feature dimension.
By varying the truncated feature dimension, we obtain the terminal performance at different machine sizes after sufficient samples and computation time.
In other words, we allow sufficient polynomial sample sizes and computation time for the algorithms to converge to their terminal performance at the designated machine size.
We additionally consider a sparse Johnson-Lindenstrauss (JL) transform as an alternative data-oblivious sketch, and, for classification, two adaptive streaming algorithms, AWM-Sketch \cite{tai2018sketching} and MISSION \cite{aghazadeh2018mission}, which adaptively identify and track important features during learning; their implementation is detailed in \Cref{sec:add-streaming}.
We focus on general-purpose algorithms with provable guarantees and leave the studies of dataset-specific heuristics to future work, as they require extensive empirical evaluation.

All algorithms are evaluated on the same tasks of predicting the label and 1D representation of a single but arbitrary test vector according to the LS-SVM decision rule and the principal component.
In particular, neither the quantum nor the classical algorithms are required to generate reusable models or output the full, memory-consuming weight vector or principal component in the original feature dimension.
}

More concretely, for each target truncated dimension $D'$, we use a random linear map $S\in \mathbb{R}^{D'\times D}$ and replace each feature vector $\vec{x}\in \mathbb{R}^{D}$ by its truncated representation $\vec{x}'=S\vec{x}\in \mathbb{R}^{D'}$.
For feature hashing, each original feature index $j\in[D]$ is assigned to one bucket $h(j)\in[D']$, and
\begin{equation}
	(S\vec{x})_b = \frac{1}{\sqrt{|B_b|}} \sum_{j:h(j)=b} x_j,
\end{equation}
where $B_b=\{j:h(j)=b\}$ is the $b$-th bucket.
For sparse JL, we use the signed version of the same construction:
\begin{equation}
	(S\vec{x})_b = \frac{1}{\sqrt{|B_b|}} \sum_{j:h(j)=b} \sigma_j x_j,
\end{equation}
where $\sigma_j\in\{\pm 1\}$ are independent random signs.
The map $h$ is chosen to be balanced, so the bucket sizes differ by at most one.
When $D'=D$, we take $S=I_D$, so the truncated dataset exactly recovers the full dataset.
The random map is specified by a seed and can be evaluated online space efficiently.
\revise{
We note that this is the sparsity $s_{\mathrm{JL}}=1$ case of general sparse JL transforms.
In \Cref{sec:add-streaming}, we conduct a systematic sweep over sparsity $s_{\mathrm{JL}}=1, 2, 4, 8$ and find that increasing the sparsity does not bring performance gain and only increases the computational burden.
Therefore, we focus on the sparsity $s_{\mathrm{JL}}=1$ in the main results.
}

We consider two tasks: binary classification and dimension reduction.
The performance of binary classification is quantified by the $5$-fold cross validation accuracy of the least-squares support vector machine (LS-SVM) with $\ell_2$ regularization.
To avoid overfitting, we use the $\ell_2$ regularization strength $\lambda=10, 200, 1, 200$ for IMDb, PBMC68k, 20Newsgroup, and Dorothea respectively.
If the dataset contains more than two categories, we average the accuracy over random pairs of categories.
In particular, 100 pairs are randomly sampled for both 20Newsgroup and PBMC68k.
For the adaptive streaming algorithms that do not have exact solutions, we report their terminal accuracy after convergence as implemented in \Cref{sec:add-streaming}.

For dimension reduction, we perform principal component analysis (PCA) on each dataset with all categories combined.
The performance is measured by how well the first principal component after randomized truncation explains the variance of the original dataset.
Concretely, for each truncation dimension $D'$, we apply a randomized feature map
$S\in \mathbb{R}^{D'\times D}$ to each feature vector (either via feature hashing or sparse JL transform), obtaining the truncated dataset
$X' = X S^T \in \mathbb{R}^{N\times D'}$.
We first compute the first principal component $\vec{w}'\in \mathbb{R}^{D'}$ of $X'$.
We then lift it back to the original feature space as
$\vec{w}'' = S^T \vec{w}'\in \mathbb{R}^{D}$ and normalize $\vec{w}''$ to unit norm.
The relative explained variance is $(\vec{w}''^{T} X^{T}X \vec{w}'')/(\vec{w}^{T} X^{T}X \vec{w})$, where $\vec{w}\in \mathbb{R}^{D}$ is the first principal component of the full dataset $X$.

\revise{
Utilizing the guarantees that the quantum algorithms provably converge to the exact LS-SVM and PCA solutions with sufficient sample size and time, we obtain their terminal performance by directly solving the relevant LS-SVM and PCA problems at different truncated feature dimensions.
This enables us to predict the terminal performance of the quantum algorithms (i.e., quantum oracle sketching and QRAM-based algorithms) without simulating deep quantum circuits on $\sim 60$ logical qubits, which would be beyond current computational capabilities.
}

We calculate the memory consumption or machine size of the algorithms as follows.
\revise{We define memory consumption as the maximal number of fundamental units required to be maintained throughout the entire computation: logical qubits for quantum processors and floating-point numbers for classical machines.
Each stored floating-point number occupies one scalar register, which we use interchangeably as the classical memory unit below.}
This is also known as the word RAM model, where at fixed numerical precision, each scalar number stored by a classical machine occupies one machine word.
As in the standard streaming-memory model~\cite{muthukrishnan2005data}, the current input sample is supplied through the input stream and is not included in the algorithm's persistent working memory; it is discarded after being processed.
For notational simplicity, in the following memory estimates, we use $D$ to denote the feature dimension after feature truncation.

\revise{
For classical sparse-matrix algorithms and QRAM-based quantum algorithms, we use the strictly conservative estimates of their actual memory consumption
\begin{equation}
	S_{\mathrm{c, sparse}}, S_{\mathrm{QRAM}} \geq N_{\mathrm{nnz}},
\end{equation}
where $N_{\mathrm{nnz}}$ is the number of non-zero elements in the datasets after feature truncation.
This is because these algorithms always store the whole sparse matrix in their memory.
For classical streaming algorithms, we use the strictly conservative estimates of their actual memory consumption
\begin{equation}
	S_{\mathrm{c,streaming}} \geq D,
\end{equation}
where $D$ is the feature dimension of the datasets after feature truncation.
This is because these algorithms proceed by storing the whole solution vector $\vec{w}\in \mathbb{R}^D$ (i.e. the weight vector of LS-SVM or the principal component in PCA) in their memory and performing a streaming update to it \cite{andoni2020streaming,mitliagkas2013memory}.
We note that these are strictly conservative estimates of the actual memory consumption of the algorithms since they may use extra memory for calculating intermediate variables.
For example, consider training LS-SVM with feature hashing and streaming stochastic gradient descent (SGD).
We truncated the dimension of each feature vector into
$\vec{z}_t\in\mathbb{R}^{D}$ using a hashing function generated from a random seed.
Then we update the weight vector as
\begin{equation}
    \vec{w}_{t+1}
    =
    (1-\eta_t\lambda)\vec{w}_t
    -
    \eta_t
    \bigl(\vec{w}_t^{\,T}\vec{z}_t-y_t\bigr)\vec{z}_t,
\end{equation}
where $\vec{w}_t\in\mathbb{R}^{D}$ is the weight vector and $\eta_t$ is the learning rate.
Each data sample $(\vec{z}_t,y_t)$ is discarded immediately after the update.
Hence, the memory consumption is given by the $D$ components of $\vec{w}_t$ plus a constant from storing the learning rate, regularization strength, random seed, etc.
Therefore, we have
\begin{equation}
    S_{\mathrm{SGD}}^{\text{LS-SVM}}
    =
    D+O(1).
\end{equation}

Streaming PCA is similar.
Consider performing PCA with streaming Oja update as
\begin{align}
    \widetilde{\vec{w}}_{t+1}
    &=
    \vec{w}_t
    +
    \eta_t\vec{z}_t
    \bigl(\vec{z}_t^{\,T}\vec{w}_t\bigr),\\
    \vec{w}_{t+1}
    &=
    \frac{\widetilde{\vec{w}}_{t+1}}
    {\lVert\widetilde{\vec{w}}_{t+1}\rVert_2}.
\end{align}
We need to store the $D$-dimensional principal-component vector and only $O(1)$ additional scalar registers.
Therefore, the total machine size is $D+O(1)$.

Overall, the plotted machine size $D$ in the figures is a conservative memory estimate for the actual memory consumption of these streaming algorithms, which is in strict favor of the classical algorithms.
Note that we are not using any theoretical lower bound from the streaming algorithm literature \cite{andoni2020streaming,mitliagkas2013memory} which was proved in the worst case and does not apply in the numerical experiments.
For the adaptive streaming algorithms, we use the measured number of stored registers, counting every feature index, weight, and sketch counter maintained, as detailed in \Cref{sec:add-streaming}.
}

For quantum oracle sketching, we have memory consumption
\begin{equation}
	S_{\mathrm{QOS}}^{\mathrm{LS-SVM}} = 2\ceil{\log_2(N+2D)} + \ceil{\log_2(s+1)} + 3 + 1,
\end{equation}
for LS-SVM prediction of a single test sample, where $N\times D$ is the dimension of the truncated data matrix $X$ and $s$ is the sparsity (i.e., the maximal number of non-zero elements in each row or column).
This is because we use quantum oracle sketching to build the following components.
\begin{enumerate}
    \item The block encoding of the augmented data matrix $\begin{pmatrix}
        X\\\sqrt{\lambda} I_D
    \end{pmatrix} \in \mathbb{R}^{(N+D)\times D}$ used in \Cref{lem:q-ridge-reg-solver}.
    Its Hermitian embedding has dimension $(N+2D)\times (N+2D)$. 
    This requires building the sparse index/element oracle for the augmented matrix, which has sparsity $s+1$.
    Hence, building its sparse index oracle requires $2\ceil{\log_2(N+2D)} + \ceil{\log_2(s+1)} + 2$ qubits, where the additional $2$ qubits are for QSVT.
    Building the sparse element oracle can reuse the same qubits, so no extra qubits are needed.
    \item The state preparation unitary of the label vector $\vec{y}\in \mathbb{R}^N$, which requires $\ceil{\log_2(N)} + 2$ qubits, where the additional $2$ qubits are for the first LCU \& QSVT and the second LCU as in \Cref{thm:q-state-sketch}. These qubits are contained in the previous count since they can be reused.
\end{enumerate}
Then we perform quantum ridge regression with amplitude amplification using QSVT-based quantum linear system solver, which requires $1$ ancilla qubit for the QSVT, contained in the previous count because we can reuse the ancilla qubit from quantum oracle sketching.
Finally, we need to perform Hadamard test to calculate the signed overlap with test state, which requires $1$ extra ancilla qubit as in \Cref{lem:interf-classical-shadow}.
The final estimate of the label is stored classically on a running average, so only $1$ extra classical scalar number register is needed.
This proves the formula for $S_{\mathrm{QOS}}^{\mathrm{LS-SVM}}$.
\revise{
Note that predicting additional test samples reruns the procedure on fresh samples within the same workspace.
Since we are focusing on the task of predicting a single but arbitrary test vector, we are not using the interferometric classical shadow procedure.
We also note that these are tailored machine size calculations for the numerical experiments, which improve over the theoretical machine size bounds presented in the theory results (e.g., \Cref{lem:block-encoding,thm:q-state-sketch}).
For example, the binary label vector state preparation unitary can be implement much more straightforwardly as compared to \Cref{thm:q-state-sketch}.
See the specific code implementation for more details.
}

Similarly, the same calculation shows that the memory consumption for PCA and dimension reduction of a single test sample is
\begin{equation}
	S_{\mathrm{QOS}}^{\mathrm{PCA}} = 2\ceil{\log_2(N+D)} + \ceil{\log_2(s)} + 3 + 1,
\end{equation}
because there is no augmentation for regularization.
\revise{
We note that these qubit consumption calculations are tailored to and confirmed by the numerical simulations, improving over the generic theoretical bounds presented in the proofs of subsequent sections.
}

\subsection{Additional studies on classical streaming algorithms}
\label{sec:add-streaming}

\revise{
The classical streaming baselines in \Cref{fig:numerics,fig:more-numerics} are built from feature hashing and sparse JL transforms.
In this section, we systematically study the robustness of the resulting classical memory-performance frontier and provide evidence for the robustness of the quantum advantages achieved by quantum oracle sketching.

First, we test several data-oblivious sketching methods, including sparse JL transforms with increasing sparsity, and find that they perform similarly to feature hashing.
Across all tested oblivious sketches, matching the performance of quantum oracle sketching still requires orders of magnitude more memory.

Second, we test more sophisticated and recent adaptive streaming algorithms \cite{tai2018sketching,aghazadeh2018mission} in the literature, which adaptively choose and change the important features to keep during the learning process, on all four datasets; their curves are shown in the classification panels of \Cref{fig:numerics,fig:more-numerics}, and \Cref{fig:survey-adaptive} shows their paired differences from feature hashing on IMDb and PBMC68k.
We find that they improve the terminal accuracy by at most $7$ percentage points at intermediate machine sizes and only $0.2$ percentage points at the largest machine sizes.
They still require orders of magnitude more memory to match the performance of quantum oracle sketching.

Third, we study how finite-sample streaming optimization approaches the exact LS-SVM and PCA solutions used to obtain the terminal points in \Cref{fig:numerics,fig:more-numerics}.
At fixed machine size, processing additional samples can reduce the optimization error of the streaming implementation toward the exact solution in its maintained feature space, but it does not change that memory-limited terminal solution.

Throughout, machine size follows the same accounting as in the previous section: the maximal number of floating-point numbers, or scalar registers, needed throughout the entire computation, including every stored feature index, weight, and sketch counter.
We ignore the hash and sign functions, which can be generated from $O(1)$ seeds, and hence the calculated machine sizes are strictly conservative estimates for the actual machine sizes.
All experiments are repeated over five independent random seeds per machine size, and, where applicable, additionally over the random category pairs for PBMC68k and 20 Newsgroups; the reported standard errors are computed over all repetitions.
}

\begin{figure}
    \centering
    \includegraphics[width=1\linewidth]{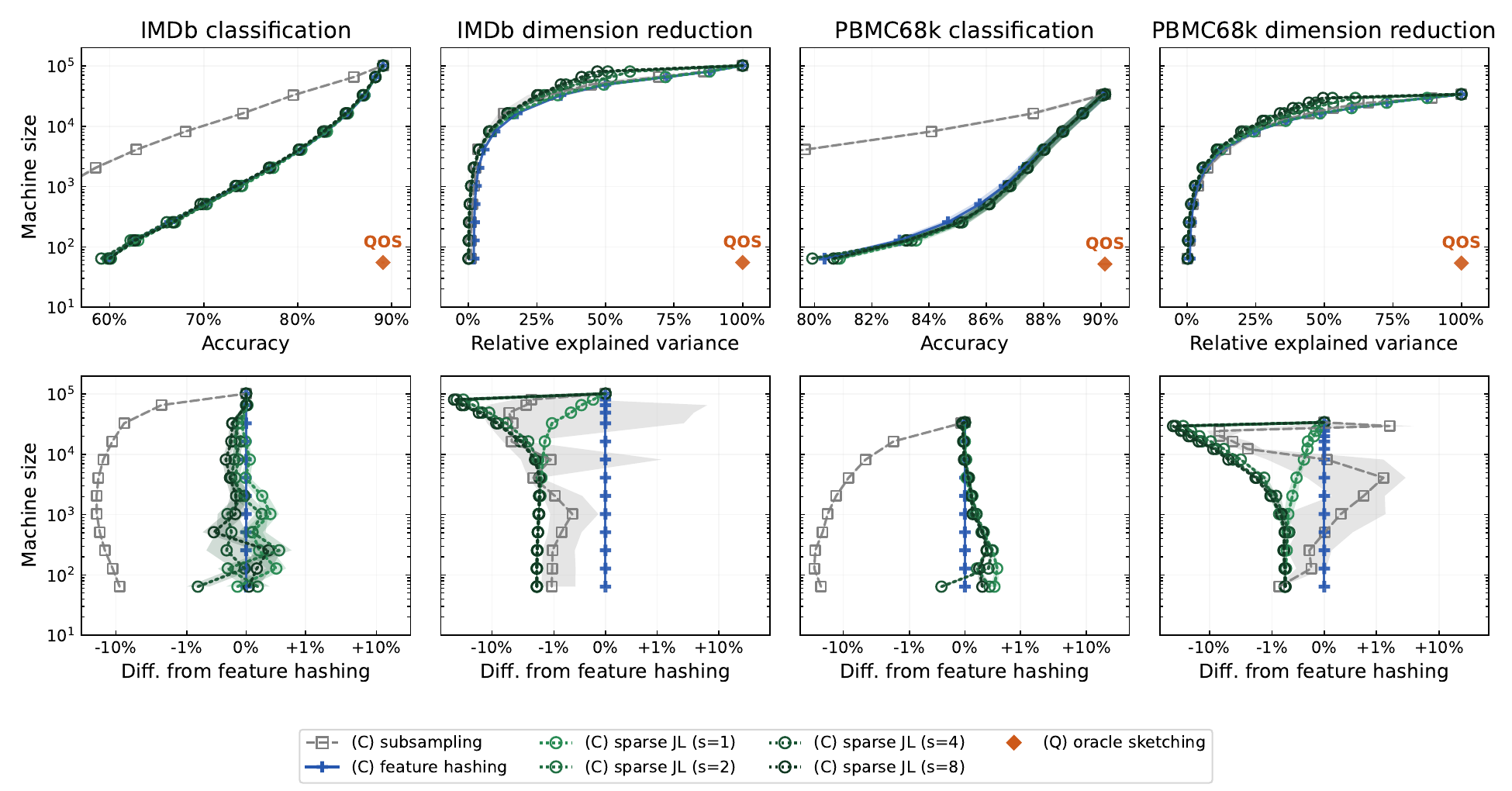}
    \caption{
    \revise{
    \textbf{Additional experiments with oblivious sketches.}
    Top row: terminal performance achieved under designated machine sizes in classification (LS-SVM) and dimension reduction (PCA) on IMDb and PBMC68k, using different oblivious randomized sketching methods: uniform coordinate subsampling, feature hashing, and signed sparse JL transforms with sparsity $s_{\mathrm{JL}}=1,2,4,8$ \cite{kane2014sparser}.
    For all oblivious sketches, we use the truncated feature dimension as a conservative estimate of their machine size.
    The orange diamond marks the terminal performance of quantum oracle sketching.
    Bottom row: performance difference of different sketches from feature hashing at equal machine size, plotted in the symmetric log scale, where the axis is linear within $\pm 1\%$; shaded regions indicate standard error.
    }
    }
    \label{fig:survey-sketches}
\end{figure}

\revise{
\textbf{Additional experiments with oblivious sketches.}
We first examine the data-oblivious sketching methods (\Cref{fig:survey-sketches}) that compress the feature space into randomized sketches.
Following the same numerical procedure as in \Cref{fig:numerics,fig:more-numerics}, we compare three families of oblivious sketching methods.
The first one is uniform coordinate subsampling, which retains a random subset of $D'$ original features:
\begin{equation}
    (S\vec{x})_b = x_{j_b},
    \qquad b\in[D'],
\end{equation}
where $j_1<\cdots<j_{D'}$ are drawn uniformly without replacement from $[D]$.
The second is feature hashing, as defined in the previous subsection and used in the main figures.
The third is the family of signed sparse JL transforms with sparsity $s_{\mathrm{JL}}=1,2,4,8$, which generalizes the sparse JL transform of the previous subsection: each feature $j\in[D]$ is assigned to $s_{\mathrm{JL}}$ distinct buckets $h_1(j),\ldots,h_{s_{\mathrm{JL}}}(j)\in[D']$, with each assignment round balanced, and
\begin{equation}
    (S\vec{x})_b
    =
    \sum_{(j, r):\, h_r(j)=b}
    \frac{\sigma_{j,r}}{\sqrt{o_b}}\, x_j,
    \qquad
    o_b = \left|\{(j, r): h_r(j)=b\}\right|,
\end{equation}
where $\sigma_{j,r}\in\{\pm 1\}$ are independent random signs and $o_b\approx s_{\mathrm{JL}} D/D'$ is the occupancy of bucket $b$.
Under this normalization, each feature carries total sketch energy $\sum_{r} 1/o_{h_r(j)}\approx D'/D$ independently of the sparsity, so the entire family shares the normalization of feature hashing and the effective regularization strength remains consistent across sparsity levels; at $s_{\mathrm{JL}}=1$, it reduces exactly to the sparse JL transform of the main figures.

As shown in \Cref{fig:survey-sketches}, for classification, the entire sparse JL family remains within one percentage point of feature hashing at every machine size on both datasets, indicating that neither the random signs nor the increased sparsity yields a material improvement.
For dimension reduction, larger sparsity is actively harmful at many truncated dimensions below the full-dimension endpoint, sacrificing tens of percentage points of relative explained variance for $s_{\mathrm{JL}}\geq 2$, as distributing the energy of each feature over multiple buckets degrades the recovery of the leading principal component.
Uniform coordinate subsampling, as a naive baseline that performs no mixing of the features, loses up to $19$ and $27$ percentage points of accuracy on IMDb and PBMC68k, respectively.
Feature hashing therefore stands as a representative and empirically near-best choice among the tested oblivious sketches.
}

\revise{
\textbf{Additional experiments with adaptive streaming algorithms.}
Second, we examine the adaptive streaming algorithms (\Cref{fig:survey-adaptive}), which devote part of their memory to identifying and exactly tracking the most informative features during learning.
All algorithms are trained on a stream of samples drawn at random from a stratified split of the dataset into $80\%$ for training and $20\%$ for testing, and minimize the same regularized objective underlying the LS-SVM of the main figures,
\begin{equation}
    \mathcal{L}(\vec{w}, b)
    =
    \frac{1}{N_{\mathrm{train}}}\sum_{i=1}^{N_{\mathrm{train}}}
    \frac{c_i}{2}
    \left(\vec{x}_i\cdot\vec{w} + b - y_i\right)^2
    +
    \frac{\lambda}{2}\|\vec{w}\|_2^2,
    \qquad
    \lambda = \frac{\alpha}{N_{\mathrm{train}}},
\end{equation}
where $N_{\mathrm{train}}$ is the number of training samples, $\alpha$ is the regularization strength of the corresponding dataset as in the previous subsection, $c_i=1$ for IMDb and 20 Newsgroups and $c_i = N_{\mathrm{train}}/(2N_{\mathrm{train},y_i})$ for PBMC68k and Dorothea, with $N_{\mathrm{train},y_i}$ the number of training samples with label $y_i$, so that the imbalanced classes are equally weighted, and the bias $b$ is unregularized and excluded from all machine-size counts as a conservative estimate.

In practice, to best ease the convergence of streaming SGD, we have made the following implementation choices to all methods and machine sizes.
For IMDb and 20 Newsgroups, the samples are drawn uniformly at random.
For PBMC68k and Dorothea, whose classes are imbalanced, we implement a class-based sampling that accommodates the class weights: the gradient are not rescaled by the class weights, and instead the samples are drawn by first drawing a class uniformly at random and then a sample uniformly within that class.
This is the importance-sampling equivalent of weighting the gradients and yields the identical expected objective.
In practice, this keeps the per-sample optimization steps from being amplified by class weights as large as $\sim 40$ on the most imbalanced pairs that destabilize SGD.

Upon receiving the sample $(\vec{x}_t, y_t)$ at step $t$, each algorithm performs the SGD update on its maintained representation $\hat{\vec{w}}_t$ of the weights,
\begin{equation}
    \vec{w}_{t+1}
    =
    \frac{\vec{w}_t}{1+\eta_t\lambda}
    -
    \eta_t\left(\vec{x}_t\cdot\hat{\vec{w}}_t + b_t - y_t\right)\vec{x}_t,
\end{equation}
together with the corresponding update of the bias, where the regularization decay is applied in its proximal form $1/(1+\eta_t\lambda)$ rather than the explicit form $1-\eta_t\lambda$ to ensure stability, under the learning rate scheduling
\begin{equation}
	\eta_t = \frac{\eta_0}{\sqrt{t}},
\end{equation}
where a single $\eta_0$ per dataset ($\eta_0=1$ for IMDb and 20 Newsgroups, $\eta_0=0.1$ for PBMC68k, and $\eta_0=0.001$ for Dorothea) is selected by the training objective on trial runs over the full-dimension problem.
Sharing a single learning rate across all methods and machine sizes isolates the effect of memory from that of hyperparameter tuning.
To keep the tuned learning rate in the same effective regime regardless of the feature scale, each run rescales its streamed features by $s=\sqrt{\rho_0/\rho}$ and the regularization strength $\lambda$ by $s^2$, where $\rho$ is the mean squared norm of the training samples in the run and $\rho_0$ that of the full-dimensional problem on which $\eta_0$ was tuned.
This is an exact reparametrization of the same objective with the same minimizer, under which the per-sample step $\eta_t\|s\vec{x}_t\|_2^2$ stays in the regime for which $\eta_0$ was tuned.
Without this reparameterization, the occupancy normalization factor in the feature hashing map would shrink the effective step by up to three orders of magnitude at small machine sizes and stall the convergence in practice.

The stopping and model-selection criteria are based solely on the training objective, so that no test information enters the training process.
Concretely, after every $N_{\mathrm{train}}/5$ processed samples, we evaluate the training objective $\mathcal{L}(\hat{\vec{w}}_t, b_t)$ of the currently maintained model, alongside its held-out accuracy recorded for reporting purposes only.
The regularization term of this objective is evaluated on the parameters that each algorithm actually stores and regularizes: the bucket weights for feature hashing, the active-set values and sketch counters for AWM-Sketch, and the count-sketch counters for MISSION.
Evaluating it on the decoded $D$-dimensional weights instead would count every sketch counter roughly $D/q$ times and be dominated by collision noise, which is not the objective the update rules minimize, and in practice leads the model selection to return premature models.
The stream stops once the lowest training objective observed so far has not decreased by a relative factor of $10^{-4}$ within a window of $N_{\mathrm{train}}$ samples, i.e., one effective pass over the training set, subject to a cap of $300N_{\mathrm{train}}$ total samples.
In practice, this cap is never hit.
Among all evaluations, we select the model attaining the minimal training objective and report its held-out accuracy.
In this way, neither the stopping rule nor the model selection ever accesses the test data.
For PBMC68k and 20 Newsgroups, we use the same $100$ random category pairs as in the main figures.
The machine sizes sweep the same grid as in \Cref{fig:numerics,fig:more-numerics}, excluding the full-dimension endpoint, at which the comparison would reflect residual optimizer and architecture effects rather than the effect of limited memory.
}

\revise{
We consider two representative adaptive streaming algorithms, AWM-Sketch \cite{tai2018sketching} and MISSION \cite{aghazadeh2018mission}, from the recent streaming machine learning literature, faithfully reimplemented from their original descriptions.
Active-set Weight-Median Sketch (AWM-Sketch) \cite{tai2018sketching} with total machine size $B$ maintains an active set $\mathcal{A}$ of the currently most significant features with capacity $B/4$, storing their indices and weights exactly, together with a depth-one signed sketch $\vec{z}\in\mathbb{R}^{B/2}$ for all remaining features, so that the weight of feature $j$ is estimated as
\begin{equation}
    \hat{w}_j
    =
    \begin{cases}
        w_j, & j\in\mathcal{A},\\
        \sigma_j z_{h(j)}, & j\notin\mathcal{A},
    \end{cases}
\end{equation}
where $h:[D]\to[B/2]$ and $\sigma:[D]\to\{\pm 1\}$ are random hash and sign functions.
The stored indices, weights, and counters total exactly $B$ registers.
After each update, a feature outside the active set is promoted whenever its estimated weight exceeds the smallest active weight in magnitude, and the evicted weight is folded back into the sketch so that no information is discarded.
The parameter choice of the active set size $B/4$ follows the best configuration reported in the original work \cite{tai2018sketching}.
}

\revise{
MISSION \cite{aghazadeh2018mission} with total machine size $B$ maintains a count sketch of the weight vector of depth three.
Let $k=\lfloor B/4\rfloor$ and $q=\lfloor(B-2k)/3\rfloor$.
The count sketch consists of three arrays of signed counters $\vec{z}_1, \vec{z}_2, \vec{z}_3\in\mathbb{R}^{q}$, equipped with independent random hash and sign functions $h_r:[D]\to[q]$ and $\sigma_r:[D]\to\{\pm 1\}$ for $r\in\{1,2,3\}$.
Every update of value $g$ to the weight of feature $j$ is recorded by adding $\sigma_r(j)\, g$ to the counter $z_{r, h_r(j)}$ for each $r\in\{1,2,3\}$, and the weight of feature $j$ is estimated as the median of the three signed counters,
\begin{equation}
    \hat{w}_j
    =
    \operatorname{median}_{r\in\{1,2,3\}}
    \sigma_{r}(j)\, z_{r, h_r(j)},
\end{equation}
which is robust against hash collisions.
In addition, MISSION exactly maintains a set $\mathcal{K}$ of the $k$ features with the largest estimated magnitudes, storing their indices and estimated weights, so that the counters, indices, and weights total $2k+3q\leq B$ registers.
Since the original work does not prescribe the division of memory between the count sketch and the exactly maintained set, we use the same even split between exact and sketched storage as in AWM-Sketch.
Its predictions use only the features in $\mathcal{K}$, forming a sparse model $\vec{x}\cdot\hat{\vec{w}} = \sum_{j\in\mathcal{K}}\hat{w}_j x_j$; each update adds the gradient into the sketch at every non-zero coordinate of the sample and refreshes the membership of $\mathcal{K}$.
As the original work optimizes the unregularized loss, we adapt it to the shared objective $\mathcal{L}$ by applying the same regularization decay as the other methods.
}

\begin{figure}
    \centering
    \includegraphics[width=1\linewidth]{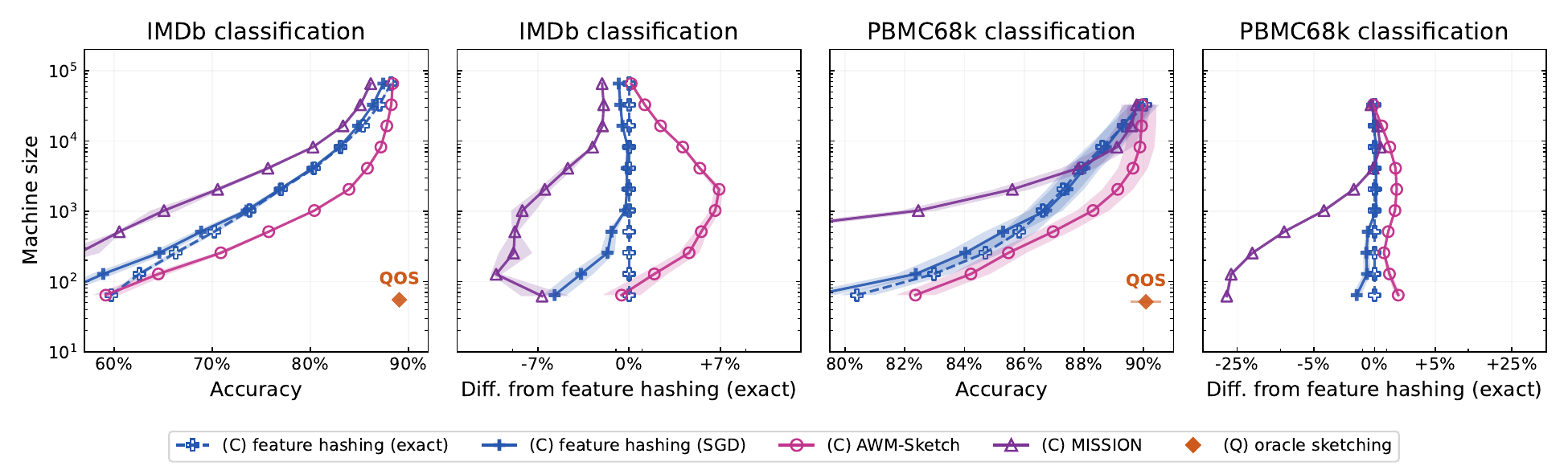}
    \caption{
    \revise{
    \textbf{Additional experiments with adaptive streaming algorithms.}
    Terminal classification accuracy achieved under designated machine sizes on IMDb and PBMC68k, using adaptive streaming algorithms trained on a stream of uniformly random training samples: Active-set Weight-Median Sketch (AWM-Sketch, magenta) \cite{tai2018sketching} and MISSION (purple) \cite{aghazadeh2018mission}, along with feature hashing baselines obtained from the exact LS-SVM solutions (dashed blue) and by streaming stochastic gradient descent (solid blue).
    The orange diamond marks the terminal performance of quantum oracle sketching.
    The difference from exact feature hashing is plotted in the symmetric log scale, where the axis is linear within $\pm 5\%$.
    Differences are evaluated at equal measured machine size by interpolating the feature hashing baseline, since the measured register counts of the adaptive algorithms can sit slightly below the designated sizes due to integer divisions.
    Shaded regions indicate standard error.
    }
    }
    \label{fig:survey-adaptive}
\end{figure}

\revise{
The reported machine sizes are measured directly from the algorithms, counting every feature index, weight, and sketch counter maintained.
To disentangle the expressivity of the maintained model from the effect of the optimizer, we evaluate feature hashing both with exactly computed terminal weights and trained by the same streaming SGD; the reference point of quantum oracle sketching is computed under the same train-test split.
}

\revise{
As shown in \Cref{fig:survey-adaptive}, the adaptive streaming algorithms do not materially change the quantum advantage conclusions.
AWM-Sketch surpasses exact feature hashing at most machine sizes on IMDb, PBMC68k, and 20 Newsgroups, by up to $6.9$, $2.0$, and $4.1$ percentage points respectively.
The largest gains are at intermediate sizes where hashing collisions hurt most, whereas at the largest sizes the gains shrink to within $0.2$ percentage points as the hashed model itself approaches the full solution.
This is expected, as its predictor extends the hashed linear model with exactly tracked weights for the heaviest features.

On Dorothea, AWM-Sketch matches feature hashing to within $0.5$ percentage points above $4\times 10^3$ registers.
Note that Dorothea only has $920$ samples in its training set, each with $10^5$ binary features.
This leads to the degenerate performance of streaming SGD on Dorothea below about $10^3$ registers for every method, and the adaptive curves enter the plotted accuracy window of \Cref{fig:more-numerics} only above this size.

On the other hand, MISSION never improves over feature hashing on IMDb and 20 Newsgroups and improves by at most $0.6$ percentage points on PBMC68k and Dorothea.
This is because its count sketch compresses the storage of the weight vector, but it only uses the recovered top-$k$ features for predictions.
In comparison, the feature hashing baseline aggregates contributions from all features through its maintained bucket weights.

These results show that, even with these more sophisticated adaptive algorithms, every classical streaming algorithm tested still needs orders of magnitude more memory (more than $10^4$ classical registers on IMDb, PBMC68k, and 20 Newsgroups, and more than $2\times 10^3$ on Dorothea) to match the terminal performance of the quantum algorithm, which only uses fewer than $60$ logical qubits.
}

\begin{figure}
    \centering
    \includegraphics[width=1\linewidth]{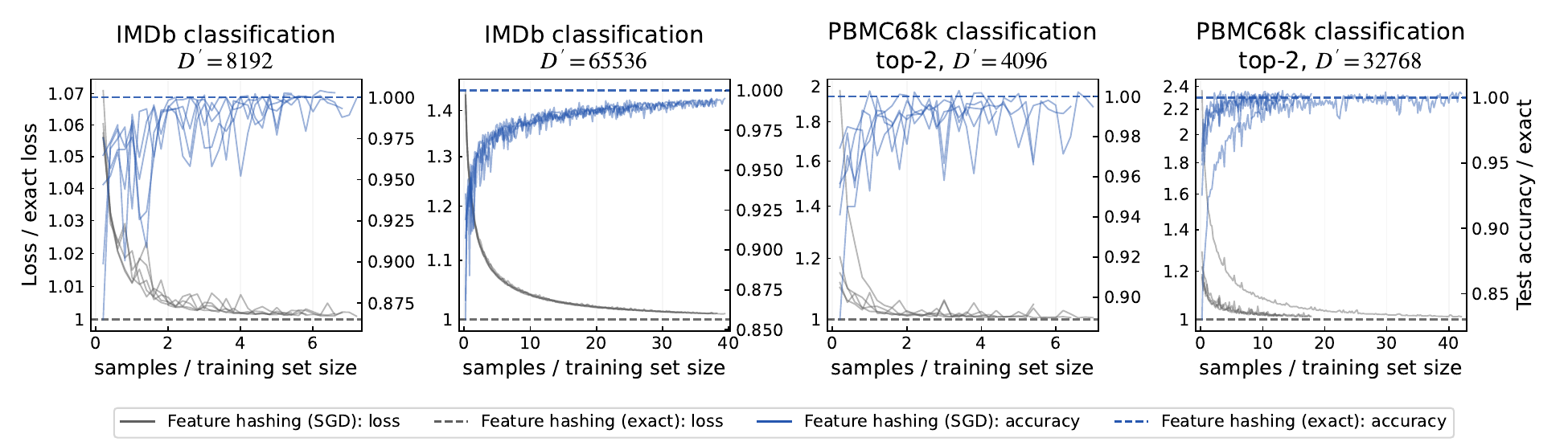}
    \caption{
    \revise{
    \textbf{Convergence of streaming algorithms to the terminal solution at fixed machine size.}
    Example training dynamics of feature-hashing stochastic gradient descent at machine sizes $D'=8192, 65536$ on IMDb, and $D'=4096, 32768$ on the PBMC68k pair of the two largest cell-type classes (CD8+ Cytotoxic T versus CD8+/CD45RA+ Naive Cytotoxic).
    Each curve represents one repetition with an independent random seed, showing the training ridge loss function relative to the exact minimizer of the same problem under feature hashing on the left axis in log scale, and the held-out accuracy relative to the same exact solution on the right axis, as functions of the number of processed samples in units of effective passes over the training set.
    As the hashed ridge loss is strictly convex, the exact solution, plotted in dashed lines, is its unique global minimum and the streaming implementation can only converge to it from above.
    }
    }
    \label{fig:survey-convergence}
\end{figure}

\revise{
\textbf{Convergence of streaming SGD.}
Third, we demonstrate that the terminal points plotted in \Cref{fig:numerics,fig:more-numerics} remove finite-sample optimization error from the classical streaming baselines by crediting each fixed feature representation with its exact converged solution.
For classification in any fixed truncated feature space, the ridge objective $\mathcal{L}$ is strictly convex, so the exact solution $\vec{w}^\star$ plotted in the figures is its unique global minimum, and every SGD iterate satisfies $\mathcal{L}(\vec{w}_t)\geq \mathcal{L}(\vec{w}^\star)$.
Processing additional samples can therefore reduce the optimization gap toward $\vec{w}^\star$, but it does not change the exact memory-limited ridge solution.
For PCA, the exact leading principal component used in the main figures likewise removes the finite-time optimization error of Oja's algorithm under the eigengap and initialization conditions of the PCA task.
Thus, the main figures compare the converged solutions associated with the fixed memory-limited feature representations, rather than finite-sample optimizer transients.
}

\revise{
In \Cref{fig:survey-convergence}, we verify this behavior directly, displaying the training dynamics of feature-hashing SGD at two machine sizes per dataset and plotting for each repetition the relative objective $\mathcal{L}(\vec{w}_t)/\mathcal{L}(\vec{w}^\star)$ and the held-out accuracy relative to that of $\vec{w}^\star$, where $\vec{w}^\star$ is the exact minimizer of the same problem after feature hashing.
To reduce computational burdens for PBMC68k, we focus on the pre-defined pair of the two largest cell-type classes, CD8+ Cytotoxic T versus CD8+/CD45RA+ Naive Cytotoxic.
As guaranteed by convexity, the observed training objective remains above and approaches the exact minimum, while the held-out accuracy stabilizes within $2\%$ of the exact solution within tens of passes over the training set.
Because held-out accuracy is not the optimized objective, it need not be monotone or upper-bounded by the accuracy of the exact ridge solution at every finite time.

We note that the number of passes the classical streaming algorithms need to converge to their terminal performance is expected to be much smaller than that of the current quantum oracle sketching methods due to the effect of shot noise and sparsity dependence.
While they can be estimated by analytical worst-case sample bounds (e.g., the $N \cdot \poly(s, 1/\gamma)$ ones presented in our theorems), these estimates are typically loose on empirical datasets where correct classification does not necessarily require accurate simulation of the perfect circuit.
A naive estimation from our theoretical bounds would yield impractical sample counts on real-world datasets and we believe that measuring the actual empirical shot requirement and convergence behavior would require end-to-end circuit simulations. 
For example, in the naive Hadamard test regime, the shot complexity scales as $O(1/\gamma^2)$, which entails a substantial overhead in data passes compared to classical streaming baselines (e.g., on the order of $10^8$ if $\gamma \sim 10^{-4}$). 
Yet being inaccurate in the estimation does not necessarily hurt the classification accuracy.
Hence a stage-separated worst-case estimate without end-to-end circuit simulations might reach misleading conclusions.
We therefore explicitly restrict our current empirical claims to the memory footprint regime and exclude discussion on the sample complexity by assuming sufficient polynomial sample sizes for both quantum and classical algorithms.
That means, with the current methods, the orders-of-magnitude memory reductions on the empirical datasets likely come with the cost of consuming more samples, which is tolerable in the limited-space with abundant data scenario considered in this work.
Circuit-level simulations for a better understanding of the end-to-end sample and time complexity of the quantum algorithms are therefore an important open direction that will make our methods more practically relevant.

Nevertheless, at fixed machine size, additional samples only reduce optimization error of the classical streaming algorithms within the maintained feature representation and cannot remove the approximation induced by that memory-limited representation.
Therefore, when both are granted sufficient samples, quantum oracle sketching converges to the exact full-data solution with fewer than $60$ logical qubits, whereas the classical streaming representations reach comparable terminal performance only after their machine size grows by orders of magnitude.
The implementation and complete experiment records of all the additional studies in this section are available in the \href{https://github.com/haimengzhao/quantum-oracle-sketching}{GitHub} repository accompanying this work.
}

\subsection{Benchmarking quantum oracle sketching}

To benchmark the performance of quantum oracle sketching, we implement the code numerically simulating it in JAX \cite{jax2018github}, which supports GPU/TPU execution and other features including automatic differentiation and vectorization.
We provide the code at \href{https://github.com/haimengzhao/quantum-oracle-sketching}{GitHub}.
We implement two versions of quantum oracle sketching: one using randomly sampled data, and one using the expected unitary.
The QSVT subroutines used in quantum oracle sketching are implemented in a memory-efficient way that takes advantage of the diagonal and direct-sum structures in quantum oracle sketching.
The QSVT rotation angles are generated via the pyqsp package \cite{martyn2021grand} and converted to circuit phases via the map in \cite{dong2021efficient}.

The implementation using randomly sampled data calculates random instances of the implemented unitary $V$.
That means when we calculate the trace distance of the output state, the sample complexity shows a $1/\epsilon^2$ scaling as predicted in \Cref{lem:sample-upper-concent} rather than the $1/\epsilon$ scaling in a quantum experimental implementation.
The $1/\epsilon$ scaling is only recovered when we calculate the error in terms of physically measurable quantities like observable expectation values or infidelity.
This feature makes the simulation extremely hard using this implementation with randomly sampled data, as it requires quadratically more samples that quickly blows up the memory for classical simulation.

To ease numerical simulation, we mainly focus on the second implementation using the expected unitary $\E[V]$.
\Cref{lem:diamond-operator-expect} guarantees that the error in diamond distance of the physical implementation is always upper bounded by the error in operator norm of the expected unitary, up to constants.
Hence, the reported error in \Cref{fig:benchmark} is a conservative upper bound on the actual implementation error using the corresponding number of samples.

We benchmark quantum oracle sketching for four kinds of oracles: Boolean functions, vectors, matrix element, and matrix index.
For Boolean function, we uniformly sample $100$ random truth tables of Boolean functions for each of $12$ dimensions ranging from $10^2$ to $10^3$ and each of $10$ sample sizes from $10^5$ to $10^8$.
We use quantum oracle sketching to assemble their phase oracles and calculate the corresponding errors in operator norm.

For vectors, we uniformly sample $10$ unit vectors for each of $12$ dimensions ranging from $10^2$ to $10^3$ and each of $10$ sample sizes from $10^5$ to $10^8$.
We use quantum oracle sketching to assemble their state preparation unitary without amplitude amplification and calculate the corresponding errors in Euclidean norm.
The norm of the resulting un-normalized quantum states are selected to be $1/(5\arcsin(1))\approx 0.127$ to ease the implementation of the $\arcsin$ function via QSVT used in quantum state sketching.

For sparse matrix element oracles, we fix the dimension of the matrices to be $100\times 100$.
Then we randomly sample $N_{\mathrm{nnz}}$ coordinates and fill them with numbers sampled uniformly from $[-1, 1]$.
We sample $200$ such random sparse matrices for each of $10$ values of $N_{\mathrm{nnz}}$ ranging from $250$ to $2000$ and each of $10$ sample sizes from $10^5$ to $10^8$.
We use quantum oracle sketching to assemble the sparse matrix element oracles and calculate the corresponding errors in operator norm.

For matrix sparse element oracles, we consider the scenario where we have access to the whole vectors of randomly sampled rows, as in binary classification or dimension reduction.
The goal is to implement the sparse row index oracle.
We fix row sparsity of the matrices to be $8$ and randomly sample $8$ columns for each row.
We fill these selected entries by values uniformly sampled from $[-1, 1]$.
We sample $5$ such random matrices for each of $6$ dimensions from $50\times 50$ to $500 \times 500$ and each of $10$ sample sizes from $10^5$ to $10^8$.
We use quantum oracle sketching to assemble the sparse row index oracle and calculate the errors in operator norm.

The results in \Cref{fig:benchmark} show that all the sample complexity scaling observed in numerical simulation agrees accurately with the theoretical prediction.
To extract the constants and exponents, we take logarithm of the sample sizes, dimensions, and errors, and perform a least-squares fit.
The residual error of the least-squares fit translates into the root-mean-squared relative errors, which are all below $3\%$.

\section{Overview and related works}
\label{sec:rel-work}

In this section, we discuss the relation between our work and the existing literature and provide an overview of our results.
We begin by providing background on the development of quantum algorithms for data processing tasks.
Then we explain the various challenges identified in the literature in reaching quantum advantages in useful classical tasks and how we tackle them.
Finally, we discuss classical space-efficient algorithms, techniques for proving space lower bounds, and how our results improve existing quantum space advantages in streaming.

\subsection{Machine learning and quantum algorithms}

\textbf{Quantum machine learning.}
Machine learning is one of the most important frontiers of modern computation, owing to the ubiquitous role of data processing and our ever-growing need for it throughout society.
Hence, it is natural to expect quantum computation, as a fundamentally new paradigm of computation, to revolutionize machine learning by providing new forms of data and new approaches to data processing \cite{biamonte2017quantum}.
Indeed, in the realm of scientific discoveries aiming to probe inherently quantum properties of Nature, people have rigorously established exponential advantages of using quantum machines to process quantum data \cite{huang2022quantum,chen2022exponential,chen2024tight,oh2024entanglement,liu2025quantum,aharonov2022quantum,allen2025quantum}.
These advantages of using quantum machines for quantum tasks are promising in facilitating science and engineering in the quantum frontier.
Yet they appear too specialized to have major impacts on our everyday life, where classical data abound and directly observable quantum effects are negligible.

In the realm of classical data processing where most real-world applications sit, whether quantum machines can deliver large advantages over classical machines is much less evident.
Existing attempts to rigorously prove quantum advantages in classical data processing rely on designing contrived classical tasks by secretly embedding problems that are inherently either quantum (e.g., entanglement, contextuality, or random circuit sampling) or cryptographic (e.g., factoring) \cite{zhao2025entanglement,gao2018quantum,gao2022enhancing,anschuetz2023interpretable,anschuetz2026arbitrary,zhang2024quantum,liu2021rigorous,gyurik2023exponential,huang2025generative}.
The limitation of this approach is that the proposed quantum algorithms are often designed for the specific contrived tasks to prove advantages and it is not clear how they apply to realistic problems. 

An alternative route is to empirically test the performance of heuristic quantum machine learning methods directly on realistic data, inspired by the empirical success of deep learning \cite{deeplearningbook}.
This approach is often referred to as variational quantum algorithms, parameterized quantum circuits, or quantum neural networks (see e.g., \cite{cerezo2021variational,cerezo2022challenges,du2025quantum,wang2023variational}).
However, extensive studies have revealed the intrinsic hardness in training such variational quantum algorithms, manifested in the forms of exponentially vanishing gradient \cite{mcclean2018barren,cerezo2021cost,wang2021noise,larocca2025barren} and exponentially many bad local minima \cite{anschuetz2022quantum}.
Furthermore, even if we can train them, their functional expressivity suffers from the same curse of dimensionality issue as classical neural network \cite{zhao2024learning}.
In fact, these discouraging results have sparked active debate on whether practical quantum advantage is even possible in classical machine learning tasks and whether trainable quantum neural networks are always classically simulatable \cite{schuld2022quantum,cerezo2023does,gil2024relation}.

\revise{Our work directly tackles the central problem of \emph{broadly applicable and provable} quantum advantage for classical data processing, by rigorously establishing exponential quantum advantages in a wide range of realistic classical machine learning tasks and providing numerical evidence for these advantages on real-world datasets.}
A promising future direction is to further enhance it with heuristic variational methods.

\vspace{1em}

\textbf{Quantum algorithms for classical tasks.}
In the quest for quantum advantages in widely-useful classical tasks, numerous quantum algorithms with rigorous performance analysis have been extensively studied.
Some prominent examples include quantum algorithms for linear systems \cite{harrow2009quantum}, recommendation systems \cite{kerenidis2017quantum}, support vector machines \cite{rebentrost2014quantum}, principal component analysis \cite{lloyd2014quantum}, generative adversarial learning \cite{lloyd2018quantum}, ordinary and partial differential equations \cite{leyton2008quantum,montanaro2016quantum}, convex optimization \cite{chakrabarti2020quantum,van2020convex}, Monte Carlo sampling \cite{montanaro2015quantum,harrow2020small}, neural network inference \cite{zlokapa2021quantum,guo2024quantum,smaldone2025hybrid}, etc.
Many of these algorithms can be understood in the unified framework of quantum linear algebra and quantum singular value transform (QSVT) \cite{gilyen2019quantum,martyn2021grand}.
See \cite{dalzell2025quantum} for a recent survey.
The performance of these quantum algorithms are usually analyzed not in an end-to-end fashion, but in terms of the number of queries they make to specific oracles that provide coherent access to the classical data.
Such oracles include quantum random access memory (QRAM), block encodings and sparse oracles of matrices, state preparation unitaries of vectors, etc.
When we take into account the cost of building such oracles, whether large quantum advantage remains becomes unclear except for a few highly specialized and structured problems \cite{clader2013preconditioned}.

\subsection{The challenging quest for quantum advantage}

\textbf{Data loading and QRAM.}
Most existing quantum algorithms for classical data processing rely on quantum random access memory (QRAM) \cite{giovannetti2008quantum}, which is a primitive that models quantum coherent access to classical data.
The quantum advantages of these algorithms survive only when QRAM can be efficiently realized \cite{aaronson2015read}.
However, how to build an efficient and fault-tolerant QRAM remains largely unclear \cite{jaques2025qram}.

To load $N$ classical bits into a quantum machine, it is unavoidable to use $\Omega(N)$ gates since each gate only has a constant number of degrees of freedom.
The idea of a QRAM is to parallelize these gates to reduce the circuit depth (and ideally the wall-clock time) to $\poly(\log N)$ by using $O(N)$ ancilla qubits.
This makes QRAM extremely memory inefficient, even worse than classical machines.
It is also unclear how to make QRAM fault tolerant with efficient control \cite{arunachalam2015robustness,hann2021resilience}, as $O(N)$ classical co-processors and classical computation might be needed to perform error correction on these $O(N)$ ancilla qubits \cite{dalzell2025distillation}.
These many classical co-processors might as well be repurposed to perform classical parallel computation and solve the target classical tasks themselves classically \cite{jaques2025qram}.

\revise{
An alternative route, called circuit QRAM or QROM, circumvents this issue by working in the standard fault-tolerant circuit model and applies $O(N)$ gates with few or no ancilla qubits \cite{babbush2018encoding}.
This makes it fault-tolerant at the expense of killing large time advantages.
Moreover, it is still space inefficient since the $O(N)$ bits specifying these gates have to be stored somewhere classically, and therefore the total space (classical and quantum) is still $O(N)$ \cite{ashhab2022quantum}, even though the qubit count may be only $\poly(\log N)$.
As a result, existing quantum algorithms that rely on QRAM have total space $O(N)$ in general.

It is possible to avoid retaining these $O(N)$ bits describing the gates only under a stronger data-access model.
For example, if an external classical procedure allows querying $f(x)$ at a chosen address $x$ at will, or can enumerate every pair $(x, f(x))$ exactly once, then one can accumulate the corresponding phase oracle online with circuit QROM using only logarithmic quantum workspace.
However, these access models cannot be easily constructed when data arrive as random samples from an underlying distribution in the standard machine learning model \cite{shalev2025online}.
For a generic massive dataset, such access models may only be reasonable when a massive data center already stores the dataset and maintains a proper data structure indexing it, which reintroduces the $O(N)$ memory cost.
}

In contrast, our quantum oracle sketching algorithm circumvents QRAM and achieves its goal of loading classical data into a quantum machine in a space-efficient and fault-tolerant way.
Our results can be viewed as a canonical data loading scheme that only consumes $\poly(\log N)$ space in total, improving exponentially over QRAM.
Moreover, in the space-efficient regime, the $\tilde O(N)$ gate complexity is optimal up to logarithmic factors due to the counting argument above.

\vspace{1em}

\textbf{Dequantization and noisy query.}
Even if we abstract away the implementation of QRAM, the mere assumption that we have efficient (i.e., $\poly(\log N)$ time) coherent access to classical data might be too strong.
An equivalent classical data access model, called sample and query access, have been shown to allow classical algorithms with only polynomial slowdown compared to the quantum algorithms.
Such classical algorithms can be systematically designed through the program of dequantization \cite{tang2023quantum,tang2019quantum,tang2021quantum}.
The existence of these quantum-inspired classical algorithms reduces many exponential quantum advantages to polynomial ones, making them less practically relevant.

Another source of quantum advantage diminishment is noise.
Despite that we can perform error correction to suppress the noise in our computing machines, we may not have full control over the entity providing the data.
As examples, one may think of typographical errors in a piece of text, the ubiquitous $1/f$ noise in electronics, or fluctuations in a magnetic field that we are trying to sense.
It has been shown that such noisy data may lead to the decrease or complete loss of quantum advantages.
For example, Heisenberg-limit quantum sensing is reduced to the standard quantum limit when noise exists parallel to the signal \cite{zhou2018achieving,huang2025vast}.
The computational speedup of Grover's algorithm for unstructured search \cite{grover1996fast} is also lost in certain noise models \cite{chen2023complexity,regev2008impossibility}.

Our results show that even with noisy data, exponential quantum space advantages persist in useful classical tasks.
Moreover, we show that any super-quadratic query separation leads to exponential space advantage.
This means that many dequantized quantum algorithms, although they might only have polynomial query speedup, they may still have exponential space advantage and hence worth revisiting.

\vspace{1em}

\textbf{Readout.}
Another bottleneck of solving classical tasks with quantum algorithms is the readout of classical results from the quantum states \cite{aaronson2015read}.
Holevo's bound asserts that one can extract at most $\poly(\log N)$ classical bits from $\poly(\log N)$ qubits \cite{holevo1973bounds}.
That means, for example, one can never hope to write out the whole solution vector $\vec{x}\in \mathbb{R}^N$ of an $N$-dimensional linear system using a $\poly(\log N)$-qubit quantum machine without repeating $\tilde{\Omega}(N)$ times.
Hence, it has been widely believed that quantum algorithms can only be used to extract certain quantum-friendly properties of the solution (e.g, a quadratic form specified by an efficiently measurable observable).

However, the takeaway drastically changes when we aim for quantum advantages in memory efficiency.
In fact, a memory-efficient classical machine with $\poly(\log N)$ bits suffers from the same ``readout issue'': one can only extract $\poly(\log N)$ bits from it.
Moreover, any property that one can efficiently extract from a $\poly(\log N)$-bit classical machine can be efficiently extracted from a $\poly(\log N)$-qubit quantum machine as well, meaning that the target properties of the solution need not be quantum-friendly anymore.
In that sense, the readout issue is no longer a caveat of quantum algorithms when we aim for memory efficiency.

For our purposes, we need to efficiently extract the properties we want without destroying sign information, which is critical in applications such as classification and dimension reduction.
To this end, we develop a technique termed \emph{interferometric classical shadow} in \Cref{lem:interf-classical-shadow}, where we combine the idea of the Hadamard test with the efficient offline prediction capability of classical shadows.
It allows us to construct a completely classical model capable of predicting any number of sparse test data.
Remarkably, this demonstrates that quantum technology enables us to construct accurate and exponentially smaller classical models out of classical data, which is provably impossible with any classical machine without exponentially larger memory.
This highly condensed classical model is only efficiently obtainable through quantum technology.
One can also view this classical model as a shadow model \cite{jerbi2024shadows} or a classical surrogate \cite{schreiber2023classical} of the quantum model with exponential quantum advantage.

\vspace{1em}

\textbf{Utility of quantum computation and provable quantum advantages.}
Building quantum computers that are useful in real-world applications is one of the central goals of quantum computation.
Yet despite decades of extensive effort, conclusive evidence of useful quantum advantages has only been established in a few specialized fields such as cryptanalysis and quantum simulation \cite{babbush2025grand}.
This is largely because known computational advantages stem from highly specialized structures in the problems (e.g., Abelian hidden subgroup structures exploited by Shor's algorithm \cite{shor1994algorithms,kitaev1995quantum} and the quantum nature of quantum simulation \cite{feynman1986quantum,lloyd1996universal}).

Computational advantages are especially hard to rigorously prove due to the notorious hardness of proving separations between computational complexity classes.
In fact, an unconditionally proved exponential quantum computational advantage directly implies $\mathsf{BPP\neq BQP}$ and hence $\mathsf{P\neq PSPACE}$, a major open conjecture in theoretical computer science.
As a result, existing useful quantum advantage claims are either proved assuming computational complexity conjectures that are widely-believed but unproven, or proved relative to the use of oracles that suffers from the data loading issue (i.e., how to instantiate the oracles), with a few relatively weaker exceptions designed based on quantum nonlocality \cite{bravyi2018quantum,bravyi2020quantum}.

Our work unconditionally proves exponential quantum space advantages in useful classical tasks, rigorously establishing memory savings as a widely-applicable utility of quantum computation.
From a fundamental perspective, these advantages persist even in the unlikely event that quantum computers turn out to be polynomially equivalent to classical computers in time (i.e., $\mathsf{BPP=BQP}$).
Our proof is information-theoretic and only relies on the exponential vastness of the Hilbert space enabled by quantum superposition.
It reveals that this quantum advantage of exponential vastness, sometimes called quantum information supremacy \cite{aaronson2024qubit}, is a generic feature of super-quadratic query separation, rather than specialized structures embedded in contrived tasks.
An experimental confirmation of our results may serve as a witness of the exponential dimension of the quantum state space \cite{huang2025vast,kretschmer2025demonstrating}, similar to how Bell inequalities witness quantum nonlocality \cite{bell1966problem}.

\subsection{Memory-efficient algorithms and quantum space advantage}

\textbf{Hitting the memory wall.}
Our burgeoning need to process massive classical data has exposed memory capacity as a bottleneck more critical than computational speed, primarily due to its significantly slower advancement.
This observation, colloquially put as ``hitting the memory wall'' \cite{wulf1995hitting}, has become increasingly relevant as large language models prosper; these models, now scaling to trillions of parameters \cite{fedus2022switch}, have grown 410-fold every two years, whereas memory capacity has only doubled over the same period \cite{gholami2024ai}.
Increasing amounts of energy are being consumed for hosting these gigantic models and maintaining massive data centers \cite{strubell2019energy,luccioni2023estimating}.
Extensive studies have been devoted to optimizing memory consumption to enable longer context window and better inference performance, even at the expense of increased computation time \cite{rajbhandari2020zero,jain2020checkmate,dao2022flashattention,dettmers2022gpt3,dettmers2023qlora,kwon2023efficient,ainslie2023gqa}.
While better models require larger training data \cite{kaplan2020scaling}, our storage capability limits the amount of data accessible, placing data such as internet traffic or global market activities beyond reach.
This memory wall also appears in many other areas of science and technology, such as large particle colliders \cite{hep2019roadmap,bejaralonso2020hllhc,khachatryan2016search}, astronomical sky surveys \cite{dewdney2009square}, single-cell RNA sequencing \cite{svensson2018exponential,lahnemann2020eleven,tsuyuzaki2020benchmarking}, the simulation of fluid dynamics \cite{hemati2014dynamic,perlman2007data} and power grids \cite{feng2008multigrid,kozhaya2002multigrid}.

\vspace{1em}

\textbf{Classical memory-efficient algorithms.}
To cope with massive classical data, algorithmic techniques such as sketching \cite{woodruff2014sketching,clarkson2009numerical,andoni2020streaming,mitliagkas2013memory}, streaming \cite{muthukrishnan2005data,alon1996space,flajolet1985probabilistic,misra1982finding,cormode2005improved}, and online learning \cite{shalev2025online} have been developed.
These techniques usually save memory by incrementally updating a model with fresh data samples without ever storing them.
Various simple statistical properties, such as frequency moments \cite{alon1996space}, distinct elements \cite{flajolet1985probabilistic}, and heavy hitters \cite{misra1982finding,cormode2005improved}, can be estimated with high probability using exponentially smaller memory.
See \cite{muthukrishnan2005data} for a survey.
This framework has been extended to numerical linear algebra via random projection techniques based on Johnson-Lindenstrauss transforms \cite{clarkson2009numerical}.
These classical sketching or streaming algorithms approximately solves matrix multiplication, linear regression, and low-rank approximation using space significantly smaller than standard classical algorithms.
See \cite{woodruff2014sketching} for a recent review.
More recent studies have proposed streaming algorithms for SVM \cite{andoni2020streaming}, PCA \cite{mitliagkas2013memory}, etc.
Similarly, online machine learning algorithms also provide memory efficiency, albeit with the main goal of minimizing regret \cite{shalev2025online}.
However, these algorithms often suffer from compromised accuracy or fail to capture hidden correlations in practice \cite{tsuyuzaki2020benchmarking,kirkpatrick2017overcoming,hemati2014dynamic}.
Their space complexity all scales at least as order $D$ for a data matrix $X\in \mathbb{R}^{N\times D}$.
A space lower bound of $\Omega(D)$ for classical machines can be proved in the streaming model via adversarially ordering of the data \cite{woodruff2014sketching}, but it is unknown in the more realistic scenario where data are randomly sampled.
In contrast, our quantum algorithms achieve space complexity $\poly(\log D)$, exponentially improving over these classical streaming or online algorithms.
Our classical lower bounds prove the $\Omega(D^{1-\zeta}), \forall \zeta>0$ lower bound for classical machines with random data access.

\vspace{1em}

\textbf{Sample-space lower bounds and communication complexity.}
Space complexity lower bounds of classical streaming algorithms are often proved via reduction to one-way communication complexity \cite{woodruff2014sketching}, where we consider the past and future of the algorithm as two parties transmitting the memory in the forward direction.
In the streaming model, we are allowed to adversarially choose the order of the data sequence, and hence we can design the hard instance by sequentially concatenating the inputs of the two parties.
This proof strategy no longer works when we have multiple passes over the data or have random sample access as in our learning setting, because the order may be randomly shuffled and one-way communication lower bounds no longer suffice.
We need communication complexity lower bounds for stronger communication models.

A rich suite of techniques has been developed in proving multi-party communication complexity lower bounds. 
One route is via information complexity \cite{bar2004information}.
But its applicability is restricted to specific problems where information quantities can be explicitly calculated \cite{dershowitz2021communication,lovett2023streaming,braverman2024new}.
An alternative approach is via query-to-communication lifting, which is a general framework that lower bounds communication complexity by classical query complexity using the density restoring partition technique \cite{goos2015deterministic,goos2016rectangles,goos2020query,anshu2020query,chattopadhyay2021query,yang2024communication,fei2025multi}.
Another related line of work directly proves classical sample-space lower bounds for various learning problems involving parity in the random query setting \cite{raz2018fast,raz2017time,garg2021memory,dinur2024time,zhan2023randomness}, which is the IID version of our data access for Boolean functions.
However, such problems is hard to quantum computers as well \cite{liu2023memory} and their relation to real-world tasks is unclear.

Contrary to these existing approaches, we consider a learning task of estimating the property of some Boolean function $f: [N]\to \bit$ using noisy random data samples.
We call this task Noisy Oracle Property Estimation (NOPE) and formally define it in \Cref{sec:cl-hard-noisy-oracle-property-est}.
The desired property is specified as a query problem of the Boolean function, while the input data are random samples of a noisy version of its oracle.
For this task, we develop a communication complexity proof technique in \Cref{sec:cl-hard-sample-space-lb} based on a modified query-to-communication lifting and density restoring partition \cite{goos2015deterministic,goos2016rectangles,goos2020query,anshu2020query,chattopadhyay2021query,yang2024communication,fei2025multi}, proving that any classical algorithm solving the task with sample size $M$ and memory $S$ must satisfy $MS\geq \Omega(NQ_C)$, where $Q_C$ is the classical query complexity of the target property.
Combined with the optimal sample complexity $M=\Theta(NQ^2)$ of quantum oracle sketching, this allows us to reveal the fundamental relation between space advantage and oracle query separation: $S\geq \Omega(Q_C/Q^2)$.
If the classical query complexity is exponential, and quantum machines can solve the query problem in $\poly(\log N)$ space with a slightly larger than quadratic query advantage (e.g., $Q_C=\Theta(N), Q=O(N^{0.49})$, which implies $S\geq \Omega(N^{0.02})$), this result unconditionally proves an exponential quantum space advantage.

\vspace{1em}

\textbf{Super-polynomial sample advantage and XOR lemmas.}
We further bootstrap our sample-space lower bound into a super-polynomial sample complexity lower bound when space is constrained.
This is proved using a hybrid approach combined with a new learning XOR lemma in \Cref{sec:cl-hard-bootstrap}, which is similar in spirit with Yao's XOR lemma \cite{yao1982theory} that amplifies hardness by taking the XOR of multiple independent problem instances.
A streaming XOR lemma was developed in \cite{assadi2021graph}, but it is not applicable to our learning setting since their proof heavily relies on the ability to adversarially choose the ordering of the data (i.e., sequentially concatenate all problem instances).
We circumvent this issue by developing a derandomization technique that allows us to prove the learning XOR lemma.
In particular, we consider a dynamic version of the NOPE task, in which we repetitively try to solve $\log^2 N$ independent instances of NOPE simultaneously.
Our learning XOR lemma gives a $1/2^{\log^2 N}=1/\mathrm{superpoly}(N)$ bound on the progress in doing so with each repetition.
Via a hybrid argument, this implies a $\mathrm{superpoly}(N)$ lower bound on the number of repetitions needed and thus the total sample complexity.
Finally, we instantiate these classical hardness results with the Forrelation property of Boolean functions \cite{aaronson2015forrelation,bansal2021kforrelation} that provides the maximal average-case query separation.

\vspace{1em}

\textbf{Lower bounds for real-world applications.}
All the above classical hardness results are proved for Boolean function problems with heavy theoretical computer science flavor.
In \Cref{sec:cl-hard-app}, we connect them to classical data processing and machine learning applications such as linear systems, binary classification, and dimension reduction, by embedding NOPE into these applications.
In particular, we construct a polynomial size quantum circuit that solves the NOPE task of estimating the Forrelation \cite{aaronson2015forrelation,bansal2021kforrelation} property.
We embed it into linear system tasks using the BQP-hardness of matrix inversion \cite{harrow2009quantum}.
We follow a similar idea to prove the BQP-hardness of binary classification and uses a modified Feynman-Kitaev circuit Hamiltonian construction \cite{kitaev2002classical} to prove the BQP-hardness of dimension reduction.
Consequently, the classical hardness results of Forrelation NOPE translates into the desired classical hardness results of the various applications.

\vspace{1em}

\textbf{Models of memory-bounded computation.}
Our model of classical learning algorithms is the same as that of branching programs, which is the standard non-uniform model for space-bounded classical computation.
It is stronger than online Turing machine and hence our classical lower bounds carries over directly.
Our formal model of quantum learning algorithms can be viewed as a model of quantum branching programs that is allowed to apply arbitrary quantum channels depending on the input data.
This model improves over existing proposals of quantum branching programs \cite{nakanishi2000ordered,ablayev2001computational,bera2023generalized} that are too restricted to be realistic, as they can only apply unitary channels and hence cannot even simulate irreversible classical computation.

\vspace{1em}

\revise{
\textbf{Quantum space advantage in streaming.}
In the standard Turing machine model, where the data storage costs are ignored, a path integral argument shows that no super-quadratic quantum space advantage exists \cite{watrous1999space,watrous2003complexity}.
In the streaming model where we take into account the costs of storing the input data, but the order of the data may be adversarially chosen, one-way quantum communication advantage can be used to prove quantum space advantage in streaming.
Examples of such one-way quantum communication advantages include the vector in subspace problem \cite{raz1999exponential,gilboa2024exponential} and the Boolean Hidden Matching problem \cite{gavinsky2007exponential}.
Such quantum communication advantages lead to quantum space advantage in some tasks such as triangle counting in graphs and Max Directed Cut \cite{le2006exponential,jain2014space,kallaugher2022quantum,kallaugher2024exponential,kallaugher2025design}.
Variants of Boolean Hidden Matching were recently demonstrated experimentally with trapped-ion systems \cite{kretschmer2025demonstrating,niroula2025realization}.
However, these tasks have limited application scope and many of their proof strategies heavily rely on the adversarial ordering of the data. 
Some of the quantum space advantages proved there \cite{kallaugher2024exponential,niroula2025realization} become classically easy when the data are randomly sampled \cite{saxena2023streaming}.
In contrast, our quantum space advantages persist when the data are randomly sampled and apply to widely-useful applications such as linear systems, binary classification, and dimension reduction.
}

\newpage

\section{Models of data access and computation}
\label{sec:model}

\begin{figure}
    \centering
    \includegraphics[width=1\linewidth]{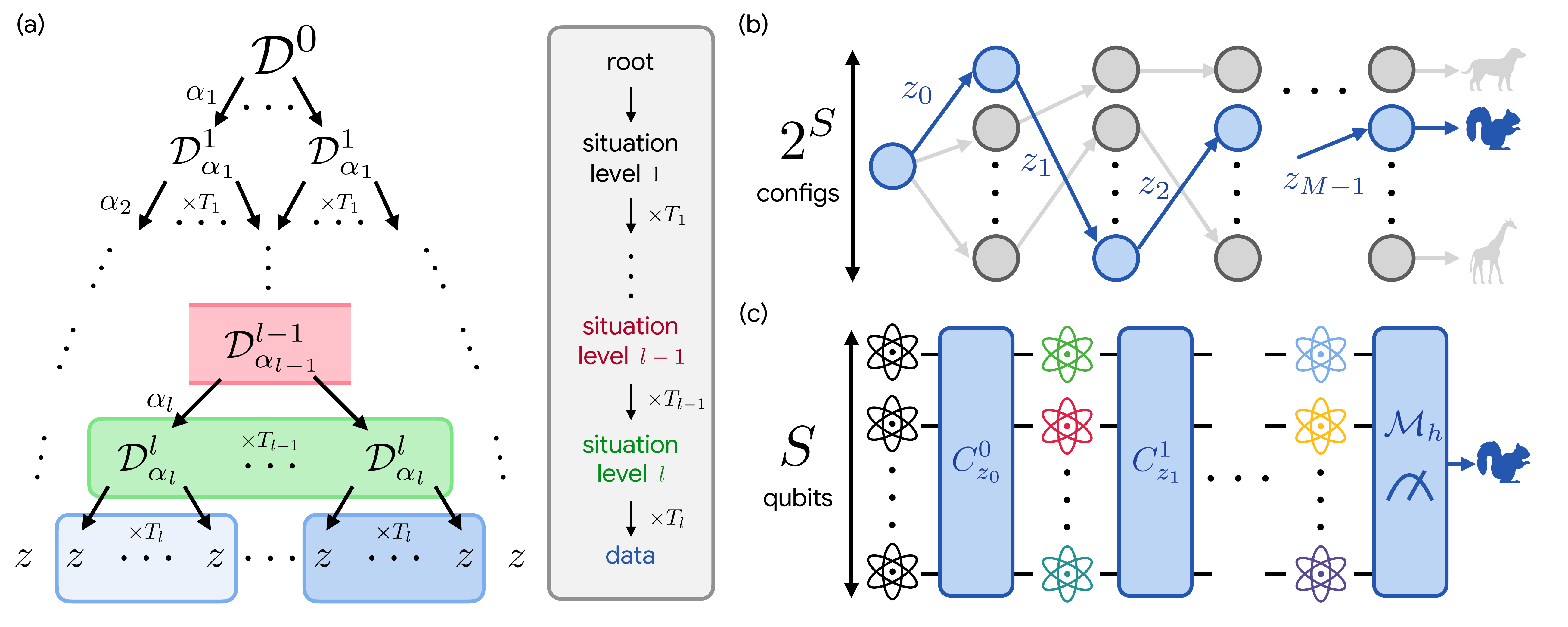}
    \caption[Overview of the models of data access and computation.]{
    \textbf{Overview of the models of data access and computation.}
    \textbf{(a)} Illustration of the tree structure of a hierarchical data generation process with $l$ situation levels, each has time scale $T_1, \ldots, T_l$.
    Random variables within the same box are IID conditioned on their shared latent situations.
    \textbf{(b)} The model of classical learning algorithms with size $S$ (i.e., $2^S$ possible configurations) and sample complexity $M$.
    The computation path when the algorithm is given a sequence of data $z_0, \ldots, z_{M-1}$ is highlighted in blue.
    \textbf{(c)} The model of quantum learning algorithms with size $S$ and sample complexity $M$.
    Upon receiving a sequence of data $z_0, \ldots, z_{M-1}$, the algorithm applies a series of quantum channels $C^0_{z_0}, \ldots, C^{M-1}_{z_{M-1}}$ and measures the final state to compute the outcome.
    }
    \label{fig:access-model}
\end{figure}

In this section, we formally introduce the models of data access and computation that we consider in this work.
They are summarized in \Cref{fig:access-model}.
In \Cref{sec:data-access}, we describe a model, dubbed hierarchical data generation processes, of how (classical) data samples are generated in the world in a dynamic and possibly correlated way.
This includes IID data as a special case, and also includes data that have time-varying features and correlation with multiple time scales (e.g., the $1/f$ noise ubiquitous in electronics, user activity that changes over time, etc.).
The goal of learning is to use these data samples to infer some property of the underlying data generation process (e.g., some rule for prediction or classification).
This general data access model will be specialized to various applications in \Cref{sec:app}.
Some useful properties of such data generation processes are proved in \Cref{sec:data-gen-property}.

To model the computation in learning, we formally define classical learning algorithms in \Cref{sec:cl-learning-alg} and quantum learning algorithms in \Cref{sec:q-learning-alg}.
We define their key properties such as size (i.e., space complexity), sample complexity, and time complexity.
In later sections, we will see that our quantum learning algorithm works for any hierarchical data generation processes and we will rigorously prove that any classical learning algorithm cannot work unless they have exponentially larger size.

\subsection{Data generation processes}
\label{sec:data-access}

The world is full of randomness and noise and in general has to be modeled probabilistically.
In a typical data processing or learning task, one collects $M$ data samples $z_1, \ldots, z_M$ from some underlying distribution in the world that generates the data.
The goal is to learn some property of the underlying distribution based on these data samples.
The data could represent a set of experimental observations for scientific discoveries, coefficients of a differential equation in simulation or real world, user data from internet activities, stock prices from financial markets, health care data from biomedical sensors, or interaction activities from an embodied robot.
The property could be as simple as some statistical property (e.g., mean and variance) of the data, or as complicated as a model of how to predict the next data sample (e.g., predicting the next word in language modeling) or a rule for classification.
As a starting point, one may model the underlying data generation process as drawing a sequence of independent and identically distributed (IID) data points
\begin{equation}
    z_i \sim \mathcal{D}, \quad i\in [M],
\end{equation}
where $\mathcal{D}$ is the underlying distribution whose property we want to learn about.
\revise{
We note that this is a standard data access model studied in online machine learning \cite{shalev2025online}.
}

We can take a binary classification task as an example.
In a typical binary classification task, one may think of the data points $z_i$ as
\begin{equation}
    z_i = (\vec{x}_i, y_i) \sim \mathcal{D},
\end{equation}
where $\vec{x}_i\in \mathbb{R}^D$ is the $D$-dimensional feature vector of a training data point that we collect from the world and $y_i\in \{\pm 1\}$ is the corresponding label.
The task is to generate a model that allows us to predict the label of another set of $m$ test data points $\vec{x}'_{j}, j\in [m]$.
For example, the simplest least-square support vector machine (LS-SVM) algorithm aggregates the training data into the form $X = (\vec{x}_1, \ldots, \vec{x}_M), \vec{y}=(y_1, \ldots, y_M)$ and makes the prediction using the following rule
\begin{equation}
    \hat{y}_j = \sgn(\vec{x}'_j \cdot (X^TX)^{-1} X^T\vec{y}), \quad j\in [m].
\end{equation}
This classification rule is the property of the underlying distribution $\mathcal{D}$ that we want to learn about.

Another cleaner, yet more theoretical example is learning properties of a Boolean function.
Consider an unknown Boolean function $f: [N]\to \bit$.
We want to learn its properties, but we do not have query access to this function.
Instead, we have a sequence of data points of the form
\begin{equation}
    z_i = (x_i, f(x_i)), \quad x_i\sim \unif([N]),
\end{equation}
that are random queries to the function $f$.
The goal is to learn some property of this function $f$, specified by say a query algorithm $\mathcal{A}(f)$.
In this example, the distribution $\mathcal{D}$ of the data samples $z_i=(x_i, f(x_i))$ is in one-to-one correspondence with the underlying function $f$.
The property of $f$ is the property of $\mathcal{D}$ that we want to learn about.

The real world is dynamic and the IID assumption is often broken by correlation among the data samples.
Such correlation could arise from a time-dependent data generation process (e.g., big things happen in the real world and the distribution changes correspondingly).
It could also come from time-correlated noise in the data that may have multiple time scales (e.g., $1/f$ noise that is ubiquitous in electronic devices).

To model data samples having noisy, time-varying features with multiple time scales, we consider the following hierarchical data generation process, where the way data are generated at each time step depends on the current \emph{situation}.
The situation changes with time and has multiple time scales specified in a hierarchical way.
We start from a root (highest level) probability distribution $\mathcal{D}^0$ supported on a finite set of possible labels $A_1$ that labels a set of probability distributions on the first level 
\begin{equation}
    \{\mathcal{D}^1_{\alpha_1}: \alpha_1\in A_1\}.
\end{equation}
We sample a random $\alpha_1\in A_1$ from $\mathcal{D}^0$, which specifies a distribution $\mathcal{D}^1_{\alpha_1}$.
We will make $T_1$ independent draws from this $\mathcal{D}^1_{\alpha_1}$ and after that we will resample a fresh instance of $\alpha_1$ from $\mathcal{D}^0$.
In this way, the parameter $T_1$ specifies the first time scale of the data generation process.

Each distribution $\mathcal{D}^1_{\alpha_1}$ on the first level is supported on another finite set of possible labels $A_2$ that labels a set of probability distributions in the second level
\begin{equation}
    \{\mathcal{D}^2_{\alpha_2}: \alpha_2\in A_2\}.
\end{equation}
Similar to the first level, now we sample a random $\alpha_2\in A_2$ from $\mathcal{D}^1_{\alpha_1}$, which specifies a distribution $\mathcal{D}^2_{\alpha_2}$.
We will make $T_2$ independent draws from this $\mathcal{D}^2_{\alpha_2}$ and after that we will resample a fresh instance of $\alpha_2$ from $\mathcal{D}^1_{\alpha_1}$.
Hence, the parameter $T_2$ characterizes a second time scale of the data generation process.

Suppose this goes on until we reach the $l$-th level and we have sampled a distribution $\mathcal{D}^l_{\alpha_l}$ that is supported on the data space $\mathcal{Z}$.
We define the current \emph{situation} to be the sampled labels $(\alpha_1, \ldots, \alpha_l)$.
For simplicity, we assume that we have included the information of the current situation (i.e., which distribution we are sampling from) into the data by including all $\alpha_1, \ldots, \alpha_l$ in $z$.
For example, $\mathcal{Z} = \mathcal{X}\times \mathcal{Y}\times A_1\times \cdots\times A_l$ and the generated samples are of the form $z_i = (x_i, y_i, \alpha_{1,i}, \ldots, \alpha_{l, i})$.

We will make $T_l$ independent draws of $z$'s from this $\mathcal{D}^l_{\alpha_l}$ and after that we will resample a fresh instance of $\alpha_l$ from $\mathcal{D}^{l-1}_{\alpha_{l-1}}$.
All together, this describes an $l$-level hierarchical data generating process with $l$ different time scales $(T_1, \ldots, T_l)$.
We use the following notation to denote this data generation process:
\begin{equation}
    \mathcal{D}=(\mathcal{D}^0 \to \mathcal{D}^1_{\alpha_1} \to^{\times T_1} \mathcal{D}^2_{\alpha_2}\to^{\times T_2} \cdots \to^{\times T_{l-1}} \mathcal{D}^l_{\alpha_l} \to^{\times T_l} z).
\end{equation}
Note that in such a hierarchical data generating process $\mathcal{D}$, the marginal distribution of each individual data sample $z_i$ is the same.
This ensures that there is a persistent signal present in the data.
Yet they are not independently distributed as there may be correlation mediated by higher levels of the hierarchy with the corresponding time scales.
This poses an apparent challenge for learning.

We can define a few characteristic properties of a hierarchical data generating process $\mathcal{D}$.
We define the \emph{refreshing time} $\tau_{\mathcal{D}}$ of $\mathcal{D}$ to be the product of all time scales
\begin{equation}
    \tau_{\mathcal{D}} = T_l\cdots T_1.
\end{equation}
$\tau_{\mathcal{D}}$ characterizes the largest timescale of correlation in the data generation process $\mathcal{D}$.
In particular, two data points that are at least $\tau_{\mathcal{D}}$ far apart from each other are independent.
We also use $\mathcal{D}\to^{\times \tau_{\mathcal{D}}} z$ to denote the process of generating a single refreshing block of data from $\mathcal{D}$.

We define the \emph{repetition number} $R_{\mathcal{D}}$ of $\mathcal{D}$ as follows
\begin{equation}
    R_{\mathcal{D}} = \max_z \left(\E\left[N_z  \middle| z_1=z\right] - \E[N_z]\right),
\end{equation}
where $N_z = \sum_{i=1}^{\tau_{\mathcal{D}}}\delta_{z_i, z}$ is the number of $z$'s in a refreshing block of data.
In other words, $R_{\mathcal{D}}$ is the expected number of extra $z$'s one gets in a refreshing time block if we condition on the first sample being $z$, maximized over all $z\in \mathcal{Z}$.
We note that it is always non-negative because by construction our data generation process can only enhance the correlation between data samples by sharing common situations (\Cref{lem:data-enhance-corr}).
When the data samples $z_i$ do not contain the full information of the situations, we need to include the situations in the condition of definition.

The repetition number $R_{\mathcal{D}}$ controls the variance of the number of samples one get on each $z$, which determines the rate of convergence in our quantum oracle sketching algorithm.
Intuitively, the more frequent the samples repeat themselves, the more redundant the data are.
Effectively, the $R_{\mathcal{D}}$ repetitive samples is only as good as a single sample, and therefore we expect that the sample complexity to blow up by a factor of $R_{\mathcal{D}}$.

To illustrate these concepts, we give two examples of data generation processes: the repetitive process $\mathcal{D}_{\mathrm{rep}}$ and the alternating process $\mathcal{D}_{\mathrm{alt}}$.
The repetitive process $\mathcal{D}_{\mathrm{rep}}$ is defined as 
\begin{equation}
    \mathcal{D}_{\mathrm{rep}} = (\mathcal{D}^0\to \mathcal{D}_{x}^1 \to^{\times N} z),
\end{equation}
where $\mathcal{D}^0 = \mathrm{Uniform}([N])$ and $\mathcal{D}^1_x$ always outputs its label $x\in [N]$ repetitively $N$ times.
We have refreshing time $\tau_{\mathcal{D}_{\mathrm{rep}}}=N$ and repetition number $R_{\mathcal{D}_{\mathrm{rep}}}=N - N\times 1/N = N-1$.
This is intuitive as $\mathcal{D}_{\mathrm{rep}}$ repeats the same output $N-1$ times.
In contrast, the alternating process is defined as
\begin{equation}
    \mathcal{D}_{\mathrm{alt}} = (\mathcal{D}^0\to \mathcal{D}_{\alpha}^1 \to^{\times N} z),
\end{equation}
where $\mathcal{D}^0 = \mathrm{Bern}(1/2)$, $\alpha\in \bit$, and $\mathcal{D}^1_\alpha$ samples $x\sim \unif([N])$ randomly $N$ times and outputs $z=(x, \alpha)$.
We still have refreshing time $\tau_{\mathcal{D}_{\mathrm{alt}}}=N$, but the repetition number is $R_{\mathcal{D}_{\mathrm{alt}}}= (1+(N-1)\times 1/N) - N\times 1/(2N)=3/2-1/N$.
This is also intuitive since the data are generated uniformly random and the expected number of repetitions is small.

In \Cref{sec:data-gen-property}, we prove some useful properties of general hierarchical data generation processes.

\subsection{Classical learning algorithms}
\label{sec:cl-learning-alg}

Now we introduce our computational model of classical learning algorithms.
In defining such learning algorithms, we need to take into account restrictions on the size of the classical machine.

We define classical learning algorithms as follows.
We use $S$ to denote the size of the classical machine (also called the space complexity), and use $M$ to denote the number of samples this learning algorithm takes in (i.e., sample complexity).
Let $\mathcal{I}$ be the set of all possible inputs to the learning algorithm at each time step.
For our data generation process defined above, $\mathcal{I} = \mathcal{Z}$.
A classical learning algorithm $\mathcal{L}$ with size $S$, sample complexity $M$, and input form $\mathcal{I}$ is defined as a directed graph with vertices arranged in $M+1$ layers ($0$ to $M$).
Each layer consists of at most $2^S$ vertices, each labeled by $S$ bits.
There is only one vertex called the root in layer $0$.
In layer $M$, each vertex $v$ has no outgoing edges and is called a leaf.
Each leaf $v$ is attached with an output $h_v$.
For each layer $i=0, \ldots, M-1$, the outgoing edge from each vertex in layer $i$ only goes to vertices in layer $i+1$.
Each vertex has $|\mathcal{I}|$ outgoing edges, labeled by each element of $\mathcal{I}$.

The computation of the learning algorithm $\mathcal{L}$ proceeds as follows.
Upon receiving a sequence of data $I_i \in \mathcal{I}, i=0, \ldots, M-1$, the algorithm starts from the root, follows the edge given by each data point $I_i$ in layer $i$ until reaching a leaf $v$ in layer $M$, and outputs $h_v$.

To keep track of the information flow during learning, we define the \emph{transcript} $\pi_\mathcal{L}(I, \alpha)$ of the learning algorithm $\mathcal{L}$ upon receiving the data sequence $I = (I_0, \ldots, I_{M-1}) \in \mathcal{I}^M$ with respect to a situation record $\alpha = (\alpha_0, \ldots, \alpha_{M-1})$ to be the concatenation of the length-$S$ bitstrings that label the vertices traversed by the computation path at layers $i$ where the situation changes $\alpha_i\neq \alpha_{i+1}$, followed by the output of $\mathcal{L}$.
For our data generation process, suppose that the situation changes in total $r$ times and the output is a single bit $h_v\in \bit$, the transcript $\pi_\mathcal{L}(I, \alpha)$ is a bitstring with total length $|\pi_\mathcal{L}(I, \alpha)|=rS+1$.

Furthermore, we define the data processing time per sample to be the time complexity of the computation per layer.
The total time complexity of the learning algorithm is equal to the sample complexity times the data processing time per sample.

Since any randomized algorithm can always be regarded as first sampling all the random numbers and then execute the corresponding deterministic algorithm (i.e., de-randomized), the deterministic definition above suffices.
We note that our definition of classical learning algorithms resemble the notion of branching programs.
They are non-uniform models of space-bounded computation (the most general form), more general than uniform ones such as online Turing machines with bounded space.
Therefore, the classical hardness results we prove also applies to uniform computational models.

\subsection{Quantum learning algorithms}
\label{sec:q-learning-alg}

Our model for quantum learning algorithms (with classical data inputs) is similar to that of classical learning algorithms, but the computation is done via a quantum circuit.
Note that since we are considering the task of processing classical data, the data that we feed into quantum learning algorithms are completely classical and the data access model is the same as that of classical learning algorithms.

In particular, we define quantum learning algorithms as follows.
We use $S$ to denote the size of the quantum machine (also called the space complexity), and use $M$ to denote the number of samples this learning algorithm takes in (i.e., sample complexity).
Let $\mathcal{I}$ be the set of all possible inputs to the quantum learning algorithm at each time step.
For our data generation process  $\mathcal{I} = \mathcal{Z}$.
A quantum learning algorithm $\mathcal{L}$ with size $S$, sample complexity $M$, and input form $\mathcal{I}$ is defined as an initial $S$-qubit quantum state $\rho_0$ and a sequence of $M$ sets of $S$-qubit quantum channels $(\mathcal{C}^{0}, \ldots, \mathcal{C}^{M-1})$, where each set $\mathcal{C}^i = \{C^i_I: I\in \mathcal{I}\}$ contains $|\mathcal{I}|$ quantum channels on $S$ qubits that are labeled by the elements of $\mathcal{I}$.
At the end, there is an $S$-qubit positive operator-valued measurement (POVM) $\{\mathcal{M}_h\}, \sum_h \mathcal{M}_h = I$ whose output labeled by $h$ is the output of the quantum learning algorithm.

The computation of the learning algorithm $\mathcal{L}$ proceeds as follows.
Upon receiving a sequence of data $I_i \in \mathcal{I}, i=0, \ldots, M-1$, the algorithm starts from the initial state $\rho_0$, sequentially applies the quantum channel $C^i_{I_i}$ given by each data point $I_i$ in step $i$ until reaching the $M$-th step. Finally, the algorithm measures $\{\mathcal{M}_h\}$ on the resulting state
\begin{equation}
    \rho_M = C^{M-1}_{I_{M-1}} \circ \cdots \circ C^0_{I_0} (\rho_0),
\end{equation}
and output the measurement outcome $h$ with probability $\tr(\mathcal{M}_h \rho_M)$ as the output of the learning algorithm.

Similar to classical learning algorithms, we define the data processing time per sample as the time complexity of implementing the quantum channel per layer.
The total time complexity is equal to the sample complexity times the data processing time per sample.

We remark that although this formal definition of quantum learning algorithm here is a non-uniform computational model, we will see that the quantum learning algorithms designed in this work all have program descriptions that can be generated by a polynomial-time classical Turing machine and therefore fit into uniform computational models.
We define (uniform) quantum learning algorithms with the additional requirement that the initial state $\rho_0$, the quantum channel sets $(\mathcal{C}^1, \ldots, \mathcal{C}^{M-1})$, and the final measurement $\{\mathcal{M}_h\}$ can be generated by a polynomial-time classical Turing machine.

\subsection{Properties of hierarchical data generation processes}
\label{sec:data-gen-property}

In this section, we prove some useful properties of the hierarchical data generation processes defined in \Cref{sec:data-access}.
We will study some fundamental properties of the repetition number and use it to characterize statistical properties of the data samples.

Let
\begin{equation}
    \mathcal{D}=(\mathcal{D}^0 \to \mathcal{D}^1_{\alpha_1} \to^{\times T_1} \mathcal{D}^2_{\alpha_2}\to^{\times T_2} \cdots \to^{\times T_{l-1}} \mathcal{D}^l_{\alpha_l} \to^{\times T_l} z)
\end{equation}
be a hierarchical data generation process with refreshing time $\tau_{\mathcal{D}}=T_l\cdots T_1$ and repetition number 
\begin{equation}
    R_{\mathcal{D}} = \max_z \left(\E\left[\sum_{i=1}^{\tau_{\mathcal{D}}}\delta_{z_i, z}  \middle| z_1=z\right] - \E\left[\sum_{i=1}^{\tau_{\mathcal{D}}}\delta_{z_i, z} \right]\right).
\end{equation}
An equivalent probability formulation of repetition number is
\begin{equation}
    R_{\mathcal{D}} = \max_z \sum_{i=1}^{\tau_{\mathcal{D}}}\left(\Pr\left[z_i=z  \middle| z_1=z\right] - \Pr[z_i=z]\right).
\end{equation}

The structure of the data generation process induces correlation between data samples.
For any two data samples $z_i, z_j$, we define their \emph{correlation depth} $K(i, j) \in \{0, \ldots, l\}$ to be the number of situation levels that they share.
In other words, $z_i, z_j$ shares the situation $(\alpha_1, \ldots, \alpha_{K(i, j)})$.
Clearly $K(i, j)=K(j, i)$.
If $K(i, j)=0$, then $z_i, z_j$ are independent and belong to different refreshing blocks of data.
The following lemma formalizes the intuition that sharing situations enhances correlation, which immediately implies $R_{\mathcal{D}}\geq 0$.

\begin{lemma}[Sharing situations enhances correlation]
\label{lem:data-enhance-corr}
    Let $z_i, z_j\in \mathcal{Z}, i, j\geq 1$ be any two random data samples from a hierarchical data generation process $\mathcal{D}$.
    Then,
    \begin{equation}
        \Pr[z_j=z|z_i=z]\geq \Pr[z_j=z]
    \end{equation}
    for any $z\in \mathcal{Z}$.
\end{lemma}

\begin{proof}
    The claim is immediate when $i=j$.
    When $i\neq j$, let $K(i, j)$ be the correlation depth between $z_i, z_j$.
    Let $\sigma = (\alpha_1, \ldots, \alpha_{K(i, j)})$ be the random shared situation.
    By construction, $z_i|\sigma$ and $z_j|\sigma$ are IID with the same distribution that we call $q_\sigma (z)$.
    Hence,
    \begin{equation}
    \begin{split}
        \Pr[z_j=z|z_i=z] &= \frac{\Pr[z_j=z, z_i=z]}{\Pr[z_i=z]} \\
        &= \frac{\E_\sigma[\Pr[z_j=z, z_i=z|\sigma]]}{\Pr[z_i=z]} \\
        &= \frac{\E_\sigma[\Pr[z_j=z|\sigma]\Pr[ z_i=z|\sigma]]}{\Pr[z_i=z]} \\
        &=\frac{\E_\sigma [q_\sigma^2(z)]}{\Pr[z_i=z]}.
    \end{split}
    \end{equation}
    Applying Jensen's inequality on $f(w)=w^2$, we have
    \begin{equation}
        \E_\sigma [q_\sigma^2(z)]\geq (\E_\sigma[q_\sigma(z)])^2 = (\E_\sigma[\Pr[z_i=z|\sigma]])^2 = (\Pr[z_i=z])^2.
    \end{equation}
    This gives us
    \begin{equation}
        \Pr[z_j=z|z_i=z] \geq \Pr[z_i=z] = \Pr[z_j=z]
    \end{equation}
    as desired, where we have used the fact that all data samples have the same marginal.
\end{proof}

Moreover, the following result shows that correlation does not decrease with correlation depth.

\begin{lemma}[Correlation is non-decreasing with correlation depth]
\label{lem:data-corr-depth}
    For any fixed $z\in \mathcal{Z}$ and four time steps $i\neq j, i'\neq j'\geq 1$, if $K(i, j)\leq K(i', j')$, then $\Pr[z_j=z|z_i=z]\leq \Pr[z_{j'}=z|z_{i'}=z]$.
\end{lemma}

\begin{proof}
    Let $k=K(i,j)$ and $k'=K(i',j')$, with $k\leq k'$.
    Let $\sigma=(\alpha_1,\ldots,\alpha_k)$ and
    $\sigma'=(\alpha_1,\ldots,\alpha_{k'})$ be naturally coupled so that $\sigma$ is the prefix of $\sigma'$.
    By construction, the marginal distribution of a situation prefix depends only on its depth, so these have the same distributions as the situations shared by $(z_i,z_j)$ and $(z_{i'},z_{j'})$, respectively.

    Let
    \begin{equation}
        q_\sigma(z)=\Pr[z_i=z\mid\sigma],
        \qquad
        q_{\sigma'}(z)=\Pr[z_{i'}=z\mid\sigma'].
    \end{equation}
    Since conditioned on their shared situation the two samples are IID, writing $p(z)$ for the common marginal gives
    \begin{equation}
        \Pr[z_j=z\mid z_i=z]
        =
        \frac{\E_\sigma[q_\sigma^2(z)]}{p(z)},
        \qquad
        \Pr[z_{j'}=z\mid z_{i'}=z]
        =
        \frac{\E_{\sigma'}[q_{\sigma'}^2(z)]}{p(z)}.
    \end{equation}
    Moreover, by the tower property,
    \begin{equation}
        q_\sigma(z)
        =
        \E[q_{\sigma'}(z)\mid\sigma].
    \end{equation}
    Hence Jensen's inequality gives
    \begin{equation}
        \E_\sigma[q_\sigma^2(z)]
        =
        \E_\sigma\left[
            \E[q_{\sigma'}(z)\mid\sigma]^2
        \right]
        \leq
        \E_{\sigma'}[q_{\sigma'}^2(z)].
    \end{equation}
    Therefore,
    \begin{equation}
        \Pr[z_j=z\mid z_i=z]
        \leq
        \Pr[z_{j'}=z\mid z_{i'}=z],
    \end{equation}
    as desired.
\end{proof}

Although the repetition number $R_{\mathcal{D}}$ is defined with a starting time of $t=1$, the following lemma shows that $R_{\mathcal{D}}$ upper bounds the repetition number of any starting time.
This property allows us to collect data starting at any time point.

\begin{lemma}[Repetition number bound is agnostic to starting time]
\label{lem:data-rep-num-start-time}
    For any $t\geq 1$, the repetition number starting at time $t$
    \begin{equation}
        R_{\mathcal{D}}(t) = \max_z \sum_{i=t}^{t+\tau_{\mathcal{D}}-1}\left(\Pr\left[z_i=z  \middle| z_t=z\right] - \Pr[z_i=z]\right)
    \end{equation}
    satisfies
    \begin{equation}
        R_{\mathcal{D}}(t)\leq R_{\mathcal{D}}.
    \end{equation}
\end{lemma}

\begin{proof}
    Note that
    \begin{equation}
        R_{\mathcal{D}}(t) = \max_z \sum_{i=1}^{\tau_{\mathcal{D}}}\left(\Pr\left[z_{t+i-1}=z  \middle| z_t=z\right] - \Pr[z_{i}=z]\right),
    \end{equation}
    where we have changed the variable and used the marginal being the same.
    We only need to show
    \begin{equation}
        \Pr[z_{t+s}=z|z_t=z]\leq \Pr[z_{1+s}=z|z_1=z]
    \end{equation}
    for any $z\in\mathcal{Z}, s\geq 0, t\geq 1 $, then we are done.
    The case $s=0$ is immediate. 
    Hence assume $s\geq 1$.
    Using \Cref{lem:data-corr-depth}, it suffices to show that the correlation depth satisfies
    \begin{equation}
        K(t, t+s)\leq K(1, 1+s), \quad \forall t\geq 1, s\geq 0.
    \end{equation}
    To show this, note that the tree structure of the data generation process implies that for any $k\in \{1, \ldots, l\}$, the data samples $(1, 1+s)$ share situations at level $k$ as long as $s< T_l\ldots T_{k}$.
    But in order for $(t, t+s)$ to share situations at level $k$, they need to satisfy extra conditions on $s, t$ (e.g., the effect of crossing boundaries, etc.).
    Therefore, any situation level shared by $(t, t+s)$ must also be shared by $(1, 1+s)$.
    In other words, we have $K(t, t+s)\leq K(1, 1+s)$ as desired.
    This proves \Cref{lem:data-rep-num-start-time}.
\end{proof}

As we will see in \Cref{sec:q-alg}, the frequency of collecting a specific data sample $z\in \mathcal{Z}$,
\begin{equation}
    m_z = \frac{1}{M} \sum_{i=t}^{t+M-1} \delta_{z_{i}, z},
\end{equation}
in a sequence of $M$ data samples $z_t, \ldots, z_{t+M-1}$ will play a crucial role in quantum oracle sketching.
In the following, we characterize the statistical properties of $m_z$ using properties of the data generation process.

Since the marginal of each data sample $p(z)$ is fixed, we always have $\E[m_z] = p(z)$.
The following lemma shows that the variance of $m_z$ is upper bounded by the repetition number $R_{\mathcal{D}}$.

\begin{lemma}[Repetition number bounds variance]
\label{lem:data-var-rep-num}
    Let $t\geq 1$ be any starting time.
    Let $z_t, \ldots, z_{t+M-1}$ be a sequence of $M$ data samples drawn from the hierarchical data generation process $\mathcal{D}$ with marginal $p(z)$ and repetition number $R_{\mathcal{D}}$.
    Let
    \begin{equation}
        m_z = \frac{1}{M} \sum_{i=t}^{t+M-1} \delta_{z_{i}, z}, \quad z\in \mathcal{Z},
    \end{equation}
    be the frequency of having the value $z$ in these data samples.
    Then, we have
    \begin{equation}
        \Var[m_z]\leq \frac{p(z)R_{\mathcal{D}}}{M},
    \end{equation}
    for any $z\in \mathcal{Z}$.
\end{lemma}

\begin{proof}
    Let the marginal be $p(z)$.
    The variance can be decomposed as
    \begin{equation}
    \begin{split}
        \Var[m_z] &= \frac{1}{M^2}\Var\left[\sum_{i=1}^M \delta_{z, z_{t+i-1}}\right] \\
        &=\frac{1}{M^2}\sum_{i, j\in[M]}\Cov(\delta_{z, z_{t+i-1}}, \delta_{z, z_{t+j-1}}).
    \end{split}
    \end{equation}
    If $z_{t+i-1}, z_{t+j-1}$ are not in the same refreshing block, they are independent and $\Cov(\delta_{z, z_{t+i-1}}, \delta_{z, z_{t+j-1}})=0$.
    On the other hand, if $z_{t+i-1}, z_{t+j-1}$ are in the same refreshing block, since each block has the same distribution, we may as well assume that they are both in the first block at time steps $i,j\in[\tau_{\mathcal{D}}]$.
    Then the covariance is equal to
    \begin{equation}
    \begin{split}
        \Cov(\delta_{z, z_{i}}, \delta_{z, z_{j}}) &= \E[\delta_{z, z_i}\delta_{z, z_j}] - \E[\delta_{z, z_i}]\E[\delta_{z, z_j}] \\
        &= \E[\delta_{z, z_i}\delta_{z, z_j}] - p^2(z) \\
        &= \Pr[z_j=z_i=z] - p^2(z) \\
        &= \Pr[z_i=z]\Pr[z_j=z|z_i=z] - p^2(z) \\
        &= p(z) (\Pr[z_j=z|z_i=z] - p(z)).
    \end{split}
    \end{equation}
    Note that from \Cref{lem:data-enhance-corr}, we have $\Cov(\delta_{z, z_{i}}, \delta_{z, z_{j}})\geq 0$.
    By construction of the data generation process $\mathcal{D}$, the joint distribution of any two distinct data samples $z_i, z_j$ only depends on their correlation depth $K(i, j)$.
    Hence, the conditional probability $\Pr[z_j=z|z_i=z]$ only depends on $K(i, j)$, and we use $q_{K(i, j)}(z)$ to denote it.
    For every $1\leq k<l$, the number of indices $j$ satisfying $K(i,j)=k$ is $T_k\cdots T_l-T_{k+1}\cdots T_l$, because $T_k\cdots T_l$ samples share the first $k$ situations, while $T_{k+1}\cdots T_l$ of them also share the $(k+1)$-st situation.
    The number of indices satisfying $K(i,j)=l$ is $T_l$, in which one is $i$ itself.
    These quantities are independent of $i$.
    Therefore, summing over all $j\in [\tau_{\mathcal{D}}]$ (which includes $i$ itself), we can regroup the sum according to the correlation depth $k$ and obtain
    \begin{equation}
    \begin{split}
        \sum_{j\in [\tau_{\mathcal{D}}]} \Pr[z_j=z|z_i=z] &= 1+\sum_{j\neq i} q_{K(i, j)}(z) = 1+\sum_{k=1}^l \sum_{j\neq i} q_{k}(z)\delta_{k, K(i, j)} \\
        &= 1+\sum_{k=1}^l q_{k}(z)\left(\sum_{j\neq i} \delta_{k, K(i, j)}\right) = \sum_{k=1}^{l-1}q_k(z)(T_k\cdots T_l-T_{k+1}\cdots T_l)+q_l(z)(T_l-1)+1,
    \end{split}
    \end{equation}
    which is independent of $i$.
    In particular, we can set $i=1$:
    \begin{equation}
        \sum_{j\in [\tau_{\mathcal{D}}]} \Pr[z_j=z|z_i=z] = \sum_{j\in [\tau_{\mathcal{D}}]} \Pr[z_j=z|z_1=z].
    \end{equation}
    
    Plugging it back into the variance expression, we obtain
    \begin{equation}
    \begin{split}
        \Var[m_z] &\leq \frac{1}{M^2}\sum_{i\in[M]}\sum_{j\in [\tau_{\mathcal{D}}]} p(z) (\Pr[z_j=z|z_i=z]-p(z)) \\
        &=\frac{1}{M^2}p(z) \sum_{i\in [M]} \sum_{j\in [\tau_{\mathcal{D}}]}(\Pr[z_j=z|z_1=z]-p(z)) \\
        &\leq \frac{1}{M^2}p(z) M R_{\mathcal{D}} \\
        &=\frac{p(z)R_{\mathcal{D}}}{M},
    \end{split}
    \end{equation}
    where the first inequality uses the non-negativity of the covariances and the second inequality uses the definition of $R_{\mathcal{D}}$.
    This proves \Cref{lem:data-var-rep-num}.
\end{proof}

When we have already processed a previous sequence of data $z_1, \ldots, z_{t-1}$ (i.e., condition on them), the distribution of $m_z$ shifts. The next lemma is a conditional version of \Cref{lem:data-var-rep-num}.

\begin{lemma}[Repetition number bounds conditional variance]
\label{lem:data-cond-var-rep-num}
    Let $t\geq 1$ and let $z_1,\ldots,z_{t-1}, z_t, \ldots, z_{t+M-1}$ be a sequence of data samples drawn from the hierarchical data generation process $\mathcal{D}$ with marginal $p(z)$ and repetition number $R_{\mathcal{D}}$.
    Let
    \begin{equation}
        m_z(t)=\frac{1}{M}\sum_{i=t}^{t+M-1}\delta_{z_i,z}, \quad z\in \mathcal{Z}
    \end{equation}
    be the frequency of having the value $z$ in this round of data processing.
    Then, we have
    \begin{equation}
        \Var[m_z(t)|z_1 \ldots, z_{t-1}]
        \leq \frac{2R_{\mathcal D}}{M} \E[m_z(t)|z_1 \ldots, z_{t-1}],
        \quad z\in\mathcal Z.
    \end{equation}
\end{lemma}

\begin{proof}
    Fix any realization of $(z_1, \ldots, z_{t-1})$ and
    $z\in\mathcal Z$.
    Since sharing situations enhances correlation, we have
    \begin{equation}
        q_s=\Pr[z_s=z|z_1, \ldots, z_{t-1}]\geq p(z),
    \end{equation}
    similar to the proof of \Cref{lem:data-enhance-corr}, as long as the situations in $z_s$ and the history are compatible.
    If not, $q_s=0$ and the corresponding covariance term in the variance expansion vanishes.
    So we only need to focus on the case where $q_s\neq 0$.
    Now for any $t\leq s<u\leq t+M-1$, assuming they are in the same refreshing block without loss of generality, every situation level shared by $z_u$ and the history $z_1, \ldots, z_{t-1}$ must also be active at time $s$.
    Therefore, conditioning on $z_s$ suffice to replace the conditioning on the history.
    In other words,
    \begin{equation}
        \Pr[z_u=z\mid z_s=z,z_1, \ldots, z_{t-1}] = \Pr[z_u=z\mid z_s=z].
    \end{equation}
    Hence, we have
    \begin{equation}
        \Pr[z_u=z\mid z_s=z,z_1, \ldots, z_{t-1}]
          -\Pr[z_u=z\mid z_1, \ldots, z_{t-1}]\leq\Pr[z_u=z\mid z_s=z]-p(z).
    \end{equation}
    This implies that
    \begin{equation}
        \Cov[\delta_{z_s, z},\delta_{z_u, z}\mid z_1, \ldots, z_{t-1}]
        \leq q_s\bigl(\Pr[z_u=z\mid z_s=z]-p(z)\bigr).
    \end{equation}
    Consequently,
    \begin{equation}
    \begin{split}
        &\Var\left[\sum_{s=t}^{t+M-1}\delta_{z_s, z}\middle|z_1, \ldots, z_{t-1}\right] \\
        &\leq \sum_{s=t}^{t+M-1}q_s(1-p(z))
        +2\sum_{s=t}^{t+M-1}q_s
          \sum_{\substack{u>s \text{ in the same refreshing block}}}
          \bigl(\Pr[z_u=z\mid z_s=z]-p(z)\bigr)\\
        &\leq 2\sum_{s=t}^{t+M-1}q_s
          \left((1-p(z))+
          \sum_{\substack{u>s \text{ in the same refreshing block}}}
          \bigl(\Pr[z_u=z\mid z_s=z]-p(z)\bigr)\right)\\
        &\leq 2R_{\mathcal D}\sum_{s=t}^{t+M-1}q_s,
    \end{split}
    \end{equation}
    where the last inequality follows from
    \Cref{lem:data-rep-num-start-time}.
    Dividing by $M^2$ yields
    \begin{equation}
        \Var[m_z(t)\mid z_1, \ldots, z_{t-1}]
        \leq \frac{2R_{\mathcal D}}{M}\cdot
        \frac{1}{M}\sum_{s=t}^{t+M-1}q_s
        =\frac{2R_{\mathcal D}}{M}
        \E[m_z(t)\mid z_1, \ldots, z_{t-1}],
    \end{equation}
    as desired.
\end{proof}

Another effect of the conditioning is that the conditional expectation of $m_z$ no longer matches $p(z)$.
Luckily, the following lemma shows that the expected drift is controlled by the repetition number $R_{\mathcal{D}}$.
Note that the bound scales as $1/M$, much better than the usual $1/\sqrt{M}$ from concentration.

\begin{lemma}[Repetition number bounds conditional drift]
\label{lem:data-cond-drift-rep-num}
    Let $t\geq 1$ be the starting time of this round of data processing.
    Let $z_1, \ldots, z_{t-1}, z_t, \ldots, z_{t+M-1}$ be a sequence of data samples drawn from the hierarchical data generation process $\mathcal{D}$ with marginal $p(z)$ and repetition number $R_{\mathcal{D}}$.
    Let
    \begin{equation}
        m_z(t) = \frac{1}{M} \sum_{i=t}^{t+M-1} \delta_{z_{i}, z}, \quad z\in \mathcal{Z},
    \end{equation}
    be the frequency of having the value $z$ in this round of data processing.
    Then, we have
    \begin{equation}
        \E_{z_1, \ldots, z_{t-1}}\left[\max_{z\in \mathcal{Z}}\Bigl|\E[m_z(t)|z_1, \ldots, z_{t-1}] - \E[m_z(t)]\Bigr|\right] \leq \frac{\sqrt{\max_{z\in \mathcal{Z}} p(z)|\mathcal{Z}|}R_{\mathcal{D}}}{M}.
    \end{equation}
\end{lemma}

\begin{proof}
    Note that the conditional independence structure in the hierarchical data generation process implies that the data samples in this round $z_t, \ldots, z_{t+M-1}$ can only correlate with the previous samples $z_1, \ldots, z_{t-1}$ through the situation $\sigma_t=(\alpha_{1,t}, \ldots, \alpha_{l,t})$ of $z_t$, which is a part of $z_t$ by construction.
    This means that
    \begin{equation}
    \begin{split}
        &\E\left[\max_z\Bigl|\E[m_z(t)|z_1, \ldots, z_{t-1}] - \E[m_z(t)]\Bigr|\right]\\
        &= \E\left[\max_z\left|\sum_{z'\in \mathcal{Z}} (\E[m_z(t)|z_1, \ldots, z_{t-1}, z_t=z'] - \E[m_z(t)]) \Pr[z_t=z'|z_1, \ldots, z_{t-1}]\right|\right] \\
        &= \E\left[\max_z\left|\sum_{z'\in \mathcal{Z}} (\E[m_z(t)|z_t=z'] - \E[m_z(t)]) \Pr[z_t=z'|z_1, \ldots, z_{t-1}]\right|\right] \\
        &\leq \sum_{z'\in \mathcal{Z}} \E \Bigl[\max_z|\E[m_z(t)|z_t=z'] - \E[m_z(t)]| \Pr[z_t=z'|z_1, \ldots, z_{t-1}]\Bigr] \\
        &=\E \left[\max_z\Bigl|\E[m_z(t)|z_t] - \E[m_z(t)]\Bigr|\right],
    \end{split}
    \end{equation}
    where we have used triangle inequality.
    Hence, it suffices to prove that
    \begin{equation}
        \E\left[\max_z|\E[m_z(t)|z_t] - \E[m_z(t)]|\right] \leq \frac{\sqrt{\max_z p(z)|\mathcal{Z}|}R_{\mathcal{D}}}{M}.
    \end{equation}

    To this end, we express this quantity in covariances as
    \begin{equation}
    \begin{split}
        &\E\left[\max_z\Bigl|\E[m_z(t)|z_t] - \E[m_z(t)] \Bigr|\right] \\
        &= \frac{1}{M} \E\left[\max_z\left|\sum_{i=0}^{M-1}(\Pr[z_{t+i}=z|z_t] - p(z))\right|\right] \\
        &=\frac{1}{M} \sum_{z'}p(z')\max_z\left|\sum_{i=0}^{M-1}(\Pr[z_{t+i}=z|z_t=z'] - p(z))\right| \\
        &=\frac{1}{M} \sum_{z'}\max_z\left|\sum_{i=0}^{M-1}(\Pr[z_{t+i}=z, z_t=z'] - p(z)p(z'))\right| \\
        &=\frac{1}{M} \sum_{z'}\max_z \left|\sum_{i=0}^{M-1}(\E[\delta_{z_{t+i}, z}\delta_{z_t, z'}] - \E[\delta_{z_{t+i}, z}]\E[\delta_{z_t, z'}])\right| \\
        &=\frac{1}{M}\sum_{z'}\max_z \left|\sum_{i=0}^{M-1} \Cov[\delta_{z_{t+i}, z}, \delta_{z_t, z'}]\right| \\
        &=\frac{1}{M}\sum_{z'}\max_z |\Sigma(z, z')|,
    \end{split}
    \end{equation}
    where we have defined the aggregated cross-covariance matrix
    \begin{equation}
        \Sigma(z, z') = \sum_{i=0}^{M-1} \Cov[\delta_{z_{t+i}, z}, \delta_{z_t, z'}].
    \end{equation}
    
    Next, we invoke the structural properties of the data generation process.
    For every $1\leq i\leq M-1$, define the conditional expectation
    \begin{equation}
        q_i(z,\sigma)
        =
        \E[
            \delta_{z_t,z}
            \mid
            \sigma_{t,t+i}=\sigma
        ],
    \end{equation}
    where $\sigma_{t,t+i}$ is the situation shared by the distinct
    samples $z_t$ and $z_{t+i}$.
    Conditioned on this shared situation, the two samples are
    independent and have the same conditional marginal.
    Hence,
    \begin{equation}
        \E[\delta_{z_t=z}|\sigma_{t, t+i}=\sigma] = \E[\delta_{z_{t+i}=z}|\sigma_{t, t+i}=\sigma] = q_i(z, \sigma).
    \end{equation}
    The law of total covariance then implies that
    \begin{equation}
    \begin{split}
        \Cov[\delta_{z_{t+i}, z}, \delta_{z_t, z'}] &= \E[\Cov[\delta_{z_{t+i}, z}, \delta_{z_t, z'}|\sigma_{t, t+i}]] + \Cov[\E[\delta_{z_{t+i}=z}|\sigma_{t, t+i}], \E[\delta_{z_t=z'}|\sigma_{t, t+i}]] \\
        &= 0 + \Cov[q_i(z, \sigma_{t, t+i}), q_i(z', \sigma_{t, t+i})] \\
        &=\Cov[q_i(z, \sigma_{t, t+i}), q_i(z', \sigma_{t, t+i})],
    \end{split}
    \end{equation}
    where we have used the fact that conditioned on the shared situation $\sigma_{t, t+i}$, the data samples are independent and their conditional covariance vanishes.
    This shows that this cross-covariance is secretly a covariance.
    Cauchy-Schwarz inequality then gives
    \begin{equation}
        |\Cov[\delta_{z_{t+i}, z}, \delta_{z_t, z'}]|=|\Cov[q_i(z, \sigma_{t, t+i}), q_i(z', \sigma_{t, t+i})]| \leq \sqrt{\Var[q_i(z, \sigma_{t, t+i})]\Var[q_i(z', \sigma_{t, t+i})]}.
    \end{equation}
    The variances can again be written as cross-covariance using the law of total covariance in reverse:
    \begin{equation}
    \begin{split}
        \Var[q_i(z, \sigma_{t, t+i})] &= \Cov[q_i(z, \sigma_{t, t+i}), q_i(z, \sigma_{t, t+i})] = \Cov[\E[\delta_{z_{t+i}=z}|\sigma_{t, t+i}], \E[\delta_{z_{t}=z}|\sigma_{t, t+i}]] \\
        &=\Cov[\delta_{z_{t+i}=z}, \delta_{z_{t}=z}] - \E[\Cov[\delta_{z_{t+i}=z}, \delta_{z_{t}=z}|\sigma_{t, t+i}]] \\
        &=\Cov[\delta_{z_{t+i}=z}, \delta_{z_{t}=z}] \\
        &=\Pr[z_{t+i}=z, z_t=z] - \Pr[z_{t+i}=z]\Pr[z_t=z] \\
        &=p(z)(\Pr[z_{t+i}=z|z_t=z] - p(z)).
    \end{split}
    \end{equation}
    Since the $i=0$ term in the repetition number is
    $1-p(z)$, \Cref{lem:data-rep-num-start-time} implies
    \begin{equation}
        \sum_{i=1}^{M-1}
        \Var[
            q_i(z,\sigma_{t,t+i})
        ]\leq
        p(z)
        \left(
            R_{\mathcal D}
            -
            (1-p(z))
        \right).
    \end{equation}
    In particular, $R_{\mathcal D}\geq 1-p(z)$ whenever $p(z)>0$.

    We now plug this back into the expression of $\Sigma$ and include the $i=0$ term.
    If $z=z'$, then
    \begin{equation}
        |\Sigma(z,z)|
        \leq
        p(z)(1-p(z))
        +
        p(z)
        \left(
            R_{\mathcal D}
            -
            (1-p(z))
        \right)=
        p(z)R_{\mathcal D}.
    \end{equation}
    If $z\neq z'$, then
    \begin{equation}
        |\Sigma(z,z')|
        \leq
        p(z)p(z')+
        \sqrt{
            p(z)
            \left(
                R_{\mathcal D}-1+p(z)
            \right)
            p(z')
            \left(
                R_{\mathcal D}-1+p(z')
            \right)
        }.
    \end{equation}
    Since $p(z)+p(z')\leq 1$, we apply Cauchy-Schwarz inequality and then apply the
    arithmetic-geometric mean inequality to get
    \begin{equation}
        \sqrt{p(z)p(z')}+
        \sqrt{
            \left(
                R_{\mathcal D}-1+p(z)
            \right)
            \left(
                R_{\mathcal D}-1+p(z')
            \right)
        }\leq
        \sqrt{
            R_{\mathcal D}-1+2p(z)
        }
        \sqrt{
            R_{\mathcal D}-1+2p(z')
        }\leq
        R_{\mathcal D}.
    \end{equation}
    Therefore, in both cases,
    \begin{equation}
        |\Sigma(z,z')|
        \leq
        R_{\mathcal D}
        \sqrt{
            p(z)p(z')
        }.
    \end{equation}
    As a result, we arrive at
    \begin{equation}
    \begin{split}
        \E\left[\max_z\Bigl|\E[m_z(t)|z_t] - \E[m_z(t)] \Bigr|\right] &\leq \frac{R_{\mathcal{D}}}{M}\sum_{z'}\max_z \sqrt{p(z)p(z')} \\
        &=\frac{\sqrt{\max_z p(z)}R_{\mathcal{D}}}{M}\sum_{z'} \sqrt{1\cdot p(z')} \\
        &\leq \frac{\sqrt{\max_z p(z)}R_{\mathcal{D}}}{M}\sqrt{\sum_{z'} 1} \sqrt{\sum_{z'}p(z')}\\
        &=\frac{\sqrt{\max_z p(z)|\mathcal{Z}|}R_{\mathcal{D}}}{M},
    \end{split}
    \end{equation}
    as desired.
    This completes the proof of \Cref{lem:data-cond-drift-rep-num}.
\end{proof}

\newpage

\section{Quantum oracle sketching}
\label{sec:q-alg}

\begin{figure}
    \centering
    \includegraphics[width=1\linewidth]{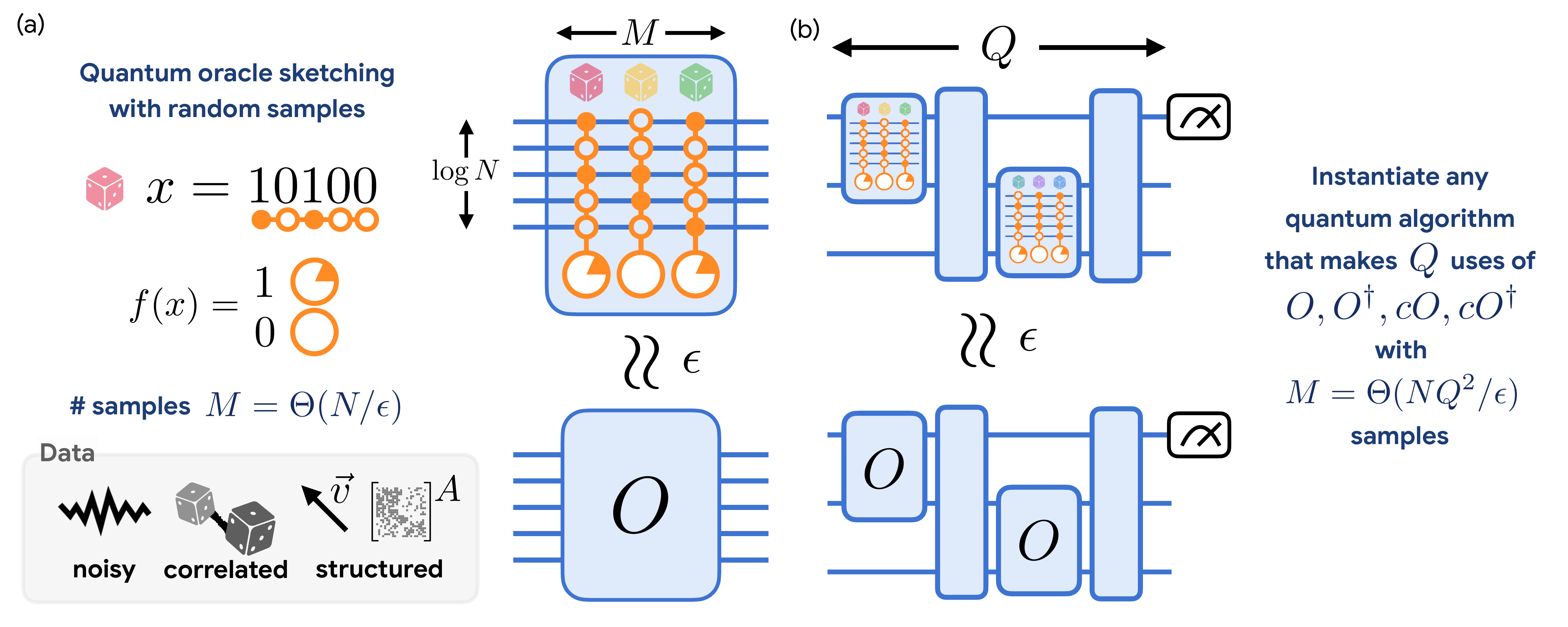}
    \caption[Schematic overview of quantum oracle sketching.]{\textbf{Schematic overview of quantum oracle sketching.}
    \textbf{(a)} Illustration of quantum oracle sketching for Boolean function data $(x, f(x)), f: [N] \to \bit$.
    We build the phase oracle $O = \sum_x (-1)^{f(x)}\ketbra{x}$ of $f$ using multi-controlled phase gates with control patterns given by $x$ and phase values given by $f(x)$ from the data.
    $M=\Theta(N/\epsilon)$ samples guarantee $\epsilon$ approximation of the phase oracle in diamond distance.
    We generalize this to accommodate noisy and correlated data inputs with generic data structures including state preparation unitaries of any vectors and block encodings of sparse matrices.
    \textbf{(b)} We use quantum oracle sketching to load data into a quantum computer and instantiate the oracle queries in any quantum algorithm.
    $M=\Theta(NQ^2/\epsilon)$ samples guarantee $\epsilon$ approximation in diamond distance of any quantum algorithm that makes $Q$ queries to the oracle $O$, its inverse $O^\dagger$, or the controlled versions $cO, cO^\dagger$.}
    \label{fig:qos}
\end{figure}

In this section, we provide a detailed description of the proposed quantum oracle sketching scheme, summarized in \Cref{fig:qos}.
In \Cref{sec:q-prelim}, we introduce necessary preliminaries including basic inequalities and techniques from quantum singular value transform (QSVT).
In \Cref{sec:q-alg-iid}, we introduce quantum oracle sketching starting from the simplest case of IID data samples of the form $(x, f(x))$, where $x\sim \unif([N])$ and $f: [N]\to \bit$ is a Boolean function whose property we want to estimate.
We show how to construct the phase oracle of $f$ using the random data samples and use it to run quantum query algorithms that estimate the properties of $f$.
Hence the name quantum oracle sketching.
Along the way, we discuss the apparent trap of decoherence that seems to rule out the possibility of quantum oracle sketching and explain how we evade decoherence.
In \Cref{sec:q-alg-opt}, we show that the sample complexity of quantum oracle sketching is optimal, by proving a matching lower bound.

Building upon this simplest case, in \Cref{sec:q-alg-ext}, we introduce several extensions of quantum oracle sketching.
In particular, we show how to handle functions with multi-bit outputs, time-varying data generation processes that are correlated and not IID, and unknown marginal distributions.
In \Cref{sec:q-alg-linear-algebra}, we use quantum oracle sketching to construct various linear algebra primitives, such as the sparse oracles and block encodings of sparse matrices, and quantum states corresponding to arbitrary vectors, all from random data samples of the matrices and vectors.
Since the state preparation algorithm requires significantly more algorithmic ingredients, we give it a standalone name of quantum state sketching.
These subroutines will be used in various applications detailed in \Cref{sec:app}.

\subsection{Preliminaries}
\label{sec:q-prelim}

\subsubsection{Inequalities}

We begin by introducing some preliminary results that we will use and reprove some of them for accessibility.
The diamond distance between unitary channels is bounded by the operator distance between the unitaries.
This is also true for channels built from isometries.

\begin{lemma}[Diamond distance and operator distance, {\cite[Lemma 3.4]{chen2021concentration}}]
\label{lem:diamond-operator}
    Let $U, V$ be isometries from $\mathbb{C}^{d_i}$ to $\mathbb{C}^{d_o}$ with $d_o\geq d_i$.
    That is, $U^\dagger U = V^\dagger V = I$.
    Let $\mathcal{U}: \rho\to U\rho U^\dagger$ and $\mathcal{V}: \rho\to V\rho V^\dagger$ be the corresponding channels.
    Then we have
    \begin{equation}
        \frac12\|\mathcal{U}-\mathcal{V}\|_\diamond\leq \|U-V\|.
    \end{equation}
\end{lemma}

\begin{proof}
    Note that to compute the diamond distance between unitary channels, it is not necessary to stabilize them with identity on an auxiliary space. 
    Meanwhile, the isometry channels $\mathcal{U}, \mathcal{V}$ can be viewed as unitary channels over an enlarged input Hilbert space.
    Therefore, we have
    \begin{equation}
    \begin{split}
        \frac12 \|\mathcal{U}-\mathcal{V}\|_\diamond &= \max_{\ket{\psi}}\frac12\|U\ket{\psi}\bra{\psi}U^\dagger - V\ket{\psi}\bra{\psi}V^\dagger\|_1 \\
        &=\max_{\ket{\psi}}\sqrt{1-|\bra{\psi}U^\dagger V \ket{\psi}|^2} \\
        &=\max_{\ket{\psi}}\sqrt{(1-|\bra{\psi}U^\dagger V \ket{\psi}|)(1+|\bra{\psi}U^\dagger V \ket{\psi}|)} \\
        &\leq \max_{\ket{\psi}}\sqrt{(1-\mathrm{Re}[\bra{\psi}U^\dagger V \ket{\psi}])(1+1)} \\
        &=\max_{\ket{\psi}}\|U\ket{\psi}-V\ket{\psi}\|_2 \\
        &=\|U-V\|. 
    \end{split}
    \end{equation}
    This proves \Cref{lem:diamond-operator}.
\end{proof}

A similar bound holds for the expectation value of random unitary channels.
This extends to channels built from isometries as well.
In the literature, this relation is often referred to as the mixing lemma \cite{hastings2017turning,chen2021concentration}.
We thank Angus Lowe and Richard Allen for pointing out a subtle mathematical flaw in the original proof presented in an earlier version of \cite[Lemma 3.4]{chen2021concentration}, and for the insightful discussion that led to the corrected proof presented here with the additional factor of $2$.

\begin{lemma}[Diamond distance and operator distance in expectation]
\label{lem:diamond-operator-expect}
Let $U$ be an isometry from $\mathbb{C}^{d_i}$ to $\mathbb{C}^{d_o}$ with $d_o \ge d_i$. That is, $U^\dagger U = I$. Let $V$ be a random isometry from $\mathbb{C}^{d_i}$ to $\mathbb{C}^{d_o}$ with $V^\dagger V = I$. Let $\mathcal{U} : \rho \to U\rho U^\dagger$ and $\mathcal{V} : \rho \to V\rho V^\dagger$ be the corresponding channels. Then we have
\begin{equation}
    \frac{1}{2}\|\mathcal{U} - \mathbb{E}[\mathcal{V}]\|_\diamond \le 2 \|U - \mathbb{E}[V]\|.
\end{equation}
\end{lemma}

\begin{proof}
Let $(p_i, V_i)$ be any random isometry ensemble. 
For any pure state $\ket{\psi}$, we expand the action of the expected channel by centering it around $\E[V]$ as
\begin{equation}
\begin{split}
    \E[V \ket{\psi}\bra{\psi} V^\dagger] &= \E[((V - \mathbb{E}[V]) + \mathbb{E}[V]) \ket{\psi}\bra{\psi} ((V - \mathbb{E}[V]) + \mathbb{E}[V])^\dagger]\\
    &= \E[V]\ket{\psi}\bra{\psi}\E[V]^\dagger + \E[(V-\E[V])\ket{\psi}\bra{\psi}(V-\E[V])^\dagger],
\end{split}
\end{equation}
where we have used $\mathbb{E}[V - \mathbb{E}[V]] = 0$ to remove the cross terms.
By triangle inequality, we have
\begin{equation}
\begin{split}
    &\frac{1}{2}\| U \ket{\psi}\bra{\psi} U^\dagger - \mathbb{E}[V \ket{\psi}\bra{\psi} V^\dagger] \|_1
    \\
    &\leq \underbrace{\frac{1}{2} \| U \ket{\psi}\bra{\psi} U^\dagger - \mathbb{E}[V] \ket{\psi}\bra{\psi} \mathbb{E}[V]^\dagger \|_1}_{\text{bias}} + \underbrace{\frac{1}{2} \| \mathbb{E}[ (V - \mathbb{E}[V]) \ket{\psi}\bra{\psi} (V - \mathbb{E}[V])^\dagger ] \|_1}_{\text{variance}}.
\end{split}
\end{equation}
The bias term is bounded as
\begin{equation}
\begin{split}
    \| U \ket{\psi}\bra{\psi} U^\dagger - \mathbb{E}[V] \ket{\psi}\bra{\psi} \mathbb{E}[V]^\dagger \|_1 &= \| (U - \mathbb{E}[V]) \ket{\psi}\bra{\psi} U^\dagger + \mathbb{E}[V] \ket{\psi}\bra{\psi} (U^\dagger - \mathbb{E}[V]^\dagger) \|_1 \\
    &\leq  \| (U - \mathbb{E}[V]) \ket{\psi}\bra{\psi} U^\dagger\|_1 + \|\mathbb{E}[V] \ket{\psi}\bra{\psi} (U^\dagger - \mathbb{E}[V]^\dagger) \|_1\\
    &\leq \| (U - \mathbb{E}[V]) \ket{\psi} \|_2 \| U\ket{\psi} \|_2 + \| \mathbb{E}[V] \ket{\psi} \|_2 \| (U - \mathbb{E}[V])\ket{\psi} \|_2 \\
    &= (\| U\ket{\psi} \|_2+\| \mathbb{E}[V] \ket{\psi} \|_2) \|(U-\E[V])\ket{\psi}\|_2,
\end{split}
\end{equation}
where we have used triangle inequality and Holder's inequality with $1/2+1/2=1$.
Since $U, V_i$ are isometries, we have $\| U\ket{\psi} \|_2 = 1$ and $\| \E[V] \ket{\psi} \|_2 \le \sum_i p_i \| V_i\ket{\psi} \|_2 = 1$.
Therefore, the bias term is bounded by
\begin{equation}
    \frac{1}{2} \| U \ket{\psi}\bra{\psi} U^\dagger - \mathbb{E}[V] \ket{\psi}\bra{\psi} \mathbb{E}[V]^\dagger \|_1 \leq \|(U-\E[V])\ket{\psi}\|_2.
\end{equation}
To bound the variance term, note that $(V - \mathbb{E}[V]) \ket{\psi}\bra{\psi} (V - \mathbb{E}[V])^\dagger$ is positive semi-definite and so is its expectation. 
Thus, its trace norm is the same as its trace. 
Using $V^\dagger V = I$, we obtain:
\begin{equation}
\begin{split}
    \| \mathbb{E}[ (V - \mathbb{E}[V]) \ket{\psi}\bra{\psi} (V - \mathbb{E}[V])^\dagger ] \|_1 &= \tr\left( \mathbb{E}[ (V - \mathbb{E}[V]) \ket{\psi}\bra{\psi} (V - \mathbb{E}[V])^\dagger ] \right) \\
    &= \mathbb{E}[ \bra{\psi} (V^\dagger - \mathbb{E}[V^\dagger])(V - \mathbb{E}[V]) \ket{\psi} ] \\
    &= \bra{\psi} (I - \mathbb{E}[V^\dagger]\mathbb{E}[V]) \ket{\psi}.
\end{split}
\end{equation}
Now we use $U^\dagger U = I$ and obtain
\begin{equation}
\begin{split}
    \bra{\psi} (U^\dagger U - \mathbb{E}[V^\dagger]\mathbb{E}[V]) \ket{\psi} &= \bra{\psi} U^\dagger(U - \mathbb{E}[V]) \ket{\psi} + \bra{\psi} (U^\dagger - \mathbb{E}[V^\dagger])\mathbb{E}[V] \ket{\psi} \\
    &\leq |\bra{\psi} U^\dagger(U - \mathbb{E}[V]) \ket{\psi}| + |\bra{\psi} (U^\dagger - \mathbb{E}[V^\dagger])\mathbb{E}[V] \ket{\psi}| \\
    &\leq \|U\ket{\psi}\|_2 \|(U-\E[V])\ket{\psi}\|_2 + \|\E[V]\ket{\psi}\|_2 \|(U-\E[V])\ket{\psi}\|_2 \\
    &\leq (1+1)\|(U-\E[V])\ket{\psi}\|_2 = 2\|(U-\E[V])\ket{\psi}\|_2,
\end{split}
\end{equation}
where we have used Cauchy-Schwarz inequality and $U, V_i$ being isometries.
Therefore, the variance term is bounded as
\begin{equation}
    \frac{1}{2} \| \E[ (V - \E[V]) \ket{\psi}\bra{\psi} (V - \E[V])^\dagger ] \|_1 \le \|(U-\E[V])\ket{\psi}\|_2.
\end{equation}
Combining the bias and variance term, we have
\begin{equation}
    \frac{1}{2}\| U \ket{\psi}\bra{\psi} U^\dagger - \E[V \ket{\psi}\bra{\psi} V^\dagger] \|_1 \leq 2 \|(U-\E[V])\ket{\psi}\|_2.
\end{equation}
Using this, we have
\begin{equation}
\begin{split}
    \frac{1}{2}\|\mathcal{U} - \E[\mathcal{V}]\|_\diamond &= \max_{\ket{\psi}} \frac{1}{2} \| (U \otimes I) \ket{\psi}\bra{\psi} (U \otimes I)^\dagger - \E[ (V \otimes I) \ket{\psi}\bra{\psi} (V \otimes I)^\dagger ] \|_1 \\
    &\leq \max_{\ket{\psi}} 2 \| (U \otimes I - \mathbb{E}[V \otimes I]) \ket{\psi} \|_2\\
    &= 2 \| (U - \mathbb{E}[V]) \otimes I \|\\
    &= 2 \| U - \mathbb{E}[V] \|,
\end{split}
\end{equation}
as desired.
\end{proof}

We will also use the subadditivity of diamond distance.
\begin{lemma}[Subadditivity of diamond distance, {\cite[Proposition 3.48]{watrous2018theory}}]
\label{lem:diamond-subadd}
    For any channels $\mathcal{C}_1, \mathcal{C}_2$ and $\mathcal{C}_1', \mathcal{C}_2'$, we have
    \begin{equation}
        \|\mathcal{C}_1'\mathcal{C}_1 - \mathcal{C}_2'\mathcal{C}_2\|_\diamond\leq \|\mathcal{C}_1'-\mathcal{C}_2'\|_\diamond + \|\mathcal{C}_1-\mathcal{C}_2\|_\diamond.
    \end{equation}
\end{lemma}

The following lemma is useful in relating trace distance and relative entropy.
We will use it in proving the optimality of quantum oracle sketching in \Cref{sec:q-alg-opt}.

\begin{lemma}[Quantum Bretagnolle-Huber inequality, {\cite[Lemma B.1]{angrisani2023unifying}, \cite{canonne2022short}}]
\label{lem:q-BH-ineq}
    Let $\rho, \sigma\in \mathbb{C}^{N\times N}$ be two quantum states. 
    Then,
    \begin{equation}
        \frac12 \|\rho-\sigma\|_1 \leq \sqrt{1-2^{-D(\rho\|\sigma)}},
    \end{equation}
    where $D(\rho\|\sigma) = \tr(\rho\log_2\rho - \rho\log_2\sigma)$ is the relative entropy between $\rho$ and $\sigma$.
\end{lemma}

We will also make use of the following moment bound for quadratic forms of random vectors \cite{diakonikolas2009bounded}.
This would be useful in bounding the error of quantum state sketching (\Cref{thm:q-state-sketch}).

\begin{lemma}[Moment bound for quadratic forms {\cite[Theorem 5.1]{diakonikolas2009bounded}}]
\label{lem:moment-quad-form}
    Let $A\in \mathbb{R}^{N\times N}$ be a symmetric matrix.
    Let $v\in \{\pm 1\}^N$ be a uniformly distributed random sign vector.
    Then for all integer $k\geq 2$, we have
    \begin{equation}
        \E[|v^T A v - \tr(A)|^k]\leq \lr{C\max\left\{\sqrt{k}\|A\|_F, k\|A\|\right\}}^k,
    \end{equation}
    where $C>0$ is a universal constant, $\|A\|_F$ is the Frobenius norm of $A$, and $\|A\|$ is the operator norm of $A$.
\end{lemma}

Another useful lemma used is the following moment bound for a sum with random signs.

\begin{lemma}[Khintchine's inequality, {\cite[Exercise 2.6.5]{vershynin2018high}}]
\label{lem:moment-inner-prod}
    Let $b_1, \ldots, b_N\in \mathbb{R}$ be fixed numbers and $v_1, \ldots, v_N\in \{\pm 1\}$ be uniformly distributed random signs.
    Then for all integers $k\geq 2$, we have
    \begin{equation}
        \E\left[\left|\sum_{j=1}^N v_j b_j \right|^k\right] \leq \lr{C\sqrt{k}\sqrt{\sum_{j=1}^Nb_j^2}}^k,
    \end{equation}
    where $C>0$ is a universal constant.
\end{lemma}

The following lemma relates the error of exchanging expectation value and matrix exponentiation on a random diagonal matrix with the variance of its matrix elements.
It will be used to bound the error of quantum oracle sketching by the variance of the data.
As a special case, we note that it can be used to improve the constant factor in the qDrift error bound analysis in \cite{chen2021concentration}.

\begin{lemma}[Expected error and variance]
\label{lem:error-var}
    Let $X = \diag(X_1, \ldots, X_d) \in \mathbb{R}^{d\times d}$ be a random diagonal matrix, where $X_1, \ldots, X_d \in \mathbb{R}$ are random variables.
    Then, we have
    \begin{equation}
        \|e^{i\E[X]} - \E[e^{iX}]\|\leq \frac{1}{2}\max_{1\leq j\leq d} \Var[X_j],
    \end{equation}
    where $\|\cdot \|$ is the operator norm.
\end{lemma}

\begin{proof}
    Since $X$ and $\E[X]$ are all diagonal matrices, they commute with each other.
    Moreover, $e^{i\E[X]} - \E[e^{iX}]$ is also a diagonal matrix with diagonal elements $e^{i\E[X_j]}-\E[e^{iX_{j}}], 1\leq j\leq d$.
    Note that for any $j\in [d]$,
    \begin{equation}
    \begin{split}
        |e^{i\E[X_j]}-\E[e^{iX_{j}}]| &= |\E[e^{i(X_{j}-\E[X_j])}]-1| \\
        &= |\E[e^{i(X_{j}-\E[X_j])}-i(X_j-\E[X_j])-1]| \\
        &\leq \E|e^{i(X_{j}-\E[X_j])}-i(X_j-\E[X_j])-1| \\
        &\leq \E\left[\frac{(X_j-\E[X_j])^2}{2}\right] \\
        &=\frac{1}{2}\Var[X_j],
    \end{split}
    \end{equation}
    where we have used triangle inequality $|\E[\cdot]|\leq \E[|\cdot|]$ and $|e^{iw}-iw-1|\leq w^2/2, \forall w\in \mathbb{R}$.
    Hence, we arrive at
    \begin{equation}
        \|e^{i\E[X]} - \E[e^{iX}]\|\leq \max_{1\leq j\leq d} |e^{i\E[X_j]}-\E[e^{iX_{j}}]| \leq \frac{1}{2} \max_{1\leq j\leq d} \Var[X_j].
    \end{equation}
    This proves \Cref{lem:error-var}.
\end{proof}

\subsubsection{Quantum singular value transform}

We will make use of the following result from quantum singular value transform (QSVT) \cite{gilyen2019quantum,martyn2021grand}.

\begin{lemma}[Quantum eigenvalue and singular value transform, {\cite[Theorem 3]{martyn2021grand}, \cite[Corollary 18, Lemma 19]{gilyen2019quantum}}]
\label{lem:qsvt}
    Let $A = \sum_\lambda \lambda \ket{\lambda}\bra{\lambda}\in \mathbb{C}^{2^n\times 2^n}$ be a Hermitian matrix with $\|A\|\leq 1$.
    Assume that we have query access to the $(n+1)$-qubit unitary $U$ and its inverse $U^\dagger$ such that $\bra{0_a}U\ket{0_a} = A$, where $\ket{0_a}$ is on the ancilla qubit $a$.
    For any $\phi\in[0, 2\pi)$, we use $\Pi_{\phi} = e^{i\phi Z_a}$ to denote the single-qubit $Z$ rotation on the ancilla $a$ and define $\tilde{\Pi}_\phi = \ketbra{0}{0}_{a'}\otimes \Pi_{\phi} + \ketbra{1}{1}_{a'}\otimes \Pi_{-\phi}$ to be a two-qubit unitary on ancilla qubits $a, a'$.
    Then, for any real polynomial $P: \mathbb{R}\to \mathbb{R}$ and even integer $d$ satisfying
    \begin{enumerate}
        \item the degree of $P$ is at most $d$,
        \item $P$ is an even function, and
        \item $|P(x)|\leq 1, \forall x\in [-1, 1]$,
    \end{enumerate}
    there exists a set of rotation angles $\phi_{i}\in [0, 2\pi), i\in [d]$, such that the $(n+2)$-qubit unitary
    \begin{equation}
        P_{\mathrm{QSVT}}(U, \phi) = H_{a'}\left[\prod_{k=1}^{d/2}\left(\tilde{\Pi}_{\phi_{2k-1}} U^\dagger \tilde{\Pi}_{\phi_{2k}} U \right)\right]H_{a'}
    \end{equation}
    satisfies
    \begin{equation}
        \bra{0_{a'}0_a} P_{\mathrm{QSVT}}(U, \phi)\ket{0_{a'}0_a} = P(A) = \sum_\lambda P(\lambda)\ket{\lambda}\bra{\lambda}.
    \end{equation}
    The same conclusion holds when $d$ is odd and $P$ is odd, using the standard odd-parity QSVT sequence, which uses $d$ alternating queries to $U$ and $U^\dagger$.

    More generally, if $U$ block encodes an arbitrary matrix $A$, the corresponding definite-parity QSVT sequence implements the singular value transformation specified by $P$. In particular, for odd $P$, it applies $P$ to the singular values while preserving the corresponding pairing of left and right singular vectors.
    When $A$ is Hermitian, this reduces to the eigenvalue
    transformation $P(A)$.
    
    Moreover, the controlled version of the constructed unitary $P_{\mathrm{QSVT}}(U, \phi)$ can be constructed by replacing $\tilde{\Pi}_\phi$ with their controlled version $c\tilde{\Pi}_{\phi} = \ket{0_{a''}}\bra{0_{a''}}\otimes I + \ket{1_{a''}}\bra{1_{a''}}\otimes\tilde{\Pi}_\phi$:
    \begin{equation}
        cP_{\mathrm{QSVT}}(U, \phi) = \ket{0_{a''}}\bra{0_{a''}}\otimes I +\ket{1_{a''}}\bra{1_{a''}}\otimes P_{\mathrm{QSVT}}(U, \phi) = H_{a'}\left[\prod_{k=1}^{d/2}\left(c\tilde{\Pi}_{\phi_{2k-1}} U^\dagger c\tilde{\Pi}_{\phi_{2k}} U \right)\right]H_{a'},
    \end{equation}
    where $a''$ is the control qubit.
    For the odd-parity sequence, the controlled version is obtained analogously, with the single unmatched query to $U$ or $U^\dagger$ also controlled.

    More generally, all the above results extend to block encodings that use more than one ancilla qubits.
\end{lemma}

We will use \Cref{lem:qsvt} to apply threshold functions, which can be approximated using a polynomial as follows.

\begin{lemma}[Polynomial approximation of threshold functions, {\cite[Page 14]{martyn2021grand}}]
\label{lem:approx-thres}
    For any $\lambda^\star\in (0, 1)$ and $\epsilon\in (0, 2\sqrt{2/(e\pi)}$, there exists a real polynomial $P: \mathbb{R}\to \mathbb{R}$ such that
    \begin{enumerate}
        \item the degree of $P$ is $O(\log(1/\epsilon)/\lambda^*)$,
        \item $P$ is an even function,
        \item $|P(x)|\leq 1, \forall x\in [-1, 1]$, and
        \item $|P(x)-1|\leq \epsilon, \forall x\in [0, \lambda^\star/2]$ and $|P(x)+1|\leq \epsilon, \forall x\in [\lambda^\star, 1]$.
    \end{enumerate}
\end{lemma}

\subsection{Quantum oracle sketching for IID data}
\label{sec:q-alg-iid}

We introduce quantum oracle sketching starting from the simplest case of learning properties of an unknown Boolean function based on its random data samples.
Let $f:[N]\to \{0, 1\}$ be a Boolean function.
Let $p: [N]\to \mathbb{R}$ be a probability distribution over $[N]$.
We show how to construct the phase oracle 
\begin{equation}
    O=\sum_{x=1}^N (-1)^{f(x)}\ket{x}\bra{x}
\end{equation}
to $\epsilon$-error in diamond distance given a sequence of data $(x_i, y_i)_{i\in \mathbb{N}}$, where $x_i$'s are sampled IID from $p(x)$ and $y_i=f(x_i)$.

In \Cref{alg:oracle-construct}, we give the simplest version of quantum oracle sketching that constructs 
\begin{equation}
    U(t) = \sum_{x=1}^N e^{i p(x)f(x) t} \ket{x}\bra{x}
\end{equation}
for arbitrary evolution time $t\geq 0$.
When the data distribution is uniform $p(x)=1/N$, we take $t=N\pi$ and obtain $U(N\pi) = O = \sum_x (-1)^{f(x)} \ket{x}\bra{x}$.

\begin{algorithm}
  \caption{Quantum oracle sketching}
  \label{alg:oracle-construct}
  \begin{algorithmic}[1]
  \Require An input state $\rho\in \mathbb{C}^{N\times N}$; a stream of $M$ data samples $(x_i, y_i)_{i=1}^M$, where $x_i\sim p(x)$ and $y_i=f(x_i)$ for some probability distribution $p:[N]\to \mathbb{R}_{\geq 0}$ and Boolean function $f: [N]\to \bit$; $t\geq 0$.
  \Ensure An output state $\left(\prod_{i=1}^M V_{i}\right)\rho \left(\prod_{i=1}^M V_{i}\right)^\dagger$.
  \For{$i=1, \ldots, M$}
    \State Get a data sample $(x_i, y_i)$ from the stream.
    \State Apply the multi-controlled phase gate $V_{i} = e^{i t y_i/M \ket{x_i}\bra{x_i}}$.
  \EndFor
  \end{algorithmic}
\end{algorithm}

Quantum oracle sketching in this case is very simple. Upon seeing a data point $(x_i, y_i)$, we apply a small, incremental quantum rotation, which is a multi-controlled phase gate
\begin{equation}
    V_i = \exp(it\frac{y_i}{M} \ketbra{x_i}),
\end{equation}
and repeat for all the $M$ data points.
We call this sequence of gates $(V_1, \ldots, V_M)$ the quantum oracle sketch of the classical data $(x_i, y_i), i=1, \ldots, M$.
These gates are applied on the fly.
That means we never store these data points after the corresponding gate application so as to minimize memory consumption.
If we want to query $U(t)$ again, we simply collect a fresh set of data samples and run \Cref{alg:oracle-construct} again.
In this case, the number of qubits used is only $\ceil{\log_2(N)}$, whereas any QRAM based algorithm stores the whole dataset and hence uses $O(N)$ memory.

\subsubsection{The apparent trap of decoherence}

This idea of applying incremental quantum rotations based on data samples naturally resembles the classical incremental updates in streaming algorithms that bypass the need to store the entire dataset. 
However, the coherence requirement of quantum computation seems to immediately rule out its plausibility: the randomness
and entropy in the data are continuously pumped into the quantum machine, causing it to decohere quickly.
This intuition is quantitatively confirmed by the error and sample complexity analysis presented below.
In particular, we will see that an analysis applicable to generic quantum rotations leads to decoherence and gives performance with no advantage over classical algorithms.
This means that the decoherence trap is real and we have to carefully design the quantum rotations for any quantum advantage to be possible.
In \Cref{thm:q-oracle-sketch-iid}, we will use the intuition provided by this failed analysis to develop a variance analysis technique that gives the optimal sample complexity and quantum advantage.

When the quantum rotations are generic, the process of quantum oracle sketching resembles the qDrift method of Hamiltonian simulation \cite{campbell2019random}, in which one randomly samples a term in the Hamiltonian and apply a short-time evolution generated by that term.
In the following, we show that the optimal qDrift analysis gives a sample complexity bound that is far too bad for any meaningful quantum advantage to exist.

More precisely, in the usual qDrift setting, we aim to simulate the time evolution $e^{iHt}$ of some target Hamiltonian $H = \sum_{x}p(x) h_x$ where each term $h_x$ has constant norm.
In the case of quantum oracle sketching, we have $h_x= f(x)\ket{x}\bra{x}$ and $\|h_x\|\leq 1$.
The total interaction strength $\lambda = \sum_x p(x) \|h_x\|=1$.
The idea of qDrift is to sample $M$ random terms of the Hamiltonian $h_{x_1}, \ldots, h_{x_M}$ according to the distribution $x_i\sim p(x)$ and apply the corresponding single-term time evolution $e^{i\lambda t h_{x_i}/M} = e^{itf(x_i)/M \ket{x_i}\bra{x_i}}$, which exactly concides with the procedure of quantum oracle sketching in this case.
Using the standard gate complexity bound for qDrift, we obtain the sample complexity bound \Cref{lem:sample-upper-qdrift}.
Note that if we take $p(x)=1/N$ and $t=N\pi$ to build a single query to the phase oracle, this bound gives us a sample complexity of $M=O(N^2)$, which is quadratic in $N$ and thus provides no quantum advantage.

To see that this offers no advantage over classical algorithms, we consider the sample complexity of classical algorithms when their memory is limited.
As one extreme, if the classical algorithm has $S_C=O(N)$ memory, it can store the entire data set using $M_C=\tilde O(N)$ samples and build a complete description of the function, which allows it to estimate any property of the function.
The other extreme is when the classical algorithm has little memory $S_C = O(1)$, it can still obtain the information of any query it wants by waiting for $\tilde{O}(N)$ samples to get lucky.
Since any property of a Boolean function can be decided with at most $N$ queries, this classical algorithm can solve any problem that requires $O(1)$ workspace using $M_C=N\cdot \tilde{O}(N)$ samples and $S_C=O(1)$ memory.
In fact, in \Cref{sec:cl-hard}, we will prove that for a classical algorithm to solve generic query problems, these two extreme cases are connected by a sample-space lower bound $M_CS_C\geq \Omega(N^2)$.

This means that if our quantum algorithm consumes $O(N^2)$ samples to make a single query, a classical machine can already solve many query problem with the same number of samples and little memory.
However, if we only use $O(N)$ samples, then the error given in \Cref{lem:sample-upper-qdrift} will be large, signifying decoherence.
In fact, Refs. \cite{chen2021concentration,kimmel2017hamiltonian} proves that this sample complexity bound in \Cref{lem:sample-upper-qdrift} is already optimal and cannot be improved for generic Hamiltonians.
In other words, decoherence is inevitable in general.

\begin{lemma}[Sample complexity upper bound via qDrift]
\label{lem:sample-upper-qdrift}
    Let $t\geq 1, 0<\epsilon\leq 1$.
    Let $f: [N]\to \bit$ be a Boolean function and $p: [N]\to \mathbb{R}_{\geq 0}$ be a probability distribution.
    Let $(x_i, y_i)_{i=1}^M$ be a sequence of IID data samples where $x_i\sim p(x)$ and $y_i = f(x_i)$.
    Let $V_i = e^{ity_i/M\ket{x_i}\bra{x_i}}$ and $U = \sum_x e^{ip(x)f(x)t}\ket{x}\bra{x}$.
    Then, 
    \begin{equation}
        M \geq \frac{8t^2}{\epsilon}
    \end{equation}
    samples suffice to guarantee that
    \begin{equation}
        \|\E[\mathcal{V}_M\cdots \mathcal{V}_{1}] - \mathcal{U}\|_\diamond \leq \epsilon,
    \end{equation}
    where $\mathcal{V}_i: \rho \to V_i\rho V_i^\dagger$ and $\mathcal{U}: \rho \to U\rho U^\dagger$ are the corresponding unitary channels.
\end{lemma}

\begin{proof}[Proof of \Cref{lem:sample-upper-qdrift}]
    Let $\epsilon\in (0, 1]$ and $t\geq 1$.
    Let $h_x = f(x)\ket{x}\bra{x}$.
    Since the samples are IID, we have
    \begin{equation}
        \E[\mathcal{V}_M\cdots \mathcal{V}_1] = \E[\mathcal{V}_M]\cdots \E[\mathcal{V}_1] = \mathcal{E}_0^M,
    \end{equation}
    where $\mathcal{E}_0$ is the average channel of a single sample:
    \begin{equation}
        \mathcal{E}_0(\rho) = \sum_x p(x) e^{ith_x/M}\rho e^{-ith_x/M} = \sum_x p(x) \sum_{j=0}^{\infty}\frac{(it/M)^j}{j!} \mathcal{L}_x^j (\rho).
    \end{equation}
    Here, we have defined $\mathcal{L}_x(\rho) = [h_x, \rho]$ with $\|\mathcal{L}_x\|_\diamond \leq 2\|h_x\|\leq 2$.
    On the other hand, the target unitary we want to implement is 
    \begin{equation}
        U = \sum_x e^{itp(x)f(x)}\ket{x}\bra{x} = e^{it\sum_x p(x)f(x)\ket{x}\bra{x}} = e^{itH},
    \end{equation}
    where $H = \sum_xp(x)f(x)\ket{x}\bra{x} = \sum_x p(x) h_x$.
    Its action can be divided into $M$ repetition of the unitary channel
    \begin{equation}
        \mathcal{U}_0(\rho) = U_0\rho U_0^\dagger =  e^{itH/M}\rho e^{-itH/M} = \sum_{j=0}^\infty \frac{(it/M)^j}{j!} \mathcal{L}^j(\rho),
    \end{equation}
    where $U_0 = e^{itH/M}, \mathcal{L}(\rho) = [H, \rho]$ with $\|\mathcal{L}\|_\diamond\leq 2\|H\|\leq 2$.
    Note that
    \begin{equation}
        \mathcal{L}(\rho) = \left[\sum_x p(x) h_x, \rho\right] = \sum_x p(x) \mathcal{L}_x(\rho).
    \end{equation}
    The diamond distance between the single-sample channels reads
    \begin{align}
        \|\mathcal{U}_0 - \mathcal{E}_0\|_\diamond &= \left\|\sum_{j=0}^\infty \frac{(it/M)^j}{j!} \mathcal{L}^j - \sum_x p(x)\sum_{j=0}^{\infty}\frac{(it/M)^j}{j!} \mathcal{L}_x^j \right\|_\diamond \\
        &=\left\|\frac{it}{M}\left(\mathcal{L}-\sum_xp(x)\mathcal{L}_x\right)+\sum_x p(x)\sum_{j=2}^{\infty}\frac{(it/M)^j}{j!} \mathcal{L}_x^j - \sum_{j=2}^\infty \frac{(it/M)^j}{j!} \mathcal{L}^j \right\|_\diamond \\
        &=\left\|\sum_x p(x)\sum_{j=2}^{\infty}\frac{(it/M)^j}{j!} \mathcal{L}_x^j - \sum_{j=2}^\infty \frac{(it/M)^j}{j!} \mathcal{L}^j \right\|_\diamond \\
        &\leq \sum_{j=2}^{\infty}\frac{(t/M)^j}{j!} \left\|\sum_x p(x)\mathcal{L}_x^j\right\|_\diamond + \sum_{j=2}^\infty \frac{(t/M)^j}{j!} \|\mathcal{L}\|_\diamond^j \\
        &\leq \left(\sum_x p(x)+1\right) \sum_{j=2}^{\infty}\frac{(2t/M)^j}{j!} \\
        &\leq 2 \cdot \frac{1}{2}\left(\frac{2t}{M}\right)^2 e^{2t/M} \\
        &=\frac{4t^2}{M^2}e^{2t/M}.
    \end{align}
    From the subadditivity of diamond distance, the overall distance is bounded by
    \begin{equation}
        \|\E[\mathcal{V}_M\cdots \mathcal{V}_{1}] - \mathcal{U}\|_\diamond=\|\mathcal{U}_0^M - \mathcal{E}_0^M\|_\diamond \leq M \|\mathcal{U}_0 - \mathcal{E}_0\|_\diamond \leq \frac{4t^2}{M}e^{2t/M}.
    \end{equation}
    To get an $\epsilon$ error, we choose $M \geq 8t^2/\epsilon$ and obtain
    \begin{equation}
        \|\E[\mathcal{V}_M\cdots \mathcal{V}_{1}] - \mathcal{U}\|_\diamond\leq \frac{1}{2}\epsilon e^{\epsilon/(4t)}\leq \frac{e^{1/4}}{2}\epsilon<\epsilon,
    \end{equation}
    where we have used $\epsilon\leq 1$ and $t\geq 1$.
    This completes the proof of \Cref{lem:sample-upper-qdrift}.
\end{proof}

\subsubsection{Evading decoherence}

As we have seen, the qDrift bound in \Cref{lem:sample-upper-qdrift} gives us an undesirable sample complexity of $O(N^2)$ when $p(x)=1/N$ and $t=N\pi$.
This signifies the inevitable decoherence for generic quantum rotations.

The mathematical origin of this decoherence is that in our case the total interaction strength $\lambda=1$ and thus the qDrift bound yields $O(\lambda^2t^2/\epsilon)=O(t^2/\epsilon)=O(N^2/\epsilon)$.
For general randomized Hamiltonian simulation problems, this bound is already optimal \cite{chen2021concentration,kimmel2017hamiltonian}.
The reason is that the total interaction strength characterizes the average effect of randomly sampling Hamiltonian terms, and the random samples are all that we have.
To go beyond this analysis, we have to carefully design the Hamiltonians and exploit their features, so that we can bypass this average effect.

The first idea that helps to go beyond this average effect is to note that our Hamiltonian terms are all diagonal in the computational basis $h_x=f(x)\ket{x}\bra{x}$, and they are orthogonal to each other.
Intuitively, as the number of samples increases, the phase accumulated on each basis $\ket{x}$ does not interfere with each other and should accumulate by themselves, following a binomial distribution with $M$ repetitions and probability $p(x)$.
From standard concentration inequalities, this phase should concentrate to $\epsilon$ error with a number of samples $M_x = O(p(x)t^2/\epsilon^2)$. 
Let $p_{\max} = \max_x p(x)$.
A union bound then implies that $M=O(p_{\max}t^2\log N / \epsilon^2)$ suffice to $\epsilon$-approximate $U(t)$ in operator norm with high probability.
When $p(x)=1/N$ and $t=N\pi$, this analysis gives a sample complexity of $M=O(N\log N/\epsilon^2)$, which bypasses the $O(N^2)$ barrier for generic Hamiltonians. 

Note that when $p_{\max}t\leq{\epsilon}/{2}$, then no samples are required.
Indeed, replacing $U$ by the identity gives $\|\mathcal{U}-\mathcal{I}\|_\diamond \leq 2\|U-I\|\leq 2p_{\max}t \leq\epsilon$.
We therefore restrict the concentration bound below to the non-trivial regime $p_{\max}t>\epsilon/2$.

\begin{lemma}[Sample complexity upper bound via concentration]
\label{lem:sample-upper-concent}
    Let $t>0, 0<\epsilon, \delta<1$.
    Let $f: [N]\to \bit$ be a Boolean function and $p: [N]\to \mathbb{R}_{\geq 0}$ be a probability distribution.
    Let $p_{\max} = \max_{x\in [N]} p(x)$ and assume $p_{\max}t>\epsilon/2$.
    Let $(x_i, y_i)_{i=1}^M$ be a sequence of IID data samples where $x_i\sim p(x)$ and $y_i = f(x_i)$.
    Let $V_i = e^{ity_i/M\ket{x_i}\bra{x_i}}$ and $U = \sum_x e^{ip(x)f(x)t}\ket{x}\bra{x}$.
    Then,
    \begin{equation}
        M \geq \frac{p_{\max} t^2}{\epsilon^2}\cdot 12\log\frac{2N}{\delta}
    \end{equation}
    samples suffice to guarantee that, with probability at least $1-\delta$, we have
    \begin{equation}
        \|\mathcal{V}_M\cdots \mathcal{V}_{1} - \mathcal{U}\|_\diamond \leq \epsilon,
    \end{equation}
    where $\mathcal{V}_i: \rho \to V_i\rho V_i^\dagger$ and $\mathcal{U}: \rho \to U\rho U^\dagger$ are the corresponding unitary channels.
\end{lemma}

\begin{proof}[Proof of \Cref{lem:sample-upper-concent}]
    For any $x\in [N]$, define $m_x = \sum_{i=1}^M \delta_{x, x_i}$ to be the number of samples $x_i$ that is equal to $x$.
    Then the constructed unitary is
    \begin{equation}
        V = V_M\cdots V_1 = \sum_x e^{itm_xf(x)/M}\ket{x}\bra{x}.
    \end{equation}
    We aim to show that $V$ approximates $U = \sum_x e^{ip(x)f(x)t}\ket{x}\bra{x}$  with high probability.

    To this end, we note that $m_x$ follows a binomial distribution with $M$ repetitions and probability $p(x)$.
        The random variable $m_x$ is binomial with mean $Mp(x)$.
    The additive Bernstein inequality gives, for every $a>0$,
    \begin{equation}
        \Pr\left[
            \left|
                \frac{m_x}{M}
                -
                p(x)
            \right|
            >
            a
        \right]
        \leq
        2\exp\left(
            -
            \frac{
                Ma^2
            }{
                2p(x)+2a/3
            }
        \right).
    \end{equation}
    We choose $a=\epsilon/(2t)$.
    Since $p_{\max}t>\epsilon/2$, we have $a<p_{\max}$.
    Therefore, $2p(x)+2a/3\leq 2p_{\max}+2p_{\max}/3\leq 3p_{\max}$.
    It follows that
    \begin{equation}
        \Pr\left[
            \left|
                \frac{tm_x}{M}
                -
                tp(x)
            \right|
            >
            \frac{\epsilon}{2}
        \right]\leq
        2\exp\left(
            -
            \frac{
                M\epsilon^2
            }{
                12p_{\max}t^2
            }
        \right)\leq
        \frac{\delta}{N},
    \end{equation}
    where the last inequality follows from
    \begin{equation}
        M
        \geq
        \frac{
            12p_{\max}t^2
        }{
            \epsilon^2
        }
        \log\frac{2N}{\delta}.
    \end{equation}
    Therefore, the union bound implies that with probability at least $1-\delta$,
    \begin{equation}
        \left|\frac{tm_xf(x)}{M} - tp(x)f(x)\right|\leq \left|\frac{tm_x}{M} - tp(x)\right|\leq \epsilon/2, \quad \forall x\in [N].
    \end{equation}
    This implies that
    \begin{equation}
    \begin{split}
        \|V-U\| &= \max_x |e^{itm_x f(x)/M}-e^{itp(x)f(x)}| \\
        &=\max_x |e^{i(tm_xf(x)/M-tp(x)f(x))}-1| \\
        &=\max_x 2\left|\sin\left(\frac{1}{2}\left(\frac{tm_xf(x)}{M}-tp(x)f(x)\right)\right)\right| \\
        &\leq\max_x \left|\frac{tm_xf(x)}{M}-tp(x)f(x)\right|\\
        &\leq \epsilon/2,
    \end{split}
    \end{equation}
    where we have used $|\sin(z)|\leq |z|, \forall z\in \mathbb{R}$.
    \Cref{lem:diamond-operator} then yields the desired result:
    \begin{equation}
        \|\mathcal{V}_M\cdots \mathcal{V}_1-\mathcal{U}\|_\diamond \leq 2\|V-U\|\leq \epsilon,
    \end{equation}
    with probability at least $1-\delta$.
\end{proof}

When $p(x)=1/N$ and $t=N\pi$, the sample complexity in \Cref{lem:sample-upper-concent} gives us a $O(N\log N/\epsilon^2)$ scaling that is better than the $O(N^2/\epsilon)$ from \Cref{lem:sample-upper-qdrift} in terms of $N$, but not in $\epsilon$.
The scaling exponent in $\epsilon$ significantly impacts the sample efficiency when we make multiple queries to the constructed oracle.
For example, when we have a query algorithm with query complexity $Q$.
In order to get a final error $\epsilon$, we need the error in each oracle to be bounded by $\epsilon/Q$.
Since we are sampling fresh data each time we query, the total sample complexity is 
\begin{equation}
    Q\cdot \tilde{O}\lr{\frac{N}{(\epsilon/Q)^2}}=\tilde{O}\lr{\frac{Q^3 N}{\epsilon^2}}
\end{equation}
incurring a cubic slowdown in $Q$.
But if the scaling is $1/\epsilon$ instead of $1/\epsilon^2$, there will only be a quadratic slowdown in $Q$.
As we will see in \Cref{sec:cl-hard}, this also makes the desired space complexity separation harder to achieve.

In the following, we introduce a second idea that improves on the $\epsilon$ dependence.
The key is to give up on having a worst-case error bound as in \Cref{lem:sample-upper-concent}, and instead only demand the random unitary channel to be close to the target in expectation as in \Cref{lem:sample-upper-qdrift}.
This leads to a quadratic suppression of error known as mixing in the literature \cite{hastings2017turning,chen2021concentration}.
However, we need to bound the error more carefully in order to recover the $p_{\max}$ factor, which is crucial in bypassing $O(N^2)$ to reach $O(N)$.

\begin{tcolorbox}
\begin{theorem}[Quantum oracle sketching for IID data]
\label{thm:q-oracle-sketch-iid}
    Let $t, \epsilon>0$.
    Let $f: [N]\to \bit$ be a Boolean function and $p: [N]\to \mathbb{R}_{\geq 0}$ be a probability distribution.
    Let $p_{\max} = \max_{x\in [N]} p(x)$.
    Let $(x_i, y_i)_{i=1}^M$ be a sequence of IID data samples where $x_i\sim p(x)$ and $y_i = f(x_i)$.
    Let $V_i = e^{ity_i/M\ket{x_i}\bra{x_i}}$ and $U = \sum_x e^{ip(x)f(x)t}\ket{x}\bra{x}$.
    Then, 
    \begin{equation}
        M \geq \frac{2p_{\max} t^2}{\epsilon}
    \end{equation}
    samples suffice to guarantee that 
    \begin{equation}
        \|\E[\mathcal{V}_M\cdots \mathcal{V}_{1}] - \mathcal{U}\|_\diamond \leq \epsilon,
    \end{equation}
    where $\mathcal{V}_i: \rho \to V_i\rho V_i^\dagger$ and $\mathcal{U}: \rho \to U\rho U^\dagger$ are the corresponding unitary channels.
    The data processing time per sample is $O(\log N)$.
\end{theorem}
\end{tcolorbox}

\revise{
Before presenting the proof, we highlight an important conceptual point regarding the error guarantee in \Cref{thm:q-oracle-sketch-iid}.
Rather than requiring every individual realization of the random unitary sequence $V_M \cdots V_1$ to be close to $U$ in operator norm (which would yield a worse $1/\epsilon^2$ sample scaling as in \Cref{lem:sample-upper-concent}), \Cref{thm:q-oracle-sketch-iid} bounds the diamond distance of the \emph{expected channel} $\E[\mathcal{V}_M \cdots \mathcal{V}_1]$.
Nevertheless, we note that this expected channel is precisely the physical quantum channel implemented on the quantum machine. 
Therefore, this error bound is still strong enough to bound the error of the quantum algorithm for the worst-case inputs.

To see this, note that each classical sample $z_i \sim p(z)$ provided by the classical data stream can be described quantum mechanically as $\rho_{\mathrm{data}} = \sum_z p(z) \ketbra{z}{z}$. 
Conditioned on sample $z_i$, we apply $V_{z_i}$ to the state of the quantum machine $\sigma$, transforming the joint state to $\sum_z p(z) \ketbra{z}{z} \otimes V_z \sigma V_z^\dagger$. 
Discarding the classical sample register corresponds to taking the partial trace over it:
\begin{equation}
    \tr_{\mathrm{data}}\left(\sum_z p(z) \ketbra{z}{z} \otimes V_z \sigma V_z^\dagger\right) = \sum_z p(z) V_z \sigma V_z^\dagger = \E[\mathcal{V}_z](\sigma).
\end{equation}
Iterating this over all $M$ samples, the state of the quantum machine at the end is exactly $\E[\mathcal{V}_M \cdots \mathcal{V}_1](\sigma)$.

Operationally, by the definition of the diamond norm, for any initial state $\sigma_0$ (possibly entangled with an ancilla) and any final measurement $\{M_y\}$ with classical measurement outcome $y$, the induced probability distribution $\tilde{P}(y)$ on the quantum machine and the ideal distribution $P(y)$ under the target unitary $\mathcal{U}$ satisfy
\begin{equation}
    d_{\mathrm{TV}}(\tilde{P}, P) = \frac{1}{2} \sum_y |\tilde{P}(y) - P(y)| \leq \frac{1}{2} \|\E[\mathcal{V}_M \cdots \mathcal{V}_1](\sigma_0) - \mathcal{U}(\sigma_0)\|_1 \leq \frac{1}{2} \|\E[\mathcal{V}_M \cdots \mathcal{V}_1] - \mathcal{U}\|_\diamond \leq \frac{\epsilon}{2}.
\end{equation}
Therefore, if an ideal algorithm succeeds (e.g., outputs an estimate satisfying $|\hat{\theta} - \theta| \leq \epsilon_{\mathrm{est}}$) with probability at least $1 - \delta_0$, the algorithm using quantum oracle sketching succeeds with probability at least $1 - \delta_0 - \epsilon/2$. 
The sketching error $\epsilon$ directly controls the overall failure probability $\delta$, rather than inflating the estimation error $\epsilon_{\mathrm{est}}$.
This property will be repeatedly used in \Cref{thm:linear-sys-upper,thm:bin-classify-upper,thm:dim-reduc-upper} to bound the failure probability of the quantum algorithms solving the relevant applications.
}

\begin{proof}[Proof of \Cref{thm:q-oracle-sketch-iid}]
    Let $h_x = f(x)\ket{x}\bra{x}$.
    Then we have $V_i = e^{ith_{x_i}/M}$ and $U = e^{it\E_x[h_x]}$, where $x\sim p(x)$.
    The gate complexity of each $V_i$ is $O(\log N)$.
    Let
    \begin{equation}
        m_x = \frac{1}{M}\sum_{i=1}^M\delta_{x, x_i}
    \end{equation}
    be the empirical frequency of a given $x$ appearing in the data.
    Note that since the $h_x$'s are orthogonal and commute with each other, we have
    \begin{equation}
        V_M\cdots V_1 = \prod_{i=1}^M e^{ith_{x_i}/M} = e^{it\frac{1}{M}\sum_{i=1}^M h_{x_i}}.
    \end{equation}
    From \Cref{lem:diamond-operator-expect}, we have
    \begin{equation}
        \frac{1}{4}\|\mathcal{U} - \E[\mathcal{V}_M\cdots \mathcal{V}_1]\|_\diamond \leq \|U - \E[V_M\cdots V_1]\| = \|e^{it\E_{x}[h_x]} - \E[e^{it\frac{1}{M}\sum_{i=1}^M h_{x_i}}]\|.
    \end{equation}
    Let $X = t\frac{1}{M}\sum_{i=1}^M h_{x_i} = t \sum_x m_xh_x$ be the random matrix in the exponent.
    Note that the linearity of expectation implies
    \begin{equation}
        \E[X] =  t\frac{1}{M}\sum_{i=1}^M \E[h_{x_i}] = t\frac{1}{M}\sum_{i=1}^M \E_x[h_x] = t\E_x[h_x],
    \end{equation}
    where we have used the fact that the marginal distribution of each individual $x_i$ is the same $x\sim p(x)$.
    Note also that the $X$ and $\E[X]$ matrices are all diagonal matrices, so they commute with each other.
    Thus we have
    \begin{align}
        &\frac{1}{4}\|\mathcal{U} - \E[\mathcal{V}_M\cdots \mathcal{V}_1]\|_\diamond \leq \|e^{i\E[X]}-\E[e^{iX}]\| 
        = \|\E[e^{i(X-\E[X])}]-1\| 
        =\|\E[e^{i(X-\E[X])} - i(X-\E[X])-1]\| \\
        &=\left\|\sum_x \E[e^{i(tm_xf(x)\ket{x}\bra{x}-\E[tm_xf(x)]\ket{x}\bra{x})} - i(tm_xf(x)\ket{x}\bra{x}-\E[tm_xf(x)]\ket{x}\bra{x})-1]\right\| \\
        &=\left\|\sum_x \ket{x}\bra{x} \E[e^{i(tm_xf(x)-\E[tm_xf(x)])} - i(tm_xf(x)-\E[tm_xf(x)])-1]\right\| \\
        &= \max_x \left|\E[e^{i(tm_xf(x)-\E[tm_xf(x)])} - i(tm_xf(x)-\E[tm_xf(x)])-1]\right| \\
        &\leq \max_x \E[|e^{i(tm_xf(x)-\E[tm_xf(x)])} - i(tm_xf(x)-\E[tm_xf(x)])-1|] \\
        &\leq \max_x \frac{1}{2}\E[(tm_xf(x)-\E[tm_xf(x)])^2] 
        =\frac{1}{2} \max_x \Var[tm_xf(x)] 
        \leq \frac{t^2}{2} \max_x \Var[m_x],
    \end{align}
    where we have used the triangle inequality, the inequality that $|e^{iw}-iw-1|\leq w^2/2, \forall w\in \mathbb{R}$, and $\{\ket{x}\}$ being orthonormal bases.
    Note that this essentially reproves \Cref{lem:error-var}, but we include it to highlight how the diagonal structure helps avoid the averaging effect in the usual qDrift analysis.

    Since the samples are IID, we have the variance
    \begin{equation}
        \Var[m_x] = \frac{1}{M^2} \Var\left[\sum_{i=1}^M \delta_{x, x_i}\right] = \frac{1}{M^2} \sum_{i=1}^M\Var[ \delta_{x, x_i}] = \frac{1}{M} p(x)(1-p(x))\leq \frac{p_{\max}}{M},
    \end{equation}
    where we have used $\delta_{x, x_i}$ being a Bernoulli variable with probability $p(x)$.

    Plugging the variance bound back in, we arrive at
    \begin{equation}
        \|\mathcal{U} - \E[\mathcal{V}_M\cdots \mathcal{V}_1]\|_\diamond\leq 2t^2 \max_x \Var[m_x]\leq \frac{2t^2p_{\max}}{M}.
    \end{equation}
    Choosing
    \begin{equation}
        M \geq \frac{2p_{\max} t^2}{\epsilon},
    \end{equation}
    we have $\|\E[\mathcal{V}_M\cdots \mathcal{V}_{1}] - \mathcal{U}\|_\diamond\leq \epsilon$.
    This concludes the proof of \Cref{thm:q-oracle-sketch-iid}.
\end{proof}

We remark that here we circumvent the averaging effect in the usual qDrift analysis by taking advantage of the fact that different $h_x$ are orthogonal to each other, similar to \Cref{lem:sample-upper-concent}.
This allows us to explicitly bound the operator norm by the maximum variance of $m_x$, which gives us the $p_{\max}$ factor.
Otherwise, all we can do is to use $\|\E[\cdot]\|\leq \E[\|\cdot\|]$ which will lead to the $\lambda = \sum_x p(x)=1$ factor in the usual qDrift analysis and a final $O(t^2/\epsilon)$ sample complexity as in \Cref{lem:sample-upper-qdrift}.

Using \Cref{thm:q-oracle-sketch-iid}, we can instantiate the quantum oracle queries in a query algorithm using a sequence of classical data samples.
In particular, we have the following result.

\begin{tcolorbox}
\begin{theorem}[Query algorithms with quantum oracle sketching]
\label{thm:q-query-alg}
    There is a classical algorithm that, for any Boolean function $f: [N]\to \bit$ and probability distribution $p: [N]\to \mathbb{R}_{\geq 0}$ with $p_{\max}= \max_{x\in [N]} p(x)$, any $t, \epsilon>0$, takes as input any quantum query algorithm $\mathcal{A}$ that queries $U = \sum_x e^{ip(x)f(x)t}\ket{x}\bra{x}, U^\dagger, cU=\ket{0}\bra{0}\otimes I+\ket{1}\bra{1}\otimes U$, or $cU^\dagger$ in total $Q$ times, and outputs a quantum learning algorithm $\mathcal{A}'$ with sample complexity
    \begin{equation}
        M =Q\ceil{\frac{2p_{\max}t^2Q}{\epsilon}} = O\lr{\frac{p_{\max}t^2Q^2}{\epsilon}}
    \end{equation}
    and input form $\mathcal{I}=[N]\times \bit$.
    Upon receiving IID data samples of the form $(x_i, y_i)_{i=1}^M$ where $x_i\sim p(x)$ and $y_i = f(x_i)$, the quantum learning algorithm $\mathcal{A}'$ satisfies
    \begin{equation}
        \|\E[\mathcal{A}'] - \mathcal{A}\|_\diamond \leq \epsilon,
    \end{equation}
    where the expectation is over random data samples.
    Meanwhile, the space complexity of $\mathcal{A}'$ is the same as that of $\mathcal{A}$ and the data processing time of $\mathcal{A}'$ is bounded by the time complexity of $\mathcal{A}$.
\end{theorem}
\end{tcolorbox}

\begin{proof}[Proof of \Cref{thm:q-query-alg}]
    We prove \Cref{thm:q-query-alg} by explicitly constructing the algorithm $\mathcal{A}'$ using \Cref{alg:oracle-construct}.
    For any unitary $U\in U(N)$, we use $\mathcal{U}: \rho\to U\rho U^\dagger$ to denote its corresponding unitary channel.
    We use $c\mathcal{U}: \rho\to cU\rho (cU)^\dagger$ to denote the unitary channel of the controlled unitary.
    Let $\mathcal{A} = \mathcal{C}_Q \mathcal{U}_Q\cdots \mathcal{C}_1 \mathcal{U}_1 \mathcal{C}_0$ be the quantum channel of the quantum query algorithm with query complexity $Q$, where $\mathcal{C}_1, \ldots, \mathcal{C}_Q$ are fixed quantum channels and $\mathcal{U}_1, \ldots, \mathcal{U}_Q\in \{\mathcal{U}, \mathcal{U}^\dagger, c\mathcal{U}, c\mathcal{U}^\dagger\}$.
    \Cref{thm:q-oracle-sketch-iid} implies that we can use $M_0 = \ceil{2p_{\max}t^2/\epsilon_1}$ samples to construct a random unitary $V\in U(N)$ by \Cref{alg:oracle-construct} such that
    \begin{equation}
        \|\E[\mathcal{V}]-\mathcal{U}\|_\diamond \leq \epsilon_1,
    \end{equation}
    where $\mathcal{V}: \rho \to V\rho V^\dagger$ is the unitary channel corresponding to $V$.
    Similarly, by changing the random unitary in \Cref{alg:oracle-construct} to its controlled version or replace $t$ by $-t$, we can implement $\mathcal{U}^\dagger, c\mathcal{U}, c\mathcal{U}^\dagger$ to $\epsilon_1$ error using the same number of samples as well.

    We construct $\mathcal{A}'$ as follows.
    We draw $M=QM_0$ samples from the data stream and use \Cref{alg:oracle-construct} to construct $V_1, \ldots, V_Q \in U(N)$.
    \Cref{thm:q-oracle-sketch-iid} guarantees that $\|\E[\mathcal{V}_i]-\mathcal{U}_i\|_\diamond\leq \epsilon_1$ for any $i\in [Q]$.
    We define
    \begin{equation}
        \mathcal{A}' = \mathcal{C}_Q \mathcal{V}_Q\cdots \mathcal{C}_1 \mathcal{V}_1 \mathcal{C}_0.
    \end{equation}
    By construction, the space complexity of $\mathcal{A}'$ is the same as that of $\mathcal{A}$.
    Since $V_1, \ldots, V_Q$ are independent because the data samples are, we have
    \begin{equation}
    \begin{split}
        \|\E[\mathcal{A}']-\mathcal{A}\|_\diamond &= \|\mathcal{C}_Q \E[\mathcal{V}_Q]\cdots \mathcal{C}_1 \E[\mathcal{V}_1] \mathcal{C}_0-\mathcal{C}_Q \mathcal{U}_Q\cdots \mathcal{C}_1 \mathcal{U}_1 \mathcal{C}_0\|_\diamond \\
        &\leq \sum_{i=1}^Q\|\E[\mathcal{V}_i]-\mathcal{U}_i\|_\diamond \\
        &\leq Q\epsilon_1,
    \end{split}
    \end{equation}
    where we have used \Cref{lem:diamond-subadd}.
    Let $\epsilon_1=\epsilon/Q$, then we arrive at
    \begin{equation}
        \|\E[\mathcal{A}']-\mathcal{A}\|_\diamond\leq \epsilon
    \end{equation}
    with 
    \begin{equation}
        M = QM_0 = Q \ceil{\frac{2p_{\max}t^2}{\epsilon_1}} = Q\ceil{\frac{2p_{\max}t^2Q}{\epsilon}}.
    \end{equation}
    This concludes the proof of \Cref{thm:q-query-alg}.
\end{proof}

\subsection{Optimality}
\label{sec:q-alg-opt}

In \Cref{thm:q-query-alg}, we have shown that we can execute any quantum query algorithm with query complexity $Q$ by performing quantum oracle sketching using classical data samples on each query.
The total number of samples has at most a quadratic slow down $\sim Q^2$.
A natural question is whether we can further improve this to linear in $Q$ and if the dependence on other parameters is tight.
One may think that it could be possible to use a ``higher-order'' version of qDrift to improve the scaling \cite{nakaji2024high}.
Yet in the following, we show that this is impossible, which also implies that the method developed in \cite{nakaji2024high} does not support sequential queries to the simulated Hamiltonian evolution.

In particular, we show in \Cref{thm:sample-lower} that the sample complexity given in \Cref{thm:q-query-alg} is tight and the quadratic slow down is necessary.
This quadratic slow down is fundamentally tied to the incoherent random sampling access to data that we have.
Combining \Cref{thm:q-query-alg,thm:sample-lower}, when
$p_{\max}=\Theta(1/N)$ and $t=\Theta(N)$, the sample complexity is $\Theta\left({NQ^2}/{\epsilon}\right)$

We also note that the assumption $p_{\max}tQ\geq 8\pi\epsilon$ in \Cref{thm:sample-lower} is indispensable.
If instead $p_{\max}tQ<8\pi\epsilon$, then we can always draw no sample at all and replace every query to $U$ by the identity.
The resulting error is bounded by $O(\|U-I\|\cdot Q)\leq O(p_{\max}tQ)\leq O(\epsilon)$.
In other words, in this case there is always an algorithm that achieves the goal without using any samples.

\begin{tcolorbox}
\begin{theorem}[Quantum oracle sketching is sample optimal]
\label{thm:sample-lower}
    Let $t>0, \epsilon\in (0, 1/12), p_{\max}\in [2/N, 1]$.
    Suppose there is a (possibly quantum) algorithm that, for any Boolean function $f: [N]\to \bit$ and probability distribution $p: [N]\to \mathbb{R}_{\geq 0}$ with $\max_{x\in [N]} p(x)\leq p_{\max}$, takes any quantum query algorithm $\mathcal{A}$ that queries $U = \sum_x e^{ip(x)f(x)t}\ket{x}\bra{x}$ in total $Q$ times, and outputs a quantum learning algorithm $\mathcal{A}'$ with input form $\mathcal{I}=[N]\times \bit$ such that
    \begin{equation}
        \|\E[\mathcal{A}'] - \mathcal{A}\|_\diamond \leq \epsilon
    \end{equation}
    upon receiving $M$ IID data samples $(x_i, y_i)_{i=1}^M$ where $x_i\sim p(x)$ and $y_i = f(x_i)$.
    Suppose $p_{\max}tQ\geq 8\pi\epsilon$, then we must have 
    \begin{equation}
        M\geq \Omega\left(\frac{p_{\max}t^2Q^2}{\epsilon}\right).
    \end{equation}
\end{theorem}
\end{tcolorbox}

\begin{proof}[Proof of \Cref{thm:sample-lower}]
    We use $M(t, Q, \epsilon)$ to denote the sample complexity, which is defined for any $\epsilon\in (0, 1)$.
    More precisely, it is the minimum sample complexity among all procedures that satisfy the guarantee in the worst case over all Boolean functions $f$, all distributions $p$ satisfying $\max_x p(x)\leq p_{\max}$ and all quantum query algorithms making at most $Q$ queries.
    We prove \Cref{thm:sample-lower} by connecting the task of oracle construction to the task of quantum state discrimination.
    We show that the oracle construction algorithm described in \Cref{thm:sample-lower} can be used to distinguish two different quantum states using $M$ copies of the states.
    Then a sample complexity lower bound for quantum state discrimination translates into the desired sample complexity bound for oracle construction.

    In particular, we note that drawing $M$ samples from the distribution $p(x)$ is weaker than having access to $M$ copies of the diagonal state $\rho = \sum_x p(x)\ket{x}\bra{x}$.
    Recall that $p_{\max}\geq 2/N$.
    Let $K = \ceil{2/p_{\max}}\in [2, N]$ and $p=1/K$.
    Note that $p=1/K\leq p_{\max}/2<p_{\max}$ and $p=1/K\geq 1/(2/p_{\max}+1)=p_{\max}/(p_{\max}+2)\geq p_{\max}/3$.
    Let $\gamma  = \pi/(2tQ)$.
    We first assume that $tQ\geq 2\pi/p_{\max}$ such that $\gamma\leq p_{\max}/4$ and hence $p+\gamma\leq 3p_{\max}/4$.

    We consider the following two states:
    \begin{equation}
        \rho_1 = \sum_{x}p_1(x)\ket{x}\bra{x} = \sum_{x=1}^K p\ket{x}\bra{x}
    \end{equation}
    and 
    \begin{equation}
        \rho_2 = \sum_{x}p_2(x)\ket{x}\bra{x} = (p+\gamma)\ket{1}\bra{1} + (p-\gamma) \ket{2}\bra{2} + \sum_{x=3}^K p\ket{x}\bra{x}.
    \end{equation}
    Indeed, we have $\max_x p_1(x)=p<p_{\max}$ and $\max_x p_2(x)=p+\gamma\leq p_{\max}$.
    The relative entropy between $\rho_1, \rho_2$ is
    \begin{equation}
        D(\rho_1\|\rho_2) = p\log\frac{p}{p+\gamma} + p\log\frac{p}{p-\gamma} = p\log\frac{1}{1-\gamma^2/p^2}\leq\frac{\gamma^2}{p}\frac{1.5}{1-\gamma^2/p^2},
    \end{equation}
    where we have used the inequality $\log(1/(1-z))\leq 1.5z/(1-z), \forall z\in [0, 1)$.
    Suppose we have $M_0$ copies of either $\rho_1$ or $\rho_2$ and we want to distinguish the two cases with success probability at least $2/3$.
    \Cref{lem:q-BH-ineq} and the operational meaning of trace distance then implies that
    \begin{equation}
        1-2\cdot \frac1{3}\leq \frac12\|\rho_1^{\otimes M_0}-\rho_2^{\otimes M_0}\|_1\leq \sqrt{1-2^{-D(\rho_1^{\otimes M_0}\|\rho_2^{\otimes M_0})}} = \sqrt{1-2^{-M_0D(\rho_1\|\rho_2)}}.
    \end{equation}
    Therefore, we have
    \begin{equation}
        M_0\geq \frac{1}{D(\rho_1\|\rho_2)}\log\frac{1}{1-(1/3)^2}\geq \frac{1}{D(\rho_1\|\rho_2)}\log\frac{9}{8}\geq \frac{2p}{3\gamma^2}\left(1-\frac{\gamma^2}{p^2}\right)\log\frac{9}{8}\geq \frac{7p}{24\gamma^2}\log\frac{9}{8},
    \end{equation}
    where we have used $\gamma/p\leq (p_{\max}/4)/(p_{\max}/3)= 3/4$.
    Since $p\geq p_{\max}/3$ and $\gamma=\pi/(2tQ)$, we arrive at
    \begin{equation}
        M_0\geq \frac{7p_{\max}t^2Q^2}{18\pi^2}\log\frac{9}{8}.
    \end{equation}
    In other words, we have proved that if $tQ\geq 2\pi/p_{\max}$, then any algorithm that can distinguish $\rho_1$ and $\rho_2$ with success probability at least $2/3$ must use $M_0$ samples.

    Now we construct an algorithm to distinguish $\rho_1$ and $\rho_2$ using the oracle construction algorithm in \Cref{thm:sample-lower}.
    We consider the Boolean function $f(x)=1, \forall x\in [N]$.
    This resembles the idea used in proving the sample complexity lower bound of quantum principle component analysis \cite{kimmel2017hamiltonian}.
    We first let $\epsilon=1/3$.
    Let $\ket{+} = (\ket{1}+\ket{2})/\sqrt{2}$ and $\ket{-} = (\ket{1}-\ket{2})/\sqrt{2}$.
    Specifically, we set the query algorithm $\mathcal{A}$ to be the simple algorithm that prepares the initial state $\ket{+}$, applies $U$ consecutively $Q$ times, and perform the two-outcome measurements $\{\Pi_0=\ket{+}\bra{+}, \Pi_1=I-\Pi_0\}$.
    If the state that generates the data stream is $\rho_1$, then 
    \begin{equation}
        U^{Q}\ket{+} = \frac{1}{\sqrt{2}}(e^{itQp}\ket{1} + e^{itQp}\ket{2})=e^{itQp}\ket{+}.
    \end{equation}
    Thus $\mathcal{A}$ will output $0$ with certainty.
    On the other hand, if the state that generates the data stream is $\rho_2$, then
    \begin{equation}
        U^{Q}\ket{+} = \frac{1}{\sqrt{2}}(e^{itQ(p+\gamma)}\ket{1} + e^{itQ(p-\gamma)}\ket{2})=e^{itQ(p+\gamma)}\frac{1}{\sqrt{2}}(\ket{1}+e^{-i2tQ\frac{\pi}{2tQ}}\ket{2}) = e^{itQ(p+\gamma)}\ket{-},
    \end{equation}
    which is orthogonal to $\ket{+}$.
    Hence $\mathcal{A}$ will output $1$ with certainty.
    That is, $\mathcal{A}$ can distinguish $\rho_1$ and $\rho_2$ with certainty.
    Meanwhile, the oracle construction algorithm in \Cref{thm:sample-lower} uses $M(t, Q, 1/3)$ samples and gives us an algorithm $\mathcal{A}'$ such that $\|\E[\mathcal{A}']-\mathcal{A}\|_\diamond\leq 1/3$.
    Since the output of $\mathcal{A}$ is a random bitstring, the diamond norm reduces to the total variation distance of the distributions of the outputs.
    This implies that $\mathcal{A}'$ can distinguish $\rho_1$ and $\rho_2$ with probability at least $2/3$.
    Therefore, we must have
    \begin{equation}
    \label{eq:sample-lower-constant-eps}
        M(t, Q, 1/3)\geq M_0 \geq \frac{7p_{\max}t^2Q^2}{18\pi^2}\log\frac{9}{8},
    \end{equation}
    if $tQ\geq 2\pi/p_{\max}$.

    For general $t, Q, \epsilon$, recall that we have $p_{\max}tQ/\epsilon\geq 8\pi$.
    Note that
    \begin{equation}
        kM(t, Q, \epsilon)\geq M(t, kQ, k\epsilon)
    \end{equation}
    for any positive integer $k$, because we can always construct $kQ$ queries to $U$ with error $k\epsilon$ by constructing $k$ chunks of $Q$ queries to $U$ with error $\epsilon$.
    Therefore, taking $k = \floor{1/(3\epsilon)}$ and using the fact that $M(t, Q, \epsilon)$ decreases with $\epsilon$, we have
    \begin{equation}
        M(t, Q, \epsilon)\geq \frac{1}{k} M(t, kQ, k\epsilon) \geq \frac{1}{k} M(t, kQ, 1/3)\geq \frac{1}{k}\frac{7p_{\max}t^2k^2Q^2}{18\pi^2}\log\frac{9}{8} \geq \frac{p_{\max}t^2Q^2}{\epsilon}\frac{7\log(9/8)}{72\pi^2},
    \end{equation}
    where we have used that $k\geq 1/(3\epsilon)-1\geq 1/(4\epsilon)$ when $\epsilon<1/12$ and $tkQ\geq tQ/(4\epsilon)\geq 2\pi/p_{\max}$ so that we can apply \Cref{eq:sample-lower-constant-eps}.
    This completes the proof of \Cref{thm:sample-lower}.
\end{proof}

\subsection{Extensions}
\label{sec:q-alg-ext}

We have shown how to quantum oracle sketching to construct a unitary $U(t)=\sum_x e^{ip(x)f(x)t}\ket{x}\bra{x}$ from a sequence of random data samples $(x_i, f(x_i))$ where $x_i\sim p(x)$ and $f: [N]\to \bit$.
When the data distribution is uniform $p(x)=1/N$, the $t=N\pi$ unitary $U(N\pi)$ gives the desired phase oracle $O=\sum_x (-1)^{f(x)}\ket{x}\bra{x}$.
We have also shown that the sample complexity for querying this oracle $Q$ times is 
\begin{equation}
    M = O\lr{\frac{p_{\max}t^2Q^2}{\epsilon}} = O\lr{\frac{NQ^2}{\epsilon}}.
\end{equation}

In this section, we discuss several extensions to this simplest version of quantum oracle sketching.
We first study the cases when we have multi-bit outputs $f: [N]\to \bit^b$, when the data are non-IID and generated from a general hierarchical data generation process with multiple time scales, and when the underlying distribution is non-uniform and unknown.

\subsubsection{Multi-bit output}
\label{sec:multibit}

We first generalize quantum oracle sketching to handle multi-bit outputs.
Let $b$ be the length of the output bitstrings.
We assume that the data samples are of the form $(x_i, y_i)$ with $y_i = f(x_i), f:[N] \to \bit^b$.
There are many equivalent ways to define oracles for functions with multi-bit outputs (e.g., the standard XOR oracle that operates on $\log N + b$ qubits and maps $\ket{x}\ket{y}$ to $\ket{x}\ket{y\oplus f(x)}$).
For our purposes, we aim to minimize the space overhead, and therefore we consider the multi-bit phase oracle 
\begin{equation}
    O: \ket{x, j} \to (-1)^{f_j(x)}\ket{x, j}, \quad \forall x\in [N], j\in [b],
\end{equation}
that operates on $\log N + \log b$ qubits.

One can equivalently consider the Boolean function 
\begin{equation}
    \hat{f}: [N]\times [b] \to \bit, \quad \hat{f}(x, j)=f_j(x)\in \bit,
\end{equation} 
whose standard phase oracle is
\begin{equation}
    O: \ket{z} \to (-1)^{\hat{f}(z)}\ket{z}, \forall z\in [N]\times [b],
\end{equation}
which is the same as the multi-bit phase oracle of $f$.
This identification immediately implies a (sample wasteful) way of using \Cref{alg:oracle-construct} to construct the multi-bit phase oracle: we simply subsample each data point $(x_i, y_i)$ with a single uniformly random coordinate $j\in [b]$.
This is the same as sampling random $(z = (x, j), \hat{f}(z))$.
This enlarges the domain from $[N]$ to $[N]\times [b]$.
\Cref{thm:q-oracle-sketch-iid} then shows that when $p(x)=1/N$ is uniform, $p(z)=1/(bN)$ and we can construct the desired oracle using $O(bN/\epsilon)$ samples.
Here, the additional factor $b$ comes from the fact that due to our wasteful use of each sample, we need $b$ samples to gather all the coordinates of each $y\in \bit^b$.
Nevertheless, as long as $b=\polylog(N)$, this sampling overhead is still tolerable.

\begin{algorithm}
  \caption{Quantum oracle construction (multi-bit version)}
  \label{alg:oracle-construct-multibit}
  \begin{algorithmic}[1]
  \Require An input state $\rho\in \mathbb{C}^{(N\times b)\times (N\times b)}$; a stream of $M$ data samples $(x_i, y_i)_{i=1}^M$, where $x_i\sim p(x)$ and $y_i=f(x_i)$ for some probability distribution $p:[N]\to \mathbb{R}_{\geq 0}$ and function $f: [N]\to \bit^b, b\in \mathbb{Z}_{\geq 1}$; $t\geq 0$.
  \Ensure An output state $\left(\prod_{i=1}^M V_{i}\right)\rho \left(\prod_{i=1}^M V_{i}\right)^\dagger$.
  \For{$i=1, \ldots, M$}
    \State Get a data sample $(x_i, y_i)$ from the stream.
    \For{$j=1, \ldots, b$}
        \State Apply the multi-controlled phase gate $V_{i} = e^{i t (y_i)_j/M \ket{x_i}\bra{x_i}\otimes \ket{j}\bra{j}}$.
    \EndFor
  \EndFor
  \end{algorithmic}
\end{algorithm}

\Cref{alg:oracle-construct-multibit} further improves the sample complexity by making use of more information in each sample.
\Cref{lem:q-oralce-sketch-multibit} shows that \Cref{alg:oracle-construct-multibit} has no sampling overhead compared to the single-bit output case.
In particular, when $p(x)=1/N$ and $t=N\pi$, the sample complexity for constructing the multi-bit phase oracle is still $O(N/\epsilon)$.

\begin{lemma}[Quantum oracle sketching for multi-bit functions]
\label{lem:q-oralce-sketch-multibit}
    Let $t, \epsilon>0$.
    Let $b\in \mathbb{Z}_{\geq 1}$
    Let $f: [N]\to \bit^b$ be a function and $p: [N]\to \mathbb{R}_{\geq 0}$ be a probability distribution.
    Let $p_{\max} = \max_{x\in [N]} p(x)$.
    Let $(x_i, y_i)_{i=1}^M$ be a sequence of IID data samples where $x_i\sim p(x)$ and $y_i = f(x_i)\in \bit^b$.
    Let $V_{ij} = e^{it(y_i)_j/M\ket{x_i}\bra{x_i}\otimes \ket{j}\bra{j}}$ for $j\in [b]$ and let $U = \sum_{x,j} e^{ip(x)f_j(x)t}\ket{x,j}\bra{x,j}$.
    Then, 
    \begin{equation}
        M \geq \frac{2p_{\max} t^2}{\epsilon}
    \end{equation}
    samples suffice to guarantee that
    \begin{equation}
        \left\|\E\left[\left(\prod_{j=1}^b\mathcal{V}_{Mj}\right)\cdots \left(\prod_{j=1}^b\mathcal{V}_{1j}\right)\right] - \mathcal{U}\right\|_\diamond \leq \epsilon,
    \end{equation}
    where $\mathcal{V}_{ij}: \rho \to V_{ij}\rho V_{ij}^\dagger$ and $\mathcal{U}: \rho \to U\rho U^\dagger$ are the corresponding unitary channels.
    The data processing time per sample is $b\cdot \polylog(N)$.
\end{lemma}

\begin{proof}[Proof of \Cref{lem:q-oralce-sketch-multibit}]
    Let $h_{x,j} = f_j(x)\ket{x}\bra{x}\otimes \ket{j}\bra{j}$.
    Note that since different $h_{x,j}$ commute with each other, we have
    \begin{equation}
        \prod_{j=1}^b V_{ij} = \prod_{j=1}^b e^{ith_{x_i,j}/M} = e^{it(\sum_{j=1}^b h_{x_i,j})/M}.
    \end{equation}
    Also note that for $x\sim p(x)$, we have $\E[\sum_{j=1}^b h_{x,j}] = \sum_{x,j}p(x)f_j(x)\ket{x,j}\bra{x,j}$ and thus $U = e^{it\E[\sum_{j=1}^b h_{x,j}]}$.
    Therefore, the rest of the proof follows verbatim as the proof of \Cref{thm:q-oracle-sketch-iid} by setting $h_x = \sum_{j=1}^b h_{x,j}$.
\end{proof}

\subsubsection{Quantum oracle sketching for correlated data}

In this section, we generalize quantum oracle sketching to handle correlated data.
In particular, we consider general hierarchical data generation processes with multiple time scales described in \Cref{sec:data-access}.
We will heavily use the statistical properties of data generation processes that we proved in \Cref{sec:data-gen-property}.
Let
\begin{equation}
    \mathcal{D}=(\mathcal{D}^0 \to \mathcal{D}^1_{\alpha_1} \to^{\times T_1} \mathcal{D}^2_{\alpha_2}\to^{\times T_2} \cdots \to^{\times T_{l-1}} \mathcal{D}^l_{\alpha_l} \to^{\times T_l} z)
\end{equation}
be a hierarchical data generation process.
For simplicity, we assume that the data samples are of the form 
\begin{equation}
    z_i = (x_i, y_i), \quad x_i\in \mathcal{X}, \quad y_i = f(x_i)\in \bit,
\end{equation}
for some Boolean function $f: \mathcal{X}\to\bit$.
Note that we include all the situations $(\alpha_{1,i}, \ldots, \alpha_{l,i})$ inside $x_i\in \mathcal{X}$.
This easily generalizes to functions with multi-bit output using the techniques from the previous section.
Since $z$ and $x$ are in one-to-one correspondence, the repetition number $R_{\mathcal{D}}$ defined with respect to $z$ in $\mathcal{D}$ is the same as that of $x$.

Our goal is to construct the unitary
\begin{equation}
    U(t): \ket{x}\to e^{ip(x)f(x)t}\ket{x},
\end{equation}
where $p(x)$ is the marginal distribution of $x$ shared by all data samples.
However, unlike the IID case, there may be correlation among data mediated by higher levels of the hierarchy.
This poses an apparent challenge to the simplest version of quantum oracle sketching described in \Cref{alg:oracle-construct}.
The main obstacle is that when we have already processes some previous data, the posterior/conditional distribution of later data samples shifts.
In order for quantum oracle sketching to work, we have to prove two guarantees: (1) after processing previous data samples, the posterior change is not too large such that we can still construct new queries using subsequent data samples; and (2) the errors of multiple queries do not correlate too much so that the total error still accumulates linearly.

Guarantee (1) is formalized in \Cref{thm:q-oracle-sketch-corr}, which shows that we can extend quantum oracle sketching to handle any hierarchical data generating process $\mathcal{D}$, with a per-query sample complexity overhead equal to its repetition number $R_{\mathcal{D}}$.
This is intuitive, since the more repetitive a data generation process is, the more samples we need to collect to gather enough information.
If every sample is repeated $R_{\mathcal{D}}$ times, we need $R_{\mathcal{D}}$ times as many samples as before.

It is informative to apply \Cref{thm:q-oracle-sketch-corr} to the two examples introduced in \Cref{sec:data-access}: the repetitive process $\mathcal{D}_{\mathrm{rep}}$ and the alternating process $\mathcal{D}_{\mathrm{alt}}$.
Recall that the repetitive process $\mathcal{D}_{\mathrm{rep}}$ is defined as 
\begin{equation}
    \mathcal{D}_{\mathrm{rep}} = (\mathcal{D}^0\to \mathcal{D}_{x}^1 \to^{\times N} z),
\end{equation}
where $\mathcal{D}^0 = \mathrm{Uniform}([N])$ and $\mathcal{D}^1_x$ always outputs its label $x\in [N]$ and repeats in total $N$ times.
We have refreshing time $\tau_{\mathcal{D}_{\mathrm{rep}}}=N$ and repetition number $R_{\mathcal{D}_{\mathrm{rep}}}\leq N$.
The alternating process is defined as
\begin{equation}
    \mathcal{D}_{\mathrm{alt}} = (\mathcal{D}^0\to \mathcal{D}_{\alpha}^1 \to^{\times N} z),
\end{equation}
where $\mathcal{D}^0 = \mathrm{Bern}(1/2)$, $\alpha\in \bit$, and $\mathcal{D}^1_\alpha$ samples $x\in [N]$ uniformly random $N$ times and outputs $z=(x, \alpha)$.
We still have refreshing time $\tau_{\mathcal{D}_{\mathrm{alt}}}=N$, but the repetition number is $R_{\mathcal{D}_{\mathrm{alt}}}\leq O(1)$.

Suppose we fix $t=N\pi$ as before in \Cref{thm:q-oracle-sketch-corr}.
Then for the alternating process $\mathcal{D}_{\mathrm{alt}}$, the oracle can still be constructed using $O(N/\epsilon)$ samples since $p_{\max} = 1/(2N)$ and $R_{\mathcal{D}_{\mathrm{alt}}}=O(1)$.
In contrast, for the repetitive process $\mathcal{D}_{\mathrm{rep}}$, we need $O(N^2/\epsilon)$ samples since $p_{\max}=1/N$ but $R_{\mathcal{D}_{\mathrm{rep}}}=N$.
This $N^2$ scaling is unavoidable, since we can only get to see all $N$ inputs after $\Omega(N^2)$ samples in the repetitive process $\mathcal{D}_{\mathrm{rep}}$.

\begin{tcolorbox}
\begin{theorem}[Quantum oracle sketching for correlated data]
\label{thm:q-oracle-sketch-corr}
    Let $t, \epsilon>0$.
    Let $\mathcal{X}$ be a finite set and let $f: \mathcal{X}\to \bit$ be a Boolean function.
    Let $\mathcal{D}$ be a hierarchical data generation process with repetition number $R_{\mathcal{D}}$ that generates a sequence of $M$ data samples $z_i=(x_i, y_i)$ where $y_i=f(x_i)$, $i=t_0, \ldots, t_0+M-1$ starting from any time step $t_0\geq 1$.
    Let $p_{\max} = \max_{x\in \mathcal{X}} p(x)$ where $p(x)$ is the marginal distribution of data.
    Let $V_{i} = e^{ity_i/M\ket{x_i}\bra{x_i}}$ and let $U = \sum_{x} e^{ip(x)f(x)t}\ket{x}\bra{x}$.
    Then, 
    \begin{equation}
        M \geq \frac{4t^2p_{\max} + 8t\sqrt{2p_{\max}|\mathcal{X}|}}{\epsilon} R_{\mathcal{D}}
    \end{equation}
    samples suffice to guarantee that
    \begin{equation}
        \E_{z_1, \ldots, z_{t_0-1}}\left\|\E\left[\mathcal{V}_{t_0+M-1}\cdots \mathcal{V}_{t_0}|z_1, \ldots, z_{t_0-1}\right] - \mathcal{U}\right\|_\diamond \leq \epsilon,
    \end{equation}
    where $\mathcal{V}_{i}: \rho \to V_{i}\rho V_{i}^\dagger$ and $\mathcal{U}: \rho \to U\rho U^\dagger$ are the corresponding unitary channels, the inner expectation is over the current $M$ samples, and the outer expectation is over the previously processed data samples $z_1, \ldots, z_{t_0-1}$.
    The data processing time per sample is $\polylog(N, 1/\epsilon)$.
    In particular, for a uniform marginal $p(x)=1/N, |\mathcal{X}|=N$ and $t=O(N)$, $M= O(NR_{\mathcal{D}}/\epsilon)$ samples suffice.
\end{theorem}
\end{tcolorbox}

\begin{proof}[Proof of \Cref{thm:q-oracle-sketch-corr}]
    The main difficulty of this extension is that the data are not IID anymore.
    When we have processed previous data samples, the (posterior/conditional) distribution of new data samples shifts due to the correlation.
    In particular, the posterior expectation of the random Hamiltonian no longer matches the target Hamiltonian.
    Therefore, we have to decompose the error into two parts: a variance part same as before, and a new bias part caused by the correlation.
    We will bound the two parts separately to prove \Cref{thm:q-oracle-sketch-corr}.

    Concretely, let $h_x = f(x)\ket{x}\bra{x}$.
    Then we have $V_i = e^{ith_{x_i}/M}$ and $U = e^{it\E_x[h_x]}$, where the expectation is over the marginal distribution $x\sim p(x)$.
    Let 
    \begin{equation}
        m_x = \frac{1}{M}\sum_{i=t_0}^{t_0+M-1} \delta_{x, x_{i}}
    \end{equation}
    be the empirical frequency of a given $x$ appearing in this block of samples $z_{t_0}, \ldots, z_{t_0+M-1}$.
    Note that since the $h_x$'s are orthogonal and commute with each other, we have
    \begin{equation}
        V_{t_0+M-1}\cdots V_{t_0} = \prod_{i=t_0}^{t_0+M-1} e^{ith_{x_i}/M} = \exp(it\frac{1}{M}\sum_{i=t_0}^{t_0+M-1} h_{x_i}).
    \end{equation}
    From \Cref{lem:diamond-operator-expect}, we have
    \begin{equation}
    \begin{split}
        &\frac{1}{4}\E\|\mathcal{U} - \E[\mathcal{V}_{t_0+M-1}\cdots \mathcal{V}_{t_0}|z_1, \ldots z_{t_0-1}]\|_\diamond \\
        &\leq \E\|U - \E[V_{t_0+M-1}\cdots V_{t_0}|z_1, \ldots z_{t_0-1}]\| \\
        &= \E\|e^{it\E_{x}[h_x]} - \E[e^{it\frac{1}{M}\sum_{i=t_0}^{t_0+M-1} h_{x_i}}|z_1, \ldots z_{t_0-1}]\|.
    \end{split}
    \end{equation}
    Let $X = t\frac{1}{M}\sum_{i=t_0}^{t_0+M-1} h_{x_i} = t \sum_x m_xh_x$ be the random matrix in the exponent.
    Note that the linearity of expectation implies
    \begin{equation}
        \E[X] =  t\frac{1}{M}\sum_{i=t_0}^{t_0+M-1} \E[h_{x_i}] = t\frac{1}{M}\sum_{i=t_0}^{t_0+M-1} \E_x[h_x] = t\E_x[h_x],
    \end{equation}
    where we have used the fact that the marginal distribution of each individual $x_i$ is the same $x\sim p(x)$.
    Now we separate the bias part and the variance part via triangle inequality
    \begin{equation}
    \begin{split}
        &\frac{1}{4}\E\|\mathcal{U} - \E[\mathcal{V}_{t_0+M-1}\cdots \mathcal{V}_{t_0}|z_1, \ldots z_{t_0-1}]\|_\diamond \\
        &\leq \E\|e^{i\E[X]} -\E[e^{iX}|z_1, \ldots, z_{t_0-1}] \| \\
        &\leq \E\|e^{i\E[X]} - e^{i\E[X|z_1, \ldots, z_{t_0-1}]}\| + \E\|e^{i\E[X|z_1, \ldots, z_{t_0-1}]} -\E[e^{iX}|z_1, \ldots, z_{t_0-1}] \| \\
        &= \E\|e^{i(\E[X] - \E[X|z_1, \ldots, z_{t_0-1}])} - 1\| + \E\|e^{i\E[X|z_1, \ldots, z_{t_0-1}]} -\E[e^{iX}|z_1, \ldots, z_{t_0-1}] \| \\
        &\leq \underbrace{\E\|\E[X] - \E[X|z_1, \ldots, z_{t_0-1}]\|}_{\mathrm{bias}} + \underbrace{\E\|e^{i\E[X|z_1, \ldots, z_{t_0-1}]} -\E[e^{iX}|z_1, \ldots, z_{t_0-1}] \|}_{\mathrm{variance}}.
    \end{split}
    \end{equation}
    where we use the fact that $X, \E[X], \E[X|z_1, \ldots, z_{t_0-1}]$ are all diagonal matrices and thus commuting, and $|e^{iw}-1|\leq |w|, \forall w\in \mathbb{R}$.

    Next, we bound the two parts separately.
    For the bias part, we have
    \begin{equation}
    \begin{split}
        \underbrace{\E\|\E[X] - \E[X|z_1, \ldots, z_{t_0-1}]\|}_{\mathrm{bias}} &= t\E\left\|\sum_x f(x)\ket{x}\bra{x} (\E[m_x]-\E[m_x|z_1, \ldots, z_{t_0-1}]) \right\| \\
        &\leq t \E \max_x \Bigl|\E[m_x]-\E[m_x|z_1, \ldots, z_{t_0-1}]\Bigr| \\
        &\leq \frac{t\sqrt{2p_{\max}|\mathcal{X}|}}{M} R_{\mathcal{D}},
    \end{split}
    \end{equation}
    where we have used \Cref{lem:data-cond-drift-rep-num} with $|\mathcal{Z}|=2|\mathcal{X}|$.
    For the variance part, we apply \Cref{lem:error-var} to the diagonal random matrix $X|z_1, \ldots, z_{t_0-1}$ with diagonal elements $t m_x f(x)$ and obtain
    \begin{equation}
    \begin{split}
        \underbrace{\E\|e^{i\E[X|z_1, \ldots, z_{t_0-1}]} -\E[e^{iX}|z_1, \ldots, z_{t_0-1}] \|}_{\mathrm{variance}} &\leq \E\left[\frac{1}{2} \max_x \Var[t m_x f(x)|z_1, \ldots, z_{t_0-1}]\right] \\
        &\leq \frac{t^2}{2} \frac{2R_{\mathcal{D}}}{M}\E[\max_x \E[m_x|z_{1}, \ldots, z_{t_0-1}]] \\
        &\leq \frac{t^2}{2} \frac{2R_{\mathcal{D}}}{M} \E[p_{\max} + \E \max_x |\E[m_x|z_1, \ldots, z_{t_0-1}]-p(x)|] \\
        &\leq \frac{t^2p_{\max}}{M}R_{\mathcal{D}} + \frac{t^2R_{\mathcal{D}}^2}{M^2}\sqrt{2p_{\max}|\mathcal{X}|},
    \end{split}
    \end{equation}
    where we have used \Cref{lem:data-cond-var-rep-num} and the bias bound (\Cref{lem:data-cond-drift-rep-num}).

    Combining the bias part and the variance part, we arrive at
    \begin{equation}
    \begin{split}
        \E\|\mathcal{U} - \E[\mathcal{V}_{t_0+M-1}\cdots \mathcal{V}_{t_0}|z_1, \ldots, z_{t_0-1}]\|_\diamond &\leq 4\left(\frac{t\sqrt{2p_{\max}|\mathcal{X}|}}{M}R_{\mathcal{D}} + \frac{t^2 p_{\max}}{M} R_{\mathcal{D}} + \frac{t^2R_{\mathcal{D}}^2}{M^2}\sqrt{2p_{\max}|\mathcal{X}|}\right) \\
        &= \frac{4t^2p_{\max} + 4t\sqrt{2p_{\max}|\mathcal{X}|}(1+tR_{\mathcal{D}}/M)}{M}R_{\mathcal{D}}
    \end{split}
    \end{equation}
    Choosing
    \begin{equation}
        M \geq \frac{4t^2p_{\max} + 8t\sqrt{2p_{\max}|\mathcal{X}|}}{\epsilon}R_{\mathcal{D}},
    \end{equation}
    we have $\E\left\|\E\left[\mathcal{V}_{t_0+M-1}\cdots \mathcal{V}_{t_0}|z_1, \ldots, z_{t_0-1}\right] - \mathcal{U}\right\|_\diamond \leq \epsilon$ as desired.

    For uniform marginal $p(x)=1/N, |\mathcal{X}|=N$ and $t=O(N)$, we have $p_{\max}=1/N$ and
    \begin{equation}
        M = \ceil{\frac{O(N)^2/N + 2O(N)\sqrt{2\cdot 1/N \cdot N}}{\epsilon}R_{\mathcal{D}}} = O\lr{\frac{NR_{\mathcal{D}}}{\epsilon}}
    \end{equation}
    samples suffice.
    This concludes the proof of \Cref{thm:q-oracle-sketch-corr}.
\end{proof}

After showing that we can still perform quantum oracle sketching with reasonable per-query sample complexity.
We move on to prove \Cref{lem:error-accumulation-time-varying} that formalizes guarantee (2): the total error still accumulates linearly.
Moreover, we show that the expected conditional errors also compound additively.
This showcases that the expected conditional error is the correct notion of error when we have correlated data and, in particular, it reduces to the diamond error of expected channel in the IID case.
Together, \Cref{lem:error-accumulation-time-varying} allows us to keep track of error accumulation easily by simply adding them up as usual, even when the data used to construct the quantum channels have correlation.

\begin{tcolorbox}
\begin{lemma}[Error accumulation for correlated quantum channels]
\label{lem:error-accumulation-time-varying}
    Let $Z_1, \ldots, Z_Q$ be $Q$ correlated random variables.
    Let $\mathcal{V}^{(1)}, \ldots, \mathcal{V}^{(Q)}$ be quantum channels, where each $\mathcal{V}^{(i)}$ depends only on $Z_i$.
    Let $\mathcal{U}^{(1)}, \ldots, \mathcal{U}^{(Q)}$ and $\mathcal{C}^{(1)}, \ldots, \mathcal{C}^{(Q)}$ be fixed quantum channels.
    Then, we have
    \begin{equation}
        \|\E[\mathcal{V}^{(Q)}\circ \mathcal{C}^{(Q)}\circ \cdots \circ \mathcal{V}^{(1)}\circ \mathcal{C}^{(1)}] - \mathcal{U}^{(Q)}\circ \mathcal{C}^{(Q)}\circ \cdots \circ \mathcal{U}^{(1)}\circ \mathcal{C}^{(1)}\|_\diamond \leq \sum_{j=1}^Q \E_{Z_1, \ldots, Z_{j-1}}\|\E[\mathcal{V}^{(j)}|Z_1, \ldots, Z_{j-1}] - \mathcal{U}^{(j)}\|_\diamond,
    \end{equation}
    where the inner expectation is over $Z_i$, and the outer expectation is over the previous $Z_1, \ldots, Z_{j-1}$.
    Moreover, for any time interval $1\leq t_0<t_1\leq Q$, we have
    \begin{equation}
    \begin{split}
        &\|\E[\mathcal{V}^{(t_1)}\circ \mathcal{C}^{(t_1)}\circ \cdots \circ \mathcal{V}^{(t_0)}\circ \mathcal{C}^{(t_0)}] - \mathcal{U}^{(t_1)}\circ \mathcal{C}^{(t_1)}\circ \cdots \circ \mathcal{U}^{(t_0)}\circ \mathcal{C}^{(t_0)}\|_\diamond \\
        &\leq \E_{Z_1, \ldots, Z_{t_0-1}}\|\E[\mathcal{V}^{(t_1)}\circ \mathcal{C}^{(t_1)}\circ \cdots \circ \mathcal{V}^{(t_0)}\circ \mathcal{C}^{(t_0)}|Z_1, \ldots, Z_{t_0-1}] - \mathcal{U}^{(t_1)}\circ \mathcal{C}^{(t_1)}\circ \cdots \circ \mathcal{U}^{(t_0)}\circ \mathcal{C}^{(t_0)}\|_\diamond \\
        &\leq \sum_{j=t_0}^{t_1}\E_{Z_1, \ldots, Z_{j-1}}\|\E[\mathcal{V}^{(j)}|Z_1, \ldots, Z_{j-1}] - \mathcal{U}^{(j)}\|_\diamond.
    \end{split}
    \end{equation}
    That is, the total error of replacing $\mathcal{U}^{(j)}$'s with $\mathcal{V}^{(j)}$'s in any time interval is bounded by the sum of individual errors, measured in expected diamond distance of conditional channels.
\end{lemma}
\end{tcolorbox}

We will also use the following immediate shared-record version of \Cref{lem:error-accumulation-time-varying}.
Suppose that the ideal channels $\mathcal{U}^{(1)}_\nu,\ldots,\mathcal{U}^{(Q)}_\nu$ depend on a shared random classical record $\nu$ that is sampled once and retained throughout the computation.
Conditioned on $\nu$, these ideal channels are fixed, so \Cref{lem:error-accumulation-time-varying} applies verbatim.
Averaging the resulting inequality over $\nu$ shows that the total error is bounded by the sum of the expected conditional diamond errors conditioned additionally on the same shared record $\nu$.

\begin{proof}[Proof of \Cref{lem:error-accumulation-time-varying}]
    We only need to prove the second claim for any time interval $1\leq t_0< t_1\leq Q$.
    The first claim follows directly when we take the whole time interval $t_0=1, t_1=Q$.

    The first inequality is a direct consequence of triangle inequality 
    \begin{equation}
        \|\E[\cdot]\|_\diamond = \|\E_{Z_1, \ldots, Z_{t_0-1}}[\E[\cdot|Z_1, \ldots, Z_{t_0-1}]]\|_\diamond\leq \E_{Z_1, \ldots, Z_{t_0-1}}\|\E[\cdot|Z_1, \ldots, Z_{t_0-1}]\|_\diamond
    \end{equation}
    To prove the second inequality, consider the hybrid quantum channels
    \begin{equation}
        \mathcal{A}_j = \E[\mathcal{U}^{(t_1)}\circ \mathcal{C}^{(t_1)}\circ \cdots \mathcal{U}^{(j)}\circ \mathcal{C}^{(j)} \circ  \mathcal{V}^{(j-1)}\circ \mathcal{C}^{(j-1)} \circ \cdots \circ \mathcal{V}^{(t_0)}\circ \mathcal{C}^{(t_0)}|Z_1,\ldots,Z_{t_0-1}], \quad t_0+1\leq j\leq t_1,
    \end{equation}
    and $\mathcal{A}_{t_1+1} = \E[\mathcal{V}^{(t_1)}\circ \mathcal{C}^{(t_1)}\circ \cdots \circ \mathcal{V}^{(t_0)}\circ \mathcal{C}^{(t_0)}|Z_1,\ldots,Z_{t_0-1}], \mathcal{A}_{t_0} =  \mathcal{U}^{(t_1)}\circ \mathcal{C}^{(t_1)}\circ \cdots \circ \mathcal{U}^{(t_0)}\circ \mathcal{C}^{(t_0)}$.
    Then we can expand the left hand side as a telescoping sum
    \begin{equation}
    \begin{split}
        &\E_{Z_1, \ldots, Z_{t_0-1}}\|\E[\mathcal{V}^{(t_1)}\circ \mathcal{C}^{(t_1)}\circ \cdots \circ \mathcal{V}^{(t_0)}\circ \mathcal{C}^{(t_0)}|Z_1,\ldots,Z_{t_0-1}] - \mathcal{U}^{(t_1)}\circ \mathcal{C}^{(t_1)}\circ \cdots \circ \mathcal{U}^{(t_0)}\circ \mathcal{C}^{(t_0)}\|_\diamond \\
        &= \E_{Z_1, \ldots, Z_{t_0-1}}\left\|\sum_{j=t_0}^{t_1}(\mathcal{A}_{j+1}-\mathcal{A}_j)\right\|_\diamond \leq \sum_{j=t_0}^{t_1}\E_{Z_1, \ldots, Z_{t_0-1}}\|\mathcal{A}_{j+1}-\mathcal{A}_j\|_\diamond.
    \end{split}
    \end{equation}
    Each term is be bounded by
    \begin{equation}
    \begin{split}
        &\E_{Z_1, \ldots, Z_{t_0-1}}\|\mathcal{A}_{j+1}-\mathcal{A}_j\|_\diamond \\
        &=\E_{Z_1, \ldots, Z_{t_0-1}}\|\mathcal{U}^{(t_1)}\circ \mathcal{C}^{(t_1)}\circ \cdots \E[(\mathcal{V}^{(j)} - \mathcal{U}^{(j)})\circ \mathcal{C}^{(j)} \circ  \mathcal{V}^{(j-1)}\circ \mathcal{C}^{(j-1)} \circ \cdots \circ \mathcal{V}^{(t_0)}\circ \mathcal{C}^{(t_0)}|Z_1,\ldots,Z_{t_0-1}]\|_\diamond \\
        &\leq \E_{Z_1, \ldots, Z_{t_0-1}}\|\E[(\mathcal{V}^{(j)} - \mathcal{U}^{(j)})\circ \mathcal{C}^{(j)} \circ  \mathcal{V}^{(j-1)}\circ \mathcal{C}^{(j-1)} \circ \cdots \circ \mathcal{V}^{(t_0)}\circ \mathcal{C}^{(t_0)}|Z_1,\ldots,Z_{t_0-1}]\|_\diamond \\
        &= \E_{Z_1, \ldots, Z_{t_0-1}}\left\|\E\Bigl[\E[\mathcal{V}^{(j)} - \mathcal{U}^{(j)}|Z_1,\ldots,Z_{j-1}]\circ \mathcal{C}^{(j)} \circ \mathcal{V}^{(j-1)}\circ \mathcal{C}^{(j-1)} \circ \cdots \circ \mathcal{V}^{(t_0)}\circ \mathcal{C}^{(t_0)}\Bigm|Z_1,\ldots,Z_{t_0-1}\Bigr]\right\|_\diamond \\
        &\leq \E_{Z_1, \ldots, Z_{t_0-1}}\left[\E\Bigl[\|\E[\mathcal{V}^{(j)} - \mathcal{U}^{(j)}|Z_1,\ldots,Z_{j-1}]\circ \mathcal{C}^{(j)} \circ \mathcal{V}^{(j-1)}\circ \mathcal{C}^{(j-1)} \circ \cdots \circ \mathcal{V}^{(t_0)}\circ \mathcal{C}^{(t_0)}\|_\diamond \Bigm| Z_1,\ldots,Z_{t_0-1} \Bigr]\right] \\
        &\leq \E_{Z_1, \ldots, Z_{j-1}}\|\E[\mathcal{V}^{(j)} - \mathcal{U}^{(j)}|Z_1,\ldots,Z_{j-1}]\|_\diamond \\
        &=\E_{Z_1, \ldots, Z_{j-1}}\|\E[\mathcal{V}^{(j)}|Z_1,\ldots,Z_{j-1}] - \mathcal{U}^{(j)}\|_\diamond,
    \end{split}
    \end{equation}
    where we have used the sub-multiplicativity of diamond norm $\|\Phi_1\circ (\cdot)\circ \Phi_2\|_\diamond \leq \|\Phi_1\|_\diamond \|\cdot\|_\diamond \|\Phi_2\|_\diamond = \|\cdot\|_\diamond$ with quantum channels $\|\Phi_1\|_\diamond = \|\Phi_2\|_\diamond=1$ twice, and the triangle inequality $\|\E[\cdot]\|_\diamond\leq \E\|\cdot\|_\diamond$.
    Plugging this back into the sum gives us the desired result.
    This completes the proof of \Cref{lem:error-accumulation-time-varying}.
\end{proof}

\subsubsection{Unknown marginal distribution}

In previous sections, we have shown how to prepare
\begin{equation}
    U(t) = \sum_x e^{ip(x)f(x)t}\ket{x}\bra{x}
\end{equation}
from any hierarchical data generation process using quantum oracle sketching.
When the data has uniform marginal $p(x)=1/N$, we can set $t=\pi N$ and obtain access to the phase oracle 
\begin{equation}
    O = \sum_x (-1)^{f(x)}\ket{x}\bra{x} = \sum_x e^{if(x)\pi}\ket{x}\bra{x}
\end{equation}
that we want.
However, the marginal distribution $p(x) \in [p_{\min}, p_{\max}]$ may be non-uniform and unknown in general. 
In this case, we need a way to remove $U(t)$'s dependence on $p(x)$ without knowing it.

One may attempt to first learn $p(x)$ (or assuming that $p(x)$ is known), and then multiply the phase rotation angle by $1/p(x)$ when performing the multi-controlled phase gate corresponding to $(x, f(x))$.
This does not work when the distribution is unknown, because learning $p(x)$ would require an exponentially large memory to store the values of $p(x)$ for all $x\in \mathcal{X}$.

In the following, we show how to use techniques from quantum singular value transform (QSVT) to accommodate unknown and non-uniform marginal with only one ancilla qubit.
Assume $p(x)\in [p_{\min}, p_{\max}]$.
The key is to note that the rotation angles in $U(t)$ are $0$ for those $x$ with $f(x)=0$, and are in $[p_{\min}, p_{\max}]$ and thus bounded away from $0$ for those $x$ with $f(x)=1$.
So all we need to do is to apply a threshold function that is equal to $1$ when the rotation angle is below a threshold value and is equal to $-1$ when the rotation angle is above the threshold.
This would give us the desired oracle $O = \sum_x (-1)^{f(x)}\ket{x}\bra{x}$.
We formalize this idea in the following lemma.
In particular, when $\mathcal{X}=[N]$, as long as the distribution $p(x)$ is not too tilted (e.g., $p_{\max}/p_{\min} \leq \polylog(N)$ and $p_{\min}\geq \Omega(1/N)$), the sample complexity is still $\tilde{O}(N)$.

\begin{tcolorbox}
\begin{lemma}[Sample complexity upper bound for unknown distributions]
\label{lem:sample-upper-unknown-dist}
    Let $\epsilon>0$.
    Let $\mathcal{X}$ be a finite set and let $f: \mathcal{X}\to \bit$ be a Boolean function.
    Let $\mathcal{D}=(\mathcal{D}^0\to \mathcal{D}^1_{\alpha_1}\to^{\times T_1}\cdots\to^{\times T_{l-1}}\mathcal{D}^l_{\alpha_l}\to^{\times T_l}z)$ be a hierarchical data generation process with repetition number $R_{\mathcal{D}}$ that generates a sequence of $M$ data samples $z_i=(x_i, y_i)$ where $y_i=f(x_i)$, $i=t_0, \ldots, t_0+M-1$ starting from any time step $t_0\geq 1$.
    Let $p_{\max} = \max_{x\in \mathcal{X}} p(x), p_{\min} = \min_{x\in \mathcal{X}} p(x)>0$, where $p(x)$ is the marginal distribution of data.
    Let $O = \sum_{x\in \mathcal{X}} (-1)^{f(x)}\ket{x}\bra{x}$ be the phase oracle of $f$.
    Then, we can use
    \begin{equation}
        M = O\left(\frac{p_{\max}\sqrt{p_{\max}|\mathcal{X}|}}{p_{\min}^2} \frac{R_{\mathcal{D}}}{\epsilon} \log^2(1/\epsilon)\right),
    \end{equation}
    samples $z_{t_0}, \ldots, z_{t_0+M-1}$ to construct a random unitary $V$ acting on the original system and two ancilla qubit $a, a'$ such that its corresponding channel
    \begin{equation}
        \mathcal{V}: \rho \to V(\ketbra{0_{a'}0_a}{0_{a'}0_a}\otimes \rho)V^\dagger
    \end{equation}
    satisfies
    \begin{equation}
        \E_{z_1, \ldots, z_{t_0-1}}\left\|\E[\mathcal{V}|z_1, \ldots, z_{t_0-1}] - \mathcal{O}\right\|_\diamond \leq \epsilon,
    \end{equation}
    where $\mathcal{O}: \rho \to \ketbra{0_{a'}0_a}{0_{a'}0_a}\otimes O\rho O^\dagger$, the inner expectation is over the current $M$ samples, and the outer expectation is over the previously processed data samples $z_1, \ldots, z_{t_0-1}$.
    The data processing time per sample is $\polylog(N)$.
    Similarly, one can implement the controlled oracle $cO = \ketbra{0}{0}\otimes I + \ketbra{1}{1}\otimes O$ with the same guarantees.
    Moreover, when the data are sampled IID, 
    \begin{equation}
        M=O\lr{\frac{p_{\max}}{p_{\min}^2}\frac{ \log^2(1/\epsilon)}{\epsilon}}
    \end{equation}
    samples suffices.
\end{lemma}
\end{tcolorbox}

\begin{proof}[Proof of \Cref{lem:sample-upper-unknown-dist}]
    We prove \Cref{lem:sample-upper-unknown-dist} by first constructing the unitary $U(t) = \sum_x e^{ip(x)f(x)t}\ket{x}\bra{x}$ using \Cref{thm:q-oracle-sketch-corr} and then applying a threshold function on the rotation angles using QSVT (\Cref{lem:qsvt,lem:approx-thres}).

    Concretely, fix $t=1/p_{\max}$.
    \Cref{thm:q-oracle-sketch-corr} asserts that we can use 
    \begin{equation}
        M_0 = O\lr{\frac{p_{\max}t^2+t\sqrt{2p_{\max}|\mathcal{X}|}}{\epsilon_1} R_{\mathcal{D}}} = O\lr{\frac{\sqrt{p_{\max}|\mathcal{X}|}}{p_{\max} \epsilon_1} R_{\mathcal{D}}}
    \end{equation}
    samples to construct one query to a (random) unitary channel $\cV^0$ such that
    \begin{equation}
        \E\| \E[\cV^0|z_{1}, \ldots, z_{t_0-1}] - \cU \|_\diamond \leq \epsilon_1,
    \end{equation}
    where $\cU$ is the unitary channel of $U = \sum_x e^{ip(x)f(x)/p_{\max}}\ket{x}\bra{x}$.
    Note that we can also implement $U^\dagger$ with the same number of samples by setting $t=-1/p_{\max}$.
    Similarly, we can implement $cU$ and $cU^\dagger$ by adding control to the multi-controlled phase gates in \Cref{thm:q-oracle-sketch-corr}.
    In the following, we will keep track of how many queries we make to $U$, $U^\dagger$, $cU$, and $cU^\dagger$ to construct the $\mathcal{V}$ that approximates $\mathcal{O}$.

    Let $\Lambda = \sum_x p(x) f(x)/p_{\max}\ket{x}\bra{x}$ and therefore $U = e^{i\Lambda}$.
    Let $S = \begin{pmatrix}
        -i &0\\
        0 &1
    \end{pmatrix}$ be a single qubit gate.
    We introduce an ancilla qubit $a$ and use two queries to construct
    \begin{equation}
        W = S_{a}X_{a}H_{a}(c_{a}U^\dagger)X_{a}(c_{a}U) H_{a},
    \end{equation}
    where the subscript indicate which qubit the gate or the control acts on.
    Then we have
    \begin{equation}
    \begin{split}
        W\ket{0_{a}} &= S_{a}X_{a}H_{a}(c_{a}U^\dagger)X_{a}(c_{a}U) \frac{1}{\sqrt{2}}(\ket{0_{a}}+\ket{1_{a}}) \\ 
        &=S_{a}X_{a}H_{a}(c_{a}U^\dagger) \frac{1}{\sqrt{2}}(\ket{1_{a}}\otimes I+\ket{0_{a}}\otimes U) \\
        &=S_{a}X_{a}H_{a}\frac{1}{\sqrt{2}}(\ket{1_{a}}\otimes U^\dagger+\ket{0_{a}}\otimes U) \\
        &=S_{a}X_{a}\frac{1}{2}(\ket{0_{a}}\otimes U^\dagger-\ket{1_{a}}\otimes U^\dagger+\ket{0_{a}}\otimes U+\ket{1_{a}}\otimes U) \\
        &=S_{a}X_{a}\left(\ket{0_{a}}\otimes \frac{U+U^\dagger}{2} + \ket{1_{a}}\otimes \frac{U-U^\dagger}{2}\right) \\
        &=\ket{1_{a}}\otimes \frac{U+U^\dagger}{2} + \ket{0_{a}}\otimes \frac{U-U^\dagger}{2i}.
    \end{split}
    \end{equation}
    Therefore, we have
    \begin{equation}
        \bra{0_{a}}W\ket{0_{a}} = \frac{U-U^\dagger}{2i} = \sin\Lambda.
    \end{equation}
    Note that we also have access to $W^\dagger$ using a similar construction.
    Since $p(x)f(x)$ is zero when $f(x)=0$ and in $[p_{\min}, p_{\max}]$ when $f(x)=1$, we know that the eigenvalues of $\sin\Lambda$ is $0$ on the basis $\{\ket{x}, f(x)=0\}$ and is in $[\sin(p_{\min}/p_{\max}), \sin(1)]$ on the basis $\{\ket{x}, f(x)=1\}$, as $\sin(x)$ is increasing on $[0, 1]$.

    Now we invoke \Cref{lem:qsvt} to apply a polynomial function $P$ to $W$.
    We choose the polynomial $P$ to approximate the threshold function with threshold value $\lambda^\star = \sin(p_{\min}/p_{\max})$ as in \Cref{lem:approx-thres}.
    In particular, \Cref{lem:approx-thres} implies that the degree of $P$ is at most $d=O(\log(1/\epsilon_2)/\lambda^\star)$ (we choose $d$ to be even), $P$ is an even function, $|P(w)|\leq 1, \forall w\in [-1, 1]$, and $|P(w)-1|\leq \epsilon_2, \forall w\in [0, \lambda^\star/2]$ and $|P(w)+1|\leq \epsilon_2, \forall w\in [\lambda^\star, 1]$.
    This means that
    \begin{equation}
        \left|P(\sin(p(x)f(x)/p_{\max})) - (-1)^{f(x)}\right|\leq \epsilon_2, \quad \forall x\in\mathcal{X}.
    \end{equation}
    Then \Cref{lem:qsvt} tells us that there exists a set of rotation angles $\phi_i\in [0, 2\pi), i\in [d]$ such that the unitary
    \begin{equation}
        P_{\mathrm{QSVT}}(W, \phi) = H_{a'}\prod_{k=1}^{d/2}(\tilde\Pi_{\phi_{2k-1}}W^\dagger \tilde\Pi_{\phi_{2k}} W)H_{a'}
    \end{equation}
    satisfies
    \begin{equation}
        \bra{0_{a'}0_a}P_{\mathrm{QSVT}}(W, \phi)\ket{0_{a'}0_a} = P(\sin\Lambda).
    \end{equation}
    The error of approximating $O$ is therefore
    \begin{equation}
        \|\bra{0_{a'}0_a}P_{\mathrm{QSVT}}(W, \phi)\ket{0_{a'}0_a} - O\| = \max_{x\in\mathcal{X}} \left|P(\sin(p(x)f(x)/p_{\max}))-(-1)^{f(x)}\right| \leq \epsilon_2.
    \end{equation}
    To convert this into a diamond norm bound, note that
    \begin{equation}
    \begin{split}
        &\|P_{\mathrm{QSVT}}(W, \phi)(\ket{0_{a'}0_a}\otimes \cdot) - \ket{0_{a'}0_a}\otimes O\| \\
        &\leq \| (I-\ketbra{0_{a'}0_a}{0_{a'}0_a} )P_{\mathrm{QSVT}}(W, \phi)\ket{0_{a'}0_a} \| + \| \bra{0_{a'}0_a}P_{\mathrm{QSVT}}(W, \phi)\ket{0_{a'}0_a} -O \| \\
        &\leq \| (I-\ketbra{0_{a'}0_a}{0_{a'}0_a} )P_{\mathrm{QSVT}}(W, \phi)\ket{0_{a'}0_a} \| + \epsilon_2.
    \end{split}
    \end{equation}
    Also,
    \begin{equation}
    \begin{split}
        &\| (I-\ketbra{0_{a'}0_a}{0_{a'}0_a} )P_{\mathrm{QSVT}}(W, \phi)\ket{0_{a'}0_a} \|^2 \\
        &= \|\bra{0_{a'}0_a}P^\dagger_{\mathrm{QSVT}}(W, \phi)(I-\ketbra{0_{a'}0_a}{0_{a'}0_a})P_{\mathrm{QSVT}}(W, \phi)\ket{0_{a'}0_a}\| \\
        &=\|1-\bra{0_{a'}0_a}P^\dagger_{\mathrm{QSVT}}(W, \phi)\ketbra{0_{a'}0_a}{0_{a'}0_a}P_{\mathrm{QSVT}}(W, \phi)\ket{0_{a'}0_a}\| \\
        &=\|O^\dagger O-\bra{0_{a'}0_a}P^\dagger_{\mathrm{QSVT}}(W, \phi)\ketbra{0_{a'}0_a}{0_{a'}0_a}P_{\mathrm{QSVT}}(W, \phi)\ket{0_{a'}0_a}\| \\
        &=\|O^\dagger (O-\bra{0_{a'}0_a}P_{\mathrm{QSVT}}(W, \phi)\ket{0_{a'}0_a}) + (O-\bra{0_{a'}0_a}P_{\mathrm{QSVT}}(W, \phi)\ket{0_{a'}0_a})^\dagger \bra{0_{a'}0_a}P_{\mathrm{QSVT}}(W, \phi)\ket{0_{a'}0_a}\| \\
        &\leq \|O\| \|O-\bra{0_{a'}0_a}P_{\mathrm{QSVT}}(W, \phi)\ket{0_{a'}0_a}\| + \|O-\bra{0_{a'}0_a}P_{\mathrm{QSVT}}(W, \phi)\ket{0_{a'}0_a}\| \|P_{\mathrm{QSVT}}(W, \phi)\| \\
        &\leq 2\epsilon_2.
    \end{split}
    \end{equation}
    Hence, the error is
    \begin{equation}
        \|P_{\mathrm{QSVT}}(W, \phi)(\ket{0_{a'}0_a}\otimes \cdot) - \ket{0_{a'}0_a}\otimes O\|\leq \sqrt{2\epsilon_2} + \epsilon_2 \leq 3\sqrt{\epsilon_2} = \epsilon/4
    \end{equation}
    when we choose $\epsilon_2 = (\epsilon/12)^2$.
    The total number of queries to $U, U^\dagger, cU$ and $cU^\dagger$ is 
    \begin{equation}
        Q = O(d) = O\left(\frac{\log(1/\epsilon_2)}{\lambda^\star}\right) = O\left(\frac{\log(1/\epsilon)}{\sin(p_{\min}/p_{\max})}\right)\leq O\left(\frac{\log(1/\epsilon)}{p_{\min}/p_{\max}}\right),
    \end{equation}
    where we have used the fact that $\sin(x)/x \geq \sin(1)/1>0.8, \forall x \in [0, 1]$.

    Finally, we construct the desired unitary $V$ by replacing all the $U, U^\dagger, cU$ and $cU^\dagger$ in $P_{\mathrm{QSVT}}(W, \phi)$ with the random unitaries constructed from samples using \Cref{thm:q-oracle-sketch-corr}.
    Since there are $Q$ queries in total, the error of approximating $P_{\mathrm{QSVT}}(W, \phi)$ is bounded by $Q\epsilon_1$, using \Cref{lem:error-accumulation-time-varying}.
    
    We set $\epsilon_1=\epsilon/(4Q)$.
    Then triangle inequality implies the desired property:
    \begin{equation}
        \E\|\E[\mathcal{V}|z_1, \ldots, z_{t_0-1}] - \mathcal{O}\|_\diamond \leq \epsilon.
    \end{equation}
    The total number of samples used is
    \begin{equation}
        M = QM_0 = O\left(\frac{\log^2(1/\epsilon)}{(p_{\min}/p_{\max})^2}\cdot \sqrt{\frac{|\mathcal{X}|}{p_{\max}}}\frac{R_{\mathcal{D}}}{\epsilon}\right) = O\left(\frac{p_{\max} \sqrt{p_{\max}|\mathcal{X}|}}{p_{\min}^2\epsilon} R_{\mathcal{D}} \log^2(1/\epsilon)\right).
    \end{equation}
    Similarly, one can construct the controlled oracle $cO$ using the controlled version of the QSVT construction as in \Cref{lem:qsvt} and the same guarantees hold.
    When the data are sampled IID, we have $M_0 =p_{\max}t^2/\epsilon_1$ from \Cref{thm:q-oracle-sketch-iid} and $M = O(p_{\max}/p_{\min}^2\cdot \log^2(1/\epsilon)/\epsilon)$.
    This completes the proof of \Cref{lem:sample-upper-unknown-dist}.
\end{proof}

\subsection{Linear algebra primitives}
\label{sec:q-alg-linear-algebra}

In this section, we utilize quantum oracle sketching to construct several useful primitives for linear algebra data (e.g., vectors, matrices, etc.).
In particular, we show how to construct the sparse oracles and block encodings for sparse matrices, and prepare quantum states corresponding to arbitrary vectors.

\subsubsection{Linear algebra data}

We begin by introducing our data access model for linear algebra data such as matrices and vectors.
For an $N$-dimensional matrix $A\in \mathbb{R}^{N\times N}$, we define its matrix data generation process $\mathcal{D}_{A}$ as a hierarchical data generation process that generates random non-zero matrix elements as data
\begin{equation}
    z = (i, j, A_{ij}), \quad (i, j)\overset{\mathrm{marginal}}{\sim} \unif(\{(i, j): A_{ij}\neq 0\}).
\end{equation}
The matrix elements $A_{ij}$ are specified by bitstrings of length $b=\polylog(N)$ to sufficient accuracy.
For simplicity, we assume that this binary representation is exact and use $A_{ij}$ to stand for the corresponding value.
Throughout this work, we also assumes that the situation information is completely specified in the data $z=(i, j, A_{ij})$ (e.g., given by the row and column indices).
From the data generation process of $A$, one can easily generate that of $A^T$ by switching $i\leftrightarrow j$ and that of the symmetrized matrix $A_{\mathrm{sym}} = \begin{pmatrix}
    0 &A\\
    A^T &0
\end{pmatrix}$ by randomly transforming $(i, j)$ to $(i, j+N)$ or $(j+N, i)$ with equal probability.
It is also straightforward to generalize to complex matrices.

We note that this encompasses rectangular matrices $A\in \mathbb{R}^{D_1\times D_2}$ by defining $N = \max(D_1, D_2)$ and embed the rectangular matrix into a larger square one with zeroes padded in.
The data generation process remains completely unchanged since zero matrix elements are never sampled.
For this reason, we always assume that $A$ is a square matrix in algorithmic constructions, because rectangular matrices are automatically handled by setting $N=\max(D_1, D_2)$.
We also assume that the number of non-zero elements $K=|\{(i,j):A_{ij}\neq 0\}|$ is known.
All sparse-oracle and block-encoding constructions below remain valid when the samples are uniform over a known padded support of size at most $Ns$, where padded positions carry a marked zero value. 
The sample-dependent operation is the identity on every padded position, and the proofs are unchanged after replacing the support size $K$ by the padded support size.

Similarly, for a $N$-dimensional vector $\vec{b} = (b_1, \ldots, b_N)^T\in \mathbb{R}^N$, we define its vector data generation process $\mathcal{D}_{\vec{b}}$ as a hierarchical data generation process that generates random components of the vector as data
\begin{equation}
    z = (j, b_j), \quad j\overset{\mathrm{marginal}}{\sim} \unif([N]).
\end{equation}
For simplicity, we also assume that the components $b_{j}$ are specified by bitstrings of length $b=\polylog(N)$ exactly and use $b_{j}$ to stand for the corresponding value.

In the following, we always assume $N=2^n$ for some integer $n$.
We can always do so, because even if $N$ is not a power of two, we can still embed the matrix/vector into a larger matrix/vector of dimension $2^{\ceil{\log N}}\in[N, 2N]$ by padding zeroes.
The matrix data generation process stays exactly the same since it only generates non-zero elements.
The block encoding of the larger matrix is also automatically a block encoding of the original, possibly rectangular matrix.
For vectors, although the vector data generation process will be diluted with a constant fraction of zeroes after the embedding, we will see that the same quantum oracle sketching algorithm works without any modification, because it does nothing upon seeing a zero component.
Therefore, we assume without loss of generality that $N=2^n$ for some integer $n$.

\subsubsection{Sparse oracles}

In this section, we show how to implement the standard sparse oracles of sparse matrices using quantum oracle sketching.
The sparse oracles of an row $s_r$-sparse and column $s_c$-sparse matrix $A$ are defined as
\begin{align}
    O_A^{\mathrm{ele}} \sth \ket{i}\ket{j}\ket{0^b} \to \ket{i}\ket{j}\ket{A_{ij}}, \quad O_A^{\mathrm{ind, row}} \sth \ket{i}\ket{k} \to \ket{i}\ket{j(i,k)}, \quad O_A^{\mathrm{ind, col}} \sth \ket{j}\ket{k} \to \ket{j}\ket{i(j,k)},
\end{align}
where $j(i,k)\in\bit^{n+1}\simeq \bit\times [N], k\in [s_r]$ represents the column index (in binary form) of the $k$-th non-zero element in row $i$ and $i(j,k)\in\bit^{n+1} \simeq \bit \times [N], k\in [s_c]$ represents the row index of the $k$-th non-zero element in column $j$. 
The additional bit at the start is meant for accommodating rows/columns with less than $s_r/s_c$ non-zero elements.
More precisely, we define $j(i, k)$ to be $0$ followed by the $k$-th non-zero column of row $i$ when $k\leq |\{j: A_{ij}\neq 0\}|$ and define it to be $1$ followed by the binary representation of $(k-1)$ when $|\{j: A_{ij}\neq 0\}|<k\leq s_r$.
In other words, the first bit signifies when the query $k$ exceeds the number of non-zero columns in this row.
We define $i(j, k)$ similarly.
The sparsity $s$ of $A$ is $s=\max(s_r, s_c)$.

In the following, we show that for $s=\polylog N$, we can implement these sparse oracles with $\polylog(N)$ qubits and $\tilde{O}(RN/\epsilon)$ samples from the matrix data generation process $\mathcal{D}_A$ with repetition number $R$.
We begin with the sparse element oracle $O_A^{\mathrm{ele}}$ that is relatively straightforward to implement.

\begin{tcolorbox}
\begin{lemma}[Sparse element oracle]
\label{lem:sparse-element-oracle}
    Let $A \in \mathbb{R}^{N\times N}$ be an $N$-dimensional, row $s_r$-sparse, and column $s_c$-sparse matrix. 
    Let $\mathcal{O}^{\mathrm{ele}}_A: \rho \to O^{\mathrm{ele}}_A \rho O^{\mathrm{ele}\dagger}_A$ be the unitary channel corresponding to its sparse element oracle $O^{\mathrm{ele}}_A$.
    Then, we can use $2\ceil{\log N}+b$ qubits and
    \begin{equation}
        M=O\lr{\frac{RN \min(s_r, s_c)}{\epsilon}}
    \end{equation}
    samples starting at any time $t_0\geq 1$ from its matrix data generation process $\mathcal{D}_A$ with repetition number $R$ to implement a random unitary channel $\mathcal{V}$ satisfying
    \begin{equation}
        \E_{z_1, \ldots, z_{t_0-1}}\|\E[\mathcal{V}|z_1, \ldots, z_{t_0-1}] - \mathcal{O}^{\mathrm{ele}}_A\|_\diamond\leq \epsilon,
    \end{equation}
    where the inner expectation is over the current $M$ samples and the outer expectation is over the previously processed data samples $z_1, \ldots, z_{t_0-1}$.
    The data processing time per sample is $\polylog(N, 1/\epsilon)$.
    The same guarantee holds for implementing $O^{\mathrm{ele}\dagger}_A$ and their controlled version $cO^{\mathrm{ele}}_A, cO^{\mathrm{ele}\dagger}_A$ with one more qubit.
\end{lemma}
\end{tcolorbox}

\begin{proof}
    Let
    $\mathcal{K}
    =
    \{(i,j):A_{ij}\neq 0\}$ be the support of the matrix and $K=|\mathcal{K}|
    \leq
    N\min(s_r,s_c)$ be the number of non-zero elements.
    For any $b$-bit string $y$, we use
    $X^y=\bigotimes_{a=1}^b X_a^{y_a}$
    to denote bitwise addition of $y$ to the $b$-qubit output register.

    Upon receiving a sample $(i,j,A_{ij})$, we apply the unitary
    \begin{equation}
        V(i,j,A_{ij})
        =
        \exp\left[
            -i\frac{\pi K}{M}
            \ketbra{i,j}
            \otimes
            \frac{I-X^{A_{ij}}}{2}
        \right].
    \end{equation}
    This gate can be implemented using
    $\polylog(N)$ gates, because $X^{A_{ij}}$ is a Pauli string
    specified by the observed $b$-bit representation of $A_{ij}$.
    Note that the generators associated with different matrix positions have orthogonal support on the $(i,j)$ register and commute.
    Moreover, $\|(I-X^{A_{ij}})/2\|\leq 1$.
    Therefore, the proof of
    \Cref{thm:q-oracle-sketch-corr} applies verbatim with
    $p_{\max}=1/K$, $|\mathcal{K}|=K$, and evolution time
    $t=\pi K$.
    The corresponding ideal unitary is
    \begin{equation}
        \exp\left[
            -i\frac{\pi}{2}
            \sum_{(i,j)\in\mathcal{K}}
            \ketbra{i,j}
            \otimes
            \left(
                I-X^{A_{ij}}
            \right)
        \right]=
        \sum_{i,j\in[N]}
        \ketbra{i,j}
        \otimes
        X^{A_{ij}}
        =
        O_A^{\mathrm{ele}},
    \end{equation}
    where we use the fact $
    \exp\left[
        -i\frac{\pi}{2}
        (I-X^y)
    \right]
    = e^{-i\pi/2}(\cos(\pi/2) + i\sin(\pi/2)X^y)
    =X^y $
    for every bit string $y$, and take $A_{ij}=0$ outside
    $\mathcal{K}$.

    Consequently, \Cref{thm:q-oracle-sketch-corr} gives
    \begin{equation}
        \E\left\|
            \E[
                \mathcal{V}
                \mid
                z_1,\ldots,z_{t_0-1}
            ]
            -
            \mathcal{O}_A^{\mathrm{ele}}
        \right\|_\diamond
        \leq
        \epsilon
    \end{equation}
    using
    \begin{equation}
        M
        =
        O\left(
            \frac{RK}{\epsilon}
        \right)
        \leq
        O\left(
            \frac{
                RN\min(s_r,s_c)
            }{\epsilon}
        \right)
    \end{equation}
    samples.
    The inverse is implemented by negating the rotation angle, and the controlled versions are implemented by adding control to each sample-dependent rotation.
\end{proof}

The column and row index oracles $O_A^{\mathrm{ind, col}}, O_A^{\mathrm{ind, row}}$ are more technically involved to implement. 
We only need to describe how to implement the row index oracle $O_A^{\mathrm{ind, row}}$, and then the implementation of the column index oracle $O_A^{\mathrm{ind, col}}$ follows verbatim by swapping the roles of $i$ and $j$.

To prepare the row index oracle, we consider an intermediate oracle that we call the \emph{cumulative counter oracle}, defined as
\begin{equation}
	O_c: \ket{i}\ket{k}\ket{l}\ket{0} \to \ket{i}\ket{k}\ket{l}\ket{1[C(i, l)< k]}, \quad i, l\in \bit^n, k\in \{1, \ldots, s_r\},
\end{equation}
where
\begin{equation}
	C(i, l) = |\{j: A_{ij}\neq 0, j<l\}|
\end{equation}
is the number of non-zero columns in row $i\in \bit^n$ that have column indices strictly smaller than $l\in \bit^n$.
The purpose of this cumulative counter oracle is to count the cumulative number of non-zero columns in a row $i$ up to a given trial column $l$.
We will use quantum oracle sketching to construct this cumulative counter oracle.
Here we use $\ceil{\log(s_r+1)}$ qubits to host the $\ket{k}$ register with $s_r$ orthogonal states via the standard encoding of $\{0, \ldots, s_r\}$.

After we have constructed the cumulative counter oracle, we can use it to perform a binary search over the non-zero columns by progressively refine our trial column $l\in \bit^n$ until it converges to the target column index $j(i, k)$.
For now let us ignore the first indicator bit of $j(i, k)$ by assuming that $k$ does not exceed the number of non-zero columns in row $i$.
We will deal with the general case later in the formal proof.
Suppose we have found the first $m\in [n]$ bits $l_m$ of the target column $j(i, k)$ and set $l=l_m00\ldots 0$.
Now we want to find the $(m+1)$-th bit.
We first flip the $(m+1)$-th bit and obtain $\ket{l_m10\ldots 0}$.
Then we apply the cumulative counter oracle $O_c$ to $\ket{i}\ket{k}\ket{l_m10\ldots 0}\ket{0}$ and get
\begin{equation}
	\ket{i}\ket{k}\ket{l_m10\ldots 0}\ket{1[C(i, l_m10\cdots 0)<k]}.
\end{equation}
Note that if the $(m+1)$-th bit of the target $j(i, k)$ is $1$ (i.e., $j(i, k)$ has prefix $l_m 1$), then there must be less than $k$ non-zero columns with prefix $l_m0$, because the $k$-th non-zero column $j(i, k)$ has prefix $l_m1$.
This means that $1[C(i, l_m10\ldots 0)<k]=1$.
On the other hand, if the $(m+1)$-th bit of $j(i, k)$ is $0$, then there must be at least $k$ non-zero columns with prefix $l_m0$ and hence $1[C(i, l_m10\ldots 0)< k]=0$.
In both cases, we have
\begin{equation}
	 1[C(i, l_m10\ldots 0)< k]=j(i, k)_{m+1},
\end{equation}
where $j(i, k)_{m+1}$ is the $(m+1)$-th bit of the target $j(i, k)$.
Therefore, the state reads
\begin{equation}
	\ket{i}\ket{k}\ket{l_m10\ldots 0}\ket{j(i, k)_{m+1}}.
\end{equation}
Finally, we flip the $(m+1)$-th bit of the $\ket{l}$ register to reset it back to $0$ and swap it with $\ket{j(i, k)_{m+1}}$.
This gives us the state
\begin{equation}
	\ket{i}\ket{k}\ket{l_mj(i, k)_{m+1}0\ldots 0}\ket{0}.
\end{equation}
Now we have found the $(m+1)$-th bit of the target $j(i, k)$.
We repeat this procedure $n$ times in total and obtain the state
\begin{equation}
	\ket{i}\ket{k}\ket{j(i, k)}\ket{0}.
\end{equation}
It remains to erase the $\ket{k}$ register.
We do so by a similar binary search using the cumulative counter oracle $O_c$ but on $\ket{k}$.
This constructs the desired row index oracle.

It remains to show how to construct the cumulative counter oracle $O_c$ with quantum oracle sketching.
We first construct an intermediate object called the cumulative counter unitary $U_c$, that has the $C(i, l)-k$ information encoded in its phase.
Then we use QSVT to apply a threshold function that calculates the binary value $1[C(i, l)< k]$ as required in the cumulative counter oracle.
In \Cref{lem:cumulat-count-unitary}, we show how to construct $U_c$ with quantum oracle sketching.
Later in \Cref{lem:sparse-index-oracle}, we show how to use $U_c$ to construct the sparse index oracles.

\begin{lemma}[Cumulative counter unitary]
\label{lem:cumulat-count-unitary}
    Let $A\in \mathbb{R}^{N\times N}$ be an $N$-dimensional and row $s_r$-sparse matrix.
    We define the cumulative counter unitary $U_c$ of $A$ as 
    \begin{equation}
        U_c: \ket{i, k, l} \to e^{i\theta(i, k, l)}\ket{i, k, l}, \quad \forall i, l\in [N], k\in [s_r],
    \end{equation}
    where the phase
    \begin{equation}
        \theta(i, k, l) = \frac{\pi}{2(s_r+1/2)} \lr{C(i,l) - k + \frac12}, \quad C(i, l) = |j\in [N]: A_{ij}\neq 0, j<l|.
    \end{equation}
    Then, we can use $2\ceil{\log(N)} + \ceil{\log(s_r+1)}+b$ qubits and
    \begin{equation}
        M = O\lr{\frac{RNs_r}{\epsilon}}
    \end{equation}
    samples starting at any time $t_0\geq 1$ from the matrix data generation process $\mathcal{D}_A$ with repetition number $R$ to implement a random unitary channel $\mathcal{V}_{c}$ satisfying
    \begin{equation}
        \E_{z_1, \ldots, z_{t_0-1}}\|\E[\cV_{c}|z_1, \ldots, z_{t_0-1}]-U_{c}(\cdot)U_{c}^\dagger\|_\diamond \leq \epsilon,
    \end{equation}
    where the inner expectation is over the current $M$ samples and the outer expectation is over the previously processed data samples $z_1, \ldots, z_{t_0-1}$.
    The data processing time per sample is $\polylog(N)$.
    The same holds for implementing $U_{c}^\dagger, cU_{c}, cU_{c}^\dagger$.
\end{lemma}

\begin{proof}
    For convenience, we define the shorthand
    \begin{align}
        a = (i, k, l)
    \end{align}
    with the corresponding set
    \begin{align}
    \label{eq:S_a}
        S_a = \left\{(i,j): A_{ij} \neq 0, j<l\right\}
    \end{align}
    and projector
    \begin{align}
        \ketbra{a} = \ketbra{i, k, l}.
    \end{align}
    Note that
    \begin{equation}
        |S_a|\leq s_r, \quad \forall a=(i, k, l),
    \end{equation}
    because $A$ is row $s_r$-sparse.
    We define the fixed offset gate
    \begin{equation}
	   U_o: \ket{i, k, l} \to e^{-i\frac{\pi}{2(s_r+1/2)}(k-1/2)} \ket{i, k, l},
    \end{equation}
    which can be easily implemented using single-qubit phase gates on each bit of $k$.
    We then rewrite $U_{c}$ as
    \begin{align}
        U_{c} = U_o\sum_a \ketbra{a} \exp[i\frac{\pi}{2(s_r+1/2)}|S_a|] = U_o\exp[i\frac{\pi}{2(s_r+1/2)}\sum_a |S_a|\ketbra{a}].
    \end{align}
    Meanwhile, the marginal distribution of $(i, j)$ from the sparse matrix stream is
    \begin{equation}
        p(i, j) = \begin{cases}
            1/K, & A_{ij}\neq 0, \\
            0, & A_{ij}=0,
        \end{cases}
    \end{equation}
    where $K=|\mathcal{K}|, \mathcal{K} = \{(i, j): A_{ij}\neq 0\}$ with $K\leq 2^n s_r$.
    Therefore, we have
    \begin{equation}
    \begin{split}
        U_o^\dagger U_{c} &= \exp[i\frac{\pi }{2(s_r+1/2)}\sum_a|S_a|\ketbra{a}] \\
        &=\exp[i\frac{\pi }{2(s_r+1/2)}\sum_{a}\sum_{(i, j)\in\mathcal{K}}1[(i, j)\in S_a]\ketbra{a}] \\
        &=\exp[i\frac{\pi }{2(s_r+1/2)}\sum_{(i, j)\in\mathcal{K}}\sum_{a}1[(i, j)\in S_a]\ketbra{a}] \\
        &=\exp[i\frac{\pi K}{2(s_r+1/2)}\sum_{(i, j)\in \mathcal{K}}p(i, j)\sum_{a: (i, j)\in S_a}\ketbra{a}]
    \end{split}
    \end{equation}
    This representation allows us to implement $U_o^\dagger U_{c}$ and hence $U_c$ using quantum oracle sketching (\Cref{thm:q-oracle-sketch-corr}).
    In particular, with each one of the $M$ samples $(i, j, A_{ij})$, we apply the unitary
    \begin{align}
        V(i,j) = \exp[\frac{i}{M}\frac{\pi K}{2(s_r+1/2)}\sum_{a\sth(i,j) \in S_a} \ketbra{a}].
    \end{align}
    This gate $V(i, j)$ can be implemented efficiently (i.e., in $\polylog N$ time) because it is a controlled phase gate where the control rule defined by $S_a$ can be efficiently checked.
    Now we follow the notations in \Cref{thm:q-oracle-sketch-corr} and define the Hamiltonian
    \begin{equation}
        h_{ij} = \frac{\pi K}{2 (s_r+1/2)} \sum_{a: (i, j)\in S_a} \ketbra{a}.
    \end{equation}
    Then $V(i, j) = e^{ih_{ij}/M}$.
    Since $h_{ij}$'s are diagonal and commute with each other, the joint action of the gates corresponding to all $M$ samples $(i_k, j_k)_{k=1}^M$, followed by a fixed gate $U_o$, is a random unitary
    \begin{equation}
        V = U_o\prod_{k=t_0}^{t_0+M-1} V(i_k, j_k) = U_o\exp[i\frac{1}{M}\sum_{k=t_0}^{t_0+M-1} h_{i_k j_k}] = U_o\exp[iX],
    \end{equation}
    where we define a random matrix $X = \frac{1}{M}\sum_{k=t_0}^{t_0+M-1} h_{i_k j_k}$ as a shorthand.
    Meanwhile, we can rewrite the target cumulative counter unitary as
    \begin{equation}
        U_{c} = U_o\exp[i\E_{i, j}[h_{ij}]] = U_o\exp[i\E[X]],
    \end{equation}
    since
    \begin{equation}
        \E[X] = \frac{1}{M}\sum_{k=t_0}^{t_0+M-1}\E[h_{i_kj_k}] = \E[h_{ij}].
    \end{equation}
    The matrix $X$ can be written in the $\ket{a}$ basis as
    \begin{equation}
    \begin{split}
        X &= \frac{1}{M}\sum_{k=t_0}^{t_0+M-1} \frac{\pi K}{2(s_r+1/2)} \sum_{a: (i_k, j_k)\in S_a} \ketbra{a} \\
        &= \frac{\pi K}{2(s_r+1/2)} \frac{1}{M}\sum_{k=t_0}^{t_0+M-1} \sum_{a} 1[(i_k, j_k)\in S_a] \ketbra{a} \\
        &= \frac{\pi K}{2(s_r+1/2)} \sum_a \ketbra{a} \frac{\sum_{k=t_0}^{t_0+M-1} 1[(i_k, j_k)\in S_a]}{M} \\
        &= t \sum_a \ketbra{a} m_a,
    \end{split}
    \end{equation}
    where we have introduced $t=\frac{\pi K}{2(s_r+1/2)}$ and random variables 
    \begin{equation}
        m_a = \frac{1}{M}\sum_{k=t_0}^{t_0+M-1} 1[(i_k, j_k)\in S_a]
    \end{equation}
    to be the empirical frequency of phase accumulation on the basis $\ket{a}$.

    To bound the error, we need to bound the variance part and bias part separately as in \Cref{thm:q-oracle-sketch-corr}.
    We first bound the variance part.
    Since $X$ is a diagonal random matrix with diagonal elements $tm_a$, \Cref{lem:error-var} implies that
    \begin{equation}
    \underbrace{\E\|\E[e^{iX}|z_1, \ldots, z_{t_0-1}] - e^{i\E[X|z_1, \ldots, z_{t_0-1}]}\|}_{\mathrm{variance}} \leq \frac{t^2}{2}\E[\max_a \Var[m_a|z_1, \ldots, z_{t_0-1}]].
    \end{equation}
    Now we define
    \begin{equation}
        m_{ij} = \frac{1}{M}\sum_{k=t_0}^{t_0+M-1} 1[(i_k, j_k)=(i, j)]
    \end{equation}
    and note that
    \begin{equation}
    \begin{split}
        m_a &= \frac{1}{M}\sum_{k=t_0}^{t_0+M-1} 1[(i_k, j_k)\in S_a]  \\
        &= \frac{1}{M}\sum_{k=t_0}^{t_0+M-1} \sum_{(i, j)\in S_a} 1[(i_k, j_k)=(i, j)]\\
        &= \sum_{(i, j)\in S_a} \frac{1}{M}\sum_{k=t_0}^{t_0+M-1} 1[(i_k, j_k)=(i, j)] \\
        &= \sum_{(i, j)\in S_a} m_{ij}
    \end{split}
    \end{equation}
    Therefore, the variance can be upper bounded by
    \begin{equation}
    \begin{split}
        \Var[m_a|z_1, \ldots, z_{t_0-1}] &= \sum_{(i, j)\in S_a}\sum_{(i', j')\in S_a}\Cov[m_{ij}, m_{i'j'}|z_1, \ldots, z_{t_0-1}] \\
        &\leq \sum_{(i, j)\in S_a}\sum_{(i', j')\in S_a}\sqrt{\Var[m_{ij}|z_1, \ldots, z_{t_0-1}]\Var[m_{i'j'}|z_1, \ldots, z_{t_0-1}]} \\
        &\leq |S_a|^2 \max_{(i, j)\in S_a}\Var[m_{ij}|z_1, \ldots, z_{t_0-1}] \\
        &\leq s_r^2 \max_{(i, j)\in S_a}\Var[m_{ij}|z_1, \ldots, z_{t_0-1}].
    \end{split}
    \end{equation}
    On the other hand, the conditional variance and bias upper bound in \Cref{lem:data-cond-var-rep-num,lem:data-cond-drift-rep-num} implies that
    \begin{equation}
        \E[\max_{i, j}\Var[m_{ij}|z_1, \ldots, z_{t_0-1}]]\leq \frac{2R_{\mathcal{D}}}{M} \E[\max_{i,j}\E[m_{ij}|z_1, \ldots, z_{t_0-1}]] \leq \frac{2R}{M}\frac{1}{K} +  \frac{2R^2}{M^2},
    \end{equation}
    where $R$ is the repetition number of the matrix data generation process.
    Thus, we have
    \begin{equation}
       \underbrace{\|\E[e^{iX}|z_1, \ldots, z_{t_0-1}] - e^{i\E[X|z_1, \ldots, z_{t_0-1}]}\|}_{\mathrm{variance}} \leq \frac{t^2}{2} s_r^2 (\frac{2R}{M K}+\frac{2R^2}{M^2}) \leq \frac{\pi^2}{4}\frac{KR}{M}(1+KR/M),
    \end{equation}
    where we have used $t=\pi K/(2(s_r+1/2))$.

    Next, we bound the bias part.
    Note that since $X, \E[X|z_1, \ldots, z_{t_0-1}]$ are diagonal and thus commuting, we have
    \begin{equation}
    \begin{split}
        \underbrace{\E\|e^{i\E[X|z_1, \ldots, z_{t_0-1}]} - e^{i\E[X]}\|}_{\mathrm{bias}} &= \E\|e^{i(\E[X]-\E[X|z_1, \ldots, z_{t_0-1}])} - 1\| \\
        &\leq \E\|\E[X]-\E[X|z_1, \ldots, z_{t_0-1}]\| \\
        &= t\E \max_a |\E[m_a|z_1, \ldots, z_{t_0-1}]-\E[m_a]| \\
        &= t\E \max_a\left|\sum_{(i, j)\in S_{a}}(\E[m_{ij}|z_1, \ldots, z_{t_0-1}]-\E[m_{ij}])\right| \\
        &\leq t s_r \E \max_{i,j}\left|\E[m_{ij}|z_1, \ldots, z_{t_0-1}]-\E[m_{ij}]\right| \\
        &\leq ts_r \frac{\sqrt{1/K \cdot K}R}{M} \\
        &= t\frac{s_rR}{M} \leq \frac{\pi K R}{2M},
    \end{split}
    \end{equation}
    where we have used \Cref{lem:data-cond-drift-rep-num}, $|S_a|\leq s_r$, and $t=\pi K/(2(s_r+1/2))$.

    Combining the variance and bias part, we have
    \begin{equation}
    \begin{split}
        &\E\|\E[V(\cdot)V^\dagger|z_1, \ldots, z_{t_0-1}] - U_{c}(\cdot)U_{c}^\dagger\|_\diamond \\
        &\leq 4\E\|\E[e^{iX}|z_1, \ldots, z_{t_0-1}] - e^{i\E[X]}\| \\
        &\leq 4\underbrace{\E\|\E[e^{iX}|z_1, \ldots, z_{t_0-1}] - e^{i\E[X|z_1, \ldots, z_{t_0-1}]}\|}_{\mathrm{variance}} + 4\underbrace{\E\|e^{i\E[X|z_1, \ldots, z_{t_0-1}]} - e^{i\E[X]}\|}_{\mathrm{bias}} \\
        &\leq 4\frac{\pi^2KR}{4M}(1+KR/M) + 4\frac{\pi K R}{2M} \leq \frac{(\pi^2+2\pi)KR}{M} (1+KR/M),
    \end{split}
    \end{equation}
    where we have invoked \Cref{lem:diamond-operator-expect}.
    
    Since $K\leq N s_r$, we conclude that
    \begin{align}
        M = 2(\pi^2+2\pi)\frac{RNs_r}{\epsilon}
    \end{align}
    samples ensures 
    \begin{align}
        \E\|\E[V(\cdot)V^\dagger|z_1, \ldots, z_{t_0-1}] - U_{c}(\cdot)U_{c}^\dagger\|_\diamond \leq \epsilon.
    \end{align}
    Note that the same analysis holds for implementing $U_{c}^\dagger, cU_{c}, cU_{c}^\dagger$ by adding additional control to the individual gates or negating the phases in them.
    This completes the proof of \Cref{lem:cumulat-count-unitary}.
\end{proof}

Now we use the cumulative counter unitary $U_c$ to build the sparse row index oracle
\begin{equation}
    O_A^{\mathrm{ind,row}} \sth \ket{i}\ket{k} \to \ket{i}\ket{j(i,k)}, \quad \forall i\in\bit^n, k\in [s_r]
\end{equation}
where $j(i,k) \in \{0,1\}^{n+1}$ is the (padded) column index of the $k$-th nonzero element of $A$ in row $i$.
The implementation of the sparse column index oracle follows similarly by swapping rows and columns.

\begin{tcolorbox}
\begin{lemma}[Sparse index oracles]\label{lem:sparse-index-oracle}
    Let $A \in \mathbb{R}^{N\times N}$ be an $N$-dimensional and row $s_r$-sparse matrix. 
    Let $\cO_A^{\mathrm{ind,row}}: \rho \to O_A^{\mathrm{ind,row}} \rho O_A^{\mathrm{ind,row}\dagger}$ be the unitary channel corresponding to the sparse row index oracle $O_A^{\mathrm{ind, row}}$.
    Then, we can use $2\ceil{\log(N)}+\ceil{\log(s_r+1)}+4+b$ qubits and
    \begin{equation}
        M = O\lr{\frac{RNs_r^3 \log^2(N)\log^2(\log(N)/\epsilon)}{\epsilon}} = \tilde{O}\lr{\frac{RNs_r^3}{\epsilon}}
    \end{equation}
    samples starting at any time $t_0\geq 1$ from the matrix data generation process $\mathcal{D}_A$ with repetition number $R$ to implement a random unitary channel $\mathcal{V}$ satisfying
    \begin{equation}
        \E_{z_1, \ldots, z_{t_0-1}}\|\E[\cV|z_1, \ldots, z_{t_0-1}] - \cO_A^{\mathrm{ind,row}}\|_\diamond\leq \epsilon,
    \end{equation}
    where the inner expectation is over the current $M$ samples and the outer expectation is over the previously processed data samples $z_1, \ldots, z_{t_0-1}$.
    The data processing time per sample is $\polylog(N, 1/\epsilon)$.
    The same guarantee holds for implementing $O^{\mathrm{ind,row}\dagger}_A$, $cO^{\mathrm{ind,row}}_A,$ and $cO^{\mathrm{ind,row}\dagger}_A$.
    Similar results apply to implementing $O^{\mathrm{ind,col}}_A$, $O^{\mathrm{ind,col}\dagger}_A$, $cO^{\mathrm{ind,col}}_A, cO^{\mathrm{ind,col}\dagger}_A$ with row $s_r$-sparse replaced by column $s_c$-sparse.
\end{lemma}
\end{tcolorbox}

\begin{proof}
    In the following we show how to implement $O^{\mathrm{ind,row}}_A$ with the claimed complexity.
    The implementation of $O^{\mathrm{ind,col}}_A$ and its cousins follows verbatim by swapping rows and columns.
    
    Recall our earlier intuition that 
    \begin{align}
        1[C(i, j(i, k)_{\leq m}10\cdots 0)-k< 0] = j(i, k)_{m+1}
    \end{align}
    for any $i\in \bit^n, k\in [s_r], m\in [n]$.
    To calculate $1[C(i, l)-k< 0]$ using the counter
    \begin{equation}
        U_c: \ket{i, k, l} \to e^{i\theta(i, k, l)}\ket{i, k, l}, \quad \theta(i, k, l) = \frac{\pi}{2(s_r+1/2)}\lr{C(i, l)-k+\frac12},
    \end{equation}
    that we have from \Cref{lem:cumulat-count-unitary}, we need to implement a threshold function that maps positive phases to $(-1)^0$ and negative phases to $(-1)^1$.
    To this end, we employ QSVT similar to the techniques used in \Cref{lem:sample-upper-unknown-dist} as follows.
    Let $S = \begin{pmatrix}
        -i &0\\
        0 &1
    \end{pmatrix}$ be a single qubit gate.
    We introduce an ancilla qubit $a$ and use two queries to construct
    \begin{equation}
        W_{c} = S_{a}X_{a}H_{a}(c_{a}U_{c}^\dagger)X_{a}(c_{a}U_{c}) H_{a},
    \end{equation}
    where the subscript $a$ indicates that the gate or the control acts on an ancilla qubit.
    Then we have
    \begin{equation}
    \begin{split}
        W_{c}\ket{0_{a}} 
        &=\ket{1_{a}}\otimes \frac{U_{c}+U_{c}^\dagger}{2} + \ket{0_{a}}\otimes \frac{U_{c}-U_{c}^\dagger}{2i}.
    \end{split}
    \end{equation}
    Therefore, we have a block encoding
    \begin{equation}
        \bra{0_{a}}W_{c}\ket{0_{a}} = \frac{U_{c}-U_{c}^\dagger}{2i} = \sum_{i, k, l} \sin(\theta(i, k, l))\ketbra{i, k, l}.
    \end{equation}
    Note that we also have access to $W_{c}^\dagger$ using a similar construction.  
    Now we apply a polynomial function $P$ to $W_{c}$.
    Since $\sin(\theta(i, k, l)) \in [\sin(\pi/(4s_r+2)), 1]$ corresponds to $1[C(i, l)< k]=0$ and $\sin(\theta(i, k, l)) \in [-1, -\sin(\pi/(4s_r+2))]$ corresponds to $1[C(i, l)< k]=1$, we choose the polynomial $P$ to $\epsilon_1$-approximate the sign function with threshold value $\lambda^\star = \sin(\pi/(4s_r+2))$.
    The degree of $P$ is at most $d=O(\log(1/\epsilon_1)/\lambda^\star)$ (we choose $d$ to be odd), $P$ is an odd function, $|P(w)|\leq 1, \forall w\in [-1, 1]$, and $|P(w)-1|\leq \epsilon_1, \forall w\in [\lambda^\star, 1]$ and $|P(w)+1|\leq \epsilon_1, \forall w\in [-1, -\lambda^\star]$.
    This means that
    \begin{equation}
        \left|P(\sin(\theta(i, k, l))) - (-1)^{1[C(i, l)< k]}\right|\leq \epsilon_1, \quad \forall i, k, l.
    \end{equation}
        Then the odd-parity part of \Cref{lem:qsvt} gives a unitary
    $P_{\mathrm{QSVT}}(W_c,\phi)$ using $d$ queries to
    $W_c,W_c^\dagger$ such that
    \begin{equation}
        \bra{0_{a'}0_a}
        P_{\mathrm{QSVT}}(W_c,\phi)
        \ket{0_{a'}0_a}
        =
        \sum_{i,k,l}
        P(\sin\theta(i,k,l))
        \ketbra{i,k,l}.
    \end{equation}
    The QSVT target is the phase oracle
    \begin{align}
        V_{c}: \ket{i, k, l} \to (-1)^{1[C(i, l)< k]}\ket{i, k, l}
    \end{align}
    The same error analysis as in \Cref{lem:sample-upper-unknown-dist} implies that
    \begin{equation}
        \|P_{\mathrm{QSVT}}(W_{c}, \phi)(\ket{0_{a'}0_a}\otimes \cdot) - \ket{0_{a'}0_a}\otimes V_{c}\|\leq \sqrt{2\epsilon_1} + \epsilon_1 \leq 3\sqrt{\epsilon_1} = \epsilon_2/4
    \end{equation}
    when we choose $\epsilon_1 = (\epsilon_2/12)^2$.
    The number of queries to $U_{c}, U^\dagger_{c}, cU_{c}$ and $cU^\dagger_{c}$ is 
    \begin{equation}
        O(d) = O\left(\frac{\log(1/\epsilon_1)}{\lambda^\star}\right) = O\left(\frac{\log(1/\epsilon)}{\sin(\pi/(4s_r+2))}\right)\leq O\left(s_r\log(1/\epsilon)\right),
    \end{equation}
    where we have used the fact that $\sin(x)/x \geq \sin(\pi/2)/(\pi/2)=2/\pi, \forall x \in [0, \pi/2]$.
    The same results hold for implementing $V_{c}^\dagger, cV_{c}, cV_{c}^\dagger$.
    We implement $U_{c}$ and its cousins using samples from the matrix data generation process as in \Cref{lem:cumulat-count-unitary}. 
    Similar to the analysis in \Cref{lem:sample-upper-unknown-dist}, we obtain an $\epsilon_2$-approximation to $V_{c}$ using a total number of samples
    \begin{equation}
        M_0 = O\lr{\frac{RNs_r}{\epsilon_2/(s_r\log(1/\epsilon_2))}\cdot (s_r\log(1/\epsilon_2))} = O\lr{\frac{RN s_r^3\log^2(1/\epsilon_2)}{\epsilon_2}}.
    \end{equation}
    
    Next, we convert this into the cumulative counter oracle
    \begin{equation}
        O_c: \ket{i, k, l}\ket{0}_o \to \ket{i, k, l}\ket{1[C(i, l)<k]}_o, \quad C(i, l) = |\{j: A_{ij}\neq 0, j< l\}|,
    \end{equation}
    via the standard phase oracle to XOR oracle conversion by introducing an output ancilla $o$:
    \begin{align}
        O_c = (H_o\otimes I)c_oV_{c}(H_o\otimes I),
    \end{align}
    where the Hadamard gate and the control are on the ancilla qubit.
    Similarly, we can implement the $\leq$ version of the cumulative counter oracle
    \begin{equation}
        O_c^{\leq}: \ket{i, k, l}\ket{0}_o \to \ket{i, k, l}\ket{1[C^{\leq}(i, l)<k]}_o, \quad C^{\leq}(i, l) = |\{j: A_{ij}\neq 0, j\leq l\}|
    \end{equation}
    with the same guarantees by changing the condition $j<l$ to $j\leq l$ in \Cref{lem:cumulat-count-unitary}.

    We now proceed to construct the sparse index oracle using the cumulative counter oracle.
    We first calculate whether $k$ exceeds the number of non-zero columns in row $i$.
    We start from $\ket{i}\ket{k}\ket{0}_{j_0}\ket{0\ldots 0}_l\ket{0}_o$, where $j_0$ represents the first indicator bit in $j(i, k)\in \bit^{n+1}$.
    This can be calculated by flipping all bits in $l$, applying $O_c^{\leq}$ to the resulting state $\ket{i, k, l=1^n}\ket{0}_{j_0}$, and flipping all bits in $l$ again.
    We use $U_{j_0}$ to denote this circuit.
    This step writes the value
    \begin{equation}
	1[C^{\leq}(i, 1^n)<k]=1[|j: A_{ij}\neq 0|<k]
    \end{equation}
    into the $j_0$ register, which exactly signals if $k$ exceeds the number of non-zero columns in row $i$, as desired.
    
    Now we execute the binary search as discussed earlier.
    When $k$ does not exceed the number of non-zero columns in this row, the $j_0$ is put into zero.
    We flip the first bit of the $l$ register, apply $O_c$, flip the first bit of $l$ again, and swap that bit with the output register $o$ of $O_c$.
    This writes the first bit of $j(i, k)$ into the first bit of the $l$ register.
    Then we proceed and repeat the same procedure to all the $n$ bits in the $l$ register.
    This results in the following circuit
    \begin{equation}
    	(\mathrm{SWAP}_{l_n, o}X_{l_n} O_c X_{l_n})\cdots (\mathrm{SWAP}_{l_1, o}X_{l_1} O_c X_{l_1})U_{j_0}: \ket{i}\ket{k}\ket{0}_{j_0}\ket{0^n}_l\ket{0}_o \to \ket{i}\ket{k}\ket{j(i, k)}_{j_0, l}\ket{0}_o
    \end{equation}
    that finds the desired column index.

    When $k$ exceeds the number of non-zero columns in this row, the $j_0$ register is put into one.
    In this case, the above binary search circuit will always write $1^n$ into $l$.
    Conditioned on $\ket{1}_{j_0}$, we apply $X^{\otimes n}$ to $l$, copy the value of $k$ into $l$, subtract one from it, and obtain $\ket{i}\ket{k}\ket{1}_{j_0}\ket{k-1}_{l}\ket{0}_o=\ket{i}\ket{k}\ket{j(i, k)}_{j_0,l}\ket{0}_o$, as desired.
    
    It remains to erase the $\ket{k}$ register.
    Conditioned on $\ket{0}_{j_0}$, we do so by a similar binary search using the cumulative counter oracle $O_c$, but over the $\ket{k}$ register.
    From the definition of $C(i, l)=|\{j: A_{ij}\neq 0, j<l\}|$, we have that
    \begin{equation}
        C(i, j(i, k)) = k-1 = k',
    \end{equation}
    where we have defined $k'=k-1$.
    Let $m=\ceil{\log(s_r+1)}$ be the binary representation length of the $\ket{k}$ register.
    We have that
    \begin{equation}
        1[C(i, j(i,k))<k'_1\ldots k'_{t-1}1_t0_{t+1}\ldots 0_m] = 1[k'<k'_1\ldots k'_{t-1}1_t0_{t+1}\ldots 0_m] = k'_t\oplus 1, \quad \forall t\in [m],
    \end{equation}
    where the subscripts mark the positions of the bits.

    We begin by subtracting $1$ from $\ket{k}$, add $1$ to $\ket{0}_o$, and obtain $\ket{i}\ket{k'}\ket{j(i, k)}\ket{1}_o$.
    Now suppose we have obtained $\ket{i}\ket{k'_1\ldots k'_t0\ldots 0}\ket{j(i, k)}\ket{1}_o$.
    We apply $\mathrm{SWAP}_{k_t, o}$ and obtain $\ket{i}\ket{k'_1\ldots k'_{t-1}1_t0\ldots 0}\ket{j(i, k)}\ket{k_t'}_o$.
    Then we apply $O_c$ to write $1[C(i, j(i,k))<k'_1\ldots k'_{t-1}1_t0_{t+1}\ldots 0_m] = k'_t\oplus 1$ into the $o$ register and obtain $\ket{i}\ket{k'_1\ldots k'_{t-1}1_t0\ldots 0}\ket{j(i, k)}\ket{1}_o$.
    Finally, we apply $X_{k_t}$ to get $\ket{i}\ket{k'_1\ldots k'_{t-1}0_t0\ldots 0}\ket{j(i, k)}\ket{1}_o$.
    Repeat this for all $m$ bits of the $k$ register gives us
    \begin{equation}
        c_{j_0=0}\left[X_o\prod_{t=1}^m (X_{k_t}O_c\mathrm{SWAP}_{k_t,o})\cdot \mathrm{SUB}^1_{k}X_{o}\right]\cdot \prod_{t=n}^1(\mathrm{SWAP}_{l_t, o}X_{l_t} O_c X_{l_t})U_{j_0}: \ket{i}\ket{k}\ket{0^{n+1}}\ket{0}_o \to \ket{i}\ket{0^m}\ket{j(i, k)}\ket{0}_o,
    \end{equation}
    where $\mathrm{SUB}^1_{k}$ means subtracting $1$ from the $k$ register and $c_{j_0=0}$ indicates controlled on $j_0=0$.

    Conditioned on $\ket{1}_{j_0}$, we know that the $l$ register stores $k-1$.
    Therefore, we subtract one from the $k$ register and apply transversal CNOTs to put $k$ into zero.
    As a result, in both cases, the output state is $\ket{i}\ket{0^m}\ket{j(i, k)}\ket{0}_o$.
    This is the desired sparse index oracle.
    
    In total, this circuit requires
    \begin{equation}
        2n + \ceil{\log(s_r+1)} + \underbrace{2}_{\text{registers $j_0, o$}} + \underbrace{2}_{\text{QSVT}} + b = 2\ceil{\log(N)} + \ceil{\log(s_r+1)} + 4 + b
    \end{equation}
    qubits and $n+\ceil{\log(s_r+1)}+1 \leq 2n+2$ uses of $cV_c$.
    We implement all these $cV_c$ using data samples.
    To obtain a final error of $\epsilon$, we set $\epsilon_2 = \epsilon/n$ and therefore the total number of samples needed is
    \begin{equation}
        M = O(n M_0) = O\lr{n\frac{R2^ns_r^3\log^2(1/(\epsilon/n))}{\epsilon/n}} = O\lr{\frac{RNs_r^3 \log^2(N)\log^2(\log(N)/\epsilon)}{\epsilon}}.
    \end{equation}
    This completes the proof of \Cref{lem:sparse-index-oracle}.
\end{proof}

\subsubsection{Block encodings}

Apart from sparse oracles, block encodings are also widely used as the default access model for matrices in many quantum algorithms.
\Cref{lem:sparse-element-oracle,lem:sparse-index-oracle} allows us to implement the sparse oracles of an $N$-dimensional and $s$-sparse matrix $A$ with error $\epsilon$ using $\tilde{O}(RNs^3/\epsilon)$ samples and $O(\log N)$ qubits.
In this section, we build on these to construct the block encoding of $A$ with constant subnormalization and essentially the same guarantees.
We use the following lemma from \cite{gilyen2019quantum} to convert sparse oracles into block encodings.

\begin{lemma}[Block encodings from sparse oracles {\cite[Lemma 48]{gilyen2019quantum}}]
\label{lem:sparse-oracle-to-block-encoding}
    Let $A\in \mathbb{C}^{2^n\times 2^n}$ be an $s$-sparse matrix where each matrix element has absolute value bounded by one.
    Then, we can implement a unitary $V_A$ such that
    \begin{equation}
        \|(\bra{0^{n+3}}\otimes I) V_A (\ket{0^{n+3}} \otimes I) - A/s\|\leq \epsilon
    \end{equation}
    with one query to each of the sparse index oracles $O_A^{\mathrm{ind,row}}$ and $O_A^{\mathrm{ind,col}}$ and sparse element oracle $O_A^{\mathrm{ele}}$ and its inverse $O_A^{\mathrm{ele}\dagger}$, using $O(n+\log^{2.5}(1/\epsilon))$ additional two qubit gates and $O(b+\log^{2.5}(1/\epsilon))$ ancilla qubits.
\end{lemma}

This gives use the following result.

\begin{tcolorbox}
\begin{theorem}[Block encodings]
\label{lem:block-encoding}
    Let $A\in \mathbb{R}^{N\times N}$ be an $N$-dimensional and $s$-sparse matrix with $\|A\|\leq 1$.
    There is a unitary $U_A$ and its corresponding channel $\mathcal{U}_A: \rho \to U_A\rho U_A^\dagger$ that block encodes $A/2$ with $\epsilon$ error and $\ceil{\log(N)}+4$ ancilla qubits 
    \begin{equation}
        \|(\langle{0^{\ceil{\log(N)}+4}}|\otimes I) U_A (|{0^{\ceil{\log(N)}+4}}\rangle \otimes I) - A/2\|\leq \epsilon,
    \end{equation}
    such that we can use $O(\log(N)+b+\log^{2.5}(s/\epsilon))$ qubits and
    \begin{equation}
        M = O\lr{\frac{RNs^5 \log^2(N)\log^4(s\log(N)/\epsilon)}{\epsilon}}
    \end{equation}
    samples starting at any time $t_0\geq 1$ from the matrix data generation process $\mathcal{D}_A$ with repetition number $R$ to implement a random unitary channel $\mathcal{V}$ satisfying
    \begin{equation}
        \E_{z_1, \ldots, z_{t_0-1}}\|\E[\mathcal{V}|z_1, \ldots, z_{t_0-1}] - \mathcal{U}_A\|_\diamond \leq \epsilon,
    \end{equation}
    where the inner expectation is over the current $M$ samples and the outer expectation is over the previously processed data samples $z_1, \ldots, z_{t_0-1}$.
    The data processing time per sample is $\polylog(N, 1/\epsilon)$.
    The same guarantee holds for implementing $U_A^\dagger, cU_A$ and $cU_A^\dagger$.
\end{theorem}
\end{tcolorbox}

Note that all downstream quantum algorithms allow a known constant block-encoding subnormalization.
The factor $2$ above changes their query and sample complexities only by universal constant factors, which we suppress throughout.
Also note that the error in the block encoding can always be absorbed into the total implementation error due to the following lemma.
Therefore, the block-encoding error can be treated as an additional implementation error with the same asymptotic resource accounting.
By decreasing the internal block-encoding accuracy by a sufficiently small universal constant factor, we henceforth use the accuracy parameter $\epsilon$ so that $\left\|\mathcal{U}_A-\widehat{\mathcal{U}}_A\right\|_\diamond\leq\epsilon$, where $\widehat{\mathcal{U}}_A$ is the channel of the exact completion.
This constant rescaling does not change any resource bound.

\begin{lemma}[Exact completion of a strictly subnormalized block encoding]
\label{lem:exact-block-completion}
Let $A$ be a matrix with norm $\|A\|\leq 1$.
Let $U$ be a unitary block encoding of $B$ satisfying $\|B-A/2\|\leq \eta\leq 1/4$.
Then there exists a unitary $\hat{U}$ that exactly block encodes $A/2$ and satisfies $\|U-\hat{U}\|\leq O(\eta)$.
The same bound holds for the inverse and controlled versions.
\end{lemma}

\begin{proof}
For any strict contraction $X$, define its canonical unitary dilation
\begin{equation}
    J(X)=
    \begin{pmatrix}
    X & \sqrt{I-XX^\dagger}\\
    \sqrt{I-X^\dagger X} & -X^\dagger
    \end{pmatrix}.
\end{equation}
Moreover, standard unitary dilation results imply that $U$ and $J(B)$ can always be connected by block-diagonal unitaries $L=I\oplus L_{\mathrm{aux}}, R=I\oplus R_{\mathrm{aux}}$ and an auxiliary unitary $W$ such that
\begin{equation}
    U
    =
    L\bigl(J(B)\oplus W\bigr)R.
\end{equation}
Here $W$ is omitted when there is no excess auxiliary space.
Since $\|A/2\|\leq 1/2$ and $\|B\|\leq 1/2+\eta\leq 3/4$, the operators $I-BB^\dagger$, $I-B^\dagger B$, $I-AA^\dagger/4$, and $I-A^\dagger A/4$ are bounded below by a positive universal constant.
The square-root map is therefore Lipschitz on their spectra, and
\begin{equation}
    \|J(B)-J(A/2)\|
    =
    O\left(
        \|B-A/2\|
    \right)
    =
    O(\eta).
\end{equation}
Define
\begin{equation}
    \hat U
    =
    L\bigl(J(A/2)\oplus W\bigr)R.
\end{equation}
Then $\hat U$ exactly block encodes $A/2$ and
\begin{equation}
    \|U-\hat U\|
    =
    \|J(B)-J(A/2)\|
    =
    O(\eta),
\end{equation}
as claimed.
Finally, $\|U^{-1}-\hat U^{-1}\|=\|U-\hat U\|$, and adding a control does not change the norm difference.
\end{proof}

\begin{proof}[Proof of \Cref{lem:block-encoding}]
    We instantiate the sparse oracles in \Cref{lem:sparse-oracle-to-block-encoding} using \Cref{lem:sparse-element-oracle,lem:sparse-index-oracle}.
    In particular, \Cref{lem:sparse-oracle-to-block-encoding} asserts that we can construct a unitary $V_A$ such that
    \begin{equation}
        \|(\bra{0^{n+3}}\otimes I) V_A (\ket{0^{n+3}} \otimes I) - A/s\|\leq \epsilon_1
    \end{equation}
    with one query to each of the sparse index oracles $O_A^{\mathrm{ind,row}}$ and $O_A^{\mathrm{ind,col}}$ and sparse element oracle $O_A^{\mathrm{ele}}$ and its inverse $O_A^{\mathrm{ele}\dagger}$, using $O(n+\log^{2.5}(1/\epsilon_1))$ additional two qubit gates and $O(b+\log^{2.5}(1/\epsilon_1))$ ancilla qubits.
    Note that $\|A\|\leq 1$ implies that each matrix element of $A$ has absolute value bounded by one, satisfying the requirement of \Cref{lem:sparse-oracle-to-block-encoding}.

    Since $\|A\|\leq 1$, every singular value of $A/s$ lies in
    $[0,1/s]$.
    By the additive-error uniform singular value amplification \cite{martyn2021grand}, there exists an odd real polynomial $P$ of degree
    $d
    =
    O\left(
        s\log\left(1/{\epsilon_2}\right)
    \right)$
    such that
    $|P(x)|\leq 1,
    \forall x\in[-1,1],$
    and
    $\left|
        P(x)-\frac{s}{2}x
    \right|
    \leq
    \epsilon_2,
    \forall |x|\leq\frac{1}{s}.$
    Applying the odd singular value transformation in
    \Cref{lem:qsvt} therefore gives a block encoding of $A/2$, which we call $U_A$.
    Since $V_A$ block encodes $A/s$ with error $\epsilon_1$, the standard robustness bound for degree-$d$ QSVT contributes $O(d\sqrt{\epsilon_1})$ additional error \cite{gilyen2019quantum}.
    We choose
    $\epsilon_1
    =
    \Theta\left(
        {\epsilon_2^2}/{d^2}
    \right)$,
    so that
    \begin{equation}
        \left\|
        (\bra{0^{n+4}}\otimes I)
        U_A
        (\ket{0^{n+4}}\otimes I)
        -
        \frac{A}{2}
        \right\|
        \leq
        O(\epsilon_2)=\epsilon,
    \end{equation}
    where we have chosen $\epsilon_2=\Theta(\epsilon)$.
    This construction queries the sparse oracles $O(d)=O(s\log(1/\epsilon))$ times and uses $O(b+\log^{2.5}(s\log(1/\epsilon)/\epsilon))\leq O(b+\log^{2.5}(s/\epsilon))$ ancilla qubits.
    
    Now, we instantiate these oracles in the construction of $U_A$ by the quantum oracle sketching using \Cref{lem:sparse-element-oracle,lem:sparse-index-oracle} with $\epsilon_3$ error in diamond distance.
    When we instantiate the sparse index oracle construction, we interpret its leading validity bit as enlarging the matrix to $A\oplus 0$.
    The sparse element oracle is extended to return zero whenever a validity bit is one.
    Thus the standard sparse block-encoding construction applies to the padded matrix, whose restriction to the valid sector is $A$.
    This yields a random unitary channel $\mathcal{V}$ satisfying
    \begin{equation}
        \E\|\E[\mathcal{V}|z_1, \ldots, z_{t_0-1}] - \mathcal{U}_A\|_\diamond \leq O(s\log(1/\epsilon) \epsilon_3) = \epsilon,
    \end{equation}
    where we choose $\epsilon_3 = \Theta(\epsilon/(s\log(1/\epsilon)))$.
    The total number of samples needed is
    \begin{equation}
    \begin{split}
        M&=O(s\log(1/\epsilon)) \cdot O\lr{\frac{R 2^n n^2 s^3 \log^2(n/\epsilon_3)}{\epsilon_3}} \\
        &= O\lr{\frac{R2^n n^2 s^5\log^2(ns^2\log(1/\epsilon)/\epsilon)\log^2(1/\epsilon)}{\epsilon}} \\
        &\leq O\lr{\frac{RNs^5 \log^2(N)\log^4(s\log(N)/\epsilon)}{\epsilon}}.
    \end{split}
    \end{equation}
    The number of qubits needed is
    \begin{equation}
        O(b+\log^{2.5}(s/\epsilon)) + O(n+b) = O(n+b+\log^{2.5}(s/\epsilon)).
    \end{equation}
    The same guarantee holds for implementing $U_A^\dagger, cU_A$ and $cU_A^\dagger$ by reversing the evolution time or adding control in each random gate.
    This completes the proof of \Cref{lem:block-encoding}.
\end{proof}

\subsubsection{Quantum state sketching}

In addition to matrices, in many applications we also need to load in vectors of the form $\vec{b}=(b_1, \ldots, b_N)^T\in \mathbb{R}^N$ as quantum state.
For example, to solve a linear system $A\vec{x}=\vec{b}$, we need to prepare the state
\begin{equation}
    \ket{b} = \frac{1}{\|\vec{b}\|_2}\sum_j b_j \ket{j},
\end{equation}
sometimes referred to as the amplitude encoding of $\vec{b}$.
Throughout this work, we assume $\vec{b}\neq 0$.
Such vectors are usually dense.
In this section, we show that we can prepare the state corresponding to any vector $\vec{b}$ using an extension of quantum oracle sketching, which we call quantum state sketching.

Before detailing our quantum state sketching algorithm, we first explain the challenge in reaching a unified state preparation procedure for any vector $\vec{b}$.
Using the techniques from previous sections, we can build the block encoding of the diagonal matrix $\diag(\vec{b}/\|\vec{b}\|_\infty)$ using $\tilde{O}(N)$ samples.
But preparing the state $\ket{b} = \sum_{j=1}^N b_j\ket{j} / \|\vec{b}\|_2$ is still hard if we only have access to this block encoding.
To see this, consider the special case of $\vec{b}=(0, \ldots, 1, \ldots, 0)$ where the position of $1$ is arbitrary yet unknown.
Then the task of preparing $\ket{b}$ using the block encoding is equivalent to Grover's unstructured search problem, which necessarily requires querying the block encoding $\Omega(\sqrt{N})$ times.
This will result in an undesirable total sample complexity of $\tilde{O}(N\cdot (\sqrt{N})^2) = \tilde{O}(N^2)$.
We note, however, that in this case $\vec{b}$ is a computational basis state and easy to prepare directly.
This incomparability between block encoding access and state preparation access was also observed and elaborated in a recent work \cite{somma2025quantum}.

On the other hand, if $\vec{b}$ is flat (i.e., $b_j\in \{\pm 1/\sqrt{N}\}$), we can directly apply $\diag(\vec{b}/\|\vec{b}\|_\infty)$ to $\ket{+^n}$ and obtain $\diag(\vec{b}/\|\vec{b}\|_\infty)\ket{+^n} = \sqrt{N}\diag(\vec{b})\cdot \frac{1}{\sqrt{N}}\sum_{j=1}^N \ket{j} = \sum_{j=1}^N b_j\ket{j} = \ket{b}$.
This means that with one query to the block encoding and thus $\tilde{O}(N)$ samples, we can prepare the state $\ket{b}$, which is in sharp contrast to the previous case.
Therefore, it is a priori unclear if there is a unified state preparation algorithm that works for any $\vec{b}$, regardless of how flat it is.

In the following, we answer this question in the affirmative and show how to prepare the state $\ket{b}$ for any vector $\vec{b}$ using $\tilde{O}(N)$ samples from its vector data generation process.
Together with the block encoding construction in the last section, this shows that vector data access is strictly stronger than both the diagonal block encoding access and access to the state preparation unitary.

\begin{tcolorbox}
\begin{theorem}[Quantum state sketching]
\label{thm:q-state-sketch}
    Let $\epsilon\in (0, 1/2)$ and $\vec{b}=(b_1, \ldots, b_N)^T\in \mathbb{R}^N$ be any $N$-dimensional non-zero vector and $\ket{b}=\sum_{j\in [N]} b_j\ket{j} / \|\vec{b}\|_2$ be its quantum state.
    There is a classical pre-processing procedure that uses
    $M_{\mathrm{pre}}=O(RN/\epsilon)$
    samples starting at any time $t_0\geq 1$ and produces an $O(\log N \cdot \log(N/\epsilon)+b)$-bit classical record $\nu$, such that for every record $\nu$, there exists a state preparation unitary $P_\nu$ on $S=O(\log N)$ qubits and its corresponding channel $\mathcal{P}_\nu:\rho\mapsto P_\nu\rho P_\nu^\dagger$ satisfying
    \begin{equation}
        P_\nu|0^S\rangle
        =
        |0^{S-\log N}\rangle\ket{b}.
    \end{equation}
    After pre-processing, starting at any time $t\geq t_0+M_{\mathrm{pre}}$, we can use $S=O(\log N)$ qubits, $O(\log N\cdot\log(N/\epsilon)+b)$ classical bits, and
    \begin{equation}
        M
        =
        O\lr{
            \frac{
                RN\log^2(N/\epsilon)\log^4(1/\epsilon)
            }{\epsilon}
        }
    \end{equation}
    samples to implement a random unitary channel
    $\mathcal{V}_\nu$ satisfying
    \begin{equation}
    \label{eq:state-sketch-composable}
        \E_{\nu,z_1,\ldots,z_{t-1}}
        \left\|
            \E\left[
                \mathcal{V}_\nu
                \middle|
                \nu,z_1,\ldots,z_{t-1}
            \right]
            -
            \mathcal{P}_\nu
        \right\|_\diamond
        \leq
        \epsilon,
    \end{equation}
    where the inner expectation is over the samples
    $z_t,\ldots,z_{t+M-1}$ and the fresh internal
    randomness used by this query and the outer expectation is over the previously processed samples $z_1,\ldots,z_{t-1}$ and the classical record $\nu$.
    Note that the record $\nu$ can be retained and reused later in downstream computation.
    With the same retained record, the same guarantee holds for implementing $P_\nu^\dagger$ and the controlled versions $cP_\nu, cP_\nu^\dagger$.
    The data processing time per sample is $\polylog(N,1/\epsilon)$.
    For one query, the total sample complexity $M_{\mathrm{pre}}+M$ has the same asymptotic bound as $M$.
\end{theorem}
\end{tcolorbox}

Note that a state preparation unitary, as used in the downstream quantum algorithms, is valid as long as its action on a specific initial state prepares the target state $\ket{b}$.
Its action on the orthogonal complement of the Hilbert space could be arbitrary.

We motivate quantum state sketching by reexamining the above naive attempt in greater details.
In the naive algorithm, we first assemble queries to the block encoding of $\diag(\vec{b}/\|\vec{b}\|_\infty)$ by approximating $\sum_j e^{it b_j/\|\vec{b}\|_\infty} \ketbra{j}, t=O(N)$ with random multi-controlled phase gates of the form $e^{it b_j/\|\vec{b}\|_\infty \ketbra{j}}$.
From \Cref{thm:q-oracle-sketch-corr}, we know that the sample complexity of doing so is controlled by the variance of the phase accumulation.
Since the phase on each basis $\ket{j}$ accumulates independently, the sample complexity is given by the sample size needed for the basis with the largest variance, which is $O(t^2(\max b_j / \|\vec{b}\|_\infty)^2 \cdot 1/N)=O(N (\max b_j / \|\vec{b}\|_\infty)^2)$.
Then we apply the block encoding to $\ket{+^n}$ and perform amplitude amplification to get $\ket{b}$.
In doing so, we need to query the block encoding $O(\sqrt{N} \|\vec{b}\|_\infty/\|\vec{b}\|_2)$ times.
Due to the quadratic slowdown, this amounts to a total sample complexity of 
\begin{equation}
    O\lr{N\lr{\frac{\max_j b_j}{\|\vec{b}\|_\infty}}^2 \cdot \lr{\sqrt{N}\frac{\|\vec{b}\|_\infty}{\|\vec{b}\|_2}}^2} = O\lr{N^2 \lr{\frac{\max_j b_j}{\|\vec{b}\|_2}}^2}.
\end{equation}
Therefore, this algorithm only works when $\vec{b}$ is flat ($\max_j b_j/\|\vec{b}\|_2=1/\sqrt{N}$) and fails when there is a large gap between the 2-norm and infinite-norm (say when $\vec{b}$ is a computational basis, $\max_j b_j/\|\vec{b}\|_2=1$).

From this detailed analysis, we see that the normalization in the block encoding $\|\vec{b}\|_\infty$ is actually not important (as long as it gives a valid block encoding) because it eventually cancels out.
The reason that this algorithm fails is because the variance is controlled by $\max_j b_j$ rather than $\|\vec{b}\|_2$, which introduces a gap that impacts the total sample complexity.
To circumvent this issue, an ideal state preparation algorithm should have its variance controlled by $\|\vec{b}\|_2$ instead, in which case the total sample complexity amounts to $O(N)$ as desired.

The key idea in achieving this goal is to perform a Hadamard transform.
Instead of preparing the state $\ket{b}$ via block encoding, we first prepare the state $\ket{b'}$ for the Hadamard transformed vector
\begin{equation}
    \vec{b}' = H^{\otimes n} \vec{b}
\end{equation}
and then apply the quantum gates $H^{\otimes n}$ to obtain $\ket{b}$.
We prepare $\ket{b'}$ by performing amplitude amplification on $\ket{+^n}$ using the block encoding of $\diag(\vec{b}'/B)$ with some normalization factor $B$.
This time, when we assemble the block encoding, we are not using random samples of the diagonal matrix elements $\vec{b}'$ any more. 
Instead, we are using random samples of $\vec{b}$, which is related to $\vec{b}'$ via a Hadamard transform.
We will show that we can still implement the block encoding using $Z$-string rotations (rather than multi-controlled phase gates), and the resulting variance is controlled by $\|\vec{b}\|_2$ as desired.
As a result, we obtain a unified state preparation algorithm with $\tilde{O}(N)$ sample complexity for any vector.

When the data samples are no longer IID, but coming from a hierarchical data generation process with repetition number $R$, the above strategy does not work anymore.
The reasons are as follows.
As we have seen in quantum oracle sketching, there are two sources of error: the variance part and the bias part.
Due to the Hadamard transformed variance structure, the variance can blow up by a factor much larger than the repetition number $R$.
Another issue comes from the different scaling of variance ($\sim t^2R/(MN)$) and bias ($\sim tR/M$) with evolution time $t$.
We need to choose $t$ at least $\sim N$ so that the bias part does not exceed the variance part too much.
But such a large $t$ may forbid $t\diag(\vec{b}')$ from being a valid block encoding.
We need $\vec{b}'$ to be roughly flat so that we can accommodate $t\sim N$ while ensuring the validity of block encoding.

With these obstacles in mind, we introduce an additional random diagonal gate $O_{h} = \sum_{j\in [N]} (-1)^{h(j)}\ketbra{j}$ defined by a random Boolean function $h: [N]\to \bit$.
In particular, we prepare the state $\ket{b'}$ for the randomized Hadamard transformed vector
\begin{equation}
    \vec{b}' = H^{\otimes n} O_h \vec{b}.
\end{equation}
We show that instantiating $O_h$ with $O(\log N)$-wise independent functions suffice to make the whole procedure efficient and we arrive at \Cref{thm:q-state-sketch}.
Intuitively, these inserted random phases decouple different data samples to keep the variance controlled, and make $\vec{b}'$ roughly flat with high probability, thereby solving the issues caused by correlated data.

Before diving into the quantum state sketching algorithm, we first introduce the standard $k$-wise independent function based on polynomials and use it to prove some useful lemmas.

\begin{lemma}[$k$-wise independent functions with polynomials]
\label{lem:state-prep-pseudorandom}
    Let $n, k$ be positive integers.
    Let $\mathbb{F}_{2^n}$ be the finite field with $2^n$ elements.
    Let $c_0, \ldots, c_{k-1}\in \mathbb{F}_{2^n}$ be $k$ uniformly random elements of $\mathbb{F}_{2^n}$.
    We define the polynomial function
    \begin{equation}
        h'(x) = \sum_{l=0}^{k-1} c_l x^l, \quad x\in \mathbb{F}_{2^n}
    \end{equation}
    to be the $k$-wise independent function seeded by $(c_0, \ldots, c_{k-1})$.
    Then, for any $k$ distinct inputs $x_1, \ldots, x_k\in \mathbb{F}_{2^n}$, the distribution of $(h'(x_1), \ldots, h'(x_k))$ is uniform over $(\mathbb{F}_{2^n})^k$.
    Consequently, if we define $h(x)\in \bit$ to be the first bit of $h'(x)\in \mathbb{F}_{2^n}$, the distribution of $(h(x_1), \ldots, h(x_k))$ is uniform over $\bit^k$.
    Moreover, given $kn$ bits that specify $c_0, \ldots, c_{k-1}$, the phase oracle of the Boolean function $h(x)$
    \begin{equation}
        O_{h}: \ket{x} \to (-1)^{h(x)}\ket{x}
    \end{equation}
    can be implemented using an $O(n)$-qubit quantum circuit with $O(kn^2)$ two-qubit gates.
\end{lemma}

\begin{proof}[Proof of \Cref{lem:state-prep-pseudorandom}]
    Note that from Lagrange interpolation, we know that the mapping from coefficients $(c_0, \ldots, c_{k-1})\in \mathbb{F}_{2^n}$ to function values $(h'(x_1), \ldots, h'(x_k))\in \mathbb{F}_{2^n}$ is a bijection for fixed, distinct inputs $(x_1, \ldots, x_k)$.
    Therefore, $(h'(x_1), \ldots, h'(x_k))$ is uniformly distributed because $(c_0, \ldots, c_{k-1})$ is.
    Consequently, the first bits of $(h'(x_1), \ldots, h'(x_k))$ are also uniformly distributed.

    Moreover, given $kn$ bits that specify $c_0, \ldots, c_{k-1}$, we can use $O(n)$ working bits (iterative, in-place computation $z\leftarrow zx+c_l$) and $O(kn^2)$ two-bit gates ($O(n^2)$ for a single multiplication) to compute the arithmetic in the polynomial.
    Promoting the classical circuit to a quantum circuit gives the desired implementation of $O_{h}$.
\end{proof}

Using this $k$-wise independent function $h$, we prove that the vector $\vec{b}'=H^{\otimes n}O_h\vec{b}$ is roughly flat with high probability.

\begin{lemma}[Flattening with seeded randomized Hadamard transform]
\label{lem:state-prep-flatten}
    Let $n$ be a positive integer and $N=2^n$.
    Let $\delta\in (0, 1/2)$ and $k=\floor{\log(N/\delta)}$.
    Let $\vec{b}\in \mathbb{R}^N$ be any vector.
    Let $h: [N]\to \bit$ be a $(2k)$-wise independent function with a length-$(2kn)$ seed uniformly distributed over $\bit^{2kn}$.
    Then, the randomized Hadamard transformed vector
    \begin{equation}
        \vec{b}' = H^{\otimes n}O_h\vec{b}
    \end{equation}
    satisfies
    \begin{equation}
        \|\vec{b}'\|_\infty \leq \|\vec{b}\|_2\sqrt{\frac{2\log (N/\delta)}{N}}
    \end{equation}
    with probability at least $1-\delta$.
\end{lemma}

\begin{proof}[Proof of \Cref{lem:state-prep-flatten}]
    Note that the components of $\vec{b}'$ reads
    \begin{equation}
        b'_l = \sum_{j\in [N]}\lr{H^{\otimes n}}_{lj} (-1)^{h(j)} b_j = \sum_{j\in [N]} (-1)^{h(j)}\frac{1}{\sqrt{N}}(-1)^{l\cdot j} b_j,
    \end{equation}
    where $l\cdot j$ is the inner product in the binary form and $h(j)$ is the only source of randomness.
    For simplicity, we define
    \begin{equation}
        a_{lj} = \frac{1}{\sqrt{N}}(-1)^{l\cdot j}b_j
    \end{equation}
    and therefore
    \begin{equation}
        b'_l = \sum_{j\in [N]} (-1)^{h(j)}a_{lj}, \quad \sum_{j=1}^N a_{lj}^2 = \frac{1}{N}\|\vec{b}\|_2^2.
    \end{equation}
    We begin by bounding the moments of $b'_l$ using the multinomial expansion:
    \begin{equation}
    \begin{split}
        \E\left[|b'_l|^{2k}\right] &= \E\left[\lr{b'_l}^{2k}\right] \\
        &= \E\left[\sum_{s_1+\cdots+s_{N}=2k} \frac{(2k)!}{s_1!\cdots s_{N}!}\prod_{j=1}^N (-1)^{s_j h(j)}a_{lj}^{s_j}\right] \\
        &= \sum_{s_1+\cdots+s_{N}=2k}\frac{(2k)!}{s_1!\cdots s_{N}!}\E\left[\prod_{j=1}^N (-1)^{s_j h(j)}a_{lj}^{s_j}\right] \\
        &=\sum_{s_1+\cdots+s_{N}=2k}\frac{(2k)!}{s_1!\cdots s_{N}!}\E\left[\prod_{j=1}^N (-1)^{s_j r_j}\right] \prod_{j=1}^N a_{lj}^{s_j}
    \end{split}
    \end{equation}
    where we have defined a truly random bitstring $r\in \bit^N$ and used the fact that $h$ is a $(2k)$-wise independent pseudorandom function (\Cref{lem:state-prep-pseudorandom}) and that within each expectation value there is at most $2k$ $h(j)$'s.
    Next, we note that in order for the expectation value to be non-zero, the $s_j$'s must be even, and in that case the expectation value is one.
    Let $s_j=2t_j$ and we have
    \begin{equation}
    \begin{split}
        \E\left[|b'_l|^{2k}\right] &= \sum_{t_1+\cdots+t_{N}=k}\frac{(2k)!}{(2t_1)!\cdots (2t_{N})!}\prod_{j=1}^N a_{lj}^{2t_j} \leq \sum_{t_1+\cdots+t_{N}=k}\frac{(2k)!}{2^{t_1}t_1!\cdots 2^{t_{N}} t_N!}\prod_{j=1}^N a_{lj}^{2t_j} \\
        &=\frac{(2k)!}{2^k}\sum_{t_1+\cdots+t_{N}=k}\frac{1}{t_1!\cdots t_N!}\prod_{j=1}^N a_{lj}^{2t_j} =\frac{(2k)!}{2^k k!}\sum_{t_1+\cdots+t_{N}=k}\frac{k!}{t_1!\cdots t_N!}\prod_{j=1}^N \lr{a_{lj}^2}^{t_j} \\
        &=\frac{(2k)!}{2^k k!} \lr{\sum_{j=1}^N a_{lj}^2}^k =\frac{(2k)!}{2^k k!} \lr{\frac{\|\vec{b}\|_2^2}{N}}^k \leq \lr{\frac{k\|\vec{b}\|_2^2}{N}}^k,
    \end{split}
    \end{equation}
    where we have used $(2t)! = (2t)\cdots (t+1)t!\geq 2^t t!$, the multinomial expansion, $\sum_j a^2_{lj}=\|\vec{b}\|_2^2/N$, and $\frac{(2k)!}{2^k k!} = \frac{(2k)\cdots (k+1)}{2^k}\leq \frac{(2k)^k}{2^k}=k^k$.

    Now, we invoke Markov's inequality:
    \begin{equation}
        \Pr[|b'_l|\geq w] = \Pr[|b_l'|^{2k}\geq w^{2k}] \leq \frac{\E[|b'_l|^{2k}]}{w^{2k}}\leq \frac{k^k \|\vec{b}\|_2^{2k}}{N^k w^{2k}}.
    \end{equation}
    Plugging in $w=\|\vec{b}\|_2\sqrt{2\log(N/\delta)/N}$ and $k=\floor{\log(N/\delta)}$ gives us
    \begin{equation}
        \Pr\left[|b'_l|\geq \|\vec{b}\|_2\sqrt{\frac{2\log(N/\delta)}{N}}\right]\leq \frac{\delta}{N}
    \end{equation}
    for each $l\in [N]$.
    Finally, a union bound over $l\in [N]$ implies that
    \begin{equation}
        \Pr\left[\|\vec{b}'\|_\infty\geq \|\vec{b}\|_2\sqrt{\frac{2\log(N/\delta)}{N}}\right]\leq N\cdot \frac{\delta}{N}\leq \delta.
    \end{equation}
    This proves \Cref{lem:state-prep-flatten}.
\end{proof}

Now we are ready to prove \Cref{thm:q-state-sketch}.

\begin{proof}[Proof of \Cref{thm:q-state-sketch}]
    We begin with a high level overview of the quantum state sketching algorithm.
    The first step is to run a space-efficient procedure to estimate $\|\vec{b}\|_2$, which will be used later to ensure that block encodings are properly normalized.
    In particular, we take $M'$ samples $z_{t_0, \ldots, t_1-1}, t_1 = t_0+M'$ and maintain an empirical average $B$ of $b_j^2$.
    We will show that, with high probability, $B$ concentrates around $\|\vec{b}\|_2^2$.
    
    Next, to prepare $\ket{b}$, we first sample a random seed $(c_0, \ldots, c_{4k-1})$ uniformly from $\bit^{4kn}$ with $k=\floor{\log(N/\delta)}$ ($\delta$ will be chosen later) and store the random seed in the classical memory, which uses $4kn=O(n\log(N/\delta))$ bits of space.
    This specifies a $(4k)$-wise independent function $h:[N]\to\bit$ that we can efficiently compute (classically or quantumly) with $O(n)$ (qu)bits and $O(kn^2)$ gates (\Cref{lem:state-prep-pseudorandom}).

    Once $B$ and the seed specifying $h$ have been generated, we define
    the retained classical pre-processing record
    \begin{equation}
        \nu=(B,h).
    \end{equation}
    The same record $\nu$ is used in every forward, inverse, and controlled state preparation query made within the same downstream computation.
    Conditioned on $\nu$, the algorithm below is fixed.
    
    Then we draw $M_0$ samples $(j_l, b_{j_l})_{l=t_1}^{t_1+M_0-1}$ from the vector stream of $\vec{b}$.
    With each sample, we compute $h(j_l)$ on the fly and apply a gate
    \begin{equation}
        V_l = \exp[i\frac{t}{M_0}(-1)^{h(j_l)}b_{j_l}\bigotimes_{w\in \mathrm{supp}(j_l)} Z_w]
    \end{equation}
    for
    \begin{equation}
        t=\frac{N}{\sqrt{4B\log(N/\delta)}},
    \end{equation}
    (cf. \Cref{lem:state-prep-flatten}), where $\mathrm{supp}(j_l) = \{w\in [n]: (j_l)_w \neq 0\}$ is the support of the bitstring $j_l$, and $Z_w$ is the Pauli $Z$ on the $w$-th qubit.
    Note that these $V_l$ gates commute with each other.
    We will show that with $M_0 = \tilde{O}(RN/\epsilon_1)$ samples, this gives us $\epsilon_1$-approximate oracle access to the unitary
    \begin{equation}
        U = \exp[i\frac{t}{\sqrt{N}}\diag(\vec{b}')], \quad \vec{b}' = H^{\otimes n}O_h\vec{b}.
    \end{equation}
    Next, we use linear combination of unitaries and QSVT to build the block encoding of $\sqrt{\frac{N}{4B\log(N/\delta)}}\diag(\vec{b}')$.
    We then apply it to $\ket{+^n}$ and perform amplitude amplification to get $\ket{b'}$.
    Finally, we apply the quantum gates $(H^{\otimes n}O_h)^\dagger = O_h H^{\otimes n}$ and arrive at $\ket{b}$.
    
    For the error analysis, we will first bound the diamond distance between the sample-instantiated state preparation circuit and the fixed ideal unitary circuit that approximately prepares $\ket{b}$, conditioned on the retained record $\nu$.
    We then modify this fixed, ideal, but approximate unitary circuit by a small unitary rotation so that it prepares $\ket{b}$ exactly on the all-zero input, and show that this correction is small.
    This implies that the error made by the sample-instantiated circuit is small and therefore yields the full-channel guarantee in \Cref{eq:state-sketch-composable}.

    \textbf{Norm estimation.}
    To show the correctness of this algorithm, we first analyze the norm estimation step, where we maintain an empirical sum
    \begin{equation}
        B = \frac{N}{M'}\sum_{l=t_0}^{t_0+M'-1} b_{j_l}^2.
    \end{equation}
    This estimator has expectation value
    \begin{equation}
        \E[B] = N\E_j[b^2_j] = \sum_{j=1}^N b^2_j = \|\vec{b}\|_2^2,
    \end{equation}
    To bound its variance, we rearrange the terms in $B$ as 
    \begin{equation}
        B = \frac{N}{M'}\sum_{l=t_0}^{t_0+M'-1} \sum_{j=1}^N 1[j_l=j]b^2_{j} = N\sum_{j=1}^N n_j b_j^2,
    \end{equation}
    where
    \begin{equation}
        n_j = \frac{1}{M'}\sum_{l=t_0}^{t_0+M'-1}1[j_l=j]
    \end{equation}
    is the frequency of seeing a particular $j$ from time $t_0$ to $t_0+M'-1$.
    According to \Cref{lem:data-var-rep-num}, the repetition number $R$ guarantees that the variance of $n_j$ is bounded by
    \begin{equation}
        \Var[n_j]\leq \frac{R}{M'N}
    \end{equation}
    with expectation value $\E[n_j]=1/N$.
    Then, the variance of $B$ is bounded by
    \begin{equation}
        \Var[B] = N^2 \sum_{j,j'=1}^N \Cov[n_j, n_{j'}] b^2_{j} b^2_{j'} \leq N^2 \sum_{j,j'=1}^N \sqrt{\Var[n_j]\Var[n_{j'}]} b^2_{j} b^2_{j'} \leq N^2 \frac{R}{M'N} \sum_{j, j'}^N b^2_{j} b^2_{j'} = \frac{RN}{M'} \|\vec{b}\|_2^4.
    \end{equation}
    Markov's inequality then implies that
    \begin{equation}
        \Pr\left[|B-\|\vec{b}\|_2^2|\geq \frac{1}{2}\|\vec{b}\|_2^2\right] = \Pr\left[|B-\E[B]|^2\geq \frac{1}{4}\|\vec{b}\|_2^4\right] \leq \frac{\Var[B]}{\|\vec{b}\|_2^4/4}\leq \frac{RN\|\vec{b}\|_2^4}{M'\|\vec{b}\|_2^4/4} = \frac{4RN}{M'}.
    \end{equation}
    We take $M'=4RN/\delta$, we have that with probability at least $1-\delta$,
    \begin{equation}
        \frac{1}{2}\|\vec{b}\|_2^2\leq B \leq \frac{3}{2}\|\vec{b}\|_2^2.
    \end{equation}
    In the following, we assume that this condition is satisfied and take the failure probability 
    \begin{equation}
        \Pr[\text{fail on norm estimation}]\leq \delta
    \end{equation}
    into account at the end.
    Note that here this failure probability is not conditioned on and therefore is averaged over previously processed data samples $z_{<t_0} = (z_1, \ldots, z_{t_0-1})$.
    We use the above bounds on $B$ only on the norm-estimation success
    event and account for its failure probability at the end.
    We do not condition the underlying data distribution on this event.
    In every conditional channel considered below, the complete data
    history before the current sample block is conditioned upon.
    Hence, the realized value of $B$, and therefore the value of $t$ that will be chosen to depend on $B$, are fixed within each conditional channel.
    We suppress this dependence on $B$ from the notation when no confusion can arise.
    To make the implemented circuit well-defined on every outcome of the pre-processing, if $B=0$ we set $t=0$ and use the corresponding fixed fallback unitary for the remainder of the construction.
    Since $\vec{b}\neq 0$, the event $B=0$ is contained in the
    norm-estimation failure event and will be covered by the failure probability above.

    \textbf{Quantum oracle sketching.}
    Next, we explain why the $V_l$ gates can be used to build the block encoding of $\diag(\vec{b}')$.
    The components of the two vectors $\vec{b}, \vec{b}'$ are connected via the following relation:
    \begin{equation}
        b'_u = \sum_j \lr{H^{\otimes n}}_{uj} \lr{O_h}_{jj} b_j = \sum_j \frac{1}{\sqrt{N}}(-1)^{u\cdot j} (-1)^{h(j)} b_j,
    \end{equation}
    where $u\cdot j$ is the bit-wise inner product of the bitstring representations of $u, j\in [N] \simeq \bit^n$.
    Note that for any basis state $\ket{u}, u\in\bit^n$,
    \begin{equation}
        \bigotimes_{w\in \mathrm{supp}(j_l)}Z_w \ket{u} = \prod_{w\in \mathrm{supp}(j_l)}(-1)^{u_w} \ket{u} = \prod_{w\in [n]} (-1)^{(j_{l})_w u_w}\ket{u} = (-1)^{j_l\cdot u}\ket{u}.
    \end{equation}
    Therefore,
    \begin{equation}
        \bigotimes_{w\in \mathrm{supp}(j_l)}Z_w = \sum_{u}(-1)^{j_l\cdot u} \ketbra{u}.
    \end{equation}
    This means that the gate $V_l$ that we apply for each sample can be rewritten as $V_l = \exp[i H_l]$ with the Hamiltonian
    \begin{equation}
    \begin{split}
        H_l &= \frac{t}{M_0}(-1)^{h(j_l)}b_{j_l}\sum_u (-1)^{j_l\cdot u}\ketbra{u} \\
        &= \frac{t}{M_0} \sum_u (-1)^{h(j_l)}b_{j_l} (-1)^{j_l\cdot u}\ketbra{u}.
    \end{split}
    \end{equation}
    The point of this rewriting is to show that the expectation value of this Hamiltonian is
    \begin{equation}
    \begin{split}
        \E_{j_l\sim \unif([N])}[H_l] &= \frac{1}{N}\sum_{j} \lr{\frac{t}{M_0} \sum_u (-1)^{h(j)} b_j (-1)^{j\cdot u} \ketbra{u}} \\
        &=\frac{t}{\sqrt{N} M_0} \sum_u \ketbra{u} \lr{\sum_{j} \frac{1}{\sqrt{N}}(-1)^{h(j)} b_j (-1)^{j\cdot u}} \\
        &=\frac{t}{\sqrt{N} M_0} \sum_u \ketbra{u} b'_u \\
        &=\frac{t}{\sqrt{N} M_0} \diag(\vec{b}'),
    \end{split}
    \end{equation}
    where we used $b'_u = \sum_j \frac{1}{\sqrt{N}} (-1)^{h(j)}b_j(-1)^{j\cdot u}$ and the expectation is taken with respect to the fixed marginal distribution $j_l\sim\unif([N])$, while holding the realized values of $B$ and $h$ fixed.
    This matches the exponent in the target unitary
    \begin{equation}
        U
        =
        \exp\left[
            i\frac{t}{\sqrt{N}}\diag(\vec{b}')
        \right],
    \end{equation}
    as defined above.
    
    Now we invoke quantum oracle sketching to show that the gate sequence we apply $V_{t_1+M_0-1}\cdots V_{t_1}$ approximates $U$ in expectation.
    For simplicity, we introduce the following notations
    \begin{equation}
        V = V_{t_1+M_0-1}\cdots V_{t_1} = \exp[i\sum_{l=t_1}^{t_1+M_0-1} H_l] = \exp[i X].
    \end{equation}
    Where the random matrix $X$ can be decomposed in the computational basis $\ket{u}, u\in [N]$ as
    \begin{equation}
    \begin{split}
        X &= \frac{t}{M_0}\sum_{l=t_1}^{t_1+M_0-1}  \sum_u \ketbra{u} (-1)^{h(j_l)} b_{j_l} (-1)^{j_l\cdot u} \\
        &=t \sum_u \ketbra{u} \frac{1}{M_0}\sum_{l=t_1}^{t_1+M_0-1} (-1)^{h(j_l)} b_{j_l} (-1)^{j_l\cdot u}\\
        &= t\sum_u \ketbra{u} m_u,
    \end{split}
    \end{equation}
    where we have defined
    \begin{equation}
        m_u = \frac{1}{M_0}\sum_{l=t_1}^{t_1+M_0-1} (-1)^{h(j_l)} b_{j_l} (-1)^{j_l\cdot u}
    \end{equation}
    to be the empirical phase accumulation on the basis $\ket{u}$.
    The approximation error can be decomposed into a variance part and a conditional bias part using
    \Cref{lem:diamond-operator-expect}:
    \begin{equation}
    \begin{split}
        &\E_{h,z_{<t_1}}
        \left\|
            \E[
                V(\cdot)V^\dagger
                \mid h,z_{<t_1}
            ]
            -
            U(\cdot)U^\dagger
        \right\|_\diamond
        \leq
        4\E_{h,z_{<t_1}}
        \left\|
            \E[e^{iX}\mid h,z_{<t_1}]
            -
            U
        \right\|\\
        &\leq
        4\E_{h,z_{<t_1}}
        \left\|
            \E[e^{iX}\mid h,z_{<t_1}]
            -
            e^{i\E[X\mid h,z_{<t_1}]}
        \right\|+
        4\E_{h,z_{<t_1}}
        \left\|
            e^{i\E[X\mid h,z_{<t_1}]}
            -
            U
        \right\|\\
        &\leq
        \underbrace{
        2\E_{h,z_{<t_1}}
        \left[
            t^2\max_u
            \Var[m_u\mid h,z_{<t_1}]
        \right]}_{\mathrm{variance}}+
        \underbrace{
        4\E_{h,z_{<t_1}}
        \left[
            t\max_u
            \left|
                \E[m_u\mid h,z_{<t_1}]
                -
                \frac{b'_u}{\sqrt{N}}
            \right|
        \right]}_{\mathrm{bias}}.
    \end{split}
    \end{equation}

    \textbf{Variance bound.}
    Now we upper bound the variance part.
    For any $u$, we first rewrite $m_u$ as 
    \begin{equation}
        m_u = \frac{1}{M_0}\sum_{l=t_1}^{t_1+M_0-1} (-1)^{h(j_l)+j_l\cdot u}b_{j_l} = \sum_{j\in [N]}(-1)^{h(j)+j\cdot u}b_j m_j, \quad m_j = \frac{1}{M_0}\sum_{l=t_1}^{t_1+M_0-1}\delta_{j_l, j}.
    \end{equation}
    For simplicity, we define a vector $v_u\in \{\pm 1\}^N$ with components $v_{uj} = (-1)^{h(j)+j\cdot u}$ and therefore $m_u = \sum_j v_{uj} b_j m_j$.
    Then we have
    \begin{equation}
        \Var[m_u|h, z_{<t_1}] = \sum_{j, j'\in[N]} v_{uj} v_{uj'} b_j b_{j'} \Cov[m_j, m_{j'}|h, z_{<t_1}] = v_u^T \Sigma v_u,
    \end{equation}
    where the matrix $\Sigma=\Sigma^T\in\mathbb{R}^{N\times N}$ has matrix elements
    \begin{equation}
        \Sigma_{jj'} = b_j b_{j'} \Cov[m_j, m_{j'}|h, z_{<t_1}].
    \end{equation}
    Note that from \Cref{lem:data-var-rep-num}, the matrix $\Sigma$ has trace
    \begin{equation}
        \tr(\Sigma) = \sum_j b_j^2 \Var[m_j|h, z_{<t_1}], \quad \E_{z_{<t_1}}[\tr(\Sigma)] \leq \sum_j b_j^2 \Var[m_j|h]\leq \|\vec{b}\|_2^2 \frac{R}{M_0N},
    \end{equation}
    and norms
    \begin{equation}
        \|\Sigma\|\leq \tr(\Sigma), \quad \|\Sigma\|_F = \sqrt{\tr(\Sigma^2)}\leq \sqrt{\|\Sigma\|\tr(\Sigma)}\leq \tr(\Sigma).
    \end{equation}
    This allows us to invoke the moment bound for quadratic forms in \Cref{lem:moment-quad-form} (the randomness is over $v_{uj}=(-1)^{h(j)+j\cdot u}$ where $h:[N]\to \bit$ is a $(4k)$-wise independent function):
    \begin{equation}
        \E_h\left[|\Var[m_u|h, z_{<t_1}]-\tr(\Sigma)|^{2k}\right]\leq \lr{C\max\left\{\sqrt{2k}\|\Sigma\|_F, 2k\|\Sigma\|\right\}}^{2k} \leq \lr{2kC\tr(\Sigma)}^{2k}
    \end{equation}
    for some universal constant $C>0$.
    This is because the $(2k)$-th moment only involves $(4k)$-order terms of $h$ and therefore the $(4k)$-wise independence of $h$ ensures that moment is the same as when $h$ is truly random, which is given by \Cref{lem:moment-quad-form}.
    For every fixed data history $z_{<t_1}$, the matrix $\Sigma$ is fixed with respect to the randomness of $h$, because $h$ is sampled independently of the data stream.
    Using $\max_u a_u\leq(\sum_u a_u^{2k})^{1/(2k)}$ and Jensen's inequality, we obtain
    \begin{equation}
    \begin{split}
        &\E_h\left[
            \max_u
            \left|
                \Var[m_u\mid h,z_{<t_1}]
                -
                \tr(\Sigma)
            \right|
        \right]\leq
        \E_h\left[
            \left(
                \sum_{u=1}^N
                \left|
                    \Var[m_u\mid h,z_{<t_1}]
                    -
                    \tr(\Sigma)
                \right|^{2k}
            \right)^{1/(2k)}
        \right]\\
        &\leq
        \left(
            \sum_{u=1}^N
            \E_h\left[
                \left|
                    \Var[m_u\mid h,z_{<t_1}]
                    -
                    \tr(\Sigma)
                \right|^{2k}
            \right]
        \right)^{1/(2k)}
        \leq
        N^{1/(2k)}\,2kC\tr(\Sigma)
        \leq
        4kC\tr(\Sigma),
    \end{split}
    \end{equation}
    where we have used $N^{1/(2k)}\leq 2$ for sufficiently large $N$.
    Therefore,
    \begin{equation}
        \E_h\left[
            \max_u
            \Var[m_u\mid h,z_{<t_1}]
        \right]
        \leq
        (1+4kC)\tr(\Sigma).
    \end{equation}

    On the norm-estimation success event, we have
    \begin{equation}
        t^2
        =
        \frac{N^2}{4B\log(N/\delta)}
        \leq
        \frac{N^2}
        {2\|\vec{b}\|_2^2\log(N/\delta)}.
    \end{equation}
    Hence, using $\E_{z_{<t_1}}[\tr(\Sigma)]\leq\frac{\|\vec{b}\|_2^2R}{M_0N}$, the contribution from the norm-estimation success event satisfies
    \begin{equation}
        2\E_{h,z_{<t_1}}\left[
            1\left[
                \frac{1}{2}\|\vec{b}\|_2^2
                \leq B
                \leq
                \frac{3}{2}\|\vec{b}\|_2^2
            \right]
            t^2
            \max_u
            \Var[m_u\mid h,z_{<t_1}]
        \right]\leq
        \frac{RN}{M_0\log(N/\delta)}
        (1+4kC)\leq
        \frac{RN}{M_0}(1+4C),
    \end{equation}
    where we have used
    $k\leq\log(N/\delta)$.

    \textbf{Bias bound.}
    Next, for any $u$, we express the bias term as
    \begin{equation}
    \begin{split}
        \underbrace{4t\E_{z_{<t_1}}\max_u\left|\E[m_{u}|h, z_{<t_1}] - \frac{b'_u}{\sqrt{N}}\right|}_{\mathrm{bias}}&= 4t\E_{z_{<t_1}}\max_u\left|\sum_{j\in [N]} (-1)^{h(j)+j\cdot u}b_j(\E[m_{j}|z_{<t_1}] - 1/N)\right| \\
        &= 4t\E_{z_{<t_1}}\max_u\left|\sum_{j\in [N]} (-1)^{h(j)+j\cdot u}b_j\Delta_j\right|,
    \end{split}
    \end{equation}
    where we have defined the random variable
    \begin{equation}
        \Delta_j = \E[m_{j}|z_{<t_1}] - 1/N.
    \end{equation}
    The $2k$-th moment over the randomness of $h$ can be bounded by \Cref{lem:moment-inner-prod} as
    \begin{equation}
        \E_h\left[\left|\sum_{j\in [N]} (-1)^{h(j)+j\cdot u}b_j\Delta_j\right|^{2k}\right]\leq \left(C'^2k\sum_{j=1}^Nb_j^2\Delta_j^2\right)^{k}\leq (\max_j |\Delta_j|C'k \|\vec{b}\|_2)^{2k}
    \end{equation}
    for some universal constant $C>0$, where we have used the fact that $h$ is $(4k)$-wise independent.
    Note also that we have
    \begin{equation}
        \E_h\left[\sum_{j\in [N]} (-1)^{h(j)+j\cdot u}b_j\Delta_j\right] = \sum_{j\in [N]} \E_h\left[(-1)^{h(j)+j\cdot u}\right]b_j\Delta_j = 0.
    \end{equation}
    The same bound that converts maximal values to moments as as used in the variance bound gives us
    \begin{equation}
    \begin{split}
        &\E_h\left[
            \max_u
            \left|
                \sum_{j\in[N]}
                (-1)^{h(j)+j\cdot u}
                b_j\Delta_j
            \right|
        \right]\leq
        \left(
            \sum_{u=1}^N
            \E_h\left[
                \left|
                    \sum_{j\in[N]}
                    (-1)^{h(j)+j\cdot u}
                    b_j\Delta_j
                \right|^{2k}
            \right]
        \right)^{1/(2k)}\\
        &\quad\leq
        N^{1/(2k)}
        C'k\|\vec{b}\|_2
        \max_j|\Delta_j|\leq
        2C'k\|\vec{b}\|_2
        \max_j|\Delta_j|.
    \end{split}
    \end{equation}
    On the norm-estimation success event,
    \begin{equation}
        t\|\vec{b}\|_2
        =
        \frac{N\|\vec{b}\|_2}
        {\sqrt{4B\log(N/\delta)}}
        \leq
        \frac{N}{\sqrt{2\log(N/\delta)}}.
    \end{equation}
    Consequently, we have
    \begin{equation}
    \begin{split}
        &4\E_{h,z_{<t_1}}\left[
            1\left[
                \frac{1}{2}\|\vec{b}\|_2^2
                \leq B
                \leq
                \frac{3}{2}\|\vec{b}\|_2^2
            \right]
            t
            \max_u
            \left|
                \E[m_u\mid h,z_{<t_1}]
                -
                \frac{b'_u}{\sqrt{N}}
            \right|
        \right]\\
        &\leq
        8C'k
        \frac{N}{\sqrt{2\log(N/\delta)}}
        \E_{z_{<t_1}}\left[
            \max_j|\Delta_j|
        \right]\leq
        8C'k
        \frac{N}{\sqrt{2\log(N/\delta)}}
        \frac{R}{M_0}\leq
        4\sqrt{2}C'
        \frac{RN\sqrt{\log(N/\delta)}}{M_0},
    \end{split}
    \end{equation}
    where we used
    \Cref{lem:data-cond-drift-rep-num} and
    $k\leq\log(N/\delta)$.

    \textbf{Variance and bias combined.}
    Combining the preceding two bounds, the contribution from the norm-estimation success event satisfies
    \begin{equation}
    \begin{split}
        &\E_{h,z_{<t_1}}\left[
            1\left[
                \frac{1}{2}\|\vec{b}\|_2^2
                \leq B
                \leq
                \frac{3}{2}\|\vec{b}\|_2^2
            \right]
            \left\|
                \E[
                    V(\cdot)V^\dagger
                    \mid h,z_{<t_1}
                ]
                -
                U(\cdot)U^\dagger
            \right\|_\diamond
        \right]\\
        &\quad\leq
        \frac{RN}{M_0}
        \left(
            1+4C
            +
            4\sqrt{2}C'
            \sqrt{\log(N/\delta)}
        \right).
    \end{split}
    \end{equation}
    We choose the deterministic sample size
    \begin{equation}
        M_0
        =
        \left\lceil
            \frac{
                C''RN\sqrt{\log(N/\delta)}
            }{\epsilon_1}
        \right\rceil
    \end{equation}
    for a sufficiently large universal constant $C''$.
    Then,
    \begin{equation}
    \begin{split}
        &\E_{h,z_{<t_1}}\left[
            1\left[
                \frac{1}{2}\|\vec{b}\|_2^2
                \leq B
                \leq
                \frac{3}{2}\|\vec{b}\|_2^2
            \right]
            \left\|
                \E[
                    V(\cdot)V^\dagger
                    \mid h,z_{<t_1}
                ]
                -
                U(\cdot)U^\dagger
            \right\|_\diamond
        \right]
        \leq
        \epsilon_1.
    \end{split}
    \end{equation}
    The same expected conditional diamond-distance guarantee holds for
    implementing $U^\dagger,cU,cU^\dagger$ by setting $t\to -t$ or
    adding control.

    \textbf{Block encoding.}
    Now, we use this $\epsilon_1$-approximation to $U=\exp(it\diag(\vec{b}')/\sqrt{N})$ to implement a block encoding of $\sqrt{\frac{N}{4B\log(N/\delta)}}\diag(\vec{b}')$.
    Thanks to \Cref{lem:error-accumulation-time-varying}, even though we have correlated data, the error still accumulates additively.
    Note that $\sqrt{\frac{N}{4B\log(N/\delta)}}\diag(\vec{b}')$ can be block encoded without further normalization with probability at least $1-\delta$ over $h$, because its norm is bounded by
    \begin{equation}
        \left\|\sqrt{\frac{N}{4B\log(N/\delta)}}\diag(\vec{b}')\right\| \leq \sqrt{\frac{N}{4B\log(N/\delta)}}\|\vec{b}'\|_\infty\leq \sqrt{\frac{N}{2\|\vec{b}\|_2^2\log(N/\delta)}} \|\vec{b}\|_2\sqrt{\frac{2\log(N/\delta)}{N}} = 1,
    \end{equation}
    where we have used $B\geq \frac{1}{2}\|\vec{b}\|_2^2$ and the fact that $\vec{b}'$ is flat with high probability (\Cref{lem:state-prep-flatten}).
    To implement the block encoding, recall that we have set 
    \begin{equation}
        t=\frac{N}{\sqrt{4B\log(N/\delta)}}.
    \end{equation}
        For brevity, define
    \begin{equation}
        D_\nu
        =
        \sqrt{
            \frac{N}{
                4B\log(N/\delta)
            }
        }
        \diag(\vec{b}').
    \end{equation}
    On the norm-estimation and flattening success event, we have
    $\|D_\nu\|\leq 1$.

    We use linear combination of unitaries to implement a block encoding of $\sin(D_\nu)$.
    This requires $O(1)$ queries to $cU,cU^\dagger$.
    Next, we use QSVT with an odd real polynomial, bounded by one on $[-1,1]$, that approximates $x\to \arcsin(x)/2$ on $[-\sin(1),\sin(1)]$.
    This costs $Q=O\left(\log(1/\epsilon_2)\right)$ queries to $cU,cU^\dagger$.
    Let $G_\nu$ denote the resulting ideal QSVT unitary and let $B_\nu$ denote its encoded block.
    By choosing the internal polynomial-approximation accuracy to be a sufficiently small universal constant multiple of $\epsilon_2$, we have $\|B_\nu-{D_\nu}/{2}\|\leq O(\epsilon_2)$.
    By \Cref{lem:exact-block-completion}, there exists a unitary $\hat{G}_\nu$ that exactly block encodes $D_\nu/2$ and satisfies $\|G_\nu-\hat{G}_\nu\|=O(\epsilon_2)$.
    The same bound holds for the inverse and controlled versions.

    The sample-instantiated QSVT circuit differs from
    $\widehat{G}_\nu$ by the sum of the oracle-instantiation error and
    the exact-completion error.
    By \Cref{lem:error-accumulation-time-varying}, its conditional
    expected diamond-norm error is at most $O(Q\epsilon_1+\epsilon_2)$.
    We choose $\epsilon_1=\Theta(\epsilon_2/Q)$ and decrease the hidden universal constants sufficiently so that
    this error is at most $\epsilon_2$.
    The same guarantee holds for the inverse and controlled versions.
    The number of samples used is
    \begin{equation}
        M_1=
        QM_0=
        O\left(
            \frac{
                RN\sqrt{\log(N/\delta)}
            }{\epsilon_2}
            Q^2
        \right)=
        O\lr{
            \frac{
                RN\log^{1/2}(N/\delta)
            }{\epsilon_2}
            \log^2(1/\epsilon_2)
        }.
    \end{equation}

    \textbf{Amplitude amplification.}
    We use the exact block encoding $\hat{G}_\nu$ only to define the fixed ideal circuit used in the analysis.
    Since $\hat{G}_\nu$ exactly block encodes $D_\nu/2$, applying its encoded block to $\ket{+^n}$ gives
    \begin{equation}
        \frac{D_\nu}{2}\ket{+^n}
        =
        \frac{1}{
            2\sqrt{4B\log(N/\delta)}
        }
        \sum_{j=1}^N
        b'_j\ket{j}=
        \frac{
            \|\vec{b}\|_2
        }{
            2\sqrt{4B\log(N/\delta)}
        }
        \ket{b'}.
    \end{equation}
    Thus the success probability is
    \begin{equation}
        \frac{
            \|\vec{b}\|_2^2
        }{
            16B\log(N/\delta)
        }.
    \end{equation}

    To prepare $\ket{b'}$ deterministically, we employ standard fixed-point amplitude amplification \cite{martyn2021grand} against the state $\ket{+^n}$, with $\epsilon_3$ error.
    Finally, we apply the gate $O_hH^{\otimes n}$ and obtain
    approximately the state $\ket{b}$.
    This is possible because we store the seed of $h$ in classical memory, and $O_h$ can be efficiently implemented on $O(n)$ qubits with $O(kn^2)=O(n^2\log(N/\delta))$ gates.
    Note that fixed-point amplitude amplification only requires a known lower bound on the initial success amplitude.
    On the norm-estimation success event,
    \begin{equation}
        \frac{
            \|\vec{b}\|_2
        }{
            2\sqrt{4B\log(N/\delta)}
        }
        \geq
        \frac{1}{
            2\sqrt{6\log(N/\delta)}
        }.
    \end{equation}
    We thus choose the fixed-point amplitude-amplification phase sequence using this lower bound, without knowing $\|\vec{b}\|_2$.
    In particular, conditioned on the retained record $\nu$ and the success event of norm-estimation and flattening, let $W_\nu$ denote the resulting complete ideal unitary circuit, using $\hat{G}_\nu$ and its inverse and controlled versions as its block-encoding queries.
    The amplitude-amplification part of $W_\nu$ uses
    \begin{equation}
        Q'=
        O\left(
            \frac{
                \log(1/\epsilon_3)
            }{
                \|\vec{b}\|_2/
                \bigl(
                    2\sqrt{4B\log(N/\delta)}
                \bigr)
            }
        \right)=
        O\left(
            \log^{1/2}(N/\delta)
            \log(1/\epsilon_3)
        \right)
    \end{equation}
    queries.
    For records on which either norm estimation or flattening fails, we choose an arbitrary exact state-preparation unitary $W_\nu$ satisfying $W_\nu|0^S\rangle=|0^{S-\log N}\rangle|b\rangle$.
    All constructions involving $\hat G_\nu$ below are used only on the norm-estimation and flattening success event.

    We fix the standard coherent phase convention for fixed-point amplitude amplification and include its known deterministic output-phase correction in $W_\nu$.
    Thus the state-preparation guarantee is phase-sensitive and holds in Euclidean norm.
    We set $\epsilon_2={\epsilon_3}/{Q'}$.
    Since $W_\nu$ uses an exact block encoding of $D_\nu/2$, its only state-preparation error is the fixed-point amplitude-amplification error.
    In particular, whenever the norm-estimation and flattening
    conditions hold,
    \begin{equation}
    \label{eq:ideal-state-prep-output}
        \left\|
            W_\nu|0^S\rangle
            -
            |0^{S-\log N}\rangle\ket{b}
        \right\|_2
        \leq
        \epsilon_3
        \leq
        2\epsilon_3.
    \end{equation}

    Now we instantiate the oracle queries with samples.
    Let $V_\nu$ denote the complete sampled unitary circuit produced by the algorithm for every retained record $\nu$.
    On the norm-estimation and flattening success event, $V_\nu$ is obtained from $W_\nu$ by replacing every ideal query to $\hat{G}_\nu$, its inverse, and its controlled versions by the corresponding sample-instantiated circuit.
    We use
    $\mathcal{W}_\nu(\rho)=W_\nu\rho W_\nu^\dagger, \mathcal{V}_\nu(\rho)=V_\nu\rho V_\nu^\dagger$
    to denote the corresponding channels.
    Applying \Cref{lem:error-accumulation-time-varying} over all block-encoding queries and averaging over the retained record gives
    \begin{equation}
        \E_{\nu,z_{<t}}\left[
            1[\text{norm estimation and flattening succeed}]
            \left\|
                \E\left[
                    \mathcal{V}_\nu
                    \middle|
                    \nu,z_{<t}
                \right]
                -
                \mathcal{W}_\nu
            \right\|_\diamond
        \right]\leq
        2Q'\epsilon_2
        =
        2\epsilon_3.
    \end{equation}
    The required number of fresh samples for this query is
    \begin{equation}
    \begin{split}
        M
        &=
        Q'M_1=
        O\lr{
            \frac{
                RN\log^{3/2}(N/\delta)
                \log^2(
                    \log^{1/2}(N/\delta)
                    \log(1/\epsilon_3)/\epsilon_3
                )
                \log^2(1/\epsilon_3)
            }{\epsilon_3}
        }\\
        &\leq
        O\lr{
            \frac{
                RN\log^2(N/\delta)\log^4(1/\epsilon_3)
            }{\epsilon_3}
        }.
    \end{split}
    \end{equation}

    The only remaining failure events are norm estimation and flattening.
    By a union bound, their total probability is at most $2\delta$.
    We now convert the approximate state preparation unitary $W_\nu$ into an exact state preparation unitary for the purpose of defining the fixed target channel.
    Note that for every record $\nu$, there exists a unitary $R_\nu$, acting non-trivially only on the subspace spanned by
    $W_\nu|0^S\rangle$ and $|0^{S-\log N}\rangle\ket{b}$, such that
    \begin{equation}
        R_\nu W_\nu|0^S\rangle
        =
        |0^{S-\log N}\rangle\ket{b}.
    \end{equation}
    Moreover, whenever
    \Cref{eq:ideal-state-prep-output} holds, this two-dimensional correction can be chosen to satisfy
    \begin{equation}
        \|R_\nu-I\|
        \leq
        2
        \left\|
            W_\nu|0^S\rangle-|0^{S-\log N}\rangle\ket{b}
        \right\|_2
        \leq
        4\epsilon_3.
    \end{equation}

    We define the exact state preparation unitary
    \begin{equation}
        P_\nu
        =
        R_\nu W_\nu,
        \qquad
        \mathcal{P}_\nu(\rho)
        =
        P_\nu\rho P_\nu^\dagger.
    \end{equation}
    By construction,
    \begin{equation}
        P_\nu|0^S\rangle
        =
        |0^{S-\log N}\rangle\ket{b}
    \end{equation}
    exactly for every $\nu$.
    Note that this rotation $R_\nu$ is only a proof device that is not implemented as circuit.

    Whenever the preceding error analysis applies,
    \Cref{lem:diamond-operator} gives
    \begin{equation}
        \|\mathcal{W}_\nu-\mathcal{P}_\nu\|_\diamond
        \leq
        2\|W_\nu-P_\nu\|
        =
        2\|I-R_\nu\|
        \leq
        8\epsilon_3.
    \end{equation}
    Using the triangle inequality and the trivial bound $2$ on the diamond distance in the failure event, we obtain
    \begin{equation}
        \E_{\nu,z_{<t}}
        \left\|
            \E\left[
                \mathcal{V}_\nu
                \middle|
                \nu,z_{<t}
            \right]
            -
            \mathcal{P}_\nu
        \right\|_\diamond\leq
        2\epsilon_3
        +
        8\epsilon_3
        +
        2\delta\cdot 2=
        10\epsilon_3+4\delta
        \leq
        \epsilon,
    \end{equation}
    where we set $\epsilon_3=\delta=\epsilon/15$.
    This concludes the error analysis.

    The pre-processing sample complexity is
    \begin{equation}
        M_{\mathrm{pre}}
        =
        M'
        =
        \frac{4RN}{\delta}
        =
        O\lr{\frac{RN}{\epsilon}},
    \end{equation}
    while each subsequent query uses
    \begin{equation}
        M
        =
        O\lr{
            \frac{
                RN\log^2(N/\epsilon)\log^4(1/\epsilon)
            }{\epsilon}
        },
    \end{equation}
    as claimed.

    For the inverse and controlled versions, we retain the same record $\nu$.
    A fresh sampled implementation of the inverse is constructed by reversing the fixed circuit architecture and replacing each inverse oracle call by its fresh sample-built implementation.
    The controlled implementations are constructed by adding control to every gate.
    Their ideal reference circuits are respectively $W_\nu^\dagger, cW_\nu, cW_\nu^\dagger$.
    The same gate-by-gate hybrid argument gives the corresponding sampled-versus-ideal diamond-distance bounds.
    Moreover, $\|P_\nu^\dagger-W_\nu^\dagger\| = \|R_\nu-I\|, \|cP_\nu-cW_\nu\|=\|P_\nu-W_\nu\|$, and the analogous identity holds for $cP_\nu^\dagger$ and $cW_\nu^\dagger$.
    Therefore, the same expected conditional diamond-distance guarantee holds for $P_\nu^\dagger, cP_\nu, cP_\nu^\dagger$ with the same retained pre-processing record $\nu$.
    This concludes the proof of \Cref{thm:q-state-sketch}.
\end{proof}

\newpage

\section{Classical hardness}
\label{sec:cl-hard}

\begin{figure}
    \centering
    \includegraphics[width=1\linewidth]{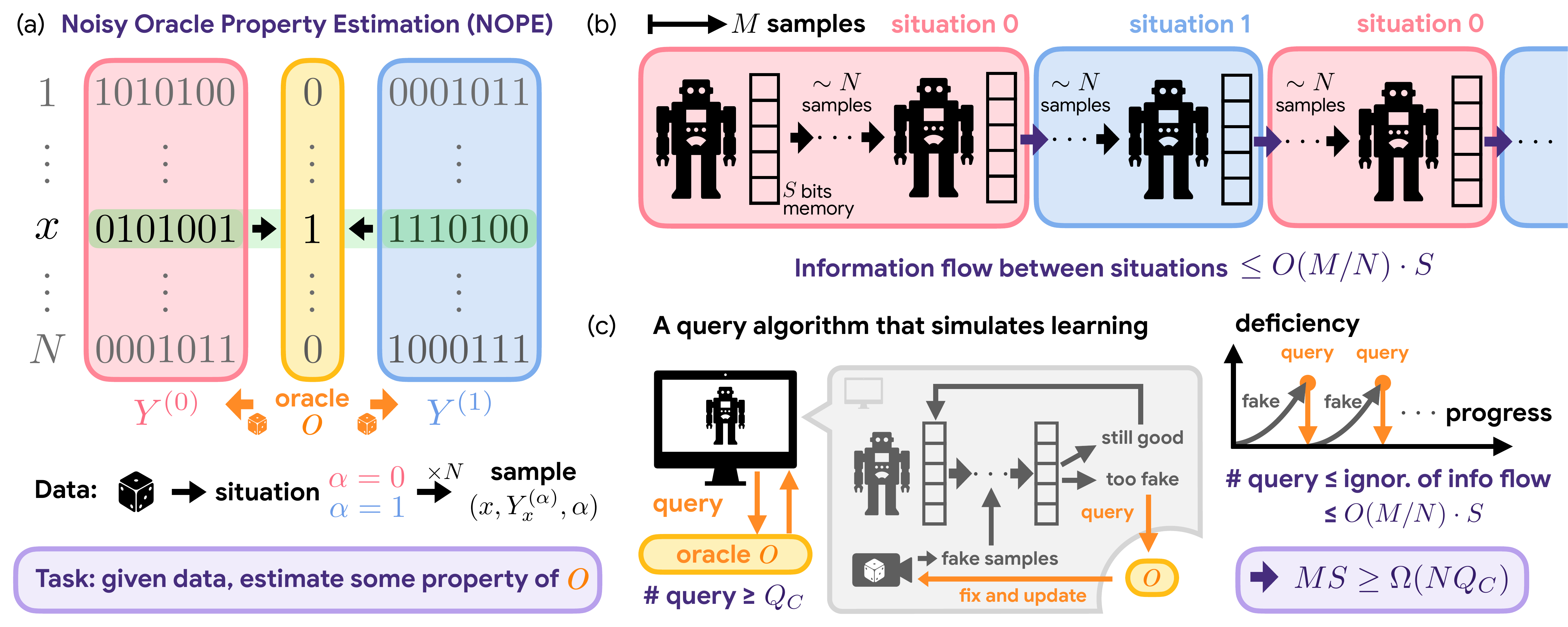}
    \caption[Overview of the classical hardness proof strategy.]{\textbf{Overview of the classical hardness proof strategy.}
    \textbf{(a)} Illustration of the Noisy Oracle Property Estimation (NOPE) task.
    We encode the truth table of an oracle $o\in \bit^N$ as noisy encodings $Y^{(0)}, Y^{(1)}$.
    The task is to estimate some property of the oracle $o$ using random query data of the noisy encoding $(x, Y^{(\alpha)}_x, \alpha)$ that depends on the current situation $\alpha$ which changes with time scale $N$.
    \textbf{(b)} Visualization of the information flow during the learning process of a classical learning algorithm that has memory size $S$ and sample complexity $M$.
    The information flow between the two situations is bounded by the number of situation changes $O(M/N)$ times $S$, the communicated number of bits per situation change.
    \textbf{(c)} Reduction from query algorithms to learning algorithms.
    We construct a query algorithm that calculates the desired oracle property by simulating a learning algorithm that solves the corresponding NOPE task.
    The query algorithm fakes data samples to feed into the learning algorithm and occasionally queries the oracle to update its data forging strategy.
    The number of queries it makes is lower bounded by the classical query complexity $Q_C$, while upper bounded by its ignorance of the information flow within the learning algorithm.
    This leads to the sample-space lower bound $MS\geq \Omega(NQ_C)$.
    }
    \label{fig:hardness}
\end{figure}

In this section, we develop the machinery for proving classical hardness in classical data processing tasks, summarized in \Cref{fig:hardness}.
We will define a learning task and prove that any classical machine that solves this task with a reasonable amount of samples must have exponentially large size, and moreover if it has slightly smaller size, it will need super-polynomially more samples.
Later in \Cref{sec:app}, we will reduce various application tasks (e.g., solving linear systems, classification, dimension reduction, etc.) to this learning task and therefore show the classical hardness of these applications.

In particular, we focus on a specific family of learning tasks called Noisy Oracle Property Estimation (NOPE).
It can be viewed as a noisy learning version of any oracle query problem.
We introduce NOPE and its data generation process in \Cref{sec:cl-hard-noisy-oracle-property-est}.
In \Cref{sec:cl-hard-comp-model}, we provide a brief recap of the formal model of classical learning algorithms that we defined in \Cref{sec:model} and will prove hardness against.

In \Cref{sec:cl-hard-sample-space-lb}, we prove a sample-space lower bound that shows any classical machine solving NOPE with a reasonable amount of data must have a size lower bounded by the classical query complexity of the oracle property estimation task, which is usually exponential.

Then, in \Cref{sec:cl-hard-bootstrap}, we build upon this sample-space lower bound and bootstrap it by adding one more time scale into the task.
In particular, we define a new task called dynamic NOPE, where the oracle changes dynamically over time and yet the property that we want to estimate remains fixed.
This allows us to prove that any classical machine with a slightly smaller size must need a super-polynomially larger sample size to solve the task.

Finally, in \Cref{sec:cl-hard-app}, we specialize NOPE tasks to the query problem of Forrelation, and prove several useful lemmas that connect NOPE to the various applications.
These results will be used in \Cref{sec:app} to reduce application tasks to NOPE and dynamic NOPE, whose classical hardness is already established in this section.
 
\subsection{Noisy oracle property estimation}
\label{sec:cl-hard-noisy-oracle-property-est}

We begin by formally introduce the Noisy Oracle Property Estimation (NOPE) task.
Intuitively, the task is to estimate some property of an oracle (i.e., a Boolean function) 
\begin{equation}
    o(x) \in \bit, \quad x\in [N],
\end{equation}
based on noisy data samples generated from it.
The oracle $o$ can be equivalently represented by its truth table 
\begin{equation}
    o \in \bit^N
\end{equation}
where $o_x = o(x), \forall x\in [N]$.
Suppose the property that we are interested in is binary and is specified by a property function
\begin{equation}
    f: \mathcal{O} \to \bit, \quad \mathcal{O}\subseteq \bit^N,
\end{equation}
that takes an oracle $o\in \bit^N$ as input and outputs its property $f(o)\in \bit$.
Here, this property function may be a partial function, meaning that it may be defined only for a subset of possible oracles $\mathcal{O}\subseteq \bit^N$.
One can equivalently think of this property function $f$ as a query algorithm that queries the oracle $o$ and computes the property $f(o)$.

In a learning task, we do not have direct knowledge of the underlying oracle $o$ and cannot make queries as we wish.
Instead, we have access to a sequence of noisy data samples $z_i$ generated according to $o$, based on which we need to estimate the property $f(o)$.
In general, the data generation process may have correlation with multiple time scales, as described by hierarchical data generation processes introduced in \Cref{sec:data-access}.
A good learning algorithm should be able to process the data samples $z_i$ and output the desired property $f(o)$ with high probability, for any oracle $o$.

To define NOPE, we consider a noisy hierarchical data generation process.
We first introduce our noise model by defining a \emph{noisy encoding} $(Y^{(0)}, Y^{(1)})$ of the oracle $o$ according to a noisy encoding function $g$.
Let $b$ be a positive integer specifying the encoding length. 
We define the noisy encoding function
\begin{equation}
    g: \bit^b \times \bit^b \to \bit
\end{equation}
to be a function that obfuscates the oracle value $o_x \in \{0, 1\}$ (i.e., the output of $g$) by providing a bipartite length-$2b$ encoding of it (i.e., the two input arguments of $g$).
This encoding function $g$ is noisy (or obfuscating), in the sense that we require $g$ to have a low discrepancy $\disc(g)\leq 2^{-\eta b}$ for some constant $\eta>0$.
This condition will be formally defined later (\Cref{def:disc}), and it intuitively means that a partial input of $g$ (e.g., its first argument) cannot reveal too much information about the value of $g$.
An example of a noisy encoding function is the inner product function.
We use 
\begin{equation}
    G=g^N: \bit^{N\times b}\times \bit^{N\times b}\to \bit^N
\end{equation}
to denote the parallel application of $g$ across all $x\in [N]$.
When $G$ maps some pair of inputs $(Y^{(0)}, Y^{(1)})$ into the oracle $o\in \bit^N$, we say that the input pair $(Y^{(0)}, Y^{(1)})$ is a noisy encoding of $o$.

Now we specify the data generation process.
Given an oracle $o\in \bit^N$, we sample a noisy encoding 
\begin{equation}
    (Y^{(0)}, Y^{(1)}) \sim \unif(G^{-1}(o))
\end{equation}
uniformly random, where $G^{-1}(o)\subseteq \bit^{N\times b}\times \bit^{N\times b}$ is the set of noisy encodings of the oracle and $Y^{(0)}, Y^{(1)}\in \bit^{N\times b}$.
This noisy encoding $(Y^{(0)}, Y^{(1)})$ is fixed from now on and is used to generate the data samples.
We consider the following hierarchical data generation process (defined in \Cref{sec:data-access})
\begin{equation}
    \mathcal{D}_{g, T}^N(o) = (\mathcal{D}^0 \to \mathcal{D}^1_{\alpha}\to^{\times T}z).
\end{equation}
where the data sample generated at each time point depends on the current binary situation $\alpha\in \bit$ sampled from $\mathcal{D}^0 = \mathrm{Bern}(1/2)$.
The situation changes slowly in time and only has a single time scale $T$.
For each situation $\alpha\in \bit$, we define $\mathcal{D}^1_\alpha$ to sample data of the form
\begin{equation}
    z=(x, y, \alpha), \quad x \sim \unif([N]), \quad y = Y^{(\alpha)}_{x} \in \bit^b
\end{equation}
where $x$ is a random query to the oracle and $y$ is a noisy encoding of the queries oracle value that depends on the situation $\alpha$.
We choose the time scale $T\geq \Omega(N)$ to avoid the situation changing too rapidly so that the learning algorithm does not have the time to gather enough information about the current situation $\alpha$ (i.e., about the noisy encoding $Y^{(\alpha)}$).
The notation $\mathcal{D}^N_{g, T}(o)$ highlights the input size $N$, noisy encoding map $g$, and time scale $T$ of this data generation process.

With the noisy data generation process $\mathcal{D}^N_{g, T}(o)$, we define the Noisy Oracle Property Estimation (NOPE) task as follows.

\begin{tcolorbox}
\begin{task}[Noisy Oracle Property Estimation (NOPE)]
\label{task:nope}
    Let $N, T, b$ be positive integers.
    Let $f: \mathcal{O} \to \bit$ be a function that specifies the target property for a set of possible oracles $\mathcal{O}\subseteq\bit^N$.
    Let $g: \bit^b \times \bit^b \to \bit$ be a noisy encoding function.
    The task of Noisy Oracle Property Estimation is to calculate the desired property $f(o)$ using data samples generated from the noisy hierarchical data generation process $\mathcal{D}^N_{g, T}(o)$ with some success probability for any $o\in \mathcal{O}$.
\end{task}
\end{tcolorbox}

In the following, we will prove classical hardness of NOPE.
In particular, we will show sample and space lower bounds for any classical learning algorithms that solves NOPE with some success probability.

\subsection{Recap of classical learning algorithms}
\label{sec:cl-hard-comp-model}

Before going into the classical hardness proofs, we first give a brief recap on the definition of classical learning algorithms that we will prove hardness for.
More details can be found in \Cref{sec:cl-learning-alg}.

Recall that in our definition of classical learning algorithms, there are two key resources that we keep track of: the size or space complexity of the classical machine $S$ and the sample complexity $M$.
We use $\mathcal{I}$ to denote the set of all possible inputs to the learning algorithm at each time step.
For NOPE defined above, $\mathcal{I} = [N]\times \bit^b\times \bit$, since the data is of the form $z=(x, y, \alpha)$.
Throughout \Cref{sec:cl-hard-sample-space-lb}, we assume such $\mathcal{I}$ unless otherwise stated.
It will change in \Cref{sec:cl-hard-bootstrap} when we introduce the dynamic version of NOPE.

A classical learning algorithm $\mathcal{L}$ with size $S$, sample complexity $M$, and input form $\mathcal{I}$ is defined as a directed graph with vertices arranged in $M+1$ layers ($0$ to $M$).
Each layer consists of at most $2^S$ vertices, each labeled by $S$ bits.
There is only one vertex called the root in layer $0$.
In layer $M$, each vertex $v$ has no outgoing edges and is called a leaf.
Each leaf $v$ is attached with an output $h_v$.
For each layer $i=0, \ldots, M-1$, the outgoing edge from each vertex in layer $i$ only goes to vertices in layer $i+1$.
Each vertex has $|\mathcal{I}|$ outgoing edges, labeled by the possible input data at this time step.
Upon receiving a sequence of data $I_i \in \mathcal{I}, i=0, \ldots, M-1$, the algorithm starts from the root, follows the edge given by each data point $I_i$ in layer $i$ until reaching a leaf $v$ in layer $M$, and outputs $h_v$.

To keep track of the information flow during learning, we define the \emph{transcript} $\pi_\mathcal{L}(I, \alpha)$ of the learning algorithm $\mathcal{L}$ upon receiving the data sequence $I = (I_0, \ldots, I_{M-1}) \in \mathcal{I}^M$ with respect to a situation record $\alpha = (\alpha_0, \ldots, \alpha_{M-1})$ to be the concatenation of the length-$S$ bitstrings that label the vertices traversed by the computation path at layers $i$ where the situation changes $\alpha_i\neq \alpha_{i+1}$, followed by the output of $\mathcal{L}$.
For our data generation process, suppose that the situation changes in total $r$ times and the output is a single bit $h_v\in \bit$, the transcript $\pi_\mathcal{L}(I, \alpha)$ is a bitstring with total length $|\pi_\mathcal{L}(I, \alpha)|=rS+1$.
Intuitively, it records a history of the state of the learning algorithm when situation changes.

Since any randomized algorithm can always be regarded as first sampling all the random numbers and then execute the corresponding deterministic algorithm, the deterministic definition above suffices.
Our definition of classical learning algorithms resemble the notion of branching programs.
They are non-uniform models of space-bounded computation, more general than uniform ones such as online Turing machines with bounded space.
Therefore, the classical hardness results we prove also applies to other weaker computational models.

\subsection{Sample-space lower bound}
\label{sec:cl-hard-sample-space-lb}

In this section, we prove a sample-space lower bound for any classical learning algorithms solving NOPE.
The key quantity that will appear in the bound is the classical query complexity $Q_C$ of the property function $f(o)$.
This relates space complexity $S$ to query complexity $Q_C$.
We will show that the higher $Q_C$ is, the larger the sample-space product $MS$ has to be to solve NOPE.

\begin{tcolorbox}
\begin{theorem}[Classical sample-space lower bound]
\label{thm:classical-lower-bound}
    Let $N$ be a large integer. 
    Let the time scale $T$ be a positive integer.
    Let $\eta\in (0, 2]$ be a constant and $c=(865/\eta^2)\log(865/\eta^2)$.
    Let $g:\bit^b\times \bit^b\to \bit$ be a noisy encoding function with encoding length $b\geq c\log N$ and discrepancy $\disc(g)\leq 2^{-\eta b}$.
    Let $f: \mathcal{O}\to \bit, \mathcal{O}\subseteq \bit^N$ be a function that specifies the target property with $\delta$-error classical randomized query complexity $Q_C^\delta$ larger than a sufficiently large constant.
    Then, any randomized classical learning algorithm $\mathcal{L}$ that solves NOPE with probability at least $1-\delta + 2^{-\eta b/8}$ must satisfy
    \begin{equation}
        MS\geq \Omega(Q_C^\delta Tb),
    \end{equation}
    where $M$ is the sample complexity of $\mathcal{L}$ and $S$ is the space complexity of $\mathcal{L}$.
\end{theorem}
\end{tcolorbox}

We prove this with a simulation argument \cite{chattopadhyay2021query}.
The intuition is as follows.
We construct a query algorithm $\mathcal{A}$ that simulates the learning algorithm $\mathcal{L}$ by randomly guessing appropriate data and feed them into the learning algorithm.
Intuitively, since the data generation process is sufficiently obfuscating, randomly guessing the data should be good enough even when we do not make queries to obtain any information about the underlying oracle.
But as the algorithm proceeds, we have forged more fake data, and the state of the simulated learning algorithm gradually drifts away from the correct state.
When the difference is significant enough, we query the oracle to fix this issue.
We will show that with this strategy, the query algorithm correctly outputs $f(o)$ with high probability, while the total query $Q$ it makes is bounded by the transcript length of the learning algorithm.
On the other hand, any query algorithm satisfies $Q\geq Q_C$.
Therefore, we have
\begin{equation}
    Q_C\leq Q \leq \tilde{O}(|\pi_\mathcal{L}(I, \alpha)|)=\tilde{O}(MS/T).
\end{equation}

In the following, we first prove some useful lemmas in \Cref{sec:cl-hard-simulation-prelim} that will be used in the simulation argument.
Then, we formalize the simulation argument in \Cref{sec:cl-hard-simulation} and detail the constructed query algorithm in \Cref{alg:simulation}.
In \Cref{sec:cl-hard-sim-correctness,sec:cl-hard-sim-complex}, we prove the correctness and the query complexity bound of the constructed query algorithm.
Finally, in \Cref{sec:cl-hard-sim-combine}, we combine these ingredients to prove \Cref{thm:classical-lower-bound}.

\subsubsection{Preliminaries}
\label{sec:cl-hard-simulation-prelim}

We use the notion of min-entropy and density \cite{goos2016rectangles,goos2020query,chattopadhyay2021query} to characterize how uniform the distribution of a random string is.
Intuitively, a random string is dense if every segment of it looks roughly uniform.

\begin{definition}[Min-entropy]
    Let $Y\in \mathcal{Y}$ be a discrete random variable. 
    The min-entropy of $Y$, denoted as $H_\infty(Y)$, is defined as
    \begin{equation}
        H_\infty(Y) = \max\{k\in \mathbb{R}: \Pr[Y=y]\leq 2^{-k}, \forall y\in \mathcal{Y}\}.
    \end{equation} 
\end{definition}

\begin{definition}[Dense random variable]
\label{def:dense}
    Let $Y\in \bit^{N\times b}$ be a random variable. 
    Let $\delta\in (0, 1]$. 
    We say $Y$ is $\delta$-dense if for every $I\subseteq [N]$, it holds that $H_\infty(Y_I)\geq \delta b|I|$.
\end{definition}

We can further keep track of finer structures of a dense random variable by the following notion of excess entropy \cite{chattopadhyay2021query}.
It is non-negative because $\Pr[Y_I=y_I]\leq 2^{-\delta b|I|}$ for any $I\subseteq [N]$, by definition of $\delta$-dense random variables.

\begin{definition}[Excess entropy]
\label{def:excess-ent}
    Let $Y\in \bit^{N\times b}$ be a $\delta$-dense random variable. 
    Let $I\subseteq [N]$ be a subset of coordinates. 
    For every $y_I\in \bit^{|I|\times b}$, we use $e_{y_I}$ to denote the non-negative number that satisfies
    \begin{equation}
        \Pr[Y_I=y_I]=2^{-\delta b|I|-e_{y_I}}.
    \end{equation}
\end{definition}

The following are some useful properties of min-entropy.

\begin{lemma}[Conditional min-entropy]
\label{lem:cond-min-entropy}
    Let $Y$ be a discrete random variable and let $\mathcal{E}$ be an event. Then,
    \begin{equation}
        H_\infty(Y|\mathcal{E}) \geq H_\infty(Y) - \log\frac{1}{\Pr[\mathcal{E}]}.
    \end{equation}
\end{lemma}

\begin{lemma}[Min-entropy of marginal]
\label{lem:marginal-min-entropy}
    Let $Y_1\in \mathcal{Y}_1, Y_2\in \mathcal{Y}_2$ be two discrete random variables. We have
    \begin{equation}
        H_\infty(Y_1)\geq H_\infty(Y_1, Y_2)-\log|\mathcal{Y}_2|.
    \end{equation}
\end{lemma}

We use discrepancy to describe how obfuscating the noisy encoding function $g: \bit^b\times\bit^b\to\bit$ is.

\begin{definition}[Discrepancy]
\label{def:disc}
    Let $\Lambda$ be a finite set. Let $g: \Lambda\times\Lambda \to \{0, 1\}$ be a function. Let $U, V$ be independent random variables uniformly sampled from $\Lambda$. Given a rectangle $R = R_1\times R_2\subseteq \Lambda\times\Lambda$, the discrepancy of $g$ with respect to $R$, denoted as $\mathrm{disc}_R(g)$, is
    \begin{equation}
        \mathrm{disc}_R(g) = |\Pr[(U, V)\in R, g(U, V)=0]-\Pr[(U, V)\in R, g(U, V)=1]|. 
    \end{equation} 
    The discrepancy of $g$, denoted as $\mathrm{disc}(g)$, is the maximum of $\mathrm{disc}_R(g)$ over all rectangles $R = R_1\times R_2\subseteq \Lambda\times\Lambda$.
\end{definition}

For any Boolean random variable $B$, we use $\mathrm{bias}(B) = |\Pr[B=0]-\Pr[B=1]|$ to denote its bias.
Given a low discrepancy map $g$, one would expect that its output is not very biased.
So if the input variables $Y^{(0)}, Y^{(1)}$ are roughly uniformly distributed, we expect the output to be roughly uniform as well.
And the more uniform $Y^{(0)}, Y^{(1)}$ are, the more uniform $g(Y^{(0)}, Y^{(1)})$ will be.
The tolerance of how biased $Y^{(0)}, Y^{(1)}$ can be such that the above holds is limited by the discrepancy of $g$.
This idea is formalized in the following lemma.

\begin{lemma}[Low discrepancy maps preserve uniformity, {\cite[Lemma 2.9]{chattopadhyay2021query}}]
\label{lem:disc-unif}
    Let $\Lambda$ be a finite set. 
    Let $g: \Lambda\times\Lambda\to \{0, 1\}$ be a function with discrepancy $\mathrm{disc}(g)\leq |\Lambda|^{-\eta}$, where $\eta>0$. 
    For any $0<\lambda\leq \eta$, let $Y^{(0)}, Y^{(1)}$ be independent random variables on $\Lambda$ with 
    \begin{equation}
        H_\infty(Y^{(0)})+H_\infty(Y^{(1)})\geq (2-\eta+\lambda)\log|\Lambda|.
    \end{equation} 
    Then
    \begin{equation}
        \mathrm{bias}(g(Y^{(0)}, Y^{(1)}))\leq |\Lambda|^{-\lambda}.
    \end{equation} 
\end{lemma}

\begin{proof}[Proof of \Cref{lem:disc-unif}]
    Let $p_u=\Pr[Y^{(0)}=u]$ and $q_v=\Pr[Y^{(1)}=v]$, and let $p_{\max}=\max_u p_u$ and $q_{\max}=\max_v q_v$.
    The min-entropy assumption implies that
    \begin{equation}
        p_{\max}q_{\max}
        =2^{-H_\infty(Y^{(0)})-H_\infty(Y^{(1)})}
        \leq |\Lambda|^{-(2-\eta+\lambda)}.
    \end{equation}
    Using $p_u=\int_0^{p_{\max}}1[p_u\geq s]\,ds$ and $q_v=\int_0^{q_{\max}}1[q_v\geq t]\,dt$, we have
    \begin{equation}
    \begin{split}
        \mathrm{bias}(g(Y^{(0)},Y^{(1)}))
        &=\left|\sum_{u,v\in\Lambda}p_uq_v(-1)^{g(u,v)}\right|
        \leq \int_0^{p_{\max}}\int_0^{q_{\max}}
        \left|\sum_{\substack{u:p_u\geq s\\v:q_v\geq t}}
        (-1)^{g(u,v)}\right|\,dt\,ds\\
        &\leq p_{\max}q_{\max}|\Lambda|^2\mathrm{disc}(g)\leq |\Lambda|^{-(2-\eta+\lambda)}|\Lambda|^2|\Lambda|^{-\eta}
        =|\Lambda|^{-\lambda},
    \end{split}
    \end{equation}
    where we have used independence, the triangle inequality, and the definition of discrepancy applied to the rectangle $\{u:p_u\geq s\}\times\{v:q_v\geq t\}$.
    This proves \Cref{lem:disc-unif}.
\end{proof}

One can also formalize this property in terms of sampling.
When $Y^{(0)}, Y^{(1)}$ are sufficiently uniform, if we sample an $Y^{(0)}=y$, then the bit $g(y, Y^{(1)})$ is roughly unbiased.

\begin{lemma}[Low discrepancy maps with random partial input preserve uniformity, {\cite[Lemma 2.10]{chattopadhyay2021query}}]
\label{lem:disc-samp}
    Let $\Lambda$ be a finite set.
    Let $g: \Lambda\times\Lambda\to \{0, 1\}$ be a function with discrepancy $\mathrm{disc}(g)\leq |\Lambda|^{-\eta}$, where $\eta>0$. 
    For any $\lambda, \gamma>0$ with $\lambda+\gamma\leq \eta$, let $Y^{(0)}, Y^{(1)}$ be independent random variables on $\Lambda$ with 
    \begin{equation}
        H_\infty(Y^{(0)})+H_\infty(Y^{(1)})\geq \left(2+\frac{1}{\log|\Lambda|}-\eta+\lambda+\gamma\right)\log|\Lambda|.
    \end{equation} 
    Then, the probability that $Y^{(0)}$ takes a value $y\in \Lambda$ such that
    \begin{equation}
        \mathrm{bias}(g(y, Y^{(1)}))> |\Lambda|^{-\lambda}
    \end{equation} 
    is less than $|\Lambda|^{-\gamma}$.
\end{lemma}

\begin{proof}[Proof of \Cref{lem:disc-samp}]
For every \(y \in \Lambda\), denote $p_y = \Pr\bigl[g(y,Y^{(1)}) = 1\bigr]$.
Our goal is to prove
\begin{equation}
  \Pr_{Y^{(0)}}[|p_{Y^{(0)}}-\tfrac12|>\tfrac12|\Lambda|^{-\lambda}]<|\Lambda|^{-\gamma}.
\end{equation}
We first show $\Pr[p_{Y^{(0)}} > \tfrac12+\tfrac12|\Lambda|^{-\lambda}]\leq |\Lambda|^{-\gamma}/2$; a symmetric argument implies $\Pr[p_{Y^{(0)}} < \tfrac12-\tfrac12|\Lambda|^{-\lambda}]\leq |\Lambda|^{-\gamma}/2$.
Then a union bound completes the proof.

Let $\mathcal{E}=\{y\in \Lambda: p_y>\frac12+\frac12 |\Lambda|^{-\lambda}\}$.
In other words, $\Pr[g(Y^{(0)}, Y^{(1)})=1|Y^{(0)}\in \mathcal{E}]>\frac12+\frac12 |\Lambda|^{-\lambda}$.
Assume, for the sake of contradiction, that $\Pr[Y^{(0)}\in\mathcal{E}]\geq \tfrac12|\Lambda|^{-\gamma}.$
This implies
\begin{equation}
\begin{split}
    H_\infty(Y^{(0)}|Y^{(0)}\in \mathcal{E})+H_\infty(Y^{(1)})&\geq H_\infty(Y^{(0)})-\log\frac{1}{\Pr[Y^{(0)}\in \mathcal{E}]}+H_\infty(Y^{(1)}) \\
    &\geq \left(2+\frac{1}{\log|\Lambda|}-\eta+\lambda+\gamma\right)\log|\Lambda| - (1+\gamma \log|\Lambda|) \\
    &=(2-\eta+\lambda)\log|\Lambda|.
\end{split}
\end{equation}
By \Cref{lem:disc-unif}, we have $\Pr[g(Y^{(0)}, Y^{(1)})=1|Y^{(0)}\in \mathcal{E}]\leq \frac12+\frac12 |\Lambda|^{-\lambda}$, which contradicts the definition of $\mathcal{E}$.
\end{proof}

Sometimes it is easier to work with the parity of the output of $g$ over some coordinates.
For any $I\subseteq [N]$, we define $g^{\oplus I}(y^0, y^1) \in \bit$ to be the parity of $g^I(y^0, y^1)$.
The discrepancy of $g^{\oplus I}$ is bounded by the discrepancy of $g$.

\begin{lemma}[Discrepancy of parity, \cite{lee2008direct}, {\cite[Theorem 2.11]{chattopadhyay2021query}}]
\label{lem:disc-parity}
    For any $I\subseteq [N]$, we have
    \begin{equation}
        (\disc(g))^{|I|}\leq \disc(g^{\oplus I})\leq (2^6\disc(g))^{|I|}.
    \end{equation}
\end{lemma}

This immediately implies properties of $g^{\oplus I}$ that are similar to \Cref{lem:disc-unif,lem:disc-samp}.

\begin{corollary}
\label{cor:disc-unif-parity}
    Let $\Lambda$ be a finite set. 
    Let $I\subseteq [N]$.
    Let $g: \Lambda\times\Lambda\to \{0, 1\}$ be a function with discrepancy $\mathrm{disc}(g)\leq |\Lambda|^{-\eta}$, where $\eta>0$. 
    For any $0<\lambda\leq \eta$, let $Y^{(0)}, Y^{(1)}$ be independent random variables on $\Lambda^I$ with 
    \begin{equation}
        H_\infty(Y^{(0)})+H_\infty(Y^{(1)})\geq \left(2+\frac{6}{\log|\Lambda|}-\eta+\lambda\right)|I|\log|\Lambda|.
    \end{equation} 
    Then
    \begin{equation}
        \mathrm{bias}(g^{\oplus I}(Y^{(0)}, Y^{(1)}))\leq |\Lambda|^{-\lambda|I|}. 
    \end{equation} 
\end{corollary}

\begin{corollary}
\label{cor:disc-samp-parity}
    Let $\Lambda$ be a finite set. 
    Let $I\subseteq [N]$.
    Let $g: \Lambda\times\Lambda\to \{0, 1\}$ be a function with discrepancy $\mathrm{disc}(g)\leq |\Lambda|^{-\eta}$, where $\eta>0$. 
    For any $\lambda, \gamma>0$ with $\lambda+\gamma\leq \eta$, let $Y^{(0)}, Y^{(1)}$ be independent random variables on $\Lambda^I$ with 
    \begin{equation}
        H_\infty(Y^{(0)})+H_\infty(Y^{(1)})\geq \left(2+\frac{7}{\log|\Lambda|}-\eta+\lambda+\gamma\right)|I|\log|\Lambda|.
    \end{equation}
    Then, the probability that $Y^{(0)}$ takes a value $y\in \Lambda$ such that
    \begin{equation}
        \mathrm{bias}(g^{\oplus I}(y, Y^{(1)}))> |\Lambda|^{-\lambda|I|} 
    \end{equation} 
    is less than $|\Lambda|^{-\gamma|I|}$.
\end{corollary}

The bias of parity is the Fourier coefficient of the distribution of random bitstrings.
In particular, for a random bitstring $Z\in \bit^m$, let $\mu: \bit^m\to \mathbb{R}$ be is probability mass function.
We consider its Fourier transform $\mu(z) = \sum_{S\subseteq [m]}\hat{\mu}(S)\chi_S(z)$, where the character $\chi_S(z) = (-1)^{\oplus_{i\in S} z_i}$ and the Fourier coefficient $\hat{\mu}(S) = \frac{1}{2^m}\sum_{z\in \bit^m}\mu(z)\chi_S(z)$.
We define $\oplus_{i\in S}z_i=0$ if $S=\emptyset$, which implies $\hat{\mu}(\emptyset) = 2^{-m}$.
We have the following lemma.

\begin{lemma}[Fourier coefficient and parity bias]
\label{lem:fourier-parity}
    Let $Z\in \bit^m$ be a random variable with probability mass function $\mu: \bit^m\to \mathbb{R}$.
    Then for any $S\subseteq [m]$, we have
    \begin{equation}
        |\hat{\mu}(S)| = \frac{1}{2^m}\mathrm{bias}(\bigoplus_{i\in S}Z_i).
    \end{equation}
\end{lemma}

\begin{proof}[Proof of \Cref{lem:fourier-parity}]
    We have $|\hat{\mu}(S)| = \frac{1}{2^m}|\sum_z \mu(z) (-1)^{\oplus_{i\in S} z_i}| = \frac{1}{2^m}|\sum_{z:\oplus_{i\in S} z_i=0}  \mu(z) - \sum_{z:\oplus_{i\in S} z_i=1}  \mu(z)| = \frac{1}{2^m}|\Pr[\oplus_{i\in S} z_i=0] - \Pr[\oplus_{i\in S} z_i=1]| = \frac{1}{2^m}\mathrm{bias}(\bigoplus_{i\in S}Z_i)$.
\end{proof}

The following lemma shows that a rapidly decaying bias in the parity of subsets of coordinates (i.e., rapidly decaying Fourier coefficients) implies a nearly uniform distribution.

\begin{lemma}[Variant of Vazirani's lemma, \cite{goos2016rectangles}, {\cite[Lemma 2.5]{chattopadhyay2021query}}]
\label{lem:vazirani}
    Let $\epsilon>0$ and let $Z\in \{0, 1\}^m$ be a random variable. If for every non-empty set $S\subseteq [m]$ we have
    \begin{equation}
        \mathrm{bias}(\oplus_{i\in S}Z_i) = |\Pr[\oplus_{i\in S}Z_i=0]-\Pr[\oplus_{i\in S}Z_i=1]|\leq \epsilon (2m)^{-|S|},
    \end{equation} 
    then for every $z\in \{0, 1\}^m$,
    \begin{equation}
        (1-\epsilon)\frac{1}{2^m}\leq \Pr[Z=z]\leq (1+\epsilon)\frac{1}{2^m}.
    \end{equation} 
\end{lemma}

\begin{proof}[Proof of \Cref{lem:vazirani}]
    Let $\mu(z) = \Pr[Z=z]$.
    Using \Cref{lem:fourier-parity}, we have
    \begin{equation}
    \begin{split}
        |\mu(z)-2^{-m}| &= \left|\sum_{S\subseteq[m], S\neq \emptyset} \hat{\mu}(S)(-1)^{\oplus_{i\in S}z_i}\right| \\
        &\leq \sum_{S\subseteq[m], S\neq \emptyset} \left|\hat{\mu}(S)\right| \\
        & = 2^{-m} \sum_{S\subseteq[m], S\neq \emptyset} \mathrm{bias}(\bigoplus_{i\in S}Z_i) \\
        &\leq 2^{-m}\sum_{S\subseteq[m], S\neq \emptyset}\epsilon(2m)^{-|S|} \\
        &=2^{-m} \sum_{i=1}^m \binom{m}{i}\epsilon(2m)^{-i} \\
        &\leq 2^{-m} \sum_{i=1}^m m^i\epsilon(2m)^{-i} \\
        &=\epsilon 2^{-m} \sum_{i=1}^m 2^{-i}\leq \epsilon 2^{-m}.
    \end{split}
    \end{equation}
    Therefore, $(1-\epsilon)\frac{1}{2^m}\leq \mu(z)\leq (1+\epsilon)\frac{1}{2^m}$.
\end{proof} 

The above lemma can also be formulated in terms of min-entropy.

\begin{lemma}[Variant of Vazirani's lemma in min-entropy, {\cite[Lemma 2.6]{chattopadhyay2021query}}]
\label{lem:vazirani-ent}
    Let $t>0$ and let $Z\in \{0, 1\}^m$ be a random variable. If for every set $S\subseteq [m]$ with $|S|\geq t$ we have 
    \begin{equation}
        \mathrm{bias}(\oplus_{i\in S}Z_i) = |\Pr[\oplus_{i\in S}Z_i=0]-\Pr[\oplus_{i\in S}Z_i=1]|\leq (2m)^{-|S|}, 
    \end{equation} 
    then
    \begin{equation}
        H_\infty(Z)\geq m-t\log m-1. 
    \end{equation} 
\end{lemma}

\begin{proof}[Proof of \Cref{lem:vazirani-ent}]
    The result holds trivially if $m=1$.
    Suppose $m\geq 2$.
    Let $\mu(z) = \Pr[Z=z]$.
    By \Cref{lem:fourier-parity}, we have
    \begin{equation}
    \begin{split}
        \mu(z)&\leq \sum_{S\subseteq[m]}|\hat{\mu}(S)| = 2^{-m}\sum_{S\subseteq[m]}\mathrm{bias}(\bigoplus_{i\in S}Z_i)\\
        &=2^{-m}\sum_{S\subseteq[m]: |S|<t}\mathrm{bias}(\bigoplus_{i\in S}Z_i) + 2^{-m}\sum_{S\subseteq[m]: |S|\geq t}\mathrm{bias}(\bigoplus_{i\in S}Z_i).
    \end{split}
    \end{equation}
    Since $t>0$, every set $S$ with $|S|\geq t$ is non-empty, and hence
    \begin{equation}
        2^{-m}\sum_{S\subseteq[m]: |S|\geq t}\mathrm{bias}(\bigoplus_{i\in S}Z_i) \leq 2^{-m}\sum_{S\subseteq[m]: |S|\geq t}(2m)^{-|S|} \leq 2^{-m}\sum_{i=1}^m \binom{m}{i}(2m)^{-i} \leq 2^{-m}\sum_{i=1}^m m^i(2m)^{-i} \leq 2^{-m}.
    \end{equation}
    For $|S|<t$, let $r=\ceil{t}-1<t$. Then
    \begin{equation}
        2^{-m}\sum_{S\subseteq[m]: |S|<t}
        \mathrm{bias}\left(\bigoplus_{i\in S}Z_i\right)
        \leq
        2^{-m}\sum_{i=0}^r\binom{m}{i}
        \leq
        2^{-m}\sum_{i=0}^r m^i.
    \end{equation}
    Therefore, for $m\geq2$,
    \begin{equation}
        \mu(z)
        \leq
        2^{-m}\left(1+\sum_{i=0}^r m^i\right)
        \leq
        2^{-m}\,2m^r
        \leq
        2^{-m+1}m^t,
    \end{equation}
    and hence $H_\infty(Z)\geq m-t\log m-1$ as desired.
\end{proof}

We also need the following definition to fix certain coordinates of our estimate of the oracle $o\in \bit^N$ when we query the true oracle.

\begin{definition}[Restriction]
    A restriction $\rho$ is a string in $\{0, 1, *\}^N$. We say that a coordinate $i\in[N]$ is free in $\rho$ if $\rho_i=*$, and otherwise we say it's fixed. 
    Given a restriction $\rho$ we use $\mathrm{free}(\rho)$ and $\mathrm{fix}(\rho)$ to denote the free and fixed coordinates in $\rho$.
    We say that a string $o\in \{0, 1\}^N$ is consistent with a restriction $\rho$ if $o_{\mathrm{fix}(\rho)}=\rho_{\mathrm{fix}(\rho)}$.
\end{definition}

Now we define a property that we want to keep during the simulation of the learning algorithm.
This property ensures that our guessed encodings of the oracle are proper encoding consistent with the queried coordinates and are jointly dense.

\begin{definition}[Structure]
\label{def:structure}
    Let $\rho\in \{0, 1, *\}^N$ be a restriction. 
    Let $\tau\in (0, 2]$. 
    Let $Y^{(0)}, Y^{(1)} \in \bit^{N\times b}$ be independent random variables. 
    Let $g:\bit^b\times \bit^b\to \bit$ be a function. 
    We say that $Y^{(0)}, Y^{(1)}$ are $(\rho, \tau)$-structured if there exist $\delta_0, \delta_1>0$ with $\delta_0+\delta_1\geq \tau$ such that
    \begin{enumerate}
        \item $Y^{(0)}_{\mathrm{free}(\rho)}$ is $\delta_0$-dense;
        \item $Y^{(1)}_{\mathrm{free}(\rho)}$ is $\delta_1$-dense;
        \item $g^N(Y^{(0)}, Y^{(1)})$ is consistent with $\rho$: $g^{\mathrm{fix}(\rho)}(Y^{(0)}_{\mathrm{fix}(\rho)}, Y^{(1)}_{\mathrm{fix}(\rho)})=\rho_{\mathrm{fix}(\rho)}$.
    \end{enumerate}
\end{definition}

The following uniform marginals lemma from \cite[Lemma 3.4]{chattopadhyay2021query} formalizes the idea that for sufficiently obfuscating encoding map $g$, randomly guessed encodings are good enough (in marginals) even when we know nothing about the true underlying oracle.

\begin{lemma}[Uniform marginals lemma, {\cite[Lemma 3.4]{chattopadhyay2021query}}]
\label{lem:uniform-marginal}
    Let $b\geq c\log N$ with some constant $c>0$.
    Let $g:\bit^b\times \bit^b\to \{0, 1\}$ be a function with $\mathrm{disc}(g)\leq 2^{-\eta b}$, and let $G=g^N: \bit^{N\times b}\times \bit^{N\times b}\to \bit^N$.
    Let $\rho\in\{0, 1, *\}^N$ be a restriction.
    Let $0<\gamma<\eta-11/c$. 
    Let $Y^{(0)}, Y^{(1)}$ be independent random variables uniformly distributed over $\mathcal{Y}^{(0)}, \mathcal{Y}^{(1)}\subseteq \bit^{N\times b}$. 
    Suppose they are $(\rho, \tau)$-structured with
    \begin{equation}
        \tau\geq 2+11/c-\eta+\gamma.
    \end{equation}
    Then, for any $o\in \bit^N$ consistent with $\rho$, consider the random variable $(\tilde{Y}^{(0)}, \tilde{Y}^{(1)})$ uniformly distributed over $G^{-1}(o)\cap (\mathcal{Y}^{(0)}\times\mathcal{Y}^{(1)})$. 
    We have that $Y^{(0)}, Y^{(1)}$ are $2^{-\gamma b}$-close to $\tilde{Y}^{(0)}, \tilde{Y}^{(1)}$ in total variation distance, respectively.
\end{lemma}

We provide a proof of \Cref{lem:uniform-marginal} with refined constants for completeness.

\begin{proof}[Proof of \Cref{lem:uniform-marginal}]
    We want to show that for any event $\mathcal{E}\subseteq \mathcal{Y}^{(0)}$, $\left|\Pr[\tilde{Y}^{(0)}\in \mathcal{E}]-\Pr[Y^{(0)}\in \mathcal{E}]\right|\leq 2^{-\gamma b}$, which implies that the total variation distance is at most $2^{-\gamma b}$. 
    Then the same holds for $Y^{(1)}$ by swapping $Y^{(0)}, Y^{(1)}$, and we are done.
    Without loss of generality, we can assume $\Pr[Y^{(0)}\in\mathcal{E}]\geq 1/2$, since otherwise we can redefine $\mathcal{E}$ to be the complement of $\mathcal{E}$.

    We first prove that when $Y^{(0)}, Y^{(1)}$ are $(\rho, \tau)$-structured, the random variable $g^{\mathrm{free}(\rho)}(Y^{(0)}_{\mathrm{free}(\rho)}, Y^{(1)}_{\mathrm{free}(\rho)})$ is roughly uniformly distributed.
    Let $I=\mathrm{free}(\rho)$.
    Let $S$ be any non-empty subset of $I$.
    Since $b\geq c\log N\geq c$, we have $\tau\geq 2+11/c-\eta+\gamma\geq 2+6/b-\eta+\gamma+2/c+3/b$. 
    By \Cref{def:structure} and \Cref{def:dense}, we have 
    \begin{equation}
        H_\infty(Y^{(0)}_S)+H_\infty(Y^{(1)}_S)\geq \tau b|S|\geq (2+6/b-\eta+\gamma+2/c+3/b)b|S|. 
    \end{equation} 
    By \Cref{cor:disc-unif-parity}, we have
    \begin{equation}
        \mathrm{bias}(g^{\oplus S}(Y^{(0)}_S, Y^{(1)}_S))\leq 2^{-(\gamma+2/c+2/b)b|S|}\leq 2^{-\gamma b|S|-2|S|}N^{-2|S|}\leq 2^{-\gamma b-3}(2|I|)^{-|S|}, 
    \end{equation} 
    where the last step follows from $N^2\geq 4N\geq 4|I|$ and $|S|\geq 1$.
    This holds for any non-empty $S\subseteq I$. 
    Therefore, by \Cref{lem:vazirani}, we have 
    \begin{equation}
        \Pr[g^I(Y^{(0)}_I, Y^{(1)}_I)=o_I]\in (1\pm 2^{-\gamma b-3})\frac{1}{2^{|I|}}.
    \end{equation}

    Next, note that by \Cref{lem:cond-min-entropy}, we have that for any $I\subseteq [N]$, $H_\infty(Y^{(0)}_I|Y^{(0)}\in \mathcal{E})\geq H_\infty(Y^{(0)}_I)-\log\frac{1}{\Pr[Y^{(0)}\in \mathcal{E}]}\geq H_\infty(Y^{(0)}_I)-1\geq H_\infty(Y^{(0)}_I)-\frac{1}{b}b|I|$. 
    Therefore, the density of $Y^{(0)}|Y^{(0)}\in \mathcal{E}$ is at most $1/b$ lower than the density of $Y^{(0)}$.
    This means that $(Y^{(0)}|Y^{(0)}\in \mathcal{E}, Y^{(1)})$ is $(\rho, \tau-1/b)$-structured.
    By the same reasoning above, we have
    \begin{equation}
        \Pr[g^I(Y^{(0)}_I, Y^{(1)}_I)=o_I|Y^{(0)}\in\mathcal{E}]\in (1\pm 2^{-\gamma b-2})\frac{1}{2^{|I|}}. 
    \end{equation} 
    
    Note that from the Bayes formula,
    \begin{equation} 
    \begin{split}
        \Pr[\tilde{Y}^{(0)}\in \mathcal{E}]&=\Pr[Y^{(0)}\in \mathcal{E}|g^N(Y^{(0)}, Y^{(1)})=o]=\frac{\Pr[g^N(Y^{(0)}, Y^{(1)})=o|Y^{(0)}\in\mathcal{E}]\Pr[Y^{(0)}\in\mathcal{E}]}{\Pr[g^N(Y^{(0)}, Y^{(1)})=o]}\\  \\
        &=\frac{\Pr[g^I(Y^{(0)}_I, Y^{(1)}_I)=o_I|Y^{(0)}\in\mathcal{E}]\Pr[Y^{(0)}\in\mathcal{E}]}{\Pr[g^I(Y^{(0)}_I, Y^{(1)}_I)=o_I]} \\
        &\leq \frac{1+2^{-\gamma b-2}}{1-2^{-\gamma b-3}}\Pr[Y^{(0)}\in\mathcal{E}] \\
        &\leq (1+2^{-\gamma b-2})(1+2^{-\gamma b-2})\Pr[Y^{(0)}\in \mathcal{E}]\\
        &=\Pr[Y^{(0)}\in\mathcal{E}] + 2^{-\gamma b}(1/2+2^{-\gamma b}/16)\Pr[Y^{(0)}\in \mathcal{E}] \\
        &\leq \Pr[Y^{(0)}\in\mathcal{E}]+2^{-\gamma b},
    \end{split}
    \end{equation}
    where we have used $1/(1-x)\leq1+2x, \forall x\in (0, 1/2]$.
    Similarly, 
    \begin{equation}
        \Pr[\tilde{Y}^{(0)}\in\mathcal{E}]\geq \frac{1-2^{-\gamma b-2}}{1+2^{-\gamma b-3}}\Pr[Y^{(0)}\in\mathcal{E}]\geq (1-2^{-\gamma b-2})(1-2^{-\gamma b-3})\Pr[Y^{(0)}\in\mathcal{E}]\geq \Pr[Y^{(0)}\in \mathcal{E}]-2^{-\gamma b}.
    \end{equation} 
    It follows that $|\Pr[Y^{(0)}\in \mathcal{E}]-\Pr[\tilde{Y}^{(0)}\in\mathcal{E}]|\leq 2^{-\gamma b}$ as required.
    This completes the proof of \Cref{lem:uniform-marginal}.
\end{proof}

\Cref{lem:uniform-marginal} ensures that as long as our guesses of the encodings are dense, we can safely use them to generate fake data and feed them into the learning algorithm without knowing anything about the oracle.
However, as the learning algorithm proceeds, the (posterior) distributions of our guesses will change since they are conditioned on the past computational path of the learning algorithm.
This will cause them to lose density, and we need the following lemma to query the oracle and restore the density of our guesses.

\begin{lemma}[Density restoring partition, {\cite[Lemma 3.5]{goos2016rectangles}}{\cite[Lemma 3.7]{chattopadhyay2021query}}]
\label{lem:density-restoring}
    Let $\delta\in (0, 1]$. 
    Let $Y\in \bit^{N\times b}$ be a random variable with support $\mathcal{Y}\subseteq \bit^{N\times b}$. 
    Then there exists a partition $\mathcal{Y} = \mathcal{Y}^1\cup\cdots\cup\mathcal{Y}^l$ where every $\mathcal{Y}^j$ is associated with a set of coordinates $I_j\subseteq [N]$ and a value $y_{I_j}\in (\bit^b)^{I_j}$ such that
    \begin{enumerate}
        \item $Y_{I_j}|Y\in\mathcal{Y}^j$ is fixed to $y_{I_j}$;
        \item $Y_{[N]-I_j}|Y\in\mathcal{Y}^j$ is $\delta$-dense.
    \end{enumerate}
    Moreover, if we use $p_{\geq j}$ to denote $\Pr[Y\in \mathcal{Y}^j\cup\cdots\cup\mathcal{Y}^l]$, then
    \begin{equation}
        H_\infty(Y_{[N]-I_j}|Y\in \mathcal{Y}^j)\geq H_\infty(Y)-\delta b |I_j|-\log\frac{1}{p_{\geq j}}.
    \end{equation}
\end{lemma}

Moreover, note that we have two variables $Y^{(0)}, Y^{(1)}$ to take care of.
When we partition $Y^{(0)}$ and fix some of its values $Y^{(0)}_I=y_I$ to restore its density, we may destroy the density of $Y^{(1)}$ because we must keep consistency with $\rho, o$ (i.e., condition on $g^I(y_I, Y^{(1)}_I)=o_I$).
Such $y_I$'s that destroy the density of the other variable are called dangerous.
We follow \cite{chattopadhyay2021query} and define them as follows.

\begin{definition}[Dangerous value]
\label{def:dangerous}
    Let $\delta\in (0, 1], \epsilon\in (0, \delta)$.
    Let $Y\in \bit^{N\times b}$ be a $\delta$-dense random variable.
    Let $g: \bit^b\times \bit^b\to \bit$ be a function. 
    We say that a value $y\in \bit^{N\times b}$ is $\epsilon$-dangerous to $Y$ if any one of the following conditions hold:
    \begin{enumerate}
        \item there exists a set of coordinates $I\subseteq [N]$ and an assignment $o_I \in \bit^{|I|}$ such that
        \begin{equation}
            \Pr[g^I(y_I, Y_I)=o_I]<2^{-|I|-1};
        \end{equation}
        \item there exists a set of coordinates $I\subseteq [N]$ and an assignment $o_I \in \{0, 1\}^{|I|}$ such that 
        \begin{equation}
            Y_{[N]-I}|(g^I(y_I, Y_I)=o_I)
        \end{equation} 
        is not $(\delta-\epsilon)$-dense.
    \end{enumerate}
\end{definition}

The following lemma shows that as long as our guesses are structured, it is very unlikely for them to take a dangerous value.

\begin{lemma}[Dangerous values are unlikely, {\cite[Lemma 3.9]{chattopadhyay2021query}}]
\label{lem:dangerous-unlikely}
    Let $b\geq c\log N$ with some constant $c\geq 4$.
    Let $g: \bit^b\times\bit^b\to \bit$ be a function with $\mathrm{disc}(g)\leq 2^{-\eta b}$. 
    Let $\tau\in (0, 2], 0<\epsilon, \gamma\leq 1$ with $\tau\geq 2+17/(c\epsilon)-\eta+\gamma$ and $\epsilon>4/b$. 
    Let $\rho$ be a restriction. 
    Let $Y^{(0)}, Y^{(1)}$ be independent variables that are $(\rho, \tau)$-structured. 
    Then the probability that $Y^{(0)}_{\mathrm{free}(\rho)}$ takes a value that is $\epsilon$-dangerous to $Y^{(1)}_{\mathrm{free}(\rho)}$ is at most $2^{-\gamma b}$.
\end{lemma}

We provide a proof of \Cref{lem:dangerous-unlikely} with refined constants for completeness.

\begin{proof}[Proof of \Cref{lem:dangerous-unlikely}]
    Without loss of generality, we assume that $\rho$ is all free.
    We introduce an auxiliary notion of $\epsilon$-biasing values such that any values that are not $\epsilon$-biasing are also not $\epsilon$-dangerous.
    We will show that $\epsilon$-biasing values are unlikely, and therefore so are $\epsilon$-dangerous values.
    Fix $\delta_0,\delta_1>0$ witnessing that $Y^{(0)},Y^{(1)}$ are $(\rho,\tau)$-structured, and write $\delta=\delta_1$.

    In particular, we say that a value $y^0\in \bit^{N\times b}$ is $\epsilon$-biasing if there exists disjoint subsets $\emptyset\neq S\subseteq [N], J\subseteq [N]-S$ and a value $y^1_J\in \bit^{|J|\times b}$ with $|S|\geq (\epsilon b |J|+e_{y^1_J}-2)/\log N$ such that
    \begin{equation}
        \mathrm{bias}(g^{\oplus S}(y^0_S, Y^{(1)}_S)|Y^{(1)}_J=y^1_J)>\frac{1}{2}(2N)^{-|S|}.
    \end{equation} 
    Here, $e_{y^1_J}$ is the excess entropy defined in \Cref{def:excess-ent}.

    We first prove that if a value $y^0$ is not $\epsilon$-biasing, then for any $I\subseteq [N]$, we have $\Pr[g^I(y^0_I, Y^{(1)})=o_I]\geq 2^{-|I|-1}$, which violates the first condition in \Cref{def:dangerous}.
    To this end, let $y^0$ be not $\epsilon$-biasing.
    We take $J=\emptyset$ in the definition of $\epsilon$-biasing, and have $\mathrm{bias}(g^{\oplus S}(y^0_S, Y^{(1)}_S))\leq\frac{1}{2}(2N)^{-|S|}$ for any non-empty $S\subseteq [N]$.
    \Cref{lem:vazirani} then implies that
    \begin{equation}
         \Pr[g^I(y^0_I, Y^{(1)}_I)=o_I]\geq \left(1-\frac{1}{2}\right) 2^{-|I|}=2^{-|I|-1}.
    \end{equation}
    Therefore, non-$\epsilon$-biasing implies violation of the first condition in \Cref{def:dangerous}.

    Next, we show that non-$\epsilon$-biasing also implies violation of the second condition in \Cref{def:dangerous}.
    Suppose not, for the sake of contradiction.
    Then there exists a set of coordinates $I\subseteq [N]$ and an assignment $o_I \in \bit^{|I|}$ such that $Y^{(1)}_{[N]-I}|(g^I(y^0_I, Y^{(1)}_I)=o_I)$ is not $(\delta-\epsilon)$-dense.
    By the definition of dense random variables and the definition of min-entropy, there exists $J\subseteq [N]-I$ and a value $y^1_J\in \bit^{|J|\times b}$ such that 
    \begin{equation}
        \Pr[Y^{(1)}_J=y^1_J|g^I(y^0_I, Y^{(1)}_I)=o_I]>2^{-(\delta-\epsilon)b|J|}.
    \end{equation}
    On the other hand, the left hand side satisfies
    \begin{equation}
    \begin{split}
        \Pr[Y^{(1)}_J=y^1_J|g^I(y^0_I, Y^{(1)}_I)=o_I] &= \frac{\Pr[g^I(y^0_I, Y^{(1)}_I)=o_I|Y^{(1)}_J=y^1_J]\Pr[Y^{(1)}_J=y^1_J]}{\Pr[g^I(y^0_I, Y^{(1)}_I)=o_I]}  \\
    &\leq\frac{\Pr[g^I(y^0_I, Y^{(1)}_I)=o_I|Y^{(1)}_J=y^1_J]2^{-\delta b |J|-e_{y^1_J}}}{2^{-|I|-1}},
    \end{split}
    \end{equation}
    where we have used the violation of the first condition that we proved above.
    Therefore, 
    \begin{equation}
    \label{eq:skewing}
        \Pr[g^I(y^0_I, Y^{(1)}_I)=o_I|Y^{(1)}_J=y^1_J]> 2^{-|I|-1+\epsilon b|J|+e_{y^1_J}}.
    \end{equation} 
    On the other hand, from the definition of non-$\epsilon$-biasing, we know that for any disjoint non-empty sets $S, J\subseteq [N]$ and value $y^1_J \in \bit^{|J|\times b}$ with $|S|\geq (\epsilon b |J|+e_{y^1_{J}}-2)/\log N$, we have
    \begin{equation}
        \mathrm{bias}(g^{\oplus S}(y^0_S, Y^{(1)}_S)|Y^{(1)}_J=y^1_J)\leq \frac12 (2N)^{-|S|}<(2N)^{-|S|}.
    \end{equation} 
    \Cref{lem:vazirani-ent} (taking $m=|I|, t = (\epsilon b |J|+e_{y^1_{J}}-2)/\log N$) then implies that
    \begin{equation}
        H_\infty(g^{I}(y^0_I, Y^{(1)}_I)|Y^{(1)}_J=y^1_J)\geq |I|-\frac{\epsilon b |J|+e_{y^1_{J}}-2}{\log N} \log|I|-1\geq |I| - \epsilon b |J|-e_{y^1_{J}}+1,
    \end{equation}
    where we have used $\log|I|\leq \log N$.
    This contradicts \Cref{eq:skewing}.
    Therefore, non-$\epsilon$-biasing also violates the second condition in \Cref{def:dangerous}.

    At this point, we have shown that non-$\epsilon$-biasing implies violation of both conditions in \Cref{def:dangerous}, which means that non-$\epsilon$-biasing implies non-$\epsilon$-dangerous.
    All we are left to show is that $\epsilon$-biasing values are unlikely.

    To prove that $\epsilon$-biasing values appear with probability at most $2^{-\gamma b}$, we use an union bound over the choice of $S, J, y^1_J$.
    We assume $J$ to be non-empty (the empty case works similarly).
    
    Let $S, J, y^1_J$ satisfy
    \begin{equation}
        |S|\geq\frac{\epsilon b|J|+e_{y^1_J}-2}{\log N}\geq \frac{\frac{1}{2}\epsilon b|J|+e_{y^1_J}}{\log N},
    \end{equation}
    where we used $\epsilon\geq 4/b$ and $|J|\geq 1$ in the second inequality.
    
    By our choice, $\delta_0+\delta_1\geq\tau$ and $Y^{(0)},Y^{(1)}$ are $\delta_0,\delta_1$-dense.
    So $H_\infty(Y^{(1)}_S)\geq \delta_1 b |S|$.
    Conditioning on $Y^{(1)}_J=y^1_J$, by \Cref{lem:cond-min-entropy}, we have
    \begin{equation}
    \begin{split}
        H_\infty(Y^{(1)}_S|Y^{(1)}_J=y^1_J)&\geq \delta_1b|S|-\log\frac{1}{\Pr[Y^{(1)}_J=y^1_J]} \\
        &=\delta_1 b|S|-(\delta_1b|J|+e_{y^1_J}) \\
        &\geq \delta_1 b|S| - b|J| - e_{y^1_J}, \quad (\delta_1\leq 1) \\
        &\geq \delta_1 b|S| -\frac{2}{\epsilon}\frac{1}{2}b\epsilon|J| - \frac{2}{\epsilon}e_{y^1_J}, \quad (\epsilon\leq 2) \\
        &\geq \delta_1 b|S| - \frac{2}{\epsilon}|S|\log N, \quad (|S|\geq (\frac12 \epsilon b |J|+e_{y^1_J})/\log N) \\
        &\geq (\delta_1 -\frac{2}{\epsilon c})b|S|, \quad (b\geq c\log N).
    \end{split}
    \end{equation}
    Therefore, 
    \begin{equation}
        H_\infty(Y^{(0)}_S) + H_\infty(Y^{(1)}_S|Y^{(1)}_J=y_J)\geq (\tau-\frac{2}{\epsilon c})b|S|\geq (2+\frac{17-2}{c\epsilon}-\eta+\gamma)b |S|= (2+\frac{15}{c\epsilon}-\eta+\gamma)b |S|.
    \end{equation}
    Then, by \Cref{cor:disc-samp-parity} with $\lambda = 3/(c\epsilon), \gamma = \gamma + 5/(c\epsilon)$, we have that the probability of $Y^{(0)}_S$ taking a value $y^0_S\in \bit^{|S|\times b}$ such that
    \begin{equation}
	\mathrm{bias}(g^{\oplus S}(y^0_S, Y^{(1)}_S)|Y^{(1)}_J=y^1_J)> 2^{-\frac{3}{c\epsilon}b|S|} 
    \end{equation}
    is at most $2^{-(\gamma+\frac{5}{c\epsilon})b|S|}$.
    The assumptions of \Cref{cor:disc-samp-parity} hold since $\tau\leq2$ implies $\eta-\gamma\geq17/(c\epsilon)$, and $b\geq c\log N\geq c\epsilon$.
    In other words, for those $S, J, y^1_J$ that satisfies $|S|\geq\frac{\epsilon b|J|+e_{y^1_J}-2}{\log N}$, the probability of $Y^{(0)}_S$ taking the above values is small.
    Moreover, since $b\geq c\log N$, $\epsilon\leq 1$, $N\geq 2$, and $|S|\geq 1$,
    \begin{equation}
        2^{-\frac{3}{c\epsilon}b|S|}
        \leq 2^{-3|S|\log N}
        \leq \frac12(2N)^{-|S|}.
    \end{equation}
    Thus the same probability bound applies to the $\epsilon$-biasing event.
    The case $J=\emptyset$ follows identically, with $e_{y^1_J}=0$.
    
    Now we take the union bound.
    First consider a fixed non-empty $S$.
    The condition $|S|\geq\frac{\epsilon b|J|+e_{y^1_J}-2}{\log N}$ implies, using $e_{y^1_J}\geq0$ and $b\geq c\log N$,
    \begin{equation}
        |J|
        \leq
        \frac{|S|\log N+2}{\epsilon b}
        \leq
        \frac{|S|}{c\epsilon}+\frac{2}{\epsilon b}
        <
        \frac{|S|}{c\epsilon}+\frac12.
    \end{equation}
    Hence, since $|J|$ is an integer,
    \begin{equation}
        |J|\leq m:=\ceil{\frac{|S|}{c\epsilon}}.
    \end{equation}
    Moreover, if there exists a non-empty $J$, the previous bound
    \begin{equation}
        |S|\geq
        \frac{\frac12\epsilon b|J|+e_{y^1_J}}{\log N}
    \end{equation}
    implies
    \begin{equation}
        \frac{|S|}{c\epsilon}
        \geq
        \frac{b}{2c\log N}
        \geq \frac12,
    \end{equation}
    and therefore
    \begin{equation}
        m\leq \frac{2|S|}{c\epsilon}.
    \end{equation}
    Thus,
    \begin{equation}
    \begin{split}
        \sum_{\substack{J\subseteq[N]\\|J|\leq m}}
        \sum_{y^1_J\in\bit^{|J|\times b}}
        2^{-(\gamma+\frac{5}{c\epsilon})b|S|}&\leq
        \sum_{j=0}^{m}
        2^{j(b+\log N)}
        2^{-(\gamma+\frac{5}{c\epsilon})b|S|}\leq
        2^{1+m(b+\log N)}
        2^{-(\gamma+\frac{5}{c\epsilon})b|S|} \\
        &\leq
        2^{\frac{3}{c\epsilon}b|S|}
        2^{-(\gamma+\frac{5}{c\epsilon})b|S|}=
        2^{-(\gamma+\frac{2}{c\epsilon})b|S|},
    \end{split}
    \end{equation}
    where we used $b\geq c\log N$, $c\geq4$, and $\frac{b|S|}{c\epsilon}\geq\frac{b}{2}\geq \frac{c}{2}\geq 2$,
    which give
    \begin{equation}
        1+m(b+\log N)
        \leq
        1+\left(2+\frac{2}{c}\right)
        \frac{b|S|}{c\epsilon}
        \leq
        \frac{3b|S|}{c\epsilon}.
    \end{equation}
    If no admissible non-empty $J$ exists, then only $J=\emptyset$ contributes, whose probability is already bounded by
    \begin{equation}
        2^{-(\gamma+\frac{5}{c\epsilon})b|S|}
        \leq
        2^{-(\gamma+\frac{2}{c\epsilon})b|S|}.
    \end{equation}
    Therefore, in all cases, for every fixed non-empty $S$ the probability of being $\epsilon$-biasing through some $J, y^1_J$ is at most $2^{-(\gamma+\frac{2}{c\epsilon})b|S|}$.
    
    Finally, taking the union bound over non-empty $S\subseteq[N]$ gives
    \begin{equation}
    \begin{split}
        \sum_{s=1}^N
        \binom Ns
        2^{-(\gamma+\frac{2}{c\epsilon})bs}
        &\leq
        \sum_{s=1}^{\infty}
        2^{-(\gamma+\frac{2}{c\epsilon})bs+s\log N}\leq
        \sum_{s=1}^{\infty}
        2^{-(\gamma+\frac{1}{c\epsilon})bs} \\
        &=
        \frac{1}{2^{(\gamma+\frac{1}{c\epsilon})b}-1}\leq
        2^{1-(\gamma+\frac{1}{c\epsilon})b}
        \leq
        2^{-\gamma b},
    \end{split}
    \end{equation}
    where we used
    $b/(c\epsilon)\geq\log(N)/\epsilon\geq1$.
    This completes the proof of \Cref{lem:dangerous-unlikely}.
\end{proof}

We will also need the following progress function called deficiency to prove query complexity bounds.

\begin{definition}[Deficiency]
    Let $Y^{(0)}, Y^{(1)}\in \bit^{N\times b}$ be two random variables.
    Given a restriction $\rho\in \{0, 1, *\}^N$, the deficiency of $Y^{(0)}, Y^{(1)}$ is defined as 
    \begin{equation}
        D_\infty(Y^{(0)}, Y^{(1)}, \rho) = 2b|\mathrm{free}(\rho)| - H_\infty(Y^{(0)}_{\mathrm{free}(\rho)})-H_\infty(Y^{(1)}_{\mathrm{free}(\rho)})\geq 0.
    \end{equation}
\end{definition}

\subsubsection{Simulation}
\label{sec:cl-hard-simulation}

We are now ready to formalize the simulation argument and prove the following result.

\begin{tcolorbox}
\begin{theorem}[Simulation]
\label{thm:simulation}
    Let $N$ be a large integer. 
    Let the time scale $T$ be a positive integer.
    Let $\eta\in (0, 2]$ be a constant and $c=(865/\eta^2)\log(865/\eta^2)$.
    Let $g:\bit^b\times \bit^b\to \bit$ be a noisy encoding function with encoding length $b\geq c\log N$ and discrepancy $\disc(g)\leq 2^{-\eta b}$.
    For any randomized classical learning algorithm $\mathcal{L}$ that has space complexity $S$ and sample complexity $M \leq 2NTb$, there exists a randomized parallel decision tree $\mathcal{A}$ with query complexity 
    \begin{equation}
        Q = O\left(\frac{MS}{Tb}+1\right)
    \end{equation}
    such that given any input $o\in \bit^N$, $\mathcal{A}$ outputs a random bitstring $\pi$ whose distribution is $2^{-\eta b/8}$-close in total variation distance with the distribution of the transcript of $\mathcal{L}$ when given data generated from $\mathcal{D}^N_{g, T}(o)$.
\end{theorem}
\end{tcolorbox}

It suffices to prove this for deterministic classical learning algorithms, since any randomized algorithm can always be regarded as first sampling all the random numbers and then execute the corresponding deterministic algorithm.
Thus we can always construct the randomized decision tree by first sampling the random numbers and execute the decision tree obtained by applying the theorem to the corresponding deterministic learning algorithm.

Given any deterministic classical learning algorithm $\mathcal{L}$, we construct the randomized decision tree $\mathcal{A}$ as in \Cref{alg:simulation}.
Intuitively, we maintain a rectangle (i.e., a Cartesian product of two subsets) $\mathcal{Y}^{(0)}\times \mathcal{Y}^{(1)} \subseteq \bit^{N\times b}\times \bit^{N\times b}$ that represents guesses of the noisy encodings of the underlying oracle.
We use the random variables $Y^{(0)}, Y^{(1)}$ uniformly distributed on $\mathcal{Y}^{(0)}\times \mathcal{Y}^{(1)}$ to generate fake data according to our guesses and feed them into the learning algorithm $\mathcal{L}$.
Whenever the situation changes, we record the state of the learning algorithm $\mathcal{L}$ and append it to the transcript $\pi$.
\Cref{lem:uniform-marginal} ensures that this strategy works as long as our guesses $Y^{(0)}, Y^{(1)}$ are dense.
When they lose density, we invoke the density restoring partition from \Cref{lem:density-restoring} to restore their density.
This requires us to query the oracle $o$ on the fixed coordinates to keep our guesses consistent with the true oracle.
During the simulation, we also keep track of several error accumulation metric ($Q, K$ and $q_{j_i}$ in \Cref{alg:simulation}) to prevent extreme cases from happening: if they happen, we halt and declare error.
We will show that such extreme cases happen rarely and this simulation outputs a transcript that is close to the correct transcript of $\mathcal{L}$ (\Cref{sec:cl-hard-sim-correctness}) with the total query complexity bounded (\Cref{sec:cl-hard-sim-complex}).
We note that by construction, $Y^{(0)}, Y^{(1)}$ are always $(\rho, \tau)$-structured at the end of each iteration.

Also note that throughout the simulation and its analysis, all probabilities are conditioned on the sampled $\alpha$ and $x$ records unless we explicitly average over them.

\begin{algorithm}
  \caption{Simulation}
  \label{alg:simulation}
  \begin{algorithmic}[1]
    \Require 
    A deterministic classical learning algorithm $\mathcal{L}$ with space complexity $S$, sample complexity $M$, and input form $\mathcal{I}=[N]\times \bit^b\times \bit$;
    the data generation process $\mathcal{D}^N_{g, T}$;
    an input $o\in\{0,1\}^N$.
    \Ensure 
    A learning algorithm transcript $\pi$.
    \State  $\epsilon\gets (96\log c)/(c\eta)$.
            $\delta\gets 1-\eta/4+\epsilon/2$.
            $\tau\gets 2\delta-\epsilon$.
    \State  Let $\pi$ be the empty string. 
            $\rho\gets\{*\}^N$.
            $\mathcal{Y}^{(0)},\mathcal{Y}^{(1)}\gets\bit^{N\times b}$.
    \State  Let $Y^{(0)} \sim \unif(\mathcal{Y}^{(0)})$, $Y^{(1)} \sim \unif(\mathcal{Y}^{(1)})$ be two independent random variables.
    \State  $v\gets$ root of $\mathcal{L}$. $K\gets 0$.
    \State Sample a full situation record $\alpha \gets (\alpha_0, \ldots, \alpha_{M-1})$ at time points $0, \ldots, M-1$ from $\mathcal{D}^N_{g, T}$.
    \State Compute the situation changing time $0=t_0<\ldots<t_{r}\leq M-1$ where $\alpha_{t_i-1}\neq \alpha_{t_i}, \forall i=1, \ldots, r$.
    \State $Q\gets 0$.
       $Q_{\max}\gets
       \ceil{
       \frac{6920}{\eta}
       \left(
       \frac{rS+4r+3b}{b}
       \right)
       }$.
    \For{$i = 0, \ldots, r-1$}
        \State  Let $A\gets \alpha_{t_i}$, $B\gets \alpha_{t_i}\oplus 1$.
        \State  $\mathcal{Y}^{(A)} \gets \mathcal{Y}^{(A)} - \{y\in \mathcal{Y}^{(A)}: y_{\mathrm{free}(\rho)} \text{ is $\epsilon$-dangerous to } Y^{(B)}_{\mathrm{free}(\rho)}\}$. 
                Let $Y^{(A)}\sim\unif(\mathcal{Y}^{(A)})$. \label{ln:remove-dangerous}
        \For{$t = t_i, \ldots, t_{i+1}-1$}
            \State  Randomly sample $x_t\sim \unif([N])$. 
                    Let $y_t \gets Y^{(A)}_{x_t} \in \bit^b$.
            \State  $v\gets$ the end point of the edge from $v$ in $\mathcal{L}$ labeled by $I_t=(x_t, y_t, A)$.
        \EndFor 
        \State  Let $\mu_i$ be the length-$S$ bitstring that labels $v$ in $\mathcal{L}$.
                Append $\mu_i$ to $\pi$.
        \State  Let $p_i$ be the probability of reaching $v$ in the above sampling process.
                $K\gets K+\log(1/p_i)$
        \State  $\mathcal{Y}^{(A)} \gets \mathcal{Y}^{(A)} - \{y\in \mathcal{Y}^{(A)}: \text{the probability of reaching $v$ conditioned on $Y^{(A)}=y$ is zero}\}$.
                Let $Y^{(A)}\sim\unif(\mathcal{Y}^{(A)})$.
        \If{$K>rS+b$} \label{ln:check-k}
          \State \textbf{halt} and declare error
        \EndIf \label{ln:check-k-end}
        \State  Find the density restoring partition of the random variable $Y^{(A)}$ supported on $\mathcal{Y}^{(A)}$: $(\mathcal{Y}^j, I_j, y_{I_j})_{j\in [l]}$. \label{ln:density-restoring}
        \State  Sample $j_i\in [l]$ with probability
               $\Pr[Y^{(A)}_{\mathrm{free}(\rho)}\in\mathcal{Y}^{j_i}]$.
        \If{$q_{j_i}=\Pr[Y^{(A)}_{\mathrm{free}(\rho)}\in \bigcup_{k\ge j_i}\mathcal{Y}^k]<\frac18 2^{-\eta b/8}\cdot\frac{1}{2Nb}$}
          \State \textbf{halt} and declare error
        \EndIf
        \State  $\mathcal{Y}^{(A)}\gets \mathcal{Y}^{(A)}-\{y\in \mathcal{Y}^{(A)}: y_{\mathrm{free}(\rho)}\notin\mathcal{Y}^{j_i}\}$. 
                Let $Y^{(A)}\sim \unif(\mathcal{Y}^{(A)})$.
        \If{$Q+|I_{j_i}|>Q_{\max}$} \label{ln:check-query}
            \State \textbf{halt} and declare error
        \EndIf \label{ln:check-query-end}
        \State Query $o_{I_{j_i}}$ and set
               $\rho_{I_{j_i}}\gets o_{I_{j_i}}$.
               $Q\gets Q+|I_{j_i}|$.    
        \State  $\mathcal{Y}^{(B)}\gets \mathcal{Y}^{(B)}-\{y^B\in \mathcal{Y}^{(B)}: g^{I_{j_i}}(y_{I_{j_i}}, y^B_{I_{j_i}})\neq \rho_{I_{j_i}}\}$. 
                Let $Y^{(B)}\sim \unif(\mathcal{Y}^{(B)})$.
    \EndFor
    \State  Let $A\gets \alpha_{t_r}$, $B\gets \alpha_{t_r}\oplus 1$
    \For{$t = t_{r}, \ldots, M-1$}
        \State  Randomly sample $x_t\sim \unif([N])$. 
                Let $y_t \gets Y^{(A)}_{x_t} \in \bit^b$.
        \State  $v\gets$ the end point of the edge from $v$ in $\mathcal{L}$ labeled by $I_t=(x_t, y_t, A)$.
    \EndFor 
    \State Append the output label $h_v\in\bit$ of leaf $v$ in $\mathcal{L}$ to $\pi$.
    \State \Return $\pi$
  \end{algorithmic}
\end{algorithm}

\subsubsection{Simulation is correct}
\label{sec:cl-hard-sim-correctness}

In this section, we prove that the output of the query algorithm $\mathcal{A}$ matches the transcript of the learning algorithm $\mathcal{L}$ in distribution.

We first calculate the number of situation changes $r$.
Note that in the data generation process, each time we sample a new situation $\alpha$, we will keep using it to generate the data for $T$ time steps.
Hence we must have $r\leq \ceil{M/T}-1\leq M/T$.

We aim to prove the correctness of \Cref{alg:simulation}: for any input $o\in \bit^N$, the distribution of $\mathcal{A}$'s output $\pi$ is $2^{-\eta b/8}$-close in total variation distance to the distribution of the transcript of $\mathcal{L}$ given data generated from $\mathcal{D}_{g, T}^N(o)$.
We follow \cite{chattopadhyay2021query} and prove this in two steps.
We define three different transcripts:
\begin{enumerate}
    \item $\pi$, the output of the query algorithm $\mathcal{A}$;
    \item $\pi'$, the output of a modified query algorithm $\mathcal{A}'$ which is the same as $\mathcal{A}$ except that it skips Lines \ref{ln:check-k}-\ref{ln:check-k-end} and Lines \ref{ln:check-query}-\ref{ln:check-query-end} in \Cref{alg:simulation}; and
    \item $\pi^\star$, the target transcript of $\mathcal{L}$ on the data generated from $\mathcal{D}_{g, T}^N(o)$.
\end{enumerate}
We will first prove that $\pi$ and $\pi'$ are $2^{-\eta b/8-1}$-close, and then prove that $\pi'$ and $\pi^\star$ are $2^{-\eta b/8-1}$-close.
This immediately implies that $\pi$ and $\pi^\star$ are $2^{-\eta b/8}$-close by triangle inequality.

We start by showing that $\pi$ and $\pi'$ are $2^{-\eta b/8-1}$-close.
We condition throughout this comparison on the sampled situation record $\alpha$, and suppress this conditioning from the notation.
Thus $r$ and $Q_{\max}$ are fixed. 
All the bounds below hold uniformly in $\alpha$, so averaging over $\alpha$ preserves the same bounds.
On the execution of $\mathcal{A}'$, let $\mathcal{E}_{K}$ be the event that $K>rS+b$ at some iteration, and let $\mathcal{E}_{Q}$ be the event that the query-budget condition in Lines \ref{ln:check-query}-\ref{ln:check-query-end} would be violated.
If $r=0$, no complete iteration is executed, both events are empty, and the claim is immediate.
In the following, assume $r\geq1$.
We couple $\pi$ and $\pi'$ by using exactly the same random choices in the two procedures.
Under this coupling, their transcripts are identical up to the first time at which either $\mathcal{E}_{K}$ or $\mathcal{E}_{Q}$ occurs.
If neither event occurs, the two complete transcripts are identical. 
Therefore,
\begin{equation}
    \Pr[\pi\neq\pi']
    \leq
    \Pr[\mathcal{E}_{K}\cup\mathcal{E}_{Q}],
\end{equation}
and the coupling characterization of total variation distance gives
\begin{equation}
    d_{\mathrm{TV}}(\pi,\pi')
    \leq
    \Pr[\mathcal{E}_{K}\cup\mathcal{E}_{Q}].
\end{equation}
To bound $\Pr[\mathcal{E}_K]$, we note that in \Cref{alg:simulation}, the value of $K$ at some iteration $i$ is calculated in the following sequential way: $\mu_0\to p_0\to j_0\to \cdots\to \mu_i\to p_i\to K$, where latter variables depend on every preceding variables.
We denote $\mu_{<i}=(\mu_0, \ldots, \mu_{i-1}), j_{<i}=(j_0, \ldots, j_{i-1})$.
If $\mathcal{A}'$ halts before $r$, we set all the subsequent $\mu_i=0^S$ and $j_i=1$.
Also note that $p_i = \Pr[\mu_i|\mu_{<i},j_{<i}]$.
Therefore, we have
\begin{equation}
\begin{split}
    \Pr[\mathcal{E}_K]
    &= \sum_{\substack{\mu_0,\ldots,\mu_{r-1}\\
                       j_0,\ldots,j_{r-1}}}
       \Pr[\mu_0,\ldots,\mu_{r-1},j_0,\ldots,j_{r-1}]
       1\left[
           \sum_{i=0}^{r-1}\log\frac{1}{p_i}>rS+b
       \right] \\
    &= \sum_{\substack{\mu_0,\ldots,\mu_{r-1}\\
                       j_0,\ldots,j_{r-1}}}
       \prod_{k=0}^{r-1}
       \Pr[\mu_k\mid\mu_{<k},j_{<k}]
       \Pr[j_k\mid\mu_{\leq k},j_{<k}]
       1\left[
           \prod_{i=0}^{r-1}
           \Pr[\mu_i\mid\mu_{<i},j_{<i}]
           <2^{-rS-b}
       \right] \\
    &\leq
       2^{-rS-b}
       \sum_{\substack{\mu_0,\ldots,\mu_{r-1}\\
                       j_0,\ldots,j_{r-1}}}
       \prod_{k=0}^{r-1}
       \Pr[j_k\mid\mu_{\leq k},j_{<k}] \\
    &= 2^{-rS-b}
       \sum_{\mu_0,\ldots,\mu_{r-1}}
       1 
    = 2^{-rS-b}(2^S)^r
    = 2^{-b}.
\end{split}
\end{equation}

Next, we bound the cumulative loss associated with the density restoring partitions.
Let $\mathcal{H}_i$ denote the complete history immediately before $j_i$ is sampled in iteration $i$ of $\mathcal{A}'$.
Conditioned on $\mathcal{H}_i$, write
\begin{equation}
    a_{i,j}
    =
    \Pr[j_i=j\mid\mathcal{H}_i],
    \qquad
    q_{i,j}
    =
    \sum_{k\geq j}a_{i,k}.
\end{equation}
Thus, the value denoted by $q_{j_i}$ in \Cref{alg:simulation} is $q_{i,j_i}$.
Recall that $l_i$ is the number of cells in the partition at iteration $i$.
If $\mathcal{A}'$ has already halted before iteration $i$, we set $l_i=1$ and $a_{i,1}=q_{i,1}=1$.
Also set $q_{i,l_i+1}=0$.

For every realization of $\mathcal{H}_i$, we have
\begin{equation}
    \E\left[
        q_{j_i}^{-1/2}
        \middle|
        \mathcal{H}_i
    \right]
    =
    \sum_{j=1}^{l_i}
    \frac{a_{i,j}}{\sqrt{q_{i,j}}}
    =
    \sum_{j=1}^{l_i}
    \frac{q_{i,j}-q_{i,j+1}}{\sqrt{q_{i,j}}}
    \leq
    2\sum_{j=1}^{l_i}
    \left(
        \sqrt{q_{i,j}}-\sqrt{q_{i,j+1}}
    \right)
    =
    2.
\end{equation}
Define
\begin{equation}
    L
    =
    \sum_{i=0}^{r-1}
    \log\frac{1}{q_{j_i}}.
\end{equation}
Iteratively taking conditional expectations gives
\begin{equation}
    \E\left[2^{L/2}\right]
    =
    \E\left[
        \prod_{i=0}^{r-1}q_{j_i}^{-1/2}
    \right]
    \leq
    2^r.
\end{equation}
Hence, defining
\begin{equation}
    \mathcal{E}_{L}
    =
    \{L>2r+2b\},
\end{equation}
Markov's inequality gives
\begin{equation}
    \Pr[\mathcal{E}_{L}]
    \leq
    \frac{2^r}{2^{r+b}}
    =
    2^{-b}.
\end{equation}

It will be shown in \Cref{sec:cl-hard-sim-complex} that if neither $\mathcal{E}_{K}$ nor $\mathcal{E}_{L}$ occurs, then the query counter never exceeds $Q_{\max}$.
Therefore,
\begin{equation}
    \mathcal{E}_{Q}
    \subseteq
    \mathcal{E}_{K}\cup\mathcal{E}_{L}.
\end{equation}
It follows that
\begin{equation}
    d_{\mathrm{TV}}(\pi,\pi')
    \leq
    \Pr[
        \mathcal{E}_{K}\cup\mathcal{E}_{Q}
    ] \leq
    \Pr[
        \mathcal{E}_{K}\cup\mathcal{E}_{L}
    ] \leq
    2^{1-b}
    \leq
    2^{-b/4-1}
    \leq
    2^{-\eta b/8-1},
\end{equation}
where the last two inequalities hold for sufficiently large $N$
since $b\geq c\log N$ and $\eta\leq2$.

Next, we show that $\pi'$ and $\pi^\star$ are $2^{-\eta b/8-1}$-close.
We first fix any $\alpha=(\alpha_0, \ldots, \alpha_{M-1})\in \bit^M$ and any $x = (x_0, \ldots, x_{M-1})\in [N]^M$.
Let $\pi'(\alpha, x)$ be the transcript of $\mathcal{A}'$ when the $\alpha$ and $x_t$'s in $\mathcal{A}'$ are replaced by corresponding entries in $\alpha$ and $x$.
Similarly, let $\pi^\star(\alpha, x)$ be the transcript of $\mathcal{L}$ when the $\alpha$ and $x_t$'s in the generated data are replaced by corresponding entries in $\alpha$ and $x$.
We will prove that $\pi'(\alpha, x)$ and $\pi^\star(\alpha, x)$ are $2^{-\eta b/8-1}$-close for any $x$.
Then averaging over random $\alpha$ and $x$ over the fixed distributions (i.e., $\mathrm{Bern}(1/2)$ and repeat $T$ times for $\alpha$, and $\unif([N])$ for $x$) immediately gives the desired result.

We prove this by going through the simulation step by step and constructing a coupling between $\pi'(\alpha, x)$ and $\pi^\star(\alpha, x)$ such that $\Pr[\pi'(\alpha, x)\neq \pi(\alpha, x)^\star]\leq 2^{-\eta b/8-1}$, which implies that $\pi'(\alpha, x)$ and $\pi^\star(\alpha, x)$ are $2^{-\eta b/8-1}$-close.
Let $(\tilde{Y}^{(0)}, \tilde{Y}^{(1)})$ be the ideal random encodings uniformly distributed over $G^{-1}(o)$.
Let $\mathcal{Y}^{(0)}_i\times \mathcal{Y}^{(1)}_i$ be the rectangle maintained in \Cref{alg:simulation} at the end of iteration $i=-1, 0, \ldots, r-1$, where $i=-1$ is the initialization $\mathcal{Y}^{(0)}_{-1}\times\mathcal{Y}^{(1)}_{-1}=\bit^{N\times b}\times\bit^{N\times b}$.
If $\mathcal{A}'$ halts early, we set subsequent $\mathcal{Y}^{(0)}_i\times \mathcal{Y}^{(1)}_i$ to be the same as the last one before halting.
Let $Y^{(0)}_i, Y^{(1)}_i$ be uniformly distributed over $\mathcal{Y}^{(0)}_i\times \mathcal{Y}^{(1)}_i$ (i.e., they are the $Y^{(0)}, Y^{(1)}$ in \Cref{alg:simulation} at the end of iteration $i$).
The idea is that, due to \Cref{lem:uniform-marginal}, we should be able to construct good coupling between $Y^{(0)}_i$ and $\tilde{Y}^{(0)}$, and also between $Y^{(1)}_i$ and $\tilde{Y}^{(1)}$.
However, we cannot use \Cref{lem:uniform-marginal} directly, because $\tilde{Y}^{(0)}$ and $\tilde{Y}^{(1)}$ may not be uniformly distributed over $G^{-1}(o)\cap (\mathcal{Y}^{(0)}_i\times \mathcal{Y}^{(1)}_i)$.
To address this issue, we introduce an intermediate random rectangle $\tilde{\mathcal{Y}}^{(0)}_i\times \tilde{\mathcal{Y}}^{(1)}_i$ coupled with $\tilde{Y}^{(0)}, \tilde{Y}^{(1)}$ such that the distribution of $\tilde{Y}^{(0)}, \tilde{Y}^{(1)}$ conditioned on a specific rectangle $\tilde{\mathcal{Y}}^{(0)}_i\times \tilde{\mathcal{Y}}^{(1)}_i$ is the uniform distribution over $G^{-1}(o)\cap (\tilde{\mathcal{Y}}^{(0)}_i\times \tilde{\mathcal{Y}}^{(1)}_i)$.
Then we show that with high probability, $\tilde{\mathcal{Y}}^{(0)}_i\times \tilde{\mathcal{Y}}^{(1)}_i$ coincides with $\mathcal{Y}^{(0)}_i\times \mathcal{Y}^{(1)}_i$.
This allows us to use \Cref{lem:uniform-marginal} and construct the desired coupling.

Concretely, we construct $\tilde{\mathcal{Y}}^{(0)}_i\times \tilde{\mathcal{Y}}^{(1)}_i$ and the relevant coupling by induction.
We set $\tilde{\mathcal{Y}}^{(0)}_{-1}\times \tilde{\mathcal{Y}}^{(1)}_{-1}=\mathcal{Y}^{(0)}_{-1}\times \mathcal{Y}^{(1)}_{-1} = \bit^{N\times b}\times \bit^{N\times b}$.
Suppose we have constructed $\tilde{\mathcal{Y}}^{(0)}_{i-1}\times \tilde{\mathcal{Y}}^{(1)}_{i-1}$ such that (1) the distribution of $\tilde{Y}^{(0)}, \tilde{Y}^{(1)}$ conditioned on a specific rectangle $\tilde{\mathcal{Y}}^{(0)}_{i-1}\times \tilde{\mathcal{Y}}^{(1)}_{i-1}$ is the uniform distribution over $G^{-1}(o)\cap (\tilde{\mathcal{Y}}^{(0)}_{i-1}\times \tilde{\mathcal{Y}}^{(1)}_{i-1})$, and (2) $\Pr[\tilde{\mathcal{Y}}^{(0)}_{i-1}\times \tilde{\mathcal{Y}}^{(1)}_{i-1}\neq \mathcal{Y}^{(0)}_{i-1}\times \mathcal{Y}^{(1)}_{i-1}]\leq 2^{-\eta b/8-1}\frac{i}{2Nb}$.
We construct $\tilde{\mathcal{Y}}^{(0)}_i\times \tilde{\mathcal{Y}}^{(1)}_i$ as follows.

Firstly, we sample $\tilde{\mathcal{Y}}^{(0)}_{i-1}\times \tilde{\mathcal{Y}}^{(1)}_{i-1}$ and $\mathcal{Y}^{(0)}_{i-1}\times \mathcal{Y}^{(1)}_{i-1}$.
If they are different, we set $\tilde{\mathcal{Y}}^{(0)}_i\times \tilde{\mathcal{Y}}^{(1)}_i = \{(\tilde{Y}^{(0)}, \tilde{Y}^{(1)})\}$ and say that the coupling fails.
Now suppose $\tilde{\mathcal{Y}}^{(0)}_{i-1}\times \tilde{\mathcal{Y}}^{(1)}_{i-1}=\mathcal{Y}^{(0)}_{i-1}\times \mathcal{Y}^{(1)}_{i-1}$.
If $\mathcal{A}'$ has halted at this point, we set $\tilde{\mathcal{Y}}^{(0)}_i\times \tilde{\mathcal{Y}}^{(1)}_i = \mathcal{Y}^{(0)}_{i-1}\times \mathcal{Y}^{(1)}_{i-1}$.
Then we proceed by following \Cref{alg:simulation} closely. Let $A, B$ be the corresponding values in iteration $i$.

In Line \ref{ln:remove-dangerous}, if either $Y^{(A)}_{i-1}$ or $\tilde{Y}^{(A)}_{i-1}$ is $\epsilon$-dangerous to $Y^{(B)}_{i-1}$, we set $\tilde{\mathcal{Y}}^{(0)}_i\times \tilde{\mathcal{Y}}^{(1)}_i = \{(\tilde{Y}^{(0)}, \tilde{Y}^{(1)})\}$.
We note that since $Y^{(A)}_{i-1}, Y^{(B)}_{i-1}$ are $(\rho, \tau)$-structured with
\begin{equation}
\begin{split}
    \tau &= 2\delta-\epsilon = 2(1-\eta/4+\epsilon/2)-\epsilon=2-\eta/2 \\
    &\geq 2+\eta/4 -\eta +\eta/8 \\
    &=2 + \frac{96\log c}{4c\epsilon}-\eta+\eta/8 \\
    &\geq 2+\frac{17}{c\epsilon} -\eta +\left(\eta/8+\frac{3\log c}{c} +\frac{4}{b}\right),
\end{split}
\end{equation}
where we have used $\epsilon=\frac{96\log c}{c\eta}$, $\epsilon\leq 1$, $b\geq c\log N\geq c$, and $\log(c)\geq \log(2)=1$.
Using \Cref{lem:dangerous-unlikely}, we have that the probability of $Y^{(A)}_{i-1}$ being $\epsilon$-dangerous to $Y^{(B)}_{i-1}$ is at most
\begin{equation}
\begin{split}
    2^{-(\eta/8+3\log(c)/c +4/b)b} &\leq \frac{1}{16}2^{-\eta b/8} 2^{-3b/c} \\
    &=\frac{1}{16}2^{-\eta b/8} 2^{-b/c - 2b/c} \\
    &\leq \frac{1}{16}2^{-\eta b/8} 2^{-\log(N) - \log(b)} \\
    & = \frac{1}{8}2^{-\eta b/8}\frac{1}{2Nb},
\end{split}
\end{equation}
where we have used $\log(c)\geq 1$, $b\geq c\log N$, and $2b\geq c\log(b)$ for large $N$.
Meanwhile, by \Cref{lem:uniform-marginal}, $Y^{(A)}_{i-1}$ is $(\frac{1}{8}2^{-\eta b/8}\frac{1}{2Nb})$-close to $\tilde{Y}^{(A)}_{i-1}$ in total variation distance.
Therefore, the total probability of this failure is at most $2\cdot\frac{1}{8}2^{-\eta b/8}\frac{1}{2Nb}$.
Suppose this failure does not happen and neither $Y^{(A)}_{i-1}$ nor $\tilde{Y}^{(A)}_{i-1}$ is $\epsilon$-dangerous to $Y^{(B)}_{i-1}$.
Since this event happens with probability larger than $1/2$, by \Cref{lem:cond-min-entropy}, conditioning random variables on this event decreases their min-entropy by at most $1$ bit and their density by at most $1/b$.
Therefore, after conditioning, $(Y^{(0)}_{i-1}, Y^{(1)}_{i-1})$ is still $(\rho, \tau-1/b)$-structured.
Note that here we are slightly abusing notation in reusing the same notation after conditioning.

Now we proceed to the steps of sampling $\mu_i$ and $j_i$.
Let $\mu_i, j_i$ be sampled based on $Y^{(A)}_{i-1}$.
Let $\tilde{\mu}_i, \tilde{j}_i$ be sampled based on $\tilde{Y}^{(A)}_{i-1}$.
Since $(Y^{(0)}_{i-1}, Y^{(1)}_{i-1})$ is still $(\rho, \tau-1/b)$-structured, by \Cref{lem:uniform-marginal}, we have that $Y^{(A)}_{i-1}$ and $\tilde{Y}^{(A)}_{i-1}$ are $(\frac{1}{8}2^{-\eta b/8}\frac{1}{2Nb})$-close to each other in total variation distance.
This implies that the sampled $(\mu_i, j_i)$ and $(\tilde{\mu}_i, \tilde{j}_i)$ are also $(\frac{1}{8}2^{-\eta b/8}\frac{1}{2Nb})$-close to each other.
Thus there exists a coupling such that
\begin{equation}
    \Pr[(\mu_i, j_i)\neq (\tilde{\mu}_i, \tilde{j}_i)]\leq \frac{1}{8}2^{-\eta b/8}\frac{1}{2Nb}.
\end{equation}
We choose the coupling so that, conditioned on the coupled values of $(\mu_i,j_i)$ and $(\tilde\mu_i,\tilde j_i)$, the variables $Y^{(A)}_{i-1}$ and $\tilde Y^{(A)}_{i-1}$ retain their original conditional distributions.
We sample from this coupling and if they differ, we set $\tilde{\mathcal{Y}}^{(0)}_i\times \tilde{\mathcal{Y}}^{(1)}_i = \{(\tilde{Y}^{(0)}, \tilde{Y}^{(1)})\}$.
Now suppose they are the same.
We proceed with \Cref{alg:simulation} and condition $Y^{(A)}_{i-1}$ and $\tilde{Y}^{(A)}_{i-1}$ on being consistent with $\mu_i, j_i$.
Lastly, if $q_{j_i}<\frac{1}{8}2^{-\eta b/8}\frac{1}{2Nb}$ (i.e., $\mathcal{A}'$ halts), we set $\tilde{\mathcal{Y}}^{(0)}_i\times \tilde{\mathcal{Y}}^{(1)}_i = \{(\tilde{Y}^{(0)}, \tilde{Y}^{(1)})\}$.
Note that the probability of this failure is at most $\frac{1}{8}2^{-\eta b/8}\frac{1}{2Nb}$ for the following reason.
Let $j_\star$ be the smallest index such that $q_{j_\star}<\frac{1}{8}2^{-\eta b/8}\frac{1}{2Nb}$, if such an index exists.
Since
\begin{equation}
    q_j
    =
    \Pr\left[
        Y^{(A)}_{\mathrm{free}(\rho)}
        \in
        \bigcup_{k\geq j}\mathcal{Y}^k
    \right]
\end{equation}
is non-increasing in $j$, the event $q_{j_i}<\frac{1}{8}2^{-\eta b/8}\frac{1}{2Nb}$ is exactly the event
$j_i\geq j_\star$. Therefore,
\begin{equation}
    \Pr\left[q_{j_i}<\frac{1}{8}2^{-\eta b/8}\frac{1}{2Nb}\right]
    =
    \Pr[j_i\geq j_\star]=
    \Pr\left[
        Y^{(A)}_{\mathrm{free}(\rho)}
        \in
        \bigcup_{k\geq j_\star}\mathcal{Y}^k
    \right]=
    q_{j_\star}
    <
    \frac{1}{8}2^{-\eta b/8}\frac{1}{2Nb}.
\end{equation}
Then we condition $Y^{(B)}_{i-1}, \tilde{Y}^{(B)}_{i-1}$ on being consistent with the oracle query.
Now we have completed a full iteration and we set $\tilde{\mathcal{Y}}^{(0)}_i$ and $\tilde{\mathcal{Y}}^{(1)}_i$ to be the support of the processed $Y^{(0)}_{i-1}$ and $Y^{(1)}_{i-1}$, respectively.
This completes the construction of $\tilde{\mathcal{Y}}^{(0)}_i, \tilde{\mathcal{Y}}^{(1)}_i$.

Now we prove that the induction properties are satisfied.
Since we have conditioned on the random variables with and without tildes being the same in every step, $\tilde{\mathcal{Y}}^{(0)}_i\times \tilde{\mathcal{Y}}^{(1)}_i =\mathcal{Y}^{(0)}_i\times \mathcal{Y}^{(1)}_i $ by construction if the coupling succeed.
The probability of failure is
\begin{equation}
\begin{split}
    \Pr[\tilde{\mathcal{Y}}^{(0)}_i\times \tilde{\mathcal{Y}}^{(1)}_i \neq\mathcal{Y}^{(0)}_i\times \mathcal{Y}^{(1)}_i]&\leq \Pr[\tilde{\mathcal{Y}}^{(0)}_{i-1}\times \tilde{\mathcal{Y}}^{(1)}_{i-1} \neq\mathcal{Y}^{(0)}_{i-1}\times \mathcal{Y}^{(1)}_{i-1}] + \left(2\cdot\frac18+\frac18+\frac18\right)2^{-\eta b/8}\frac{1}{2Nb} \\
    &\leq \left(\frac{i}{2} + \frac{1}{2}\right)2^{-\eta b/8}\frac{1}{2Nb} \\
    &=2^{-\eta b/8-1}\frac{i+1}{2Nb}
\end{split}
\end{equation}
as desired.
Finally, we show that conditioned on $\tilde{\mathcal{Y}}^{(0)}_i\times \tilde{\mathcal{Y}}^{(1)}_i$, $\tilde{Y}^{(0)}, \tilde{Y}^{(1)}$ are uniform on $G^{-1}(o)\cap(\tilde{\mathcal{Y}}^{(0)}_i\times \tilde{\mathcal{Y}}^{(1)}_i)$.
Note that when the coupling fails, this is automatically true since we have set $\tilde{\mathcal{Y}}^{(0)}_i\times \tilde{\mathcal{Y}}^{(1)}_i=\{(\tilde{Y}^{(0)}, \tilde{Y}^{(1)})\}$.
When the coupling succeed, since all the coupling does to $\tilde{Y}^{(0)}, \tilde{Y}^{(1)}$ is to condition them on being in $\tilde{\mathcal{Y}}^{(0)}_i\times \tilde{\mathcal{Y}}^{(1)}_i$.
Therefore, the induction hypothesis implies that the desired property is still true.
This completes the induction and we have proven that for all $i=0, \ldots, r-1$, it holds true that (1) the distribution of $\tilde{Y}^{(0)}, \tilde{Y}^{(1)}$ conditioned on a specific rectangle $\tilde{\mathcal{Y}}^{(0)}_{i}\times \tilde{\mathcal{Y}}^{(1)}_{i}$ is the uniform distribution over $G^{-1}(o)\cap (\tilde{\mathcal{Y}}^{(0)}_{i}\times \tilde{\mathcal{Y}}^{(1)}_{i})$, and (2) $\Pr[\tilde{\mathcal{Y}}^{(0)}_{i}\times \tilde{\mathcal{Y}}^{(1)}_{i}\neq \mathcal{Y}^{(0)}_{i}\times \mathcal{Y}^{(1)}_{i}]\leq 2^{-\eta b/8-1}\frac{i+1}{2Nb}$.

Finally, we compare the data generated in the last situation block $t_r,\ldots,M-1$.
Condition on all preceding couplings succeeding and on
$\tilde{\mathcal{Y}}^{(0)}_{r-1}
\times
\tilde{\mathcal{Y}}^{(1)}_{r-1}
=
\mathcal{Y}^{(0)}_{r-1}
\times
\mathcal{Y}^{(1)}_{r-1}$,
where for $r=0$ we use the initial rectangles indexed by $-1$.
Let $A=\alpha_{t_r}$ be the situation in the last block.
By \Cref{lem:uniform-marginal}, conditioned on this common rectangle,
the current fake encoding $Y^{(A)}$ and the corresponding ideal encoding $\tilde{Y}^{(A)}$ are $\frac{1}{8}2^{-\eta b/8}\frac{1}{2Nb}$-close in total variation distance.
Hence, they can be coupled such that $\Pr[Y^{(A)}\neq\tilde{Y}^{(A)}]\leq\frac{1}{8}2^{-\eta b/8}\frac{1}{2Nb}$.
If this final coupling succeeds, then, since $\alpha$, $x$, and the learning algorithm are fixed, the fake and ideal procedures feed exactly the same samples in the last block and produce the same final output.
Therefore,
\begin{equation}
    \Pr[
        \pi^\star(\alpha,x)
        \neq
        \pi'(\alpha,x)
    ]
    \leq
    2^{-\eta b/8-1}
    \frac{r}{2Nb}
    +
    \frac{1}{8}
    2^{-\eta b/8}
    \frac{1}{2Nb}\leq
    2^{-\eta b/8-1},
\end{equation}
where we used $r\leq \ceil{M/T}-1\leq 2Nb-1$.
Averaging over the fixed distributions of $\alpha$ and $x$ gives
\begin{equation}
    d_{\mathrm{TV}}(
        \pi',
        \pi^\star
    )
    \leq
    2^{-\eta b/8-1}.
\end{equation}
Together with $d_{\mathrm{TV}}(\pi,\pi')\leq2^{-\eta b/8-1}$, this completes the proof that $\pi$ and $\pi^\star$ are $2^{-\eta b/8}$-close in total variation distance.

\subsubsection{Simulation does not query too much}
\label{sec:cl-hard-sim-complex}

In this section, we show that the query $\mathcal{A}$ makes is bounded by the sample-space product.
To bound the query complexity of \Cref{alg:simulation}, we keep track of the deficiency of $Y^{(0)}, Y^{(1)}$ through out the simulation.
Initially, since $\mathcal{Y}^{(0)}=\mathcal{Y}^{(1)}=\bit^{N\times b}$ and $\rho = *^N$, we have $D_\infty(Y^{(0)}, Y^{(1)}, \rho) = 2bN-bN-bN=0$.

Now we go through the steps of \Cref{alg:simulation}.
In Line \ref{ln:remove-dangerous}, we condition $Y^{(A)}$ on being not $\epsilon$-dangerous to $Y^{(B)}$, which changes the distribution of $Y^{(A)}$.
In \Cref{sec:cl-hard-sim-correctness}, we have shown that the probability of not being dangerous is at least $1/2$.
Thus by \Cref{lem:cond-min-entropy}, the min-entropy of $Y^{(A)}$ decreases by at most $1$ bit.
Then we condition $Y^{(A)}$ on being consistent with reaching vertex $v$ labeled by $\mu_i$, which happens with probability $p_i$.
This decreases the min-entropy of $Y^{(A)}$ by at most $\log(1/p_i)$.
Therefore, up to this point, the deficiency increases by at most $\log(1/p_i)+1$.

Next, we proceed to the density restoring step.
We first condition $Y^{(A)}$ on being in the selected subset $\mathcal{Y}^{j_i}$.
From \Cref{lem:density-restoring}, we know that this conditioning decreases the min-entropy of $Y^{(A)}$ by at most $\delta b |I_{j_i}|+\log(1/q_{j_i})$ bits.
Then we query the oracle on $I_{j_i}$ and fix the corresponding coordinates in $\rho, Y^{(A)}, Y^{(B)}$.
This decreases $|\mathrm{free}(\rho)|$ by $|I_{j_i}|$.
Also, from \Cref{lem:marginal-min-entropy}, the min-entropy of $Y^{(B)}$ decreases by at most $b|I_{j_i}|$ while the min-entropy of $Y^{(A)}$ does not change since it is already fixed on $I_{j_i}$.
Finally, we condition $Y^{(B)}$ on being consistent with $Y^{(A)}_{I_{j_i}}$ and $\rho_{I_{j_i}}$ and change the distribution of $Y^{(B)}$.
Since $Y^{(A)}_{\mathrm{free}(\rho)}$ is not $\epsilon$-dangerous to $Y^{(B)}_{\mathrm{free}(\rho)}$, by definition we have $\Pr[g^{I_{j_i}}(y_{I_{j_i}}, y^B_{I_{j_i}})= \rho_{I_{j_i}}]\geq 2^{-|I_{j_i}|-1}$.
Thus this conditioning decreases the min-entropy of $Y^{(B)}$ by at most $|I_{j_i}|+1$ bits.
This completes the density restoring step and the decrease in deficiency is at least
\begin{equation}
    -(\delta b |I_{j_i}|+\log(1/q_{j_i}))
    +2b|I_{j_i}|
    -b|I_{j_i}|
    -( |I_{j_i}|+1) 
    =
    \left(
        1-\delta-\frac{1}{b}
    \right)
    b|I_{j_i}|
    -
    \log\frac{1}{q_{j_i}}
    -1.
\end{equation}
Since $b\geq c$ and $\eta>0$, we have
\begin{equation}
    1-\delta-\frac{1}{b}
    \geq
    1-\delta-\frac{8}{c}-\frac{\eta}{8}.
\end{equation}
Moreover, using $\delta=1-\eta/4+\epsilon/2$, $\epsilon=(96\log c)/(c\eta)$, $0<\eta\leq2$, and $c\geq2$, we have
\begin{equation}
\begin{split}
    1-\delta-\frac{8}{c}-\frac{\eta}{8}
    &=
    \frac{\eta}{4}
    -
    \frac{\epsilon}{2}
    -
    \frac{8}{c}
    -
    \frac{\eta}{8} \\
    &=
    \frac{\eta}{8}
    -
    \frac{48\log c}{c\eta}
    -
    \frac{8}{c} \\
    &\geq
    \frac{\eta}{8}
    -
    \frac{48\log c+16}{c\eta} \\
    &\geq
    \frac{\eta}{8}
    -
    \frac{64\log c}{c\eta}
    \geq
    \frac{\eta}{6920},
\end{split}
\end{equation}
if we choose
$c=\frac{865}{\eta^2}\log(\frac{865}{\eta^2})$.
Therefore, the decrease in deficiency during the density restoring
step is at least
\begin{equation}
    \frac{\eta b}{6920}|I_{j_i}|
    -
    \log\frac{1}{q_{j_i}}
    -1.
\end{equation}
Note that this bound also holds when $I_{j_i}=\emptyset$.

To summarize, in a single complete iteration, the deficiency is increased by at most
\begin{equation}
    \log\frac{1}{p_i}
    +
    \log\frac{1}{q_{j_i}}
    +
    2
    -
    \frac{\eta b}{6920}|I_{j_i}|.
\end{equation}
For the analysis only, let
\begin{equation}
    K_s
    =
    \sum_{i=0}^{s-1}\log\frac{1}{p_i},
    \qquad
    L_s
    =
    \sum_{i=0}^{s-1}\log\frac{1}{q_{j_i}},
    \qquad
    Q_s
    =
    \sum_{i=0}^{s-1}|I_{j_i}|,
\end{equation}
for every $s$ such that the first $s$ iterations of $\mathcal{A}'$ are completed.
Since the initial deficiency is zero, after any such $s$ iterations the deficiency is upper bounded by
\begin{equation}
    D_\infty
    \leq
    K_s+L_s+2s
    -
    \frac{\eta b}{6920}Q_s.
\end{equation}
Suppose neither $\mathcal{E}_{K}$ nor $\mathcal{E}_{L}$ occurs.
Since $K_s,L_s$ are non-decreasing and $s\leq r$, from the definition of $\mathcal{E}_{K}$ and $\mathcal{E}_{L}$, we have
\begin{equation}
    K_s\leq rS+b,
    \qquad
    L_s\leq2r+2b.
\end{equation}
Therefore,
\begin{equation}
    0
    \leq
    D_\infty
    \leq
    rS+4r+3b
    -
    \frac{\eta b}{6920}Q_s,
\end{equation}
and hence
\begin{equation}
    Q_s
    \leq
    \frac{6920}{\eta}
    \left(
        \frac{rS+4r+3b}{b}
    \right)
    \leq
    Q_{\max}.
\end{equation}
Thus, if neither $\mathcal{E}_{K}$ nor $\mathcal{E}_{L}$ occurs, the
query-budget condition in
Lines \ref{ln:check-query}-\ref{ln:check-query-end} is never violated.
This proves
\begin{equation}
    \mathcal{E}_{Q}
    \subseteq
    \mathcal{E}_{K}\cup\mathcal{E}_{L},
\end{equation}
as used in \Cref{sec:cl-hard-sim-correctness}.

By construction, the worst-case query complexity of
$\mathcal{A}$ is at most
\begin{equation}
\begin{split}
    Q_{\max}
    &=
    \ceil{
    \frac{6920}{\eta}
    \left(
        \frac{rS+4r+3b}{b}
    \right)
    } \\
    &=
    O\left(
        \frac{rS}{b}
        +
        \frac{r}{b}
        +
        1
    \right) \\
    &=
    O\left(
        \frac{MS}{Tb}
        +
        1
    \right),
\end{split}
\end{equation}
where we used $r\leq M/T$ and $S\geq1$.
This completes the proof of \Cref{thm:simulation}.

\subsubsection{Proof of \texorpdfstring{\Cref{thm:classical-lower-bound}}{Theorem \ref{thm:classical-lower-bound}}}
\label{sec:cl-hard-sim-combine}

Finally, we are ready to prove \Cref{thm:classical-lower-bound}.
\begin{proof}[Proof of \Cref{thm:classical-lower-bound}]
    Let $\mathcal{L}$ be any randomized classical learning algorithm with space complexity $S$, sample complexity $M$, and input form $\mathcal{I} = [N]\times \bit^b\times \bit$ such that given data generated from $\mathcal{D}_{g, T}^N(o)$, $\mathcal{L}$ outputs $f(o)$ with probability at least $1-\delta + 2^{-\eta b/8}$.
    Since $f:\bit^N\to \bit$, its query complexity satisfies $Q_C^\delta \leq N$.
    If $M>2NTb$, then we immediately have 
    \begin{equation}
        MS\geq M>2NTb\geq 2Q_C^\delta Tb = \Omega( Q_C^\delta T b)
    \end{equation}
    as required.

    Suppose otherwise that $M\leq 2NTb$.
    From \Cref{thm:simulation}, we know that there exists a randomized parallel decision tree $\mathcal{A}$ with query complexity $Q=O(MS/(Tb)+1)$ such that given any input $o\in \bit^N$, $\mathcal{A}$ outputs a random bitstring $\pi$ whose distribution is $2^{-\eta b/8}$-close in total variation distance to the distribution of the transcript of $\mathcal{L}$ when given data generated from $\mathcal{D}^N_{g, T}(o)$.
    In particular, let $a$ be the last bit of the transcript of $\mathcal{L}$, which is the output of $\mathcal{L}$.
    Then the last bit of $\pi$ must be equal to $a$ with probability at least $1-2^{-\eta b/8}$.
    On the other hand, $\Pr[a=f(o)]\geq 1-\delta+2^{-\eta b/8}$ from the guarantee of $\mathcal{L}$.
    Thus we have that the last bit of $\pi$ is equal to $f(o)$ with probability at least $1-\delta+2^{-\eta b/8} - 2^{-\eta b/8}=1-\delta$ by the union bound.
    This gives us a query algorithm (executing $\mathcal{A}$ and outputting the last bit of its output) with query complexity $Q=O(MS/(Tb)+1)$ that outputs $f(o)$ with probability at least $1-\delta$.
    From the definition of $\delta$-error classical randomized query complexity, we have
    \begin{equation}
        Q_C^\delta \leq Q = O\left(\frac{MS}{Tb}+1\right).
    \end{equation}
    Since $Q_C^\delta$ is larger than a sufficiently large constant, we arrive at $MS\geq \Omega(Q_C^\delta T b)$, as required.
    This completes the proof of \Cref{thm:classical-lower-bound}.
\end{proof}

\subsection{Bootstrap with one more time scale}
\label{sec:cl-hard-bootstrap}

In this section, we build upon \Cref{thm:classical-lower-bound} and bootstrap it to a much stronger lower bound by introducing a new task called dynamic Noisy Oracle Property Estimation (dynamic NOPE).
We design this new task by adding one more time scale into the original NOPE.
We call it dynamic NOPE, because the oracle becomes dynamic: it changes over time, but the property that we want to estimate stays fixed.

Recall that our interested oracle property is given by a property function $f: \mathcal{O} \to \bit, \quad \mathcal{O}\subseteq \bit^N$.
To make the oracle dynamic, we consider randomly sampled oracles from two distributions $p_0, p_1$ supported on $\mathcal{O}\subseteq \bit^N$, where $o\sim p_B$ satisfies $f(o)=B$ for $B\in\bit$.
The relevant notion of query complexity in this case is the distributional query complexity with a uniform prior on the two distributions.
More precisely, the $\epsilon$-error classical distributional query complexity with respect to $p_0,p_1$, denoted by $Q_C^\epsilon$, is the minimum number of queries made by a randomized classical query algorithm $\mathcal{A}$ satisfying
\begin{equation}
    \Pr_{\substack{\gamma\sim\unif(\bit)\\o\sim p_\gamma}}
    [\mathcal{A}(o)=\gamma]
    \geq 1-\epsilon,
\end{equation}
where the probability also includes the internal randomness of
$\mathcal{A}$.
We write $Q_C=Q_C^{1/3}$.
We introduce noise in the same way as before.
We consider a noisy encoding function $g:\bit^b\times\bit^b\to\bit$ with encoding length $b=\ceil{c\log N}$ and discrepancy $\disc(g)\leq 2^{-\eta b}$, where $c=(865/\eta^2)\log(865/\eta^2)$ for some constant $\eta\in (0.1, 2]$.
We will call such encoding functions \emph{good noisy encoding functions}.

In a dynamic setting, we introduce two time scales $T_1, T_2$ and another parameter $L$ that characterize how dynamic the oracle can be.
They are positive integers and their relation with $N$ will be specified later.
We consider a $2$-level hierarchical data generation process $\mathcal{D}_{g, f}^{N}(B)$ that depends on the binary property $B\in\bit$ we want to estimate:
\begin{equation}
    \mathcal{D}_{g, f}^{N, T_1}(B) = (\mathcal{D}^0_B\to \mathcal{D}_{(\gamma_j, o_j, Y^{(0,j)}, Y^{(1,j)})_{j=1}^{L}}^1 \to^{\times T_1} \mathcal{D}^2_{(\beta, \alpha, Y^{(\alpha,\beta)})} \to^{\times T_2} z),
\end{equation}
where we have $L$ sampled oracles $o_j\in \bit^N, j\in [L]$, each with its own property $\gamma_j =f(o_j)\in \bit$ and noisy encoding $Y^{(0,j)}, Y^{(1,j)}\in\bit^{N\times b}$.
The situation $\beta\in [L]$ labels which oracle we are currently getting samples from, and $\alpha\in \bit$ specifies which part of the noisy encoding we are looking at.
The data sample is of the form
\begin{equation}
    z_i=(x_i, y_i, \alpha_i, \beta_i), \quad x_i\in [N], \quad y_i\in \bit^b,
\end{equation}
where $x_i$ is a random query and $y_i$ is the corresponding noisy oracle value.

We set $L=\ceil{\log^2 N}\geq 5$ and $T_2=N$.
$T_1$ is tunable but scales as $T_1=\polylog(N)$.
The sampling distributions are defined as follows.
$\mathcal{D}^0_{B}$ samples a length-$L$ bitstring $\gamma$ with parity 
\begin{equation}
    \bigoplus_{j=1}^{L}\gamma_{j}=B
\end{equation} 
uniformly random.
This means that the final property that we want to estimate is the XOR of the properties of all the $L$ oracles.
For each $j\in [L]$, we set $\gamma_j$ to be the $j$-th bit of the sampled bitstring $\gamma$, sample a random oracle $o_j\sim p_{\gamma_j}$ and sample a noisy encoding pair $(Y^{(0, j)}, Y^{(1, j)})\sim \unif((g^N)^{-1}(o_j))$.
These meta-data $(\gamma_j, o_j, Y^{(0,j)}, Y^{(1,j)})_{j=1}^{L}$ will be used to generate the data samples.
Now we define $\mathcal{D}^1_{(\gamma_j, o_j, Y^{(0,j)}, Y^{(1,j)})_{j=1}^{L}}$.
We first sample a uniformly random oracle label $\beta\sim\unif([L])$ and a random bit $\alpha\sim\mathrm{Bern}(1/2)$.
Then we pick out $Y^{(\alpha, \beta)}$ and use it to generate the data samples.
In particular, we define $\mathcal{D}^2_{(\beta, \alpha, Y^{(\alpha, \beta)})}$ as sampling a random query $x\sim \unif([N])$ and its corresponding noisy oracle value $y = Y^{(\alpha, \beta)}_{x}\in \bit^b$.
This gives us a data sample $z = (x, y, \alpha, \beta)$.

The task of dynamic NOPE, is to estimate the binary property $B$ using these data samples.

\begin{tcolorbox}
\begin{task}[Dynamic Noisy Oracle Property Estimation (dynamic NOPE)]
\label{task:dynamic-nope}
    Let $N, T_1$ be positive integers.
    Let $f: \mathcal{O} \to \bit$ be a function that specifies the target property for a set of possible oracles $\mathcal{O}\subseteq\bit^N$.
    Let $g: \bit^b \times \bit^b \to \bit$ be a noisy encoding function.
    The task of dynamic Noisy Oracle Property Estimation is to calculate the desired property $B\in \bit$ using data samples generated from the noisy hierarchical data generation process $\mathcal{D}^{N, T_1}_{g, f}(B)$, for any $B\in \bit$.
\end{task}
\end{tcolorbox}

Intuitively, in dynamic NOPE, the target property $B$ that we are trying to estimate is the parity of the length-$L$ bitstring $\gamma$.
In order to predict the parity of $\gamma$, we effectively need to predict all its $L$ components $\gamma_j, j\in [L]$ simultaneously.
Meanwhile, information about $\gamma_j$ is encoded as the property of the generated oracle $o_j$.
Hence, we are trying to solve $L$ independent instances of NOPE described in \Cref{thm:classical-lower-bound}.

Dynamic NOPE is hard to solve, because \Cref{thm:classical-lower-bound} tells us that if the classical learning algorithm does not have enough memory size (determined by $N, Q_C$), its success probability on a single NOPE instance is at most $2/3+o(1)\leq 1/2+\delta$ for some $\delta$.
We will build upon this and show that its success probability of solving dynamic NOPE is at most $1/2 + (2\delta)^{L} = 1/2 + N^{-\Omega(\log(N))}$.
Finally, we use a hybrid argument to convert this success probability upper bound into a sample complexity lower bound of $N^{\Omega(\log N)}=N^{\omega(1)}$.
Together, this means that any classical learning algorithm with insufficient memory would need a tremendous amount of samples to solve dynamic NOPE.

Formally, we prove the following two classical hardness results.
They hold for any $T_1$ of our choice.
The first result shows that given only the data from one refreshing block (i.e., $M = \tau_{\mathcal{D}}$ where $\tau_{\mathcal{D}}=T_1T_2$ is the refreshing time of $\mathcal{D}^{N, T_1}_{g, f}(B)$), any classical learning algorithm with insufficient size cannot perform much better than random guessing at dynamic NOPE.

\begin{tcolorbox}
\begin{theorem}[Classical single-block hardness of dynamic NOPE]
\label{thm:classical-lower-bound-single-block}
    Let $N, T_1$ be large integers.
    Let $f: \mathcal{O}\to \bit, \mathcal{O}\subseteq \bit^N$ be a function that specifies the target property with $(1/3)$-error classical distributional query complexity $Q_C\geq \Omega(T_1^2 \log^2(N)\log\log(N))$.
    Let $g$ be a good noisy encoding function.
    Then, for any randomized classical learning algorithm $\mathcal{L}$ with sample complexity $M=\tau_{\mathcal{D}}$, if its space complexity
    \begin{equation}
        S\leq o\lr{\frac{Q_C}{T_1^2 \log^2(N)}},
    \end{equation}
    its success probability of solving dynamic NOPE is at most
    \begin{equation}
        \frac{1}{2} + \frac{1}{N^{\omega(1)}}.
    \end{equation}
\end{theorem}
\end{tcolorbox}

The second result shows that if we want to solve dynamic NOPE with high probability, but our classical machine does not have sufficient size, we must collect a super-polynomial amount of samples.

\begin{tcolorbox}
\begin{theorem}[Classical sample complexity of dynamic NOPE]
\label{thm:classical-lower-bound-superpoly-sample}
    Let $N, T_1$ be large integers.
    Let $f: \mathcal{O}\to \bit, \mathcal{O}\subseteq \bit^N$ be a function that specifies the target property with $(1/3)$-error classical distributional query complexity $Q_C\geq \Omega(T_1^2 \log^2(N)\log\log(N))$.
    Let $g$ be a good noisy encoding function.
    Then, for any randomized classical learning algorithm $\mathcal{L}$ that solves dynamic NOPE with probability at least $2/3$, if its space complexity
    \begin{equation}
        S\leq o\lr{\frac{Q_C}{T_1^2 \log^2(N)}},
    \end{equation}
    it must have sample complexity
    \begin{equation}
        M\geq \tau_{\mathcal{D}}N^{\omega(1)},
    \end{equation}
    where $\tau_{\mathcal{D}}=T_1N$ is the refreshing time of dynamic NOPE.
\end{theorem}
\end{tcolorbox}

The remaining parts of this section are devoted to prove \Cref{thm:classical-lower-bound-single-block,thm:classical-lower-bound-superpoly-sample}.
Since we are randomly sampling oracles, we first prove a distributional version of the sample-space lower bound in \Cref{sec:cl-hard-bootstrap-distributional}.
Then, in \Cref{sec:cl-hard-bootstrap-xor,sec:cl-hard-bootstrap-xor-derandom,sec:cl-hard-bootstrap-xor-derandom-xor,sec:cl-hard-bootstrap-xor-finish}, we use a derandomization technique to prove a learning XOR lemma that suppresses the advantage of any algorithm exponentially.
In \Cref{sec:cl-hard-bootstrap-adv-to-sample}, we use a hybrid argument to show that an super-polynomially small advantage leads to a super-polynomial sample complexity.
Finally, we complete the proof in \Cref{sec:cl-hard-bootstrap-finish}.

\subsubsection{Distributional sample-space lower bound}
\label{sec:cl-hard-bootstrap-distributional}

We begin by noting that the sample-space lower bound we prove for NOPE (\Cref{thm:classical-lower-bound}) applies only to algorithms that are required to work for any input oracle $o\in\bit^N$.
But in dynamic NOPE, the learning algorithm only needs to succeed on average for random oracles drawn from $p_0, p_1$.
Nevertheless, we can still prove a sample-space lower bound as follows by replacing randomized query complexity with distributional query complexity.

\begin{lemma}[Distributional sample-space lower bound]
\label{lem:classical-lower-bound-dist}
    Let $N$ be a large integer. 
    Let the time scale $T$ be a positive integer.
    Let $\eta\in (0, 2]$ be a constant and $c=(865/\eta^2)\log(865/\eta^2)$.
    Let $b\geq c\log N$.
    Let $g:\bit^b\times \bit^b\to \bit$ be an encoding map with $\disc(g)\leq 2^{-\eta b}$.
    Let $f: \mathcal{O}\to \bit, \mathcal{O}\subseteq \bit^N$ be any function with $\delta$-error classical distributional query complexity $Q_C^\delta$ with respect to the distributions $p_0, p_1$ supported on $\mathcal{O}$ that is larger than a sufficiently large constant.
    Let $\mathcal{L}$ be any randomized classical learning algorithm with space complexity $S$, sample complexity $M$, and input form $\mathcal{I} = [N]\times \bit^b\times \bit$ such that, given
    $\gamma\sim\unif(\bit)$, $o\sim p_\gamma$, and data generated from
    $\mathcal{D}_{g,T}^N(o)$, $\mathcal{L}$ outputs $\gamma$ with probability at least $1-\delta + 2^{-\eta b/8}$.
    Then, $\mathcal{L}$ must satisfy
    \begin{equation}
        MS\geq \Omega(Q_C^\delta Tb).
    \end{equation}
\end{lemma}

\begin{proof}[Proof of \Cref{lem:classical-lower-bound-dist}]
    The proof resembles the proof of \Cref{thm:classical-lower-bound}.
    Let $\mathcal{L}$ be any randomized classical learning algorithm with space complexity $S$, sample complexity $M$, and input form $\mathcal{I} = [N]\times \bit^b\times \bit$ such that, given
    $\gamma\sim\unif(\bit)$, $o\sim p_\gamma$, and data generated from
    $\mathcal{D}_{g,T}^N(o)$, $\mathcal{L}$ outputs $\gamma$ with probability at least $1-\delta + 2^{-\eta b/8}$.
    Since $f:\bit^N\to \bit$, its distributional query complexity satisfies $Q_C^\delta \leq N$.
    If $M>2NTb$, then we immediately have $MS\geq M>2NTb\geq 2Q_C^\delta Tb = \Omega(Q_C^\delta T b)$ as required.

    Suppose otherwise that $M\leq 2NTb$.
    From \Cref{thm:simulation}, we know that there exists a randomized parallel decision tree $\mathcal{A}$ with query complexity $Q=O(MS/(Tb)+1)$ such that given any input $o\in \bit^N$, $\mathcal{A}$ outputs a random bitstring $\pi$ whose distribution is $2^{-\eta b/8}$-close in total variation distance to the distribution of the transcript of $\mathcal{L}$ when given data generated from $\mathcal{D}^N_{g, T}(o)$.
    In particular, let $a$ be the last bit of the transcript of $\mathcal{L}$, which is the output of $\mathcal{L}$.
    Then the last bit of $\pi$ must be equal to $a$ with probability at least $1-2^{-\eta b/8}$.
    On the other hand, $\Pr[a=\gamma]\geq 1-\delta+2^{-\eta b/8}$.
    Thus we have that the last bit of $\pi$ is equal to $\gamma$ with probability at least $1-\delta+2^{-\eta b/8} - 2^{-\eta b/8}=1-\delta$.
    Averaging over $\gamma\sim\unif(\bit)$, this gives us a query algorithm (executing $\mathcal{A}$ and outputting the last bit of its output) with query complexity $Q=O(MS/(Tb)+1)$ that satisfies
    \begin{equation}
        \Pr_{\substack{\gamma\sim\unif(\bit)\\o\sim p_\gamma}}
        [\mathcal{A}(o)=\gamma]
        \geq
        1-\delta.
    \end{equation}
    From the definition of $\delta$-error classical distributional query complexity, we have
    \begin{equation}
        Q_C^\delta \leq Q = O\left(\frac{MS}{Tb}+1\right).
    \end{equation}
    Since $Q_C^\delta$ is larger than a sufficiently large constant, we arrive at $MS\geq \Omega(Q_C^\delta T b)$, as required.
\end{proof}

In the following, we focus on the case of $T=N$ as in dynamic NOPE.

\subsubsection{Learning XOR Lemma}
\label{sec:cl-hard-bootstrap-xor}

\Cref{lem:classical-lower-bound-dist} shows that if our classical machine does not have enough size, it cannot solve a single problem instance of NOPE.
In dynamic NOPE, we further reduce the advantage (i.e., success probability increase over random guessing) exponentially by requiring the algorithm to predict the XOR of $L$ instances.
This is formalized in the following result that we call the learning XOR lemma.

\begin{tcolorbox}
\begin{lemma}[Learning XOR lemma]
\label{lem:xor-lemma}
    Let $N$ be a large integer. 
    Let $T_1, L$ be positive integers.
    Let $\eta\in (0, 2]$ be a constant and $c=(865/\eta^2)\log(865/\eta^2)$.
    Let $g:\bit^b\times \bit^b\to \bit$ be a noisy encoding function with encoding length $b\geq c\log N$ and discrepancy $\disc(g)\leq 2^{-\eta b}$.
    Let $f: \mathcal{O}\to \bit, \mathcal{O}\subseteq\bit^N$ be any function.
    Suppose that any randomized classical learning algorithm $\mathcal{L}$ with space complexity $T_1L(S+\ceil{\log L})$, sample complexity $T_1N$, and input form $\mathcal{I}=[N]\times \bit^b\times \bit$, given data generated from $\mathcal{D}^N_{g, N}(o), o\sim p_\gamma, \gamma\sim \unif(\bit)$, cannot output $\gamma\in\bit$ with success probability more than $1/2+\delta$.
    Then, any randomized classical learning algorithm $\mathcal{L}^\oplus$ with space complexity $S$, sample complexity $T_1N$, and input form $\mathcal{I}\times [L]$, given data generated from $(\mathcal{D}^1_{(\gamma_j, o_j, Y^{(0, j)}, Y^{(1, j)})_{j=1}^L}\to^{\times T_1} \mathcal{D}^2_{(\beta, \alpha, Y^{(\alpha, \beta)})}\to^{\times N} z)$ as in dynamic NOPE, cannot output $B=\bigoplus_{j=1}^L\gamma_j\in\bit$ with success probability more than
    \begin{equation}
        \frac{1}{2} + \frac{(2\delta)^L}{2}
    \end{equation}
\end{lemma}
\end{tcolorbox}

There are two challenges in proving \Cref{lem:xor-lemma}.
The first challenge is that the random sampling of $\beta$ may allow the algorithm to see more samples that belong to the same problem instance labeled by $\beta$.
For example, if two consecutive $\beta$'s are the same, then the algorithm effectively have twice as many samples, and thus have a higher success probability than usual.
The second challenge is that the $L$ problem instances are actually not independent, because the algorithm can carry information through the computation and may be able to solve several instances in a joint way.
In other words, the algorithm does not have to approach each problem instances independently.
The random sampling of $\beta$'s, moreover, allows the algorithm to introduce more correlation between problem instances, as compared to a case where streams of different instances are concatenated sequentially in an adversarial order (cf. the streaming XOR lemma as in \cite[Theorem 1]{assadi2021graph}).

To overcome these challenges, we proceed in two steps to prove \Cref{lem:xor-lemma}.
Firstly, we derandomize $\beta$ by showing that any learning algorithm $\mathcal{L}^\oplus$ can be used to construct a learning algorithm $\mathcal{L}'$ with at least the same success probability, but increased space complexity $S+\ceil{\log L}$ and sample complexity $T_1NL$, for a different data generation process where $\beta$ appears in a fixed ordering $\beta=1, \ldots, L, 1, \ldots, L, \ldots$ rather than randomly sampled (there are still $N$ samples under each $\beta$).
Next, we show that the learning algorithm $\mathcal{L}'$ cannot have success probability more than $1/2+(2\delta)^L/2$ even if it can correlate different instances, by analyzing a corresponding communication problem that takes into account the correlation between different problem instances generated by the algorithm.
Together, this shows that the success probability of $\mathcal{L}^{\oplus}$ cannot exceed $1/2+(2\delta)^L/2$.

\subsubsection{Derandomization}
\label{sec:cl-hard-bootstrap-xor-derandom}

In the first step of proving \Cref{lem:xor-lemma}, we consider the data generation process $\mathcal{D}^{\mathrm{order}}_{(\gamma_j)_{j=1}^L}$ that generates $T_1NL$ samples $z_1, \ldots, z_{T_1NL}$ as follows.
Instead of sampling random $\beta\in[L]$ and $T_1N$ samples as in $(\mathcal{D}^1_{(\gamma_j, o_j, Y^{(0, j)}, Y^{(1, j)})_{j=1}^L}\to^{\times T_1} \mathcal{D}^2_{(\beta, \alpha, Y^{(\alpha, \beta)})}\to^{\times N} z)$, we sample $T_1N$ samples $z_1^\beta, \ldots, z_{T_1N}^\beta$ in the same way but for every $\beta\in [L]$.
This means that for a given $\beta$, $z^\beta_1, \ldots, z^\beta_{T_1N}$ have the same $\beta$ values, while their $\alpha$ values are re-sampled after each $N$ consecutive samples.
This gives us $LT_1N$ samples in total. 
We order them in a round-robin way (in terms of $\beta$) such that the $\beta$'s appear sequentially as
\begin{equation}
\label{eq:derandom-beta}
    \beta = \underbrace{\underbrace{1, \ldots, 1}_{N}, \underbrace{2, \ldots, 2}_N, \ldots, \underbrace{L, \ldots, L}_N}_{NL}, \underbrace{1, \ldots, 1}_N, \ldots
\end{equation}
and within the data samples that have the same $\beta$ value, their ordering stays the same.
More formally, we define 
\begin{equation}
    z_{(a-1)NL+(b-1)N+c} = z^b_{(a-1)N+c}, \quad \forall a\in[T_1], b\in[L], c\in [N].
\end{equation}
This defines the data generation process $\mathcal{D}^{\mathrm{order}}_{(\gamma_j)_{j=1}^L}$ with derandomized $\beta$.

Now, we use $\mathcal{L}^\oplus$ to construct $\mathcal{L}'$ that process data from $\mathcal{D}^{\mathrm{order}}_{(\gamma_j)_{j=1}^L}$ and prove the following lemma.

\begin{lemma}[Derandomize $\beta$]
\label{lem:derandomize-beta}
    Let $\mathcal{L}^\oplus$ be any randomized classical learning algorithm with space complexity $S$ and sample complexity $T_1N$ that given data generated from $(\mathcal{D}^1_{(\gamma_j, o_j, Y^{(0, j)}, Y^{(1, j)})_{j=1}^L}\to^{\times T_1} \mathcal{D}^2_{(\beta, \alpha, Y^{(\alpha, \beta)})}\to^{\times N} z)$ as in dynamic NOPE, outputs $B=\bigoplus_{j=1}^L \gamma_j \in\bit$ with probability $p_{\mathrm{succ}}$.
    Then, there exists a randomized classical learning algorithm $\mathcal{L}'$ with space complexity $S+\ceil{\log L}$ and sample complexity $T_1NL$ that given data generated from $\mathcal{D}^{\mathrm{order}}_{(\gamma_j)_{j=1}^L}$, outputs $B=\bigoplus_{j=1}^L \gamma_j \in\bit$ with the same probability $p_{\mathrm{succ}}$.
\end{lemma}

\begin{proof}[Proof of \Cref{lem:derandomize-beta}]
    We prove this lemma by explicitly constructing the learning algorithm $\mathcal{L}'$ from $\mathcal{L}^\oplus$.
    The constructed algorithm $\mathcal{L}'$ process the data in chunks of size $NL$.
    At the beginning of each chunk, $\mathcal{L}'$ randomly selects a $\beta\sim \unif([L])$ and keeps it in its memory.
    This uses $\ceil{\log L}$ bits of memory.
    It waits until the data block of size $N$ that corresponds to this particular $\beta$ arrives, and then feeds these $N$ samples sequentially into $\mathcal{L}^\oplus$, which operates on the remaining memory of size $S$.
    After processing these $N$ samples, $\mathcal{L}'$ waits until the end of this $NL$-size chunk and erases its selection of $\beta$.
    Then it proceeds to the next $NL$-size chunk and randomly selects a new $\beta$.
    After processing all the $T_1$ chunks of size $NL$, $\mathcal{L}'$ has already given $T_1 N$ data samples to $\mathcal{L}^{\oplus}$ and $\mathcal{L}^{\oplus}$ generates an output bit in return, which is then outputted by $\mathcal{L}'$ as the final outcome.
    Note that since the $\beta$'s are selected uniformly, the data samples that are fed into $\mathcal{L}^\oplus$ follow exactly the same distribution as $(\mathcal{D}^1_{(\gamma_j, o_j, Y^{(0, j)}, Y^{(1, j)})_{j=1}^L}\to^{\times T_1} \mathcal{D}^2_{(\beta, \alpha, Y^{(\alpha, \beta)})}\to^{\times N} z)$, by definition of $\mathcal{D}^{\mathrm{order}}_{(\gamma_j)_{j=1}^L}$ in \Cref{eq:derandom-beta}.
    This implies that the success probability of $\mathcal{L}'$ is the same as that of $\mathcal{L}^\oplus$.
\end{proof}

\subsubsection{Derandomized learning XOR lemma}
\label{sec:cl-hard-bootstrap-xor-derandom-xor}

In the second step of proving \Cref{lem:xor-lemma}, we show that the success probability of $\mathcal{L}'$ cannot exceed $1/2+(2\delta)^L/2$.
The proof idea follows that of \cite[Theorem 1]{assadi2021graph}, but the structure of our data generation processes are very different from the adversarial ordering there.
Therefore, we adopt a different technical construction.
In particular, their data streams are concatenated sequentially whereas ours are interleaved as in \Cref{eq:derandom-beta}.
This interleaved structure is inherent in our learning setting, because our problem instance labels $\beta$ are sampled randomly.

\begin{lemma}[Derandomized learning XOR lemma]
\label{lem:xor-lemma-derandom}
    Suppose that any randomized classical learning algorithm $\mathcal{L}$ with space complexity $T_1LS$ and sample complexity $T_1N$, given data generated from $\mathcal{D}^N_{g, N}(o), o\sim p_\gamma, \gamma\sim \unif(\bit)$, cannot output $\gamma\in\bit$ with success probability more than $1/2+\delta$.
    Then, any randomized classical learning algorithm $\mathcal{L}'$ with space complexity $S$ and sample complexity $T_1NL$, given data generated from $\mathcal{D}^{\mathrm{order}}_{(\gamma_j)_{j=1}^L}$, cannot output $B=\bigoplus_{j=1}^L\gamma_j\in\bit$ with success probability more than $1/2+(2\delta)^L/2$.
\end{lemma}

To prove \Cref{lem:xor-lemma-derandom}, we consider an $L$-player communication game.
It suffices to prove it when $B\sim\unif(\bit)$.
Under this choice, the procedure that first samples $B$ and then samples $\gamma\in\bit^L$ uniformly conditioned on $\bigoplus_{\beta=1}^L\gamma_\beta=B$ is equivalent to sampling $\gamma\sim\unif(\bit^L)$ directly.
Therefore, the pairs $(\gamma_\beta,z^\beta)$, $\beta\in[L]$, are mutually independent before the protocol starts.
In particular, there are $L$ players $Q_1, \ldots, Q_L$ where the player $Q_\beta$ receives the data $z_1^\beta, \ldots, z^\beta_{T_1N}$ in $\mathcal{D}^{\mathrm{order}}_{(\gamma_j)_{j=1}^L}$.
Their goal is to communicate and output a final bit $\hat{B}\in\bit$ that matches $B = \bigoplus_{j=1}^L \gamma_j$.
Recall that $z_1^\beta, \ldots, z^\beta_{T_1N}$ are the data samples corresponding to the problem instance $\beta\in [L]$.
Intuitively, these players will execute a communication protocol $\pi$ based on the learning algorithm $\mathcal{L}'$, and the communication between different players will characterize the correlation between problem instances induced by the learning algorithm.
Without loss of generality, we assume that the learning algorithm $\mathcal{L}'$ and the communication protocol $\pi$ are both deterministic, since there always exists a way of fixing the random numbers in a randomized algorithm such that the resulting success probability is at least that of the randomized algorithm.

We work in the blackboard model for communication in this game.
In other words, the players each write a message on the blackboard in the order $Q_1, \ldots, Q_L$ and the messages written on the blackboard are visible to all players afterwards.
This constitutes one round of communication, and the messages are never erased.
The next round starts again from $Q_1$.
We use $\mathcal{M}^j_i$ to denote the message written by player $Q_i$ in round $j$.
Let $\mathcal{B}^j_i$ be the content of the blackboard before player $Q_i$ communicates in round $j$, and let $\mathcal{B}^j$ be the content of the blackboard after round $j$ completes. 
Suppose that there are $r$ rounds in total.

For any $\beta\in [L]$ and final blackboard content $\mathcal{B}$, we define
\begin{equation}
    \mathrm{bias}_\pi (\beta, \mathcal{B}) = |\Pr[\gamma_\beta = 0|\mathcal{B}^r=\mathcal{B}] - \Pr[\gamma_\beta = 1|\mathcal{B}^r=\mathcal{B}]|
\end{equation}
to be the bias of $\gamma_\beta$ conditioned on the final blackboard displaying $\mathcal{B}$.
This intuitively characterizes the progress $\pi$ makes in solving the $\beta$-th problem instance.
For a partial blackboard content $\mathcal{B}$, we use the same notation $\mathrm{bias}_\pi(\beta,\mathcal{B})$ to denote the largest bias with which $\gamma_\beta$ can be predicted from that point onward, when all future messages not written by $Q_\beta$ may be chosen arbitrarily and adaptively depending on the current blackboard content, while the future messages written by $Q_\beta$ are generated
from $z^\beta$ according to the original protocol $\pi$.
For a final blackboard content, this agrees with the definition above.
Similarly, we define
\begin{equation}
    \mathrm{bias}_\pi (\mathcal{B}) = \left|\Pr\left[\bigoplus_{\beta=1}^L\gamma_\beta = 0|\mathcal{B}^r=\mathcal{B}\right] - \Pr\left[\bigoplus_{\beta=1}^L\gamma_\beta = 1|\mathcal{B}^r=\mathcal{B}\right]\right|
\end{equation}
to be the bias of $\bigoplus_{\beta=1}^L\gamma_\beta$ conditioned on the final blackboard displaying $\mathcal{B}$.
This characterizes the progress $\pi$ makes in solving the XOR problem.
In particular, the success probability (i.e., the output $\hat{B}$ of $\pi$ matches $B=\bigoplus_{j=1}^L\gamma_j$)
\begin{equation}
\begin{split}
    \Pr[\hat{B}=B] &= \E_\mathcal{B}\left[\Pr[\hat{B}=B|\mathcal{B}^r=\mathcal{B}]\right]\\
    &\leq \E_\mathcal{B}\left[\max_{\theta\in\bit}\Pr[B=\theta|\mathcal{B}^r=\mathcal{B}]\right] \\
    &= \E_\mathcal{B}\left[\frac{1+\mathrm{bias}_\pi(\mathcal{B})}{2}\right] = \frac{1}{2} + \frac{1}{2}\E_\mathcal{B}[\mathrm{bias}_\pi(\mathcal{B})],
\end{split}
\end{equation}
where we have used the fact that the final output $\hat{B}$ conditioned on the final blackboard content $\mathcal{B}^r=\mathcal{B}$ is a deterministic value $\theta$, since $\pi$ is deterministic.

Now we focus on a specific communication protocol $\pi$ given by the learning algorithm $\mathcal{L}'$ with at least the same success probability.
We will describe a randomized protocol, but we make it deterministic by fixing the randomness as described earlier.
The communication protocol proceeds as follows.
The player $Q_1$ feeds its data $z^1_1, \ldots, z^1_N$ sequentially into $\mathcal{L}'$, which produces a memory state of $\mathcal{L}'$ that the player $Q_1$ writes as the message $\mathcal{M}^1_1$ onto the blackboard.
Then player $Q_2$ reads the message $\mathcal{M}^1_1$ and use it as the memory state of $\mathcal{L}'$ after one data block of size $N$ is fed in.
$Q_2$ now feeds its data $z^2_1, \ldots, z^2_N$ into $\mathcal{L}'$, which again produces a memory state of $\mathcal{L}'$ that the player $Q_2$ writes as the message $\mathcal{M}^1_2$ onto the blackboard.
We proceed like this and when player $Q_L$ writes its message $\mathcal{M}^1_L$, the algorithm $\mathcal{L}'$ has received the first data chunk of size $NL$ (compare with \Cref{eq:derandom-beta}) and we proceed to the next round to read in the next data chunk of size $NL$.
After all the $T_1$ data chunks of size $NL$ are fed into $\mathcal{L}'$, it is now round $r=T_1$ and player $Q_L$ will write down the final output $\hat{B}$ of $\mathcal{L}'$, which we define as the output of the communication protocol $\pi$.
Since the data fed into $\mathcal{L}'$ follows the distribution $\mathcal{D}^{\mathrm{order}}_{(\gamma_j)_{j=1}^L}$, the success probability of $\pi$ is that of the corresponding
fixed-randomness version of $\mathcal L'$, and hence is at least that of the original randomized algorithm.
In the following, we focus on this communication protocol $\pi$ (after fixing its internal randomness appropriately).

We prove that this protocol $\pi$ cannot make good progress on any of the individual problem instances.

\begin{lemma}[Hardness of each individual problem instance]
\label{lem:hard-each-instance}
    $\E_{\mathcal{B}}[\mathrm{bias}_\pi(\beta, \mathcal{B})]\leq 2\delta, \quad \forall \beta\in[L]$, where $\mathcal{B}$ is the random variable of the partial or final blackboard content at any time step.
\end{lemma}

\begin{proof}[Proof of \Cref{lem:hard-each-instance}]
    We prove it by constructing a learning algorithm $\mathcal{L}$ with space complexity $T_1LS$ and sample complexity $T_1N$ that can output $\gamma_\beta\in\bit$ with decent probability given data generated from $\mathcal{D}^N_{g, N}(o), o\sim p_{\gamma_\beta}$, which contradicts our assumption in \Cref{lem:xor-lemma-derandom}.

    Fix any time step in the protocol and let $\mathcal{B}$ be the random partial blackboard content at that time step.
    Suppose for the sake of contradiction that there is a $\beta\in [L]$ such that $\E_{\mathcal{B}}[\mathrm{bias}_\pi(\beta, \mathcal{B})]> 2\delta$.
    For every possible value of $\mathcal{B}$, fix a continuation of the communication protocol and
    a final prediction $\hat{\theta}_{\mathcal{B}}$ that achieve
    $\mathrm{bias}_\pi(\beta,\mathcal{B})$ in its definition.
    Then, by the definition of $\mathrm{bias}_\pi(\beta, \mathcal{B})$, we have
    \begin{equation}
        \E_{\mathcal{B}}\left[\Pr[\gamma_\beta=\hat{\theta}_{\mathcal{B}}|\mathcal{B}]\right] >\frac{1+2\delta}{2} = \frac{1}{2}+\delta.
    \end{equation}

    We use $z^\beta = (z^\beta_1, \ldots, z^\beta_{T_1N})$ to denote that data samples corresponding the $\beta$-th problem instance.
    By the averaging argument, we know that there is a way of fixing all the other random data samples $z^1, \ldots, z^{\beta-1}, z^{\beta+1}, \ldots, z^L$ to some $z^{1*}, \ldots, z^{\beta-1*}, z^{\beta+1*}, \ldots, z^{L*}$ such that
    \begin{equation}
        \Pr[\gamma_\beta = \hat{\theta}_{\mathcal{B}^*}] >\frac{1}{2}+\delta,
    \end{equation}
    where $\mathcal{B}^*=\mathcal{B}^*(z^{1*}, \ldots, z^{\beta-1*}, z^\beta, z^{\beta+1*}, \ldots, z^{L*})$ is the random variable of the blackboard content at the fixed time step when the data is $z^{1*}, \ldots, z^{\beta-1*}, z^\beta, z^{\beta+1*}, \ldots, z^{L*}$.
    Note that it is a random variable with randomness induced by $z^\beta$ only, since the communication protocol is deterministic.

    We now construct the learning algorithm $\mathcal{L}$ by hard-coding $z^{1*}, \ldots, z^{\beta-1*}, z^{\beta+1*}, \ldots, z^{L*}$.
    Specifically, $\mathcal{L}$ proceeds as follows.
    Let $z^\beta$ be the input data sequence of $\mathcal{L}$.
    At the beginning, $\mathcal{L}$ first feeds $z^{1*}_1, \ldots, z^{1*}_N, \ldots,  z^{\beta-1*}_1, \ldots, z^{\beta-1*}_N$ into the communication protocol $\pi$ (and thus $\mathcal{L}'$), and calculate all the messages $\mathcal{M}^1_1, \mathcal{M}^1_{\beta-1}$.
    Recall that these messages are the memory states of $\mathcal{L}'$ after processing corresponding data blocks of size $N$.
    When the first data sample $z^\beta_1$ arrives, $\mathcal{L}$ feeds it into $\mathcal{L}'$, obtain the memory state of $\mathcal{L}'$.
    Now $\mathcal{L}$ writes the memory state and all the previous messages $\mathcal{M}^1_1, \mathcal{M}^1_{\beta-1}$ into its memory, which uses $(1+(\beta-1))S$ bits of memory.
    In a similar way, $\mathcal{L}$ moves on and feeds $z^\beta_2, \ldots, z^\beta_N$ sequentially into $\mathcal{L}'$.
    After processing $z^\beta_N$, it computes $\mathcal{M}^1_\beta$ and also put this message into its memory.
    Then, it feeds the rest of the first data chunk $z^{\beta+1*}_1, \ldots, z^{\beta+1*}_N, \ldots,  z^{L*}_1, \ldots, z^{L*}_N$ into $\mathcal{L}'$, compute the messages $\mathcal{M}^1_{\beta+1}, \ldots, \mathcal{M}^1_L$ and store them inside its memory.
    In other words, after processing this first data chunk of size $NL$, the algorithm $\mathcal{L}$ now stores in its memory all the messages of the first round $\mathcal{M}^1_1, \ldots, \mathcal{M}^1_L$, using $LS$ bits of memory.
    This completes the processing of the first $N$ samples from $z^\beta$.
    Now, $\mathcal{L}$ moves on to the next $N$ samples of $z^\beta$ and process them in a similar way.
    It never erases its memory and keeps storing more messages.

    When the fixed time step in the protocol is reached, $\mathcal{L}$ observes the resulting partial blackboard content $\mathcal{B}$ and follows the fixed continuation associated with this $\mathcal{B}$.
    In this continuation, the messages of $Q_\beta$ are generated from the genuine remaining samples of $z^\beta$, while all other messages are generated by the hard-coded adaptive continuation.
    At the end, $\mathcal{L}$ outputs the final prediction prescribed by the continuation.
    This means that the success probability of $\mathcal{L}$ is $\Pr[\gamma_\beta = \hat{\theta}_{\mathcal{B}^*}] >1/2+\delta$.
    Since the only things $\mathcal{L}$ writes into its memory are all the messages $\mathcal{M}^1_1, \ldots, \mathcal{M}^{T_1}_L$, which correspond to $T_1L$ snapshots of the memory states of $\mathcal{L}'$, the space complexity of $\mathcal{L}$ is $T_1LS$.
    The number of samples $\mathcal{L}$ uses is $T_1N$.
    Also, since the data of different $\beta$'s are generated independently, the input distribution is the same as $\mathcal{D}^N_{g, N}(o), o\sim p_{\gamma_{\beta}}$.
    As a result, $\mathcal{L}$ is a learning algorithm with space complexity $T_1LS$ and sample complexity $T_1N$, that given data generated from $\mathcal{D}^N_{g, N}(o), o\sim p_{\gamma_{\beta}}$, outputs $\gamma_\beta$ with success probability more than $1/2+\delta$, violating the assumption of \Cref{lem:xor-lemma-derandom}.
    This proves \Cref{lem:hard-each-instance}.
\end{proof}

Now that we have proved that $\pi$ cannot make good progress on each individual problem instance, we move on to show that it cannot make good progress on the final XOR problem.
To do this, we need to show that different $\gamma_\beta$'s are independent even when conditioned on the final blackboard content $\mathcal{B}^r$, which can be used by $\pi$ to make predictions.

We make use of the following fact from information theory.

\begin{lemma}[{\cite[Proposition A.4]{assadi2021graph}}]
\label{lem:mutual-info-cond-indep}
    For any random variables $A, B, C, D$, if $A$ and $D$ are independent conditioned on both $B$ and $C$, then
    \begin{equation}
        I(A; B|CD)\leq I(A; B|C).
    \end{equation}
\end{lemma}
\begin{proof}
    Note that since $A$ and $D$ are independent conditioned on $B, C$, we have
    \begin{equation}
        H(A|BCD) = H(AD|BC) - H(D|BC) = H(A|BC) + H(D|BC) - H(D|BC) = H(A|BC).
    \end{equation}
    Therefore, 
    \begin{equation}
        I(A;B|CD) = H(A|CD) - H(A|BCD) = H(A|CD) - H(A|BC) \leq H(A|C) - H(A|BC) = I(A;B|C),
    \end{equation}
    where we have used the fact that conditioning reduces entropy $H(A|CD)\leq H(A|C)$.
\end{proof}

We use this to prove the following lemma that shows the conditional independence of problem instances.
This property resembles the rectangle property of usual communication protocols on independent inputs.

\begin{lemma}[Conditional independence of problem instances]
\label{lem:cond-independence}
    For any index $\beta\in[L]$, let $z^{-\beta} = (z^1, \ldots, z^{\beta-1}, z^{\beta+1}, \ldots, z^L)$ and $\gamma^{-\beta} = (\gamma^1, \ldots, \gamma^{\beta-1}, \gamma^{\beta+1}, \ldots, \gamma^L)$. Then, for any partial or final blackboard content $\mathcal{B}$, $(\gamma^\beta, z^\beta)$ and $(\gamma^{-\beta}, z^{-\beta})$ are independent conditioned on $\mathcal{B}$.
\end{lemma}

\begin{proof}[Proof of \Cref{lem:cond-independence}]
    We prove this by induction.
    First note that at the beginning of the game, the blackboard is empty and we have that $(\gamma^\beta, z^\beta)$ and $(\gamma^{-\beta}, z^{-\beta})$ are independent, because the input data are independent.
    Next, assume that for some $0\leq k\leq r-1$, $(\gamma^\beta, z^\beta)$ and $(\gamma^{-\beta}, z^{-\beta})$ are independent conditioned on $\mathcal{B}^k=\mathcal{B}, \forall \mathcal{B}$.
    We will prove that $z^\beta$ and $z^{-\beta}$ are independent conditioned on $\mathcal{B}^{k+1}=\mathcal{B}, \forall \mathcal{B}$.
    As a shorthand, we will simply write $(\gamma^\beta, z^\beta)$ and $(\gamma^{-\beta}, z^{-\beta})$ as $z^\beta, z^{-\beta}$.

    To prove the induction step, we look at the conditional mutual information $I(z^\beta; z^{-\beta}|\mathcal{B}^{k+1})$.
    We know from the inductive hypothesis that $I(z^\beta; z^{-\beta}|\mathcal{B}^{k})=0$ because $z^\beta$ and $z^{-\beta}$ are independent conditioned on $\mathcal{B}^k$.
    On the other hand, note that $\mathcal{B}^{k+1} = \mathcal{B}^k \mathcal{M}^{k+1}_1 \ldots \mathcal{M}^{k+1}_L = \mathcal{B}^{k+1}_{\beta+1}\mathcal{M}^{k+1}_{\beta+1}\ldots\mathcal{M}^{k+1}_{L}$, and $\mathcal{M}^{k+1}_{\beta+1},\ldots, \mathcal{M}^{k+1}_{L}$ are deterministic functions of $\mathcal{B}^{k+1}_{\beta+1}$ and $z^{-\beta}$ because $\pi$ is deterministic.
    Hence, $\mathcal{M}^{k+1}_{\beta+1},\ldots, \mathcal{M}^{k+1}_{L}$ are deterministic and thus independent from $z^\beta$, conditioned on $\mathcal{B}^{k+1}_{\beta+1}$ and $z^{-\beta}$.
    This allows us to invoke \Cref{lem:mutual-info-cond-indep} and obtain
    \begin{equation}
    \label{eq:mutual-info-beta}
        I(z^\beta; z^{-\beta}|\mathcal{B}^{k+1}) = I(z^\beta; z^{-\beta}|\mathcal{B}^{k+1}_{\beta+1}\mathcal{M}^{k+1}_{\beta+1}\ldots\mathcal{M}^{k+1}_{L})\leq I(z^\beta;z^{-\beta}|\mathcal{B}^{k+1}_{\beta+1}).
    \end{equation}
    Next, note that $\mathcal{B}^{k+1}_{\beta+1}=\mathcal{B}^{k+1}_{\beta}\mathcal{M}^{k+1}_\beta$, and $\mathcal{M}^{k+1}_\beta$ is a deterministic function of $\mathcal{B}^{k+1}_{\beta}$ and $z^\beta$.
    Hence, $\mathcal{M}^{k+1}_\beta$ is deterministic and thus independent from $z^{-\beta}$, conditioned on $\mathcal{B}^{k+1}_\beta$ and $z^\beta$.
    We again invoke \Cref{lem:mutual-info-cond-indep} and obtain
    \begin{equation}
        I(z^\beta; z^{-\beta}|\mathcal{B}^{k+1})\leq I(z^\beta;z^{-\beta}|\mathcal{B}^{k+1}_{\beta+1}) = I(z^\beta;z^{-\beta}|\mathcal{B}^{k+1}_{\beta}\mathcal{M}^{k+1}_{\beta})\leq I(z^\beta;z^{-\beta}|\mathcal{B}^{k+1}_{\beta}).
    \end{equation}
    Finally, we have $\mathcal{B}^{k+1}_\beta = \mathcal{B}^k \mathcal{M}^{k+1}_1\mathcal{M}^{k+1}_{\beta-1}$.
    Similar to \Cref{eq:mutual-info-beta}, we have $\mathcal{M}^{k+1}_{1},\ldots, \mathcal{M}^{k+1}_{\beta-1}$ are deterministic and thus independent from $z^\beta$, conditioned on $\mathcal{B}^k$ and $z^{-\beta}$.
    Therefore,
    \begin{equation}
        I(z^\beta; z^{-\beta}|\mathcal{B}^{k+1})\leq I(z^\beta;z^{-\beta}|\mathcal{B}^{k+1}_{\beta}) = I(z^\beta;z^{-\beta}|\mathcal{B}^{k}\mathcal{M}^{k+1}_1\ldots\mathcal{M}^{k+1}_{\beta-1})\leq I(z^\beta;z^{-\beta}|\mathcal{B}^{k})=0.
    \end{equation}
    Due to the non-negativity of conditional mutual information, we have
    \begin{equation}
        I(z^\beta; z^{-\beta}|\mathcal{B}^{k+1})=0
    \end{equation}
    and thus $z^\beta$ and $z^{-\beta}$ are independent conditioned on $\mathcal{B}^{k+1}$.
    The same argument applies when the induction is stopped immediately after any individual message, and hence proves the claim for every partial blackboard content.
    This proves \Cref{lem:cond-independence}.
\end{proof}

Meanwhile, for independent random bits, the bias of their XOR is dampened exponentially.

\begin{lemma}[XOR of independent random bits, {\cite[Prop. A.9]{assadi2021graph}}]
\label{lem:xor-indep}
    Let $X_1, \ldots, X_L$ be independent random bits.
    We have
    \begin{equation}
        \mathrm{bias}\left(\bigoplus_{j=1}^L X_j\right) = \prod_{j=1}^L \mathrm{bias}(X_j).
    \end{equation}
\end{lemma}

\begin{proof}[Proof of \Cref{lem:xor-indep}]
    We only need to show the case of $L=2$, and the general case follows directly by recursively grouping random bits.
    Let $X_1, X_2$ be two independent random bits.
    Let $\beta_1=\mathrm{bias}(X_1), \beta_2=\mathrm{bias}(X_2)$ and $b_1=\argmax_{x\in\bit}\Pr[X_1=x], b_2=\argmax_{x\in\bit}\Pr[X_2=x]$.
    Then $\Pr[X_1=b_1]=(1+\beta_1)/2, \Pr[X_2=b_2]=(1+\beta_2)/2$.
    From the independence of $X_1, X_2$, we have
    \begin{equation}
    \begin{split}
        \Pr[X_1\oplus X_2=b_1\oplus b_2] &= \Pr[X_1=b_1, X_2=b_2] + \Pr[X_1=1-b_1, X_2=1-b_2] \\
        &=\Pr[X_1=b_1]\Pr[X_2=b_2] + \Pr[X_1=1-b_1]\Pr[X_2=1-b_2] \\
        &=\frac{1+\beta_1}{2}\frac{1+\beta_2}{2}+\frac{1-\beta_1}{2}\frac{1-\beta_2}{2} \\
        &=\frac{1}{2} (1+\beta_1\beta_2).
    \end{split}
    \end{equation}
    Therefore, we have $\mathrm{bias}(X_1\oplus X_2) = \beta_1\beta_2 = \mathrm{bias}(X_1)\mathrm{bias}(X_2)$.
    This completes the proof of \Cref{lem:xor-indep}.
\end{proof}

By combining \Cref{lem:hard-each-instance,lem:cond-independence,lem:xor-indep}, we prove the derandomized learning XOR lemma \Cref{lem:xor-lemma-derandom}.

\begin{proof}[Proof of \Cref{lem:xor-lemma-derandom}]
    Recall that the success probability of the communication protocol $\pi$ is the same as that of the learning algorithm $\mathcal{L}'$, which we use $p_{\mathrm{succ}} = \Pr[\hat{B}=B] \leq \frac{1}{2}+\frac{1}{2}\E_{\mathcal{B}}[\mathrm{bias}_\pi[\mathcal{B}]]$ to denote.
    Our goal is to prove an upper bound on $\E_{\mathcal{B}}[\mathrm{bias}_\pi[\mathcal{B}]]$ and therefore $p_{\mathrm{succ}}$.

    To this end, we fix any final blackboard content $\mathcal{B}$.
    Note that \Cref{lem:cond-independence} implies that the $\gamma_\beta$'s are independent from each other even conditioned on $\mathcal{B}^r=\mathcal{B}$, because any $\gamma_\beta$ is only correlated with $z^\beta$ and not $z^{-\beta}$.
    This allows us to invoke \Cref{lem:xor-indep} and obtain
    \begin{equation}
        \mathrm{bias}_\pi(\mathcal{B}) = \mathrm{bias}\left(\bigoplus_{\beta=1}^L\gamma_\beta\middle|\mathcal{B}^r=\mathcal{B}\right) = \prod_{\beta=1}^L \mathrm{bias}(\gamma_\beta|\mathcal{B}^r=\mathcal{B}) = \prod_{\beta=1}^L \mathrm{bias}_\pi(\beta, \mathcal{B}).
    \end{equation}

        On the other hand, applying
    \Cref{lem:hard-each-instance} to the empty blackboard content gives
    \begin{equation}
        \mathrm{bias}_\pi(\beta,\emptyset)
        \leq 2\delta,
        \qquad \forall \beta\in[L].
    \end{equation}

    We now expose the blackboard messages one at a time.
    Suppose that the current partial blackboard content is
    $\mathcal{B}^j_i=\mathcal{B}$ and that the next message
    $\mathcal{M}^j_i$ is written by $Q_i$.
    By the definition of $\mathrm{bias}_\pi(i,\mathcal{B})$, we may first observe the value of $\mathcal{M}^j_i$ and then, for each possible value $m$, use a continuation and a final prediction that achieve $\mathrm{bias}_\pi(i,\mathcal{B}m)$.
    The final prediction may depend on $m$: it predicts $0$ if $0$ is more likely in that branch and predicts $1$ otherwise.
    This is a valid continuation starting from $\mathcal{B}$, and hence
    \begin{equation}
        \E\left[
            \mathrm{bias}_\pi
            (i,\mathcal{B}\mathcal{M}^j_i)
            \middle|
            \mathcal{B}^j_i=\mathcal{B}
        \right]
        \leq
        \mathrm{bias}_\pi(i,\mathcal{B}).
    \end{equation}

    For every $\beta\neq i$, \Cref{lem:cond-independence} implies that $\mathcal{M}^j_i$ is independent of $(\gamma_\beta,z^\beta)$ conditioned on $\mathcal{B}^j_i=\mathcal{B}$.
    Moreover, since this message $\mathcal{M}^j_i$ is not written by $Q_\beta$, any particular value of it is a particular choice of the continuation and may be hard-coded into the continuation in the definition of $\mathrm{bias}_\pi(\beta,\mathcal{B})$.
    Hence, for every possible message $m$, we have
    \begin{equation}
        \mathrm{bias}_\pi(\beta,\mathcal{B}m)
        \leq
        \mathrm{bias}_\pi(\beta,\mathcal{B}),
        \qquad \forall \beta\neq i.
    \end{equation}

    It follows that
    \begin{equation}
        \E\left[
            \prod_{\beta=1}^L
            \mathrm{bias}_\pi
            (\beta,\mathcal{B}\mathcal{M}^j_i)
            \middle|
            \mathcal{B}^j_i=\mathcal{B}
        \right] \leq
        \left(
            \prod_{\substack{\beta=1\\\beta\neq i}}^L
            \mathrm{bias}_\pi(\beta,\mathcal{B})
        \right)
        \E\left[
            \mathrm{bias}_\pi
            (i,\mathcal{B}\mathcal{M}^j_i)
            \middle|
            \mathcal{B}^j_i=\mathcal{B}
        \right]\leq
        \prod_{\beta=1}^L
        \mathrm{bias}_\pi(\beta,\mathcal{B}).
    \end{equation}
    Iterating over all the blackboard messages, we obtain
    \begin{equation}
        \E_{\mathcal{B}}\left[
            \prod_{\beta=1}^L
            \mathrm{bias}_\pi(\beta,\mathcal{B})
        \right]
        \leq
        \prod_{\beta=1}^L
        \mathrm{bias}_\pi(\beta,\emptyset)
        \leq
        (2\delta)^L,
    \end{equation}
    where $\mathcal{B}$ now denotes the final blackboard content.
    Therefore,
    \begin{equation}
        \E_{\mathcal{B}}[\mathrm{bias}_\pi(\mathcal{B})] = \E_{\mathcal{B}}\left[\prod_{\beta=1}^L \mathrm{bias}_\pi(\beta, \mathcal{B}) \right]\leq (2\delta)^L.
    \end{equation}
    Hence, we arrive at
    \begin{equation}
        p_{\mathrm{succ}} \leq \frac{1}{2} + \frac{1}{2}\E_{\mathcal{B}}[\mathrm{bias}_\pi(\mathcal{B})]\leq \frac{1}{2} + \frac{1}{2}(2\delta)^L.
    \end{equation}
    This completes the proof of \Cref{lem:xor-lemma-derandom}.
\end{proof}

\subsubsection{Proof of \texorpdfstring{\Cref{lem:xor-lemma}}{Lemma \ref{lem:xor-lemma}}}
\label{sec:cl-hard-bootstrap-xor-finish}

We are now ready to prove the original learning XOR lemma (\Cref{lem:xor-lemma}).
Recall that the general proof idea is the following.
We first use the learning algorithm $\mathcal{L}^\oplus$ to construct an algorithm $\mathcal{L}'$ with at least the same success probability but learns from the derandomized distribution $\mathcal{D}^{\mathrm{order}}_{(\gamma_j)_{j=1}^L}$ as in \Cref{lem:derandomize-beta}.
Then, we invoke \Cref{lem:xor-lemma-derandom} to upper bound the success probability of $\mathcal{L}'$, which implies an upper bound the success probability of $\mathcal{L}^\oplus$.

\begin{proof}[Proof of \Cref{lem:xor-lemma}]
    Let $\mathcal{L}^\oplus$ be any randomized classical learning algorithm with space complexity $S$ and sample complexity $T_1N$, such that given data generated from $(\mathcal{D}^1_{(\gamma_j, o_j, Y^{(0, j)}, Y^{(1, j)})_{j=1}^L}\to^{\times T_1} \mathcal{D}^2_{(\beta, \alpha, Y^{(\alpha, \beta)})}\to^{\times N} z)$ as in dynamic NOPE, outputs $B=\bigoplus_{j=1}^L\gamma_j\in\bit$ with success probability $p_{\mathrm{succ}}$.
    \Cref{lem:derandomize-beta} implies that there exists a randomized classical learning algorithm $\mathcal{L}'$ with space complexity $S+\ceil{\log L}$ and sample complexity $T_1NL$ that given data generated from $\mathcal{D}^{\mathrm{order}}_{(\gamma_j)_{j=1}^L}$, outputs $B=\bigoplus_{j=1}^L \gamma_j \in\bit$ with probability $p_{\mathrm{succ}}$.
    Now we invoke \Cref{lem:xor-lemma-derandom}, which shares the assumption of \Cref{lem:xor-lemma}, and asserts that the success probability of $\mathcal{L}'$ is at most $1/2+(2\delta)^L/2$.
    Therefore, we have $p_{\mathrm{succ}}\leq 1/2+(2\delta)^L/2$, concluding the proof of \Cref{lem:xor-lemma}.
\end{proof}

\subsubsection{Low advantage leads to large sample complexity}
\label{sec:cl-hard-bootstrap-adv-to-sample}

The learning XOR lemma (\Cref{lem:xor-lemma}) shows that the advantage of predicting $B$ within a single $\gamma\in\bit^L$ instance decays exponentially with $L$.
In this section, we translate this exponentially decaying advantage into a blow up in sample complexity using a hybrid argument.

\begin{tcolorbox}
\begin{lemma}[Low advantage leads to large sample complexity]
\label{lem:adv-to-sample}
    Let $\tau$ be a positive integer and $\mathcal{Z}$ be a finite set.
    Let $\mathcal{D}_0, \mathcal{D}_1$ be two distributions on $\mathcal{Z}^{\tau}$.
    Suppose that any randomized classical learning algorithm with space complexity $S$, sample complexity $\tau$, and input form $\mathcal{Z}$, given one length-$\tau$ data block distributed according to $\mathcal{D}_B$, cannot output $B$ with probability more than $1/2+\delta$.
    Then, any randomized classical learning algorithm with space complexity $S$, sample complexity $M$, and input form $\mathcal{Z}$, that receives a sequence obtained by concatenating independent length-$\tau$ blocks drawn from $\mathcal{D}_B, B\in\bit$ and outputs $B$ with probability at least $2/3$, must satisfy
    \begin{equation}
        M\geq \left(\frac{1}{6\delta}-1\right)\tau.
    \end{equation}
\end{lemma}
\end{tcolorbox}

\begin{proof}[Proof of \Cref{lem:adv-to-sample}]
    We prove \Cref{lem:adv-to-sample} using a hybrid argument.
    Let $\mathcal{L}$ be a randomized classical learning algorithm with space complexity $S$, sample complexity $M$, and input form $\mathcal{Z}$, that receives a sequence obtained by concatenating IID length-$\tau$ blocks from $\mathcal{D}_B, B\in\bit$ and outputs $\hat{B}=B$ with probability at least $2/3$.
    Let $r=\ceil{M/\tau}$ and $M' = r\tau \in [M, M+\tau]$.
    We define a new learning algorithm $\mathcal{L}'$ with sample complexity $M'$ (now a multiple of $\tau$) by executing $\mathcal{L}$ on the first $M$ samples and discard the rest.
    Clearly, it has the same space complexity and success probability as $\mathcal{L}$.
    
    Now we define $r+1$ hybrid probability distributions
    \begin{equation}
        \mathcal{H}_{i} = (\mathcal{D}_1)^{\otimes i} \otimes (\mathcal{D}_0)^{\otimes (r-i)}, \quad 0\leq i\leq r.
    \end{equation}
    In particular, $M'=r\tau$ data samples drawn from $\mathcal{D}_0$ follows the distribution $\mathcal{H}_0 = \mathcal{D}_0^{\otimes r}$, while $M'$ data samples drawn from $\mathcal{D}_1$ follows the distribution $\mathcal{H}_r = \mathcal{D}_1^{\otimes r}$.
    We use $Z = (Z_1, \cdots, Z_r)$ to denote the data samples.
    Note that here each $Z_i\in \mathcal{Z}^\tau$ consists of $\tau$ samples that may have correlation among them.
    Then, the learning guarantee of $\mathcal{L}'$ reads
    \begin{equation}
        \left|\Pr_{Z\sim \mathcal{H}_r}[\hat{B}=1] - \Pr_{Z\sim \mathcal{H}_0}[\hat{B}=1]\right|\geq \frac{2}{3} - \frac{1}{3} = \frac{1}{3}.
    \end{equation}
    
    On the other hand, triangle inequality implies that
    \begin{equation}
    \begin{split}
        \frac{1}{3}\leq \left|\Pr_{Z\sim \mathcal{H}_r}[\hat{B}=1] - \Pr_{Z\sim \mathcal{H}_0}[\hat{B}=1]\right| &= \left|\sum_{i=1}^r\left(\Pr_{Z\sim \mathcal{H}_{i}}[\hat{B}=1] - \Pr_{Z\sim \mathcal{H}_{i-1}}[\hat{B}=1]\right)\right| \\
        &\leq \sum_{i=1}^r\left|\Pr_{Z\sim \mathcal{H}_{i}}[\hat{B}=1] - \Pr_{Z\sim \mathcal{H}_{i-1}}[\hat{B}=1]\right| \\
        &\leq r \max_{i\in [r]}\left|\Pr_{Z\sim \mathcal{H}_{i}}[\hat{B}=1] - \Pr_{Z\sim \mathcal{H}_{i-1}}[\hat{B}=1]\right|.
    \end{split}
    \end{equation}
    Let $i^* = \argmax_{i\in [r]}\left|\Pr_{Z\sim \mathcal{H}_{i}}[\hat{B}=1] - \Pr_{Z\sim \mathcal{H}_{i-1}}[\hat{B}=1]\right|$ be any $i\in [r]$ that maximizes the value.
    Then,
    \begin{equation}
    \label{eq:hybrid-i-star}
        \left|\Pr_{Z\sim \mathcal{H}_{i^*}}[\hat{B}=1] - \Pr_{Z\sim \mathcal{H}_{i^*-1}}[\hat{B}=1]\right|\geq \frac{1}{3r}.
    \end{equation}

    Now we use the learning algorithm $\mathcal{L}'$ to construct another learning algorithm $\mathcal{L}_0$ with space complexity $S$, sample complexity $\tau$, and input form $\mathcal{Z}$, that given one length-$\tau$ data block distributed according to $\mathcal{D}_B, B\in\bit$, outputs $\hat{B}_0=B$ with probability at least $1/2+1/(6r)$.
    Note that if this is true, \Cref{lem:adv-to-sample} follows immediately.
    This is because the assumption implies that 
    \begin{equation}
        \frac{1}{2} + \frac{1}{6r}\leq \frac{1}{2}+\delta
    \end{equation} 
    and therefore
    \begin{equation}
        \frac{M}{\tau}+1\geq \ceil{\frac{M}{\tau}}= r\geq \frac{1}{6\delta},
    \end{equation}
    which yields the desired result
    \begin{equation}
        M\geq \left(\frac{1}{6\delta}-1\right)\tau
    \end{equation}
    and completes the proof.

    To construct $\mathcal{L}_0$, we hard-code $i^\star$, $\mathcal{L}'$, and $\mathcal{D}_0, \mathcal{D}_1$ into the construction as follows.
    $\mathcal{L}_0$ first draws $i^*-1$ IID data blocks of size $\tau$ from $\mathcal{D}_1$ and feeds them into $\mathcal{L}'$.
    Next, $\mathcal{L}_0$ reads in the actual input data of size $\tau$ and feeds it into $\mathcal{L}'$.
    Then, it draws $r-i^*$ IID data blocks from $\mathcal{D}_0$ and feeds them into $\mathcal{L}'$. 
    Finally, we hard-code the sign of $\Pr_{Z\sim\mathcal{H}_{i^\star}}[\hat{B}=1]-\Pr_{Z\sim\mathcal{H}_{i^\star-1}}[\hat{B}=1]$.
    If this sign is nonnegative, $\mathcal{L}_0$ outputs the bit produced by $\mathcal{L}'$.
    If it is negative, $\mathcal{L}_0$ outputs its complement.
    
    Note that if the input of $\mathcal{L}_0$ is drawn from $\mathcal{D}_0$, then the $M'=r\tau$ data samples fed into $\mathcal{L}'$ follows the distribution $\mathcal{H}_{i^\star-1}$.
    On the other hand, if the input of $\mathcal{L}_0$ is drawn from $\mathcal{D}_1$, then the data samples fed into $\mathcal{L}'$ follows $\mathcal{H}_{i^\star}$.
    Hence, \Cref{eq:hybrid-i-star} implies that $\mathcal{L}_0$ has success probability at least
    \begin{equation}
        \frac{1 + \left|\Pr_{Z\sim \mathcal{H}_{i^*}}[\hat{B}=1] - \Pr_{Z\sim \mathcal{H}_{i^*-1}}[\hat{B}=1]\right|}{2}\geq \frac{1}{2} + \frac{1}{6r}.
    \end{equation}
    Moreover, by construction, $\mathcal{L}_0$ indeed has space complexity $S$ and sample complexity $\tau$.
    This concludes the proof of \Cref{lem:adv-to-sample}.
\end{proof}

\subsubsection{Proof of \texorpdfstring{\Cref{thm:classical-lower-bound-superpoly-sample,thm:classical-lower-bound-single-block}}{Theorems \ref{thm:classical-lower-bound-superpoly-sample,thm:classical-lower-bound-single-block}}}
\label{sec:cl-hard-bootstrap-finish}

We are now ready to prove the classical single-block hardness and sample complexity of dynamic NOPE (\Cref{thm:classical-lower-bound-single-block,thm:classical-lower-bound-superpoly-sample}).
We will make use of the distributional sample-space lower bound (\Cref{lem:classical-lower-bound-dist}), the learning XOR lemma (\Cref{lem:xor-lemma}), and the sample complexity blow-up from low advantage (\Cref{lem:adv-to-sample}).

\begin{proof}[Proof of \Cref{thm:classical-lower-bound-superpoly-sample,thm:classical-lower-bound-single-block}]
    Let $\mathcal{L}$ be any randomized classical learning algorithm with space complexity $S$, sample complexity $M$, and input form $\mathcal{I} = [N]\times \bit^b\times \bit \times [L]$ such that it solves dynamic NOPE with probability at least $p_{\mathrm{succ}}$.
    That is, for any $B\in\bit$, given data generated from $\mathcal{D}_{g, f}^{N, T_1}(B)$, $\mathcal{L}$ outputs $B$ with probability at least $p_{\mathrm{succ}}$.
    Define $S_1 = T_1L(S+\ceil{\log L})$ and $M_1 = T_1N$ as in the learning XOR lemma (\Cref{lem:xor-lemma}).
    Note that the assumption
    \begin{equation}
        S=o\lr{\frac{Q_C}{T_1^2\log^2(N)}}
    \end{equation}
    implies that
    \begin{equation}
        S_1M_1 = NT_1^2L (S+\ceil{\log L}) \leq O\left(NT_1^2L\left(\frac{Q_C}{T_1^2\log^2(N)} + \log\log N \right)\right) \leq O(NQ_C) = o(NQ_C b),
    \end{equation}
    where we have used $Q_C\geq \Omega(T_1^2L\log\log N)$, $L=\ceil{\log^2(N)}$, and $b=\ceil{c\log N}$.
    Then the distributional sample-space lower bound (\Cref{lem:classical-lower-bound-dist}) asserts that any randomized classical learning algorithm with space complexity $S_1$, sample complexity $M_1$, and input form $\mathcal{I}_1 = [N]\times \bit^b\times \bit$, given $\gamma\sim\unif(\bit)$, $o\sim p_\gamma$, and data generated from $\mathcal{D}_{g,N}^N(o)$, cannot output $\gamma$ with probability more than
    \begin{equation}
        1-\frac{1}{3} + 2^{-\eta b/8}\leq \frac{1}{2} + \frac{1}{6} + 2^{-\eta \cdot 210/8}\leq \frac{1}{2}+\frac{1}{6} + 2^{-21/8}<\frac{1}{2} + \frac{1}{3},
    \end{equation}
    where we have used $Q_C$ being the $(1/3)$-error query complexity and $b>c\log N\geq c \geq 865/\eta^2 >210$ since $\eta\in (0.1, 2]$.
    This gives us the assumption we need for invoking the learning XOR lemma (\Cref{lem:xor-lemma}) with $\delta=1/3$, which implies that any randomized classical learning algorithm $\mathcal{L}^\oplus$ with space complexity $S$, sample complexity $T_1N=\tau_{\mathcal{D}}$, and input form $\mathcal{I}_1\times [L] = [N] \times \bit^b\times \bit\times [L]=\mathcal{I}$, given data generated from $(\mathcal{D}^1_{(\gamma_j, o_j, Y^{(0, j)}, Y^{(1, j)})_{j=1}^L}\to^{\times T_1} \mathcal{D}^2_{(\beta, \alpha, Y^{(\alpha, \beta)})}\to^{\times N} z)$, cannot output $B=\bigoplus_{j=1}^L\gamma_j\in\bit$ with success probability more than 
    \begin{equation}
        \frac{1}{2} + \frac{(2/3)^L}{2} = \frac{1}{2}+\frac{1}{N^{\omega(1)}}.
    \end{equation}
    This proves \Cref{thm:classical-lower-bound-single-block}.
    
    This is also exactly the assumption we need for invoking \Cref{lem:adv-to-sample} that shows low advantage leads to sample complexity blow up with $\tau = T_1N=\tau_{\mathcal{D}}, \delta = (2/3)^L/2, \mathcal{Z} = \mathcal{I}$.
    Applying \Cref{lem:adv-to-sample} to $\mathcal{L}$ gives us that if its success probability is at least $p_{\mathrm{succ}}\geq 2/3$, then
    \begin{equation}
        M \geq \left(\frac{1}{3(2/3)^L}-1\right)\tau_{\mathcal{D}} \geq \frac{1}{6(2/3)^L}\tau_{\mathcal{D}} \geq \tau_{\mathcal{D}} \Omega\left((1.5)^{\log^2 N}\right) \geq \tau_{\mathcal{D}} N^{\Omega(\log N)}\geq \tau_{\mathcal{D}}N^{\omega(1)},
    \end{equation}
    where we have used $L\geq 5>\log_{3/2}(6)$ such that $1\leq 1/(6(2/3)^L)$.
    This concludes the proof of \Cref{thm:classical-lower-bound-superpoly-sample}.
\end{proof}

\subsection{Connect to applications}
\label{sec:cl-hard-app}

In this section, we develop the necessary tools that connect dynamic NOPE to the various application tasks that we will explore in \Cref{sec:app}, such as solving linear systems, binary classification, dimension reduction, etc.

We begin by focusing on a specific oracle property called Forrelation \cite{aaronson2015forrelation,bansal2021kforrelation} and take the inner product function as the noisy encoding function.
We design a distributional version of Forrelation, show that inner product is a good noisy encoding function, and prove some useful lemmas in \Cref{sec:cl-hard-app-prelim}.

In \Cref{sec:cl-hard-app-embed}, we construct an efficient quantum circuit that solves the dynamic NOPE task, which we then embed into various applications.
In particular, in \Cref{sec:conn-linear-sys,sec:conn-bin-classify,sec:conn-dim-reduc}, we embed this circuit into the tasks of solving linear systems, binary classification, and dimension reduction.
This shows that if we can solve the application tasks, we can solve dynamic NOPE, which is hard.
Hence the application tasks must also be hard to solve.
The full hardness proofs will be given in \Cref{sec:app} for the various application tasks.

\subsubsection{Forrelation and inner product}
\label{sec:cl-hard-app-prelim}

We first introduce the oracle property function $f$ that we will use in dynamic NOPE.
In particular, we consider the following oracle query problem called Forrelation.

\begin{definition}[$(\delta, K, n)$-Forrelation, \cite{aaronson2015forrelation,bansal2021kforrelation}]
\label{def:forrelation}
    Let $n$ be a positive integer and $N=2^n$.
    Let $K\geq 2$ be an integer and let $\delta\in (0, 1)$.
    We define the partial Boolean function $\mathrm{forr}^n_{\delta, K}: \bit^{KN}\to \bit$ as
    \begin{equation}
        \mathrm{forr}^n_{\delta, K}(o) = \begin{cases}
            1, \quad \mathrm{forr}^n_{K}(o)\geq \delta, \\
            0, \quad |\mathrm{forr}^n_{K}(o)|\leq \delta/2,
        \end{cases}
    \end{equation}
    where
    \begin{equation}
        \mathrm{forr}^n_K(o) = \left(\frac{(-1)^{o_1}}{\sqrt{N}}\right)^T H^{\otimes n} \diag((-1)^{o_2})H^{\otimes n} \cdots \diag((-1)^{o_{K-1}})H^{\otimes n}\left(\frac{(-1)^{o_K}}{\sqrt{N}}\right),
    \end{equation}
    $H = \frac{1}{\sqrt{2}}\begin{pmatrix}
        1 &1\\
        1 &-1
    \end{pmatrix}$ is the Hadamard gate, and $o=(o_1, \ldots, o_K)\in \bit^{KN}$ with $o_i\in \bit^N$.
\end{definition}

Forrelation has a $O_\epsilon(1)$ versus $\Omega(N^{1-\epsilon})$ quantum-classical query complexity separation for any arbitrarily small constant $\epsilon>0$.
Specifically, \cite{bansal2021kforrelation} showed that there are two distributions $p_0', p_1'$ over $\{0, 1\}^{KN}$ such that the following holds.
Note that we are using the $\bit$ notation for oracles whereas \cite{bansal2021kforrelation} used $\{\pm 1\}$.

\begin{lemma}[Oracle distributions for Forrelation, {\cite[Figure 1, Theorem 3.1, Section 5.4]{bansal2021kforrelation}}]
\label{lem:forrelation-distribution}
    Let $K\geq 2, n, N=2^n$ be positive integers and $\delta=2^{-5K}$.
    There are two distributions $p_0', p_1'$ supported on $\bit^{KN}$ such that
    \begin{enumerate}
        \item $\Pr_{o\sim p_0'}[\mathrm{forr}^n_{\delta, K}(o)=0]\geq 1-4/(\delta^2N)$ and $\Pr_{o\sim p_1'}[\mathrm{forr}^n_{\delta, K}(o)=1]\geq 6\delta$;
        \item Any classical query algorithm $\mathcal{A}$ that queries $o\in \bit^{KN}$ and outputs $\mathcal{A}(o)\in \bit$ such that 
        \begin{equation}
            |\E_{o\sim p_1'}[\mathcal{A}(o)] - \E_{o\sim p_0'}[\mathcal{A}(o)]|>\frac{\delta}{4}
        \end{equation}
        must make at least
        \begin{equation}
            \Omega\left(\frac{1}{K^{28}}\left(\frac{N}{\log(KN)}\right)^{1-1/K}\right)
        \end{equation}
        queries to $o$.
    \end{enumerate}
\end{lemma}

Recall that in dynamic NOPE, we need two oracle distributions $p_\gamma$ supported
entirely on the two promise classes of $\mathrm{forr}^n_{\delta,K}$.
Rather than using the distributions $p_0',p_1'$ in \Cref{lem:forrelation-distribution} that does not always give oracles with the correct property, we construct the required distributions directly from the worst-case randomized query lower bound using Yao's minimax principle.
This allows us to obtain the following distributional query complexity separation of Forrelation.

\begin{lemma}[Distributional query complexity separation of Forrelation]
\label{lem:forrelation-query-separation}
    Let $n$ be a large integer and $N=2^n$. Let $K\geq 2$ be an integer and let $\delta=2^{-5K}$.
    Let $\eta\in (0, 1/3]$.

    Then, there exist two distributions $p_0,p_1$ on $\bit^{KN}$ such
    that
    \begin{equation}
        \mathrm{supp}(p_\gamma)
        \subseteq
        \left\{
            o:
            \mathrm{forr}^n_{\delta,K}(o)=\gamma
        \right\},
        \quad \forall \gamma\in\bit.
    \end{equation}
    Given $o\sim p_\gamma$ for any $\gamma\in\bit$, there exists a
    quantum algorithm with
    $\log N+O(K)$ space complexity,
    $O(K2^{10K}\log N\log(1/\eta))$ gate complexity, and
    $O(K2^{10K}\log(1/\eta))$ query complexity that outputs $\gamma$
    with success probability at least $1-\eta$.

    Meanwhile, any randomized classical query algorithm $\mathcal{A}$
    satisfying
    \begin{equation}
        \Pr_{\substack{\gamma\sim\unif(\bit)\\o\sim p_\gamma}}
        [\mathcal{A}(o)=\gamma]
        \geq \frac{2}{3},
    \end{equation}
    where the probability also includes the internal randomness of
    $\mathcal{A}$, must make
    \begin{equation}
        \Omega\left(
            \frac{1}{K^{29}}
            \left(
                \frac{N}{\log(KN)}
            \right)^{1-1/K}
        \right)
    \end{equation}
    queries.
\end{lemma}

\begin{proof}[Proof of \Cref{lem:forrelation-query-separation}]
    Taking the advantage parameter in
    \cite[Theorem 1.3]{bansal2021kforrelation} to be $\xi=1/6$, the theorem states that there exists a constant $c>0$ such that every randomized classical algorithm that computes $\mathrm{forr}^n_{\delta,K}$ with error at most $1/2-\xi/2=5/12$
    on every promised input must make at least
    \begin{equation}
        \frac{c}{K^{29}}
        \left(
            \frac{N}{\log(KN)}
        \right)^{1-1/K}
    \end{equation}
    queries.

    Define
    \begin{equation}
        q_0
        =
        \left\lfloor
            \frac{c}{2K^{29}}
            \left(
                \frac{N}{\log(KN)}
            \right)^{1-1/K}
        \right\rfloor.
    \end{equation}
    By Yao's minimax principle \cite{yao1977probabilistic}, there exists
    a distribution $\mu$ supported on the domain of $\mathrm{forr}^n_{\delta,K}$ such that every deterministic classical algorithm $D$ making at most $q_0$ queries
    satisfies
    \begin{equation}
        \Pr_{o\sim\mu}[D(o)\neq \mathrm{forr}^n_{\delta,K}(o)]
        \geq
        \frac{5}{12}.
    \end{equation}

    Let
    \begin{equation}
        \lambda_\gamma
        =
        \Pr_{o\sim\mu}[\mathrm{forr}^n_{\delta,K}(o)=\gamma].
    \end{equation}
    Suppose the deterministic classical algorithm $D$ makes no query and always outputs $0$ or $1$, then the error probability bound becomes a bound on $\lambda_\gamma$, which implies
    \begin{equation}
        \frac{5}{12}
        \leq
        \lambda_\gamma
        \leq
        \frac{7}{12},
        \qquad \forall\gamma\in\bit.
    \end{equation}
    We define
    \begin{equation}
        p_\gamma(o)
        =
        \begin{cases}
            \mu(o)/\lambda_\gamma,
            & \mathrm{forr}^n_{\delta,K}(o)=\gamma,\\
            0,
            & \text{otherwise}.
        \end{cases}
    \end{equation}

    Suppose, for the sake of contradiction, that a randomized classical query
    algorithm $\mathcal{A}$ makes at most $q_0$ queries and satisfies
    \begin{equation}
        \Pr_{\substack{\gamma\sim\unif(\bit)\\o\sim p_\gamma}}
        [\mathcal{A}(o)=\gamma]
        \geq
        \frac{2}{3}.
    \end{equation}
    For every $\gamma\in\bit$, let
    \begin{equation}
        e_\gamma
        =
        \Pr_{o\sim p_\gamma}
        [\mathcal{A}(o)\neq\gamma],
    \end{equation}
    where the probability includes the internal randomness of
    $\mathcal{A}$.
    The assumed success probability implies
    \begin{equation}
        e_0+e_1
        \leq
        \frac{2}{3}.
    \end{equation}

    Let $\mathcal{A}_r$ be the deterministic classical algorithm obtained by fixing the internal randomness of $\mathcal{A}$ to $r$.
    Then,
    \begin{equation}
        \E_r\left[
            \Pr_{o\sim\mu}
            [\mathcal{A}_r(o)\neq \mathrm{forr}^n_{\delta,K}(o)]
        \right]
        =
        \lambda_0e_0+\lambda_1e_1 \leq
        \frac{7}{12}(e_0+e_1) \leq
        \frac{7}{18}
        <
        \frac{5}{12}.
    \end{equation}
    Hence, there exists a fixing $r$ such that the deterministic classical algorithm $\mathcal{A}_r$ makes at most $q_0$ queries and has error probability strictly smaller than $5/12$ under $\mu$, contradicting the defining property of $\mu$.
    Therefore, every randomized classical query algorithm with success probability at least $2/3$ must make
    \begin{equation}
        \Omega\left(
            \frac{1}{K^{29}}
            \left(
                \frac{N}{\log(KN)}
            \right)^{1-1/K}
        \right)
    \end{equation}
    queries, as desired.

    It remains to prove the quantum upper bound.
    By \cite[Corollary 1.4]{bansal2021kforrelation}, there exists a quantum circuit acting on $\log N+O(K)$ qubits that computes $\mathrm{forr}^n_{\delta,K}(o)$ with error at most $1/3$ on every promised input, using $O(K2^{10K})$ queries and $O(K2^{10K}\log N)$ gates.

    Applying standard coherent threshold amplification based on quantum singular value transformation \cite{martyn2021grand} to this constant-error circuit gives a unitary quantum circuit $\mathcal{A}^O$, with no intermediate measurements, whose designated output qubit is equal to $\mathrm{forr}^n_{\delta,K}(o)$ with probability at least $1-\eta$.
    Indeed, the two promise cases have acceptance probabilities at most $1/3$ and at least $2/3$, respectively, so a bounded polynomial threshold transformation of degree $O(\log(1/\eta))$ separates these two constant-gap intervals.
    This coherent amplification uses $O(\log(1/\eta))$ calls to the constant-error circuit and its inverse and $O(1)$ additional qubits.
    In particular, $(\mathcal{A}^O)^\dagger$ is implemented by
    reversing the gates of $\mathcal{A}^O$.
    This gives the claimed space, gate, and query complexities.
    Since $p_\gamma$ is supported entirely on inputs satisfying $\mathrm{forr}^n_{\delta,K}(o)=\gamma$, the output qubit of $\mathcal{A}^O$ equals $\gamma$ with probability at least $1-\eta$ whenever $o\sim p_\gamma$.
    This completes the proof of \Cref{lem:forrelation-query-separation}.
\end{proof}

Next, we introduce the inner product function as our noisy encoding function.

\begin{definition}[Inner product noisy encoding function]
\label{def:inner-prod}
    Let $b$ be a positive integer.
    We define the inner product noisy encoding function $g: \bit^b\times \bit^b \to \bit$ as
    \begin{equation}
        g(y^{(0)}, y^{(1)}) = \bigoplus_{i=1}^b \lr{ y^{(0)}_i\cdot y^{(1)}_i}.
    \end{equation}
\end{definition}

The following standard proof shows that the inner product noisy encoding function $g$ has low discrepancy (defined in \Cref{def:disc}) $\disc(g)\leq 2^{-\eta b}$ with $\eta=1/2$.

\begin{lemma}[Inner product has low discrepancy]
\label{lem:innder-prod-disc}
    Let $b$ be a positive integer and $g: \bit^b\times \bit^b \to \bit$ be the inner product noisy encoding function.
    Then,
    \begin{equation}
        \disc(g)\leq 2^{-b/2}.
    \end{equation}
\end{lemma}

\begin{proof}[Proof of \Cref{lem:innder-prod-disc}]
    Let $R=S\times T$ be any rectangle where $S, T\subseteq \bit^b$.
    Let $U, V\sim \unif(\bit^b)$.
    From \Cref{def:disc}, we have
    \begin{equation}
    \begin{split}
        \disc_R(g) &= \left|\Pr[(U, V)\in R, g(U, V)=0] - \Pr[(U, V)\in R, g(U, V)=1]\right| \\
        &=\left|\frac{1}{2^b}\cdot\frac{1}{2^b}\sum_{u\in S, v\in T}(-1)^{g(u, v)}\right| \\
        &=\frac{1}{2^{3b/2}}|s^TH t|,
    \end{split}
    \end{equation}
    where $s, t\in\bit^{2^b}$ are length-$2^b$ Boolean vectors such that $s_u = 1[u\in S]$ and $t_v = 1[v\in T]$ for all $u, v\in \bit^b$, and $H\in \{\pm 1\}^{2^b\times 2^b}$ such that $H_{uv}=(-1)^{g(u, v)}/2^{b/2}$.
    Note that there we are identifying length-$b$ bitstrings with elements from $[2^b]$.

    Next, we prove that $H$ is an orthogonal matrix.
    Indeed, the matrix element reads
    \begin{equation}
    \begin{split}
        (H^TH)_{uv} &= \sum_{w\in \bit^b} H_{wu}H_{wv} \\
        &= \frac{1}{2^b}\sum_{w\in \bit^b} (-1)^{g(w, u)\oplus g(w, v)} \\
        &= \frac{1}{2^b}\sum_{w\in \bit^b} (-1)^{\bigoplus_{j=1}^b w_j(u_j\oplus v_j)} \\
        &= \frac{1}{2^b}\prod_{j=1}^b \sum_{w_j\in\bit} (-1)^{w_j(u_j\oplus v_j)} \\
        &= \frac{1}{2^b}\prod_{j=1}^b (2\cdot 1[u_j\oplus v_j=0]) \\
        &= \delta_{u, v}.
    \end{split}
    \end{equation}
    Hence, $H^TH = I$, and we arrive at
    \begin{equation}
        \disc_R(g) = \frac{1}{2^{3b/2}}|s^TH t|\leq \frac{1}{2^{3b/2}} \|s\| \|Ht\| = \frac{1}{2^{3b/2}} \|s\| \|t\| \leq \frac{1}{2^{3b/2}} \sqrt{2^b} \sqrt{2^b} = 2^{-b/2}
    \end{equation}
    for any rectangle $R$.
    Therefore, $\disc(g)\leq 2^{-b/2}$.  
\end{proof}

We will embed quantum circuits into application tasks, which naturally involve complex numbers.
However, in most classical data processing applications, we work in real numbers.
The following standard lemma provides a canonical way to realify complex matrices and vectors.

\begin{lemma}[Realification]
\label{lem:realification}
    For any complex vector $\vec{v}\in \mathbb{C}^d$, we define its realification $\mathcal{R}[\vec{v}]\in \mathbb{R}^{2d}$ as 
    \begin{equation}
        \mathcal{R}[\vec{v}] = \begin{pmatrix}
            \Re[\vec{v}] \\
            \Im[\vec{v}]
        \end{pmatrix}.
    \end{equation}
    For any complex matrix $A\in \mathbb{C}^{d\times d}$, we define its realification $\mathcal{R}[A]\in \mathbb{R}^{2d\times 2d}$ as
    \begin{equation}
        \mathcal{R}[A] = \begin{pmatrix}
            \Re[A] &-\Im[A] \\
            \Im[A] &\Re[A]
        \end{pmatrix}.
    \end{equation}
    Then, the following properties hold
    \begin{enumerate}
        \item (Isometry): $\|\mathcal{R}[\vec{v}]\|_2 = \|\vec{v}\|_2, \forall \vec{v}\in \mathbb{C}^d$.
        \item (Inner product): $\mathcal{R}[\vec{u}]^T\mathcal{R}[\vec{v}] = \Re[\vec{u}^\dagger \vec{v}], \forall \vec{u},\vec{v}\in \mathbb{C}^d$.
        \item (Linearity): $\mathcal{R}[A\vec{v}]=\mathcal{R}[A]\mathcal{R}[\vec{v}], \forall A\in \mathbb{C}^{d\times d},  \vec{v}\in \mathbb{C}^d$.
        \item (Matrix operations): for any $A, B\in \mathbb{C}^{d\times d}$, $\mathcal{R}[AB] = \mathcal{R}[A]\mathcal{R}[B]$, $\mathcal{R}[A^\dagger] = \mathcal{R}[A]^T$. If $A$ is invertible, $\mathcal{R}[A]^{-1}=\mathcal{R}[A^{-1}]$.
        \item (Singular values): For any $A\in \mathbb{C}^{d\times d}$, the singular values of $\mathcal{R}[A]$ are the same as those of $A$, with doubled multiplicity. Consequently, the norm and condition number of $\mathcal{R}[A]$ are the same as those of $A$.
        \item (Structured matrices): The realification of Hermitian matrices are symmetric. The realification of unitary matrices are orthogonal. The realification of $s$-sparse matrices are $2s$-sparse.
        \item (Eigenvectors): For any Hermitian matrix $A\in \mathbb{C}^{d\times d}$ with the largest eigenvalue $\lambda$ and a unique corresponding eigenvector $\vec{v}$, its realification $\mathcal{R}[A]$ is symmetric with a two-fold degenerate largest eigenvalue $\lambda$ corresponding to two eigenvectors $\mathcal{R}[\vec{v}], \mathcal{R}[i\vec{v}]$.
    \end{enumerate}
\end{lemma}

The above standard realification doubles the eigenvalue degeneracy of a matrix.
To avoid that, we use the following realification of quantum circuits \cite{bernstein1993quantum,aharonov2003simple}.

\begin{lemma}[Realification of quantum circuits]
\label{lem:realification-circuit}
    Let $U_T\cdots U_1, U_i\in U(2^n)$ be a sequence of $T$ unitaries on $n$ qubits.
    Let $\ket{\psi}\in \mathbb{C}^{2^n}$ be an $n$-qubit quantum state.
    Then, $\mathcal{R}[U_i], i\in [T]$ are $2^{n+1}$-dimensional real, orthogonal matrices and therefore unitaries on $n+1$ qubits.
    Similarly, $\mathcal{R}[\ket{\psi}]$ is a quantum state on $n+1$ qubits with real components.
    In particular, $\mathcal{R}[\ket{x}]=\ket{0}\ket{x}, \forall x\in \bit^n$.
\end{lemma}

\begin{proof}
    This follows directly from Items 1 and 6 of \Cref{lem:realification}.
\end{proof}

\Cref{lem:realification-circuit} avoids the degeneracy problem because of the simple observation: the eigenspace of $\mathcal{R}[\ketbra{\psi}]$ with eigenvalue one is (doubly) degenerate, while that of $\mathcal{R}[\ket{\psi}]\mathcal{R}[\ket{\psi}]^T$ is not.

\subsubsection{Quantum circuit for dynamic NOPE}
\label{sec:cl-hard-app-embed}

In this section, we construct a quantum circuit that efficiently solves the dynamic NOPE task in \Cref{sec:cl-hard-bootstrap} for the oracle property Forrelation with inner product noisy encoding.
Then we embed this quantum circuit into various application tasks that will be used in \Cref{sec:app}.

The following lemma gives us the desired circuit.
For simplicity, throughout the hardness constructions below, we implicitly round $K, L$ upward to powers of two and choose $b$ such that $b+1$ is a power of two. 
All displayed choices of these parameters are understood up to this constant-factor rounding, which does not affect any asymptotic bounds.

\begin{lemma}[Quantum circuit for dynamic NOPE]
\label{lem:encode-q-circ}
    Let $n, N=2^n, T_1$ be large integers.
    Let $K\geq 2$ be a constant integer.
    Let $g:\bit^b\times \bit^b\to \bit$ be the inner product noisy encoding function with encoding length $b=\ceil{40678\log(KN)}$.
    Consider the oracle property function $f=\mathrm{forr}^n_{2^{-5K}, K}: \bit^{KN} \to \bit$.
    Then, there exists a $(\log N+O(\log\log N))$-qubit quantum circuit $C$ consisting of $O(\log^3 N\log\log N)$ fixed two-qubit gates and $O(\log^3 N \log\log N)$ diagonal gates satisfying 
    \begin{equation}
        O_0: \ket{x, \alpha, \beta, k} \to (-1)^{\left(Y^{(\alpha, \beta)}_x\right)_k}\ket{x, \alpha, \beta, k}, \quad \forall x\in [KN], \alpha\in \bit, \beta\in [L], k\in [b],
    \end{equation}
    or its controlled version $cO_0$, where $Y^{(\alpha, \beta)}\in \bit^{KN\times b}$ are any noisy encodings that can be generated in a single refreshing block of $\mathcal{D}^{KN, T_1}_{g, f}(B)$, such that
    \begin{equation}
        \left\|
            C\ket{0}
            -
            \ket{B}_a\ket{0}
        \right\|_2
        \leq
        \frac{1}{20},
    \end{equation}
    where the second $\ket{0}$ denotes all qubits other than the output qubit $a$.
    In particular, measuring $C\ket{0}$ in the computational basis returns $B$ on the output qubit and $0$ on all the remaining qubits with probability at least $0.99$.
\end{lemma}

\begin{proof}[Proof of \Cref{lem:encode-q-circ}]
    Recall that the data generation process, as defined in \Cref{sec:cl-hard-bootstrap}, reads
    \begin{equation}
        \mathcal{D}_{g, f}^{KN, T_1}(B) = (\mathcal{D}^0_B\to \mathcal{D}_{(\gamma_j, o_j, Y^{(0,j)}, Y^{(1,j)})_{j=1}^{L}}^1 \to^{\times T_1} \mathcal{D}^2_{(\beta, \alpha, Y^{(\alpha,\beta)})} \to^{\times T_2} z),
    \end{equation}
    where $\gamma_j \in\bit$, $o_j\in \bit^{KN}$, $Y^{(0,j)}, Y^{(1,j)}\in\bit^{KN\times b}$, $\beta\in [L]$, $\alpha\in \bit$, and $z_i=(x_i, y_i, \alpha_i, \beta_i)$ with $x_i\in [KN]$ and $y_i\in \bit^b$.
    Here, $L=\ceil{\log^2(KN)}\geq 5$.
    To prove \Cref{lem:encode-q-circ}, we construct a quantum query algorithm that queries the unitary
    \begin{equation}
        O_0: \ket{x, \alpha, \beta, k}\to (-1)^{\left(Y^{(\alpha, \beta)}_x\right)_k} \ket{x, \alpha, \beta, k}, \quad \forall x\in [KN], \alpha\in \bit, \beta\in [L], k\in [b],
    \end{equation}
    and its controlled version $cO_0$, and predicts $B$ with success probability at least $2/3$.
    Here, we take $O_0$ to act as identity on all other input states.
    Note that the number of qubits that $O_0$ acts on is 
    \begin{equation}
        n_1 = \ceil{\log(KN)} + 1 + \ceil{\log(L)} + \ceil{\log(b+1)} = \log N + O(\log\log N),
    \end{equation}
    because $L=\ceil{\log^2 (KN)}$ and $b = \ceil{40678\log(KN)}$.
    Here we are hosting the $\ket{k}$ register with $\ceil{\log(b+1)}$ qubits since it will be allowed to take values $k=0, \ldots, b$ later.

    To construct the quantum query algorithm that predicts $B$, we start by computing the oracle values $o_\beta\in \bit^{KN}, \beta\in [L]$ from $Y^{(\alpha, \beta)}\in \bit^{KN\times b}, \alpha\in \bit$ using the inner product noisy encoding map $g: \bit^{b}\times \bit^b\to \bit$.
    In particular, we introduce an ancilla qubit $a$ and note that
    \begin{equation}
        O_0^{\oplus} = H_a c_aO_0 H_a: \ket{x, \alpha, \beta, k}\ket{\chi}_a \to \ket{x, \alpha, \beta, k}\ket{\chi \oplus \left(Y^{(\alpha, \beta)}_{x}\right)_k}_a
    \end{equation}
    gives us the standard XOR oracle on $n_1+1$ qubits using two Hadamard gates.
    Now, we introduce three ancilla qubits $a, a_1, a_2$ and do the following steps.
    We start from 
    \begin{equation}
        \ket{x, 0, \beta, k}\ket{\chi}_a\ket{0}_{a_1}\ket{0}_{a_2}
    \end{equation}
    and apply $O^\oplus_0$ with ancilla $a_1$, yielding 
    \begin{equation}
        \ket{x, 0, \beta, k}\ket{\chi}_a\ket{\left(Y^{(0, \beta)}_{x}\right)_k}_{a_1}\ket{0}_{a_2}.
    \end{equation}
    Then we apply a NOT gate on the $\alpha$ register and apply $O^\oplus_0$ with ancilla $a_2$, obtaining 
    \begin{equation}
        \ket{x, 1, \beta, k}\ket{\chi}_a\ket{\left(Y^{(0, \beta)}_{x}\right)_k}_{a_1}\ket{\left(Y^{(1, \beta)}_{x}\right)_k}_{a_2}.
    \end{equation}
    Next, we reversibly compute the AND of $a_1, a_2$ via a Toffoli gate and add the result in $a$.
    This gives us 
    \begin{equation}
        \ket{x, 1, \beta, k}\ket{\chi\oplus \left(Y^{(0, \beta)}_{x}\right)_k\cdot \left(Y^{(1, \beta)}_{x}\right)_k}_a\ket{\left(Y^{(0, \beta)}_{x}\right)_k}_{a_1}\ket{\left(Y^{(1, \beta)}_{x}\right)_k}_{a_2}.
    \end{equation}
    Finally, we uncompute everything.
    Note that the inverse unitaries used in uncomputation are the same as the original unitaries $O_0^\dagger = O_0$.
    This constructs the unitary
    \begin{equation}
        O^{\oplus, \mathrm{prod}}_0: \ket{x, \beta, k}\ket{\chi}_{a} \to \ket{x, \beta, k}\ket{\chi\oplus \left(Y^{(0, \beta)}_{x}\right)_k\cdot \left(Y^{(1, \beta)}_{x}\right)_k}_{a}
    \end{equation}
    using only two ancilla qubits and $O(1)$ queries to $O_0^\oplus$.
    From $O^{\oplus, \mathrm{prod}}_0$, we compute the inner product as follows.
    We start from $\ket{x, \beta, 1}\ket{\chi}_a$ and apply $O^{\oplus, \mathrm{prod}}_0$, giving us
    \begin{equation}
        \ket{x, \beta, 1}\ket{\chi\oplus \left(Y^{(0, \beta)}_{x}\right)_1\cdot \left(Y^{(1, \beta)}_{x}\right)_1}_{a}.
    \end{equation}
    Then we add one to the $k$ register and apply $O^{\oplus, \mathrm{prod}}_0$.
    We repeat this until $k=b$ and add another one to it to take $k$ back to $0$.
    Here all additions on the $k$ register are performed modulo $b+1$.
    The final state is
    \begin{equation}
        \ket{x, \beta, 0}\ket{\chi\oplus \bigoplus_{k=1}^b\left(\left(Y^{(0, \beta)}_{x}\right)_k\cdot \left(Y^{(1, \beta)}_{x}\right)_k\right)}_{a} = \ket{x, \beta, 0}\ket{\chi\oplus g(Y^{(0, \beta)}_{x}, Y^{(1, \beta)}_{x})}_a = \ket{x, \beta, 0}\ket{\chi\oplus (o_\beta)_x}_a.
    \end{equation}
    In this way, we have constructed the unitary
    \begin{equation}
        O^\oplus: \ket{x, \beta}\ket{\chi}_a \to \ket{x, \beta} \ket{\chi\oplus (o_\beta)_x}
    \end{equation}
    using $O(1)$ ancilla and $O(b)$ queries to $O^\oplus_0$.
    Lastly, we turn this into the phase oracle
    \begin{equation}
        O: \ket{x, \beta} \to (-1)^{(o_\beta)_x}\ket{x, \beta}, \quad \forall x\in[NK], \beta\in [L].
    \end{equation}
    with one more ancilla (e.g., in $\ket{-}$).
    As a result, we have constructed the phase oracle for solving any $\beta\in [L]$ instance of the Forrelation problem (see \Cref{def:forrelation}) using $O(1)$ ancilla and $O(b)$ queries to $O_0$.
    Note that here we are computing the inner product bit by bit at the cost of an additional factor of $b$ in query complexity, as compared to use $O(b)$ ancilla and compute the inner product directly.
    We choose this way of computing the inner product to keep the space complexity small: $\log N + O(\log\log N)$ rather than $O(b) = c\log N$ with a large constant $c$.
    This will be crucial in determining the size of the linear system in \Cref{lem:embed-qc-into-linear-sys} and proving the final classical hardness result with the optimal exponent.

    Now, we invoke \Cref{lem:forrelation-query-separation}, which gives us a quantum algorithm $\mathcal{A}^O$ with $\log N + O(K)$ space complexity, $O(K 2^{10K}\log N \log(1/\eta))=O(\log N\log(1/\eta))$ gate complexity, making $O(K 2^{10K}\log(1/\eta))=O(\log(1/\eta))$ queries to $O$ (with $\beta$ register set to $\beta$) that outputs $\gamma_\beta$ with success probability at least $1-\eta$, when $o_\beta \sim p_{\gamma_\beta}$.
    Using $\mathcal{A}^O$, we compute the parity $B = \bigoplus_{\beta=1}^L \gamma_\beta$ as follows.
    We introduce an ancilla qubit $a$ to record the result.
    We first initialize the $\beta$ register to $\ket{1}$, execute $\mathcal{A}^O$, and apply a CNOT gate to copy the output qubit to the ancilla $a$.
    Then we run $\mathcal{A}^O$ in reverse to uncompute, add one to $\beta$, execute $\mathcal{A}^O$ again (now for $\beta=2$), and copy out the result with a CNOT gate.
    We repeat the above procedure until $\beta=L$.
    This is the quantum circuit $C$ that we construct.

    We now bound the coherent error of this computation.
    Fix any $\beta\in[L]$ and any $\chi\in\bit$.
    Since $\mathcal{A}^O$ outputs $\gamma_\beta$ with probability atleast $1-\eta$, the component of $\mathcal{A}^O\ket{\beta}\ket{0}$ on which its output qubit is different from $\gamma_\beta$ has Euclidean norm at most $\sqrt{\eta}$.
    On the component with output $\gamma_\beta$, copying the output to $a$ has exactly the desired effect $\ket{\chi}_a\mapsto\ket{\chi\oplus\gamma_\beta}_a$.
    On the remaining component, the difference between the actual and desired operations has norm at most twice the norm of that component.
    Therefore,
    \begin{equation}
        \left\|
            (\mathcal{A}^O)^\dagger
            \operatorname{CNOT}_{\mathrm{out},a}
            \mathcal{A}^O
            \ket{\beta}\ket{0}\ket{\chi}_a
            -
            \ket{\beta}\ket{0}
            \ket{\chi\oplus\gamma_\beta}_a
        \right\|_2 \leq
        2\sqrt{\eta}.
    \end{equation}
    Applying this bound successively for $\beta=1,\ldots,L$ and using the triangle inequality after each step gives
    \begin{equation}
        \left\|
            C\ket{0}
            -
            \ket{B}_a\ket{0}
        \right\|_2
        \leq
        2L\sqrt{\eta},
        \quad
        B=\bigoplus_{\beta=1}^L\gamma_\beta,
    \end{equation}
    where after the last execution we return the $\beta$ register to $\ket{0}$ using fixed gates.

    We choose $\eta=\frac{1}{1600L^2}$ and therefore
    \begin{equation}
        \left\|
            C\ket{0}
            -
            \ket{B}_a\ket{0}
        \right\|_2
        \leq
        \frac{1}{20}.
    \end{equation}
    Consequently,
    \begin{equation}
        \left|
            \bra{B}_a\bra{0}C\ket{0}
        \right|^2
        \geq
        1-(1/20)^2
        >
        0.99.
    \end{equation}
    Thus, measuring all qubits in the computational basis returns $B$ on the output qubit and $0$ on every remaining qubit with probability at least $0.99$.
    The circuit has space complexity $n_1+O(1)=\log N + O(\log\log N)$, gate complexity $O(L\log N \log(1/\eta))=O(L\log N \log\log N)=O(\log^3 N \log\log N)$, and queries $O_0$
    \begin{equation}
        Q=O(bL\log L) = O(\log^3 N\log\log N)
    \end{equation}
    times.
    This completes the proof of \Cref{lem:encode-q-circ}.
\end{proof}

\subsubsection{Connect to linear systems}
\label{sec:conn-linear-sys}

Next, we embed this quantum circuit into a sparse and well-conditioned matrix following the idea of \cite{harrow2009quantum}, such that solving the linear system given by this matrix is as hard as executing the quantum circuit.
This will be used to prove the classical hardness of linear system in \Cref{sec:linear-system}.

\begin{lemma}[Embed a quantum circuit into a real linear system]
\label{lem:embed-qc-into-linear-sys}
    Let $n$ be a large integer and $T, s$ be positive integers.
    Let $U_T\cdots U_1$ be an $n$-qubit quantum circuit composed of $T$ $n$-qubit unitaries $U_i, i\in [T]$ that are all $s$-sparse.
    Define a $3T2^n$ dimensional unitary $U$ as 
    \begin{equation}
        U = \sum_{t=1}^T\left(\ket{t+1}\bra{t}\otimes U_t + \ket{t+T+1}\bra{t+T}\otimes I + \ket{t+2T+1~\mathrm{mod}~ 3T}\bra{t+2T}\otimes U^\dagger_{T+1-t}\right).
    \end{equation}
    Let
    \begin{equation}
    \label{eq:B}
        B_c = \begin{pmatrix}
        I_{3T2^n} &0 \\
        0 &I_{3T2^n} - Ue^{-1/T}
    \end{pmatrix} \in \mathbb{C}^{6T2^n\times 6T2^n} = \mathbb{C}^{2\times 2}\otimes \mathbb{C}^{3T2^n\times 3T2^n}, 
    \end{equation}
    where the basis vectors in the first part of the factorization $\mathbb{C}^{2\times 2}$ are labeled by $\{\ket{0_a}, \ket{1_a}\}$.
    Let $P\in \mathbb{C}^{6T2^n\times 6T2^n}$ be the basis permutation matrix that swaps $\ket{0_a}$ and $\ket{1_a}$ while keeping the rest unchanged.
    Similarly, define $R\in \mathbb{C}^{6T2^n\times 6T2^n}$ to be the basis permutation matrix that swaps the basis vectors $\ket{1_a}\ket{t}\ket{0}\ket{\chi}$ and $\ket{0_a}\ket{t}\ket{0}\ket{\chi}$ for all $t\in [T+1, 2T], \chi\in\bit^{n-1}$ and keeps the rest unchanged.
    Consider
    \begin{equation}
    \label{eq:A}
        A_c = \frac{1}{2}\begin{pmatrix}
            0 &P^\dagger B_c R \\
            R^\dagger B_c^\dagger P &0
        \end{pmatrix}
        \in \mathbb{C}^{12T2^n \times 12T2^n}, \quad \vec{b}_c = \begin{pmatrix}
            1 \\
            0 \\
            \vdots \\
            0
        \end{pmatrix}
        \in \mathbb{R}^{12T2^n},
    \end{equation}
    whose realification
    \begin{equation}
        A=\mathcal{R}[A_c]\in \mathbb{R}^{24T2^n\times 24T2^n}, \quad \vec{b} = \mathcal{R}[\vec{b}_c] \in \mathbb{R}^{24T2^n}
    \end{equation}
    define a real linear system $A\vec{x} = \vec{b}$ of dimension $24T2^n$ for $\vec{x}\in\mathbb{R}^{24T2^n}$.
    
    Then, $A$ is real, symmetric and $O(s)$-sparse with operator norm $\|A\|\leq 1$ and condition number $\kappa = \|A^{-1}\|\|A\|\leq 4T$.
    Moreover, let
    \begin{equation}
        \mathcal{M}_c = I_2 \otimes \begin{pmatrix}
            1 &0\\
            0 &0
        \end{pmatrix} \otimes I_{3T2^n}, \quad \mathcal{M}=\mathcal{R}[\mathcal{M}_c] \in \mathbb{R}^{24T2^n\times 24T2^n}.
    \end{equation}
    The solution vector $\vec{x}$ satisfies
    \begin{equation}
    \label{eq:quad-form-vs-decision-bit}
        \frac{\vec{x}^T\mathcal{M}\vec{x}}{\|\vec{x}\|_2\|\mathcal{M}\|\|\vec{x}\|_2}  = \frac{e^{-2}}{1+e^{-2}+e^{-4}}\|\braket{0_1}{\psi}\|^2,
    \end{equation}
    where $\ket{\psi} = U_T\cdots U_1\ket{0}$ is the $n$-qubit output state of the circuit and $\ket{0_1}$ is the zero state on the first qubit of the circuit.
\end{lemma}

We rescale the resulting coefficient matrix as $A\mapsto A/\|A\|$ such that $\|A\|=1$.
The normalization factor depends only on $T$.
Thus, we can apply this rescaling to every matrix-element sample using the public circuit description.
This gives $\|A\|=1$ while preserving the sparsity, condition number, and normalized quadratic form.

\begin{proof}[Proof of \Cref{lem:embed-qc-into-linear-sys}]
    First note that $A_c$ is clearly Hermitian $A_c^\dagger = A_c$ by construction.
    Therefore, its realification $A$ is symmetric.
    Its sparsity depends on $U$.
    Since for any basis state with some clock value $\ket{t}$, there is only one term in $U$ that acts on it, changes the clock value to $\ket{t+1}$, and applies an $s$-sparse unitary $U_t$ (or $I, U^\dagger_{T+1-t}$), thus the unitary $U$ is also $s$-sparse.
    Therefore, $A_c$ and its realification $A=\mathcal{R}[A_c]$ are both $O(s)$ sparse.
    To compute its condition number, note that $A_c^\dagger A_c = \frac{1}{4}\begin{pmatrix}
            P^\dagger B_c B_c^\dagger P &0\\
            0 &R^\dagger B_c^\dagger B_c R
        \end{pmatrix}$.
    Since $P, R$ are basis permutation matrix, the singular values of $A_c$ are the same as those of $B_c/2$, and thus $A, A_c, B_c$ share the same condition number $\kappa$.
    $B_c$ is clearly invertible by construction.
    Moreover, $\|B_c\| \leq 1+e^{-1/T}<2$ and 
    \begin{equation}
        \|B_c^{-1}\|\leq \frac{1}{1-e^{-1/T}}\leq \frac{1}{1/(2T)}\leq 2T
    \end{equation}
    because $1-e^{-w}\geq w/2, \forall w\in (0, 1]$.
    Therefore, we have condition number $\kappa = \|B_c\|\|B^{-1}_c\|\leq 4T$ and norm $\|A\|=\|B_c\|/2\leq 1$ as desired.

    Next, we define the complex solution vector $\vec{x}_c = A^{-1}_c\vec{b}_c$ that satisfies
    \begin{equation}
        \vec{x} = A^{-1}\vec{b} = \mathcal{R}[A^{-1}_c]\mathcal{R}[\vec{b}_c] = \mathcal{R}[A^{-1}_c\vec{b}_c] = \mathcal{R}[\vec{x}_c].
    \end{equation}
    We calculate the complex solution vector 
    \begin{equation}
        \frac{1}{2}\vec{x}_c = \frac{1}{2}A^{-1}_c\vec{b}_c = \begin{pmatrix}
            0 & P^\dagger (B_c^{-1})^\dagger R \\
            R^\dagger B_c^{-1} P &0
        \end{pmatrix}\vec{b} = \vec{0}\oplus \left(R^\dagger B^{-1} P\begin{pmatrix}
            1 \\
            0 \\
            \vdots \\
            0
        \end{pmatrix}\right)
        = \ket{1_b} \otimes \left(R^\dagger B_c^{-1} P\ket{0_a}\ket{t=1}\ket{0^n}\right),
    \end{equation}
    where the first equality can be verified by direct calculation $A^{-1}A=I$ and we are using $\ket{0_b}, \ket{1}_b$ to label the basis vectors corresponding to the two blocks in the block matrix $A$.
    Now we plug in the expressions for $B_c, P$ and obtain
    \begin{equation}
    \begin{split}
        \frac{1}{2}\vec{x}_c &= \ket{1_b}\otimes (R^\dagger \underbrace{(\ket{0_a}\bra{0_a}\otimes I_{3T2^n} + \ket{1_a}\bra{1_a}\otimes (I_{3T2^n}-Ue^{-1/T})^{-1})}_{B_c^{-1}}\underbrace{\ket{1_a}\ket{t=1}\ket{0^n}}_{P\ket{0_a}\ket{t=1}\ket{0^n}}) \\
        &=\ket{1_b}\otimes R^\dagger \ket{1_a} (I_{3T2^n}-Ue^{-1/T})^{-1} \ket{t=1}\ket{0^n}.
    \end{split}
    \end{equation}
    Note that $P$ is introduced here to transform the bias vector $\vec{b}_c$ into the nontrivial subspace of $B_c$.
    Using the formula for geometric series, we have
    \begin{equation}
    \begin{split}
        (I_{3T2^n}-Ue^{-1/T})^{-1} \ket{t=1}\ket{0^n} &= \sum_{k\geq 0} U^k e^{-k/T}\ket{t=1}\ket{0^n} \\
        &= (1+e^{-3} +e^{-6} +\cdots)\sum_{k=0}^{3T-1} U^k e^{-k/T}\ket{t=1}\ket{0^n} \\
        &= \frac{1}{1-e^{-3}}\left(\sum_{k=0}^{3T-1} e^{-k/T} \ket{k+1} U_k\cdots U_1 \ket{0^n}\right),
    \end{split}
    \end{equation}
    where we have used $U^{3T}=I$ and the shorthand $U_{t}=I$ for $t\in [T+1, 2T]$ and $U_{t}=U^\dagger_{3T+1-t}$ for $t\in [2T+1, 3T]$.
    Recall that we use $\ket{0_1}, \ket{1_1}$ to denote the basis vector of the first qubit in the $n$-qubit circuit.
    Therefore, the solution vector satisfies
    \begin{equation}
    \begin{split}
        \frac{1}{2}(1-e^{-3})\vec{x}_c &= \ket{1_b}\sum_{k=0}^{3T-1} e^{-k/T} R^\dagger \ket{1_a} \ket{k+1}U_k\cdots U_1\ket{0^n} \\
        &= \ket{1_b}\sum_{k=0}^{3T-1} e^{-k/T} R^\dagger \ket{1_a} \ket{k+1}(\ket{0_1}\bra{0_1}U_k\cdots U_1\ket{0^n} + \ket{1_1}\bra{1_1}U_k\cdots U_1\ket{0^n}) \\
        &=\ket{1_b}\ket{0_a}\sum_{k=T}^{2T-1}e^{-k/T}\ket{k+1}\ket{0_1}\bra{0_1}U_k\cdots U_1\ket{0^n} \\
        &+ \ket{1_b}\ket{1_a}\sum_{k\in [0, T-1]\cup[2T, 3T-1]}e^{-k/T}\ket{k+1}\ket{0_1}\bra{0_1}U_k\cdots U_1\ket{0^n} \\
        &+\ket{1_b}\ket{1_a}\sum_{k=0}^{3T-1}e^{-k/T}\ket{k+1}\ket{1_1}\bra{1_1}U_k\cdots U_1\ket{0^n}.
    \end{split}
    \end{equation}
    This implies that
    \begin{equation}
    \begin{split}
        \frac{1}{4}\vec{x}^\dagger_c \mathcal{M}_c \vec{x}_c &= \frac{1}{4}\vec{x}^\dagger_c \ket{0_a}\bra{0_a} \vec{x}_c \\
        &= \frac{1}{(1-e^{-3})^2}\left\|\sum_{k=T}^{2T-1}e^{-k/T}\ket{k+1}\bra{0_1}U_k\cdots U_1\ket{0^n} \right\|^2 \\
        &= \frac{1}{(1-e^{-3})^2}\left\|\sum_{k=T}^{2T-1}e^{-k/T}\ket{k+1}\braket{0_1}{\psi} \right\|^2 \\
        &=\frac{1}{(1-e^{-3})^2}\sum_{k=T}^{2T-1}e^{-2k/T} \|\braket{0_1}{\psi}\|^2 \\
        &=\frac{e^{-2}}{(1-e^{-3})^2}\|\braket{0_1}{\psi}\|^2\sum_{k=0}^{T-1}e^{-2k/T},
    \end{split}
    \end{equation}
    where we have used the fact that $U_k\cdots U_1\ket{0^n} = U_T\cdots U_1\ket{0^n} = \ket{\psi}$ for $k\in [T, 2T-1]$.
    On the other hand, the norm of $\vec{x}$ reads
    \begin{equation}
        \frac{1}{4}\|\vec{x}_c\|_2^2 = \frac{1}{(1-e^{-3})^2}\sum_{k=0}^{3T-1} e^{-2k/T} = \frac{1}{(1-e^{-3})^2} (1+ e^{-2} + e^{-4})\sum_{k=0}^{T-1}e^{-2k/T}.
    \end{equation}
    This gives us 
    \begin{equation}
        \frac{\vec{x}^\dagger_c}{\|\vec{x}_c\|_2} \mathcal{M}_c \frac{\vec{x}_c}{\|\vec{x}_c\|_2} = \frac{e^{-2}}{1+e^{-2}+e^{-4}}\|\braket{0_1}{\psi}\|^2.
    \end{equation}
    Taking the realification via \Cref{lem:realification} and noting that $\|\mathcal{M}\|=\|\mathcal{M}_c\|=1$ completes the proof of \Cref{lem:embed-qc-into-linear-sys}.
\end{proof}

\subsubsection{Connect to binary classification}
\label{sec:conn-bin-classify}

Then, we move on to the application of binary classification.
We consider binary classification with least-squares SVM.
In a least-squares SVM, we consider a binary-label training dataset of $N$ samples $(\vec{x}_i, y_i)_{i=1}^N$, each specified by a $D$-dimensional feature vector $\vec{x}_i\in \mathbb{R}^D$ and a label $y_i \in \{\pm1\}$.
We use a matrix $X = (\vec{x}_1, \ldots, \vec{x}_N)^T\in \mathbb{R}^{N\times D}$ and a vector $\vec{y} = (y_1, \ldots, y_N)^T\in \{\pm 1\}^N$ to represent the training set.
The goal is to use the training set to classify a new test point $\vec{x}\in \mathbb{R}^D$ via $\hat{y} = \sgn(\vec{x}^T (X^T X+\lambda I_D)^{-1} X^T\vec{y})$, where $\lambda\geq 0$ is the $\ell_2$ regularization strength.

We show below that any quantum circuit $U_T\cdots U_1$ of size $T$ can be embedded into a training dataset $(X, \vec{y})$ with feature length $D=N$ and a sparse test vector $\vec{x}$ such that calculating $\hat{y}=\sgn(\vec{x}^T (X^T X)^{-1} X^T \vec{y})$ with no regularization suffices to determine $\sgn(\bra{\psi}Z_1\ket{\psi})$, where $\ket{\psi}=U_T\cdots U_1\ket{0}$ is the output state of the circuit.
One can interpret this result as proving the BQP hardness of binary classification.
This will be used to prove classical hardness of binary classification in \Cref{sec:binary-classification}.

\begin{lemma}[Embed a quantum circuit into an SVM]
\label{lem:embed-qc-into-svm}
    Let $n$ be a large integer and $T, s$ be positive integers.
    Let $T'=2T+n+1$.
    Let $U_T\cdots U_1$ be an $n$-qubit quantum circuit composed of $T$ $n$-qubit unitaries $U_i, i\in [T]$ that are all $s$-sparse.
    Let $A_c\in \mathbb{C}^{12T'2^n\times 12T'2^n}=\mathbb{C}^{2}\otimes \mathbb{C}^{2}\otimes \mathbb{C}^{3T'}\otimes \mathbb{C}^{2^n}$ be the complex matrix specified by the unitary sequence $U_1^\dagger \cdots U_T^\dagger Z_1U_T\cdots U_1H_n\cdots H_1$ in \Cref{lem:embed-qc-into-linear-sys}, where $Z_1$ is the Pauli $Z$ gate on the first qubit of the circuit and $H_i$ is the Hadamard gate on the $i$-th qubit.
    We use $b, a$ to label the two binary registers corresponding to the two layers of blocks inside $A_c$ as defined in \Cref{lem:embed-qc-into-linear-sys}.
    Let
    \begin{equation}
        X_c = \begin{pmatrix}
            A_c &0\\
            0 &A_c
        \end{pmatrix}\in \mathbb{C}^{24T'2^n\times 24T'2^n}, \quad \vec{y}_c = \begin{pmatrix}
            \vec{y}_1\\
            \vec{y}_2
        \end{pmatrix}\in \{\pm 1\}^{24T'2^n},
    \end{equation}
    where
    \begin{equation}
    \begin{split}
        \vec{y}_1 &= (1, \ldots, 1)^T = \sum_{b_b, b_a\in \bit}\sum_{t=1}^{3T'} \ket{b_b}\ket{b_a}\ket{t}\sqrt{2^n}\ket{+^n}\in \{\pm 1\}^{12T'2^n}, \\
        \vec{y}_2 &= \left(2\ket{0_b}\ket{0_a}\ket{t=1}-\sum_{b_b, b_a\in \bit}\sum_{t=1}^{3T'} \ket{b_b}\ket{b_a}\ket{t}\right)\sqrt{2^n}\ket{+^n} \in\{\pm 1\}^{12T'2^n}.
    \end{split}
    \end{equation}
    Let
    \begin{equation}
        \vec{x}_c = (\ket{0_c}+\ket{1_c})\ket{1_b}\ket{0_a}\ket{T'+1}\ket{0^n} \in \mathbb{C}^{24T'2^n} = \mathbb{C}^{2}\otimes \mathbb{C}^{2}\otimes \mathbb{C}^{2}\otimes \mathbb{C}^{3T'}\otimes \mathbb{C}^{2^n}
    \end{equation}
    be a complex $2$-sparse test vector, where we use $c$ to label the binary register corresponding to the two blocks of $X_c$.
    Let the realification
    \begin{equation}
        X = \mathcal{R}[X_c]/\|\mathcal{R}[X_c]\|\in \mathbb{R}^{48T'2^n\times 48T'2^n}, \quad \vec{y} = \mathcal{R}[(1+i)\vec{y}_c] \in \{\pm 1\}^{48T'2^n}, \quad \vec{x} =  \mathcal{R}[\vec{x}_c] \in \mathbb{R}^{48T'2^n}
    \end{equation}
    define a real training dataset of size $48T'2^n$ and feature dimension $48T'2^n$ and a $4$-sparse test vector.
    Then, $X$ is real, symmetric, and $O(s)$-sparse with norm $\|X\|= 1$ and condition number $\kappa = \|X^{-1}\|\|X\|\leq 4T'$, and
    \begin{equation}
        \vec{x} \cdot \frac{(X^T X)^{-1}X^T\vec{y}}{\|(X^T X)^{-1}X^T\vec{y}\|} = C \bra{\psi}Z_1\ket{\psi},
    \end{equation}
    where
    \begin{equation}
        \frac{1}{50T'^{3/2}}\leq C\leq \frac{2}{5T'^{1/2}}
    \end{equation}
    and $\ket{\psi} = U_T\cdots U_1\ket{0}$ is the $n$-qubit output state of the circuit and $Z_1$ is the Pauli $Z$ observable on the first qubit of the circuit.
    In particular, we have
    \begin{equation}
    \label{eq:class-vs-decision-bit}
        \sgn(\vec{x}^T (X^T X)^{-1} X^T \vec{y}) = \sgn(\bra{\psi}Z_1\ket{\psi}).
    \end{equation}
\end{lemma}

Note that the normalization factor $\|\mathcal{R}[X_c]\|=\|X_c\|$ is a known function of $T'$ and is independent of the encoded oracle instance. Hence, every sampled row in the SVM application setting can be rescaled online by this known factor without additional sample or space complexity.

\begin{proof}[Proof of \Cref{lem:embed-qc-into-svm}]
    We use $V_{T'}\cdots V_{1} = U_1^\dagger \cdots U_T^\dagger Z_1U_T\cdots U_1 H_n\cdots H_1$ to denote the embedded unitary sequence that has length $T'=2T+n+1$.
    From \Cref{lem:embed-qc-into-linear-sys}, we know that $A_c$ is Hermitian, $O(s)$ sparse, and has norm $\|A_c\|\leq 1$ and condition number at most $4T'=4(2T+n+1)$.
    This implies that both $X_c, X$ are $O(s)$ sparse with norm $\leq 1$ and condition number $\kappa\leq 4T'=4(2T+n+1)$.
    We also have
    \begin{equation}
        (X_c^\dagger X_c)^{-1}X_c^\dagger\vec{y}_c = X_c^{-1}\vec{y}_c = \begin{pmatrix}
            A_c^{-1} &0\\
            0 & A_c^{-1}
        \end{pmatrix}\begin{pmatrix}
            \vec{y}_1 \\
            \vec{y}_2
        \end{pmatrix} = \begin{pmatrix}
            A_c^{-1}\vec{y}_1 \\
            A_c^{-1}\vec{y}_2
        \end{pmatrix}.
    \end{equation}
    Let $\vec{x}_0 = \ket{1_b}\ket{0_a}\ket{T'+1}\ket{0^n} \in \mathbb{C}^{12T'2^n}$ such that
    \begin{equation}
        \vec{x}_c = \begin{pmatrix}
            \vec{x}_0\\
            \vec{x}_0
        \end{pmatrix}.
    \end{equation}
    Let 
    \begin{equation}
        \vec{b}_c = \frac{\vec{y}_1+\vec{y}_2}{2} = \ket{0_b}\ket{0_a}\ket{t=1}\sqrt{2^n}\ket{+^n} \in \{0, \pm 1\}^{12T'2^n}.
    \end{equation}
    Then we have
    \begin{equation}
        \vec{x}_c^\dagger (X_c^\dagger X_c)^{-1}X_c^\dagger \vec{y}_c = \begin{pmatrix}
            \vec{x}_0^\dagger &\vec{x}_0^\dagger
        \end{pmatrix}
        \begin{pmatrix}
            A^{-1}_c\vec{y}_1\\
            A^{-1}_c\vec{y}_2
        \end{pmatrix} = \vec{x}^\dagger_0 A^{-1}_c(\vec{y}_1+\vec{y}_2) = 2\vec{x}_0^\dagger A^{-1}_c\vec{b}_c = 2\bra{1_b}\bra{0_a}\bra{T'+1}\bra{0^n}A^{-1}_c\vec{b}_c.
    \end{equation}
    The calculations in the proof of \Cref{lem:embed-qc-into-linear-sys} shows that
    \begin{equation}
    \begin{split}
        \frac{1}{2}(1-e^{-3})A^{-1}_c\vec{b}_c &= \ket{1_b}\ket{0_a}\sum_{k=T'}^{2T'-1}e^{-k/T'}\ket{k+1}\ket{0_1}\bra{0_1}V_k\cdots V_1\sqrt{2^n}\ket{+^n} \\
        &+ \ket{1_b}\ket{1_a}\sum_{k\in [0, T'-1]\cup[2T', 3T'-1]}e^{-k/T'}\ket{k+1}\ket{0_1}\bra{0_1}V_k\cdots V_1\sqrt{2^n}\ket{+^n} \\
        &+\ket{1_b}\ket{1_a}\sum_{k=0}^{3T'-1}e^{-k/T'}\ket{k+1}\ket{1_1}\bra{1_1}V_k\cdots V_1\sqrt{2^n}\ket{+^n}.
    \end{split}
    \end{equation}
    Plugging this in, we have
    \begin{equation}
    \begin{split}
        \vec{x}_c^\dagger (X_c^\dagger X_c)^{-1}X_c^\dagger \vec{y}_c &= \frac{4\sqrt{2^n}}{1-e^{-3}} \bra{0^n} e^{-1}\ket{0_1}\bra{0_1}V_{T'}\cdots V_1\ket{+^n} \\
        &=\frac{4\sqrt{2^n}e^{-1}}{1-e^{-3}} \bra{0^n}V_{T'}\cdots V_1\ket{+^n} \\
        &=\frac{4\sqrt{2^n}e^{-1}}{1-e^{-3}} \bra{0^n}U_1^\dagger \cdots U_T^\dagger Z_1U_T\cdots U_1H_n\cdots H_1\ket{+^n} \\
        &=\frac{4\sqrt{2^n}e^{-1}}{1-e^{-3}} \bra{\psi}Z_1\ket{\psi},
    \end{split}
    \end{equation}
    where $\ket{\psi} = U_T\cdots U_1\ket{0^n}$ is the output state of the $n$-qubit circuit.
    This implies that $\vec{x}_c^\dagger (X_c^\dagger X_c)^{-1}X_c^\dagger \vec{y}_c$ is real and
    \begin{equation}
        \sgn(\Re[\vec{x}_c^\dagger (X_c^\dagger X_c)^{-1}X_c^\dagger \vec{y}_c]) = \sgn(\bra{\psi}Z_1\ket{\psi}).
    \end{equation}
    
    Finally, we bound the norm of $\vec{x}_c$ and $(X_c^\dagger X_c)^{-1}X_c^\dagger\vec{y}_c$.
    We have
    \begin{equation}
        \|\vec{x}_c\|_2 = \sqrt{2}, \quad \|\vec{y}_c\|_2 = \sqrt{24T'2^n},
    \end{equation}
    and
    \begin{equation}
        \|(X_c^\dagger X_c)^{-1}X_c^\dagger\vec{y}_c\|_2 =\|X_c^{-1} \vec{y}_c\|_2 \in \left[\frac{\|\vec{y}_c\|_2}{\|X_c\|}, \frac{\kappa\|\vec{y}_c\|_2}{\|X_c\|}\right] \subseteq \left[\sqrt{24T'2^n}, 4T'\sqrt{24T'2^n}\right].
    \end{equation}
    Therefore, we have
    \begin{equation}
        \frac{\vec{x}_c^\dagger}{\|\vec{x}_c\|}\frac{(X_c^\dagger X_c)^{-1}X_c^\dagger\vec{y}_c}{\|(X_c^\dagger X_c)^{-1}X_c^\dagger\vec{y}_c\|} = C \bra{\psi}Z_1\ket{\psi},
    \end{equation}
    where
    \begin{equation}
        C \in \left[\frac{e^{-1}}{1-e^{-3}}\frac{1}{\sqrt{48T'^3}}, \frac{e^{-1}}{1-e^{-3}}\frac{4}{\sqrt{48T'}}\right]\subseteq \left[\frac{1}{50T'^{3/2}}, \frac{2}{5T'^{1/2}}\right].
    \end{equation}

    The real label vector used in the final training problem is $\vec{y}=\mathcal{R}[(1+i)\vec{y}_c]$.
    By the realification identity,
    \begin{equation}
        \vec{x}\cdot (X^TX)^{-1}X^T\vec{y}=
        \|\mathcal{R}[X_c]\|
        \Re\left[
            (1+i)\vec{x}_c^\dagger
            (X_c^\dagger X_c)^{-1}
            X_c^\dagger\vec{y}_c
        \right] = \|\mathcal{R}[X_c]\| \vec{x}_c^\dagger
        (X_c^\dagger X_c)^{-1}
        X_c^\dagger\vec{y}_c,
    \end{equation}
    since we have shown that $\vec{x}_c^\dagger(X_c^\dagger X_c)^{-1}X_c^\dagger\vec{y}_c$ is real.
    On the other hand,
    \begin{equation}
        \|(X^TX)^{-1}X^T\vec{y}\|=\|\mathcal{R}[X_c]\|
        \left\|
            \mathcal{R}\left[
                (1+i)
                (X_c^\dagger X_c)^{-1}
                X_c^\dagger\vec{y}_c
            \right]
        \right\|_2=
        \sqrt{2}\|\mathcal{R}[X_c]\|
        \left\|
            (X_c^\dagger X_c)^{-1}
            X_c^\dagger\vec{y}_c
        \right\|_2.
    \end{equation}
    Combining these calculations with $\|\vec{x}_c\|_2=\sqrt{2}$, we obtain
    \begin{equation}
        \vec{x}\cdot
        \frac{
            (X^TX)^{-1}X^T\vec{y}
        }{
            \|(X^TX)^{-1}X^T\vec{y}\|_2
        }
        =
        C\bra{\psi}Z_1\ket{\psi}.
    \end{equation}
    This completes the proof of \Cref{lem:embed-qc-into-svm}.
\end{proof}

\subsubsection{Connect to dimension reduction}
\label{sec:conn-dim-reduc}

Finally, we consider the application of dimension reduction.
We focus on the task of performing principal component analysis (PCA) on a data matrix $X\in \mathbb{R}^{N\times D}$, which reduces the data to one dimension.
That means finding the eigenvector $\vec{w}\in \mathbb{R}^D$ of $X^TX\in \mathbb{R}^{D\times D}$ with the largest eigenvalue $\lambda_{\max}$ and projecting the test vector $\vec{x}\in \mathbb{R}^D$ to that direction as $\vec{x}\cdot \vec{w}$.
This reduces the test vector $\vec{x}$ to a scalar $\vec{x}\cdot \vec{w}$, its 1D representation.

In the following, we show that any quantum circuit $U_T\cdots U_1$ of size $T$ can be embedded into a data matrix $X$ with feature length $D=N$ and a sparse test vector $\vec{x}$ such that calculating the 1D representation $\vec{x}\cdot \vec{w}$ to $1/\poly(T, \log(N))$ error suffices to determine the output of that circuit.
We also show that there is a sparse guiding vector that have good overlap with the principal component.
One can interpret this result as proving the BQP hardness of dimension reduction given a good guiding vector.
This will be used to prove the classical hardness of dimension reduction in \Cref{sec:dimension-reduction}.

\begin{lemma}[Embed a quantum circuit into PCA]
\label{lem:embed-qc-into-pca}
    Let $n$ be a large integer and $T, s$ be positive integers.
    Let $U_T\cdots U_1$ be an $n$-qubit quantum circuit composed of $T$ $n$-qubit unitaries $U_i, i\in [T]$ that are all $s$-sparse.
    Let
    \begin{equation}
    \begin{split}
        H_{\mathrm{circ}} &= \ketbra{0}\otimes (I_{2^{n+1}}-\ketbra{0^{n+1}}) \\
        &+ \sum_{t=1}^T \frac{1}{2} \left( \ketbra{t-1} \otimes I_{2^{n+1}} + \ketbra{t} \otimes I_{2^{n+1}} - \ketbra{t}{t-1} \otimes \mathcal{R}[U_t] - \ketbra{t-1}{t} \otimes \mathcal{R}[U_t^\dagger] \right) \\
        &\in \mathbb{R}^{(T+1)2^{n+1}\times (T+1)2^{n+1}},
    \end{split}
    \end{equation}
    where $\mathcal{R}[\cdot]$ is the realification defined in \Cref{lem:realification-circuit}.
    Let 
    \begin{equation}
        X = I_{(T+1)2^{n+1}} - \frac{H_{\mathrm{circ}}}{\|H_{\mathrm{circ}}\|} \in \mathbb{R}^{(T+1)2^{n+1}\times (T+1)2^{n+1}}
    \end{equation}
    define a real dataset of size $(T+1)2^{n+1}$ and feature dimension $(T+1)2^{n+1}$.
    Consider the $1$-sparse guiding vector $\vec{g}$ and test vector $\vec{x}$ defined as
    \begin{equation}
        \vec{g} = \ket{0}\ket{0^{n+1}} \in \mathbb{R}^{(T+1)2^{n+1}}, \quad \vec{x} = \ket{T}\ket{0^{n+1}} \in \mathbb{R}^{(T+1)2^{n+1}}.
    \end{equation}
    Then, the data matrix $X$ is real, symmetric, and $O(s)$-sparse with norm $\|X\|=1$ and gap 
    \begin{equation}
        \Delta = \lambda_{\max}(X^TX) - \lambda_{\mathrm{sec}}(X^TX) \geq \Omega\lr{\frac{1}{T^3}},
    \end{equation}
    where $\lambda_{\max}(X^TX), \lambda_{\mathrm{sec}}(X^TX)$ are the largest and second largest eigenvalues of $X^TX$.
    Moreover, the guiding vector have overlap 
    \begin{equation}
        \vec{g}\cdot \vec{w} = \frac{1}{\sqrt{T+1}}, \quad \vec{w} = \underset{\|\vec{w}\|_2=1}{\argmax}~\vec{w}^TX^TX\vec{w},
    \end{equation}
    with the principal component $\vec{w}$ of $X$, and the test vector $\vec{x}$ has 1D representation
    \begin{equation}
        \xi(\vec{x}) = \vec{x}\cdot \vec{w} = \frac{1}{\sqrt{T+1}}\Re[\expval{0^{n}|U_T\cdots U_1|0^n}]
    \end{equation}
\end{lemma}

We note that if we do not require the guiding vector $\vec{g}$ to be sparse, we can pad the circuit with $\Theta(T)$ identity gates in the beginning and choose $\vec{g}$ to be an equal superposition of all the clock time with identity gates tensor product with $\ket{0^{n+1}}$.
This will give a guiding vector with constant overlap $\vec{g}\cdot \vec{w}=\Theta(1)$ but sparsity $O(T)$.

\begin{proof}[Proof of \Cref{lem:embed-qc-into-pca}]
    Since each $U_t$ is $O(s)$-sparse, it is immediate by construction that $X$ is real, symmetric,and $O(s)$ sparse.
    In addition, since the $U_t$'s are unitary, we have that $\|H_{\mathrm{circ}}\|\leq 1+ T\cdot O(1)= O(T)$ by triangle inequality.
    It is also straightforward to show that $H_{\mathrm{circ}}\geq 0$.
    Therefore, we have $0\leq X\leq I_{(T+1)2^{n+1}}$.
    We prove the remaining properties by analyzing the spectrum of $H_{\mathrm{circ}}$ following \cite[Section 14.4]{kitaev2002classical}.
    Consider the basis rotation unitary
    \begin{equation}
        W = \sum_{t=0}^T \ketbra{t} \otimes \mathcal{R}[U_t \cdots U_1].
    \end{equation}
    We have that
    \begin{equation}
        W^T H_{\mathrm{circ}} W =  \ketbra{0}\otimes (I_{2^{n+1}}-\ketbra{0^{n+1}}) + \begin{pmatrix}
        \frac{1}{2} & -\tfrac12 & 0 & \cdots & 0\\
        -\tfrac12 & 1 & -\tfrac12 & \ddots & \vdots\\
        0 & -\tfrac12 & 1 & \ddots & 0\\
        \vdots & \ddots & \ddots & \ddots & -\tfrac12\\
        0 & \cdots & 0 & -\tfrac12 & \frac{1}{2}
        \end{pmatrix} \otimes I_{2^{n+1}}.
    \end{equation}
    Define 
    \begin{equation}
	E_0=\begin{pmatrix}
    \frac12 & -\tfrac12 & 0 & \cdots & 0\\
    -\tfrac12 & 1 & -\tfrac12 & \ddots & \vdots\\
    0 & -\tfrac12 & 1 & \ddots & 0\\
    \vdots & \ddots & \ddots & \ddots & -\tfrac12\\
    0 & \cdots & 0 & -\tfrac12 & \frac12
    \end{pmatrix}, \quad E_1 = \begin{pmatrix}
    \frac12+1 & -\tfrac12 & 0 & \cdots & 0\\
    -\tfrac12 & 1 & -\tfrac12 & \ddots & \vdots\\
    0 & -\tfrac12 & 1 & \ddots & 0\\
    \vdots & \ddots & \ddots & \ddots & -\tfrac12\\
    0 & \cdots & 0 & -\tfrac12 & \frac12
    \end{pmatrix} \in \mathbb{R}^{(T+1)\times (T+1)}.
    \end{equation}
    Elementary calculations show that $E_1$ has minimal eigenvalue $\Omega(1/T^2)$ and $E_0$ has minimal eigenvalue $0$ with non-degenerate eigenvector
    \begin{equation}
        \vec{w'} = \frac{1}{\sqrt{T+1}}(1, \ldots, 1)^T
    \end{equation}
    and second minimal eigenvalue $\Omega(1/T^2)$.
    Since $W^TH_{\mathrm{circ}}W$ is equal to $E_0$ when the qubit register is $\ket{0^{n+1}}$ and equal to $E_1\otimes I_{2^{n+1}}$ when the qubit register is orthogonal to $\ket{0^{n+1}}$, we know that $W^TH_{\mathrm{circ}}W$ has minimal eigenvalue $0$ and second minimal eigenvalue $\Omega(1/T^2)$, and so does $H_{\mathrm{circ}}$.
    The non-degenerate eigenvector with minimal eigenvalue $0$ of $H_{\mathrm{circ}}$ is therefore
    \begin{equation}
        \vec{w} = W(\vec{w}'\otimes \ket{0^{n+1}}) = \frac{1}{\sqrt{T+1}} \sum_{t=0}^{T} \ket{t}\otimes \mathcal{R}[U_t\cdots U_1]\ket{0^{n+1}}.
    \end{equation}
    Since $X = I_{(T+1)2^{n+1}} - H_{\mathrm{circ}}/\|H_{\mathrm{circ}}\|$, the vector $\vec{w}$ is indeed the principal component of $X$, with eigenvalue $\lambda_{\max}(X) = 1-0 = 1$.
    Therefore, the norm $\|X\|=1$ and the gap of $X$ is
    \begin{equation}
        \Delta = \lambda_{\max}^2(X) - \lambda_{\mathrm{sec}}^2(X) \geq \lambda_{\max}(X) (\lambda_{\max}(X) - \lambda_{\mathrm{sec}}(X)) \geq 1\cdot \frac{\Omega(1/T^2)}{\|H_{\mathrm{circ}}\|}\geq \Omega\lr{\frac{1}{T^3}},
    \end{equation}
    because $\|H_{\mathrm{circ}}\|\leq O(T)$.
    Moreover, the guiding vector $\vec{g} = \ket{0}\ket{0^{n+1}}$ has overlap
    \begin{equation}
        \vec{g}\cdot \vec{w} = \frac{1}{\sqrt{T+1}}\expval{0^{n+1}|0^{n+1}} = \frac{1}{\sqrt{T+1}}.
    \end{equation}
    The 1D representation of the test vector $\vec{x} = \ket{T}\ket{0^{n+1}}$ is 
    \begin{equation}
        \xi(\vec{x}) = \vec{x}\cdot \vec{w} = \frac{1}{\sqrt{T+1}}\bra{0^{n+1}}\mathcal{R}[U_T\cdots U_1]\ket{0^{n+1}} = \frac{1}{\sqrt{T+1}}\Re[\expval{0^{n}|U_T\cdots U_1|0^n}],
    \end{equation}
    as desired.
    This completes the proof of \Cref{lem:embed-qc-into-pca}.
\end{proof}

\newpage

\section{Applications}
\label{sec:app}

In this section, we apply quantum oracle sketching to various applications, including solving linear systems (\Cref{sec:linear-system} and \Cref{fig:linear-sys}), binary classification (\Cref{sec:binary-classification} and \Cref{fig:binary-classification}), and dimension reduction (\Cref{sec:dimension-reduction}).
We show that we can solve these useful classical data processing tasks in an end-to-end fashion with a small quantum computer.
Then, we use techniques developed in \Cref{sec:cl-hard} to rigorously prove the classical hardness of solving these tasks.
In particular, we prove that, using the same number of samples, any classical machine needs exponentially larger size to solve these tasks.
Moreover, in a dynamic setting, if the classical machine does not have enough size, it would need super-polynomially more samples.
Throughout this section, we use $\ket{x}=\sum_j x_j\ket{j}/\|\vec{x}\|_2$ to denote the real quantum state corresponding to any vector $\vec{x}=(x_1, \ldots, x_N)^T\in \mathbb{R}^N$.
In dynamic tasks, we calculate quantum oracle sketching guarantees conditioned on the sampled datasets, and suppress that conditioning for simplicity.

\subsection{Linear system}
\label{sec:linear-system}

\begin{figure}
    \centering
    \includegraphics[width=1\linewidth]{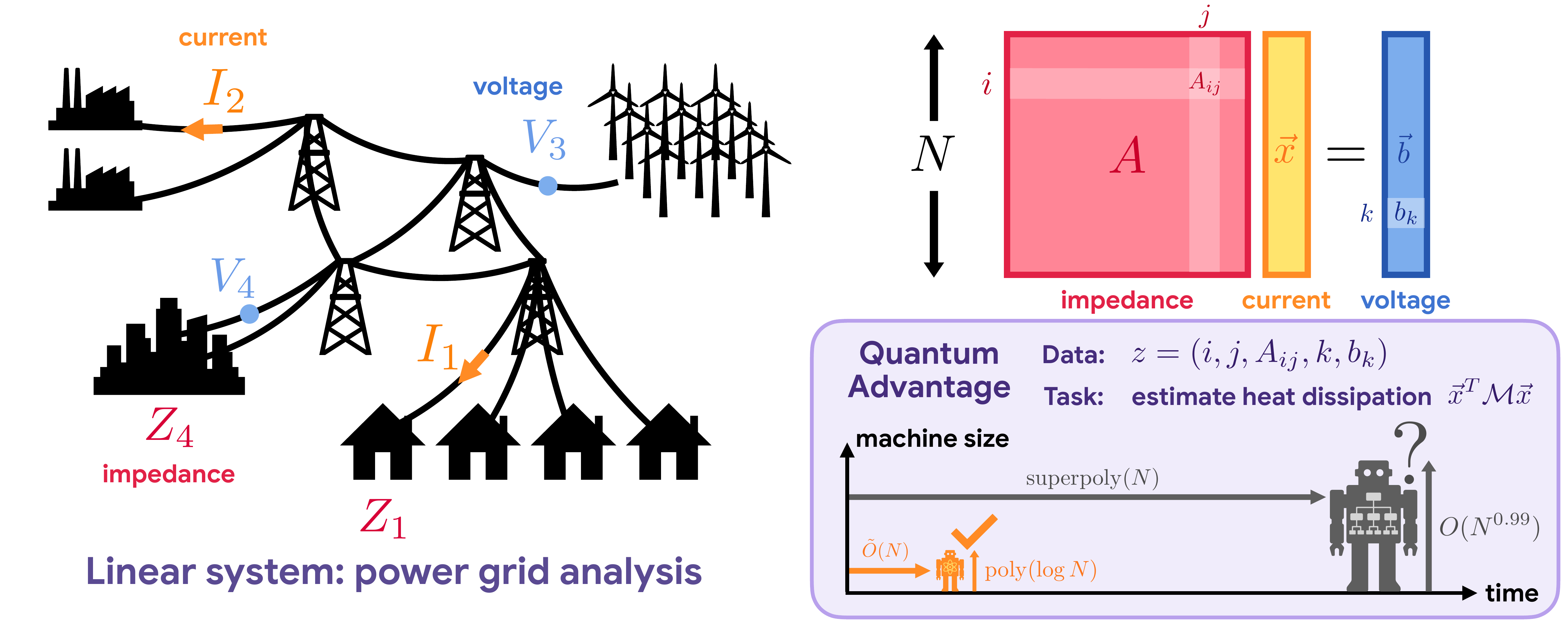}
    \caption[Overview of the linear system task.]{\textbf{Overview of the linear system task.}
    We illustrate the linear system task with a particular real-world application scenario in power grid analysis.
    In power grid analysis, samples are obtained by performing measurements on a power grid to read off the impedance values and the voltage values of random locations that are changing dynamically.
    We may want to estimate the heat dissipation of some critical junctions, given by a quadratic form of the current.
    This reduces to solving a high-dimensional linear system $A\vec{x}=\vec{b}$.
    Our results show that a small quantum machine with size $\poly(\log N)$ can solve this task with $\tilde{O}(N)$ samples, whereas any classical machine with exponentially larger size $O(N^{0.99})$ cannot solve the task unless it uses a sample size at least super-polynomial in $N$.
    }
    \label{fig:linear-sys}
\end{figure}

We begin with the fundamental primitive of solving linear systems, summarized in \Cref{fig:linear-sys}.
The solution of large linear systems serves as the computational bottleneck in a vast array of scientific and engineering disciplines. 
It is a key subroutine in many problems including regression, optimization, and solving differential equations that governs chemicals, materials, biological systems, fluid, and mechanical systems, etc.
As we increase the dataset size, sampling rate, or precision, the dimension $N$ of these systems grows rapidly.
Consequently, storing the data for processing becomes a bottleneck.
In this section, we show that a small quantum machine can solve linear systems better than exponentially larger classical machines.

Formally, we consider $N$-dimensional linear systems of the form
\begin{equation}
    A\vec{x} = \vec{b}, \quad A\in \mathbb{R}^{N\times N}, \quad \vec{x}, \vec{b}\in \mathbb{R}^N,
\end{equation}
where $A$ is a symmetric, sparse and well-conditioned matrix with operator norm $\|A\|= 1$ after appropriate rescaling.
The assumption that $A$ is symmetric incurs no loss of generality; any general linear system can always be embedded into a symmetric formulation $A_{\mathrm{sym}}\vec{x}_{\mathrm{sym}} = \vec{b}_{\mathrm{sym}}$, where $A_{\mathrm{sym}} = \begin{pmatrix}
    0 &A\\
    A^T &0
\end{pmatrix} \in \mathbb{R}^{2N\times 2N}$ and $\vec{b}_{\mathrm{sym}} = (\vec{b}, \vec{0})^T \in \mathbb{R}^{2N}$ with the solution $\vec{x}_{\mathrm{sym}} = (\vec{0}, \vec{x})^T$.
This embedding preserves the sparsity and condition number.

We characterize the tractability of the linear system with two parameters: the sparsity $s$ (the maximal number of nonzero elements per row or column) and condition number $\kappa = \|A^{-1}\|\|A\|$.
We focus on the regime where the dimension $N$ is very large, and yet it remains sparse and well-conditioned:
\begin{equation}
    s, \kappa \leq \poly(\log N).
\end{equation}

\subsubsection{Problem formulation}

Now we formally define our data processing task of solving linear systems, summarized in \Cref{task:linear-sys,task:linear-sys-dynamic}.
In particular, we specify the data generation process and the goal of the task.
We consider a data generation process where we randomly get a non-zero matrix element of coefficient matrix $A$ and a random component of the bias vector $\vec{b} = (b_1, \ldots, b_N)^T$.

Recall that in \Cref{sec:q-alg-linear-algebra}, we define the matrix data generation process of the matrix $A$ as a hierarchical data generation process that generates random non-zero matrix elements as data
\begin{equation}
    z^{\mathrm{coeff}} = (i, j, A_{ij}), \quad (i, j)\overset{\mathrm{marginal}}{\sim} \unif(\{(i, j): A_{ij}\neq 0\}).
\end{equation}
We assume that the linear system is properly normalized such that $\|A\|\leq 1$.
The vector data generation process of $\vec{b}$ is a hierarchical data generation process that generates random components of the vector as data
\begin{equation}
    z^{\mathrm{bias}} = (k, b_k), \quad k\overset{\mathrm{marginal}}{\sim} \unif([N]).
\end{equation}
The matrix elements $A_{ij}$ and vector components $b_k$ are specified by bitstrings of length $b=\poly(\log N)$ to sufficient accuracy.
For simplicity, we assume that these binary representations are exact and use $A_{ij}, b_k$ to stand for the corresponding values.

In the task of solving linear systems, we consider any hierarchical data generation process $\mathcal{D}_{\mathrm{LS}}(A, \vec{b})$ with bounded repetition number $R$ that generates data samples of the form
\begin{equation}
    z = (z^{\mathrm{coeff}}, z^{\mathrm{bias}}),
\end{equation}
where
\begin{equation}
    z^{\mathrm{coeff}} = (i, j, A_{ij}), \quad (i, j)\overset{\mathrm{marginal}}{\sim} \unif(\{(i, j): A_{ij}\neq 0\}), \quad z^{\mathrm{bias}} = (k, b_k), \quad k\overset{\mathrm{marginal}}{\sim} \unif([N])
\end{equation}
gives us random non-zero matrix elements of the coefficient matrix $A$ and random components of the bias vector $\vec{b}$.
Recall that the repetition number $R$ characterizes the correlation in the data, and is defined as
\begin{equation}
    R_{\mathrm{coeff/bias}} = \max_{z^{\mathrm{coeff/bias}}} \left(\E\left[N_{z^{\mathrm{coeff/bias}}}  \middle| z_1^{\mathrm{coeff/bias}}={z^{\mathrm{coeff/bias}}}\right] - \E[N_{z^{\mathrm{coeff/bias}}}]\right),
\end{equation}
where $N_z = \sum_{i=1}^{\tau}\delta_{z_i, z}$ is the number of repeating $z$'s in a refreshing block of data and $\tau$ is refreshing time that bounds the correlation time scale.
For simplicity, we use a single repetition number $R=\max\{R_{\mathrm{coeff}}, R_{\mathrm{bias}}\}$ and assume that,
conditioned on the fixed pair $(A,\vec b)$, the coefficient and bias data streams are independent hierarchical data generation processes.
Throughout this work, we also assumes that the situation information of each stream is completely specified in the data $z$ (e.g., given by the row and column indices).
The specific way of sampling these data can be arbitrary as long as it satisfies the above form.
Our discussion generalizes straightforwardly to the alternative scenario where each time we can choose to get either $z^{\mathrm{coeff}}$ or $z^{\mathrm{bias}}$.

Next, we specify the goal of solving a high-dimensional linear system.
The standard task of calculating the whole solution vector $\vec{x}=A^{-1}\vec{b}$ is undesirable when the dimension is very large, since writing down the solution vector already takes $O(N)$ memory.
Instead, in many cases, our goal is to estimate some property of the solution vector.
In particular, we consider the prototypical task of estimating the normalized quadratic form value
\begin{equation}
    \frac{\vec{x}^T\mathcal{M}\vec{x}}{\|\vec{x}\|\|\mathcal{M}\|\|\vec{x}\|} 
\end{equation}
to $\epsilon$ additive error for some symmetric matrix $\mathcal{M}\in \mathbb{R}^{N\times N}$ with known norm $\|\mathcal{M}\|$.
We assume that $\mathcal{M}$ can be efficiently measured on a quantum machine.
More precisely, by saying that $\mathcal{M}$ can be efficiently measured, we mean that there exists a coherent unitary circuit $W_{\mathcal{M}}$, together with its inverse and controlled versions, such that for every input state $\ket{\psi}$, measuring a designated output qubit of $W_{\mathcal{M}}\ket{0}\ket{\psi}$ returns $1$ with probability $\frac{1}{2}(1+\bra{\psi}\mathcal{M}\ket{\psi}/\|\mathcal{M}\|)$.
The circuit uses $S_{\mathcal{M}}=\poly(\log N)$ qubits and $\poly(\log N)$ gates.
This assumption can be relaxed if we allow the quantum machine to run longer.
Throughout this work, we assume $\vec{b}\neq 0$ and $\mathcal{M}\neq 0$.

This motivates our definition of the linear system task as follows.

\begin{tcolorbox}
\begin{task}[Linear system task]
\label{task:linear-sys}
    Let $N, R$ be integers and $\epsilon\in (0, 1]$.
    Consider any symmetric matrix $\mathcal{M}\in \mathbb{R}^{N\times N}$ that specifies a quadratic form of our interest and can be  efficiently measured.
    The linear system task is to estimate the normalized quadratic form value 
    \begin{equation}
    \frac{\vec{x}^T\mathcal{M}\vec{x}}{\|\vec{x}\|\|\mathcal{M}\|\|\vec{x}\|} 
    \end{equation}
    to $\epsilon$ additive error using data samples from any data generation process $\mathcal{D}_{\mathrm{LS}}(A, \vec{b})$ defined above with repetition number at most $R$ that corresponds to a linear system $A\vec{x}=\vec{b}, A\in \mathbb{R}^{N\times N}, \vec{b}\in \mathbb{R}^N$, where the coefficient matrix $A$ has $\|A\|=1$ norm, $\poly(\log N)$, sparsity and $\poly(\log N)$ condition number.
\end{task}
\end{tcolorbox}

Our quantum algorithm is flexible enough to handle correlated data with time-varying features.
This allows us to further consider a dynamic scenario, where the linear system changes over time, but the property we want to estimate remains approximately fixed.
This resembles the batch processing strategy common in modern large-scale data analysis, where we keep processing batches of data that share a common underlying property that we want to learn.
We use $\theta \in \mathbb{R}$ to denote that underlying property value.
In particular, we consider any hierarchical data generation process $\mathcal{D}_{\mathrm{DLS}}(\theta, \epsilon)$ with bounded repetition number $R$ and refreshing time $\tau$ (defined for each sampled linear system) of the form
\begin{equation}
    \mathcal{D}_{\mathrm{DLS}}(\theta, \epsilon) = (\mathcal{D}_{\theta, \epsilon}^0 \to \mathcal{D}_{\mathrm{LS}}(A, \vec{b})\to^{\times \tau} z).
\end{equation}
In other words, the linear system $A\vec{x}=\vec{b}$ changes every $\tau$ time steps and we keep getting random non-zero matrix elements of $A$ and random components of $\vec{b}$ of the current linear system as the data.
We require all linear systems sampled from $\mathcal{D}_{\theta, \epsilon}^0 $ to have $\|A\|=1$ norm, $\poly(\log N)$ sparsity, $\poly(\log N)$ condition number, and the desired property:
\begin{equation}
    \left|\frac{\vec{x}^T\mathcal{M}\vec{x}}{\|\vec{x}\|\|\mathcal{M}\|\|\vec{x}\|} - \theta\right|\leq \epsilon, \quad \vec{x}=A^{-1}\vec{b}.
\end{equation}
The specific way that $\mathcal{D}_\theta^0$ samples the linear system can be arbitrary as long as it satisfies the above requirements.
Formally, we define the dynamic linear system task as follows.

\begin{tcolorbox}
\begin{task}[Dynamic linear system task]
\label{task:linear-sys-dynamic}
    Let $N, R, \tau$ be integers and $\epsilon\in (0, 1]$.
    Consider any symmetric matrix $\mathcal{M}\in \mathbb{R}^{N\times N}$ that specifies a quadratic form of our interest and can be efficiently measured.
    The dynamic linear system task is to estimate the underlying normalized quadratic form value $\theta$ to $2\epsilon$ additive error using data samples from any $\mathcal{D}_{\mathrm{DLS}}(\theta, \epsilon)$ defined above with repetition number at most $R$ and refreshing time $\tau$.
\end{task}
\end{tcolorbox}

In the following, we first state our main results on quantum advantage in solving linear systems.
Then we prove the quantum easiness and classical hardness in subsequent sections.

\subsubsection{Main results}

Our first result shows that given the same amount of samples, a small quantum machine can solve the linear system task better than an exponentially larger classical machine.
Note that the scaling $N^{1-\zeta}$ is effectively $N$ since it holds for any constant $\zeta>0$.

\begin{tcolorbox}
\begin{theorem}[Quantum advantage in solving linear systems]
\label{thm:q-adv-linear-sys}
    Consider the linear system task with dimension $N$ and repetition number $R$ defined in \Cref{task:linear-sys}.
    Using $\tilde O(RN)$ samples, a quantum machine with $\poly(\log N)$ size can solve it with $1/\poly(\log N)$ error and high success probability, while any classical machine with $o(N^{1-\zeta})$ size for any constant $\zeta>0$ cannot solve it with a sufficiently small constant error and success probability more than $1/2 + 1/N^{\omega(1)}$.
    Moreover, the data processing time per sample of the quantum machine is $\poly(\log N)$.
\end{theorem}
\end{tcolorbox}

The second result shows that if the size of the classical machine is slightly smaller than $o(N)$, it would need super-polynomially more samples than a small quantum machine to solve the dynamic linear system task.
In the context of batch processing, this means that quantum machines can solve the task with a few batches of data, whereas sub-exponential size classical machines require super-polynomially many batches.

\begin{tcolorbox}
\begin{theorem}[Quantum advantage in solving dynamic linear systems]
\label{thm:q-adv-linear-sys-dynamic}
    Consider the dynamic linear system task with dimension $N$, repetition number $R$, and sufficient refreshing time $\tau=\tilde \Theta(RN)$ defined in \Cref{task:linear-sys-dynamic}.
    A quantum machine with $\poly(\log N)$ size can use $\tilde O(RN)$ samples to solve it with a sufficiently small constant error parameter and high success probability, while any classical machine with $o(N^{1-\zeta})$ size for any constant $\zeta>0$ that solves the same task with probability at least $2/3$ must collect at least $R N^{\omega(1)}$ samples.
    Moreover, the data processing time per sample of the quantum machine is $\poly(\log N)$.
\end{theorem}
\end{tcolorbox}

Together, \Cref{thm:q-adv-linear-sys,thm:q-adv-linear-sys-dynamic} establish unconditional and exponential quantum advantages in the foundational task of solving linear systems that appears in almost every aspect of science and engineering.

\subsubsection{Quantum algorithm}

Here, we prove the quantum algorithm parts of \Cref{thm:q-adv-linear-sys,thm:q-adv-linear-sys-dynamic}.
We construct an algorithm that solves a high-dimensional linear system on a small quantum computer.
We do so by combining quantum oracle sketching (\Cref{thm:q-oracle-sketch-corr}), quantum state sketching (\Cref{thm:q-state-sketch}), and standard quantum linear system solvers.
We use the Heisenberg-scaling quantum amplitude estimation to measure $\mathcal{M}$, which gives us the optimal $1/\epsilon^2$ sample complexity after the quadratic slowdown.
One may also use standard quantum limit measurements that will lead to a $1/\epsilon^4$ dependence on the error $\epsilon$.

\begin{tcolorbox}
\begin{theorem}[Solving linear systems with quantum oracle sketching]
\label{thm:linear-sys-upper}
    Let $\epsilon, \delta \in (0, 1)$.
    Let $N$ be a large integer and $A\vec{x}=\vec{b}$ be an $N$-dimensional linear system.
    Here, $A\in \mathbb{R}^{N\times N}$ is an $s$-sparse matrix with norm $\|A\|=1$ and condition number at most $\kappa$.
    $\vec{x}, \vec{b}\in \mathbb{R}^{N}$ are $N$-dimensional vectors.
    Consider any symmetric matrix $\mathcal{M}\in \mathbb{R}^{N\times N}$ that specifies a quadratic form of our interest and can be efficiently measured using $S_{\mathcal{M}}$ qubits.
    Then, there exists a quantum algorithm that produces an estimate $\hat{\theta}\in [-1, 1]$ satisfying
    \begin{equation}
        \left|\hat{\theta} - \frac{\vec{x}^T\mathcal{M}\vec{x}}{\|\vec{x}\|\|\mathcal{M}\|\|\vec{x}\|}\right| \leq \epsilon
    \end{equation}
    with probability at least $1-\delta$, using 
    \begin{equation}
        S=O\left(b+\log^2(N)\log^{2.5}\left(\frac{s\kappa}{\epsilon\delta}\right) + S_{\mathcal{M}}\right)
    \end{equation}
    qubits and 
    \begin{equation}
        M = O\lr{\frac{RN\log^2(N)s^5\kappa^2\log^{8}\left(\frac{\log(N)s\kappa}{\epsilon\delta}\right)}{\epsilon^2\delta}}
    \end{equation}
    samples from the data generation process $\mathcal{D}_{\mathrm{LS}}(A, \vec{b})$ with repetition number $R$.
    
    In particular, we have 
    \begin{equation}
	S = \poly(\log N), \quad M = \tilde{O}(RN),
    \end{equation}
    when $s, \kappa, b, \epsilon^{-1}, \delta^{-1}, S_{\mathcal{M}}\leq \poly(\log N)$.
    The data processing time per sample is $\poly(\log N)$.
\end{theorem}
\end{tcolorbox}

We note that \Cref{thm:linear-sys-upper} immediately implies the quantum algorithm part of \Cref{thm:q-adv-linear-sys} by definition of the linear system task.
It also proves the quantum algorithm part of \Cref{thm:q-adv-linear-sys-dynamic} by taking $\tau = \tilde{O}(RN)$ satisfying $\tau\geq M = \tilde{O}(RN)$ and noting that
\begin{equation*}
    \left|\hat{\theta} - \theta\right| \leq \left|\hat{\theta} - \frac{\vec{x}^T\mathcal{M}\vec{x}}{\|\vec{x}\|\|\mathcal{M}\|\|\vec{x}\|}\right| + \left|\frac{\vec{x}^T\mathcal{M}\vec{x}}{\|\vec{x}\|\|\mathcal{M}\|\|\vec{x}\|} - \theta\right| \leq \epsilon+\epsilon=2\epsilon
\end{equation*}
with probability at least $1-\delta$, as required.

We use the following quantum linear system solver as the backbone of our quantum algorithm and instantiate the oracle queries with quantum oracle sketching.
We refer interested readers to a recent survey \cite{morales2024quantum} for an overview on alternative quantum linear system solvers.

\begin{lemma}[Quantum linear system solver {\cite{costa2022optimal}}]
\label{lem:q-linear-sys-solver}
    Let $n$ be an integer and $N=2^n$.
    Let $A\vec{x}=\vec{b}$ be a linear system, where $A\in \mathbb{C}^{N\times N}, \vec{x}, \vec{b}\in \mathbb{C}^{N}$ with $\|A\|=1$ and condition number at most $\kappa$.
    There exists a quantum algorithm that executes a unitary circuit $V_{\mathrm{LS}}$ satisfying
    \begin{equation}
        \|V_{\mathrm{LS}}\ket{0}-\ket{x}\ket{0}\|_2\leq \epsilon, \quad \ket{x} = \frac{A^{-1}\ket{b}}{\|A^{-1}\ket{b}\|},
    \end{equation}
    using $O(n)$ qubits, $O(\kappa \log(1/\epsilon))$ gates, and $O(\kappa \log(1/\epsilon))$ queries to the block encoding of $A$ and the state preparation unitary of $\ket{b}$ and their inverse and controlled versions.
\end{lemma}

Recall that in \Cref{lem:block-encoding}, we have shown that $\tilde{O}(RNs^5)$ samples suffice to implement the block encoding of an $s$-sparse matrix $A$.
In addition, \Cref{thm:q-state-sketch} shows that we can use $\tilde{O}(RN)$ samples to prepare the quantum state $\ket{b} = \sum_{j=1}^N b_j \ket{j} / \|\vec{b}\|$ of the vector $\vec{b} = (b_1, \ldots, b_N)^T$.
Together, these primitives allow us to implement all the necessary subroutines that we need in quantum linear system solvers.
In particular, we combine \Cref{lem:q-linear-sys-solver} with \Cref{lem:block-encoding} and \Cref{thm:q-state-sketch} to prove \Cref{thm:linear-sys-upper}.

\begin{proof}[Proof of \Cref{thm:linear-sys-upper}]
    Let $n = \ceil{\log_2(N)}$ and embed $A, \vec{b}$ into $2^n \in [N, 2N]$ dimension to apply quantum oracle and state sketching.
    \Cref{lem:q-linear-sys-solver} states that there is a quantum linear system solver $V_{\mathrm{LS}}$ that produces a state $\rho$ satisfying
    \begin{equation}
        \|V_{\mathrm{LS}}\ket{0}-\ket{x}\ket{0}\|_2\leq \epsilon/4
    \end{equation}
    using $O(n)$ qubits and 
    \begin{equation}
        Q_0=O(\kappa \log(1/\epsilon))
    \end{equation}
    queries to the block encoding of $A$ and the state preparation unitary of $\ket{b}$ and their inverse and controlled versions.

    Our proof proceeds in two steps.
    We first use this quantum linear system solver to construct a query algorithm that estimates the quadratic form value to $\epsilon$ error with probability at least $1-\delta/2$.
    Then, we instantiate the queries with quantum oracle and state sketching.
    The instantiation error is chosen to be $\delta/2$ such that the total variation distance error on the final estimate is $\delta/2$.
    This immediately implies that the final estimate is accurate up to $\epsilon$ error with probability at least $1-\delta/2-\delta/2 = 1-\delta$.

    As the first step, we construct the query algorithm that estimates the quadratic form value.
    The quantum linear system solver produces a state that is $\epsilon/2$ close to $\ket{x}$ using $O(n)$ qubits and $Q_0=O(\kappa \log(1/\epsilon))$ queries to $A$ and $\ket{b}$.
    Note that we can also execute the inverse of the quantum linear system solver with the same query complexity.
    This allows us to use standard quantum amplitude estimation \cite{montanaro2015quantum,brassard2000quantum} to measure $\mathcal{M}/\|\mathcal{M}\|$ on $\rho$, the reduced state on the output qubits of $V_{\mathrm{LS}}\ket{0}\bra{0}V_{\mathrm{LS}}^\dagger$, and get a classical estimate $\hat{\theta}_0$ satisfying
    \begin{equation}
        \left|\hat{\theta}_0 - \tr(\rho \mathcal{M}/\|\mathcal{M}\|)\right| \leq \epsilon/2
    \end{equation}
    with probability at least $1-\delta/2$ using $O(\log(1/\delta)/\epsilon)$ queries to the quantum linear system solver and its inverse, along with additional $S_{\mathcal{M}}$ qubits and $O(T_{\mathcal{M}}\log(1/\delta)/\epsilon)$ time, where $T_{\mathcal{M}}=\poly(\log N)$ is the time complexity of measuring $\mathcal{M}$.
    The resulting estimation error of the quadratic form value is
    \begin{eqsplit}
        \left|\hat{\theta}_0 - \frac{x^T\mathcal{M}x}{\|x\|\|\mathcal{M}\| \|x\|}\right| &= \left|\hat{\theta}_0 - \tr(\ketbra{x}  \mathcal{M}/\|\mathcal{M}\|)\right| \\
        &\leq \left|\hat{\theta}_0 - \tr(\rho \mathcal{M}/\|\mathcal{M}\|)\right| + \left|\tr(\rho \mathcal{M}/\|\mathcal{M}\|) - \tr(\ketbra{x} \mathcal{M}/\|\mathcal{M}\|)\right| \\
        &\leq \epsilon/2 + \|\rho - \ketbra{x}\|_1 \frac{\|\mathcal{M}\|}{\|\mathcal{M}\|} \\
        &\leq \epsilon/2 + 2\cdot \epsilon/4 = \epsilon.
    \end{eqsplit}
    The total number of queries to $A$ and $\ket{b}$ amounts to
    \begin{equation}
        Q = O(Q_0\log(1/\delta)/\epsilon) = O(\kappa \log(1/\epsilon)\log(1/\delta)/\epsilon).
    \end{equation}
    We view this whole quantum algorithm as a query algorithm that produces a classical random variable $\hat{\theta}_0$ which is $\epsilon$ close to the quadratic form value with probability at least $1-\delta/2$.
    
    The second step is to instantiate the queries to $A$ and $\ket{b}$ in the query algorithm using quantum oracle and state sketching.
    We first replace all queries to the block encoding of $A$ and its inverse and controlled versions to their $\epsilon_1$-approximate versions in \Cref{lem:block-encoding}.
    This incurs an error of at most $E_1 = Q\epsilon_1 = \delta/6$, where we set $\epsilon_1 = \delta/(6Q) = \Theta(\epsilon\delta/(\kappa\log(1/\epsilon)\log(1/\delta)))$.
    
    Next, we replace the queries to $\epsilon_1$-approximate versions of block encoding of $A$ and its inverse and controlled versions by the random unitary channel that we build from samples in \Cref{lem:block-encoding}.
    This incurs an additional error of at most $E_2 = Q\epsilon_1 = \delta/6$, and uses $O(n+b+\log^{2.5}(s/\epsilon_1)) = O(n+b+\log^{2.5}(\frac{s\kappa}{\epsilon\delta}))$ qubits and 
    \begin{equation}
        M_A = Q\cdot O\lr{\frac{R2^nn^2s^5\log^4(ns/\epsilon_1)}{\epsilon_1}} \leq O\lr{\frac{R2^nn^2s^5\kappa^2\log^8(\frac{ns\kappa}{\epsilon\delta})}{\delta \epsilon^2}}
    \end{equation}
    samples from the data generation process.
    Note that here we only use the coefficient data and throw away the bias data in each sample.

    Finally, we run the preprocessing procedure in
    \Cref{thm:q-state-sketch} once for $\vec{b}$ and retain the resulting record $\nu_b$.
    Conditioned on $\nu_b$, \Cref{thm:q-state-sketch} provides a fixed state preparation unitary $P_{\nu_b}$ for $\ket{b}$.
    We implement these queries to the state preparation unitary, its inverse, and their controlled versions by the corresponding $\epsilon_1$-error random unitary channels constructed from fresh sample blocks according to \Cref{thm:q-state-sketch}.
    Applying the shared-record version of
    \Cref{lem:error-accumulation-time-varying} conditioned on $\nu_b$ and then averaging over $\nu_b$ gives an additional error
    \begin{equation}
        E_3
        =
        Q\epsilon_1
        =
        \delta/6.
    \end{equation}
    This uses $O(n\log(N/\epsilon_1)) \leq O(n^2\log(\frac{\kappa}{\epsilon\delta}))$ qubits and 
    \begin{equation}
        M_b = Q\cdot O\lr{\frac{R2^n n^2\log^4(1/\epsilon_1)}{\epsilon_1}} \leq O\lr{\frac{R2^nn^2\kappa^2\log^{8}(\frac{\kappa}{\epsilon\delta})}{\delta \epsilon^2}}
    \end{equation}
    samples from the data generation process, where the one-time pre-processing cost is absorbed into the stated sample bound.
    Here we only use the bias data and throw away the coefficient data in each sample.
    
    According to \Cref{lem:error-accumulation-time-varying}, the total error in instantiating the query algorithm is bounded by
    \begin{equation}
        E_1+E_2+E_3 = 3\cdot \delta/6 = \delta/2.
    \end{equation}
    This means that the output of the query-instantiated quantum algorithm, which is a classical random variable $\hat{\theta}$, has a distribution that is $\frac{1}{2}\cdot \delta/2=\delta/4$ close to $\hat{\theta}_0$ in total variation distance.
    In particular, this implies that the success probability
    \begin{equation}
        \Pr\left[\left|\hat{\theta} - \frac{x^T\mathcal{M}x}{\|x\|\|\mathcal{M}\| \|x\|}\right| \leq \epsilon\right] \geq \Pr\left[\left|\hat{\theta}_0 - \frac{x^T\mathcal{M}x}{\|x\|\|\mathcal{M}\| \|x\|}\right| \leq \epsilon\right] - \delta/4 \geq 1-\delta/2-\delta/4\geq 1-\delta,
    \end{equation}
    as desired.
    The total number of qubits used is
    \begin{equation}
        O(n)+S_{\mathcal{M}} + O\left(n+b+\log^{2.5}\left(\frac{s\kappa}{\epsilon\delta}\right)\right)+O\left(n^2\log(\frac{\kappa}{\epsilon\delta})\right) \leq O\left(b+n^2\log^{2.5}\left(\frac{s\kappa}{\epsilon\delta}\right) + S_{\mathcal{M}}\right).
    \end{equation}
    The total number of samples is
    \begin{equation}
        M=M_A+M_b \leq O\lr{\frac{R2^n n^2 s^5 \kappa^2\log^{8}(\frac{ns\kappa}{\epsilon\delta})}{\epsilon^2\delta}}.
    \end{equation}
    This completes the proof of \Cref{thm:linear-sys-upper}.
\end{proof}

\subsubsection{Classical hardness}

In this section, we prove the two classical hardness results in \Cref{thm:q-adv-linear-sys,thm:q-adv-linear-sys-dynamic}.
We prove them by constructing a specific linear system using \Cref{lem:embed-qc-into-linear-sys}.
Solving it amounts to solving the (dynamic) Noisy Oracle Property Estimation (NOPE) task defined in \Cref{sec:cl-hard} (\Cref{task:nope,task:dynamic-nope}), whose classical hardness we have already proved in \Cref{thm:classical-lower-bound-single-block,thm:classical-lower-bound-superpoly-sample}.

This gives us two classical hardness results.
The first result follows from \Cref{thm:classical-lower-bound-single-block} and shows that any classical learning algorithm that wants to perform better than random guessing in solving a linear system with the same number of samples quantum algorithms need must have $\Omega(N^{1-\zeta})$ size for any constant $\zeta>0$.
The second result follows from \Cref{thm:classical-lower-bound-superpoly-sample} and shows that when the linear system is dynamic, any classical learning algorithm will need a number of samples super-polynomial in $N$ if it has size $o(N^{1-\zeta})$ for any constant $\zeta>0$.
Note that it suffices to consider $0<\zeta<1$, since for $\zeta\geq1$ the claimed space lower bounds are vacuous.

The general idea of proving these results is to embed the quantum circuit that solves dynamic NOPE from \Cref{lem:encode-q-circ} into a linear system via \Cref{lem:embed-qc-into-linear-sys}.
The normalized quadratic form value $\frac{\vec{x}^T\mathcal{M} \vec{x}}{\|\vec{x}\|\|\mathcal{M}\|\|\vec{x}\|}$ encodes the oracle property $B\in\bit$ that we want to estimate in (dynamic) NOPE.
Then we show that given any classical algorithm that solves the linear system task of this specific linear system, we can use it to construct a classical algorithm that solves (dynamic) NOPE.
In the (dynamic) NOPE task, we take the oracle property function $f$ to be $K$-Forrelation (\Cref{def:forrelation}) and the noisy encoding function $g$ to be the inner product (\Cref{def:inner-prod}).

In the following, we construct the data generation process $\mathcal{D}^{N,K, R}_{\mathrm{LS}}(B), B\in\bit$ for linear systems, where $N$ is the dimension of the linear system, $K=\Theta(1)$ is the constant in the Forrelation that we will embed, $R$ is an integer that specifies the repetition number, and $B$ indicates whether the quadratic form of the solution vector has a large or small value (see \Cref{eq:quad-form-vs-decision-bit}).
The goal is to solve for $B$.
It helps to compare this construction to the dynamic NOPE data generation process defined in \Cref{sec:cl-hard-bootstrap}.

Given $N$, we define another large integer $N'$ as follows.
Let $T = \Theta(\log^3 N' \log\log N')$ be the total number of two-qubit gates and diagonal gates in the quantum circuit from \Cref{lem:encode-q-circ}.
Let $n' = \log N' + O(\log\log N')$ be the number of qubits in that circuit from \Cref{lem:encode-q-circ}.
To properly embed that circuit into the $N$-dimensional linear system, we define $N'$ such that $N = 24T 2^{n'} = N' \cdot \polylog N'$ as in \Cref{lem:embed-qc-into-linear-sys}.
This implies $N' = N / \polylog N$.
The resulting linear systems have sparsity $s=O(1)$ and condition number $\kappa = O(T) = O(\log^3N\log\log N)$ from \Cref{lem:embed-qc-into-linear-sys}.
Let $L=\ceil{\log^2 (KN')}\geq 5$ be the number of independent oracle instances that we have in dynamic NOPE.

We define the data generation process as
\begin{equation}
    \mathcal{D}^{N, K, R}_{\mathrm{LS}}(B) = (\mathcal{D}^0_B\to \mathcal{D}^1_{A}\to^{\times T_1} \mathcal{D}^2_{(\alpha, \beta)}\to^{\times T_2} z=(i, j, A_{ij}, l, b_l) \to^{\times T_3} z),
\end{equation}
where $T_3=\Theta(R)$ is chosen such that the repetition number of the resulting process is at most $R$, $T_2 = KN'$, $T_1 = \ceil{M_Q/(T_2T_3)} = \polylog(KN')$, $M_Q=RN\polylog(N)$ is the number of samples quantum machines need in \Cref{thm:linear-sys-upper}, $A\in \mathbb{R}^{N\times N}$ is an $N$-dimensional, symmetric, $O(1)$-sparse matrix with condition number $\kappa=O(T)=O(\log^3 N\log\log N)$, $\alpha\in\bit, \beta\in[L] $ label which part of the matrix $A$ that we are currently collecting matrix element data from, $i, j\in [N]$ labels the row/column of the matrix elements $A_{ij}$, and $l\in [N], b_l$ is a random component of a fixed bias vector $\vec{b}$.

The data are sampled in the following way that resembles dynamic NOPE in \Cref{sec:cl-hard-bootstrap}.
$\mathcal{D}^0_{B}$ samples a length-$L$ bitstring $\gamma$ with parity 
\begin{equation}
    \bigoplus_{j=1}^{L}\gamma_{j}=B
\end{equation} 
uniformly random.
For each $j\in [L]$, we sample a random oracle $o_j\sim p_{\gamma_j}$, where $p_0, p_1$ are the distributions of Forrelation defined in \Cref{lem:forrelation-query-separation}.
Then we sample a noisy encoding pair $(Y^{(0, j)}, Y^{(1, j)})\sim \unif((g^{KN'})^{-1}(o_j))$ using the inner product noisy encoding function $g$ defined in \Cref{def:inner-prod}.

Next, note that $(Y^{(0,j)}, Y^{(1,j)})_{j=1}^{L}$ specifies an $n'$-qubit quantum circuit $C$ with $T=O(\log^3N' \log\log N')$ gates via \Cref{lem:encode-q-circ}, and the quantum circuit $C$ gives a linear system $A\vec{x}=\vec{b}$ with dimension $24T 2^{n'}=N$ via \Cref{lem:embed-qc-into-linear-sys}.
Here, $A$ is indeed symmetric and $O(1)$-sparse with condition number $\kappa = O(T)=O(\log^3 N\log\log N)$.
Now we define $\mathcal{D}^1_{A}$.
We first sample a uniformly random coordinate $\beta\sim \unif([L])$ and a random bit $\alpha\sim\mathrm{Bern}(1/2)$ as in dynamic NOPE.
Then we pick out $Y^{(\alpha, \beta)}$ and use it to generate the data samples.

In particular, we define $\mathcal{D}^2_{(\alpha,\beta)}$ as follows.
For every fixed $(\alpha,\beta)$, the support pattern of the
corresponding rows of $A$ is completely determined by the public circuit layout and is independent of $Y^{(\alpha,\beta)}$.
Indeed, every oracle-dependent factor is nonzero for either value of the encoded bit.
Conditioned on the selected $(\alpha,\beta)$, we sample a row $i$ with probability proportional to its number of nonzero columns, and then sample one of its nonzero columns uniformly.
Consequently, for every nonzero position $(i,j)$ in the selected $(\alpha,\beta)$ slice,
\begin{equation}
    \Pr[(i,j)\mid\alpha,\beta]
    =
    \frac{
        |\{j':A_{ij'}\neq0\}|
    }{
        \sum_{i'\text{ selected}}
        |\{j':A_{i'j'}\neq0\}|
    }
    \cdot
    \frac{1}{
        |\{j':A_{ij'}\neq0\}|
    }=
    \frac{1}{
        \sum_{i'\text{ selected}}
        |\{j':A_{i'j'}\neq0\}|
    }.
\end{equation}
The different $(\alpha,\beta)$ slices have the same public support pattern.
Since $(\alpha,\beta)$ is sampled uniformly, the resulting marginal distribution of $(i,j)$ is uniform over all nonzero positions of $A$.
After the position $(i,j)$ is selected, we calculate $A_{ij}$ exactly from \Cref{eq:A,eq:B} via the known embedding, where the only data dependence is through $(-1)^{(Y^{(\alpha,\beta)}_x)_k}$.
This gives the required sample $(i,j,A_{ij})$.
We randomly sample a bias vector component $(l\sim \unif([N]), b_l)$ and obtain the full data sample $z=(i, j, A_{ij}, l, b_l)$.
We repeat this sample $T_3$ times, completing the data generation process.

This data generation process $\mathcal{D}_{\mathrm{LS}}^{N, K, R}$ is a valid data generation process of dynamic linear systems.
It produces matrix element data uniformly distributed over the non-zero elements of the matrix $A$.
We define the underlying value according to \Cref{lem:embed-qc-into-linear-sys} as
\begin{equation}
    \theta_B = c_{\mathrm{LS}}(1-B), \quad c_{\mathrm{LS}}=\frac{e^{-2}}{1+e^{-2}+e^{-4}},
\end{equation}
which satisfies $|\theta_B-c_{\mathrm{LS}}q_B|\leq 0.1c_{\mathrm{LS}}$, where $q_B$ is the probability of measuring $0$ on the embedded circuit given by \Cref{lem:encode-q-circ} when the underlying oracle property is $B$.
\Cref{lem:encode-q-circ} ensures that $q_1\leq 0.1, q_0\geq 0.9$.
The repetition number is $O(T_3)$ because, conditioned on the current higher-level situation, only the group containing the first sample and the remaining draws in the same $(\alpha, \beta)$ block can make an excess contribution. 
The former contributes at most $T_3$, while the latter contributes $T_3(T_2-1)O(1/T_2)=O(T_3)$, since every possible unrepeated sample has probability $O(1/T_2)$.
Thus $R_{\mathcal D}=O(T_3)$.
We choose $T_3=\Theta(R)$ so that $R_{\mathcal D}\leq R$ and suppress this constant rescaling below.
The refreshing time is $\tau = T_1T_2T_3 = O(M_Q)=\tilde O(RN)$, satisfying the requirements in \Cref{thm:q-adv-linear-sys,thm:q-adv-linear-sys-dynamic}.

This data generation process for linear systems $\mathcal{D}^{N, K, R}_{\mathrm{LS}}(B)$ is designed to reduce to the dynamic NOPE data $\mathcal{D}^{KN', T_1}_{g, f}(B)$ in \Cref{sec:cl-hard-bootstrap} via \Cref{lem:encode-q-circ,lem:embed-qc-into-linear-sys}.
Using this data generation process, we prove the following two results.

\begin{tcolorbox}
\begin{theorem}[Classical hardness of solving linear systems]
\label{thm:classical-hardness-linear-sys}
    Let $\zeta>0$ be any constant.
    Let $N$ be the dimension of a linear system task and $R$ be its repetition number.
    Using $\tilde{O}(RN)$ samples, any randomized classical learning algorithm with space complexity 
    \begin{equation}
        S\leq o(N^{1-\zeta})
    \end{equation}
    cannot solve the linear system task with $\epsilon=c_{\mathrm{LS}}/8$ error and success probability more than $1/2+1/N^{\omega(1)}$.
\end{theorem}
\end{tcolorbox}

\begin{tcolorbox}
\begin{theorem}[Classical hardness of solving dynamic linear systems]
\label{thm:classical-hardness-linear-sys-dynamic}
    Let $\zeta>0$ be any constant.
    Let $N$ be the dimension of a dynamic linear system task and $R, \tau=\tilde{\Theta}(RN)$ be its repetition number and refreshing time.
    Any randomized classical learning algorithm that solves the task with error parameter $\epsilon=c_{\mathrm{LS}}/8$ and success probability at least $2/3$ must have sample complexity
    \begin{equation}
        M\geq RN^{\omega(1)}
    \end{equation}
    if its space complexity
    \begin{equation}
        S \leq o(N^{1-\zeta}).
    \end{equation}
\end{theorem}
\end{tcolorbox}

\Cref{thm:classical-hardness-linear-sys} immediately implies the classical hardness part of \Cref{thm:q-adv-linear-sys} because the first $\tilde{O}(RN)$ samples from the constructed dynamic linear system data belongs to the same linear system and therefore is a valid sequence of non-dynamic linear system data.
\Cref{thm:classical-hardness-linear-sys-dynamic} directly implies the classical hardness part of \Cref{thm:q-adv-linear-sys-dynamic}.
Together with \Cref{thm:linear-sys-upper}, this completes the proof of the quantum advantage claims in \Cref{thm:q-adv-linear-sys,thm:q-adv-linear-sys-dynamic}.

\begin{proof}[Proof of \Cref{thm:classical-hardness-linear-sys,thm:classical-hardness-linear-sys-dynamic}]
    We prove \Cref{thm:classical-hardness-linear-sys,thm:classical-hardness-linear-sys-dynamic} by showing that given any classical learning algorithm that can estimate the normalized quadratic form value to $\epsilon=c_{\mathrm{LS}}/8$ error in the constructed linear system task, we can use it to construct an algorithm that decides the oracle property $B\in \bit$ in dynamic NOPE, which we have proved to be hard in \Cref{thm:classical-lower-bound-single-block,thm:classical-lower-bound-superpoly-sample}.

    First note that from \Cref{lem:embed-qc-into-linear-sys}, we have that the normalized quadratic form value of the linear systems in $\mathcal{D}_{\mathrm{LS}}^{N, K, R}$ is
    \begin{equation}
        \frac{\vec{x}^T\mathcal{M}\vec{x}}{\|\vec{x}\|\|\mathcal{M}\|\|\vec{x}\|} = c_{\mathrm{LS}}q_B,
    \end{equation}
    where $q_B$ is the probability of measuring $0$ on the embedded circuit when the underlying oracle property is $B$.
    \Cref{lem:encode-q-circ} ensures that $q_1\leq 0.1, q_0\geq 0.9$.
    That means, if we can estimate the normalized quadratic form value to $\epsilon=c_{\mathrm{LS}}/8$ error in the static case and $2\epsilon=c_{\mathrm{LS}}/4$ in the dynamic case, we can decide the value of $B\in \bit$.
    Also note that in the dynamic case, we have $|\theta_B-c_{\mathrm{LS}}q_B|\leq 0.1c_{\mathrm{LS}}\leq \epsilon$, satisfying the definition of the dynamic task.

    We choose $K=\ceil{1.001/\zeta}$ such that 
    \begin{equation}
        \frac{N'^{1-1/K}}{\polylog N'} \geq \frac{N^{1-\zeta+\zeta/1001}}{\polylog N}\geq \Omega(N^{1-\zeta}).
    \end{equation}
    For the sake of contradiction, we suppose we have a randomized classical learning algorithm $\mathcal{L}$ with space complexity
    \begin{equation}
        S \leq o(N^{1-\zeta}) \leq o\left(\frac{N'^{1-1/K}}{\polylog N'}\right)
    \end{equation}
    and sample complexity $M$ that, given data drawn from $\mathcal{D}^{N,K,R}_{\mathrm{LS}}(B)$, solves the corresponding static or dynamic linear-system task and hence decides $B$ with probability $p_{\mathrm{succ}}$.
    In the following, we design a classical learning algorithm $\mathcal{L}'$ that decides $B$ in dynamic NOPE using data from $\mathcal{D}^{KN', T_1}_{g, f}(B)$.

    The first step of $\mathcal{L}'$ is to generate data samples distributed exactly according to $\mathcal{D}^{N,K,R}_{\mathrm{LS}}(B)$ from the data stream $\mathcal{D}^{KN',T_1}_{g,f}(B)$.
    For each group of $T_3$ repeated target samples, $\mathcal{L}'$ reads one source sample $z=
    \left(
        x,
        Y^{(\alpha,\beta)}_x,
        \alpha,
        \beta
    \right)$
    from $\mathcal{D}^{KN',T_1}_{g,f}(B)$.
    It then uses the same public uniform-nonzero-position sampling rule as in the definition of
    $\mathcal{D}^{N,K,R}_{\mathrm{LS}}(B)$.
    More explicitly, it partitions the public support in the selected $(\alpha,\beta)$ slice into fixed-gate positions and oracle-dependent positions and chooses between the two classes with probability proportional to their cardinalities.
    If a fixed-gate position is selected, it samples uniformly from the fixed positions and calculates the corresponding entry from the public circuit description; the values $x$ and $Y^{(\alpha,\beta)}_x$ in the source sample are simply ignored.
    If an oracle-dependent position is selected, the uniformly random source coordinate $x$ is used as the corresponding oracle coordinate.
    The algorithm samples $k\sim\unif(\{0, \ldots, b\})$ and all remaining public block, clock, realification, and working-register coordinates uniformly.
    It then calculates the selected matrix entry exactly according to \Cref{eq:A,eq:B}.
    In particular, it only depends on the data via $(Y^{(\alpha,\beta)}_x)_k$, via the known embedding, permutations, and realification.
    Independently, $\mathcal{L}'$ samples a uniformly random component $(l,b_l)$ of the fixed bias vector.
    It repeats the resulting sample $z_{\mathrm{LS}}=(i,j,A_{ij},l,b_l)$ exactly $T_3$ times.
    Because the two support classes are selected proportionally to their cardinalities and each position is uniform inside its class, this procedure produces exactly the distribution $\mathcal{D}^{N,K,R}_{\mathrm{LS}}(B)$.

    After sampling a linear system data point $z_{\mathrm{LS}}=(i, j, A_{ij}, l, b_l)$, we feed it into the learning algorithm $\mathcal{L}$ for linear systems.
    We repeat this $M$ times so that $\mathcal{L}$ receives $M$ samples whose distribution matches that of $\mathcal{D}^{N, K, R}_{\mathrm{LS}}(B)$, produces an estimate of the normalized quadratic form value, and based on that estimate the value of the underlying bit $B$.
    We use the estimated bit of $\mathcal{L}$ as the final output of $\mathcal{L}'$.

    Note that since the data generation does not require knowledge of the previous data samples from $\mathcal{D}^{KN', T_1}_{g, f}(B)$, it can be performed online and thus the space complexity of $\mathcal{L}'$ is $S'=S$, the same as that of $\mathcal{L}$.
    Moreover, one source sample is used for every group of $T_3$ repeated target samples, so
    $M'\leq\lceil M/T_3\rceil\leq M/T_3+1.$
    The success probability of $\mathcal{L}'$ is $p_{\mathrm{succ}}$, the same as that of $\mathcal{L}$.

    Finally, we invoke \Cref{thm:classical-lower-bound-single-block,thm:classical-lower-bound-superpoly-sample}.
    Note that for the inner product $g$, we have $\eta=1/2$, $c = (865/\eta^2)\log(865/\eta^2) = (865\times 4)\log(865\times 4) \approx 40677.68$, and therefore the choice of $b=\ceil{40678\log(KN')}$ satisfies the requirement.
    For the Forrelation $f$ we use, \Cref{lem:forrelation-query-separation} implies that the $(1/3)$-error classical distributional query complexity is (using $K=\Theta(1)$)
    \begin{equation}
        Q_C \geq \Omega\left(\frac{N'^{1-1/K}}{\polylog N'}\right).
    \end{equation}
    Therefore, we have
    \begin{equation}
        S'=S \leq o\left(\frac{N'^{1-1/K}}{\polylog N'}\right)\implies S'\leq  o\lr{\frac{Q_C}{T_1^2 L}},
    \end{equation}
    satisfying the condition of \Cref{thm:classical-lower-bound-single-block,thm:classical-lower-bound-superpoly-sample}.
    Therefore, if the $M=T_1T_2T_3=\tilde{O}(RN)$ samples are drawn from the first refreshing block only, then $M'\leq M/T_3 = T_1KN'$ and \Cref{thm:classical-lower-bound-single-block} implies that 
    \begin{equation}
        p_{\mathrm{succ}}\leq \frac12+\frac{1}{(KN')^{\omega(1)}} \leq \frac12+\frac{1}{N^{\omega(1)}},
    \end{equation}
    proving \Cref{thm:classical-hardness-linear-sys}.
    Similarly, \Cref{thm:classical-lower-bound-superpoly-sample} implies that 
    \begin{equation}
        M\geq T_3(M'-1)\geq R(KN')^{\omega(1)}= RN^{\omega(1)},
    \end{equation}
    proving \Cref{thm:classical-hardness-linear-sys-dynamic}.
    This completes the proof of \Cref{thm:classical-hardness-linear-sys,thm:classical-hardness-linear-sys-dynamic} and, together with the quantum algorithm result \Cref{thm:linear-sys-upper}, proves \Cref{thm:q-adv-linear-sys,thm:q-adv-linear-sys-dynamic}.
\end{proof}

\newpage

\subsection{Binary classification}
\label{sec:binary-classification}

\begin{figure}
    \centering
    \includegraphics[width=1\linewidth]{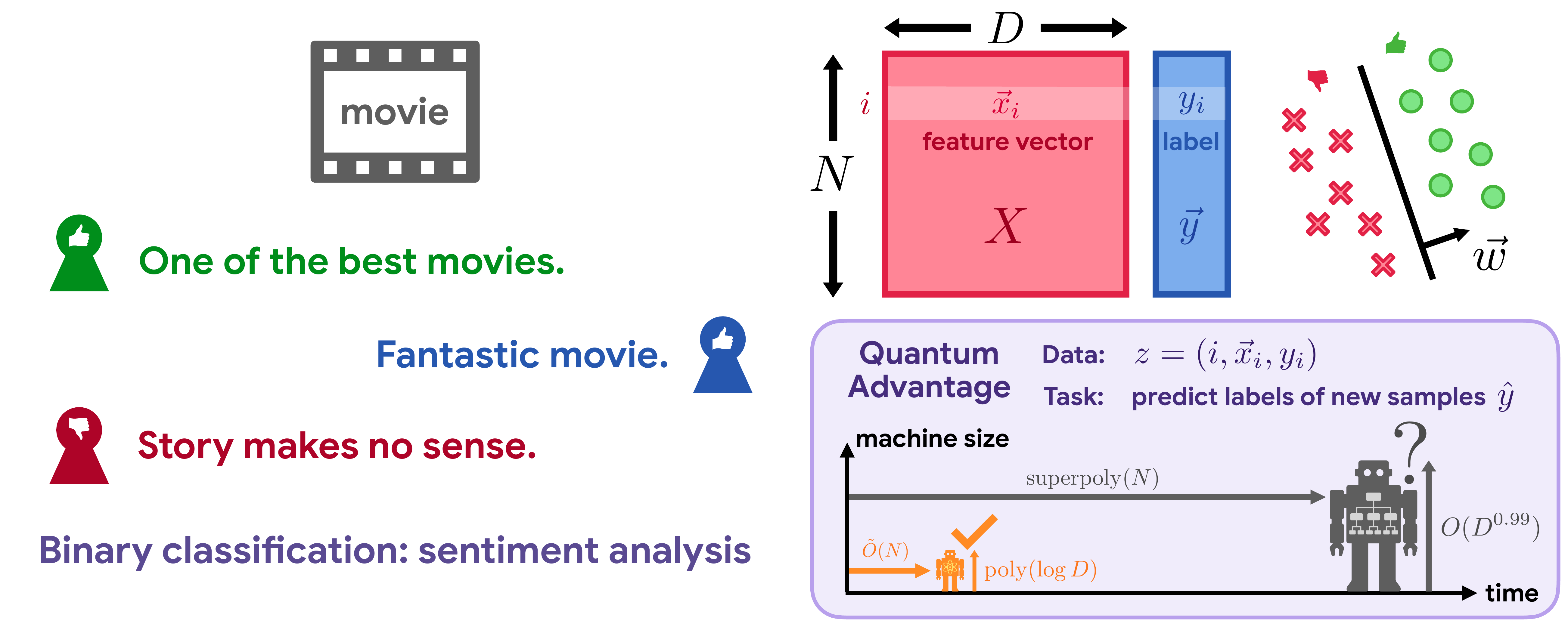}
    \caption[Overview of the binary classification task.]{\textbf{Overview of the binary classification task.}
    We illustrate the binary classification task with a particular real-world application scenario in sentiment analysis of movie reviews.
    We collect positive and negative reviews of movies from users, encode them into feature vectors $\vec{x}_i\in \mathbb{R}^D$, and use them to predict labels of new samples.
    A sample is the feature vector of a review $\vec{x}_i$ and its label $y_i$ (positive or negative).
    A canonical way of binary classification is the support vector machine (SVM), which is trained on the high-dimensional data matrix $X\in \mathbb{R}^{N\times D}$ and label vector $\vec{y}$ and produces a decision boundary represented by a weight vector $\vec{w}\in \mathbb{R}^D$ that classifies new samples. 
    Our results show that a small quantum machine with size $\poly(\log D)$ can solve this task with $\tilde{O}(N)$ samples, whereas any classical machine with exponentially larger size $O(D^{0.99})$ cannot solve the task unless it uses a sample size at least super-polynomial in $N$.
    }
    \label{fig:binary-classification}
\end{figure}

In this section, we consider the fundamental task of classification in data science and machine learning.
We focus on the prototypical example of binary classification.
In a binary classification task, we are given $N$ data samples of the form
\begin{equation}
    z_i = (i, \vec{x}_i, y_i), \quad \vec{x}_i \in \mathbb{R}^D, y_i\in \{\pm1\}, i\in [N],
\end{equation}
where $\vec{x}_i$ is a vector of $D$ features and $y_i\in \{\pm 1\}$ is its binary label.
This forms a training dataset of size $N$, represented by
\begin{equation}
    X = (\vec{x}_1, \ldots, \vec{x}_N)^T = \begin{pmatrix}
        \text{---}\vec{x}_1^T\text{---} \\
        \vdots \\
        \text{---}\vec{x}_N^T\text{---}
    \end{pmatrix} \in \mathbb{R}^{N\times D}, \quad \vec{y} = \begin{pmatrix}
        y_1\\
        \vdots\\
        y_N
    \end{pmatrix} \in \{\pm 1\}^N,
\end{equation}
where $X$ is the feature matrix and $\vec{y}$ is the label vector. 
Here, we adopt the convention used in data science, where each row represents a data sample and each column is a feature.
The goal of binary classification is to use this training dataset to predict the labels of future unseen test data samples $\vec{x}'$.

A standard method for performing binary classification is support vector machine (SVM).
The most basic version of SVM is the least-squares SVM (LS-SVM), where we aim to find a weight vector $\vec{w}\in \mathbb{R}^D$ that explains the dataset by minimizing the quadratic loss function
\begin{equation}
    \min_{\vec{w}\in \mathbb{R}^D} \|X\vec{w} - \vec{y}\|^2_2 + \lambda \|\vec{w}\|_2^2
\end{equation}
with $\ell_2$ regularization strength $\lambda\geq 0$.
This method is also known as ridge classifier.
Generalizations to other regularization methods such as $\ell_1$ (lasso) are straightforward.
We assume that the data matrix has been properly normalized such that $\|X\|= 1$ and therefore all matrix elements $|X_{ij}|\leq 1$.
The closed form solution of the weight vector reads
\begin{equation}
    \vec{w} = (X^TX+\lambda I_D)^{-1}X^T \vec{y},
\end{equation}
where $I_D$ is the $D\times D$ identity matrix.
We predict the label of a test vector $\vec{x}'$ according to the rule
\begin{equation}
    \hat{y} = \sgn(\vec{x}'\cdot  \vec{w}) = \sgn(\vec{x}'^T (X^TX+\lambda I_D)^{-1}X^T\vec{y}).
\end{equation}
We assume that the test vector is also properly normalized such that all its components have magnitude less than one: $\|\vec{x}'\|_\infty\leq 1$.
The classifiability of a test vector $\vec{x}'$ can be characterized by the margin $\gamma(\vec{x}')$, defined as
\begin{equation}
    \gamma(\vec{x}') = \frac{|\vec{x}'\cdot \vec{w}|}{\|\vec{w}\|_2},
\end{equation}
which is the magnitude of the projection of $\vec{x}'$ on to the vector $\vec{w}$, indicating how far the test point is from the decision boundary.
A classifiable test vector should have a margin that is not too small (e.g., $\gamma(\vec{x}')\geq 1/\poly(\log N, \log D)$).
We additionally assume that the regularized linear model provides a nontrivial fit to the training labels.
That is, we assume that the normalized optimal regularized residual
\begin{equation}
    r_{\mathrm{reg}}(X,\vec{y})
    =
    \frac{
        \min_{\vec{w}\in\mathbb{R}^D}
        \left(
            \|X\vec{w}-\vec{y}\|_2^2
            +
            \lambda\|\vec{w}\|_2^2
        \right)
    }{
        \|\vec{y}\|_2^2
    }
    \leq 0.99,
\end{equation}
where this constant $0.99$ can be any constant less than the maximal residual $1$.
When this condition is satisfied, we say that the LS-SVM is non-trivial.

We characterize the tractability of the binary classification task by three parameters: the sparsity $s$, the regularized condition number $\kappa_{\mathrm{reg}}$, and the classifiability of the test vector $\gamma(\vec{x}')$.
The sparsity $s$ is the maximal number of non-zero elements in $X$ per row or column, which bounds the number of non-zero features per sample and the number of samples each feature is presented in.
Such sparsity appears in many natural datasets (e.g., natural language data with common words trimmed off.)
The regularized condition number is the square root of the condition number of the matrix that we invert in LS-SVM, $X^TX+\lambda I_D$, which is equal to
\begin{equation}
    \kappa_{\mathrm{reg}} = \sqrt{\frac{\sigma_{\max}^2(X)+\lambda}{\sigma_{\min}^2(X)+\lambda}} \leq \min\left\{\kappa, \sqrt{1+\frac{1}{\lambda}}\right\},
\end{equation}
where $\sigma_{\max}(X), \sigma_{\min}(X)$ are the maximal and minimal singular value of $X$ satisfying $\sigma_{\min}^2(X)=\lambda_{\min}(X^TX)$ and $\kappa = \sigma_{\max}(X)/\sigma_{\min}(X)$ is the condition number of $X$.
When no regularization is imposed ($\lambda=0$), the regularized condition number $\kappa_{\mathrm{reg}}$ is the same as the condition number $\kappa$.
In natural dataset, often only the largest few singular values matter for classification, but they are buried in a long tail of small singular values that are not important.
Choosing a regularization strength $\lambda$ that effectively truncates the long tail provides a balance between good regularized condition number $\kappa_{\mathrm{reg}}$ and good classification performance.
We focus on the regime where the number of samples $N$ and feature dimension $D$ are very large, and yet the binary classification task remains sparse, well-conditioned, and classifiable:
\begin{equation}
    s, \kappa_{\mathrm{reg}}\leq \poly(\log N, \log D), \quad \gamma(\vec{x}'_j)\geq \frac{1}{\poly(\log N, \log D)},
\end{equation}
for every test vector $\vec{x}'_j, j=1, \ldots m$.
We call such test vectors classifiable vectors.
We take $N$ and $D$ to count only nonzero training samples and nonzero features.
Indeed, an all-zero row of $X$ does not affect the weight vector $(X^TX+\lambda I)^{-1}X^T\vec{y}$, and removing such rows cannot increase the normalized optimal residual.
Similarly, an all-zero column does not affect any prediction and may be discarded.
Note that we always have $D=\tilde{\Theta}(N)$ when the sparsity $s\leq \poly(\log N, \log D)$.

\subsubsection{Problem formulation}

Now we formally define our data processing task of binary classification, summarized in \Cref{task:bin-classify,task:bin-classify-dynamic}.
In particular, we specify the data generation process and the goal of the task.
We consider a data generation process where we randomly get a sparse feature vector $(i, \vec{x}_i)$ from the sparse feature matrix $X$ and its corresponding label $y_i$.
The components of $\vec{x}_i$ are specified by bitstrings of length $b=\poly(\log N, \log D)$ to sufficient accuracy.
For simplicity, we assume that this binary representation is exact and use $\vec{x}_i$ to stand for the corresponding values.

Specifically, we consider any hierarchical data generation process $\mathcal{D}_{\mathrm{BC}}(X, \vec{y})$ with bounded repetition number $R$ that generates data samples of the form
\begin{equation}
    z = (i, \vec{x}_i, y_i), \quad i\overset{\mathrm{marginal}}{\sim} \unif([N]),
\end{equation}
where $\vec{x}_i$ is a random entry of feature vector in the feature matrix $X$ and $y_i$ is the label of that training data point.
Recall that the repetition number $R$ characterize the correlation in the data, and is defined as
\begin{equation}
    R = \max_z \left(\E\left[N_z  \middle| z_1=z\right] - \E[N_z]\right),
\end{equation}
where $N_z = \sum_{i=1}^{\tau}\delta_{z_i, z}$ is the number of repeating $z$'s in a refreshing block of data and $\tau$ is refreshing time that bounds the correlation time scale.
Throughout this work, we also assumes that the situation information is completely specified in the data $z$ (e.g., given by the row index).
The specific way of sampling these data can be arbitrary as long as it satisfies the above form.
Without loss of generality, we represent every sparse row by
$s$ slots, padding unused slots with marked zero-valued dummy entries.
Our results generalize straightforwardly to the more demanding scenario, where we only get a random non-zero subset of the feature vector $\vec{x}_i$, or a random and possibly unmatched label $(j, y_j)$.

Next, we specify the goal of the binary classification task.
We aim to classify all possible sparse and classifiable test vectors $\vec{x}'_j, j=1, \ldots, m$ that are properly normalized $\|\vec{x}'_j\|_\infty\leq 1$.
In other words, we want to calculate the prediction
\begin{equation}
	\hat{y}_j = \sgn(\vec{x}'_j \cdot \vec{w}), \quad \vec{w} = (X^TX+\lambda I_D)^{-1} X^T \vec{y},
\end{equation}
for any $s$-sparse test vector $\vec{x}'_j$ that is classifiable.
That is, 
\begin{equation}
    \gamma(\vec{x}'_j) = \frac{|\vec{x}'_j\cdot \vec{w}|}{\|\vec{w}\|_2}\geq \frac{1}{\poly(\log N, \log D)}, \quad \forall j\in [m].
\end{equation}

This motivates our definition of the binary classification task as follows.

\begin{tcolorbox}
\begin{task}[Binary classification task]
\label{task:bin-classify}
    Let $N, D, R$ be integers and $\lambda\geq 0$ be any $\ell_2$ regularization strength.
    The binary classification task is to predict the label $\hat{y}_j$ of any $\poly(\log N, \log D)$-sparse and classifiable test vector $\vec{x}'_j$ in a test set $\{\vec{x}'_j\}_{j=1}^m$ of any size $m$, according to a non-trivial LS-SVM rule
    \begin{equation}
	   \hat{y}_j = \sgn(\vec{x}'_j \cdot \vec{w}), \quad \vec{w} = (X^TX+\lambda I_D)^{-1} X^T \vec{y},
    \end{equation}
    using data samples from any data generation process $\mathcal{D}_{\mathrm{BC}}(X, \vec{y})$ defined above with repetition number at most $R$ that corresponds to the normalized feature matrix $X\in \mathbb{R}^{N\times D}, \|X\|= 1$ with label vector $\vec{y}\in \mathbb{R}^N$, where the feature matrix $X$ has $\poly(\log N, \log D)$ sparsity and $\poly(\log N, \log D)$ regularized condition number.
\end{task}
\end{tcolorbox}

Our quantum algorithm is flexible enough to handle correlated data with time-varying features.
This allows us to further consider a dynamic scenario, where the training data $(X, \vec{y})$ changes over time, but the labels of the test vectors remain fixed.
This resembles the batch training strategy common in modern large-scale machine learning, where we train our model on batches of data that are different but the desired labels of the test vectors are fixed.
We use $\{\hat{y}_j\}_{j=1}^m$ to denote that fixed set of labels of the test vectors $\vec{x}_j'$.
We consider any hierarchical data generation process $\mathcal{D}_{\mathrm{DBC}}(\{(\vec{x}_j', \hat{y}_j)\}_{j=1}^m)$ with bounded repetition number $R$ and refreshing time $\tau$ (defined for each sampled dataset) of the form
\begin{equation}
    \mathcal{D}_{\mathrm{DBC}}(\{(\vec{x}_j', \hat{y}_j)\}_{j=1}^m) = (\mathcal{D}^0_{\{(\vec{x}_j', \hat{y}_j)\}_{j=1}^m} \to \mathcal{D}_{\mathrm{BC}}(X, \vec{y}) \to^{\times \tau} z).
\end{equation}
In other words, the training set $(X, \vec{y})$ changes every $\tau$ time steps, and we keep getting random feature vectors and corresponding labels of the current training dataset.
We require all training set $(X, \vec{y})$ sampled from $\mathcal{D}^0_{\{(\vec{x}_j', \hat{y}_j)\}_{j=1}^m}$ to have $\poly(\log N, \log D)$ sparsity, $\poly(\log N, \log D)$ regularized condition number, and fixed non-trivial classification rule:
\begin{equation}
    \hat{y}_j = \sgn(\vec{x}_j'^T (X^TX+\lambda I_D)^{-1}X^T\vec{y}).
\end{equation}
The specific way that $\mathcal{D}^0_{\{(\vec{x}_j', \hat{y}_j)\}_{j=1}^m}$ samples the training set can be arbitrary as long as it satisfies the above requirements.
\revise{
In other words, the labels of the test points stay the same, while the training examples the machine sees drift over time.
}
Formally, we define the dynamic binary classification task as follows.

\begin{tcolorbox}
\begin{task}[Dynamic binary classification task]
\label{task:bin-classify-dynamic}
    Let $N, D, R, \tau$ be integers and $\lambda\geq 0$ be any $\ell_2$ regularization strength.
    The dynamic binary classification task is to predict the label $\hat{y}_j$ of any $\poly(\log N, \log D)$-sparse and classifiable test vector $\vec{x}'_j$ in a test set $\{\vec{x}'_j\}_{j=1}^m$ of any size $m$ using data samples from any $\mathcal{D}_{\mathrm{DBC}}(\{(\vec{x}_j', \hat{y}_j)\}_{j=1}^m)$ defined above with repetition number at most $R$ and refreshing time $\tau$.
\end{task}
\end{tcolorbox}

In the following, we first state our main results on quantum advantage in binary classification.
Then we prove the quantum easiness and classical hardness in subsequent sections.

\subsubsection{Main results}

Our first result shows that given the same amount of samples, a small quantum machine can solve the binary classification task better than an exponentially larger classical machine.
This means that using a quantum machine, we can build a better model with exponentially smaller size.
Note that the scaling $D^{1-\zeta}$ is effectively $D$ since it holds for any constant $\zeta>0$.

\begin{tcolorbox}
\begin{theorem}[Quantum advantage in binary classification]
\label{thm:q-adv-bin-classify}
    Consider the binary classification task with sample dimension $N$, feature dimension $D$ and repetition number $R$ defined in \Cref{task:bin-classify}.
    Using $\tilde O(RN)$ samples, a quantum machine with $\poly(\log D)$ size can solve it with high success probability, while any classical machine with $o(D^{1-\zeta})$ size for any constant $\zeta>0$ cannot solve it with success probability more than $1/2 + 1/N^{\omega(1)}$.
    Moreover, the data processing time per sample of the quantum machine is $\poly(\log D)$.
\end{theorem}
\end{tcolorbox}

The second result shows that if the size of the classical machine is slightly smaller than $o(D)$, it would need super-polynomially more samples than a small quantum machine to solve the dynamic binary classification task.
In the context of batch training, this means that quantum machines can solve the task with a few batches of training data, whereas sub-exponential size classical machines require super-polynomially many batches.

\begin{tcolorbox}
\begin{theorem}[Quantum advantage in dynamic binary classification]
\label{thm:q-adv-bin-classify-dynamic}
    Consider the dynamic binary classification task with sample dimension $N$, feature dimension $D$, repetition number $R$, and sufficient refreshing time $\tau=\tilde \Theta(RN)$ defined in \Cref{task:bin-classify-dynamic}.
    A quantum machine with $\poly(\log D)$ size can use $\tilde O(RN)$ samples to solve it with high success probability, while any classical machine with $o(D^{1-\zeta})$ size for any constant $\zeta>0$ that solves it with probability at least $2/3$ must collect at least $R N^{\omega(1)}$ samples.
    Moreover, the data processing time per sample of the quantum machine is $\poly(\log D)$.
\end{theorem}
\end{tcolorbox}

Together, \Cref{thm:q-adv-bin-classify,thm:q-adv-bin-classify-dynamic} establish unconditional and exponential quantum advantages in the foundational task of binary classification.

\subsubsection{Quantum algorithm}

Here, we prove the quantum algorithm parts of \Cref{thm:q-adv-bin-classify,thm:q-adv-bin-classify-dynamic}.
We construct an algorithm that runs high-dimensional SVM on a small quantum computer.
We do so by combining quantum oracle sketching (\Cref{thm:q-oracle-sketch-corr}), quantum state sketching (\Cref{thm:q-state-sketch}), and standard quantum ridge regression solver that prepares the weight vector $\ket{w}$ of the LS-SVM.
We use a variant of Clifford classical shadow \cite{huang2020predicting}, which we call interferometric classical shadow (\Cref{lem:interf-classical-shadow}), to readout the weight vector $\ket{w}$ into a classical representation, which is used for predicting the labels of all possible test vectors.

\begin{tcolorbox}
\begin{theorem}[Binary classification with quantum oracle sketching]
\label{thm:bin-classify-upper}
    Let $N, D$ be large integers, $\delta\in (0, 1)$, and $\lambda\geq 0$ be the $\ell_2$ regularization strength.
    Let $X = (\vec{x}_1, \ldots, \vec{x}_N)^T \in \mathbb{R}^{N\times D}, \vec{y} = (y_1, \ldots, y_N)^T \in \mathbb{R}^N$ be a training dataset where $X$ is $s$-sparse with norm $\|X\|= 1$ and regularized condition number $\kappa_{\mathrm{reg}} = \sqrt{(\sigma^2_{\max}(X)+\lambda)/(\sigma^2_{\min}(X)+\lambda)}$.
    Assume that the normalized optimal regularized residual satisfies $r_{\mathrm{reg}}(X,\vec{y})\leq 0.99$.
    Consider a set of $s$-sparse test vectors $\{\vec{x}_j'\}_{j=1}^m, \|\vec{x}_j'\|_\infty\leq 1$, of any size $m$ that all have margin $\gamma(\vec{x}'_j)\geq \gamma_{\mathrm{test}}, \forall j\in [m]$.
    There exists a quantum algorithm that can output the prediction
    \begin{equation}
        \hat{y}_j = \sgn(\vec{x}'_j \cdot \vec{w}), \quad \vec{w} = (X^TX+\lambda I_D)^{-1} X^T \vec{y}, 
    \end{equation}
    for any $j\in [m]$ with probability at least $1-\delta$ using 
    \begin{equation}
        S=\tilde{O}\lr{b + \frac{s\log^2(D)(\min\{s\log(D), \log(m)\}+\log^{2.5}(1/\delta))}{\gamma_{\mathrm{test}}^2}+\log(sD)\log(\kappa_{\mathrm{reg}}) +\log^{2.5}(\kappa_{\mathrm{reg}}) }
    \end{equation}
    qubits and
    \begin{equation}
        M=\tilde{O}\lr{\frac{RN s^{7}\kappa_{\mathrm{reg}}^4}{\gamma_{\mathrm{test}}^4\delta}\min\{s^2\log^2(D), \log^2(m)\}}
    \end{equation}
    samples from the data generation process $\mathcal{D}_{\mathrm{BC}}(X, \vec{y})$ with repetition number $R$.

    In particular, we have
    \begin{equation}
        S = \poly(\log D), \quad M = \tilde{O}(RN),
    \end{equation}
    when $s, \kappa_{\mathrm{reg}}, b, \gamma_{\mathrm{test}}^{-1}, \delta^{-1}\leq \poly(\log N, \log D)$.
    The data processing time per sample is $\poly(\log D)$.
\end{theorem}
\end{tcolorbox}

We note that \Cref{thm:bin-classify-upper} immediately implies the quantum algorithm part of \Cref{thm:q-adv-bin-classify} by definition of the binary classification task.
It also proves the quantum algorithm part of \Cref{thm:q-adv-bin-classify-dynamic} by taking $\tau = \tilde{O}(RN)$ satisfying $\tau\geq M = \tilde{O}(RN)$.

We use the following simple quantum ridge regression solver as the backbone of our quantum algorithm and instantiate its oracle queries with quantum oracle sketching.
The solver applies QSVT to construct the pseudo-inverse of an augmented matrix and then uses standard fixed-point amplitude amplification.
This is a straightforward solver that suffices for our purposes.

\begin{lemma}[Quantum ridge regression solver]
\label{lem:q-ridge-reg-solver}
    Let $\epsilon\in(0,1/2]$.
    Let $N, D$ be large integers.
    Let $\lambda\geq 0$ be the normalized $\ell_2$ regularization strength.
    Let $X\in \mathbb{R}^{N\times D}, \vec{y}\in \mathbb{R}^{N}$ be the training dataset with norm $\|X\|= 1$ and regularized condition number $\kappa_{\mathrm{reg}}=\sqrt{(\sigma^2_{\max}(X)+\lambda)/(\sigma^2_{\min}(X)+\lambda)}$.
    Assume that the normalized optimal regularized residual satisfies $r_{\mathrm{reg}}(X,\vec{y})\leq 0.99$.
    There exists a quantum algorithm that applies a unitary $V$ such that
    \begin{equation}
        \|V|0^S\rangle
        -
        |0^{S-\log(D)}\rangle \ket{w}\|_2
        \leq \epsilon,
        \qquad
        \ket{w}
        =
        \frac{(X^TX+\lambda I_D)^{-1}X^T\vec{y}}
        {\|(X^TX+\lambda I_D)^{-1}X^T\vec{y}\|},
    \end{equation}
    using $S=O\left(\log(\max\{N,D\})\right)$
    qubits and $O\left(\kappa_{\mathrm{reg}}^2\log^2\left(\frac{\kappa_{\mathrm{reg}}}{\epsilon}\right)\right)$
    gates and queries to the block encoding of $X$ and the state preparation unitary of
    $\ket{y}=\frac{1}{\|\vec{y}\|}\sum_i y_i\ket{i},$
    and their inverse and controlled versions.
\end{lemma}

\begin{proof}[Proof of \Cref{lem:q-ridge-reg-solver}]
    Define the normalized augmented matrix and input state
    \begin{equation}
        A_\lambda
        =
        \frac{1}{\sqrt{1+\lambda}}
        \begin{pmatrix}
            X\\
            \sqrt{\lambda}I_D
        \end{pmatrix},
        \qquad
        \ket{b}
        =
        \frac{1}{\|\vec{y}\|}
        \begin{pmatrix}
            \vec{y}\\
            0_D
        \end{pmatrix}.
    \end{equation}
    The singular values of $A_\lambda$ are
    \begin{equation}
        \sigma_i(A_\lambda)
        =
        \sqrt{
        \frac{\sigma_i^2(X)+\lambda}
        {1+\lambda}
        }
        \in
        \left[
        \frac{1}{\kappa_{\mathrm{reg}}},
        1
        \right].
    \end{equation}
    Moreover,
    \begin{equation}
    \begin{split}
        A_\lambda^+\ket{b}
        &=
        \frac{\sqrt{1+\lambda}}{\|\vec{y}\|}
        (X^TX+\lambda I_D)^{-1}X^T\vec{y},
    \end{split}
    \end{equation}
    whose normalized state is exactly $\ket{w}$.

    Let $A_\lambda=\sum_i \sigma_i\ket{u_i}\bra{v_i}$ be its singular value decomposition, and decompose
    \begin{equation}
        \ket{b}
        =
        \sum_i\beta_i\ket{u_i}
        +
        \ket{b_\perp},
        \qquad
        \ket{b_\perp}\in\ker(A_\lambda^\dagger).
    \end{equation}
    Since $A_\lambda A_\lambda^+$ is the projector onto
    $\operatorname{col}(A_\lambda)$, the regularized residual can be written as
    \begin{equation}
        r_{\mathrm{reg}}(X,\vec{y})
        =
        \left\|
        (I-A_\lambda A_\lambda^+)\ket{b}
        \right\|_2^2
        =
        \|\ket{b_\perp}\|_2^2.
    \end{equation}
    Therefore, the non-trivialness of the residual implies that
    \begin{equation}
        \sum_i|\beta_i|^2
        =
        1-r_{\mathrm{reg}}(X,\vec{y})
        \geq 0.01.
    \end{equation}

    By the standard QSVT pseudo-inverse construction
    \cite{gilyen2019quantum,martyn2021grand}, there is an odd
    polynomial $P$ of degree
    \begin{equation}
        O\left(
        \kappa_{\mathrm{reg}}
        \log\left(\frac{\kappa_{\mathrm{reg}}}{\eta}\right)
        \right)
    \end{equation}
    satisfying $|P(x)|\leq 1$ on $[-1,1]$ and
    \begin{equation}
        \left|
        P(x)-\frac{1}{2\kappa_{\mathrm{reg}}x}
        \right|
        \leq \eta,
        \qquad
        x\in
        \left[
        \frac{1}{\kappa_{\mathrm{reg}}},
        1
        \right].
    \end{equation}
    Define $P^{(\mathrm{SV})}(A_\lambda^\dagger)=\sum_i P(\sigma_i)\ket{v_i}\bra{u_i}$.
    Applying singular value transformation to a block encoding of $A_\lambda^\dagger$ therefore produces, in the success branch,
    the unnormalized state
    \begin{equation}
        \ket{\widetilde z}
        =
        P^{(\mathrm{SV})}(A_\lambda^\dagger)\ket{b},
    \end{equation}
    which approximates the ideal unnormalized state
    \begin{equation}
        \ket{z}
        =
        \frac{1}{2\kappa_{\mathrm{reg}}}
        A_\lambda^+\ket{b}.
    \end{equation}
    Note that here we apply QSVT to $A_\lambda^\dagger$ rather than $A_\lambda$, because the input is in the left singular-vector space and the desired output is in the right singular-vector space.
    Also note that a block-encoding unitary for $A_\lambda^\dagger$, together with its adjoint and controlled versions, can be implemented using $O(1)$ queries to the corresponding versions of the block encoding of $X$, since the identity block is known and the two row blocks can be combined using a single selector qubit.
    
    Since $P$ is odd, $P(0)=0$, and hence the component
    $\ket{b_\perp}\in\ker(A_\lambda^\dagger)$ makes no contribution
    to the success branch. Therefore,
    \begin{equation}
    \begin{split}
        \left\|\ket{\widetilde z}-\ket{z}\right\|_2^2
        &=
        \sum_i |\beta_i|^2
        \left|
        P(\sigma_i)
        -
        \frac{1}{2\kappa_{\mathrm{reg}}\sigma_i}
        \right|^2\\
        &\leq
        \eta^2\sum_i|\beta_i|^2
        \leq \eta^2.
    \end{split}
    \end{equation}
    
    Let $a=\|\ket{z}\|_2, \widetilde a=\|\ket{\widetilde z}\|_2$.
    The ideal success amplitude satisfies
    \begin{equation}
        a=
        \frac{\|A_\lambda^+\ket{b}\|_2}
        {2\kappa_{\mathrm{reg}}}=
        \frac{1}{2\kappa_{\mathrm{reg}}}
        \sqrt{
        \sum_i\frac{|\beta_i|^2}{\sigma_i^2}
        }\geq
        \frac{1}{2\kappa_{\mathrm{reg}}}
        \sqrt{\sum_i|\beta_i|^2}=
        \frac{\sqrt{1-r_{\mathrm{reg}}(X,\vec{y})}}
        {2\kappa_{\mathrm{reg}}}\geq
        \frac{1}{20\kappa_{\mathrm{reg}}},
    \end{equation}
    where we have used $\sigma_i\leq 1$.
    
    We choose the approximation error $\eta=\frac{\epsilon}{80\kappa_{\mathrm{reg}}}$.
    The actual success amplitude is bounded below by
    \begin{equation}
        \widetilde a
        \geq a-\eta
        \geq
        \frac{1}{20\kappa_{\mathrm{reg}}}
        -
        \frac{1}{80\kappa_{\mathrm{reg}}}
        \geq
        \frac{1}{40\kappa_{\mathrm{reg}}}.
    \end{equation}
    Moreover, defining
    $\ket{\widetilde w}
    =
    {\ket{\widetilde z}}/{\widetilde a}$,
    we have
    \begin{equation}
        \|\ket{\widetilde w}-\ket{w}\|_2
        =
        \left\|
        \frac{\ket{\widetilde z}}{\widetilde a}
        -
        \frac{\ket{z}}{a}
        \right\|_2\leq
        \frac{
        \|\ket{\widetilde z}-\ket{z}\|_2
        +
        |a-\widetilde a|
        }{a}\leq
        \frac{2\eta}{a}\leq
        40\kappa_{\mathrm{reg}}\eta\leq
        \frac{\epsilon}{2}.
    \end{equation}
    
    Standard fixed-point amplitude amplification with target error
    $\epsilon/2$ therefore prepares $\ket{\widetilde w}$ using
    \begin{equation}
        O\left(
        \kappa_{\mathrm{reg}}\log(1/\epsilon)
        \right)
    \end{equation}
    applications of the QSVT circuit and its inverse. Combining the
    amplitude-amplification error with
    $\|\ket{\widetilde w}-\ket{w}\|_2\leq\epsilon/2$ shows that the
    final state is $\epsilon$-close to $\ket{w}$.
    
    With the above choice of $\eta$, the total query and gate complexities are consequently
    \begin{equation}
        O\left(
        \kappa_{\mathrm{reg}}^2
        \log^2\left(
        \frac{2\kappa_{\mathrm{reg}}}{\epsilon}
        \right)
        \right),
    \end{equation}
    as desired.
\end{proof}

We use Clifford classical shadow tomography to readout the weight vector and make predictions.

\begin{lemma}[Clifford classical shadow, \cite{huang2020predicting}]
\label{lem:cliffod-classical-shadow}
    Let $\rho$ be any $D$-dimensional quantum state and $O_1, \ldots, O_m \in \mathbb{C}^{D\times D}$ be observables.
    Let $\epsilon, \delta \in (0, 1)$.
    Then, there is a quantum algorithm that can predict $\hat{o}_i$ such that
    \begin{equation}
        |\hat{o}_i-\tr(\rho O_i)|\leq \epsilon, \quad \forall i\in [m],
    \end{equation}
    with probability at least $1-\delta$, using
    \begin{equation}
        O\lr{\max_{i\in [m]} \tr(O_i^2) \cdot \frac{\log(m/\delta)}{\epsilon^2}}
    \end{equation}
    copies of $\rho$.
    Moreover, the time complexity of predicting any $\hat{o}_i$ is $\polylog(D)$ if the expectation value of $O_i$ on any stabilizer state can be calculated in $\polylog(D)$ time.
\end{lemma}

To make predictions on the labels, we need to estimate the sign of $\expval{x|w}$.
Measuring the observable $\ketbra{x}$ on $\ket{w}$ only gives us the magnitude $|\expval{x|w}|^2$.
The standard way of keeping the sign information is via Hadamard test, in which we introduce an ancilla, prepare the state $\frac{1}{\sqrt{2}}(\ket{0}\ket{x} + \ket{1}\ket{w})$, and measure the ancilla.
However, this contradicts our goal of classical shadow, where we want to collect a universal set of data from $\ket{w}$ that can be used to predict many different $\ket{x}$ afterwards.
To address this issue, we develop the following technique that we call interferometric classical shadow.

\begin{tcolorbox}
\begin{lemma}[Interferometric classical shadow]
\label{lem:interf-classical-shadow}
    Let $\ket{w}, \ket{x_1}, \ldots, \ket{x_m}\in \mathbb{C}^D$ be $D$-dimensional pure states.
    Assume that we have complete classical descriptions of $\ket{x_1}, \ldots, \ket{x_m}$.
    Let $\epsilon, \delta\in (0, 1)$.
    Then, there is a quantum algorithm that can predict $\hat{o}_i$ such that
    \begin{equation}
        |\hat{o}_i - \Re[\expval{x_i|w}]|\leq \epsilon, \quad \forall i\in [m],
    \end{equation}
    with probability at least $1-\delta$, using
    \begin{equation}
        O\lr{\frac{\log(m/\delta)}{\epsilon^2}}
    \end{equation}
    queries of the controlled state preparation unitary of $\ket{w}$.
    Moreover, the time complexity of predicting any $\hat{o}_i$ is $\polylog(D)$ if the overlap between $\ket{x_i}$ and any stabilizer state can be calculated in $\polylog(D)$ time.
\end{lemma}
\end{tcolorbox}

\begin{proof}[Proof of \Cref{lem:interf-classical-shadow}]
    We apply Clifford classical shadow (\Cref{lem:cliffod-classical-shadow}) on the state
    \begin{equation}
        \ket{\tilde{w}} = \frac{1}{\sqrt{2}}(\ket{0}\ket{0}+\ket{1}\ket{w}),
    \end{equation}
    which can be prepared with a single query of the controlled state preparation of $\ket{w}$ on $\frac{1}{\sqrt{2}}(\ket{0}+\ket{1})\ket{0}$.
    Consider the observables $O_i = \ketbra{x_i+} - \ketbra{x_i-}$, where
    \begin{equation}
    \begin{split}
        \ket{x_i\pm} = \frac{1}{\sqrt{2}}(\ket{0}\ket{0} \pm \ket{1}\ket{x_i} )
    \end{split}
    \end{equation}
    satisfy $\expval{x_i+|x_i-}=0$.
    Note that $\tr(O_i^2) = 2$ and we have
    \begin{equation}
    \begin{split}
        \tr(O_i\ketbra{\tilde w}) &= |\expval{x_i+|\tilde w}|^2 - |\expval{x_i-|\tilde w}|^2 \\
        &= \left|\frac{1+\expval{x_i|w}}{2}\right|^2 - \left|\frac{1-\expval{x_i|w}}{2}\right|^2 \\
        &=\Re[\expval{x_i|w}].
    \end{split}
    \end{equation}
    Moreover, evaluating the expectation value of $O_i$ on any stabilizer state reduces to evaluating the overlap between $\ket{x_i}$ and any stabilizer state.
    \Cref{lem:interf-classical-shadow} then follows directly from \Cref{lem:cliffod-classical-shadow}.
\end{proof}

Recall that in \Cref{lem:block-encoding}, we have shown that $\tilde{O}(RNs^5)$ samples suffice to implement the block encoding of an $s$-sparse matrix.
In addition, \Cref{thm:q-state-sketch} shows that we can use $\tilde{O}(RN)$ samples to prepare the quantum state of any vector.
We combine \Cref{lem:q-ridge-reg-solver} with \Cref{lem:block-encoding} and \Cref{thm:q-state-sketch} to prove \Cref{thm:bin-classify-upper}.

\begin{proof}[Proof of \Cref{thm:bin-classify-upper}]
    Let $n = \ceil{\log_2(\max(N, D))}\leq \min\{O(\log(sD)), O(\log(sN))\}$ and embed $X, \vec{y}$ into $2^n$ dimension to apply quantum oracle and state sketching.
    \Cref{lem:q-ridge-reg-solver} states that there is a quantum ridge regression solver, which is a unitary $V$ that prepares the state $\ket{w}$ with $\epsilon/2$ error using $O(n)$ qubits and 
    \begin{equation}
        Q_0=O\left(
        \kappa_{\mathrm{reg}}^2
        \log^2\left(
        \frac{\kappa_{\mathrm{reg}}}{\epsilon}
        \right)
        \right)
    \end{equation}
    queries to the block encoding of $X$ and the state preparation unitary of $\ket{y}$ and their inverse and controlled versions.
    By replacing all gates and queries in $V$ with their controlled versions, we obtain a controlled state preparation unitary $cV$ that prepares $\ket{w}$ with error $\epsilon/2$.

    Similar to the proof of \Cref{thm:linear-sys-upper}, our proof of \Cref{thm:bin-classify-upper} proceeds in two steps.
    We first use this quantum ridge regression solver to construct a query algorithm that can predict $\hat{y}_j$ correctly for any $j\in [m]$ with probability at least $1-\delta/2$.
    Then, we instantiate the queries with quantum oracle and state sketching.
    The instantiation error is chosen to be $\delta/2$ such that the total variation distance error on the final prediction is $\delta/2$.
    This immediately implies that the final prediction is correct with probability at least $1-\delta/2-\delta/2 = 1-\delta$.

    As the first step, we construct the query algorithm that makes correct predictions.
    The quantum ridge regression solver gives us a controlled state preparation unitary $cV$ that prepares $\ket{w}$ to $\epsilon/2$ using $O(n)$ qubits and $Q_0=O(\kappa_{\mathrm{reg}}^2\log^2(\kappa_{\mathrm{reg}}/\epsilon))$ queries to $X$ and $\ket{y}$.
    To readout the weight vector well enough that we can predict any $s$-sparse test vector $\vec{x}'_j$, we imagine that we have an $\epsilon_0$-covering net $\mathcal{N}$ over the set
    \begin{equation}
        \mathcal{X}_{\mathrm{test}} = \{\vec{u}\in \mathbb{R}^{D}: \|\vec{u}\|_0\leq s, \|\vec{u}\|_2 = 1\}
    \end{equation}
    in $\|\cdot \|_2$.
    Note that this covering net is only a analysis tool and it is not used in the algorithm.
    The covering net has size
    \begin{equation}
        |\mathcal{N}|\leq \binom{D}{s}\left( \frac{3}{\epsilon_0} \right)^{O(s)} \leq 2^{O(s\log(D/\epsilon_0))}
    \end{equation}
    We run interferometric classical shadow (\Cref{lem:interf-classical-shadow}) using the controlled state preparation unitary $cV$ with the test states $\ket{x'}$, $\vec{x}'\in \mathcal{N}$, error $\epsilon_0$, and success probability $\delta/2$.
    This consumes
    \begin{equation}
        O\lr{\frac{\log(|\mathcal{N}|/\delta)}{\epsilon_0^2}} = O\lr{\frac{s\log(D/\epsilon_0) + \log(1/\delta)}{\epsilon_0^2}}
    \end{equation}
    queries to $cV$ and guarantees that the prediction of the overlap is $\epsilon_0$ accurate on the $\epsilon_0$-covering net $\mathcal{N}$ of $\mathcal{X}_{\mathrm{test}}$.
    For every $\vec{u}\in\mathcal{X}_{\mathrm{test}}$, fix $\widetilde{\vec{u}}\in\mathcal{N}$ satisfying $\|\vec{u}-\widetilde{\vec{u}}\|_2\leq\epsilon_0$, and use the classical-shadow estimate for $\widetilde{\vec{u}}$ as the estimate for $\vec{u}$. 
    Note that the net need not be explicitly enumerated or stored because $\widetilde{\vec{u}}$ can be computed on demand by rounding the nonzero coordinates of the sparse test vector.
    On the simultaneous-success event,
    \begin{equation}
    \left|
    \hat{o}(\widetilde{\vec{u}})
    -
    \vec{u}\cdot\vec{w}
    \right|
    \leq
    \left|
    \hat{o}(\widetilde{\vec{u}})
    -
    \widetilde{\vec{u}}\cdot\vec{w}
    \right|
    +
    \left|
    (\widetilde{\vec{u}}-\vec{u})\cdot\vec{w}
    \right|\leq
    \epsilon_0+
    \|\widetilde{\vec{u}}-\vec{u}\|_2
    \leq
    2\epsilon_0=E_0.
    \end{equation}
    For every classifiable test vector $\vec{x}'\neq 0$, its normalized vector $\vec{x}'/\|\vec{x}'\|_2$ belongs to $\mathcal{X}_{\mathrm{test}}$, and multiplication by the positive scalar $\|\vec{x}'\|_2$ does not change the predicted sign.
    We choose $\epsilon_0$ such that $\epsilon/2=E_0=O(\epsilon_0)$.
    Combining the error of the state preparation unitary, we know that the produced estimator is an $E_0+\epsilon/2 = \epsilon$ accurate estimate of $\expval{x'|w}$ for all $\vec{x}' \in \mathcal{X}_{\mathrm{test}}$.
    Meanwhile, if the size of test vectors $m$ is small, then $O(\log(m/\delta)/\epsilon^2)$ queries suffice.
    Hence the number of queries to $cV$ that we need is
    \begin{equation}
        O\lr{\frac{\min \left\{ s\log(D/\epsilon) \right., \log(m)\}+\log(1/\delta)}{\epsilon^2}}.
    \end{equation}

    From the assumption, we know that the margin of each test vector $\vec{x}'_j$ satisfies
    \begin{equation}
        \gamma(\vec{x}'_j) = \frac{|\vec{x}'_j \cdot \vec{w}|}{\|\vec{w}\|_2} = \|\vec{x}'_j\|_2 |\expval{x'_j|w}| \geq \gamma_{\mathrm{test}}.
    \end{equation}
    This implies that
    \begin{equation}
        |\expval{x'_j|w}| \geq \frac{\gamma_{\mathrm{test}}}{\|\vec{x}'\|_2}\geq \frac{\gamma_{\mathrm{test}}}{\sqrt{s}},
    \end{equation}
    because $\vec{x}'_j$ is $s$-sparse and hence $\|\vec{x}'_j\|_2 \leq \sqrt{s}\|\vec{x}_j'\|_\infty\leq \sqrt{s}$.
    Therefore, estimating $\expval{x'_j|w}$ to $\gamma_{\mathrm{test}}/(3\sqrt{s})$ suffices to determine
    \begin{equation}
        \hat{y}_j = \sgn(\vec{x}'_j\cdot \vec{w}) = \sgn\left(\expval{x'_j|w}\right).
    \end{equation}
    We therefore set
    \begin{equation}
        \epsilon = \frac{\gamma_{\mathrm{test}}}{3\sqrt{s}}.
    \end{equation}
    This gives us a quantum query algorithm that makes 
    \begin{equation}
    \begin{split}
        Q
        &=
        Q_0\cdot
        O\lr{
        \frac{
        \min\{s\log(D/\epsilon_0),\log(m)\}
        +
        \log(1/\delta)
        }{
        \epsilon_0^2
        }
        }\\
        &\leq
        O\lr{
        \frac{
        s\kappa_{\mathrm{reg}}^2
        \log^2(s\kappa_{\mathrm{reg}}/\gamma_{\mathrm{test}})
        \left(
        \min\left\{
        s\log\left(\frac{sD}{\gamma_{\mathrm{test}}}\right),
        \log(m)
        \right\}
        +
        \log(1/\delta)
        \right)
        }{
        \gamma_{\mathrm{test}}^2
        }
        }\\
        &\leq
        \tilde{O}\lr{
        \frac{
        s\kappa_{\mathrm{reg}}^2
        \left(
        \min\{s\log(D),\log(m)\}
        +
        \log(1/\delta)
        \right)
        }{
        \gamma_{\mathrm{test}}^2
        }
        }.
    \end{split}
    \end{equation}
    queries to $X, \ket{y}$ and can output the prediction $\hat{y}_j, \forall j\in [m]$ (a classical random variable) with probability at least $1-\delta/2$.
    Moreover, the space complexity is
    \begin{equation}
    \begin{split}
        S_0 &= O\lr{n} + \log^2(D)\cdot O\lr{\frac{\min\{s^2\log(sD/\gamma_{\mathrm{test}}), s\log(m)\}+s\log(1/\delta)}{\gamma_{\mathrm{test}}^2}} + O(\log^2(D)) \\
        &\leq \tilde{O}\lr{\frac{1}{\gamma_{\mathrm{test}}^2}s\log^2(D)(\min\{s\log(D), \log(m)\}+\log(1/\delta))},
    \end{split}
    \end{equation}
    where the first term comes from the ridge regression solver, the second comes from the classical shadow data, and the third one comes from classical simulation of Clifford circuits.
    
    The second step is to instantiate the queries to $X$ and $\ket{y}$ in the query algorithm using quantum oracle and state sketching.
    We first replace all queries to the block encoding of $X$ and its inverse and controlled versions to their $\epsilon_1$-approximate versions in \Cref{lem:block-encoding}.
    This incurs an error of at most $E_1 = Q\epsilon_1 = \delta/6$, where we set $\epsilon_1 = \delta/(6Q)$.
    
    Next, we replace the queries to $\epsilon_1$-approximate versions of block encoding of $X$ and its inverse and controlled versions by the random unitary channel that we build from samples in \Cref{lem:block-encoding}.
    This incurs an additional error of at most $E_2 = Q\epsilon_1 = \delta/6$, and uses $O(n+b+\log^{2.5}(s/\epsilon_1)) = O(n+b+\log^{2.5}(sQ/\delta))$ qubits and 
    \begin{equation}
        M_X = Q\cdot O\lr{\frac{R2^nn^2s^5\log^4(ns/\epsilon_1)}{\epsilon_1}} \leq O\lr{\frac{R2^nn^2s^5Q^2\log^4(\frac{nsQ}{\delta})}{\delta}}
    \end{equation}
    samples from the data generation process.
    Note that here we only use the feature data and throw away the label data in each sample.
    Also note that here, given a uniformly random row sample, we independently sample one of its padded $s$ slots uniformly. 
    Hence the resulting real-or-dummy matrix-element slot is uniform over the $Ns$ padded slots, so we may invoke the padded-support version of \Cref{lem:block-encoding}.
    The fresh slot label is retained in the transformed sample, so this subsampling changes the repetition number by at most a constant factor, which is absorbed into $R$.

    Finally, we run the preprocessing procedure in
    \Cref{thm:q-state-sketch} once for $\vec{y}$ and retain the resulting record $\nu_y$.
    Conditioned on $\nu_y$, \Cref{thm:q-state-sketch} provides a fixed state preparation unitary $P_{\nu_y}$ for $\ket{y}$.
    We implement these queries to the state preparation unitary, its inverse, and their controlled versions by the corresponding $\epsilon_1$-error random unitary channels constructed from fresh sample blocks according to \Cref{thm:q-state-sketch}.
    Applying the shared-record version of
    \Cref{lem:error-accumulation-time-varying} conditioned on $\nu_y$ and then averaging over $\nu_y$ gives an additional error
    \begin{equation}
        E_3
        =
        Q\epsilon_1
        =
        \delta/6.
    \end{equation}
    This uses $O(n\log(N/\epsilon_1)) \leq O(n\log(\frac{NQ}{\delta}))$ qubits and 
    \begin{equation}
        M_y = Q\cdot O\lr{\frac{R2^n n^2\log^4(1/\epsilon_1)}{\epsilon_1}} \leq O\lr{\frac{R2^nn^2Q^2\log^{4}(\frac{Q}{\delta})}{\delta}}
    \end{equation}
    samples from the data generation process, where the one-time pre-processing cost is absorbed into the stated sample bound.
    Here we only use the label data and throw away the feature data in each sample.
    
    According to \Cref{lem:error-accumulation-time-varying}, the total error in instantiating the query algorithm is bounded by
    \begin{equation}
        E_1+E_2+E_3 = 3\cdot \delta/6 = \delta/2.
    \end{equation}
    This means that the output of the query-instantiated quantum algorithm, which is a classical random variable must be correct with probability at least $1-\delta/2-\frac{1}{2}\delta/2\geq 1-\delta$.
    The total number of qubits used is
    \begin{equation}
    \begin{split}
        &S_0 + O(n+b+\log^{2.5}(sQ/\delta)) + O(n\log(NQ/\delta))\\
        &\leq \tilde{O}\lr{b + \frac{s\log^2(D)(\min\{s\log(D), \log(m)\}+\log^{2.5}(1/\delta))}{\gamma_{\mathrm{test}}^2}+\log(sD)\log(\kappa_{\mathrm{reg}}) +\log^{2.5}(\kappa_{\mathrm{reg}}) }
    \end{split}
    \end{equation}
    The total number of samples is
    \begin{equation}
        M=M_X+M_y
        \leq
        O\lr{
        \frac{
        R2^n n^2s^5Q^2
        \log^{4}(\frac{nsQ}{\delta})
        }{
        \delta
        }
        }
        \leq
        \tilde{O}\lr{
        \frac{
        RN s^{7}\kappa_{\mathrm{reg}}^4
        }{
        \gamma_{\mathrm{test}}^4\delta
        }
        \min\{s^2\log^2(D), \log^2(m)\}
        }.
    \end{equation}
    This completes the proof of \Cref{thm:bin-classify-upper}.
\end{proof}

\subsubsection{Classical hardness}

In this section, we prove the two classical hardness results in \Cref{thm:q-adv-bin-classify,thm:q-adv-bin-classify-dynamic}.
We fix the $\ell_2$ regularization strength $\lambda=0$ throughout the proof so that the regularized condition number is the same as the condition number of $X$: $\kappa_{\mathrm{reg}} = \kappa$.
We will construct a (dynamic) binary classification task with $N=D$.
Hence, for simplicity, we will use $N$ alone throughout this section.
We prove the classical hardness results by constructing a specific binary classification task using \Cref{lem:embed-qc-into-svm}.
Solving it amounts to solving the (dynamic) Noisy Oracle Property Estimation (NOPE) task defined in \Cref{sec:cl-hard} (\Cref{task:nope,task:dynamic-nope}), whose classical hardness we have already proved in \Cref{thm:classical-lower-bound-single-block,thm:classical-lower-bound-superpoly-sample}.

This gives us two classical hardness results.
The first result follows from \Cref{thm:classical-lower-bound-single-block} and shows that any classical learning algorithm that wants to perform better than random guessing in binary classification with the same number of samples quantum algorithms need must have $\Omega(D^{1-\zeta})$ size for any constant $\zeta>0$.
The second result follows from \Cref{thm:classical-lower-bound-superpoly-sample} and shows that when the binary classification task is dynamic, any classical learning algorithm will need a number of samples super-polynomial in $N$ if it has size $o(D^{1-\zeta})$ for any constant $\zeta>0$.
Note that it suffices to consider $0<\zeta<1$, since for $\zeta\geq1$ the claimed space lower bounds are vacuous.

The general idea of proving these results is to embed the quantum circuit that solves dynamic NOPE from \Cref{lem:encode-q-circ} into a binary classification task via \Cref{lem:embed-qc-into-svm}.
The label $\hat{y}$ of a fixed, sparse test vector $\vec{x}$ encodes the oracle property $B\in\bit$ that we want to estimate in (dynamic) NOPE.
Then we show that given any classical algorithm that predicts the label $\hat{y}$, we can use it to construct a classical algorithm that solves (dynamic) NOPE.
In the (dynamic) NOPE task, we take the oracle property function $f$ to be $K$-Forrelation (\Cref{def:forrelation}) and the noisy encoding function $g$ to be the inner product (\Cref{def:inner-prod}).

In the following, we construct the data generation process $\mathcal{D}^{N,K, R}_{\mathrm{BC}}(B), B\in\bit$ for binary classification, where $N$ is both the sample dimension and the feature dimension (i.e., $D=N$), $K=\Theta(1)$ is the constant in the Forrelation that we will embed, $R$ is an integer that specifies the repetition number, and $B$ indicates whether the label of that fixed test vector is $+1$ or $-1$ (see \Cref{lem:embed-qc-into-svm}).
The goal is to solve for $B$.
It helps to compare this construction to the dynamic NOPE data generation process defined in \Cref{sec:cl-hard-bootstrap}.

Given $N$, we define another large integer $N'$ as follows.
Let $T = \Theta(\log^3 N' \log\log N')$ be the total number of two-qubit gates and diagonal gates given in \Cref{lem:encode-q-circ} and let $n' = \log N' + O(\log\log N')$ be the number of qubits in \Cref{lem:encode-q-circ}.
Let $T'=2T+n'+1$ be the length of the enlarged circuit in \Cref{lem:embed-qc-into-svm}.
To properly embed that circuit into the $N$-dimensional training dataset, we define $N'$ such that $N = 48T' 2^{n'} = N' \cdot \polylog N'$ as in \Cref{lem:embed-qc-into-svm}.
This implies $N' = N / \polylog N$.
The resulting feature matrix $X$ have sparsity $s=O(1)$ and condition number $\kappa = O(T') = O(\log^3N\log\log N)$ from \Cref{lem:embed-qc-into-svm}.
We fix the label vector to be the $\vec{y}$ in \Cref{lem:embed-qc-into-svm}.
Let $L=\ceil{\log^2 (KN')}\geq 5$ be the number of independent oracle instances that we have in dynamic NOPE.

Now we define the data generation process
\begin{equation}
    \mathcal{D}^{N, K, R}_{\mathrm{BC}}(B) = (\mathcal{D}^0_B\to \mathcal{D}^1_{X}\to^{\times T_1} \mathcal{D}^2_{(\alpha, \beta)}\to^{\times T_2} z=(i, \vec{x}_i, y_i) \to^{\times T_3} z),
\end{equation}
where $T_3=\Theta(R)$ is chosen such that the repetition number of the resulting process is at most $R$, $T_2 = KN'$, $T_1 = \ceil{M_Q/(T_2T_3)} = \polylog(KN')$, $M_Q=RN\polylog(N)$ is the number of samples quantum machines need in \Cref{thm:bin-classify-upper}, $X\in \mathbb{R}^{N\times N}$ is an $N$-dimensional, symmetric, $O(1)$-sparse feature matrix with condition number $\kappa=O(T)=O(\log^3 N\log\log N)$, $\alpha\in\bit, \beta\in[L] $ label which part of the matrix $X$ that we are currently collecting matrix element data from, $i\in [N]$ labels the $i$-th training data point $\vec{x}_i$ (the $i$-th row of the matrix $X$), and $y_i$ is the corresponding label in $\vec{y}$.

The data are sampled in the following way that resembles dynamic NOPE in \Cref{sec:cl-hard-bootstrap}.
$\mathcal{D}^0_{B}$ samples a length-$L$ bitstring $\gamma$ with parity 
\begin{equation}
    \bigoplus_{j=1}^{L}\gamma_{j}=B
\end{equation} 
uniformly random.
For each $j\in [L]$, we sample a random oracle $o_j\sim p_{\gamma_j}$, where $p_0, p_1$ are the distributions of Forrelation defined in \Cref{lem:forrelation-query-separation}.
Then we sample a noisy encoding pair $(Y^{(0, j)}, Y^{(1, j)})\sim \unif((g^{KN'})^{-1}(o_j))$ using the inner product noisy encoding function $g$ defined in \Cref{def:inner-prod}.

Next, note that $(Y^{(0,j)}, Y^{(1,j)})_{j=1}^{L}$ specifies an $n'$-qubit quantum circuit $C$ with $T=O(\log^3N' \log\log N')$ gates via \Cref{lem:encode-q-circ}, and the quantum circuit $C$ gives a training dataset $(X, \vec{y})$ with dimension $48T' 2^{n'}=N$ via \Cref{lem:embed-qc-into-svm}.
Here, $X$ is indeed $O(1)$-sparse with condition number $\kappa = O(T)=O(\log^3 N\log\log N)$.
Now we define $\mathcal{D}^1_{X}$.
We first sample a uniformly coordinate $\beta\sim \unif([L])$ and a random bit $\alpha\sim\mathrm{Bern}(1/2)$ as in dynamic NOPE.
Then we pick out $Y^{(\alpha, \beta)}$ and use it to generate the data samples.

In particular, we define $\mathcal{D}^2_{(\alpha, \beta)}$ in the following way.
We first sample a random row of the matrix $X$ as follows.
We note that after uniformly sampling the realification blocks given in \Cref{lem:realification}, the linear space corresponding to the matrix $X$ has a particular factorization given by \Cref{lem:embed-qc-into-svm}.
We sample the nested blocks of $X$ uniformly randomly.
Then we are left with the subspace $\ket{t}\ket{\psi}$ where $\ket{t}$ is the clock register and $\ket{\psi}$ is the $n'$-qubit register that the quantum circuit runs on.
We sample a clock time $t\sim \unif([3T'])$ and the matrix is reduced to a specific gate in the $n'$-qubit subspace (either a fixed two qubit gate or a diagonal gate that depends on $Y^{(\alpha, \beta)}$).
The remaining subspace further factorizes into $\ket{x, \alpha, \beta, k}$ and the rest of the working qubits.
We sample a random basis of this $n'$-qubit subspace by plug in the specific $(\alpha, \beta)$ that we have already sampled, sample $x \sim \unif([KN']), k\sim \unif(\{0, \ldots, b\})$, and sample a computational basis of the rest of the working qubits uniformly random.
This together specifies and thus samples a row $i$ of the matrix $X$.
Note that the marginal distribution of $i$ is uniform over $[N]$ (because $\alpha, \beta$ are sampled uniformly), though there are correlations between consecutive samples of $i$ since they share the same set of $(\alpha, \beta)$.

Note that by construction of the matrix $X$ as in \Cref{lem:embed-qc-into-svm}, the picked out row vector $\vec{x}_i$ is $O(1)$ sparse, and has components that, before the normalization, is the real or imaginary part of either $1$, from the identity matrices in \Cref{eq:A,eq:B}, or a matrix element of a fixed two qubit gates, or a value depends on $(Y^{(\alpha,\beta)}_x)_k$ via the known embedding, permutations, and realification (see \Cref{eq:B} and \Cref{lem:embed-qc-into-svm}) which is solely specified by $Y^{(\alpha, \beta)}$.
This gives us the sample $z=(i, \vec{x}_i, y_i)$ and we repeat this sample $T_3$ times, completing the data generation process.

This data generation process $\mathcal{D}_{\mathrm{BC}}^{N, K, R}$ is a valid data generation process of dynamic binary classification.
It produces a random row sample uniformly distributed over the rows of $X$.
The test vector is a single fixed $O(1)$-sparse vector $\vec{x}'$ specified in \Cref{lem:embed-qc-into-svm}, which satisfies
\begin{equation}
    \frac{\vec{x}'\cdot \vec{w}}{\|\vec{w}\|_2} = C (2q_B-1), \quad \vec{w} = (X^TX)^{-1}X^T\vec{y}, \quad C\geq \frac{1}{200(2T+n+1)^{3/2}} = \frac{1}{\polylog(N)}
\end{equation}
where $q_B$ is the probability of measuring $0$ on the embedded circuit given by \Cref{lem:encode-q-circ} when the underlying oracle property is $B$.
\Cref{lem:encode-q-circ} ensures that $q_1\leq 0.1, q_0\geq 0.99\geq 0.9$.
Therefore, we have the label
\begin{equation}
    \hat{y} = \sgn(\vec{x}'\cdot \vec{w}) = (-1)^B,
\end{equation}
and the margin
\begin{equation}
    \gamma(\vec{x}') = C|2q_B-1|\geq 1/\polylog(N),
\end{equation}
as required.
Moreover, since $\lambda=0$ and $X$ is square and invertible, $r_{\mathrm{reg}}(X,\vec{y})=0$, so this hard instance indeed satisfies the non-trivial residual requirement.
The repetition number is $O(T_3)$ because, conditioned on the current higher-level situation, only the group containing the first sample and the remaining draws in the same $(\alpha, \beta)$ block can make an excess contribution. 
The former contributes at most $T_3$, while the latter contributes $T_3(T_2-1)O(1/T_2)=O(T_3)$, since every possible unrepeated sample has probability $O(1/T_2)$.
Thus $R_{\mathcal D}=O(T_3)$.
We choose $T_3=\Theta(R)$ so that $R_{\mathcal D}\leq R$ and suppress this constant rescaling below.
The refreshing time is $\tau = T_1T_2T_3 = O(M_Q)=\tilde O(RN)$, satisfying the requirements in \Cref{thm:q-adv-bin-classify,thm:q-adv-bin-classify-dynamic}.

This data generation process for binary classification $\mathcal{D}^{N, K, R}_{\mathrm{BC}}(B)$ is designed to reduce to the dynamic NOPE data $\mathcal{D}^{KN', T_1}_{g, f}(B)$ in \Cref{sec:cl-hard-bootstrap} via \Cref{lem:encode-q-circ,lem:embed-qc-into-svm}.
Using this data generation process, we prove the following two results.

\begin{tcolorbox}
\begin{theorem}[Classical hardness of binary classification]
\label{thm:classical-hardness-bin-classify}
    Let $\zeta>0$ be any constant.
    Let $N, D$ be the sample and feature dimension of a binary classification task and $R$ be its repetition number.
    Using $\tilde{O}(RN)$ samples, any randomized classical learning algorithm with space complexity 
    \begin{equation}
        S\leq o(D^{1-\zeta})
    \end{equation}
    cannot solve the binary classification task with success probability more than $1/2+1/N^{\omega(1)}$.
\end{theorem}
\end{tcolorbox}

\begin{tcolorbox}
\begin{theorem}[Classical hardness of dynamic binary classification]
\label{thm:classical-hardness-bin-classify-dynamic}
    Let $\zeta>0$ be any constant.
    Let $N, D$ be the sample and feature dimension of a dynamic binary classification task and $R, \tau=\tilde{\Theta}(RN)$ be its repetition number and refreshing time.
    Any randomized classical learning algorithm that solves the task with success probability at least $2/3$ must have sample complexity
    \begin{equation}
        M\geq RN^{\omega(1)}
    \end{equation}
    if its space complexity
    \begin{equation}
        S \leq o(D^{1-\zeta}).
    \end{equation}
\end{theorem}
\end{tcolorbox}

\Cref{thm:classical-hardness-bin-classify} immediately implies the classical hardness part of \Cref{thm:q-adv-bin-classify} because the first $\tilde{O}(RN)$ samples from the constructed binary classification data belongs to the same training dataset and therefore is a valid sequence of non-dynamic binary classification data.
\Cref{thm:classical-hardness-bin-classify-dynamic} directly implies the classical hardness part of \Cref{thm:q-adv-bin-classify-dynamic}.
Together with the quantum algorithm result \Cref{thm:bin-classify-upper}, this completes the proof of the quantum advantage claims in \Cref{thm:q-adv-bin-classify,thm:q-adv-bin-classify-dynamic}.

\begin{proof}[Proof of \Cref{thm:classical-hardness-bin-classify,thm:classical-hardness-bin-classify-dynamic}]
    Recall that $N=D$ in the task that we construct.
    For simplicity, we will use $N$ throughout the proof.
    We prove \Cref{thm:classical-hardness-bin-classify,thm:classical-hardness-bin-classify-dynamic} by showing that given any classical learning algorithm that can predict the label $\hat{y}$ of the test vector $\vec{x}'$, we can use it to construct an algorithm that decides $B$ from $\mathcal{D}^{KN', T_1}_{g, f}(B)$, which we have proved to be hard in \Cref{thm:classical-lower-bound-single-block,thm:classical-lower-bound-superpoly-sample}.
    
    First note that, since
    \begin{equation}
        \hat{y} = (-1)^B,
    \end{equation}
    if we can predict the label $\hat{y}$, we can indeed decide $B$.

    We choose $K=\ceil{1.001/\zeta}$ such that 
    \begin{equation}
        \frac{N'^{1-1/K}}{\polylog N'} \geq \frac{N^{1-\zeta+\zeta/1001}}{\polylog N}\geq \Omega(N^{1-\zeta}).
    \end{equation}
    For the sake of contradiction, suppose we have a randomized classical learning algorithm $\mathcal{L}$ with space complexity
    \begin{equation}
        S \leq o(N^{1-\zeta}) \leq o\left(\frac{N'^{1-1/K}}{\polylog N'}\right)
    \end{equation}
    and sample complexity $M$ that given a sequence of data samples drawn from $\mathcal{D}^{N, K, R}_{\mathrm{BC}}(B)$, predicts $\hat{y}$ and hence decides $B$ with probability $p_{\mathrm{succ}}$.
    In the following, we design a classical learning algorithm $\mathcal{L}'$ that decides $B$ using data from $\mathcal{D}^{KN', T_1}_{g, f}(B)$.

    The first step of $\mathcal{L}'$ is to generate data samples distributed exactly according to $\mathcal{D}^{N,K,R}_{\mathrm{BC}}(B)$ from the source stream $\mathcal{D}^{KN',T_1}_{g,f}(B)$.
    
    For every group of $T_3$ repeated target samples, $\mathcal{L}'$ reads one source sample $z=
    (
        x,
        Y^{(\alpha,\beta)}_x,
        \alpha,
        \beta
    )$.
    It then uses the same public row-sampling procedure as in the definition of $\mathcal{D}^{N,K,R}_{\mathrm{BC}}(B)$.
    In particular, it uses the sampled values of $\alpha,\beta,x$ and samples all remaining public realification, block, clock, and working-register coordinates according to the original construction.
    If the selected clock time corresponds to a fixed gate, the resulting row is calculated from the public circuit description, and $Y^{(\alpha,\beta)}_x$ is ignored.
    If it corresponds to an oracle gate, the algorithm samples $k\sim\unif(\{0,\ldots,b\})$; for $k=0$ the oracle acts as the identity, while for $k\in[b]$ the row is calculated from $(Y^{(\alpha,\beta)}_x)_k$ via the known embedding.
    In either case, it applies the public permutations and realification map and divides the resulting row by the known normalization factor $\|\mathcal{R}[X_c]\|$.
    The corresponding label $y_i$ is determined by the fixed public label vector in \Cref{lem:embed-qc-into-svm}.
    The algorithm repeats the resulting sample $z_{\mathrm{BC}}=(i,\vec{x}_i,y_i)$ exactly $T_3$ times.
    Note that this procedure generates data samples that exactly matches $\mathcal{D}^{N, K, R}_{\mathrm{BC}}(B)$ by construction.

    After sampling a training data point $z_{\mathrm{BC}}=(i, \vec{x}_i, y_i)$, we feed it into the learning algorithm for binary classification $\mathcal{L}$.
    We repeat this $M$ times so that $\mathcal{L}$ receives $M$ samples whose distribution matches that of $\mathcal{D}^{N, K, R}_{\mathrm{BC}}(B)$ and produces a prediction of the label $\hat{y}$ that provides a prediction of $B$.
    We use this predicted bit of $\mathcal{L}$ as the final output of $\mathcal{L}'$.

    Note that since the data generation does not require knowledge of the previous data samples from $\mathcal{D}^{KN', T_1}_{g, f}(B)$, it can be performed online and thus the space complexity of $\mathcal{L}'$ is $S'=S$.
    Moreover, one source sample is used for every group of $T_3$ repeated target samples, so
    $M'\leq\lceil M/T_3\rceil\leq M/T_3+1$.
    The success probability of $\mathcal{L}'$ is $p_{\mathrm{succ}}$, the same as that of $\mathcal{L}$.

    Finally, we invoke \Cref{thm:classical-lower-bound-single-block,thm:classical-lower-bound-superpoly-sample}.
    Note that for the inner product $g$, we have $\eta=1/2$, $c = (865/\eta^2)\log(865/\eta^2) = (865\times 4)\log(865\times 4) \approx 40677.68$, and therefore the choice of $b=\ceil{40678\log(KN')}$ satisfies the requirement.
    For the Forrelation $f$ we use, \Cref{lem:forrelation-query-separation} implies that the $(1/3)$-error classical distributional query complexity is (using $K=\Theta(1)$)
    \begin{equation}
        Q_C \geq \Omega\left(\frac{N'^{1-1/K}}{\polylog N'}\right)
    \end{equation}
    and therefore
    \begin{equation}
        S'=S \leq o\left(\frac{N'^{1-1/K}}{\polylog N'}\right)\implies S'\leq o\lr{\frac{Q_C}{T_1^2 L}}
    \end{equation}
    satisfying the condition of \Cref{thm:classical-lower-bound-single-block,thm:classical-lower-bound-superpoly-sample}.
    Therefore, if the $M=T_1T_2T_3=\tilde{O}(RN)$ samples are drawn from the first refreshing block only, then $M'\leq M/T_3 = T_1KN'$ and \Cref{thm:classical-lower-bound-single-block} implies that
    \begin{equation}
        p_{\mathrm{succ}}\leq \frac12+\frac{1}{(KN')^{\omega(1)}}\leq \frac12+\frac{1}{N^{\omega(1)}},
    \end{equation}
    proving \Cref{thm:classical-hardness-bin-classify}.
    On the other hand, \Cref{thm:classical-lower-bound-superpoly-sample} implies that 
    \begin{equation}
        M\geq T_3(M'-1)\geq R(KN')^{\omega(1)}= RN^{\omega(1)},
    \end{equation}
    proving \Cref{thm:classical-hardness-bin-classify-dynamic}.
    This completes the proof of \Cref{thm:classical-hardness-bin-classify,thm:classical-hardness-bin-classify-dynamic} and, together with the quantum algorithm result \Cref{thm:bin-classify-upper}, proves \Cref{thm:q-adv-bin-classify,thm:q-adv-bin-classify-dynamic}.
\end{proof}

\newpage

\subsection{Dimension reduction}
\label{sec:dimension-reduction}

\begin{figure}
    \centering
    \includegraphics[width=1\linewidth]{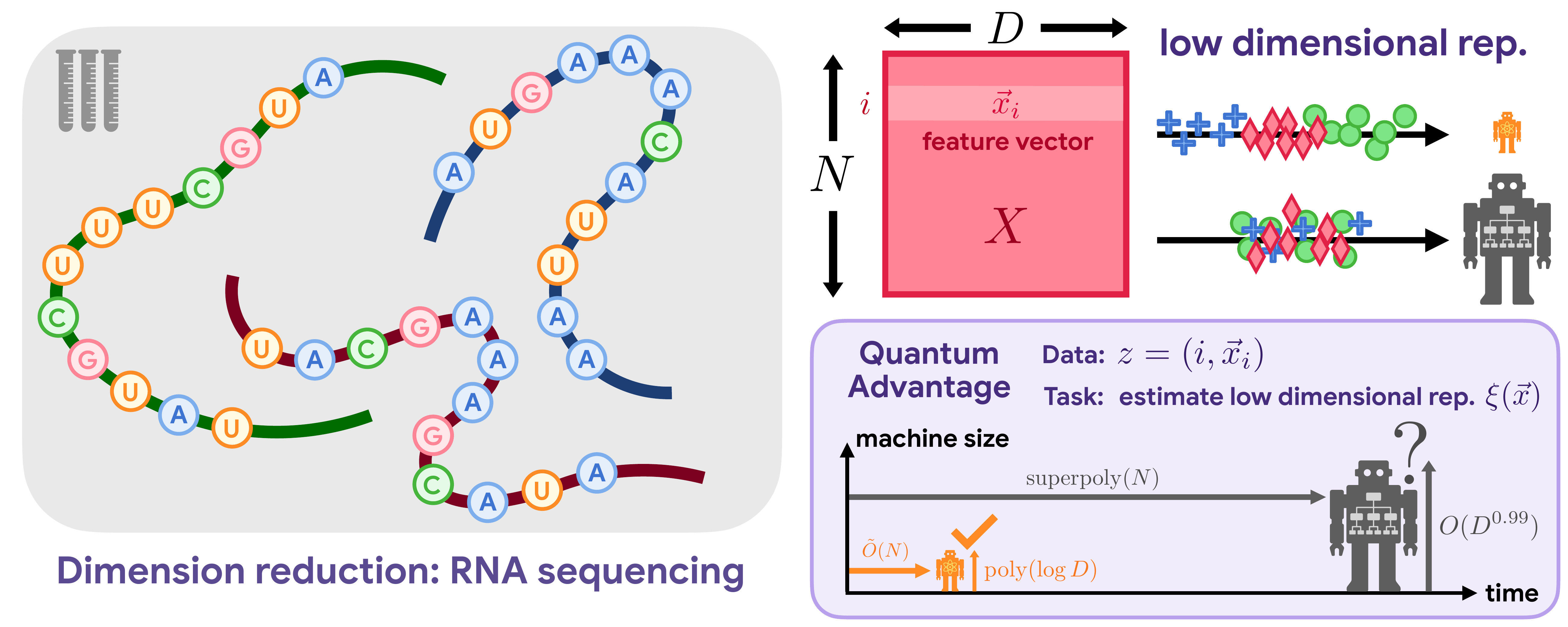}
    \caption[Overview of the dimension reduction task.]{\textbf{Overview of the dimension reduction task.}
    We illustrate the dimension reduction task with a particular real-world application scenario in single cell RNA sequencing (scRNA-seq).
    We conduct RNA sequencing experiments to obtain gene sequences in cell samples, which are represented as feature vectors $\vec{x}_i \in \mathbb{R}^D$ (e.g., via frequency counting of $k$-mer representations).
    A sample is the feature vector $\vec{x}_i$ of a single RNA sequence.
    A canonical way of dimension reduction is principal component analysis (PCA), which is performed on the high-dimensional data matrix $X\in \mathbb{R}^{N\times D}$ and produces a principal component vector $\vec{w}\in \mathbb{R}^D$ that represents the most important feature combination. 
    We obtain the low dimension representation $\xi(\vec{x})$ of a sample $\vec{x}$ by projecting it onto the principal component $\vec{w}$.
    Our results show that a small quantum machine with size $\poly(\log D)$ can solve this task with $\tilde{O}(N)$ samples, whereas any classical machine with exponentially larger size $O(D^{0.99})$ cannot solve the task unless it uses a sample size at least super-polynomial in $N$.}
    \label{fig:dim-reduct}
\end{figure}

In this section, we consider the fundamental task of dimension reduction in data science and machine learning.
We focus on principal component analysis (PCA), the standard method for dimension reduction.
For simplicity, we consider the task of reducing the dimension to one. 
It is straightforward to generalize this to more dimensions.

In PCA, we are given a dataset $X = (\vec{x}_1, \ldots, \vec{x}_N)^T$ of size $N$, where each data sample $\vec{x}_i \in \mathbb{R}^D$ is a $D$-dimensional feature vector and is properly normalized such that $\|X\|= 1$.
This normalization is without loss of generality, because multiplying $X$ by a positive scalar does not change its principal component. 
The gap parameter $\Delta$ below is always defined after this normalization.

The principal component of this dataset $X$ is a unit vector $\vec{w}\in \mathbb{R}^D$, the direction of which explains the most variance of the data.
It is the top eigenvector of the covariance matrix $X^TX\in \mathbb{R}^{D\times D}$ with the largest eigenvalue $\lambda_{\max} = \sigma_{\max}^2$, where $\sigma_{\max}$ is the largest singular value of $X$.
The principal component $\vec{w}$ provides us with a way to perform dimension reduction on any set of test vector $\{\vec{x}'_j\}_{j=1}^m$.
We reduce each test vector $\vec{x}'_j$ to one dimension by projecting it to the direction of $\vec{w}$:
\begin{equation}
    \xi(\vec{x}'_j) = \vec{x}'_j\cdot \vec{w}, \quad \vec{w} = \underset{\|\vec{w}\|_2=1}{\argmax}~\vec{w}^TX^TX\vec{w}.
\end{equation}
We call $\xi(\vec{x}'_j)$ the one-dimensional ($1D$) representation of the test vector $\vec{x}_j'$.
The goal of dimension reduction is to find the 1D representation $\xi(\vec{x}'_j)$ of any test vector $\vec{x}'_j$, given the dataset $X$.

It is often the case that we can obtain a good initial guess $\vec{g}\in \mathbb{R}^D, \|\vec{g}\|_2=1,$ of the principal component.
We call this initial guess $\vec{g}$ the guiding vector.
The quality of a guiding vector $\vec{g}$ is given by the overlap between $\vec{g}$ and the true principal component $\vec{w}$:
\begin{equation}
	\vec{g}\cdot \vec{w},
\end{equation}
and a good guiding vector should have a large overlap with the principal component, as compared to a random vector that has overlap $\sim 1/\sqrt{D}$.
For example, when there is a prominent feature in the dataset, the corresponding basis vector is usually a good guiding vector.
Another way to obtain a good guiding vector, is to first conduct a small-scale PCA on a sub-sampled dataset, and use the resulting principal component as the guiding vector for the full PCA.
Since the spectral gap satisfies $\Delta>0$, the principal component is unique up to an overall sign.
Throughout this work, we fix its orientation by requiring $\vec{g}\cdot\vec{w}>0$.
All occurrences of the 1D representation $\xi(\vec{x})=\vec{x}\cdot\vec{w}$ use this convention.

We characterize the tractability of the dimension reduction task by three parameters: the sparsity $s$, the spectral gap $\Delta$, and the quality $\chi$ of the guiding vector $\vec{g}$ defined below.
The sparsity $s$ is the maximal number of non-zero elements in $X$ per row or column, which bounds the number of non-zero features per sample  and the number of samples each feature is presented in.
Such sparsity appears in many natural datasets (e.g., natural language data with common words trimmed off.)
The spectral gap $\Delta$ is the gap between the largest eigenvalue $\lambda_{\max}(X^TX)$ of $X^TX$ and the second largest eigenvalue $\lambda_{\mathrm{sec}}(X^TX)$ 
\begin{equation}
	\Delta = \lambda_{\max}(X^TX) - \lambda_{\mathrm{sec}}(X^TX) = \sigma_{\max}^2(X) - \sigma_{\mathrm{sec}}^2(X),
\end{equation}
where $\sigma_{\max}(X), \sigma_{\mathrm{sec}}(X)$ are the largest and second largest singular values of $X$.
$\Delta$ captures the gap between the signal (the principal component) and the noise (the other singular vectors) in the data.
A large gap means a distinguishable signal in the data.
This corresponds to the spike in the spiked covariance model, a standard model for the covariance structure of high-dimensional data \cite{johnstone2001distribution}.
Under the power law distribution of eigenvalues usually seen in realistic datasets \cite{newman2005power}, with the normalization $\|X\|=1$, the gap is a constant $1-2^{-\alpha}=O(1)$, where $\alpha$ is the exponent in the power law.
Even when the spectrum is indeed continuous, the parameter $\Delta$ serves as a cutoff scale at which we wish to distinguish the principal component from competing modes.
Our algorithm will return a vector $\vec{w}$ that is a linear combination of all singular vectors that have singular values close to the principal component within $\Delta$.
These singular vectors are roughly equally well up to precision $\Delta$.
Hence this is sufficient for the purpose of dimension reduction.
The quality of the guiding vector $\chi$ represents how good our initial guess is.
We define the quality parameter $\chi$ by
\begin{equation}
	\vec{g}\cdot \vec{w} \geq \tilde{\Omega}\lr{\frac{1}{D^{(1-\chi)/2}}},
\end{equation}
where $\chi=0$ means that the guiding vector is no better than a random guess, and $\chi=1$ means that the guiding vector already has constant overlap with the principal component.
We assume that $\chi=\Theta(1)$ is a constant.

We focus on the regime where the number of samples $N$ and feature dimension $D$ are very large, and yet the dimension reduction task remains sparse, gapped, and well-guided:
\begin{equation}
	s\leq \poly(\log N, \log D), \quad \Delta \geq \frac{1}{\poly(\log N, \log D)}, \quad \vec{g}\cdot\vec{w}\geq \tilde{\Omega} \left( \frac{1}{D^{(1-\chi)/2}} \right).
\end{equation}
Note that all-zero rows do not affect $X^TX$, and all-zero columns correspond to features with zero principal-component weight.
We therefore discard them and take $N,D$ to count only nonzero samples and features.
Consequently, we always have $D=\tilde{\Theta}(N)$ when the sparsity $s\leq \poly(\log N, \log D)$.

\subsubsection{Problem formulation}

Now we formally define our data processing task of dimension reduction, summarized in \Cref{task:dim-reduc,task:dim-reduc-dynamic}.
In particular, we specify the data generation process and the goal of the task.
We assume that a guiding vector $\vec{g}$ with quality $\chi$ is known to us and the state $\ket{g}$ can be prepared in $\poly(\log D)$ time (e.g., say it is sparse and we can directly prepare the state $\ket{g}$).
We consider a data generation process where we randomly get a sparse feature vector $(i, \vec{x}_i)$ from the sparse feature matrix $X$.
The components of $\vec{x}_i$ are specified by bitstrings of length $b=\poly(\log N, \log D)$ to sufficient accuracy.
For simplicity, we assume that this binary representation is exact and use $\vec{x}_i$ to stand for the corresponding values.

Specifically, we consider any hierarchical data generation process $\mathcal{D}_{\mathrm{DR}}(X)$ with bounded repetition number $R$ that generates data samples of the form
\begin{equation}
    z = (i, \vec{x}_i), \quad i\overset{\mathrm{marginal}}{\sim} \unif([N]),
\end{equation}
where $\vec{x}_i$ is a random entry of feature vector in the feature matrix $X$.
Recall that the repetition number $R$ characterize the correlation in the data, and is defined as
\begin{equation}
    R = \max_z \left(\E\left[N_z  \middle| z_1=z\right] - \E[N_z]\right),
\end{equation}
where $N_z = \sum_{i=1}^{\tau}\delta_{z_i, z}$ is the number of repeating $z$'s in a refreshing block of data and $\tau$ is refreshing time that bounds the correlation time scale.
Throughout this work, we also assumes that the situation information is completely specified in the data $z$ (e.g., given by the row index).
The specific way of sampling these data can be arbitrary as long as it satisfies the above form.
Without loss of generality, we represent every sparse row by
$s$ slots, padding unused slots with marked zero-valued dummy entries.
Our results generalize straightforwardly to the more demanding scenario, where we only get a random non-zero subset of the feature vector $\vec{x}_i$.

Next, we specify the goal of the dimension reduction task.
We aim to estimate the 1D representation $\xi(\vec{x}'_j)$ of all possible sparse test vectors $\vec{x}_j', j=1, \ldots, m$ that are properly normalized $\|\vec{x}_j'\|_\infty\leq 1$.
In other words, we want to estimate the 1D representation
\begin{equation}
	\xi(\vec{x}'_j)=\vec{x}'_j \cdot \vec{w}, \quad \vec{w} = \underset{\|\vec{w}\|_2=1}{\argmax}~\vec{w}^TX^TX\vec{w},
\end{equation}
for any $s$-sparse test vector $\vec{x}'_j$.

This motivates our following definition of the dimension reduction task.

\begin{tcolorbox}
\begin{task}[Dimension reduction task]
\label{task:dim-reduc}
    Let $N, D, R$ be integers and $\epsilon\in (0, 1]$.
    The dimension reduction task is to estimate the 1D representation
    \begin{equation}
        \xi(\vec{x}_j') = \vec{x}'_j\cdot \vec{w}, \quad \vec{w} = \underset{\|\vec{w}\|_2=1}{\argmax}~\vec{w}^TX^TX\vec{w},
    \end{equation}
    of any $\poly(\log N, \log D)$-sparse test vector $\vec{x}'_j$ in a test set $\{\vec{x}_j'\}_{j=1}^m$ of any size $m$ to $\epsilon$ additive error, using data samples from any data generation process $\mathcal{D}_{\mathrm{DR}}(X)$ defined above with repetition number at most $R$ that corresponds to the normalized data matrix $X\in \mathbb{R}^{N\times D}, \|X\|= 1$, which has $\poly(\log N, \log D)$ sparsity and $1/\poly(\log N, \log D)$ gap, given a guiding vector $\vec{g}$ with quality $\chi\in [0, 1]$.
\end{task}
\end{tcolorbox}

We also consider the dynamic scenario where we have correlated data with time-varying features.
That means the data matrix $X$ changes over time, but the 1D representations $\xi(\vec{x}'_j)$ of the test vectors $\vec{x}'_j$ remain approximately the same.
One can also think of this as having a time-varying data matrix $X$ with its principal component roughly fixed.
This resembles the batch processing strategy common in modern large-scale data mining, where we analyze the data on batches of data that are different but the desired 1D representations are fixed.
We use $\{\xi_j\}_{j=1}^m$ to denote that fixed set of 1D representations of the test vectors $\vec{x}_j'$.
We consider any hierarchical data generation process $\mathcal{D}_{\mathrm{DDR}}(\{(\vec{x}_j', \xi_j)\}_{j=1}^m)$ with bounded repetition number $R$ and refreshing time $\tau$ (defined for each sampled dataset) of the form
\begin{equation}
	\mathcal{D}_{\mathrm{DDR}}(\{(\vec{x}_j', \xi_j)\}_{j=1}^m, \epsilon) = (\mathcal{D}^0_{\{(\vec{x}_j', \xi_j)\}_{j=1}^m, \epsilon} \to \mathcal{D}_{\mathrm{DR}}(X) \to^{\times \tau} z).
\end{equation}
In other words, the data matrix $X$ changes every $\tau$ time steps, and we keep getting random feature vectors of the current training dataset.
We require all data matrices $X$ sampled from $\mathcal{D}^0_{\{(\vec{x}_j', \xi_j)\}_{j=1}^m, \epsilon}$ to have norm $\|X\|=1$, $\poly(\log N, \log D)$ sparsity, at least $1/\poly(\log N, \log D)$ gap, and give roughly the same 1D representations:
\begin{equation}
	|\xi_j - \vec{x}'_j\cdot \vec{w}|\leq \epsilon, \quad \vec{w} = \underset{\|\vec{w}\|_2=1}{\argmax}~\vec{w}^TX^TX\vec{w}.
\end{equation}
We also assume that the guiding vector $\vec{g}$ has quality at least $\chi$ with respect to all data matrices $X$.
The specific way that $\mathcal{D}^0_{\{(\vec{x}_j', \xi_j)\}_{j=1}^m, \epsilon}$ samples the data matrix $X$ can be arbitrary as long as it satisfies the above requirements.
Formally, we define the dynamic dimension reduction task as follows.

\begin{tcolorbox}
\begin{task}[Dynamic dimension reduction task]
\label{task:dim-reduc-dynamic}
    Let $N, D, R, \tau$ be integers and $\epsilon\in (0, 1]$.
    The dynamic dimension reduction task is to estimate the underlying 1D representation $\xi_j$ of any $\poly(\log N, \log D)$-sparse test vector $\vec{x}_j'$ in a test set $\{\vec{x}'_j\}_{j=1}^m$ of any size $m$ to $2\epsilon$ error using data samples from any $\mathcal{D}_{\mathrm{DDR}}(\{(\vec{x}_j', \xi_j)\}_{j=1}^m, \epsilon)$ defined above with repetition number at most $R$ and refreshing time $\tau$, given a guiding vector $\vec{g}$ with quality at least $\chi\in [0, 1]$.
\end{task}
\end{tcolorbox}

In the following, we first state our main results on quantum advantage in dimension reduction.
Then we prove the quantum easiness and classical hardness in subsequent sections.

\subsubsection{Main results}

Our first result shows that given the same amount of samples, a small quantum machine can solve the dimension reduction task better than an exponentially larger classical machine.
Moreover, the better the quality of the guiding vector is, the larger the quantum advantage is.
This means that using a quantum machine, we can build a better model with exponentially smaller size, as long as there is a guiding vector slightly better than random guessing.
Note that the scaling $D^{(1-\zeta)\chi}$ is effectively $D^{\chi}$ since it holds for any constant $\zeta>0$.

\begin{tcolorbox}
\begin{theorem}[Quantum advantage in dimension reduction]
\label{thm:q-adv-dim-reduc}
    Consider the dimension reduction task with sample dimension $N$, feature dimension $D$, repetition number $R$, and guiding vector quality $\chi\in (0, 1]$ defined in \Cref{task:dim-reduc}.
    Using $\tilde O(RND^{1-\chi})$ samples, a quantum machine with $\poly(\log D)$ size can solve it with $1/\poly(\log(D))$ error and high success probability, while any classical machine with $o(D^{(1-\zeta)\chi})$ size for any constant $\zeta>0$ cannot solve it with $1/\poly(\log(D))$ error and success probability more than $0.67$.
    Moreover, the data processing time per sample of the quantum machine is $\poly(\log D)$.
\end{theorem}
\end{tcolorbox}

The second result shows that if the size of the classical machine is slightly smaller than $o(D^{2\chi-1})$, it would need super-polynomially more samples than a small quantum machine to solve the dynamic dimension reduction task.
This separation starts as soon as we have a guiding state of quality $\chi>1/2$ (i.e., $\vec{g}\cdot \vec{w}\geq \tilde{\Omega}(1/D^{1/4})$).
In the context of batch processing, this means that quantum machines can solve the task with a few batches of training data, whereas sub-exponential size classical machines require super-polynomially many batches.

\begin{tcolorbox}
\begin{theorem}[Quantum advantage in dynamic dimension reduction]
\label{thm:q-adv-dim-reduc-dynamic}
    Consider the dynamic dimension reduction task with sample dimension $N$, feature dimension $D$, repetition number $R$, guiding vector quality $\chi\in (1/2, 1]$, and sufficient refreshing time $\tau=\tilde \Theta(RND^{1-\chi})$ defined in \Cref{task:dim-reduc-dynamic}.
    A quantum machine with $\poly(\log D)$ size can use $\tilde O(RN D^{1-\chi})$ samples to solve it with $1/\poly(\log(D))$ error and high success probability, while any classical machine with $o(D^{(1-\zeta)(2\chi-1)})$ size for any constant $\zeta>0$ that solves it with $1/\poly(\log(D))$ error and probability at least $2/3$ must collect at least $R N^{\omega(1)}$ samples.
    Moreover, the data processing time per sample of the quantum machine is $\poly(\log D)$.
\end{theorem}
\end{tcolorbox}

Together, \Cref{thm:q-adv-dim-reduc,thm:q-adv-dim-reduc-dynamic} establish unconditional and exponential quantum advantages in the foundational task of dimension reduction.

\subsubsection{Quantum algorithm}

Here, we prove the quantum algorithm parts of \Cref{thm:q-adv-dim-reduc,thm:q-adv-dim-reduc-dynamic}.
We construct an algorithm that performs high-dimensional PCA on a small quantum computer.
We do so by combining quantum oracle sketching (\Cref{thm:q-oracle-sketch-corr}), quantum state sketching (\Cref{thm:q-state-sketch}), and a ground state preparation algorithm \cite{lin2020near}.
We again use the interferometric classical shadow developed in the last section (\Cref{lem:interf-classical-shadow}) to readout the principal component $\ket{w}$ into a classical representation, which is used for predicting the 1D representations of all possible test vectors.

\begin{tcolorbox}
\begin{theorem}[Dimension reduction with quantum oracle sketching]
\label{thm:dim-reduc-upper}
    Let $N, D$ be large integers, $\epsilon, \delta \in (0, 1)$, and $\chi \in [0, 1]$ be the quality parameter.
    Let $X = (\vec{x}_1, \ldots, \vec{x}_N)\in \mathbb{R}^{N\times D}$ be a dataset matrix that is $s$-sparse with norm $\|X\|= 1$ and gap $\Delta = \lambda_{\max}(X^TX)-\lambda_{\mathrm{sec}}(X^TX)$.
    Consider a set of $s$-sparse test vectors $\{\vec{x}'_j\}_{j=1}^m, \|\vec{x}'_j\|_\infty\leq 1$, of any size $m$.
    There exists a quantum algorithm that can output an $\epsilon$-approximate estimate $\hat{\xi}_j$ of the 1D representation
    \begin{equation}
        |\hat{\xi}_j - \xi(\vec{x}'_j)|\leq \epsilon, \quad \xi(\vec{x}'_j) = \vec{x}'_j\cdot \vec{w}, \quad \vec{w} = \underset{\|\vec{w}\|_2=1}{\argmax}~\vec{w}^TX^TX\vec{w},
    \end{equation}
    for any $j\in [m]$ with probability at least $1-\delta$ using 
    \begin{equation}
        S = \tilde{O}\lr{b + S_{\vec{g}} + \frac{s}{\epsilon^2}\log^{2.5}(D)(\min\{s\log(D), \log(m)\}+\log^{2.5}(1/\delta))+\log^{2.5}(1/\Delta)}
    \end{equation}
    qubits and
    \begin{equation}
        M = \tilde{O}\lr{\frac{RND^{1-\chi}s^{7}}{\Delta^2\epsilon^4\delta}\min\{s^2\log^2(D), \log^2(m)\}}.
    \end{equation}
    samples from the data generation process $\mathcal{D}_{\mathrm{DR}}(X)$ with repetition number $R$, given a guiding vector $\vec{g}$ that can be efficiently prepared in $S_{\vec{g}}$ space and has overlap $\vec{g}\cdot \vec{w}\geq \tilde{\Omega}(1/D^{(1-\chi)/2})$.
    In particular, we have
    \begin{equation}
    	S = \poly(\log D), \quad M = \tilde{O}(RND^{1-\chi}),
    \end{equation}
    when $s, \Delta^{-1}, b, \epsilon^{-1}, \delta^{-1}, S_{\vec{g}}\leq \poly(\log N, \log D)$.
    The data processing time per sample is $\poly(\log D)$.
\end{theorem}
\end{tcolorbox}

We note that \Cref{thm:dim-reduc-upper} immediately implies the quantum algorithm part of \Cref{thm:q-adv-dim-reduc} by definition of the dimension reduction task.
It also proves the quantum algorithm part of \Cref{thm:q-adv-dim-reduc-dynamic} by taking $\tau = \tilde{O}(RND^{1-\chi})$ satisfying $\tau\geq M = \tilde{O}(RND^{1-\chi})$ and noting that
\begin{equation*}
    |\hat{\xi}_j - \xi_j| \leq |\hat{\xi}_j - \xi(\vec{x}'_j)| + |\xi(\vec{x}'_j) - \xi_j| \leq \epsilon+\epsilon=2\epsilon
\end{equation*}
with probability at least $1-\delta$, as required.

We use the following quantum ground state preparation algorithm as the backbone of our quantum algorithm and instantiate the oracle queries with quantum oracle sketching.
We note that for coherent usage, the failure probability originally stated in \cite{lin2020near} is absorbed into the error parameter $\epsilon$ by replacing standard amplitude amplification with fixed-point amplitude amplification.
We bound the error in $2$-norm to avoid ambiguity in the global phase.

\begin{lemma}[Quantum ground state preparation {\cite[Corollary 9]{lin2020near}}]
\label{lem:q-ground-state-prep}
    Let $D$ be a large integer. 
    Let $H\in \mathbb{R}^{D\times D}$ be a real, symmetric matrix with the smallest eigenvalue $\lambda_1$ and corresponding unit eigenvector $\vec{w} \in \mathbb{R}^D$ and second smallest eigenvalue $\lambda_2$ such that $\lambda_2-\lambda_1\geq \Delta$.
    Assume that a number $\mu$ satisfying
    $\lambda_1
    \leq
    \mu
    \leq
    \lambda_1+{\Delta}/{4}$
    is known classically.
    Let $\vec{g}\in \mathbb{R}^D$ be a unit vector such that $\vec{g}\cdot \vec{w}\geq q$.
    There exists a quantum algorithm that applies a unitary $V$ such that
    \begin{equation}
        \|V|0^S\rangle - |0^{S-\log(D)}\rangle \ket{w}\|_2\leq \epsilon,
    \end{equation}
    using $S=O(\log(D) + \log(1/q))$ qubits, $\tilde{O}\lr{\frac{1}{q\Delta}\log(1/\epsilon)}$ gates, and $\tilde{O}\lr{\frac{1}{q\Delta}\log(1/\epsilon)}$ queries to the block encoding of $H$ and the state preparation unitary of $\ket{g}$ and their inverse and controlled versions.
\end{lemma}

Note that in the algorithm described in \cite{lin2020near}, for a fixed value of $\mu$, the filtering and fixed-point
amplitude-amplification phases define a fixed coherent unitary.
Its inverse and controlled versions are obtained by reversing or controlling the same circuit.

Recall that in \Cref{lem:block-encoding}, we have shown that $\tilde{O}(RNs^5)$ samples suffice to implement the block encoding of an $s$-sparse matrix.
In addition, \Cref{thm:q-state-sketch} shows that we can use $\tilde{O}(RN)$ samples to prepare the quantum state of any vector.
We combine \Cref{lem:q-ground-state-prep} with \Cref{lem:block-encoding}, \Cref{thm:q-state-sketch}, and interferometric classical shadow (\Cref{lem:interf-classical-shadow}) to prove \Cref{thm:dim-reduc-upper}.

\begin{proof}[Proof of \Cref{thm:dim-reduc-upper}]
Let $n = \ceil{\log_2(\max(N, D))}\leq \min\{O(\log(sD)), O(\log(sN))\}$ and embed $X$ into $2^n$ dimension to apply quantum oracle and state sketching.
We define $H = -X^TX\in \mathbb{R}^{D\times D}$, which is real and symmetric.
The smallest eigenvalue of $H$ corresponds to the largest singular value of $X$, and therefore the ground state of $H$ is indeed the principal component $\vec{w}$ of $X$.
Since $\|X\|=1$, the smallest eigenvalue of $H$ is exactly $\lambda_1(H)=-1$.
Therefore, the required ground-energy upper bound $\mu$ in \Cref{lem:q-ground-state-prep} can be chosen as $\mu=-1$.
The gap between the smallest and second smallest eigenvalues of $H$ is $\Delta$ by definition.
The block encoding of $H$ can be implemented with two queries to the block encoding of $X$ and its inverse.
We use $q = \vec{g}\cdot \vec{w} \geq \tilde{\Omega}(1/D^{(1-\chi)/2})$ to denote the overlap between the guiding state $\vec{g}$ and the principal component $\vec{w}$.

\Cref{lem:q-ground-state-prep} states that there is a quantum ground state preparation algorithm, which is a unitary $V$ that prepares the state $\ket{w}$ with $\epsilon'/2$ error using $O(n+\log(1/q)) = O(n + (1-\chi)\log(D))$ qubits and 
\begin{equation}
    Q_0=\tilde O\lr{\frac{1}{q\Delta}\log(1/\epsilon')} = \tilde O\lr{\frac{D^{(1-\chi)/2}}{\Delta}\log(1/\epsilon')}
\end{equation}
queries to the block encoding of $X$ and the state preparation unitary of $\ket{g}$ and their inverse and controlled versions.
By replacing all gates and queries in $V$ with their controlled versions, we obtain a controlled state preparation unitary $cV$ that prepares $\ket{w}$ with error $\epsilon'/2$.

Similar to the proof of \Cref{thm:linear-sys-upper,thm:bin-classify-upper}, our proof of \Cref{thm:dim-reduc-upper} proceeds in two steps.
We first use this quantum ground state preparation algorithm to construct a query algorithm that can estimate the 1D representation $\xi(\vec{x}'_j)$ to $\epsilon$ error for any $j\in [m]$ with probability at least $1-\delta/2$.
Then, we instantiate the queries with quantum oracle and state sketching.
The instantiation error is chosen to be $\delta/2$ such that the total variation distance error on the final estimate is $\delta/2$.
This immediately implies that the final prediction is $\epsilon$-accurate with probability at least $1-\delta/2-\delta/2 = 1-\delta$.

As the first step, we construct the query algorithm that estimates the 1D representation.
The quantum ground state preparation algorithm gives us a controlled state preparation unitary $cV$ that prepares $\ket{w}$ to $\epsilon'/2$ using $O(n+(1-\chi)\log(D))$ qubits and $Q_0$ queries to $X$ and $\ket{g}$.
To readout the weight vector well enough that we can predict any $s$-sparse test vector $\vec{x}'_j$, we imagine that we have an $\epsilon_0$-covering net $\mathcal{N}$ over the set
\begin{equation}
    \mathcal{X}_{\mathrm{test}} = \{\vec{u}\in \mathbb{R}^{D}: \|\vec{u}\|_0\leq s, \|\vec{u}\|_2 = 1\}
\end{equation}
in $\|\cdot \|_2$.
Note that this covering net is only a analysis tool and it is not used in the algorithm.
The covering net has size
\begin{equation}
    |\mathcal{N}|\leq \binom{D}{s}\left( \frac{3}{\epsilon_0} \right)^{O(s)} \leq 2^{O(s\log(D/\epsilon_0))}
\end{equation}
We run interferometric classical shadow (\Cref{lem:interf-classical-shadow}) using the controlled state preparation unitary $cV$ with the test states $\ket{x'}$, $\vec{x}'\in \mathcal{N}$, error $\epsilon_0$, and success probability $\delta/2$.
This consumes
\begin{equation}
    O\lr{\frac{\log(|\mathcal{N}|/\delta)}{\epsilon_0^2}} = O\lr{\frac{s\log(D/\epsilon_0) + \log(1/\delta)}{\epsilon_0^2}}
\end{equation}
queries to $cV$ and guarantees that the prediction of the overlap is $\epsilon_0$ accurate on the $\epsilon_0$-covering net $\mathcal{N}$ of $\mathcal{X}_{\mathrm{test}}$.
For every $\vec{u}\in\mathcal{X}_{\mathrm{test}}$, fix $\widetilde{\vec{u}}\in\mathcal{N}$ satisfying $\|\vec{u}-\widetilde{\vec{u}}\|_2\leq\epsilon_0$, and use the classical-shadow estimate for $\widetilde{\vec{u}}$ as the estimate for $\vec{u}$. 
Note that the net need not be explicitly enumerated or stored because $\widetilde{\vec{u}}$ can be computed on demand by rounding the nonzero coordinates of the sparse test vector.
On the simultaneous-success event,
\begin{equation}
\left|
\hat{o}(\widetilde{\vec{u}})
-
\vec{u}\cdot\vec{w}
\right|
\leq
\left|
\hat{o}(\widetilde{\vec{u}})
-
\widetilde{\vec{u}}\cdot\vec{w}
\right|
+
\left|
(\widetilde{\vec{u}}-\vec{u})\cdot\vec{w}
\right|\leq
\epsilon_0+
\|\widetilde{\vec{u}}-\vec{u}\|_2
\leq
2\epsilon_0=E_0.
\end{equation}
We choose $\epsilon_0$ such that $\epsilon'/2=E_0=O(\epsilon_0)$.
For any nonzero sparse test vector $\vec{x}'_j$, we apply the overlap estimator to $\vec{x}'_j/\|\vec{x}'_j\|_2$ and output
\begin{equation}
    \hat{\xi}_j=\|\vec{x}'_j\|_2\cdot \hat{o}(\vec{x}'_j/\|\vec{x}'_j\|_2).
\end{equation}
The zero vector is handled separately by outputting zero.
Combining the error of the state preparation unitary, we know that the produced estimator $\hat{o}(\vec{x}'_j/\|\vec{x}'_j\|_2)$ is an $E_0+\epsilon'/2 = \epsilon'$ accurate estimate of $\expval{x'|w}$ for all $\vec{x}'_j/\|\vec{x}'_j\|_2 = \ket{x_j'} \in \mathcal{X}_{\mathrm{test}}$.
Meanwhile, if the size of test vectors $m$ is small, then $O(\log(m/\delta)/\epsilon'^2)$ queries suffice.
Hence, with 
\begin{equation}
    O\lr{\frac{\min \left\{ s\log(D/\epsilon') \right., \log(m)\}+\log(1/\delta)}{\epsilon'^2}}
\end{equation}
queries to $cV$, we can produce an estimator $\hat{o}(\vec{x}'_j/\|\vec{x}'_j\|_2)$ satisfying
\begin{equation}
	|\hat{o}(\vec{x}'_j/\|\vec{x}'_j\|_2) - \expval{x_j'|w}|\leq \epsilon'
\end{equation}
for any $j\in [m]$ with probability at least $1-\delta/2$.

Furthermore, the 1D representation is given by
\begin{equation}
	\xi(\vec{x}_j') = \vec{x}_j' \cdot \vec{w} = \|\vec{x}_j'\|_2\cdot \expval{x_j'|w}.
\end{equation}
Therefore,
\begin{equation}
	|\hat{\xi}_j - \xi(\vec{x}_j')| = \|\vec{x}_j'\|_2 \cdot |\hat{o}(\vec{x}'_j/\|\vec{x}'_j\|_2) - \expval{x_j'|w}|\leq \|\vec{x}_j'\|_2\epsilon'\leq \sqrt{s}\epsilon' = \epsilon,
\end{equation}
where we have chosen $\epsilon' = \epsilon/\sqrt{s}$ and used the fact that $\vec{x}'_j$ is $s$-sparse and hence $\|\vec{x}'_j\|_2\leq \sqrt{s}\|\vec{x}'_j\|_\infty\leq \sqrt{s}$.
This gives us the desired quantum query algorithm that makes 
\begin{equation}
\begin{split}
    Q &= Q_0\cdot O\lr{\frac{\min\{s\log(D/\epsilon'), \log(m)\}+\log(1/\delta)}{\epsilon'^2}} \\
    &= \tilde{O}\lr{\frac{sD^{(1-\chi)/2}}{\Delta \epsilon^2}(\min\{s\log(D), \log(m)\}+\log(1/\delta))}
\end{split}
\end{equation}
queries to $X$ and can output the $\epsilon$-error prediction $\hat{\xi}_j, \forall j\in [m]$ (a classical random variable) with probability at least $1-\delta/2$.
Moreover, the space complexity is
\begin{equation}
\begin{split}
    S_0 &= O\lr{n+(1-\chi)\log(D)} + \log^2(D)\cdot \tilde{O}\lr{\frac{\min\{s^2\log(D), s\log(m)\}+s\log(1/\delta)}{\epsilon^2}} + O(\log^2(D)) \\
    &\leq \tilde{O}\lr{\frac{s}{\epsilon^2}\log^2(D)(\min\{s\log(D), \log(m)\}+\log(1/\delta))},
\end{split}
\end{equation}
where the first term comes from the ground state preparation algorithm, the second comes from the classical shadow data, and the third one comes from classical simulation of Clifford circuits.

The second step is to instantiate the queries to $X$ in the query algorithm using quantum oracle sketching.
We first replace all queries to the block encoding of $X$ and its inverse and controlled versions to their $\epsilon_1$-approximate versions in \Cref{lem:block-encoding}.
This incurs an error of at most $E_1 = Q\epsilon_1 = \delta/4$, where we set $\epsilon_1 = \delta/(4Q)$.

Next, we replace the queries to $\epsilon_1$-approximate versions of block encoding of $X$ and its inverse and controlled versions by the random unitary channel that we build from samples in \Cref{lem:block-encoding}.
This incurs an additional error of at most $E_2 = Q\epsilon_1 = \delta/4$, and uses $O(n+b+\log^{2.5}(s/\epsilon_1)) = O(n+b+\log^{2.5}(sQ/\delta))$ qubits and 
\begin{equation}
\begin{split}
    M &= Q\cdot O\lr{\frac{R2^nn^2s^5\log^4(ns/\epsilon_1)}{\epsilon_1}} \leq O\lr{\frac{R2^nn^2s^5Q^2\log^4(\frac{nsQ}{\delta})}{\delta}} \\
    &\leq \tilde{O}\lr{\frac{RND^{(1-\chi)}s^{7}}{\Delta^2\epsilon^4\delta}\min\{s^2\log^2(D), \log^2(m)\}}
\end{split}
\end{equation}
samples from the data generation process.
Note that here, given a uniformly random row sample, we independently sample one of its padded $s$ slots uniformly. 
Hence the resulting real-or-dummy matrix-element slot is uniform over the $Ns$ padded slots, so we may invoke the padded-support version of \Cref{lem:block-encoding}.
The fresh slot label is retained in the transformed sample, so this subsampling changes the repetition number by at most a constant factor, which is absorbed into $R$.

According to \Cref{lem:error-accumulation-time-varying}, the total error in instantiating the query algorithm is bounded by
\begin{equation}
    E_1+E_2 = 2\cdot \delta/4 = \delta/2.
\end{equation}
This means that the output of the query-instantiated quantum algorithm, which is a classical random variable must be correct with probability at least $1-\delta/2-\frac{1}{2}\delta/2\geq 1-\delta$.
The total number of qubits used is
\begin{equation}
\begin{split}
    &S_0 + S_{\vec{g}} + O(n+b+\log^{2.5}(sQ/\delta))\\
    &\leq \tilde{O}\lr{b + S_{\vec{g}} + \frac{s}{\epsilon^2}\log^{2.5}(D)(\min\{s\log(D), \log(m)\}+\log^{2.5}(1/\delta))+\log^{2.5}(1/\Delta)}.
\end{split}
\end{equation}
The total number of samples is
\begin{equation}
    M=\tilde{O}\lr{\frac{RND^{(1-\chi)}s^{7}}{\Delta^2\epsilon^4\delta}\min\{s^2\log^2(D), \log^2(m)\}}.
\end{equation}
This completes the proof of \Cref{thm:dim-reduc-upper}.
\end{proof}

\subsubsection{Classical hardness}

In this section, we prove the two classical hardness results in \Cref{thm:q-adv-dim-reduc,thm:q-adv-dim-reduc-dynamic}.
We will construct a (dynamic) dimension reduction task with $N=D$.
Hence, for simplicity, we will use $N$ alone throughout this section.
We prove the classical hardness results by constructing a specific dimension reduction task using \Cref{lem:embed-qc-into-pca}.
Solving it amounts to solving the (dynamic) Noisy Oracle Property Estimation (NOPE) task defined in \Cref{sec:cl-hard} (\Cref{task:nope,task:dynamic-nope}), whose classical hardness we have already proved in \Cref{lem:classical-lower-bound-dist} and \Cref{thm:classical-lower-bound-superpoly-sample}.

This gives us two classical hardness results.
The first result (\Cref{thm:classical-hardness-dim-reduc}) follows from the distributional sample-space lower bound of NOPE (\Cref{lem:classical-lower-bound-dist}) and shows that any classical learning algorithm that wants to achieve success probability $0.67$ in dimension reduction with the same number of samples quantum algorithms need must have $\Omega(D^{(1-\zeta)\chi})$ size for any constant $\zeta>0$.
The second result (\Cref{thm:classical-hardness-dim-reduc-dynamic}) follows from \Cref{thm:classical-lower-bound-superpoly-sample} and shows that when the dimension reduction task is dynamic, any classical learning algorithm will need a number of samples super-polynomial in $N$ if it has size $o(D^{(1-\zeta)(2\chi-1)})$ for any constant $\zeta>0$.
Note that it suffices to consider $0<\zeta<1$, since for $\zeta\geq1$ the claimed space lower bounds are vacuous.
The dynamic result follows a similar reasoning as the linear system and binary classification case.
We therefore prove it first.
The non-dynamic result requires a slight modification, which we will detail at the end of this section.

The general idea of proving these results is to embed the quantum circuit that solves dynamic NOPE from \Cref{lem:encode-q-circ} into a dimension reduction task via \Cref{lem:embed-qc-into-pca}.
The 1D representation $\xi(\vec{x})$ of a fixed, sparse test vector $\vec{x}$ encodes the oracle property $B\in\bit$ that we want to estimate in (dynamic) NOPE.
Then we show that given any classical algorithm that solves the dimension reduction task, we can use it to construct a classical algorithm that solves (dynamic) NOPE.
In the (dynamic) NOPE task, we take the oracle property function $f$ to be $K$-Forrelation (\Cref{def:forrelation}) and the noisy encoding function $g$ to be the inner product (\Cref{def:inner-prod}).

In the following, we construct the data generation process $\mathcal{D}^{N,K, R}_{\mathrm{DR}}(B), B\in\bit$ for dimension reduction, where $N$ is both the sample dimension and the feature dimension (i.e., $D=N$), $K=\Theta(1)$ is the constant in the Forrelation that we will embed, $R$ is an integer that specifies the repetition number, and $B$ indicates whether the 1D representation of that fixed test vector has a large or small value (see \Cref{lem:embed-qc-into-pca}).
The goal is to solve for $B$.
It helps to compare this construction to the dynamic NOPE data generation process defined in \Cref{sec:cl-hard-bootstrap}.

Given $N$, we define another large integer $N'$ as follows.
Let $T = \Theta(\log^3 N' \log\log N')$ be the total number of two-qubit gates and diagonal gates given in \Cref{lem:encode-q-circ} and let $n' = \log N' + O(\log\log N')$ be the number of qubits in \Cref{lem:encode-q-circ}.
To properly embed that circuit into the $N$-dimensional training dataset, we define $N'$ such that $N = (T+1) 2^{n'+1} = N' \cdot \polylog N'$ as in \Cref{lem:embed-qc-into-pca}.
This implies $N' = N / \polylog N$.
The resulting data matrix $X$ have sparsity $s=O(1)$ and gap $\Delta \geq \Omega(1/T^3)=\tilde{\Omega}(1/\log^9 N')$ from \Cref{lem:embed-qc-into-pca}.
We fix the guiding vector to be the $\vec{g}$ in \Cref{lem:embed-qc-into-pca} that has overlap $\vec{g}\cdot \vec{w} = 1/\sqrt{T+1} = \Omega(1/\polylog(N))$, which means that it already has maximal quality $\chi=1$.
Hence this guiding vector has quality at least $\chi$ for any $\chi\in [0, 1]$.
Let $L=\ceil{\log^2 (KN')}\geq 5$ be the number of independent oracle instances that we have in dynamic NOPE.

Now we define the data generation process
\begin{equation}
    \mathcal{D}^{N, K, R}_{\mathrm{DR}}(B) = (\mathcal{D}^0_B\to \mathcal{D}^1_{X}\to^{\times T_1} \mathcal{D}^2_{(\alpha, \beta)}\to^{\times T_2} z=(i, \vec{x}_i) \to^{\times T_3} z),
\end{equation}
where $T_3=\Theta(R)$ is chosen such that the repetition number of the resulting process is at most $R$, $T_2 = KN'$, $T_1 = \ceil{M_Q/(T_2T_3)} = (KN')^{1-\chi}\polylog(KN')$, $M_Q=RND^{1-\chi}\polylog(N)$ is the number of samples quantum machines need in \Cref{thm:dim-reduc-upper}, $X\in \mathbb{R}^{N\times N}$ is an $N$-dimensional, symmetric, $O(1)$-sparse feature matrix with gap $\Delta\geq \tilde{\Omega}(1/\log^9 N)$, $\alpha\in\bit, \beta\in[L] $ label which part of the matrix $X$ that we are currently collecting data from, and $i\in [N]$ labels the $i$-th training data point $\vec{x}_i$ (the $i$-th row of the matrix $X$).

The data are sampled in the following way that resembles dynamic NOPE in \Cref{sec:cl-hard-bootstrap}.
$\mathcal{D}^0_{B}$ samples a length-$L$ bitstring $\gamma$ with parity 
\begin{equation}
    \bigoplus_{j=1}^{L}\gamma_{j}=B
\end{equation} 
uniformly random.
For each $j\in [L]$, we sample a random oracle $o_j\sim p_{\gamma_j}$, where $p_0, p_1$ are the distributions of Forrelation defined in \Cref{lem:forrelation-query-separation}.
Then we sample a noisy encoding pair $(Y^{(0, j)}, Y^{(1, j)})\sim \unif((g^{KN'})^{-1}(o_j))$ using the inner product noisy encoding function $g$ defined in \Cref{def:inner-prod}.

Next, note that $(Y^{(0,j)}, Y^{(1,j)})_{j=1}^{L}$ specifies an $n'$-qubit quantum circuit $C$ with $T=O(\log^3N' \log\log N')$ gates via \Cref{lem:encode-q-circ}, and the quantum circuit $C$ gives a data matrix $X$ with dimension $(T+1) 2^{n'+1}=N$ via \Cref{lem:embed-qc-into-pca}.
Here, $X$ is indeed $O(1)$-sparse with gap $\Delta\geq \tilde{\Omega}(1/\log^9 N)$.
Now we define $\mathcal{D}^1_{X}$.
We first sample a uniformly coordinate $\beta\sim \unif([L])$ and a random bit $\alpha\sim\mathrm{Bern}(1/2)$ as in dynamic NOPE.
Then we pick out $Y^{(\alpha, \beta)}$ and use it to generate the data samples.

In particular, we define $\mathcal{D}^2_{(\alpha, \beta)}$ in the following way.
We first sample a random row of the matrix $X$ as follows.
Note that the linear space in which the matrix $X$ lives factorizes as $\ket{t}\ket{\psi}$ where $\ket{t}$ is the clock register and $\ket{\psi}$ is the $(n'+1)$-qubit register that the realified quantum circuit runs on.
We sample a clock time $t\sim \unif(\{0, \ldots, T\})$ and the matrix is reduced to a specific gate in the $(n'+1)$-qubit subspace (either a fixed two qubit gate or a diagonal gate that depends on $Y^{(\alpha, \beta)}$).
The remaining subspace further factorizes into $\ket{x, \alpha, \beta, k}$ and the rest of the working qubits.
We sample a random basis of this $(n'+1)$-qubit subspace by plug in the specific $(\alpha, \beta)$ that we have already sampled, sample $x \sim \unif([KN']), k\sim \unif(\{0, \ldots, b\})$, and sample a computational basis of the rest of the working qubits uniformly random.
This together specifies and thus samples a row $i$ of the matrix $X$.
Note that the marginal distribution of $i$ is uniform over $[N]$ (because $\alpha, \beta$ are sampled uniformly), though there are correlations between consecutive samples of $i$ since they share the same set of $(\alpha, \beta)$.

Note that by construction of the matrix $X$ as in \Cref{lem:embed-qc-into-pca}, the picked out row vector $\vec{x}_i$ is $O(1)$ sparse, and has components that is the real or imaginary part of either $1$, from the identity matrices in \Cref{lem:embed-qc-into-pca}, or a matrix element of a fixed two qubit gates, a value that only depends on the data via $(Y^{(\alpha,\beta)}_x)_k$, via the known embedding, permutations, and realification.
This gives us the sample $z=(i, \vec{x}_i)$ and we repeat this sample $T_3$ times, completing the data generation process.

This data generation process $\mathcal{D}_{\mathrm{DR}}^{N, K, R}$ is a valid data generation process of dynamic dimension reduction.
It produces a random row sample uniformly distributed over the rows of $X$.
The test vector is a single fixed $O(1)$-sparse vector $\vec{x}'$ specified in \Cref{lem:embed-qc-into-pca}, which satisfies
\begin{equation}
    \xi(\vec{x}')
    =
    \vec{x}'\cdot\vec{w}
    =
    \frac{1}{\sqrt{T+1}}
    \Re\left[
        \bra{0}C\ket{0}
    \right],
    \quad
    \vec{w}
    =
    \underset{\|\vec{w}\|_2=1}{\argmax}
    ~\vec{w}^TX^TX\vec{w}.
\end{equation}
By \Cref{lem:encode-q-circ}, when $B=0$,
$\Re\left[
    \bra{0}C\ket{0}
\right] = \frac{2-\|C\ket{0}-\ket{0}\|_2^2}{2}
\geq
1-\frac{1}{2}(1/20)^2
>
0.99$,
whereas when $B=1$,
$\left|
    \Re\left[
        \bra{0}C\ket{0}
    \right]
\right|
\leq
\frac{1}{20}$.
Therefore,
\begin{equation}
    B=0
    \quad\Longrightarrow\quad
    \xi(\vec{x}')
    >
    \frac{0.99}{\sqrt{T+1}},
    \qquad
    B=1
    \quad\Longrightarrow\quad
    \left|\xi(\vec{x}')\right|
    \leq
    \frac{0.05}{\sqrt{T+1}}.
\end{equation}
We define the underlying value in the dynamic task as 
\begin{equation}
	\xi_B = \frac{1-B}{\sqrt{T+1}},
\end{equation}
such that $|\xi_B-\xi(\vec{x}')|\leq 0.05/\sqrt{T+1}$.
The quality of the guiding vector $\vec{g}$ given in \Cref{lem:embed-qc-into-pca} is maximal ($\chi=1$) and hence is indeed at least $\chi$ for any $\chi\in [0, 1]$.
The repetition number is $O(T_3)$ because, conditioned on the current higher-level situation, only the group containing the first sample and the remaining draws in the same $(\alpha, \beta)$ block can make an excess contribution. 
The former contributes at most $T_3$, while the latter contributes $T_3(T_2-1)O(1/T_2)=O(T_3)$, since every possible unrepeated sample has probability $O(1/T_2)$.
Thus $R_{\mathcal D}=O(T_3)$.
We choose $T_3=\Theta(R)$ so that $R_{\mathcal D}\leq R$ and suppress this constant rescaling below.
The refreshing time is $\tau = T_1T_2T_3 = O(M_Q)=\tilde O(RND^{1-\chi})$, satisfying the requirements in \Cref{thm:q-adv-dim-reduc,thm:q-adv-dim-reduc-dynamic}.

This data generation process for dimension reduction $\mathcal{D}^{N, K, R}_{\mathrm{DR}}(B)$ is designed to reduce to the dynamic NOPE data $\mathcal{D}^{KN', T_1}_{g, f}(B)$ in \Cref{sec:cl-hard-bootstrap} via \Cref{lem:encode-q-circ,lem:embed-qc-into-pca}.
Using this data generation process, we prove the following result.

\begin{tcolorbox}
\begin{theorem}[Classical hardness of dynamic dimension reduction]
\label{thm:classical-hardness-dim-reduc-dynamic}
    Let $\zeta>0$ be any constant.
    Let $N, D$ be the sample and feature dimension of a dimension reduction task and $R, \tau=\tilde{\Theta}(RND^{1-\chi})$ be its repetition number and refreshing time.
    Given a guiding vector of constant quality at least $\chi>1/2$, any randomized classical learning algorithm that solves the task with some error parameter $\epsilon=\tilde \Theta(1/\log^{1.5}(N))$ and success probability at least $2/3$ must have sample complexity
    \begin{equation}
        M\geq RN^{\omega(1)}
    \end{equation}
    if its space complexity
    \begin{equation}
        S \leq o(D^{(1-\zeta)(2\chi-1)}).
    \end{equation}
\end{theorem}
\end{tcolorbox}

\Cref{thm:classical-hardness-dim-reduc-dynamic} directly implies the classical hardness part of \Cref{thm:q-adv-dim-reduc-dynamic}.
Together with the quantum algorithm result \Cref{thm:dim-reduc-upper}, this completes the proof of the quantum advantage claim in \Cref{thm:q-adv-dim-reduc-dynamic}.
However, to obtain the claim in \Cref{thm:q-adv-dim-reduc} where $2\chi-1$ is improved to $\chi$, we need to design a different data generation process based on the distributional sample-space lower bound (\Cref{lem:classical-lower-bound-dist}) that we will detail later.

\begin{proof}[Proof of \Cref{thm:classical-hardness-dim-reduc-dynamic}]
    Recall that $N=D$ in the task that we construct.
    For simplicity, we will use $N$ throughout the proof.
    We prove \Cref{thm:classical-hardness-dim-reduc-dynamic} by showing that given any classical learning algorithm that can estimate the 1D representation $\xi(\vec{x}')$ of the test vector $\vec{x}'$, we can use it to construct an algorithm that decides $B$ from $\mathcal{D}^{KN', T_1}_{g, f}(B)$, which we have proved to be hard in \Cref{thm:classical-lower-bound-superpoly-sample}.

    First note that from
    \Cref{lem:embed-qc-into-pca,lem:encode-q-circ}, the 1D
    representation satisfies
    \begin{equation}
        B=0
        \quad\Longrightarrow\quad
        \xi(\vec{x}')
        >
        \frac{0.99}{\sqrt{T+1}},
        \qquad
        B=1
        \quad\Longrightarrow\quad
        \left|\xi(\vec{x}')\right|
        \leq
        \frac{0.05}{\sqrt{T+1}}.
    \end{equation}
    We set
    \begin{equation}
        \epsilon
        =
        \frac{1}{16\sqrt{T+1}}
        =
        \widetilde{\Theta}\left(
            \frac{1}{\log^{3/2}N}
        \right).
    \end{equation}
    Moreover,
    \begin{equation}
        |\xi_B-\xi(\vec{x}')|
        \leq
        \frac{0.05}{\sqrt{T+1}}
        <
        \epsilon,
    \end{equation}
    so the constructed process satisfies the promise of the dynamic task.
    Furthermore, any solver for the dynamic task produces an estimate $\hat{\xi}$ satisfying
    \begin{equation}
        |\hat{\xi}-\xi_B|
        \leq
        2\epsilon
        =
        \frac{1}{8\sqrt{T+1}}.
    \end{equation}
    Therefore, thresholding $\hat{\xi}$ at $\frac{1}{2\sqrt{T+1}}$ determines $B$.

    We choose $K=\ceil{\frac{1.001}{\zeta(2\chi-1)}}$ such that 
    \begin{equation}
        \frac{N'^{1-1/K}}{\polylog N'} \geq \frac{N^{1-\zeta(2\chi-1)+\zeta(2\chi-1)/1001}}{\polylog N}\geq \Omega(N^{1-\zeta(2\chi-1)}).
    \end{equation}
    For the sake of contradiction, suppose we have a randomized classical learning algorithm $\mathcal{L}$ with space complexity
    \begin{equation}
        S \leq o(N^{(1-\zeta)(2\chi-1)}) = o(N^{(2\chi-2) +1- \zeta(2\chi-1)}) \leq o\left(N^{2\chi-2}\cdot \frac{N'^{1-1/K}}{\polylog N'}\right) = o\left(\frac{N'^{2\chi-1-1/K}}{\polylog N'}\right)
    \end{equation}
    and sample complexity $M$ that given a sequence of data samples drawn from $\mathcal{D}^{N, K, R}_{\mathrm{DR}}(B)$, solves the dynamic dimension-reduction task with parameter $\epsilon$ and probability $p_{\mathrm{succ}}$.
    In the following, we design a classical learning algorithm $\mathcal{L}'$ that decides $B$ using data from $\mathcal{D}^{KN', T_1}_{g, f}(B)$.

    The first step of $\mathcal{L}'$ is to generate data samples that look like $\mathcal{D}^{N, K, R}_{\mathrm{DR}}(B)$ from $\mathcal{D}^{KN', T_1}_{g, f}(B)$.
    For every group of $T_3$ repeated target samples, $\mathcal{L}'$ reads one source sample $z=(x, Y_x^{(\alpha,\beta)}, \alpha, \beta)$ from $\mathcal{D}^{KN',T_1}_{g,f}(B)$.
    It then follows exactly the same public row-sampling procedure as in the definition of $\mathcal{D}^{N,K,R}_{\mathrm{DR}}(B)$:
    it samples the clock basis from $\{0,\ldots,T\}$ and all remaining public basis coordinates uniformly.
    If all nonzero entries of the selected row are fixed, the value $Y_x^{(\alpha,\beta)}$ is ignored.
    Otherwise, the oracle-dependent entries are calculated using $(-1)^{(Y_x^{(\alpha,\beta)})_k}$ via the known embedding.
    The algorithm then repeats the resulting sample $z_{\mathrm{DR}}=(i,\vec{x}_i)$ exactly $T_3$ times.
    Note that this procedure generates data samples that exactly matches $\mathcal{D}^{N, K, R}_{\mathrm{DR}}(B)$ by construction.

    After sampling a training data point $z_{\mathrm{DR}}=(i, \vec{x}_i)$, we feed it into dimension reduction solver $\mathcal{L}$.
    We repeat this $M$ times so that $\mathcal{L}$ receives $M$ samples whose distribution matches that of $\mathcal{D}^{N, K, R}_{\mathrm{DR}}(B)$ and produces an estimate of the fixed underlying representation $\xi_B$, from which $\mathcal{L}'$ determines $B$ using the threshold above.
    We use this predicted bit of $\mathcal{L}$ as the final output of $\mathcal{L}'$.

    Note that since the data generation does not require knowledge of the previous data samples from $\mathcal{D}^{KN', T_1}_{g, f}(B)$, it can be performed online and thus the space complexity of $\mathcal{L}'$ is $S'=S$.
    Moreover, one source sample is used for every group of $T_3$ repeated target samples, so $M'\leq\lceil M/T_3\rceil\leq M/T_3+1$.
    The success probability of $\mathcal{L}'$ is $p_{\mathrm{succ}}$, the same as that of $\mathcal{L}$.

    Finally, we invoke \Cref{thm:classical-lower-bound-superpoly-sample}.
    Note that for the inner product $g$, we have $\eta=1/2$, $c = (865/\eta^2)\log(865/\eta^2) = (865\times 4)\log(865\times 4) \approx 40677.68$, and therefore the choice of $b=\ceil{40678\log(KN')}$ satisfies the requirement.
    For the Forrelation $f$ we use, \Cref{lem:forrelation-query-separation} implies that the $(1/3)$-error classical distributional query complexity is (using $K=\Theta(1)$)
    \begin{equation}
        Q_C \geq \Omega\left(\frac{N'^{1-1/K}}{\polylog N'}\right).
    \end{equation}
    Together with $T_1 = N'^{1-\chi} \polylog(KN')$, we have
    \begin{equation}
        S'=S \leq o\left(\frac{N'^{2\chi-1-1/K}}{\polylog N'}\right) \implies S'\leq o\lr{\frac{Q_C}{T_1^2 L}}
    \end{equation}
    satisfying the condition of \Cref{thm:classical-lower-bound-superpoly-sample}.
    Therefore, \Cref{thm:classical-lower-bound-superpoly-sample} implies that 
    \begin{equation}
        M\geq T_3(M'-1)\geq R(KN')^{\omega(1)}= RN^{\omega(1)},
    \end{equation}
    proving \Cref{thm:classical-hardness-dim-reduc-dynamic}.
    Together with the quantum algorithm result \Cref{thm:dim-reduc-upper}, this proves the quantum advantage claim in dynamic dimension reduction (\Cref{thm:q-adv-dim-reduc-dynamic}).
\end{proof}

Finally, we prove the following result that implies the classical hardness part of \Cref{thm:q-adv-dim-reduc}.

\begin{tcolorbox}
\begin{theorem}[Classical hardness of dimension reduction]
\label{thm:classical-hardness-dim-reduc}
    Let $\zeta>0$ be any constant.
    Let $N, D$ be the sample and feature dimension of a dimension reduction task and $R$ be its repetition number.
    Given a guiding vector of constant quality $\chi\in (0, 1]$, using $\tilde{O}(RND^{1-\chi})$ samples, any randomized classical learning algorithm with space complexity
    \begin{equation}
        S\leq o(D^{(1-\zeta)\chi})
    \end{equation}
    cannot solve the dimension reduction task with error $\epsilon=\tilde \Theta(1/\log^{1.5}(N))$ and success probability more than $0.67$.
\end{theorem}
\end{tcolorbox}

The origin of the difference between the $D^{\chi}$ exponent and the $D^{2\chi-1}$ exponent is as follows.
The ground state preparation algorithm has query complexity $Q\sim 1/(\vec{g}\cdot \vec{w})\sim D^{(1-\chi)/2}$.
Due to the quadratic slowdown of quantum oracle sketching (caused by the incoherent random sampling of classical data), we have sample complexity $M\sim NQ^2 \sim ND^{1-\chi}$.
The classical sample-space lower bound in \Cref{lem:classical-lower-bound-dist} then provides the classical space lower bound $S\gtrsim DQ_C/M \sim D\cdot N/(ND^{1-\chi}) \sim D^\chi$ as claimed.
This is the intuition behind \Cref{thm:classical-hardness-dim-reduc}.
As for the dynamic version (\Cref{thm:classical-hardness-dim-reduc-dynamic}), the learning XOR lemma (\Cref{lem:xor-lemma}) incurs an additional $T_1\sim M/N\sim D^{1-\chi}$ factor in the space lower bound due to interleaving problem instances.
This leads to a final space lower bound of $S\gtrsim D^{\chi}/T_1 \sim D^{\chi-(1-\chi)} = D^{2\chi-1}$ as in \Cref{thm:classical-hardness-dim-reduc-dynamic}.

\begin{proof}[Proof of \Cref{thm:classical-hardness-dim-reduc}]
    We modify the dynamic data generation process for dimension reduction $\mathcal{D}_{\mathrm{DR}}^{N, K, R}(B)$ such that (1) all the $L$ problem instances are the same (i.e., $\gamma_1=\cdots=\gamma_L = \gamma_0\in \bit$ and $o_\beta, Y^{(\alpha, \beta)}$ are the same for different $\beta\in [L]$) and (2) $L$ is rounded to the smallest odd number.
    This means that the decision bit
    \begin{equation}
        B = \bigoplus_{j=1}^L \gamma_j = \gamma_0.
    \end{equation}
    With this modification, the data generation process of the dimension reduction task is reduced to mimic that of the non-dynamic NOPE $\mathcal{D}_{g, KN'}^{KN'}(o), o\sim p_{\gamma_0}$ from \Cref{sec:cl-hard-noisy-oracle-property-est} with the Forrelation property $f$ and each sample repeated $\Theta(R)$ times.
    We call this new data generation process $\tilde{\mathcal{D}}_{\mathrm{DR}}^{N, K, R}(B)$.

    Now we consider the same reduction as in the proof of \Cref{thm:classical-hardness-dim-reduc-dynamic}.
    We have that estimating the 1D representation to $\epsilon=\tilde \Theta(1/\log^{1.5}(N))$ error suffice to determine the value of $B\in \bit$ and hence $\gamma_0$.
    We choose $K=\ceil{\frac{1.001}{\zeta\chi}}$ such that 
    \begin{equation}
        \frac{N'^{1-1/K}}{\polylog N'} \geq \frac{N^{1-\zeta\chi+\zeta\chi/1001}}{\polylog N}\geq \Omega(N^{1-\zeta\chi}).
    \end{equation}
    For the sake of contradiction, suppose we have a randomized classical learning algorithm $\mathcal{L}$ with space complexity
    \begin{equation}
        S \leq o(N^{(1-\zeta)\chi}) = o(N^{\chi-1+1-\zeta\chi}) \leq o\left(N^{\chi-1}\cdot \frac{N'^{1-1/K}}{\polylog N'}\right) = o\left(\frac{N'^{\chi-1/K}}{\polylog N'}\right)
    \end{equation}
    and sample complexity $M$ that given a sequence of data samples drawn from the modified $\tilde{\mathcal{D}}_{\mathrm{DR}}^{N, K, R}(B)$, estimates the 1D representation to $\epsilon$ error and hence decides $\gamma_0\in \bit$ with probability $p_{\mathrm{succ}}\geq 0.67$.
    With the same data reduction detailed in the proof of \Cref{thm:classical-hardness-dim-reduc-dynamic}, $\mathcal{L}$ can be used to construct a classical learning algorithm $\mathcal{L}'$ that decides $\gamma_0$ with probability $p_{\mathrm{succ}}\geq 0.67$ using the same space complexity $S$ and $M'\leq\lceil M/T_3\rceil\leq O(M/R+1)$ samples from $\mathcal{D}^{KN'}_{g, KN'}(o), o\sim p_{\gamma_0}$.

    Finally, we invoke \Cref{lem:classical-lower-bound-dist}.
    Note that for the inner product $g$, we have $\eta=1/2$, $c = (865/\eta^2)\log(865/\eta^2) = (865\times 4)\log(865\times 4) \approx 40677.68$, and therefore the choice of $b=\ceil{40678\log(KN')}$ satisfies the requirement.
    For the Forrelation $f$ we use, \Cref{lem:forrelation-query-separation} implies that the $(1/3)$-error classical distributional query complexity is (using $K=\Theta(1)$)
    \begin{equation}
        Q_C \geq \Omega\left(\frac{N'^{1-1/K}}{\polylog N'}\right).
    \end{equation}
    Meanwhile, the success probability satisfies
    \begin{equation}
        p_{\mathrm{succ}}=0.67 > 2/3 + 2^{-\eta b/8} = 2/3 + 2^{-\Theta(\log N')}
    \end{equation}
    for large enough $N'$, satisfying the conditions in \Cref{lem:classical-lower-bound-dist}.
    Therefore, \Cref{lem:classical-lower-bound-dist} implies that 
    \begin{equation}
    \begin{split}
        S&\geq \Omega\lr{\frac{Q_C KN' \log(KN')}{M'}} = \Omega\lr{\frac{N'^{1-1/K}\cdot N'}{M/R+1} \frac{1}{\polylog(N')}} \\
        &= \Omega\lr{\frac{N'^{2-1/K}}{N^{2-\chi}\polylog(N')}} = \Omega\lr{\frac{N'^{\chi-1/K}}{\polylog(N')}},
    \end{split}
    \end{equation}
    where we have used $M=\tilde{O}(RND^{1-\chi}) = \tilde{O}(RN^{2-\chi})$.
    This contradicts $S\leq o\lr{\frac{N'^{\chi-1/K}}{\polylog(N')}}$.
    Therefore, no such classical learning algorithm exists.
    This completes the proof of \Cref{thm:classical-hardness-dim-reduc}.
    Together with the quantum algorithm result \Cref{thm:dim-reduc-upper}, this proves the quantum advantage claim in non-dynamic dimension reduction (\Cref{thm:q-adv-dim-reduc}).
\end{proof}

\section{Evidence for superpolynomial time advantage}
\label{app:best-known-classical}

\revise{
In the preceding sections, we have rigorously proved superpolynomial sample and time advantages for the dynamic tasks.
For the static tasks, our unconditional result is instead an exponential space advantage under the same number of samples.
In this section, we provide complexity-theoretic evidence that a superpolynomial time advantage persists in the IID static task when both quantum and classical machines are restricted to $\polylog(N)$ space.

In particular, we will first review the generic low-space classical algorithm for solving linear systems obtained from parallel matrix inversion, then improve its runtime using the well-conditioning promise of our task.
These are the best known classical algorithms that can solve the linear systems in $\polylog(N)$ space, and yet they have $\mathrm{superpoly}(N)$ runtime.
In comparison, quantum oracle sketching solves the task with $\polylog(N)$ space and $\tilde O(N)$ time, which is superpolynomially faster.
We then conclude with a brief discussion of the complexity-theoretic evidence supporting this superpolynomial time separation.

We note that the results in this section are separate from the hard instances based on Forrelation-NOPE that we constructed in previous sections for \textit{unconditionally} proving classical hardness.
In particular, those hard instances can indeed be solved with $\poly(\log N)$ space and $\tilde O(N^2)$ samples.
Our results in this section show that the situation is different in general.
To avoid confusion for careful readers: the unconditional sample and time separation stated in the main text are proved for dynamic tasks and we do not claim to unconditionally prove superpolynomial time advantage in the static tasks; we only provide complexity-theoretic evidence for it.

\subsection{Space-efficient classical linear system solvers}

We start with the classic result on solving matrix inversion with classical parallel computation.
For simplicity, we focus on real and symmetric matrices.
Csanky showed that matrix inversion and solving linear systems can be performed in $O(\log^2 N)$ parallel time using a polynomial number of processors, and Berkowitz gave a uniform polynomial-size, $O(\log^2 N)$-depth construction for the closely related determinant, characteristic polynomial, and adjoint problems
\cite{csanky1976fast,berkowitz1984computing}.
In standard complexity terminology, these results show that these primitive linear algebra problems are in $\mathsf{NC}^2$.

There is a well-established relation between circuit depth and sequential space complexity
\cite{borodin1977relating}.
In particular, a uniform polynomial-size circuit of depth $d$ can be evaluated recursively using space essentially proportional to its depth, up to logarithmic bookkeeping factors.
Applying this to the $\mathsf{NC}^2$ algorithms above gives a classical simulation using
\begin{equation}
S_{\mathrm{C}}=\polylog(N)
\end{equation}
space.
However, the time and query complexity of this method grows exponentially with depth.
This is because due to space constraints, it cannot store many intermediate values in the computation and therefore has to recalculate them whenever they are needed.
This is similar to the path integral idea used in proving $\mathsf{BQPSPACE}=\mathsf{PSPACE}$ \cite{watrous1999space,watrous2003complexity}.
This leads to the time and complexity
\begin{equation}
T_{\mathrm{C}}
=
2^{O(\log^2 N)}
=
N^{O(\log N)}.
\end{equation}

We can then estimate the normalized quadratic form required in \Cref{task:linear-sys} as follows.
Note that the individual components of the numerator and denominator in the quadratic form can be calculated and accumulated sequentially without storing the full solution vector.
Enumerating polynomially many indices only require an additional polynomial factor in time and $O(\log N)$ additional space for indexing.
Given an observable $\mathcal{M}$ in \Cref{task:linear-sys} that can be evaluated with a $\polylog(N)$-size circuit, we can also estimate it with the same low-space simulation algorithm.
This gives a classical algorithm that runs in $\polylog(N)$ space and
\begin{equation}
T_{\mathrm{C}}=\mathrm{superpoly}(N)
\end{equation}
time.

Note that this algorithm is a classical query algorithm that makes query to the entries of $A$ and $\vec{b}$.
To turn it into a learning algorithm that works for randomly sampled data, one can just wait for $\tilde O(N)$ time to obtain any specific query according to standard probability theory.
For simplicity, we ignore the technical complications caused by the zero elements of the matrix.
Therefore, a classical low-space algorithm making $Q$ queries can always be converted into a learning algorithm with $\tilde O(QN)$ sample complexity.

For the low-space algorithm solving $\mathsf{NC}^2$ problems above, we have $Q=N^{O(\log N)}$, and therefore the sample complexity reads
\begin{equation}
M_{\mathrm{C}}
=
N\cdot N^{O(\log N)} = \mathrm{superpoly}(N),
\end{equation}
which is superpolynomially larger than that of quantum oracle sketching.

\subsection{Improvement for well-conditioned linear systems}

Note that the above algorithm does not yet make use of the condition number promise $\kappa=\polylog(N)$.
Here we show that a good condition number can be used to improve the runtime, but yet still runs in $\mathrm{superpoly}(N)$ time.

To construct the improved algorithm, we utilize the well known polynomial approximation of $1/x, |x|\in [1/\kappa, 1]$ \cite{gilyen2019quantum}, which states that there is an odd polynomial $p_d$ of degree
\begin{equation}
d
=
O\left(
\kappa\log\frac{\kappa}{\eta}
\right)
\end{equation}
such that
\begin{equation}
\left|p_d(x)-\frac{1}{x}\right|
\leq \eta, \quad |x|\in [1/\kappa, 1].
\end{equation}
For $\kappa,\eta^{-1}=\polylog(N)$, we have
\begin{equation}
d=\polylog(N).
\end{equation}

The solution vector can be approximated using the polynomial as
\begin{equation}
\tilde{\vec{x}}
=
p_d(A)\vec{b}.
\end{equation}
The error bound reads
\begin{equation}
\|\tilde{\vec{x}}-\vec{x}\|
\leq
\eta\|\vec{b}\|
\leq
\eta\|\vec{x}\|,
\qquad
\vec{x}=A^{-1}\vec{b},
\end{equation}
where the last inequality follows from $\|A\|\leq1$.
Thus $\eta=1/\polylog(N)$ indeed suffices for our purposes.

Using the similar recursive recomputation idea, $p_d(A)\vec{b}$ can be evaluated space-efficiently as follows.
To calculate any entry of a matrix power, we use
\begin{equation}
(A^r)_{ij}
=
\sum_{k=1}^{N}
(A^{\lfloor r/2\rfloor})_{ik}
(A^{\lceil r/2\rceil})_{kj},
\end{equation}
to recursively compute $(A^r)_{ij}$.
Note that this recursion finishes in depth $O(\log r)$.
At any level in the recursion, the algorithm only needs to store the current indices, the exponent value and the accumulating value.
Hence it consumes
\begin{equation}
S_{\mathrm{C}}
=
\polylog(N)
\end{equation}
space.
Its runtime satisfy the recursive relation
\begin{equation}
T(r)
\leq
O(N)T(\lceil r/2\rceil),
\end{equation}
and therefore
\begin{equation}
T(r)
=
N^{O(\log r)}.
\end{equation}
Since every term in $p_d(A)$ has power
$r\leq d=\polylog(N)$, the runtime reads
\begin{equation}
T_{\mathrm{C}}
=
N^{O(\log\log N)}.
\end{equation}
Multiplying by $\vec b$, computing the normalization, and evaluating the quadratic form only adds polynomial factors in $N$ and do not change this superpolynomial scaling.
As above, every query to $A_{ij}$ or $b_j$ can be implemented from the random data samples by waiting $\tilde O(N)$ samples, which gives the total sample and time complexity
\begin{equation}
M_{\mathrm{C}},T_{\mathrm{C}}
=
N^{O(\log\log N)}.
\end{equation}

For the most general efficiently measurable observable $\mathcal{M}$ allowed in \Cref{task:linear-sys}, recursively simulating its $\polylog(N)$-size measurement circuit may contribute an additional $\exp(\polylog N)$ factor.
Thus the best known $\polylog(N)$-space classical algorithm has runtime
\begin{equation}
T_{\mathrm{C}}=\mathrm{superpoly}(N),
\end{equation}
which remains superpolynomial under the well-conditioning promise.
For the simple observables relevant to the standard linear-system primitive and to the hardness constructions in this work, the explicit inversion procedure above gives the sharper
$N^{O(\log\log N)}$ bound.

An alternative classical algorithm is given by the low-space simulations of quantum circuits in \cite{aaronson2016complexity}.
Aaronson and Chen showed that an $n$-qubit quantum circuit of depth $d$ can be simulated using $O(n\log d)$ classical space and
\begin{equation}
O\left(n(2d)^{n+1}\right)
=
d^{O(n)}
\end{equation}
time
\cite{aaronson2016complexity}.
For $n=O(\log N)$ and $d=\polylog(N)$, this again gives runtime
\begin{equation}
N^{O(\log\log N)} = \mathrm{superpoly}(N).
\end{equation}
This provides an alternative interpretation of the same bottleneck faced by all these best known classical algorithms: storing only logarithmically many amplitudes reduces memory consumption, but at the cost of a blowup in time complexity because one has to recursively recompute sums over exponentially many intermediate configurations, in line with the path integral intuition.

\subsection{Relation to \texorpdfstring{$\mathsf{NC}$, $\mathsf{SC}$}{NC, SC}, and directed connectivity}

The difficulty of improving these bounds to polynomial time is closely related to long-standing questions about space-time tradeoffs in complexity theory

Recall that
$\mathsf{NC}
=
\bigcup_{k\geq1}\mathsf{NC}^k$
consists of problems admitting uniform polynomial-size circuits of polylogarithmic depth, whereas
$\mathsf{SC}$ consists of problems admitting classical algorithms that simultaneously uses polynomial time and polylogarithmic space.
Intuitively, circuit size corresponds to time complexity, while circuit depth can be converted into space complexity via the established recursive recomputation method used above
\cite{borodin1977relating}.
However, whether this low-space simulation can be done without blowing up the time exponentially has long been unknown.
In particular, whether general $\mathsf{NC}$ computations can be carried out simultaneously in polynomial time and polylogarithmic space (i.e., in $\mathsf{SC}$) is a long-standing open problem.
In fact, it has been conjectured to be impossible
\cite{papakonstantinou2014depth}.

Matrix inversion sits directly in this landscape.
General matrix inversion is complete for the class $\mathsf{DET}$ under log-space reductions, with
\begin{equation}
\mathsf{DET}
\subseteq
\mathsf{NC}^2
\subseteq
\mathsf{DSPACE}(O(\log^2 N))
\end{equation}
\cite{ta2013inverting}.
Thus, for unrestricted matrix inversion, obtaining a polynomial-time algorithm while retaining only polylogarithmic space would imply $\mathsf{DET}$ in $\mathsf{SC}$.
More broadly, a generic polynomial-time, polylogarithmic-space simulation of arbitrary $\mathsf{NC}^2$ circuits would establish $\mathsf{NC}^2\subseteq\mathsf{SC}$, which would constitute substantial progress in complexity theory.

Condition number promise in \Cref{task:linear-sys} means that the standard $\mathsf{DET}$-completeness result for unrestricted matrix inversion does not directly give $\mathsf{NC}^2$ or $\mathsf{DET}$ reductions for our static task.
Therefore, the above discussion only provides complexity-theoretic evidence for the classical hardness of the problem rather than a rigorous reduction.
Nevertheless, the best known classical algorithm in $\polylog(N)$-space has runtime superpolynomial in $N$, even after exploiting the well-conditioning promise as above.
In contrast, quantum oracle sketching solves the static linear-system task using $\polylog(N)$ space and $\tilde O(N)$ total time.
This gap motivates the evidence for the superpolynomial quantum time advantage stated in the main text.

In fact, one can relax the $\polylog(N)$ space restriction even further to sublinear space (e.g., $O(N^{0.99})$). 
Currently, there is no known classical algorithm for solving general linear systems with $O(N^{0.99})$ space and $\poly(N)$ time.
In fact, it has been shown that such a linear system solver with multiplicative guarantees would imply a $\mathrm{poly}(N)$-time and $O(N^{0.99})$-space algorithm for directed $s$-$t$ connectivity (a canonical $\mathsf{NL}$-complete problem), which remains open \cite{liu2026numerical}.
It is known that directed connectivity can be solved using Savitch's algorithm in $O(\log^2 N)$ space and $N^{O(\log N)}$ time, whereas standard graph search runs in polynomial time using $\tilde O(N)$ space.
Barnes, Buss, Ruzzo, and Schieber obtained a polynomial-time algorithm using
$\frac{N}{2^{\Theta(\sqrt{\log N})}}
=
N^{1-o(1)}$
space, but no polynomial-time algorithm using $O(N^{1-\zeta})$ space is known for any constant $\zeta>0$
\cite{savitch1970relationships,barnes1998sublinear}.

We note that these do not directly apply to our task since we have additional promises on condition number and sparsity.
Yet they provides complexity-theoretic evidence that the exponential space advantage and superpolynomial time advantage are not restricted to the dynamic tasks, but already exist in the static tasks.
}

\end{document}